\pdfoutput=1

\documentclass[a4paper,12pt,twoside,colorlinks]{report}

\usepackage{palatino} 			
\usepackage[hyperindex=true,colorlinks=true,linkcolor=black,citecolor=black,urlcolor=black,filecolor = black]{hyperref}
\usepackage{fancyhdr} 			
\usepackage{graphicx} 			
\usepackage{rotating}			
\usepackage[utf8]{inputenc} 	
\usepackage[T1]{fontenc}
\usepackage[francais]{babel}  
\usepackage{minitoc} 				
\usepackage{color}
\usepackage{lettrine}				
\usepackage{xcolor}
\usepackage{caption}
\usepackage{journal_font}  		
\usepackage{natbib}
\usepackage{eso-pic}
\usepackage{tocloft}
\usepackage{appendix}
\usepackage{amssymb}
\usepackage{amsmath}
\usepackage{afterpage}
\usepackage{pdfpages}
\usepackage{perpage}
\usepackage{pxfonts}			
\usepackage{multirow}

\definecolor{rouge}{rgb}{0.7,0.,0.}

\definecolor{bleu_chapitre}{rgb}{0.,0.,0.7}
\makeatletter   
\def\thickhrulefill{\leavevmode \leaders \hrule height 1ex \hfill \kern \z@}		
\def\@makechapterhead#1{%
  {\parindent \z@ \raggedright
    \reset@font \color{bleu_chapitre}
    \vspace*{10\p@}%
    \par
    \Large \scshape \@chapapp{} \Huge\bfseries \thechapter
    \par\nobreak
    \vspace*{10\p@}%
    \hrule
    \par
    \vspace*{1\p@}%
    \hrule
    \vspace*{20\p@}
    \Huge \bfseries #1\par\nobreak
    \vskip 70\p@
  }}

\mtcselectlanguage{french} 		

\newcommand{\malettrine}[1]{
  \lettrine[lines=2,lhang=0.33,loversize=0.33]{#1} 	
}

\makeatletter
\let\LaTeX@startsection\@startsection
\renewcommand{\@startsection}[6]{\LaTeX@startsection%
{#1}{#2}{#3}{#4}{#5}{\color{bleu_chapitre}\raggedright \large{#6}}}
\makeatother

\makeatletter
\renewcommand                	
{\section}{\@startsection{section}{2}{0mm}  
{1.\baselineskip}{.2\baselineskip}  
{\bf\large}}
\makeatother 

\makeatletter
\renewcommand                	
{\subsection}{\@startsection{subsection}{2}{10mm}  
{.3\baselineskip}{.2\baselineskip}  
{\bf\large}}
\makeatother 

\makeatletter
\renewcommand              	 
{\subsubsection}{\@startsection{subsubsection}{2}{20mm}
{.3\baselineskip}{.2\baselineskip}  
{\bf\large}}
\makeatother

\makeatletter
\renewcommand              	 
{\paragraph}{\@startsection{paragraph}{2}{0mm}
{.3\baselineskip}{.2\baselineskip}  
{\bf\large}}
\makeatother

\newcommand{\refbleu}{
\hypersetup
{
colorlinks=true, 
linkcolor=bleu_chapitre, 
}}

\newcommand{\refnoir}{
\hypersetup
{
colorlinks=true, 
linkcolor=black 
}}

\def\arcmin{\hbox{$^\prime$}}
\def\arcsec{\hbox{$^{\prime\prime}$}}

\newcommand{\compactlist}{\setlength{\itemsep}{0pt} \setlength{\parskip}{0pt} \setlength{\leftskip}{1cm}}

\newcommand{\nocontentsline}[3]{}
\newcommand{\tocless}[2]{\bgroup\let\addcontentsline=\nocontentsline#1{#2}\egroup}

\def\tablefootmark#1{$^{#1}$\,\ignorespaces}
\def\tablefoottext#1#2{$^{(#1)}$~#2}

\def\ciap{\c ci-après\ }

\title{
  \textcolor{black}
  {
  \vspace*{-5cm} \\ 
  \normalsize Université Pierre et Marie Curie-Paris 6  \\[.1cm]
  \Large{\textbf{École Doctorale d'Île-de-France} } \\[1cm]
  \Huge THÈSE \\[.4cm]
   \normalsize pour obtenir le titre de \\[.4cm]
   \Large \textbf{Docteur en Sciences} \\[.4cm]
   \normalsize de l' Université Paris VI - Pierre et Marie Curie \\[.5cm]
     \normalsize Spécialité : Astronomie \& d'Astrophysique  \\[1.cm]
   \normalsize Présentée et soutenue par\\
   \Large{Alexandre GALLENNE} \\ [1.5cm]
   \LARGE{\textbf{LES CÉPHÉIDES À HAUTE RÉSOLUTION ANGULAIRE : ENVELOPPE CIRCUMSTELLAIRE ET PULSATION}} \\[1.5cm]
   \large{Soutenue publiquement le 19 Octobre 2011 \\
   devant le jury composé de : }\\[1.cm]
   \normalsize
   \begin{center}
	\begin{tabular}{llcl}
		\textit{Président :}			& Patrick \textsc{Boissé}				& - & Institut d'Astrophysique de Paris				\\[.2cm]	
		\textit{Examinateur :} 	& Nicolas \textsc{Nardetto}		    & - & Observatoire de la Côte d'Azur					\\[.2cm]
		\textit{Examinateur :} 	& Yann \textsc{Clénet}				    & - & LESIA (CNRS - Observatoire de Paris)			\\[.2cm]
		\textit{Examinateur :} 	& Pascal \textsc{Bordé}				    & - & Institut d'Astrophysique Spatiale				\\[.2cm]
		\textit{Rapporteur :}		& Denis \textsc{Mourard}				& - & Observatoire de la Côte d'Azur					\\[.2cm]
		\textit{Rapporteur :}		& Pascal \textsc{Fouqué}				& - & Laboratoire Astrophysique de Toulouse		\\[.2cm]
      	\textit{Directeur :}			& Pierre \textsc{Kervella}				& - & Observatoire de Paris								\\[.2cm]
      	\textit{Co-directeur :}	& Antoine \textsc{Mérand}			& - & European Southern Observatory				\\[.2cm]
	\end{tabular} \\[2.5cm]
	\end{center}
	\begin{center}
	\normalsize{Travaux effectués à l'Observatoire de Paris - LESIA, 5 place Jules Janssen,
    92195 Meudon et à l'ESO, Alonso de C\'ordova 3107, Vitacura, Santiago de Chile}
    \end{center}
   \date{}
  }  
}

\MakePerPage[1]{footnote}  

\begin{document}

\maketitle

\tightmtctrue   

\renewcommand\contentsname{\textcolor{bleu_chapitre}{\emph{Sommaire}}}
\renewcommand\listfigurename{\textcolor{bleu_chapitre}{\emph{Table des figures}}}
\renewcommand\listtablename{\textcolor{bleu_chapitre}{\emph{Liste des tableaux}}}
\renewcommand\bibname{\thispagestyle{empty}\textcolor{bleu_chapitre}{\emph{Références}}}

\cleardoublepage  
\thispagestyle{empty}

\begin{figure}[!p]
\centering\includegraphics[width = .6\linewidth]{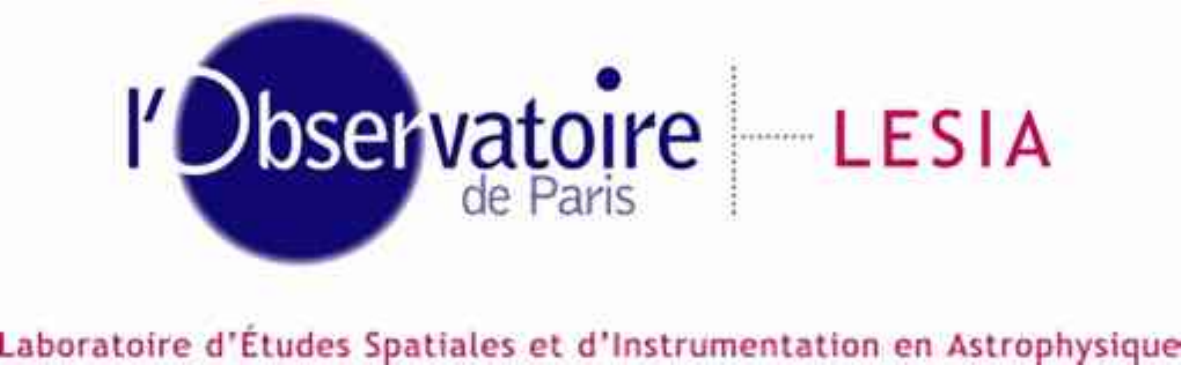}\hspace{.5cm}
\end{figure}
\begin{figure}[!p]
\centering\includegraphics[width = .5\linewidth]{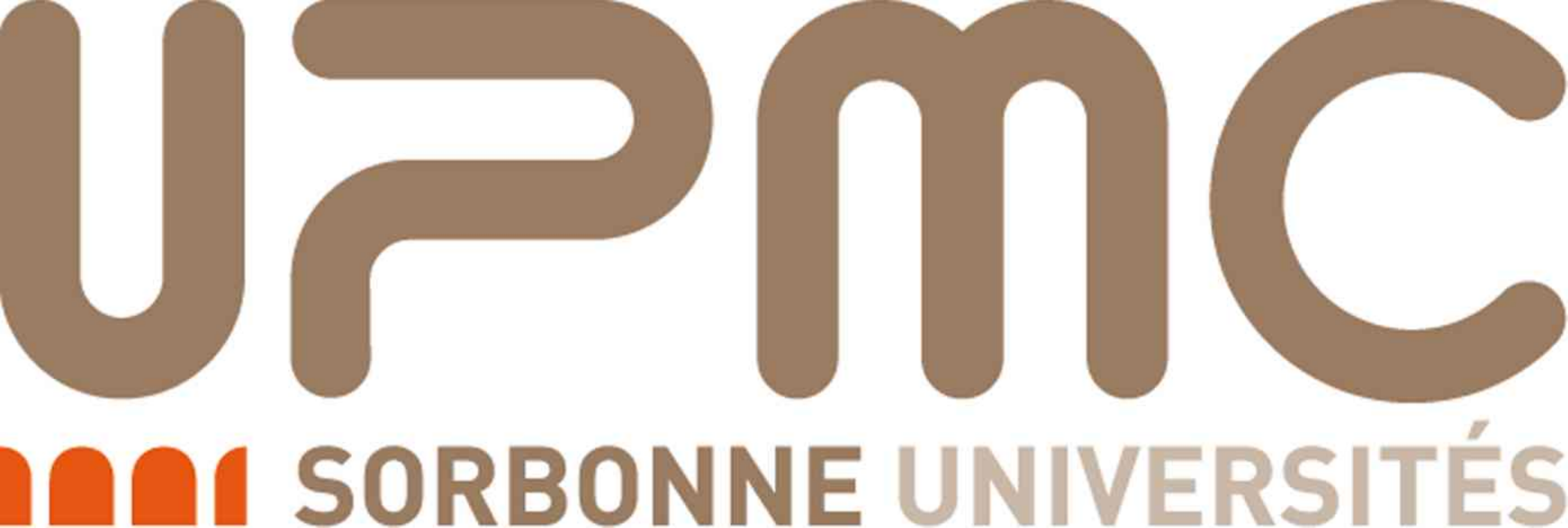}\hspace{.5cm}
\end{figure}
\begin{figure}[!p]
\centering\includegraphics[width = .3\linewidth]{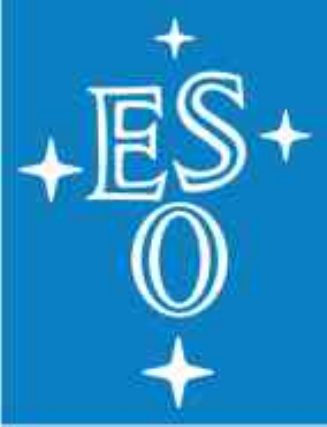}
\end{figure}

\thispagestyle{empty}		
\cleardoublepage

%
%
\pagestyle{empty}
\cleardoublepage  

\chapter*{\textcolor{bleu_chapitre}{\emph{Remerciements}}}
\thispagestyle{empty}

L'objectif fixé lors de ma reprise des études universitaires est maintenant atteint. Le chemin fut difficile mais j'y suis arrivé grâce à une motivation constante, au soutient de mes amis, de ma famille et de mes collaborateurs. 

Cette thèse est le résumé d'un travail de recherche effectué au sein de deux instituts, l'Observatoire Européen Austral (ESO) et le Laboratoire d'Études Spatiales et d'Instrumentation en Astrophysique (LESIA). J'ai passé mes deux premières années à l'ESO, à Santiago du Chili, sous un climat ensoleillé et chaleureux, et ma dernière année au LESIA à Meudon, où le soleil était malheureusement beaucoup moins présent.

Mes premiers remerciements vont tout naturellement à Pierre Kervella, qui m'a d'abord accueilli en stage de master 2 en 2008, puis par la suite encouragé à commencer mes travaux de recherche au sein de l'ESO. Bien que le contact direct pendant les deux premières années était compliqué, son encadrement, son soutient et sa confiance sont restés sans faille tout au long de ces 3 ans. La dernière année en France m'a permis d'avoir plus interactions avec lui, à la fois sur le plan scientifique et humain. Je le remercie de m'avoir encouragé à postuler pour une bourse de thèse à l'ESO au Chili, car cela a complètement changé ma vie. En bref, ces années de thèse ont été enrichissantes, et parfois même amusantes, grâce notamment aux nombreuses nuits passées au Mont Wilson (avec les ours et les pumas).

La transition avec le Chili n'aurait jamais pu se faire sans la co-direction d'Antoine Mérand. Je le remercie pour les connaissances qu'il m'a transmises en astrophysique, en instrumentation et en programmation avec Yorick. Son arrivée à Santiago de seulement quelques mois avant moi a fait que nous avions tout les deux une même étape à franchir : l'apprentissage de l'espagnol. J'ai apprécié les nombreuses discussions très instructives que nous avons pu avoir, et ceux malgré ses devoirs à Paranal et ses "descanso".

Un travail de thèse se doit d'être examiné par un jury, je tiens donc à remercier les membres de mon jury qui ont répondu présent. Merci pour le soin que vous avez apporté à l'examen de mon travail.

Je remercie également les deux instituts qui m'ont accueillis dans leurs locaux. Tout d'abord l'ESO, qui m'a offert un financement de 2 ans à Santiago, permettant ainsi un contact plus direct avec les télescopes et les instruments, et surtout, la rencontre de chercheurs/ingénieurs dans des domaines astrophysiques variés. Je remercie particulièrement J. B. Lebouquin, J. Girard, G. Montagnier et F. Patru pour les discussions nombreuses et variées. Je remercie également tout le personnel de l'ESO, notamment M. West, P. Jiron, H. Schmidt, S. Silva et I. Riveros pour leur travail dans mon intégration à l'ESO et au Chili. De retour en France en octobre 2010 (un peu nostalgique), je suis prêt pour ma dernière année au sein du LESIA de l'Observatoire à Meudon. Mon insertion dans l'équipe de haute résolution angulaire du bâtiment 5 s'est faite de manière simple et chaleureuse, notamment grâce aux personnes que j'avais déjà rencontrées lors de mon stage de master 2. Je remercie toutes les personnes des bâtiments 5 et 6 pour avoir fait de cette dernière année (stressante), une année pleine d'humour et de sympathie. La bataille des places de bureau est rude dans ces deux bâtiments, je remercie donc particulièrement Pierre Léna qui m'a gentiment prêté le sien. Enfin, merci aux ITA de l'Observatoire de Meudon pour leur aide informatique et logistique.

Ces trois ans ont été riches en expériences humaines, principalement grâce à la collocation dans la fameuse casa de la rue Ricardo Matte Perez. Mes colocataires, les gens de passage à la maison, les nombreux asados, les Miercoles Po' et autres sorties ont fait de mon séjour en Amérique du Sud une expérience inoubliable. Merci à tout ceux qui ont contribué à ces années extraordinaires : Pedro (el crespo), Pedro (el pelado), Bene, Svante, Sky, Michel (el suizo), Michel (el gato), Thomas, Olivia, Claire, Ignacio, les suisses, Boris, Alex, Karen, Victor et Pao, ... et toutes les autres personnes formidables que j'aurais pu oublier de citer. Une pensée particulière pour l'un de mes coloc, Vincent, avec lequel s'est crée une réelle complicité tout au long de ces deux années. Voici une petite anecdote incroyable sur la naissance de cette amitié : Vincent et moi étions dans le même vol d'arrivée Paris-Santiago (sans se connaître) ; à peine sortie de l'aéroport, un peu dépaysé et troublé par tout ces gens parlant espagnol, que j'entends déjà deux personnes parler français, et de plus attendant le même taxi que moi. Viens alors le premier contact : "vous êtes Français?", "non, nous sommes suisses!", disent-ils. De là a démarré une conversation dans laquelle j'ai appris que Vincent, l'un des deux suisses, allait rester au minimum 2 ans au Chili (il était technicien à la Silla pour le télescope suisse) et qu'il allait vivre dans la même maison que moi. C'est quand même incroyable : même vol, même taxi, même durée du séjour, même maison et en plus on parle tout les deux la même langue! Ce séjour a donc commencé d'une manière assez inattendue, et cette rencontre a probablement facilité mon insertion dans ce pays merveilleux qui m'était complètement inconnu.

Même à des milliers de kilomètres, mes amis de France étaient toujours présents. Mon retour m'a permis de passer plus de temps avec eux et de partager mon expérience transatlantique. Je les remercie des bons moments passer en leur compagnie. Merci à Chris, Fredo, Bruno, Bubu, Audrey, Sam, Amélie, Adeline, ... et les autres.

Merci ma belle-famille, Cesar, Irene, Danha et Mathias (el diablito), qui m'ont accueillis en leur sein. J'ai appris énormément de choses sur la culture chilienne grâce à eux, notamment leur sens de l'hospitalité et leur gentillesse. Merci également à ma famille, mon père, ma mère, Vincent, Mégane, Alexandra, et une nouvelle arrivée, Mélinda, qui ont toujours cru en moi et dont le soutient moral a toujours été présent.

Todo esto no tendría mucha importancia sin mi mujer, Jerusnha, siempre a mi lado. Su encuentro en la Nona cambió completamente mi vida. Mi estancia en su país fue excepcional gracias a ella, tan maravilloso que decidimos casarnos en Chile. Ella hizo el esfuerzo por alejarse de su familia y pasar mi último año de tesis junto a mi en Francia, y le agradezco por esto. Gracias por haberse quedada a mi lado, sé que no fue fácil para ti en Francia, lejos de tu familia y de tus amigos, y además un marido muy ocupado para terminar su tesis. Gracias por haberme sostenido y animado siempre.

\begin{flushright}
\vspace*{2cm}
\textit{À ma famille ...}

\textit{y a mi compañera de vida, Jerusnha}
\end{flushright}

\pagestyle{plain}
\pagenumbering{roman} \setcounter{page}{1}		

\cleardoublepage

\dominitoc
\tableofcontents

\cleardoublepage
\thispagestyle{empty}		
\listoffigures
\cleardoublepage
\thispagestyle{empty}		
\listoftables
\cleardoublepage
\thispagestyle{empty}		

\pagenumbering{arabic} \setcounter{page}{1}		


\cleardoublepage     

\pagestyle{fancy}
\fancyhf{}
\lhead[\nouppercase{\emph{\thepage}}]{\nouppercase{\emph{Introduction}}}
\rhead[\nouppercase{\emph{Introduction}}]{\nouppercase{\emph{\thepage}}}
\newpage

\chapter*{\textcolor{bleu_chapitre}{\emph{Introduction}}}
\addstarredchapter{Introduction}

\thispagestyle{empty}


\refnoir

\malettrine{C}{'}est souvent par une belle soirée d'été que commence la passion de l'astronomie. Une nuit claire, sans Lune, éloigné de toute lumière artificielle, allongé dans l'herbe, relaxé, détendu, rêveur, les yeux perdus dans la profondeur du ciel. Commence ensuite la recherche de constellations, ces ensembles d'étoiles reliées par des lignes imaginaires pour former des figures caractéristiques. L'observation du ciel nous offre alors un premier contact avec l'univers, une première vision d'un espace infini.

Ces constellations, au nombre de 88, sont composées d'étoiles brillantes qui nous apparaissent comme regroupées, situées à une même distance de la Terre. C'est ce que pensaient auparavant les astronomes en situant les étoiles à une même distance, appartenant à une même sphère céleste. Nous savons maintenant que les étoiles d'une constellation ne sont pas nécessairement liées physiquement, et peuvent même être distantes de plusieurs années lumières.

L'estimation des distances Galactiques et extragalactiques est l'une des questions majeures en astronomie. Sans cette variable, il est impossible d'estimer certains paramètres tels que la taille ou la masse d'un objet astrophysique. Comment connaître la distance d'une étoile, d'une galaxie, ... ? Mesurer la brillance d'un astre n'est pas vraiment fiable puisqu'un objet peu lumineux et proche peut nous sembler identique à un autre astre très brillant et distant. On a alors recours à d'autres méthodes en fonction de la proximité des objets astrophysiques. Pour les astres proches, nous nous servons de leur mouvement apparent par rapport aux étoiles plus lointaines, pour les astres plus éloignés, nous utilisons les étoiles Céphéides et leur relation période--luminosité, tandis que pour des objets encore plus lointains, on se sert par exemple des galaxies ou des quasars et leur décalage vers le rouge (effet Doppler), car les étoiles ne sont plus visibles individuellement.

Dans ce manuscrit, je m'intéresse aux étoiles de type Céphéide, des étoiles pulsantes connues comme "chandelles standards" pour l'échelle des distances dans l'univers. L'estimation des distances par les Céphéides passe par l'utilisation d'une relation liant la luminosité intrinsèque de l'étoile à sa période de pulsation. Cette loi n'est pas parfaite, elle a besoin d'être étalonnée et corrigée des éventuelles sources de biais.

Le sujet porte plus particulièrement sur les enveloppes circumstellaires de Céphéides qui ont été récemment découvertes. L'existence de ces enveloppes soulève des questions sur 1) la perte de masse des Céphéides et 2) le biais pouvant être causé sur la mesure de distance par leur présence. Cette thématique étant nouvelle dans le cadre des Céphéides, les premiers travaux consistent en l'exploration de diverses techniques d'observations dans différentes gammes de longueurs d'onde afin de favoriser leur détection et permettre la caractérisation de certains paramètres physiques tels que la taille, la géométrie ou la température. L'impact exact des enveloppes circumstellaires sur l'estimation des distances n'est pas connu à ce jour. Le biais n'est probablement que de quelques pour cent mais une quantification précise est nécessaire si l'on souhaite une détermination de la distance avec une précision inférieure à 1\,\%.

Je m'intéresse également dans ce manuscrit à la mesure de distance via l'utilisation de la méthode interférométrique de Baade--Wesselink. Cet outil puissant permet d'avoir une estimation de la distance indépendamment de la relation période--luminosité, permettant par la suite son étalonnage. 

L'utilisation des Céphéides comme indicateurs de distance a eu un rôle essentiel dans la mesure des distances Galactiques et extragalactiques. Au début du XXe siècle, elles permirent notamment de placer le Soleil en dehors du centre Galactique, d'estimer la taille de la Voie Lactée, ou encore d'estimer la distance de la galaxie Andromède. De nos jours, l'usage des Céphéides est devenu un échelon essentiel dans l'échelle des distances extragalactiques et nous permet de concevoir l'immensité de l'univers.

\cleardoublepage  

\pagestyle{fancy}
\fancyhf{}
\lhead[\nouppercase{\emph{\thepage}}]{\nouppercase{\emph{\rightmark}}}
\rhead[\nouppercase{\emph{\leftmark}}]{\nouppercase{\emph{\thepage}}}
\newpage

\chapter[Mesurer l'univers: les Céphéides]{\emph{Mesurer l'univers: les Céphéides}}
\label{chapitre__mesurer_l_univers_les_cephéides}

\thispagestyle{empty}

\vspace*{-1cm}

\refbleu 
\textcolor{bleu_chapitre}{\minitoc}
\refnoir 

\section{Bref historique}

\malettrine{E}{}lles tiennent leur nom de l'étoile $\delta$~Cephei, l'archétype des étoiles variables Céphéides découvert en 1784 par John Goodricke \citep{Goodricke-1786-}. Contrairement à ce que l'on pourrait penser, ce n'est pas le prototype de cette classe d'étoile car la première à être découverte fut $\eta$~Aql quelques mois plus tôt par l'astronome amateur Edward Piggot. Ces étoiles très brillantes aux variations périodiques de luminosité apparente intriguent de plus en plus les astronomes, entraînant des observations intensives au fil des années. Environ un siècle après la découverte de E. Piggot, l'astronome russe Aristarkh Belopol’skii \citep{Belopolsky-1896-02} découvrit le décalage des raies spectrales de $\delta$~Cep en fonction de la période (la période étant définie comme l'intervalle de temps entre deux phases de même luminosité) et plus particulièrement que le maximum de la courbe de vitesse radiale correspondait au minimum de luminosité. Les astronomes de l'époque expliquèrent ce phénomène observé comme étant lié à des variations orbitales de l'étoile causées par un compagnon proche et suggérèrent que les étoiles de type Céphéide pourraient simplement être des binaires spectroscopiques. 

\begin{figure}[!t]
  	\centering\includegraphics[width = .55\linewidth]{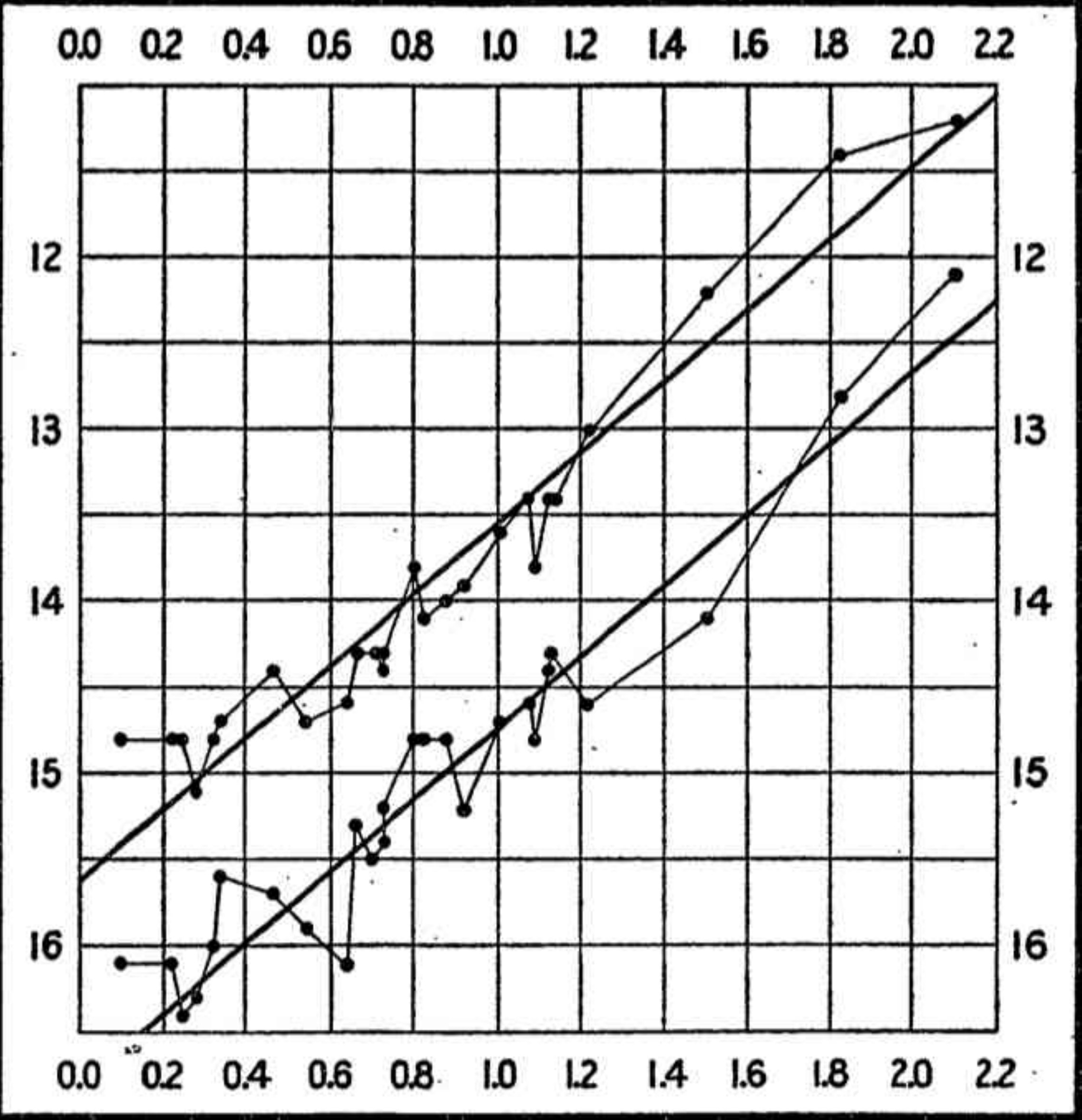}
 	\caption[Première relation P--L]{\textbf{Première relation P--L} : Première courbe période--luminosité tracé par H. Leavitt 				\citep{Leavitt-1912-03} pour 25 étoiles variables situées dans le Petit Nuage de Magellan. Le logarithme de la période est en abscisse et la magnitude photographique en ordonnée. Les deux courbes représentent la relation P--L pour les magnitudes maximales et minimales.}
	\label{image__leavitt_PL}
\end{figure}

Alors que les astronomes cherchaient désespérément des effets de binarité, la collecte des données sur les étoiles variables augmentait, jusqu'à en découvrir dans d'autres galaxies. Ce furent notamment les travaux de Henrietta Leavitt, qui catalogua des étoiles variables situées dans les Nuages de Magellan \citep{Leavitt-1908-}. Son échantillon comprenait 1777 étoiles variables mais seulement 17 disposaient d'une mesure de période. En ordonnant ces étoiles par période croissante, elle fit la découverte que les étoiles variables les plus brillantes avaient les plus longues périodes. En 1912 elle détermina la période et la magnitude de 8 variables supplémentaires et confirma sa découverte précédente \citep{Leavitt-1912-03}. Elle fait également l'hypothèse que toutes ces étoiles sont à la même distance de la Terre reliant ainsi les magnitudes apparentes aux magnitudes intrinsèques : c'est la naissance de la relation période--luminosité (P--L). La première courbe P--L fut obtenue en 1912 avec 25 étoiles variables (Fig.~\ref{image__leavitt_PL}). Cette relation linéaire de la forme $M = a\log{P} + b$ est aujourd'hui l'une des plus utilisée pour l'estimation des distances dans l'univers et je développerai son utilité dans la Section~\ref{section__la_relation_periode_luminosite}.

Cependant le point zéro $b$ de cette relation ne fut pas déterminé à l'époque car la distance du Petit Nuage de Magellan (SMC) n'était pas connue. Cette relation non étalonnée n'était donc pas directement utilisable.

En 1913, Ejnar Hertzsprung identifia certaines étoiles observées par H. Leavitt comme étant des Céphéides et mesura leur distance par la méthode de la parallaxe statistique (définition dans la Section~\ref{subsection__les_echelles_de_distance}) pour estimer le point zéro. Bien qu'imprécise, il combina son estimation de $b$ et la pente $a$ déterminée à partir des données de H. Leavitt pour obtenir une relation P--L étalonnée. L'équation qu'il obtint lui permis d'estimer la distance du SMC. Nous savons maintenant que l'estimation était fausse mais ce résultat était pour l'époque extraordinaire. En 1918 Harlow Shapley améliora l'étalonnage de la relation précédente basée sur des Céphéides du SMC. Il remarqua également que la pente de la relation P--L pour les étoiles variables des amas globulaires était identique à celle des étoiles variables du SMC et décida donc de regrouper toutes les Céphéides dans une relation P--L unique. Sans le savoir il venait de rajouter un type de Céphéides différent de celui utilisé par H. Leavitt en 1912. De nos jours ces deux types sont connus sous le nom de Céphéides classiques (type I) et Céphéides de type II, j'en parlerai plus en détail en Section~\ref{section__les_differents_types}. C'est seulement en 1956 que Walter Baade proposa d'utiliser deux relations P--L distinctes \citep{Baade-1956-02}. Malgré cette erreur, H. Shapley basa la plupart de ses travaux sur cette relation unique et l'utilisa, entre autres, pour étudier la structure de notre Galaxie et mesurer les distances d'amas globulaires. En 1924, l'impact de cette relation sur la communauté scientifique fut énorme quand ils réalisèrent, notamment grâce à Edwin Hubble et ses mesures de distances extragalactiques, tout le potentiel de cette relation linéaire.

Alors que des travaux observationnels étaient en cours pour un meilleur étalonnage de cette loi, les travaux théoriques de 
Sir Arthur Eddington en 1927 furent le point de départ d'une théorie expliquant la pulsation des variables céphéides \citep{Eddington-1927-05}. Les Céphéides sont des étoiles subissant des oscillations radiales causées par la force de gravitation agissant vers l'intérieur et la force de pression du gaz agissant vers l'extérieur. Des théories plus élaborées et une meilleure compréhension des processus physiques liés au mécanisme de pulsation suivirent quelques années plus tard avec les travaux notamment de Sergei Zhevakin en 1953, John Cox en 1963 \citep{Cox-1963-06} et Robert Christy \citep{Christy-1966-}. Je parlerai brièvement du mécanisme de pulsation dans la Section~\ref{section__mecanisme_de_pulsation}.

À la même époque des astronomes se penchèrent sur le problème de la dispersion de la relation P--L. Certains évoquèrent une mauvaise correction de l'extinction interstellaire et d'autres préférèrent l'utilisation d'une relation du type période--luminosité--couleur (P--L--C). \citet{Fernie-1967-04} expliqua qu'une relation P--L--C n'est pas utile si les magnitudes sont bien corrigées de l'extinction. De plus, \citet{Stothers-1988-06} montra également, que cette relation est dépendante de la métallicité de l'étoile. 

La relation P--L est donc loin d'être parfaite et nécessite d'être peaufinée pour atteindre des mesures de distances avec une bonne précision. Ce bref historique n'est bien sûr pas exhaustif et je renvoie le lecteur aux articles publiés sur ce sujet pour une revue plus complète et détaillée \citep[par exemple][]{Fernie-1969-12,Feast-1999-07}.

\section*{De nos jours ...}

Les travaux sur les Céphéides sont toujours d'actualité, tant sur le plan observationnel que théorique. Cette relation P--L (encore appelée loi de Leavitt) est un outil puissant pour la détermination des distances cosmologiques et l'estimation de la constante de Hubble $H_0$ via la loi du même nom. Dans le milieu des années 80, l'observation de Céphéides pour déterminer $H_0$ avec une précision de $\pm 10\,\%$ est désignée comme l'un des trois projets clés du télescope spatial Hubble \citep[voir par exemple][]{Freedman-2010-09}. L'étalonnage précis de la relation P--L est nécessaire et divers travaux sont effectués dans ce but via l'utilisation de diverses techniques, comme par exemple \citet{Gieren-1998-03} qui utilisèrent la brillance de surface ou encore \citet{Kervella-2004-08} grâce aux mesures de diamètres par interférométrie.

Les Céphéides sont également un bon laboratoire d'étude dans le cadre de l'évolution et de la pulsation stellaire, fournissant des informations fondamentales sur les étoiles de masses intermédiaires. Les données récoltées sont utilisées pour contraindre et améliorer les modèles d'évolution stellaire.

Notre connaissance sur ce type d'étoile variable s'est améliorée au fil des années grâce aux avancées technologiques. Par l'utilisation de diverses techniques (Section~\ref{image__les_echelles_de_distance}) nous pouvons par exemple pour les plus proches, estimer leur distance et leur diamètre avec une bonne précision, mesurer leur variation de diamètre, ou encore évaluer leur profil d'intensité. Il a même été détecté récemment une enveloppe circumstellaire autour de certaines Céphéides \citep{Kervella-2006-03,Merand-2006-07,Merand-2007-08}. L'environnement autour des Céphéides, et principalement des Céphéides classiques, constitue le sujet de cette thèse et je parlerai tout au long de ce manuscrit des diverses techniques que j'ai utilisées pour leur détection et caractérisation. Avant cela il est nécessaire d'avoir une certaine compréhension des Céphéides elles-mêmes ainsi que leur utilisation.

\section{Les différents types de Céphéides}
\label{section__les_differents_types}

Les Céphéides sont des étoiles supergéantes très brillantes et sont donc observables à de grandes distances (par exemple le rayon de $\delta$~Cep fait environ 40 fois celui du Soleil). Elles ont une période de pulsation comprise entre 1 et 150 jours, une amplitude allant de $10$ à $20\,\%$ en diamètre et peuvent atteindre une variation photométrique de l'ordre de 2 magnitudes. Le type spectral varie également avec la pulsation entre les types F et K. 

\begin{figure}[!t]
	\centering\includegraphics[width = .9\linewidth]{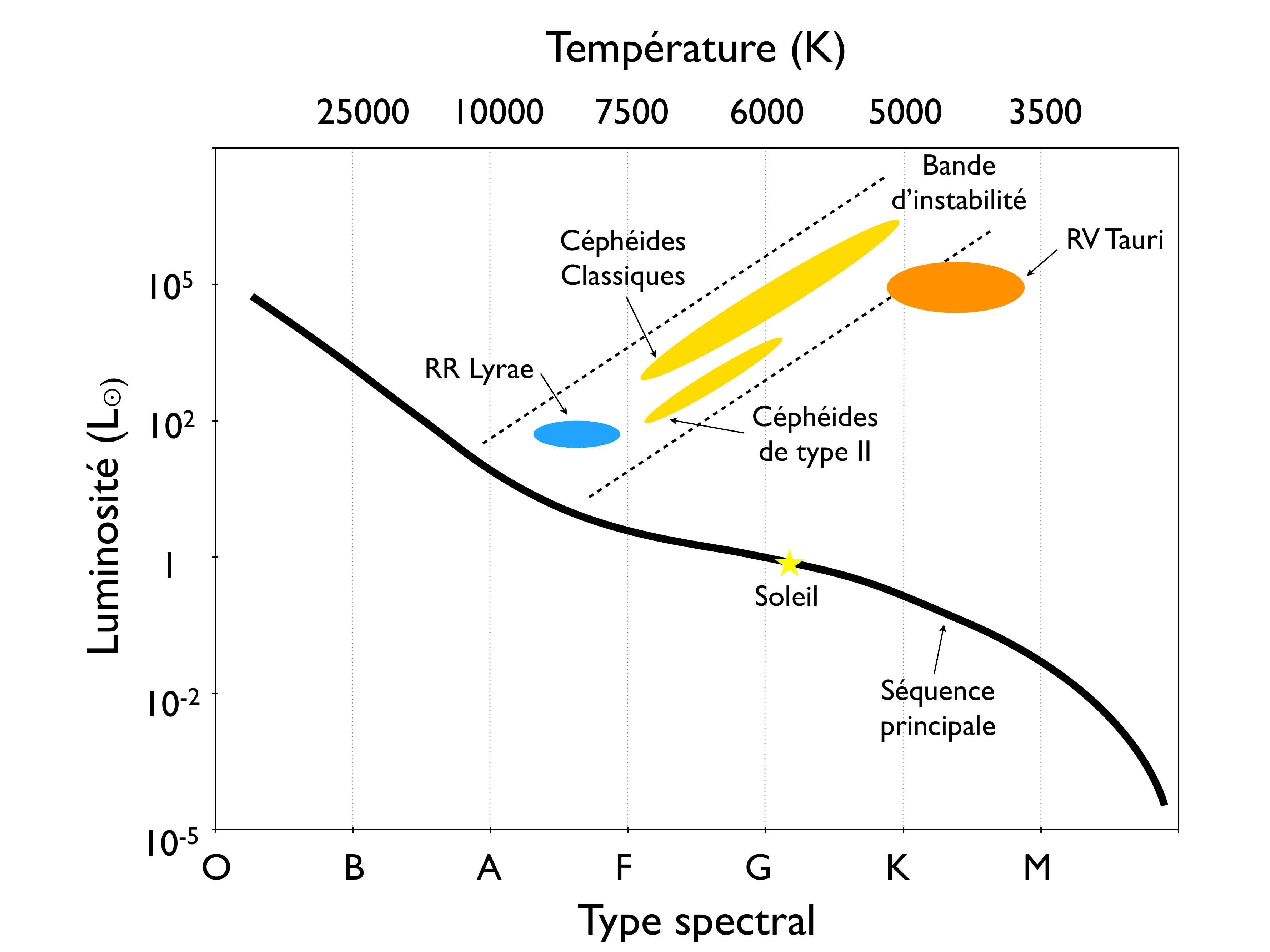}
	\caption[Diagramme H--R]{\textbf{Diagramme H--R} : diagramme luminosité--température de quelques étoiles variables dans la bande d'instabilité.}
	\label{image__diagramme_hr}
\end{figure}

\begin{figure}[!p]
	\centering\includegraphics[width = .8\linewidth]{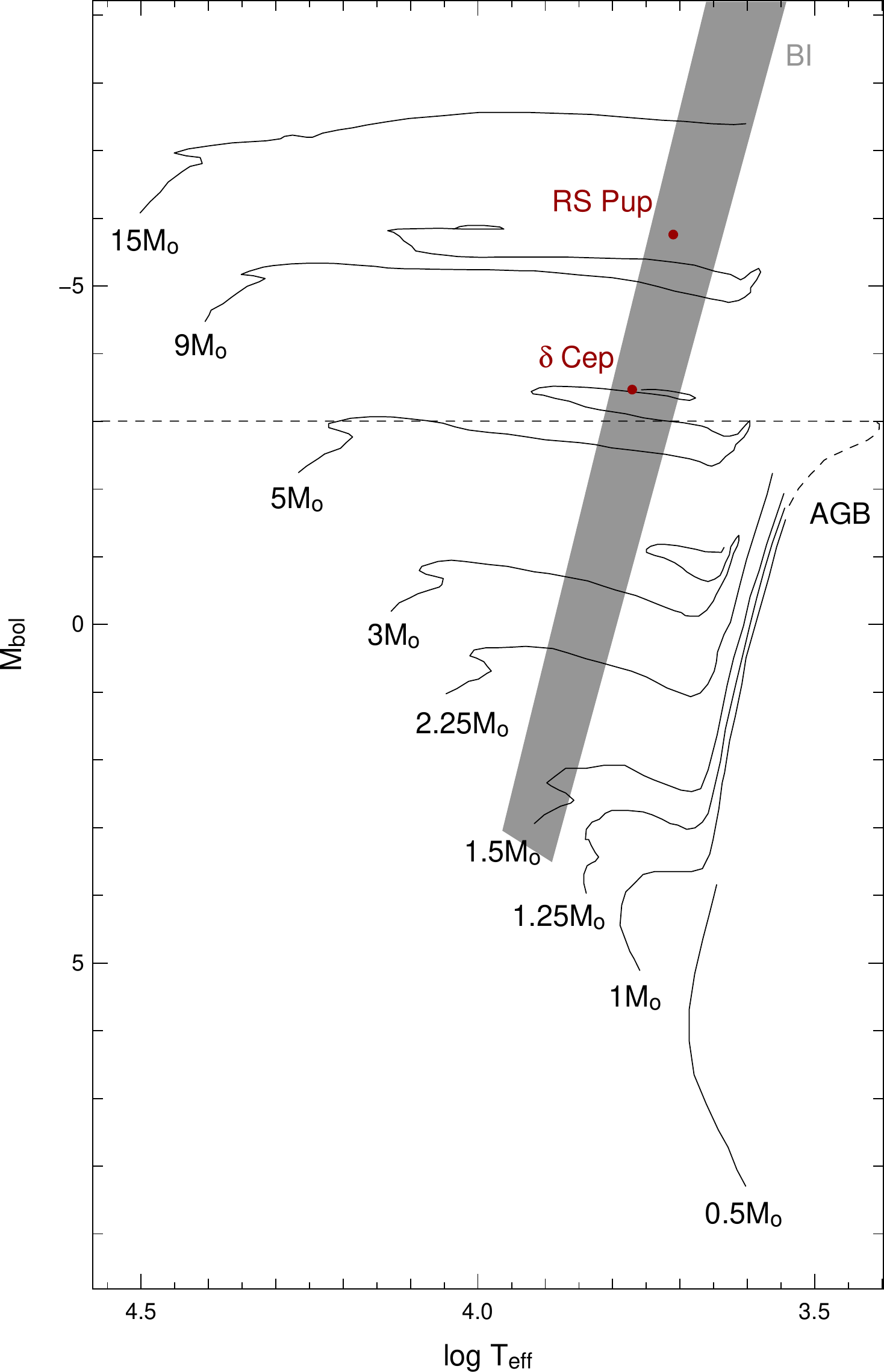}
	\caption[Évolution des étoiles en fonction de leur masse]{\textbf{Évolution des étoiles en fonction de leur masse} :  parcours d'une étoile dans le diagramme H--R en fonction de sa masse \citep[d'après][]{Kaler-1997-03}. La bande d'instabilité est notée $BI$. On remarque que les étoiles les plus massives font plusieurs passages dans la bande d'instabilité.}
	\label{image__evolution}
\end{figure}

Dans le diagramme de Hertzsprung--Russell (H--R) ces étoiles se situent dans ce que l'on appelle la bande d'instabilité (Fig~\ref{image__diagramme_hr}). Cette région étroite et presque verticale (en échelle logarithmique) contient plusieurs types d'étoiles variables comme par exemple les étoiles RR Lyrae, RV Tau, ... et s'étend de la séquence principale jusqu'à la branche des géantes rouges (voir l'annexe~\ref{annexe__evolution_des_etoile_de_masse_intermediaire} pour connaitre les diverses phases de l'évolution d'une étoile). Au cours de leur évolution, les étoiles traversent cette bande d'instabilité assez rapidement et peuvent, pour les plus massives, la traverser plusieurs fois en fonction de leur stade d'évolution. Sur la figure~\ref{image__evolution} est tracée l'évolution d'étoiles de différentes masses \citep[d'après][]{Kaler-1997-03}. On s'aperçoit notamment que la masse a une influence sur le ou les passages de l'étoile dans cette bande, et que l'étoile passe au stade Céphéide à différentes phases de son évolution, suggérant l'existence de différents groupes de Céphéides distinguables par une composition chimique différente. 

Avant 1956 les Céphéides étaient réunies en un seul groupe malgré la position différente de certaines étoiles dans le diagramme H--R. Les astronomes de l'époque et notamment Baade et Hubble ont remarqué une différence d'environ 1.5 magnitudes entre la relation P--L utilisant les Céphéides Galactiques et celle utilisant les Céphéides des amas globulaires. Pour expliquer cela Baade suggéra qu'il s'agissait peut-être de populations distinctes de Céphéides. Nous savons maintenant que c'est effectivement le cas et que les Céphéides sont principalement divisées en deux groupes, les Céphéides de type I et de type II.

\subsection{Les types I}

Ces étoiles plus connues sous le nom de Céphéides classiques, de variables de type $\delta$~Cep ou aussi parfois pas abus de langage simplement nommées Céphéides, sont des étoiles de population I, c'est à dire de la population d'étoile la plus jeune de l'univers ($\lesssim 9$ milliards d'années) et sont par conséquent riches en métaux. Elles sont généralement assimilées à des étoiles de masse intermédiaire ($3$--$10\,M_\odot$) post-séquence principale, en phase de combustion centrale de l'hélium.

\begin{figure}[!p]
	\begin{minipage}[h]{\linewidth}
		\centering\includegraphics[width = .95\linewidth]{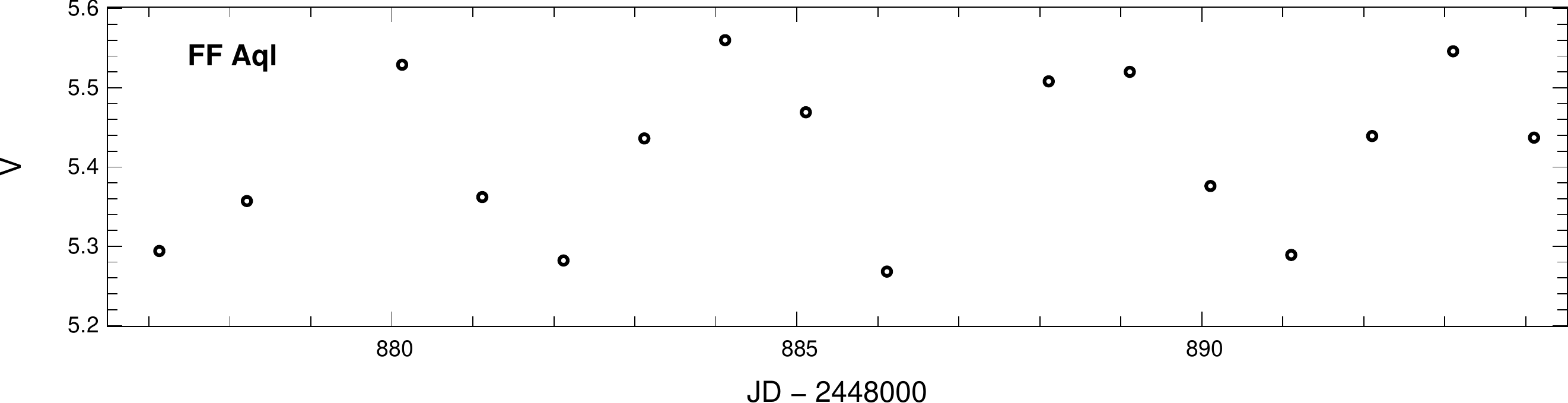}
	\end{minipage}
	\vfill
	\vspace{.2cm}
	\begin{minipage}[h]{\linewidth}
		\centering\includegraphics[width = .95\linewidth]{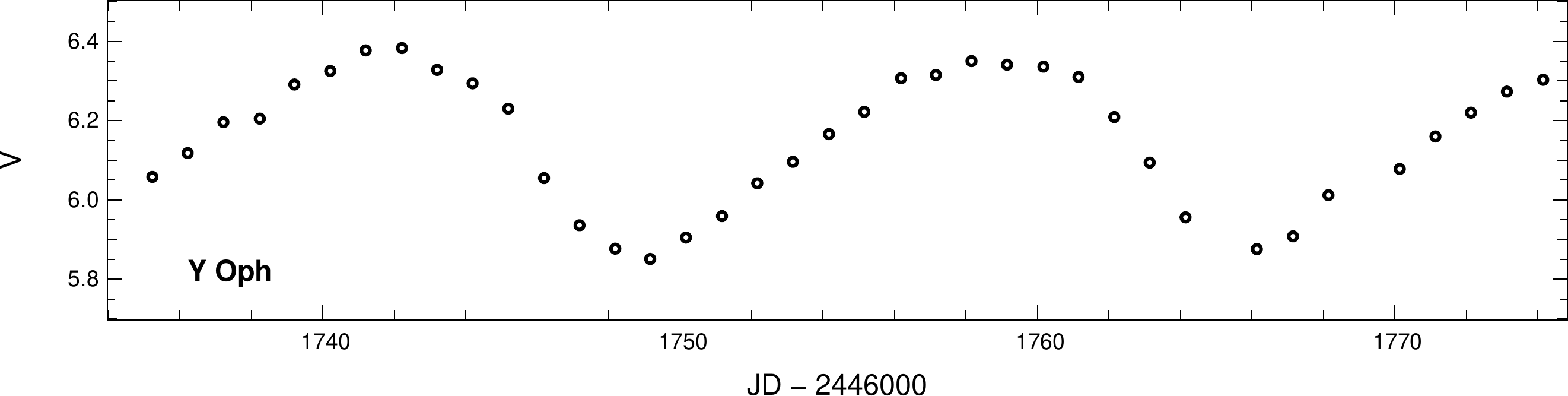} 
	\end{minipage}
	\vfill
	\vspace{.2cm}
	\begin{minipage}[h]{\linewidth}
		\centering\includegraphics[width = .95\linewidth]{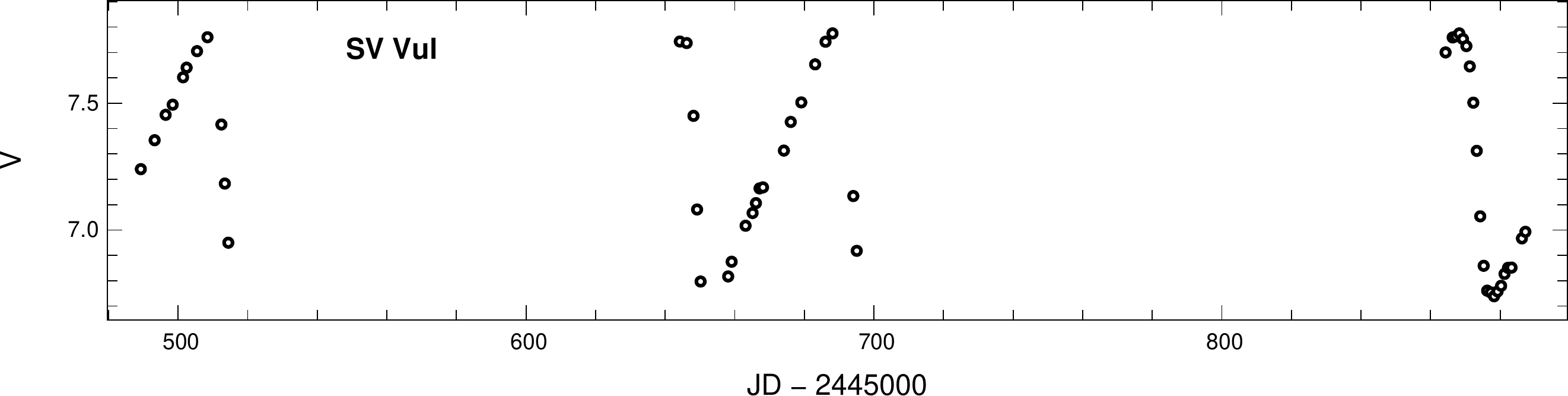}
	\end{minipage}
	\caption[Courbes de lumière de FF~Aql, Y~Oph et SV~Vul]{\textbf{Courbes de lumière de FF~Aql, Y~Oph et SV~Vul} : graphe de la magnitude apparente en fonction du temps. Ces étoiles ont des périodes respectives de $4.47\,\mathrm{j}, 17.12\,\mathrm{j}$ et $45.01\,\mathrm{j}$ \citep[d'après les données photométriques en bande $V$ de][]{Berdnikov-2008-04}.}
	\label{image__courbe_de_lumiere_type_I}
\end{figure}

Elles pulsent avec une période très régulière comprise entre 2 et 100\,jours avec une courbe de lumière très régulière avec le temps. Je présente comme exemples sur la figure~\ref{image__courbe_de_lumiere_type_I}, les courbes de lumière pour les Céphéides FF~Aql, Y~Oph et SV~Vul ayant des périodes respectives de $4.47\,\mathrm{j}, 17.12\,\mathrm{j}$ et $45.01\,\mathrm{j}$.

Elles sont principalement concentrées dans le plan galactique, ce qui rend délicates les mesures photométriques due à l'absorption interstellaire. J'aborderai ce point dans le Chapitre~\ref{chapitre__etude_d_exces_infrarouge_par_photometrie}. 

Durant ma thèse j'ai principalement travaillé sur des Céphéides de ce type car elles ont un rôle fondamental dans l'étalonnage de l'échelle des distances Galactiques et extragalactiques.

\begin{figure}[!p]
	\begin{minipage}[h]{\linewidth}
		\centering\includegraphics[width = .95\linewidth]{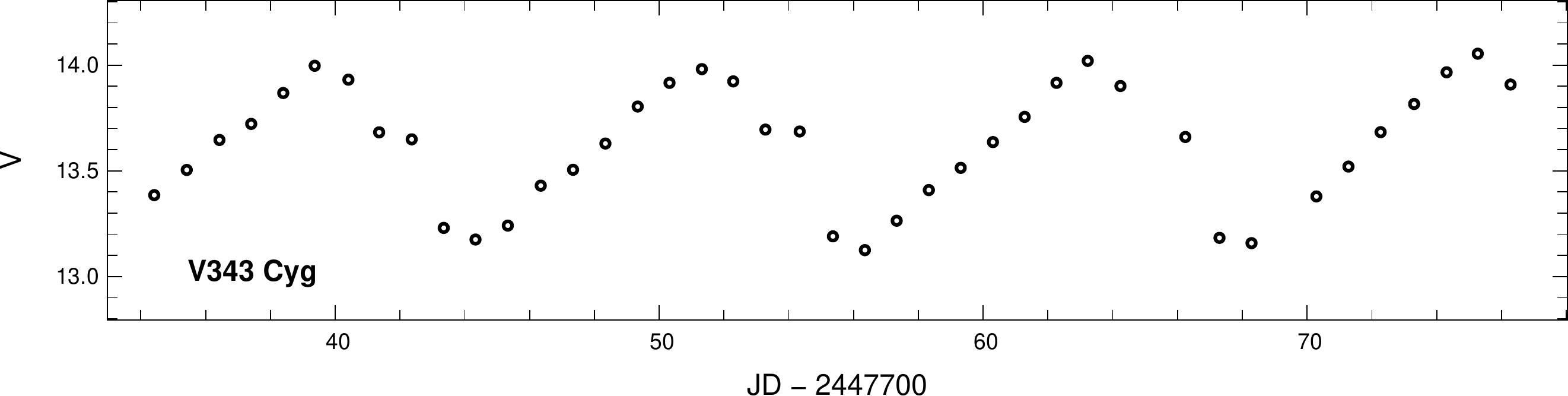} 
	\end{minipage}
	\vfill
	\vspace{.2cm}
	\begin{minipage}[h]{\linewidth}
		\centering\includegraphics[width = .95\linewidth]{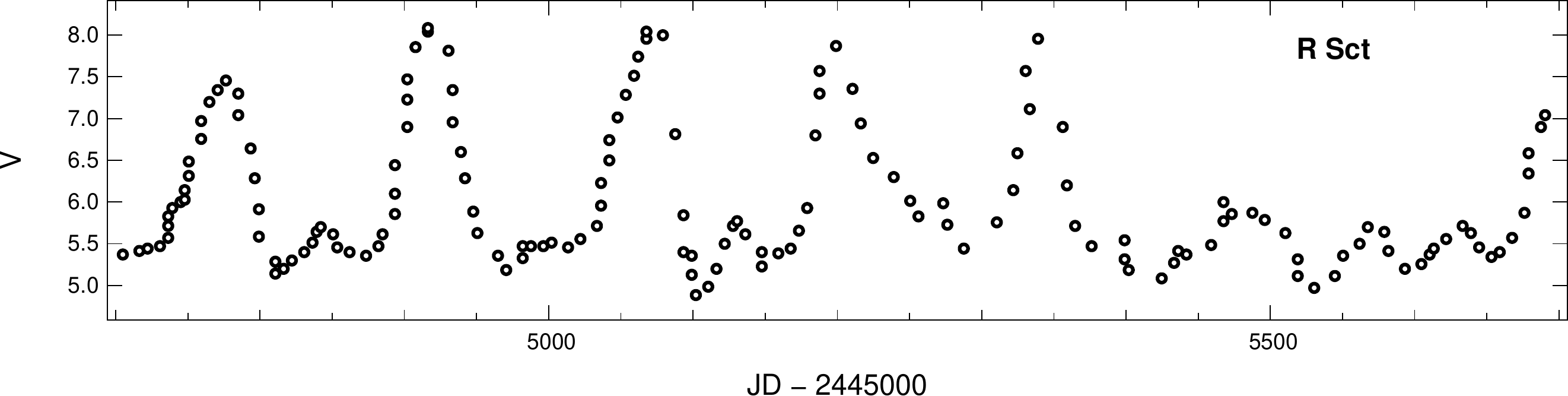}
	\end{minipage}
	\caption[Courbes de lumière de V343~Cyg et R~Sct]{\textbf{Courbes de lumière de V343~Cyg et R~Sct} : graphe de la magnitude apparente en fonction du temps. Ces étoiles ont des périodes respectives de $11.93\,\mathrm{j}$ et $146.50\,\mathrm{j}$ (d'après les données photométriques de \citet{Berdnikov-2008-04} et \citet{Matsuura-2002-06} en bande $V$).}
	\label{image__courbe_de_lumiere_type_II}
\end{figure}

\subsection{Les types II}

Elles sont plus vieilles et moins brillantes que les Céphéides classiques et se trouvent généralement dans les halos galactiques et les amas globulaires. Ce sont des étoiles de population II, c'est à dire des étoiles anciennes ($\gtrsim 9$ milliards d'années) assez pauvres en métaux. On les associe généralement à un stade d'évolution plus avancé que celui des Type I, c'est à dire à une phase postérieure à la combustion de l'hélium dans le c\oe ur.

Elles ont une masse de l'ordre de quelques $M_\odot$ et une période allant de un à plusieurs dizaines de jours. La courbe de lumière en fonction du temps pour les longues périodes sont très irrégulières et permettent, en plus de la métallicité, de les distinguer des Céphéides classiques. Sur la figure~\ref{image__courbe_de_lumiere_type_II} est représenté comme exemples les courbes de lumière des étoiles R~Sct et V343~Cyg, de période respective $146.50\,\mathrm{j}$ et $11.93\,\mathrm{j}$.

Les Céphéides de type II sont elles-mêmes divisées en trois sous-groupes en fonction de leur période et considérées comme étant à différents stades d'évolution :

\begin{itemize}
	\compactlist
	\item \textit{BL~Her} : étoiles de courte période ($P\,\lesssim\,4\,\mathrm{j}$) qui évoluent de la branche horizontale vers la phase AGB
	\item \textit{W~Vir} : étoiles de période intermédiaire ($4\,\mathrm{j}\,\lesssim\ P\,\lesssim\,20\,\mathrm{j}$) traversant la bande d'instabilité pendant sa phase AGB
	\item \textit{RV~Tau} : étoiles de longue période ($P\,\gtrsim\,20\,\mathrm{j}$) évoluant de la phase AGB vers la phase des naines blanches
\end{itemize}

Bien que moins brillantes que les Céphéides classiques, ces étoiles sont également utilisées pour l'estimation des distances Galactiques et extragalactiques.

Mon travail de thèse a principalement porté sur l'étude des Céphéides classiques Galactiques, par conséquent j'omettrai désormais, sauf mention contraire, le terme Classique pour désigner ce type d'étoile.

\section{Mécanisme de pulsation}
\label{section__mecanisme_de_pulsation}

Le principal mécanisme de pulsation des Céphéides a été découvert par E. Eddington et est connu sous le nom de mécanisme $\kappa$. Il est relatif au comportement de l'opacité dans les couches extérieures de l'étoile et plus précisément au changement d'opacité des couches d'hélium.

La compression de l'étoile sous l'effet de la gravitation entraîne une augmentation de la pression et de la température des couches externes. En se contractant, l'hélium s'ionise et devient opaque aux radiations. Cela bloque le flux radiatif des couches profondes, l'énergie s'accumule, la température et la pression interne de l'étoile augmente et les deux forces parviennent à s'équilibrer. Les forces de pression deviennent ensuite supérieures à la gravitation et ont pour effet d'élever les couches de gaz situées au dessus : l'étoile est en expansion. En se dilatant, les couches de gaz se refroidissent et l'hélium se recombine devenant ainsi transparent aux radiations. Sans cette source de chaleur additionnelle, l'expansion s'arrête et la force de gravitation reprend le dessus : l'étoile se contracte. En se contractant, la pression et la température augmentent et l'hélium s'ionise à nouveau. Le milieu redevient opaque aux rayonnements et l'expansion recommence.

Alors que certaines zones de l'atmosphère contribuent à l'excitation, d'autres contribuent à son amortissement. Toutefois cet amortissement est faible et permet à l'étoile d'osciller de façon très périodique. La période caractéristique des oscillations est directement liée à la densité moyenne de l'étoile :
\begin{equation}
	P \propto \rho^{-\frac{1}{2}}
	\label{equation_p_rho}
\end{equation}
Ceci peut se comprendre en raisonnant sur le temps caractéristique dit de "chute libre". La phase de compression de l'étoile correspond à une phase d'effondrement ayant comme temps caractéristique:
\begin{displaymath}
	t_\mathrm{ff} = \sqrt{ \frac{R^3}{GM} } \propto \sqrt{ \frac{1}{\rho\,G} } \propto \rho^{-\frac{1}{2}}
\end{displaymath}

Ainsi les Céphéides de faible densité ont une plus longue période de pulsation que celles de densité élevée. Cette observable $P$ est le paramètre principal pour une Céphéide et est de plus mesurable avec une bonne précision.

Notons également qu'il existe des modes propres de pulsation dépendant du profil de densité dans l'étoile. Il y a un mode dit fondamental noté $P_0$ et les modes harmoniques, dont la fréquence est multiple de celle de $P_0$ et noté $P_1$, $P_2$, ...

\section{Intérêt et utilisation des Céphéides: mesure de distance}

Les Céphéides ont deux propriétés qui font d'elles des chandelles standards pour les mesures de distance. Premièrement elles sont très brillantes et sont donc observables à de grandes distances et deuxièmement, comme mentionné plus tôt, la période fondamentale $P_0$ des Céphéides est directement liée à la luminosité grâce à la relation P--L.

Cette relation est un outil puissant dans l'estimation des distances cosmiques car c'est un pont reliant les distances locales à celles cosmologiques. Voyons plus précisément où se place cette relation parmi l'échelle des distances dans l'univers.

\subsection{L'échelle des distances}
\label{subsection__les_echelles_de_distance}

\begin{figure}[!p]
	\centering\includegraphics[width = \linewidth]{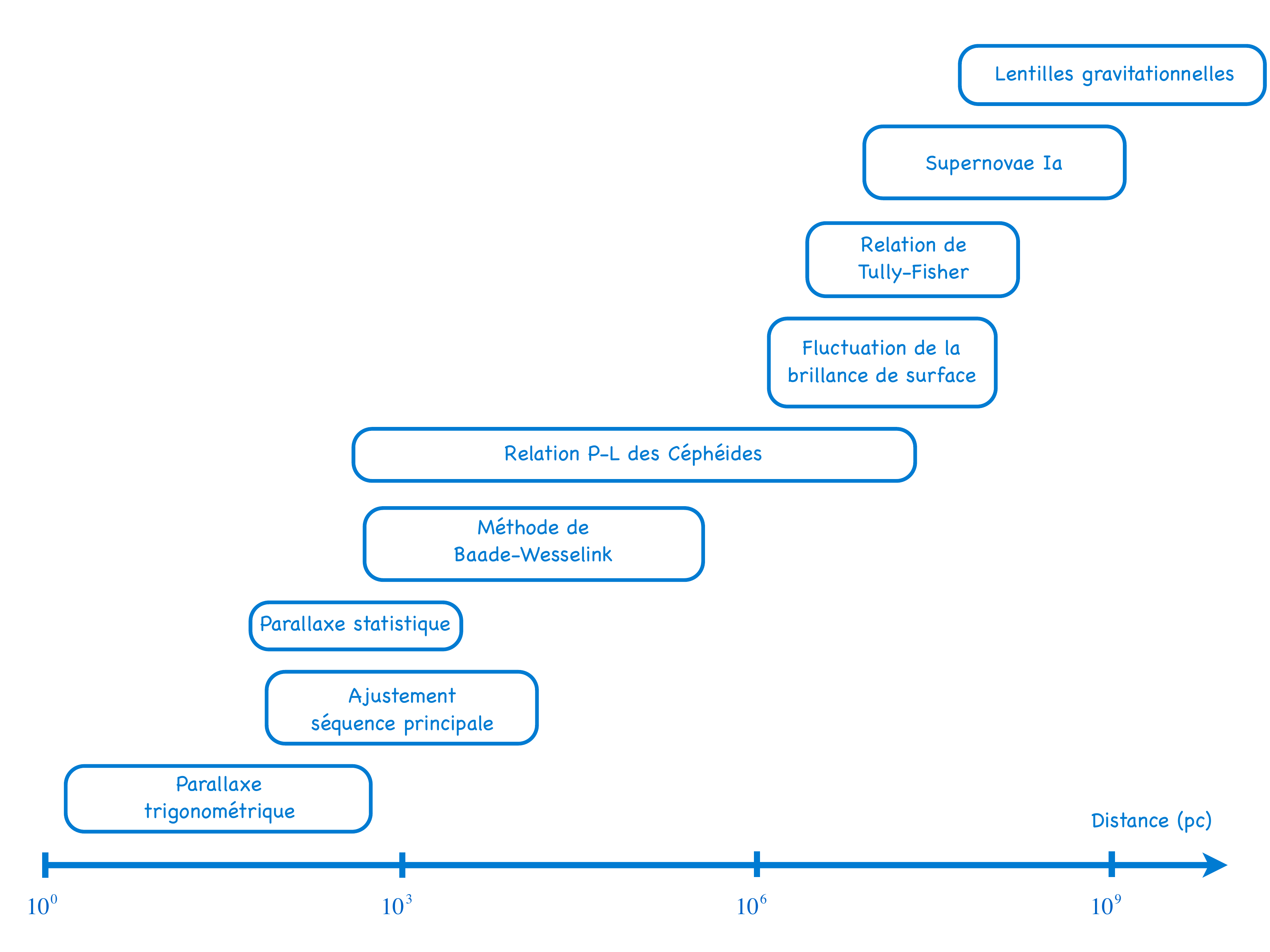}
	\caption[L'échelle des distances]{\textbf{L'échelle des distances} : différentes méthodes utilisées pour la mesure des distances. Toutes les techniques ne sont pas indiquées ici, il en existe beaucoup d'autres, mais celles là sont les plus utilisées.}
	\label{image__les_echelles_de_distance}
\end{figure}

L'estimation des distances est faite d'empilement de techniques pour une échelle de distance donnée. Je présente sur la Fig.~\ref{image__les_echelles_de_distance} quelques méthodes utilisées pour estimer les distances. La problématique est que l'utilisation d'une technique nécessite l'étalonnage des distances par la méthode précédente. La difficulté réside donc dans la propagation des incertitudes. On s'aperçoit que la relation P--L a une position importante, à la fois en tant qu'indicateur primaire de distance dans le groupe local (contenant environ 40 galaxies sur un diamètre d'environ 3 millions de parsecs) mais aussi pour l'étalonnage des indicateurs secondaires (relation de Tully-Fisher, ...).

Voici quelques méthodes utilisées en fonction de la distance supposée de l'objet astrophysique :

\begin{itemize}
	\compactlist
	\item \textbf{la parallaxe trigonométrique :} elle consiste à observer le changement de position d'une étoile proche au cours de l'année par rapport à des objets extrêmement éloignés. La trajectoire de l'étoile paraît décrire une petite ellipse. La parallaxe $\pi$ est le demi-grand axe de cette ellipse et est définie comme égale au rapport du rayon de l'orbite terrestre à la distance de l'étoile.

	\item \textbf{l'ajustement de la séquence principale :} la magnitude apparente d'étoiles dans un amas est mesurée puis tracée en fonction de l'excès de couleur ($B - V$). Comme ces étoiles sont approximativement à la même distance, le graphe obtenu est équivalent à un diagramme H--R uniquement translaté en ordonnée d'un coefficient proportionnel à la distance (la magnitude apparente est en ordonnée au lieu de la magnitude absolue). Il suffit d'ajuster la séquence principale du graphe obtenu avec un diagramme H--R connu pour obtenir la distance.

	\item \textbf{la parallaxe statistique :} l'idée est la même que la parallaxe trigonométrique excepté que maintenant on observe le mouvement d'un groupe d'étoiles fixes (supposées physiquement associées) par rapport à la position du soleil avec son environnement. La vitesse angulaire apparente de ce groupe d'étoile est directement liée à sa distance. L'utilisation de cette méthode se fait par des mesures s'étalant sur plusieurs années.

	\item \textbf{la méthode de Baade Wesselink :} C'est la plus utilisée des techniques ; elle permet de calculer la distance d'une Céphéides en reliant les variations de diamètre linéaire aux variations de diamètre angulaire. Je parlerai plus en détail de cette méthode dans la Section~\ref{section__la_methode_de_baade_wesselink}.

	\item \textbf{la relation période--luminosité des Céphéides:} une loi linéaire (en log) relie la période fondamentale de pulsation à la luminosité intrinsèque de l'étoile. Cette relation sera détaillée dans la Section~\ref{section__la_relation_periode_luminosite}

	\item \textbf{les fluctuations de la brillance de surface des galaxies:} cette méthode utilise le fait que plus une galaxie est éloignée, moins les détails à l'intérieur sont perceptibles. L'étude des fluctuations d'un pixel à un autre sera différente si l'on résout la galaxie ou pas. Par exemple une galaxie éloignée aura des fluctuations d'un pixel à un autre plus régulières qu'une galaxie proche.

	\item \textbf{la relation de Tully-Fisher :} c'est une relation linéaire (en log) reliant la luminosité intrinsèque d'une galaxie spirale à sa vitesse de rotation. Si l'on mesure la vitesse de rotation (par spectroscopie), on peut avec cette relation obtenir la magnitude absolue qui, en comparant avec la magnitude apparente, donne une mesure de la distance.

	\item \textbf{les supernovae de type Ia :} cette méthode est basée sur la similarité du profil des courbes de luminosité absolue des supernovae Ia. La mesure du profil de luminosité apparente permet l'estimation des distances.

	\item \textbf{les lentilles gravitationnelles :} quand un quasar est observé à travers une lentille gravitationnelle, des images multiples du même objet sont observées. En mesurant le décalage de temps d'arrivée des photons entre ces images multiples et en connaissant les angles de déviation, il est possible d'estimer la distance. Cependant il est nécessaire de connaître auparavant la distance relative entre le quasar et la lentille ainsi que des informations sur la lentille elle-même (masse, distribution, ...).

\end{itemize}

Il existe bien d'autres techniques pour estimer les distances que je ne détaillerai pas ici, citons par exemple les échos de lumière, la relation de Faber-Jackson, les supernovae de type II, l'effet Sunyaev-Zeldovich, ... L'un des objectifs final de cet échafaudage de techniques est la détermination précise de la constante de Hubble $H_0$, nécessaire pour contraindre les modèles cosmologiques.

La relation P--L est donc un maillon central dans l'estimation des distances dans l'univers. Cependant elle a besoin des techniques précédentes pour être étalonnée (généralement celle de Baade-Wesselink). Elle permet également l'étalonnage des autres indicateurs de distance.

\subsection{La relation Période--Luminosité}
\label{section__la_relation_periode_luminosite}

La relation entre la période et la luminosité observée par H. Leavitt implique une corrélation entre la luminosité et la densité tel que :
\begin{displaymath}
	L \propto \rho^\alpha
\end{displaymath}

\noindent où $\alpha$ est une constante et la relation~\ref{equation_p_rho} implique :
\begin{displaymath}
	P \propto L^{-\frac{1}{2\alpha} }
\end{displaymath}

En utilisant la définition de la magnitude absolue $M_\lambda = -2.5\,\log{L_\lambda} + cste$ on obtient la fameuse relation P--L :
\begin{equation}
	M_\lambda = a_\lambda\,\log{P} + b_\lambda
	\label{equation_p_l}
\end{equation}
où $a_\lambda$ et $b_\lambda$ sont des constantes dépendantes de la longueur d'onde. Si l'on connaît la magnitude apparente moyenne, nous pouvons déduire grâce au module de distance $\mu = m_\lambda - M_\lambda = 5\,\log{d} - 5$ la distance de la Céphéide et de la galaxie hôte (si la taille de la galaxie est petite devant sa distance).

\subsection{En pratique ... }

Des observations photométriques d'une Céphéide sont effectuées afin de fournir la magnitude apparente de l'étoile. Nous traçons ensuite la magnitude observée en fonction du temps pour obtenir une courbe de lumière (comme par exemple sur les Fig.~\ref{image__courbe_de_lumiere_type_I} et \ref{image__courbe_de_lumiere_type_II}). De cette courbe sont déterminés deux paramètres, la magnitude apparente moyenne de l'étoile et sa période. Si $a_\lambda$ et $b_\lambda$ sont connues, la magnitude absolue est déterminée via la relation P--L. Un fois la magnitude apparente et absolue déterminées, il ne reste qu'à déterminer la distance de l'étoile via l'utilisation du module de distance.

La précision de la distance dépend donc de la qualité des données photométriques et de la connaissance des paramètres $a_\lambda$ et $b_\lambda$ de la relation P--L. Les télescopes et instruments actuels nous permettent d'obtenir des courbes de lumière assez précises et ne limitent donc pas l'estimation des distances. Les principales sources d'incertitudes sont l'ajustement de la pente, $a_\lambda$, du point zéro, $b_\lambda$ et la correction de l'absorption interstellaire.

\subsection{L'étalonnage}

Pour être fiable, l'équation~\ref{equation_p_l} a besoin d'être étalonnée, c'est à dire qu'il faut déterminer $a_\lambda$ et $b_\lambda$. En général le paramètre $a_\lambda$ est assez bien déterminé en utilisant des Céphéides du Grand Nuage de Magellan (LMC). Le point zéro $b_\lambda$ est un peu plus délicat puisqu'il est nécessaire de connaître la distance d'un échantillon de Céphéides de manière précise et indépendante.

Lors des premiers étalonnages de la relation, l'erreur estimée sur les distances étaient grandes et atteignait 32\,\% \citep{1958ApJ...127..513S}. Onze ans plus tard, \citet{Sandage-1969-08} étalonnent une relation P--L--C et réduisent l'incertitude à 26\,\%. Cette relation s'affine au fur et à mesure des années, notamment en remarquant qu'il existe une loi P--L différente pour chaque type de Céphéides. Je ne présenterai pas ici d'historique sur l'étalonnage, mais pour le lecteur intéressé, des revues détaillées peuvent-être trouvées dans \citet{Madore-1991-09}, \citet{Sandage-2006-12}, \citet{Fouque-2007-12} et \citet{Barnes-2009-09}. 

\begin{figure}[!t]
	\begin{minipage}[h]{.45\linewidth}
  		\centering\includegraphics[width = \linewidth]{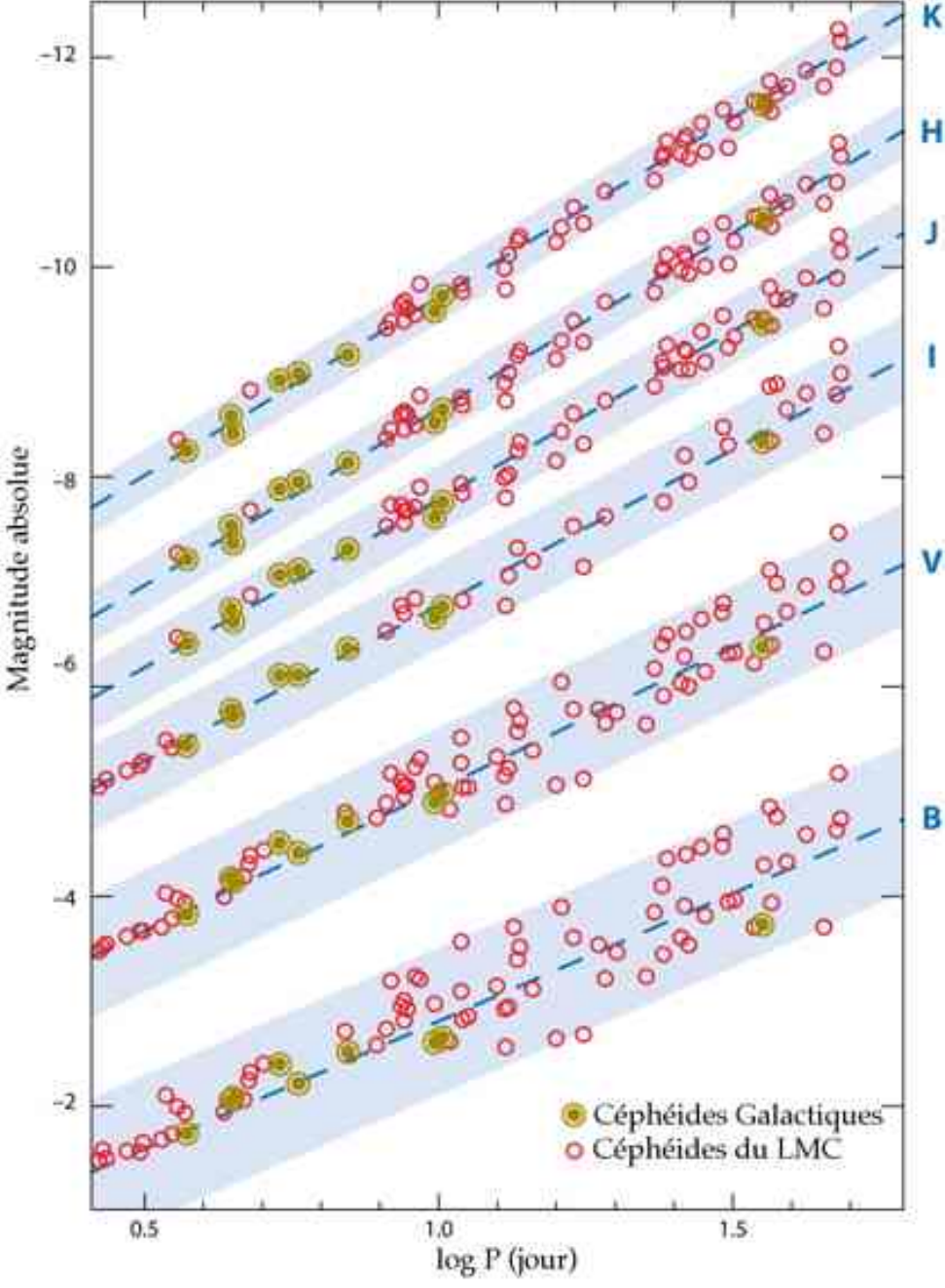}
	\end{minipage}
	\hfill
	\hspace{.2cm}
	\begin{minipage}[h]{.5\linewidth}
 		\caption[Relation P--L en bande $B, V, I, J, H, K$]{\textbf{Relation P--L en bande $B, V, I, J, H, K$} : la pente est étalonnée avec des Céphéides du LMC et le point zéro avec des Céphéides Galactiques. On s'aperçoit que la pente augmente avec la longueur d'onde alors que sa dispersion diminue \citep[d'après][]{Freedman-2010-09}.}
  		\label{image__calibration_p_l}
	\end{minipage}
\end{figure}

La pente est généralement bien définie grâce aux milliers de Céphéides identifiées dans le LMC qui permettent une bonne étude statistique. Cependant $a_\lambda$ dépend de la longueur d'onde d'observation. Comme le montre la Fig.~\ref{image__calibration_p_l}, la pente augmente avec les longueurs d'onde croissantes alors que sa dispersion diminue. Cette diminution de la dispersion peut s'expliquer par la décroissance de l'extinction interstellaire en fonction de la longueur d'onde, mais surtout par la diminution de l'effet dû à la température. De ce fait, la majorité des relations P--L aujourd'hui sont des relations infrarouges. Cependant comme nous le verrons dans la suite, l'utilisation d'observations en infrarouge peut créer une nouvelle source d'incertitude.

À ce jour, l'incertitude est de l'ordre de 3--5\,\% \citep{Benedict-2007-04,Riess-2011-04} et est principalement due à l'estimation du point zéro. Le plus souvent des Céphéides Galactiques sont utilisées pour déterminer $b_\lambda$ car il est plus facile de mesurer leur distance par des méthodes directes (parallaxe, ...). Certaines de ces méthodes ont été détaillées en Section~\ref{subsection__les_echelles_de_distance}. L'un des problèmes est que la majorité des Céphéides (classiques) sont localisées dans le plan Galactique, les poussières présentent sur la ligne de visée décroissent la luminosité apparente de l'étoile et par conséquent induisent une erreur sur la distance. Néanmoins, l'extinction interstellaire peut être diminuée en observant dans l'infrarouge où les effets sont moindres, mais pas négligeables.

La relation la plus précise à ce jour est une relation en bande $K$, étalonnée par 10 Céphéides Galactiques grâce à des mesures précises de parallaxes avec le HST et une bonne détermination de l'extinction \citep{Benedict-2007-04} :
\begin{displaymath}
	M_K = - 3.32_{\,\pm\,0.12}\,(\log{P} - 1) - 5.71_{\,\pm\,0.03}
\end{displaymath}

Nous noterons également que \citet{Storm-2011-10} convergent vers une relation P--L similaire ($M_K = - 3.30_{\,\pm\,0.06}\,(\log{P} - 1) - 5.65_{\,\pm\,0.02}$).

Depuis 2006, une nouvelle source d'incertitude est apparue via la découverte d'enveloppes circumstellaires autour de certaines Céphéides Galactiques \citep{Kervella-2006-03,Merand-2006-07,Merand-2007-08,Barmby-2010-11}. Avant cela, seul une Céphéide, RS~Puppis, était connue pour être entourée d'un environnement de gaz et de poussière. L'échantillon de Céphéides actuel semble montrer qu'elles possèdent peut-être toutes une composante circumstellaire. L'existence de ces enveloppes soulève non seulement des questions sur leur processus de formation mais également concernant l'impact sur l'estimation de $b_\lambda$ via les mesures "directes" de distance (directe dans le sens où l'on utilise pas la relation P--L). Plus particulièrement, la méthode de Baade-Wesselink dont je parlerai dans la Section~\ref{section__la_methode_de_baade_wesselink} utilise des mesures photométriques, spectroscopiques et/ou interférométriques de diamètre pour l'estimation des distances et ces mesures pourraient être biaisées par l'émission infrarouge de l'environnement circumstellaire. Une erreur sur la mesure de diamètre se répercute donc sur l'étalonnage de la relation P--L, puis sur les indicateurs de distance secondaires puis sur $H_0$.

Il est donc nécessaire d'étudier ces enveloppes afin de quantifier un éventuel biais sur l'étalonnage via les mesures de diamètres. À ce jour peu d'enveloppes ont été détectées car selon les études récentes \citep{Kervella-2006-03,Merand-2006-07,Merand-2007-08} leur taille angulaire ne serait que de quelques rayons stellaires (soit quelques millisecondes d'angle) et leur détection nécessite donc l'utilisation des techniques de haute résolution angulaire. C'est ce dont je parlerai tout au long de ce manuscrit.

Avant cela regardons la méthode la plus utilisées pour l'étalonnage de la relation P--L, qui pourrait être biaisée par la présence d'enveloppes circumstellaires.

\section{La méthode de Baade-Wesselink}
\label{section__la_methode_de_baade_wesselink}

Le principe de cette technique est de comparer les variations linéaires du diamètre stellaire (parallèle à la ligne de visée) aux variations angulaires du diamètre (perpendiculaire à la ligne de visée). On suppose généralement que les Céphéides oscillent radialement. La distance est calculée géométriquement par la relation :
\begin{displaymath}
	d = \frac{2\,\Delta R}{\Delta \theta}
\end{displaymath}

$\Delta R$ est évalué en intégrant la courbe de vitesse radiale mesurée par spectroscopie à haute résolution et $\Delta \theta$ est estimé soit par photométrie, on parle alors de la méthode de brillance de surface, soit par interférométrie, on parle alors de la méthode de Baade-Wesselink interferométrique (IBWM).

En fait cette relation n'est pas si simple car il faut tenir compte de certains autres paramètres. Le premier paramètre le plus important est le facteur de projection $p$. La spectroscopie ne mesure pas directement la vitesse de pulsation $v_p$ mais plutôt la vitesse de pulsation projetée et intégrée sur toute la surface visible de l'étoile. On définit donc la vitesse de pulsation comme $v_\mathrm{p}\,=\,p\,v_\mathrm{r}$. Ce facteur $p$ n'est pas très bien connu, la recherche de sa variation avec la pulsation est très active, mais nous n'avons toujours pas d'estimation précise. Sa valeur serait comprise entre 1.27 et 1.36 \citep{Merand-2006-,Nardetto-2007-07,Groenewegen-2007-11}.

Le second paramètre concerne le profil d'assombrissement centre-bord de l'étoile, c'est un effet de projection du disque stellaire où la lumière provenant du centre montre des couches plus profondes et plus chaude que le limbe. Cet effet, contenu dans un paramètre noté $k$, rentre en jeu dans les mesures de diamètres par interférométrie. On note généralement $\theta_\mathrm{UD}\,=\,k\,\theta_\mathrm{LD}$ où $k$ est déterminé à partir de modèles hydrostatiques d'atmosphères. Je parlerai plus précisément de ce paramètre au Chapitre~\ref{chapitre__acces_a_la_haute_resolution_angulaire_interferometrie}.

L'équation précédente devient plus précisément :
\begin{displaymath}
\theta_\mathrm{UD}(T) - \theta_\mathrm{UD}(0) = -2 \frac{kp}{d} \int_0^T v_\mathrm{r}(t)dt
\end{displaymath}

Si l'on utilise la technique de brillance de surface, on récupère directement $\theta_\mathrm{LD}$, le paramètre $k$ n'intervient donc pas.

\subsection{La méthode de brillance de surface}

Certaines des relations exposées ici sont présentées plus en détails dans le Chapitre~\ref{chapitre__etude_d_exces_infrarouge_par_photometrie}.

La puissance absolue rayonnée par une étoile de rayon $R$ et de température effective $T_\mathrm{eff}$ est donnée par la relation $L_\lambda = 4\pi R^2 F_\lambda$. Le flux mesuré sur Terre est relié à la distance de l'étoile par la relation $f_\lambda = L_\lambda/(4\pi d^2)$. En supposant que l'étoile rayonne comme un corps noir, on a $F_\lambda = \pi B_\lambda(T_\mathrm{eff})$, où $B_\lambda(T_\mathrm{eff})$ représente la fonction de Planck. La combinaison de ces équations donne :
\begin{equation}
	\label{equation__F_lambda}
	f_\lambda = \frac{\pi \theta_\mathrm{LD}^2}{4} B_\lambda(T_\mathrm{eff})
\end{equation}

En reliant cette équation à la magnitude apparente $m_\lambda$, on obtient :
\begin{displaymath}
	m_{\lambda_0} = - 5\,\log{\theta_\mathrm{LD}} - 2.5\,\log{B_\lambda(T_\mathrm{eff})} +cst
\end{displaymath}
où l'indice $0$ signifie que les magnitudes sont corrigées de l'extinction interstellaire. La brillance de surface $S_\lambda$ est définie comme :
\begin{displaymath}
	- 2.5\,\log{B_\lambda(T_\mathrm{eff})} +cst = S_\lambda = m_{\lambda_0} + 5 \log{\theta_\mathrm{LD}}
\end{displaymath}

$S_\lambda$ est indépendante de la distance et en faisant l'approximation que l'étoile rayonne comme un corps noir, elle est uniquement liée à la température. Il existe donc une relation avec l'indice de couleur (différence entre deux magnitudes apparentes obtenues dans deux bandes spectrales) telle que :
\begin{displaymath}
	S_{\lambda_1} = c_1\,(m_{\lambda_1} - m_{\lambda_2})_0 + c_2
\end{displaymath}
 Les diamètres angulaires peuvent donc être prédits par la combinaison des deux relations précédentes :
\begin{displaymath}
	\log{\theta_\mathrm{LD}} = a_1\,(m_{\lambda_1} - m_{\lambda_2})_0 + a_2 - 0.2\,m_{\lambda_{1,0}}
\end{displaymath}
où j'ai noté $a_1\ =\ 0.2\,c_1$ et $a_2\ =\ -0.2\,c_2$.

L'étalonnage de cette relation nécessite des mesures photométriques et des diamètres angulaires connus avec une bonne précision. La relation de brillance de surface moyenne pour les Céphéides a été interpolée par \citet{Kervella-2004-03} sur un échantillon de 9 étoiles. La dispersion est minimale avec l'utilisation de l'indice $V - K$ :
\begin{displaymath}
	S_V =  -0.1336_{\,\pm\,0.0008}(V - K)_0 + 3.9530_{\,\pm\,0.0006}
\end{displaymath}

Notons que \citet{Fouque-1997-04} avaient déduit une relation similaire pour les étoiles géantes et supergéantes en accord avec cette relation. À une phase donnée, une Céphéide peut donc être comparée à une étoile supergéante non pulsante.

Il est maintenant possible d'estimer les variations de diamètres (photométriques) de l'étoile en mesurant les variations de magnitudes apparentes corrigées de l'extinction interstellaire :
\begin{equation}
	\Delta\,\log{\theta_\mathrm{LD}}  =  -0.2672\,\Delta (V - K)_0 - 0.2\,\Delta V_0
	\label{equation__BS}
\end{equation}

On utilise le plus souvent une relation de brillance de surface infrarouge (IRSB), car comme je l'ai exposé précédemment, la relation P--L est moins dispersée. De plus la correction des effets de l'extinction y est plus faible.

\bigskip
\noindent \textbf{Limitations de la méthode :}

La première limitation est bien sûr le facteur $p$ qui n'est pas très bien estimé pour le moment. Certains pensent que sa dépendance avec la phase de pulsation causerait une erreur sur la distance de l'ordre de $4$--$5\,\%$ \citep{Sabbey-1995-06} alors que d'autres \citep{Nardetto-2004-12} affirment une erreur de seulement $0.2\,\%$. Il n'existe pas de valeur optimale et il dépend probablement de la Céphéide étudiée. Il a été mesuré par \citet{Merand-2006-} pour l'étoile $\delta$~Cep : $p\,=\,1.27\,\pm\,0.06$. Il existerait également une loi facteur de projection--période \citep{Nardetto-2009-05} : $p\,=\,-0.08_{\,\pm\,0.05}\,\log{P}\,+\,1.31_{\,\pm\,0.06}$.

Une autre limitation concerne la correction de l'absorption interstellaire par une bonne estimation de l'extinction $E(B - V)$ et d'une loi de rougissement appropriée (je renvoie le lecteur au Chapitre~\ref{chapitre__etude_d_exces_infrarouge_par_photometrie} pour plus de détails sur l'absorption interstellaire). En utilisant l'équation~\ref{equation__BS}, une erreur de $\sigma_\mathrm{V_0}\,=\,0.2\,\mathrm{mag}$ sur la magnitude $V_0$ entraînerait une erreur relative minimale qui n'est pas négligeable :
\begin{displaymath}
	\frac{\sigma_\theta}{\theta}\ \geqslant \ \sigma_\mathrm{V_0}\,\sqrt{0.1336^2 + 0.2^2}\,\sim\,5\,\%
\end{displaymath}

La présence d'une enveloppe circumstellaire causerait également une erreur sur l'estimation du diamètre. Si cet environnement (noté CSE çi-après) contribue à l'émission photosphérique de l'étoile en bande $K$ tel que $F_\mathrm{cse}/F_\star\,=\,5\,\%$, alors l'erreur sur le diamètre serait :
\begin{displaymath}
	\log\,\left(\frac{\theta_\mathrm{\star\,+\,\mathrm{cse}} }{\theta_\star } \right) = 2.5 \times 0.2672\,\log \left(\frac{F_\mathrm{\star\,+\,cse}}{F_\star} \right)
\end{displaymath}
\begin{displaymath}
	\Longleftrightarrow \log\,\left(\frac{\theta_\star + d\theta}{\theta_\star } \right) = 2.5 \times 0.2672\,\log \left(\frac{F_\star +F_\mathrm{cse}}{F_\star} \right)
\end{displaymath}
\begin{displaymath}
	\Rightarrow \frac{d\theta}{\theta} = \left( 1\,+\,\frac{F_\mathrm{cse}}{F_\star} \right)^{0.0668} - 1\,\sim\,1.0\,\%
\end{displaymath}

Ces deux incertitudes, liées à l'extinction et à la présence d'une enveloppe, se propageraient donc sur l'estimation de la distance avec le même ordre de grandeur.

La présence d'un compagnon est également une source d'erreur non négligeable. Cela aurait un effet sur la mesure photométrique de la Céphéide (elle apparaîtrait plus brillante) et sur la vitesse radiale mesurée (via la vitesse orbitale). L'impact est d'autant plus important que la différence de magnitude entre les deux objets est faible. On doit donc également tenir compte du phénomène de binarité de certaines Céphéides.

\subsection{La méthode interférométrique}

\begin{figure}[!p]
  \centering\includegraphics[width = .85\linewidth]{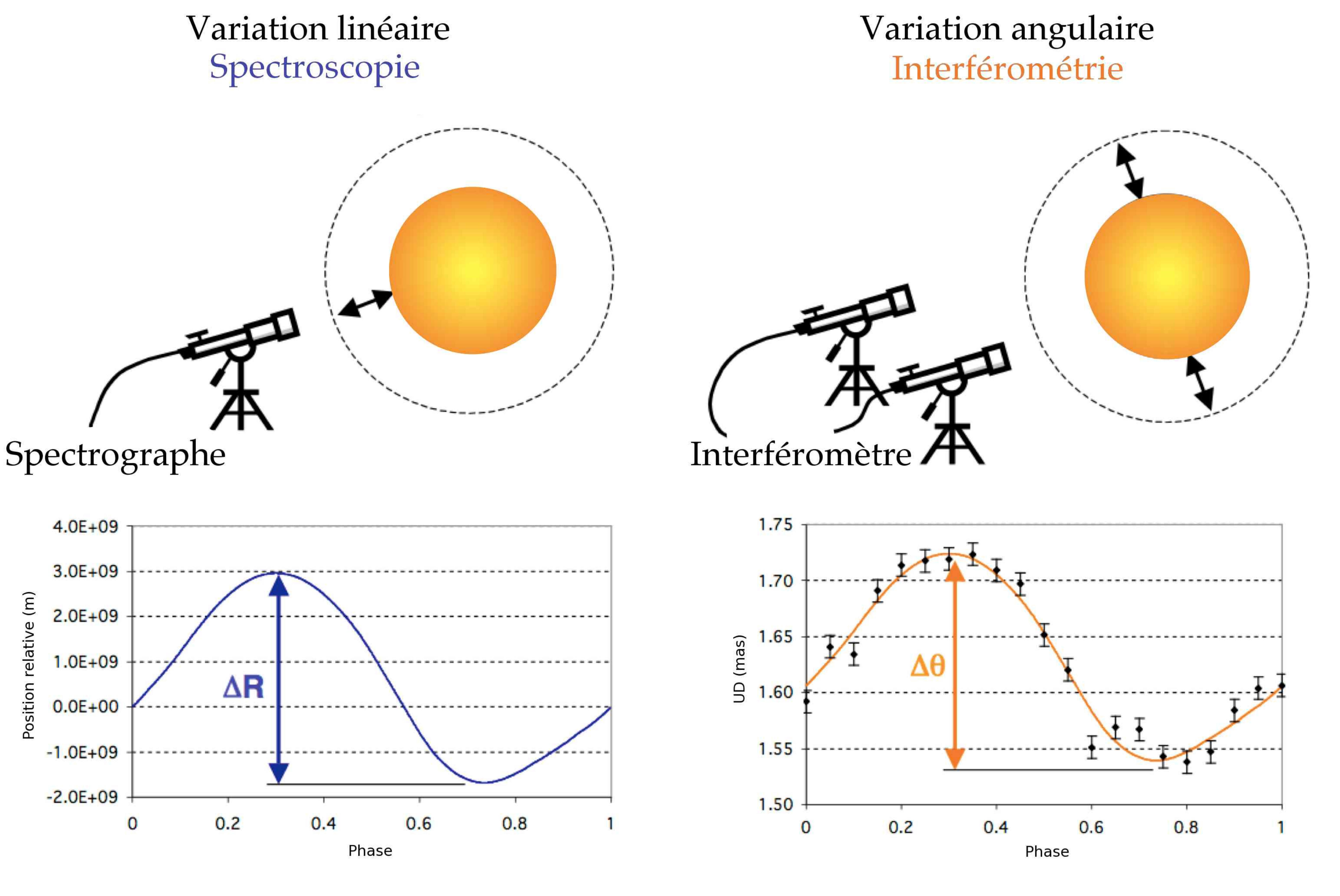}
  \caption[Méthode IBWM]{\textbf{Méthode IBWM} : Les deux techniques utilisées sont la spectroscopie haute résolution et l'interférométrie. La première donne accès aux variations de vitesse radiale et la deuxième aux variations de diamètre angulaire \citep[d'après][]{Kervella-2004-09}.}
  \label{image__baade_wesselink}
\end{figure}

Cette technique, aussi appelé parallaxe de pulsation, est plus directe pour la détermination des diamètres stellaires. Comme nous le verrons dans le Chapitre~\ref{chapitre__acces_a_la_haute_resolution_angulaire_interferometrie}, l'interférométrie permet une mesure directe du diamètre angulaire d'une étoile. En échantillonnant les mesures sur le cycle de pulsation de l'étoile il est possible de mesurer les variations de diamètre. La Fig.~\ref{image__baade_wesselink} représente une vue schématique de la IBWM.

\begin{figure}[!p]
	\centering\includegraphics[width = .76\linewidth]{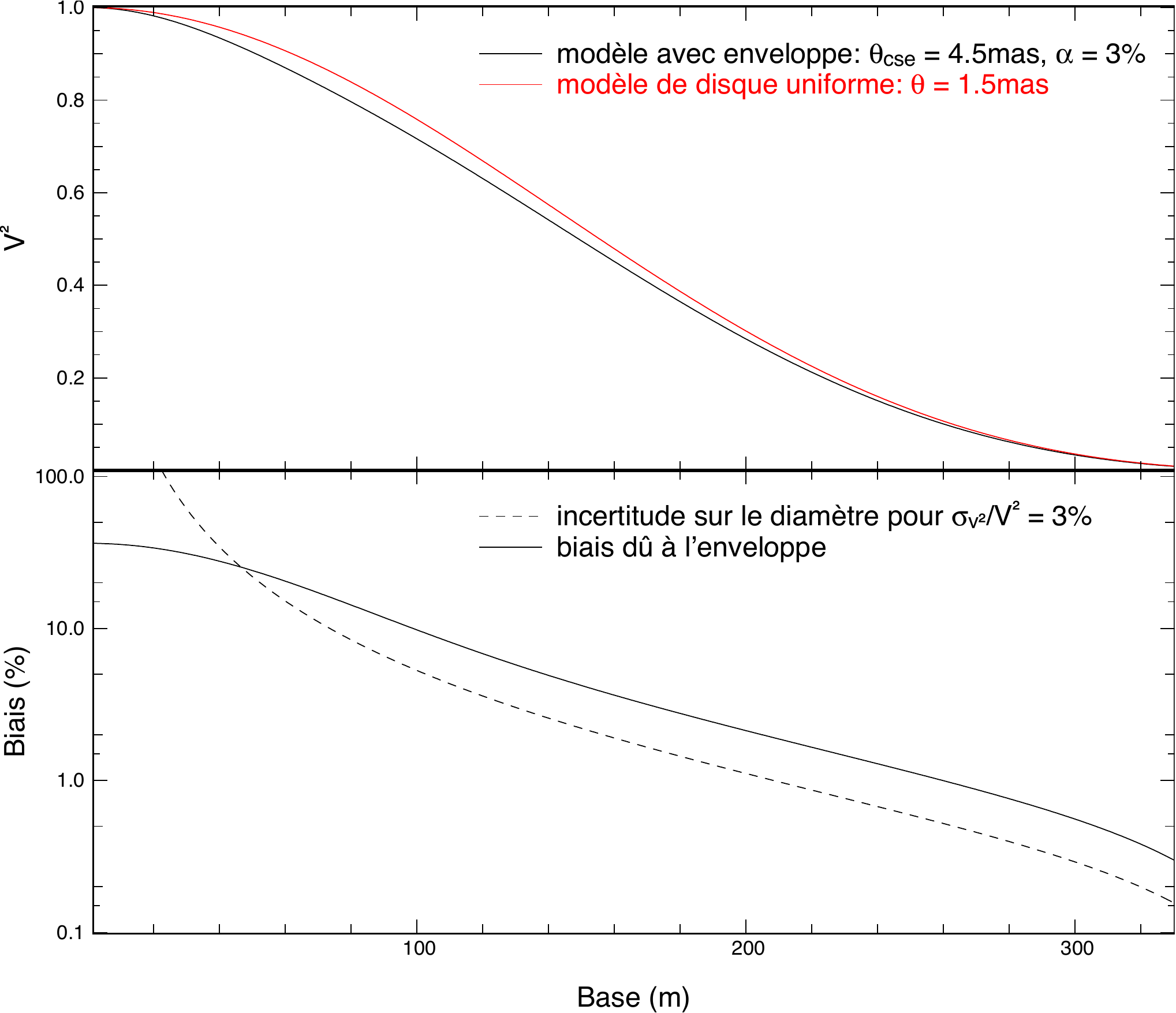}
  	\caption[Biais sur l'estimation du diamètre lié à la présence d'une enveloppe]{\textbf{Biais sur l'estimation du diamètre lié à la présence d'une enveloppe} : dans la fenêtre du haut est tracée la visibilité pour un modèle de disque uniforme (rouge) et un modèle étoile+enveloppe (noir). Dans la fenêtre du bas est représenté le biais sur le diamètre angulaire due à la présence de l'enveloppe en fonction de la base (trait plein), ainsi que l'incertitude sur la mesure de visibilité (tirets).}
  	\label{image__bias_visibility}
\end{figure}

C'est la seule méthode à ce jour qui donne accès à une précision de quelques pour cent sur les distances à partir du sol. Notons toutefois que les mesures de diamètre angulaire par interférométrie sont toujours dépendantes des modèles que nous utilisons.

\bigskip
\noindent \textbf{Limitations de la méthode :}

Le facteur $p$ reste la principale source d'incertitude. Il peut être estimé pour certaines étoiles qui disposent déjà d'une mesure de distance précise \citep{Merand-2006-,Groenewegen-2007-11}.

Un modèle analytique d'assombrissement centre-bord doit également être utilisé. Une loi généralement adoptée est une loi de puissance de type $I(\mu)/I(0) = \mu^\alpha$, avec $\mu = \cos{\theta}$ et où $\theta$ est l'angle entre la surface stellaire et la ligne de visée. Le paramètre $\alpha$ peut être ajusté aux données ou fixé à partir de modèles d'atmosphère \citep{Claret-2000-11}. Je reviendrai sur l'assombrissement centre-bord au Chapitre~\ref{chapitre__acces_a_la_haute_resolution_angulaire_interferometrie}.

La présence d'une enveloppe est également une source d'erreur puisque le diamètre mesuré apparaîtra plus grand que le diamètre stellaire réel et par conséquence la distance sera sous-estimée. Cependant cette erreur dépend de la base (distance entre les télescopes) et des modèles de visibilité $V^2$ utilisés. Adoptant la même démarche que \citet{Merand-2006-07}, on peut définir :
\begin{displaymath}
A_\mathrm{f}(\theta ,B,\lambda) = \frac{\partial V^2(\theta ,B,\lambda)}{\partial \theta}\frac{\theta}{V^2(\theta ,B,\lambda)}
\end{displaymath}

En première approximation, nous pouvons écrire : 
\begin{equation}
\frac{\theta_\mathrm{cse} - \theta_\star}{\theta_\star} = \frac{V^2_\mathrm{cse} - V^2_\star}{V^2_\star}\,\frac{1}{A_\mathrm{f}}
\label{equation__biais}
\end{equation}

Pour une étoile, on a comme modèle de visibilité le plus simple celui d'un disque uniforme $V^2_\star(\theta ,B,\lambda) = (2\,J_1(\pi \theta_\star B/\lambda)/(\pi \theta_\star B/\lambda))^2$, où $J_1$ représente la fonction de Bessel de première espèce d'ordre 1, $B$ la base et $\lambda$ la longueur d'onde d'observation. Modélisons par exemple l'enveloppe par une fonction Gaussienne de largeur à mi-hauteur $\theta_\mathrm{cse}$ contribuant au flux de l'étoile tel que $F_\mathrm{cse} = \alpha F_\star$, on obtient alors comme modèle de visibilité :
\begin{displaymath}
V^2 = \left( \frac{V_\star + \alpha V_\mathrm{cse}}{1 + \alpha} \right)^2 = \left( \frac{V_\star + \alpha e^{-a_1\theta_\mathrm{cse}^2}}{1 + \alpha} \right)^2 
\end{displaymath}

J'ai représenté sur la Fig.~\ref{image__bias_visibility} (fenêtre du haut) ce modèle de visibilité ainsi que celui d'un disque uniforme pour comparaison avec $\theta_\star = 1.5\,\mathrm{mas}$, $\theta_\mathrm{cse} = 3\,\theta_\star$ et un flux relatif de l'enveloppe $\alpha = 3\,\%$.
Dans la fenêtre du bas de la Fig.~\ref{image__bias_visibility} est représenté le biais sur l'estimation du diamètre du à la présence de l'enveloppe (Eq~\ref{equation__biais}) ainsi que l'incertitude $\sigma_\theta/\theta = \sigma_\mathrm{V^2}/V^2\,A_\mathrm{f}^{-1}$ sur la mesure de la visibilité (pour une précision $\sigma_\mathrm{V^2}/V^2 = 3\,\%$ typique de l'instrument FLUOR). On remarque que pour une base de 100\,m, le biais est de l'ordre de $10\,\%$ et n'est donc pas négligeable.

À l'issue des exemples précédents, on s'aperçoit alors qu'il est important d'étudier l'environnement des Céphéides. Quelle que soit la méthode utilisée,  la présence de ces enveloppes a un impact sur l'estimation des distances via les mesures de diamètre angulaire par photométrie ou interférométrie.

\section{Environnement des Céphéides: mécanismes de formation}

L'existence de ces enveloppes circumstellaires n'est pas bien comprise à ce jour. Leur taille de quelques rayons stellaires ($\sim2$--$3\,R_\star$) et leur faible luminosité ($\sim1$--$10\,\%$ du flux photosphérique) les rendent difficiles à détecter. L'utilisation des techniques de la haute résolution angulaire devient alors nécessaire.

Leur formation est encore une énigme. La théorie la plus envisagée et simple à concevoir est que l'étoile perdrait ou aurait perdu de la masse pendant son évolution via la combinaison de plusieurs mécanismes (vents radiatifs, pulsation, ...). L'hypothèse d'un vestige de la nébuleuse d'origine n'est pas envisageable compte tenu de leur faible dimension spatiale, excepté peut-être pour la Céphéides RS~Puppis dont sa nébuleuse a une taille angulaire de $\sim2\arcmin$ en bande $V$. Je reparlerai de cette étoile plus en détail dans le Chapitre~\ref{chapitre__imagerie_a_haute_resolution_spatiale_optique_adaptative_et_lucky_imaging}.

Une théorie probable serait donc la perte de masse des Céphéides, où l'accumulation de particules s'échappant de l'attraction gravitationnelle formerait l'enveloppe circumstellaire. Un aspect intéressant concerne certaines recherches récentes \citep{Neilson-2011-04,Neilson-2010-11} qui semblent indiquer que la perte de masse pourrait également avoir un rôle significatif dans le problème de la divergence de masse des Céphéides. Ce problème connu depuis des années concerne les masses estimées à partir de modèles d'évolution stellaire (utilisant $T_\mathrm{eff}$ et $L$) et celles estimées à partir de modèles de pulsation (utilisant $P$ et l'amplitude des pulsations), où les masses basées sur ce dernier modèle ont tendance à être plus petites que celles basées sur un modèle d'évolution. \citet{Neilson-2010-11} suggèrent que ce problème peut-être résolu en incluant dans les modèles un mécanisme de perte de masse lorsque l'étoile se trouve dans la bande d'instabilité.

Voyons les divers mécanismes favorisant l'éjection de matière et vérifions s'ils sont possibles dans le cas des Céphéides.

\subsection{La rotation}

La rotation d'une étoile ajoute une force supplémentaire (la force centrifuge) pouvant rompre son équilibre hydrostatique, c'est à dire l'équilibre entre la gravité (force vers l'intérieur de l'étoile) et les forces de pression (matière et rayonnement, vers l'extérieur). Sous l'effet de la rotation, l'étoile se déforme (s'aplatit) et éjecte de la matière par l'équateur si la force centrifuge domine la force de gravitation. Pour un point à la surface de l'étoile, on définit la vitesse critique $v_\mathrm{c}$ telle que :
\begin{displaymath}
\frac{GM}{R^2_\mathrm{eq}} = \frac{v_\mathrm{c}^2}{R_\mathrm{eq}} \quad \Longrightarrow \quad \frac{v_\mathrm{c}^2}{v_\mathrm{c_\odot}^2} = \left( \frac{M}{M_\odot} \right) \left( \frac{R}{R_\odot} \right)^{-1}
\end{displaymath}
où $R_\mathrm{eq}$ représente le rayon équatorial. Pour le Soleil par exemple, $v_\mathrm{c_\odot} \sim 430\,\mathrm{km\,s^{-1}}$ et sa vitesse de rotation mesurée est $v \sim 2\,\mathrm{km\,s^{-1}}$. Il tourne très lentement et n'éjecte donc pas de matière sous l'effet de sa rotation.

Les Céphéides ont également une faible vitesse de rotation. D'après un modèle simple appliqué à un échantillon de Céphéides, \citet{Nardetto-2006-07} ont trouvé que les courtes périodes tournent plus rapidement que les longues périodes mais que l'on a toujours $v\,\sin i\,<\,20\,\mathrm{km\,s^{-1}}$. On peut donner un ordre de grandeur de la vitesse critique de cet échantillon en estimant $R$ à partir d'une relation période--rayon \citep[$\log{(R/R_\odot)} = 0.750\,\log{P} + 1.075$,][]{Gieren-1998-03} et $M$ avec une relation période--masse--rayon \citep[$M/M_\odot = (40\,P)^{-1.49}\,(R/R_\odot)^{2.54}$,][]{Fricke-1972-02}. La Table~\ref{table__vitesse_critique} liste les vitesses critiques théoriques de ces Céphéides et l'on constate que $v\,\sin i\,\ll\,v_c$. L'éjection de matière sous l'effet de la rotation n'est donc pas un mécanisme envisageable pour les Céphéides.

Toutefois, si le progéniteur est un rotateur rapide, la formation d'une enveloppe via l'éjection de matière pourrait se faire avant le stade Céphéide.

\begin{table}[!t]
\centering
\begin{tabular}{cccccccc} 
\hline
\hline
Noms	  			& $\log{P}$	& 	$v\,\sin i$ 		& $M$ 				& $R$				& $v_c$  				& $\log{(L/L_\odot)}$	&	$\Gamma_e$	\\
						&					&	($\mathrm{km\,s^{-1}}$)	&	($M_\odot$)	&	($R_\odot$)	& ($\mathrm{km\,s^{-1}}$)	&	&	\\
\hline
R~Tra				&  0.530		& 	15					&	3.7				&	29.7				&	149.9				&	2.992	&	0.008	\\
S~Cru				&  0.671		& 	10					&	4.2				&	37.9				&	143.6				&	3.156	&	0.010	\\
Y~Sgr				&  0.761		& 	16					&	4.6				&	44.3				&	137.6				&	3.230	&	0.011	\\
$\beta$~Dor		&  0.993		& 	6						&	5.7				&	66.0				&	126.0				&	3.528	&	0.018	\\
$\zeta$~Gem	&  1.006		& 	6						&	5.8				&	67.6				&	124.5				&	3.544	&	0.018	\\
Y~Oph				&  1.234		& 	4						&	7.2				&	100.1			&	114.6				&	3.808	&	0.028	\\
RZ~Vel				&  1.310		& 	3						&	7.7				&	114.1			&	111.2				&	3.896	&	0.031	\\
$\ell$~Car		&  1.551		& 	7						&	9.7				&	173.0			&	100.7				&	4.176	&	0.046	\\
RS~Pup				&  1.617		& 	$<\,1$				&	10.3				&	193.9			&	98.2					&	4.256	&	0.053	\\
\hline
\end{tabular}
\caption[Paramètres physiques d'un échantillon de Céphéides]{\textbf{Paramètres physiques d'un échantillon de Céphéides} : $v\,\sin i$ de quelques Céphéides \citep{Nardetto-2006-07} comparés à la vitesse de rotation critique théorique, ainsi que la facteur d'Eddington.}
\label{table__vitesse_critique}
\end{table}

\subsection{Les vents radiatifs}
La perte de masse liée aux vents stellaires joue un rôle important pour les étoiles chaudes. Or les Céphéides sont des étoiles plutôt froides avec une température effective $4000\,\lesssim\,T_\mathrm{eff}\,\lesssim\,6500\,\mathrm{K}$, les vents radiatifs n'interviennent donc probablement pas dans notre cas. 

Faisons cependant une analyse simple. La perte de masse due aux vents stellaires intervient lorsque la force créée par la pression de radiation devient plus forte que la force de gravitation. En faisant l'hypothèse que la force radiative est liée à la diffusion des photons par les électrons libres, on a :
\begin{displaymath}
F_\mathrm{rad} = \frac{\sigma_T L_\star}{4\pi r^2c}
\end{displaymath}

Il est intéressant d'introduire le rapport entre cette force radiative $F_\mathrm{rad}$ et la force gravitationnelle $F_\mathrm{grav} = Gm_pM_\star/r^2$, appelé facteur d'Eddington :
\begin{displaymath}
\Gamma_e = \frac{F_\mathrm{rad}}{F_\mathrm{grav}} \simeq 3\times10^{-5}\ \left(\frac{L_\star}{L_\odot}\right) \left(\frac{M_\star}{M_\odot}\right)^{-1}
\end{displaymath}

Si $\Gamma_e\,<\,1$, la pression de radiation n'est pas assez forte pour former un vent stellaire. Le cas $\Gamma_e\,=\,1$ correspond au point critique où l'on définit la luminosité limite dite d'Eddington ($L_\mathrm{edd}$). Une étoile dont la luminosité est supérieure à $L_\mathrm{edd}$ expulsera de la matière sous l'effet de la pression de radiation.

En reprenant l'échantillon de Céphéides précédent et en estimant leur luminosité à partir de la magnitude absolue\footnote{d'après le David Dunlap Observatory Catalogue of Galactic Classical Cepheids : \url{http://www.astro.utoronto.ca/DDO/research/cepheids/cepheids.html}} \citep{Fernie-1995-01}, nous pouvons calculer $\Gamma_e$. Ces valeurs sont reportées dans la Table~\ref{table__vitesse_critique}. On constate que $\Gamma_e\,\ll\,1$ pour toutes les étoiles de l'échantillon. Une perte de masse par le seul mécanisme des vents radiatifs est donc exclue pour les Céphéides. Cependant un couplage avec un autre mécanisme d'éjection (pulsation) est tout à fait envisageable, le gaz déjà en déplacement sous l'effet du premier mécanisme peut ensuite être accéléré vers l'extérieur par la pression de radiation.

Les oscillations de la photosphère pourraient permettre la formation des environnements circumstellaires. Cependant l'amplitude des pulsations doit être importante pour vaincre à elle seul les effets de la gravité. Les ondes de choc se produisant dans l'atmosphère peuvent contribuer à l'éjection de masse. Ces chocs peuvent émerger par exemple lorsqu'une couche supérieure de la photosphère en contraction rencontre une couche inférieure toujours en expansion. 

Des oscillations non-radiales peuvent également contribuer à la perte de masse de manière non-uniforme. Cependant il n'existe pas de preuves observationnelles sur de telles oscillations pour les Céphéides \citep[d'après][ certaines Céphéides du LMC seraient excitées par des modes non-radiaux, mais cela n'a pas été confirmé]{Moskalik-2004-05}. Elles sont donc toujours considérées comme étant soumises à des oscillations radiales. \citet{Mulet-Marquis-2009-} ont reconsidéré la théorie des pulsations non-radiales et arrivent à la conclusion que pour des étoiles de $T_\mathrm{eff}\,>\,6200\,\mathrm{K}$, seules des oscillations non-radiales devraient exister.

L'effet de la pulsation et des chocs couplées à la pression de radiation est le phénomène de perte de masse le plus envisagé actuellement \citep{Neilson-2008-09}.  \citet{Neilson-2009-09,Neilson-2011-04} ont montré que la perte de masse via ces processus est une explication raisonnable à l'existence des enveloppes autour des Céphéides.

\subsection{Le magnétisme}
La présence d'une activité magnétique est assez incertaine pour les étoiles Céphéides. Si le champ est assez fort, il peut avoir une influence sur l'éjection de matière, mais cela n'a toujours pas été détecté, encourageant donc l'hypothèse d'un faible champ magnétique. Très peu de données existent à ce sujet et les mesures existantes pointent vers une activité magnétique faible, c'est à dire $<\,20\,\mathrm{G}$ \citep{Borra-1981-07,Borra-1984-09,Wade-2002-09}. 

Le magnétisme semble donc peu probable comme mécanisme d'éjection de matière mais peut cependant intervenir dans la réorganisation de la matière éjectée par l'étoile. Toutefois, le manque de résolution spatiale actuel ne nous permet pas de vérifier cette possibilité.

\subsection{La binarité}

Le phénomène d'étoiles doubles ou multiples est très commun dans l'univers. Plus de $60\,\%$ des Céphéides classiques appartiennent à un système binaire ou multiple \citep{Szabados-1995-}. Bien que souvent négligé, l'effet dû à la présence d'un compagnon sur les mesures (photométrie, spectroscopie, ...) ne doit pas être oublié. L'une des difficultés de leur détection est due au fait que les Céphéides sont assez éloignées et de surcroît très brillantes par rapport à leurs compagnons. De plus la séparation entre les deux objets est en général assez petite, impliquant que l'imagerie directe est impossible pour le moment. Par exemple pour FF~Aql et W~Sgr la séparation est $a = 12.8\,\pm\,0.9\,\mathrm{mas}$ et $a = 12.9\,\pm\,0.3\,\mathrm{mas}$ respectivement \citep{Benedict-2007-04}.

La formation d'une enveloppe circumstellaire par binarité n'est possible que dans certains cas. Si l'une des étoiles remplit entièrement son lobe de Roche, alors de la matière est transférée via le point de Lagrange $L_1$ vers l'autre composante plus massive (voir Annexe~\ref{section__potentiel_de_gravitation_dans_un_systeme_binaire}). Comme le système est en rotation autour du centre de gravité, la matière "tombe" de façon spirale et forme alors un disque d'accrétion. Dans notre cas, cette hypothèse nécessite une composante compacte arrachant la matière de la Céphéide, mais à ce jour il n'existe pas d'indices observationnels pour confirmer cela. De plus, les systèmes binaires connus impliquant des Céphéides semblent être assez  éloignés l'un de l'autre pour être des systèmes détachés, rendant cette hypothèse peu probable \citep[][pour un exemple de Céphéide double dans un système détaché dans le LMC]{Pietrzynski-2010-12}.

Une autre possibilité serait liée à l'excentricité de l'orbite faisant qu'à chaque passage au périastre, le compagnon arracherait un peu de matière à la Céphéide. Comme le système n'est pas résolu, il est difficile de confirmer cette hypothèse.

\section{Objectifs de la thèse : étude des enveloppes circumstellaires}
\label{section__objectifs_de_la_thèse}

Le statut de "chandelle standard" des Céphéides donne une importance supplémentaire à l'étude de leurs enveloppes circumstellaires. Le but de cette thèse est d'apporter des solutions observationnelles pour caractériser les paramètres physiques, identifier un éventuel rapport avec la pulsation et estimer leur influence sur l'étalonnage de la relation période--luminosité.

C'est un sujet nouveau dans le cadre des étoiles Céphéides puisque ces environnements circumstellaires n'ont été détectés que récemment \citep{Kervella-2006-03}. Cela m'a donc permis d'orienter mes recherches vers plusieurs techniques d'observation et diverses gammes de longueur d'onde. À la vue des premiers résultats montrant une taille de quelques rayons stellaires, les techniques de la haute résolution angulaire semblent nécessaires à l'étude de ces environnements. De plus, pour réduire le contraste entre l'étoile et l'enveloppe, des longueurs d'onde supérieures à $2\,\mu\mathrm{m}$ sont préférables.

Dans un premier temps je parlerai d'observations de l'enveloppe étendue de la Céphéide RS~Puppis en imagerie par optique adaptative avec \emph{NACO}, couplée à un mode d'observation dit "cube". Cette étude m'a permis de déduire le rapport de flux entre l'enveloppe et la photosphère de l'étoile dans deux bandes étroites centrées sur $\lambda = 2.18\,\mu\mathrm{m}$ et $\lambda = 1.64\,\mu\mathrm{m}$. De plus grâce au mode cube, j'ai également pu effectuer une étude statistique du bruit de speckle me permettant d'étudier une éventuelle asymétrie.

Dans un second temps, j'utilise des données \emph{VISIR} pour étudier la distribution d'énergie spectrale d'un échantillon de Céphéides. Ces images, qui sont limitées par la diffraction, m'ont permis d'effectuer une photométrie précise dans la bande $N$ et de mettre en évidence un excès infrarouge lié à la présence d'une composante circumstellaire. D'autre part en appliquant une analyse de Fourier j'ai montré que certaines de ces composantes sont résolues.

Je poursuis ensuite l'étude de ces enveloppes en passant à une plus haute résolution spatiale grâce à l'interférométrie. J'ai exploré la bande $K\arcmin$ avec l'instrument de recombinaison \emph{FLUOR} pour un échantillon de Céphéides. Les données étant récentes (dernières observations en août 2011), elles sont toujours en cours de réduction et d'analyse, et je ne présenterai donc que certains premiers résultats. Grâce à de nouvelles données sur l'étoile Y~Oph, j'ai approfondi l'étude de son enveloppe circumstellaire. En utilisant un modèle d'étoile entourée d'une couronne sphérique, j'ai déterminé une taille angulaire de $4.54\,\pm\,1.13$\,mas et une profondeur optique de $0.011\,\pm\,0.006$. Pour deux autres Céphéides, U~Vul et S~Sge, j'ai appliqué la méthode de Baade--Wesselink afin d'estimer une première mesure directe de leur distance. J'ai trouvé une distance de $d = 647\,\pm\,45$\,pc et $d = 661\,\pm\,57$\,pc , respectivement pour U~Vul et S~Sge, ainsi qu'un rayon linéaire moyen $R = 53.4\,\pm\,3.7\,R_\odot$ et $R = 57.5\,\pm\,4.9\,R_\odot$ respectivement. L'étude interférométrique se conclut par une estimation du diamètre angulaire de la Céphéide R~Sct.

Je terminerai ce manuscrit par des conclusions liées à l'application des diverses techniques d'observations et j'exposerai les perspectives possibles sur les Céphéides et leur enveloppe circumstellaire.

\cleardoublepage     

\pagestyle{fancy}
\fancyhf{}
\lhead[\nouppercase{\emph{\thepage}}]{\nouppercase{\emph{\rightmark}}}
\rhead[\nouppercase{\emph{\leftmark}}]{\nouppercase{\emph{\thepage}}}
\newpage

\chapter[Imagerie à haute résolution spatiale: Optique adaptative et "lucky-imaging"]{\emph{Imagerie à haute résolution spatiale: Optique adaptative et "lucky-imaging"}}
\label{chapitre__imagerie_a_haute_resolution_spatiale_optique_adaptative_et_lucky_imaging}

\thispagestyle{empty}

\vspace*{-1cm}

\refbleu
\textcolor{bleu_chapitre}{\minitoc}
\refnoir

\section{Introduction}

\malettrine{L}{'}observation depuis l'espace est idéale en astronomie, mais malheureusement un télescope spatial coûte cher et on observe donc le plus souvent depuis le sol. Le handicap pour un télescope terrestre est la turbulence atmosphérique qui a pour effet de limiter la résolution spatiale, rendant un télescope de 10\,m pas plus efficace qu'un télescope de quelques centimètres de diamètre. Ce phénomène de turbulence est observable à l'\oe il nu lorsqu'on contemple le ciel nocturne, et se traduit par une impression de scintillement des étoiles.

La turbulence dans l'atmosphère est liée aux mouvements de masses d'air de différentes températures. Ces mouvements aléatoires produisent des variations d'indice de réfraction de l'air qui altèreront légèrement la trajectoire des photons, provoquant une dégradation de l'image. On surmonte ce problème de dégradation en corrigeant les fluctuations avec un système d'optique adaptative. L'idée originelle d'un tel système provient de \citet{Babcock-1953-10} qui affirma qu'il était possible de corriger la déformation d'une image si l'on connaît la manière dont elle est déformée. Ce n'est que vers la fin des années 80 que le premier système d'optique adaptative (OA) appliqué à l'astronomie voit le jour. Le premier résultat prometteur est présenté par \citet{Rousset-1990-04} et rapportait la résolution d'un système binaire.

Une onde plane entrant dans l'atmosphère se voit déformée après sa traversée. Un système d'OA mesure cette déformation causée par la turbulence et la corrige en temps réel pour obtenir une meilleure qualité d'image. Les éléments clefs d'un tel système sont l'analyseur de front d'onde, le miroir déformable et le calculateur. Bien que performants de nos jours, ces sous-systèmes ont leurs limitations. Ils dépendent fortement de la vitesse caractéristique de la turbulence et actuellement, la correction n'est possible que dans le visible et le proche infrarouge, où les effets de la turbulence sont moindres. La source observée doit également être brillante pour permettre une bonne correction, ce qui n'est malheureusement pas souvent le cas. On alors recours à une étoile prise comme référence (naturelle ou artificielle) pour mesurer la déformation du front d'onde.

Bien qu'efficace, une OA n'est pas parfaite et la correction de la déformation n'est pas toujours optimale. Les variations temporelles de la turbulence (à une échelle de quelques millisecondes) produisent des variations de qualité de la correction. Ainsi le front d'onde sera d'autant mieux corrigé que la turbulence sera faible. Lors d'une observation à travers une atmosphère turbulente, cette variation est moyennée sur le temps d'acquisition, menant à une résolution spatiale inférieure à la limite de résolution du télescope. Ceci peut être amélioré en enregistrant des images de temps d'exposition assez court (pas trop quand même pour permettre à l'OA de fonctionner), dans le but de "figer" la turbulence atmosphérique. Ce procédé est connu sans utilisation d'une OA comme "lucky-imaging". Avec cette technique, nous pouvons sélectionner les images les moins altérées par la turbulence. Il est également possible de combiner cette technique à un système d'OA, dont on peut choisir les meilleures images corrigées. La résolution atteinte est proche de la limite de résolution du télescope et meilleure qu'avec une image longue pose avec OA.

En raison de sa petite taille angulaire, une haute résolution spatiale est nécessaire pour obtenir une image de l'enveloppe circumstellaire autour d'une Céphéide. Une image dégradée par la turbulence peut facilement empêcher la détection d'une enveloppe. D'un autre coté, la correction partielle d'une OA peut également entraîner la formation d'un "halo" qui peut, pour des scientifiques moins expérimentés dans ce domaine, facilement être confondu avec un environnement circumstellaire.

J'ai été amené durant mon travail de thèse à réduire les données de l'instrument \emph{NACO} du VLT. Cet instrument est doté d'une optique adaptative \citep[NAOS,][]{Rousset-2003-02} et permet plusieurs modes d'observations. Mes données concernent la Céphéide RS~Pup, et ont été obtenues en mode "cube" aux longueurs d'onde de $1.64\,\mu\mathrm{m}$ et $2.18\,\mu\mathrm{m}$. Dans un premier temps, je parlerai d'une manière non-exhaustive de l'optique adaptative, en prenant comme exemple le système NAOS, le but étant d'exposer les paramètres fondamentaux et le principe de fonctionnement. La technique de "lucky-imaging" sera ensuite introduite et testée sur quelques exemples. La Section~\ref{observation_de_rs_puppis_avec_naco} concerne l'application simultanée de l'OA et du "lucky-imaging" à la Céphéide RS~Pup. Je développerai mes méthodes d'analyses et mes résultats qui ont donné lieu à une publication dans la revue Astronomy \& Astrophysics (Annexes~\ref{article__naco}).

\section{Principe de l'optique adaptative}

Dans cette partie j'expose brièvement les bases de l'optique adaptative. Cette section n'a pas pour but de redémontrer la théorie de la turbulence, mais plutôt de se faire une idée sur le principe général et les paramètres de base. Je renvoie le lecteur vers des revues comme \citet{Roddier-2004-11} ou \citet{Beckers-1993-} pour une théorie plus exhaustive.

\subsection{Fonction d'étalement de point idéale}

\begin{figure}[!p]
	\resizebox{\hsize}{!}{
  		\centering\includegraphics[width= .9\linewidth]{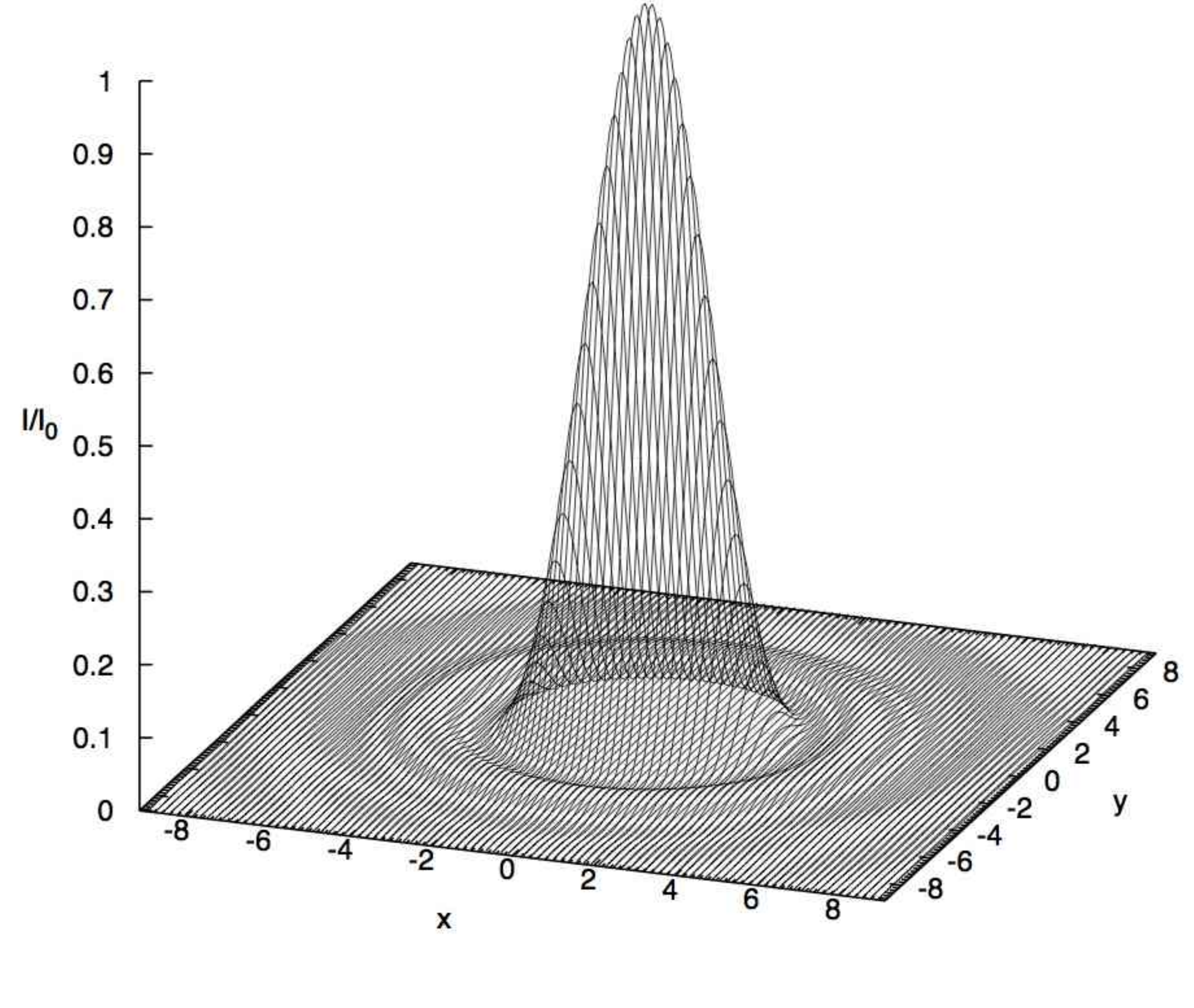} \hspace{.2cm}
  		\centering\includegraphics[width= .78\linewidth]{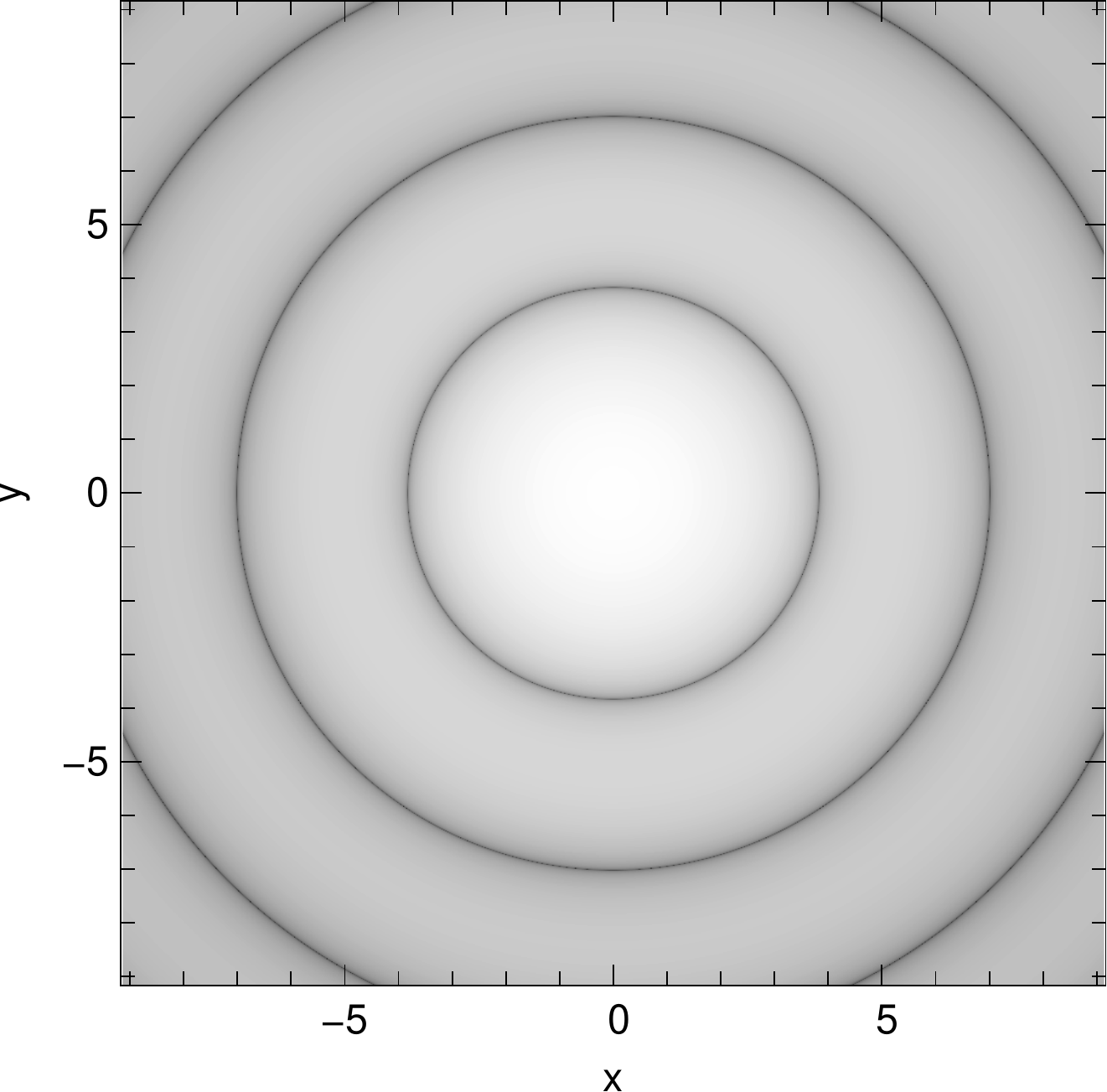} \hspace{.2cm}
  		\centering\includegraphics{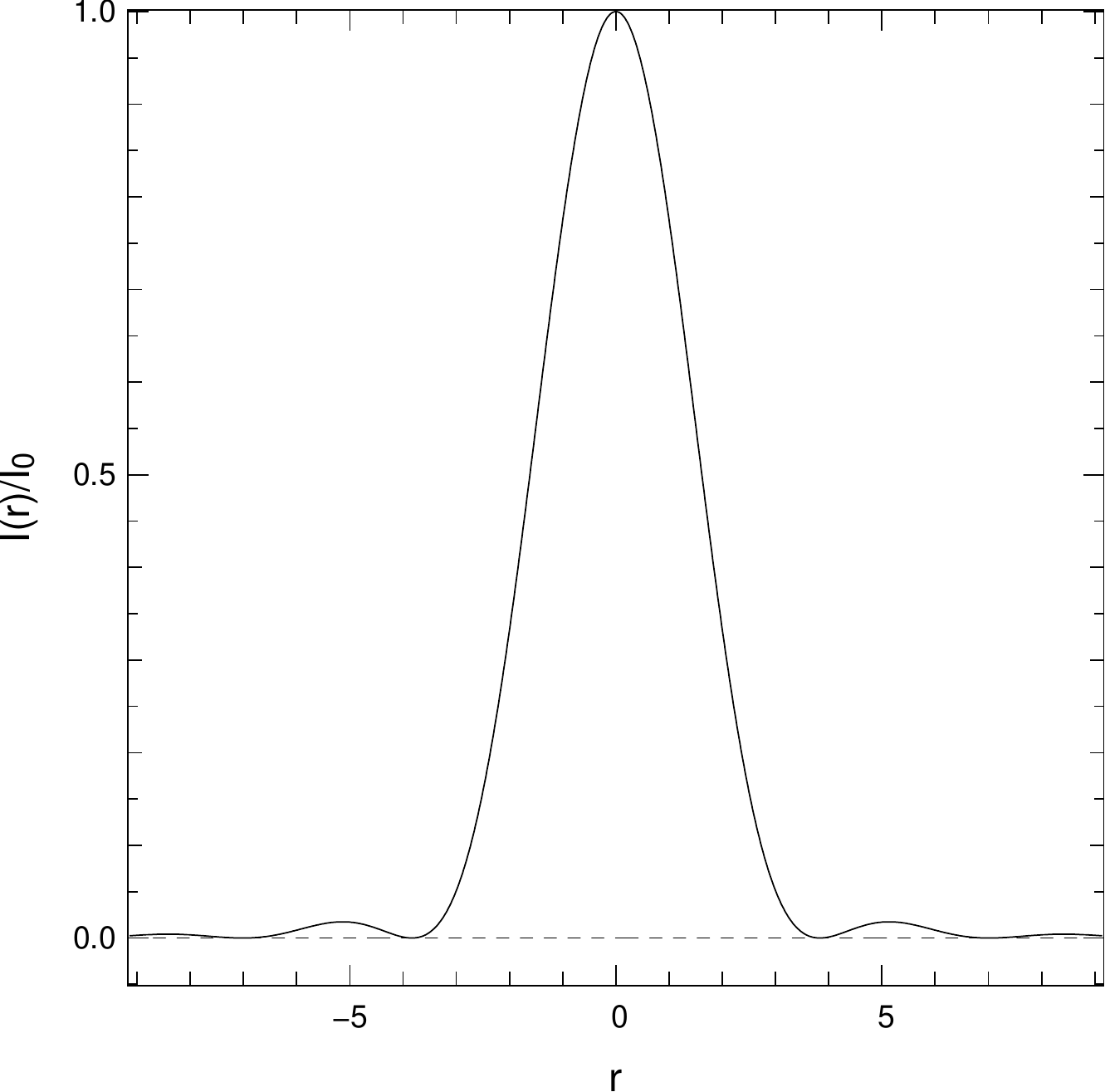} \hspace{.2cm}
	}
	\caption[Fonction d'étalement de point idéale]{\textbf{Fonction d'étalement de point idéale} : réponse impulsionnelle d'un télescope parfait en 3, 2 et 1 dimension. Le graphe à 2 dimensions a été tracé en échelle logarithmique afin de mieux discerner les anneaux.}
  	\label{image__airy}
\end{figure}

La résolution d'un télescope est imposée par la nature ondulatoire de la lumière. L'image d'un point source est donnée par la théorie de la diffraction et est nommée fonction d'étalement de point (FEP, ou PSF en anglais). Considérons une onde plane arrivant sur un télescope composée d'un miroir circulaire, l'intensité observée au foyer, appelée figure de diffraction, est :
\begin{displaymath}
I(r) = I_0 \left( \frac{2J_1(r)}{r} \right)^2
\end{displaymath}
où $r = \pi \sqrt{x^2 + y^2} D/(\lambda f)$, $(x, y)$ les coordonnées au foyer, $D$ le diamètre du télescope, $\lambda$ la longueur d'onde et $f$ la distance focale. La FEP, également appelée fonction d'Airy, est tracée sur la Fig.~\ref{image__airy} pour un télescope de diamètre $D = 8\,\mathrm{m}$, $\lambda = 2.2\,\mu\mathrm{m}$ et $f = 100\,\mathrm{m}$.

\begin{figure}[!p]
  		\centering\includegraphics[width= .25\linewidth]{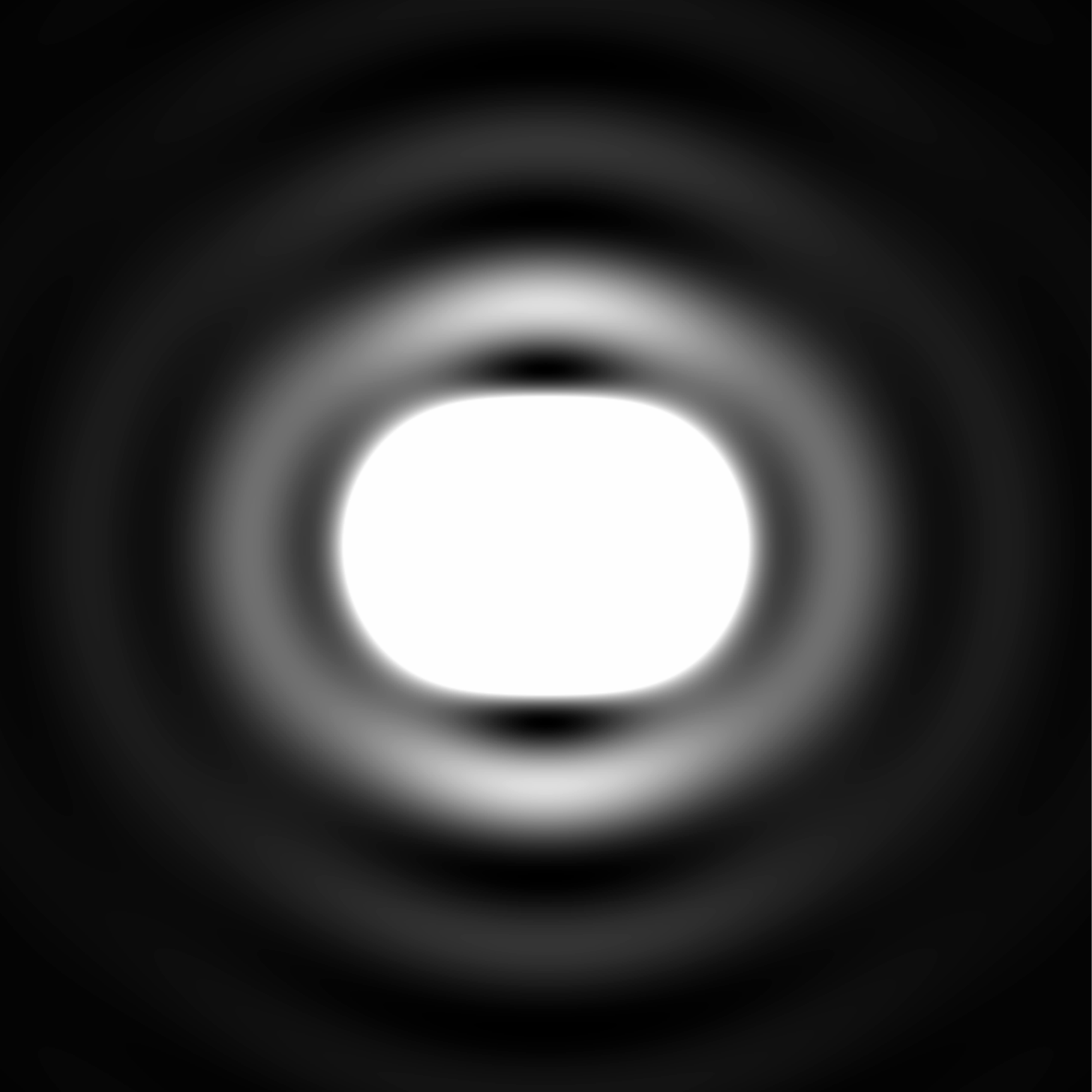} \hspace{.5cm}
  		\centering\includegraphics[width= .25\linewidth]{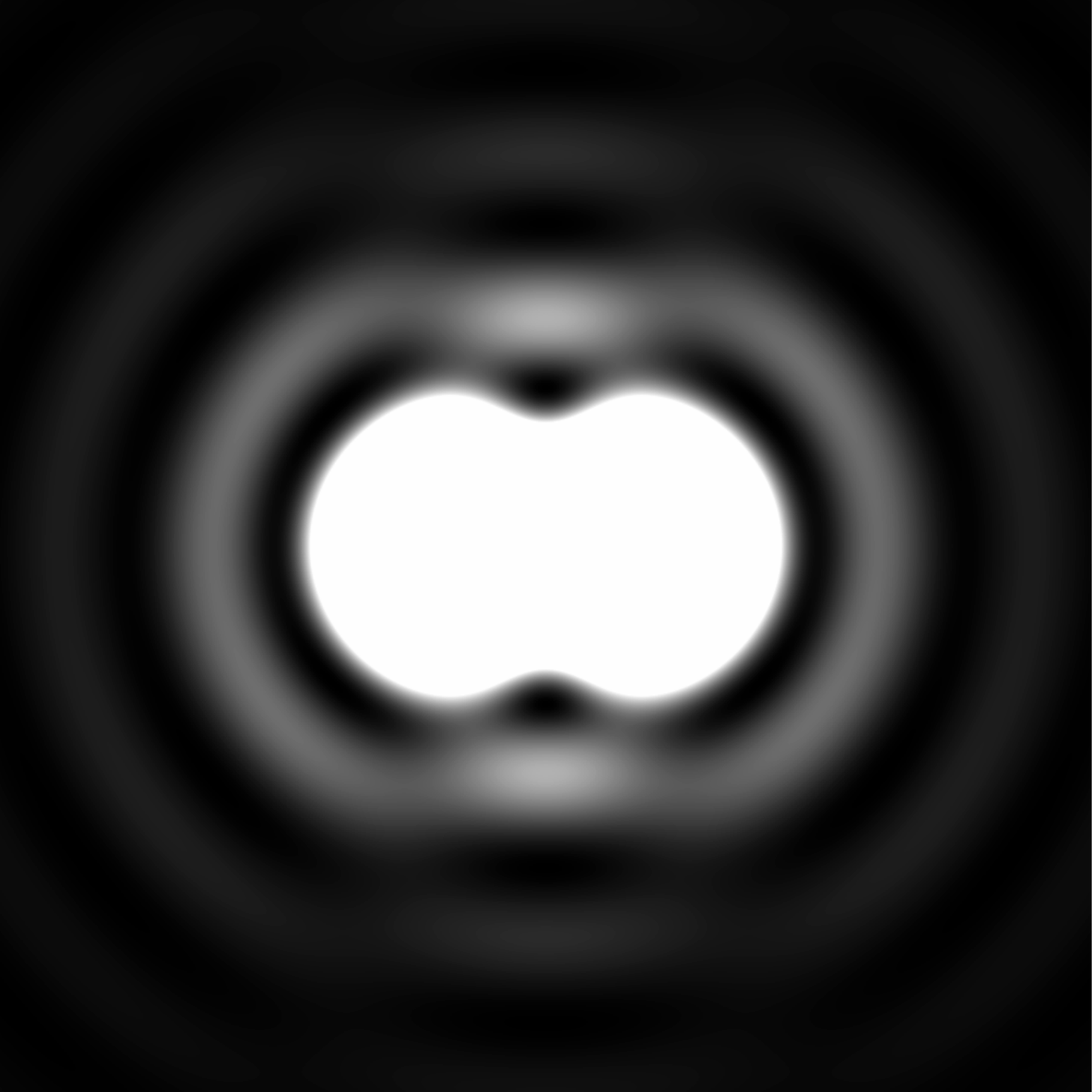} \hspace{.5cm}
  		\centering\includegraphics[width= .25\linewidth]{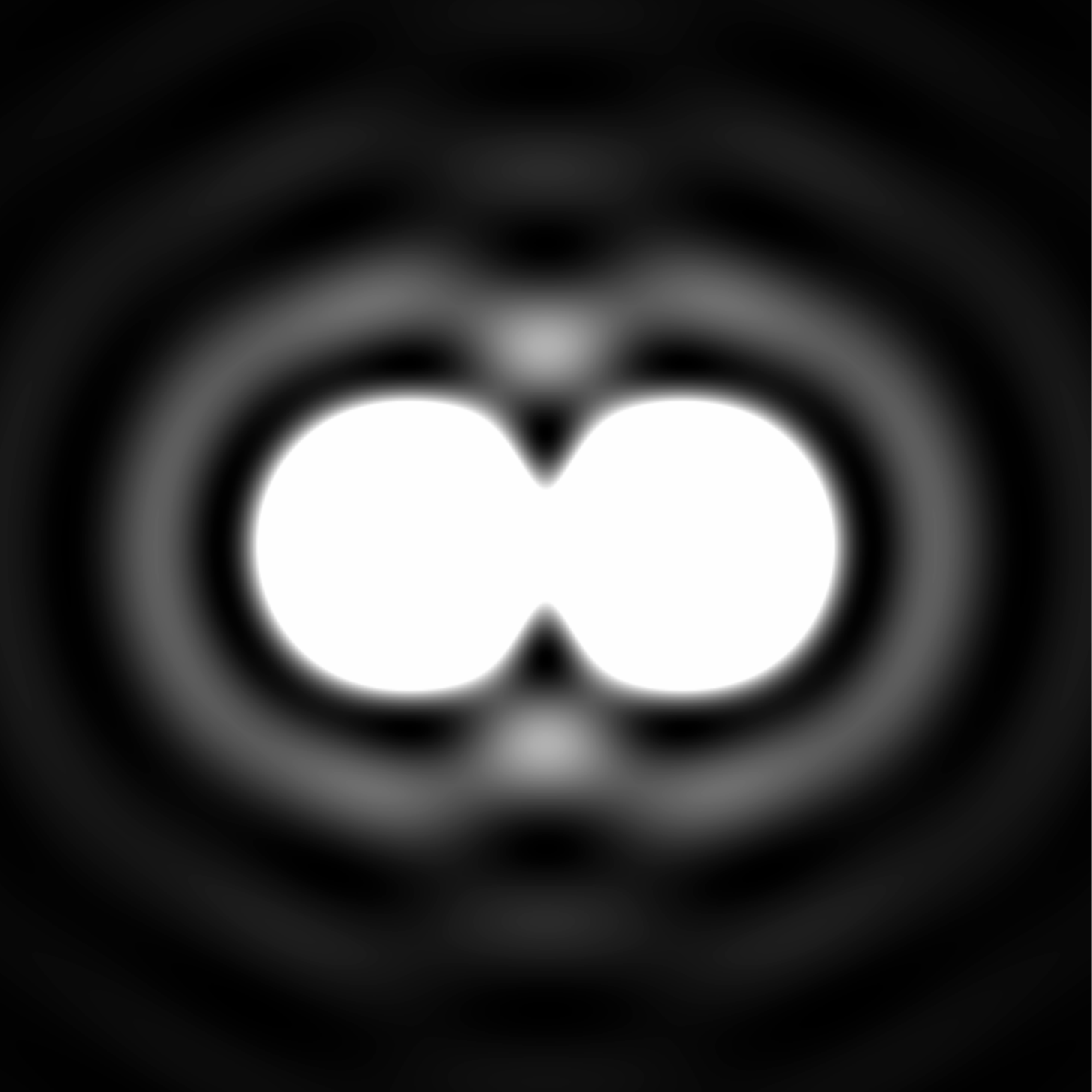}

  		\centering\includegraphics[width= .25\linewidth]{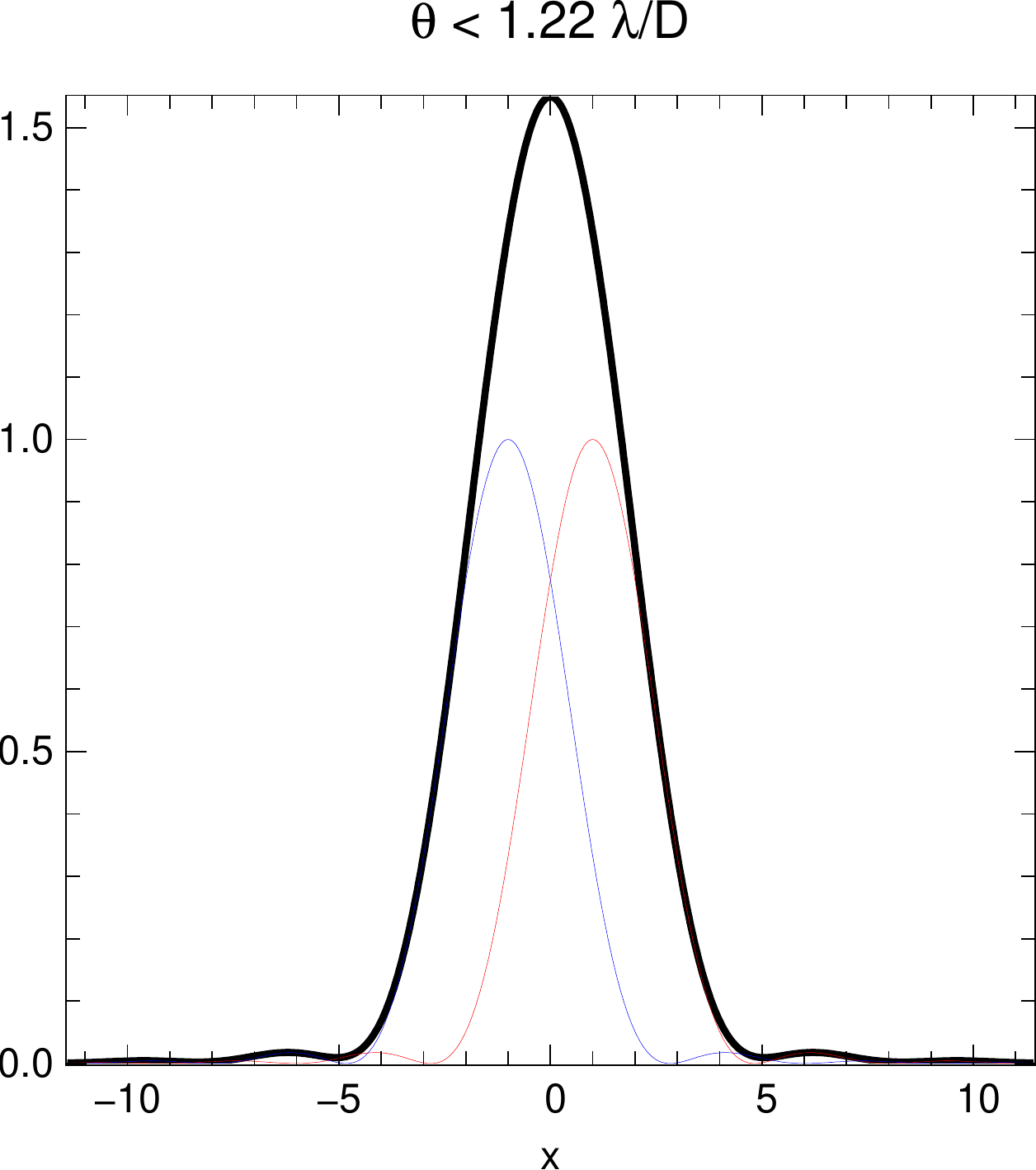} \hspace{.5cm}
  		\centering\includegraphics[width= .25\linewidth]{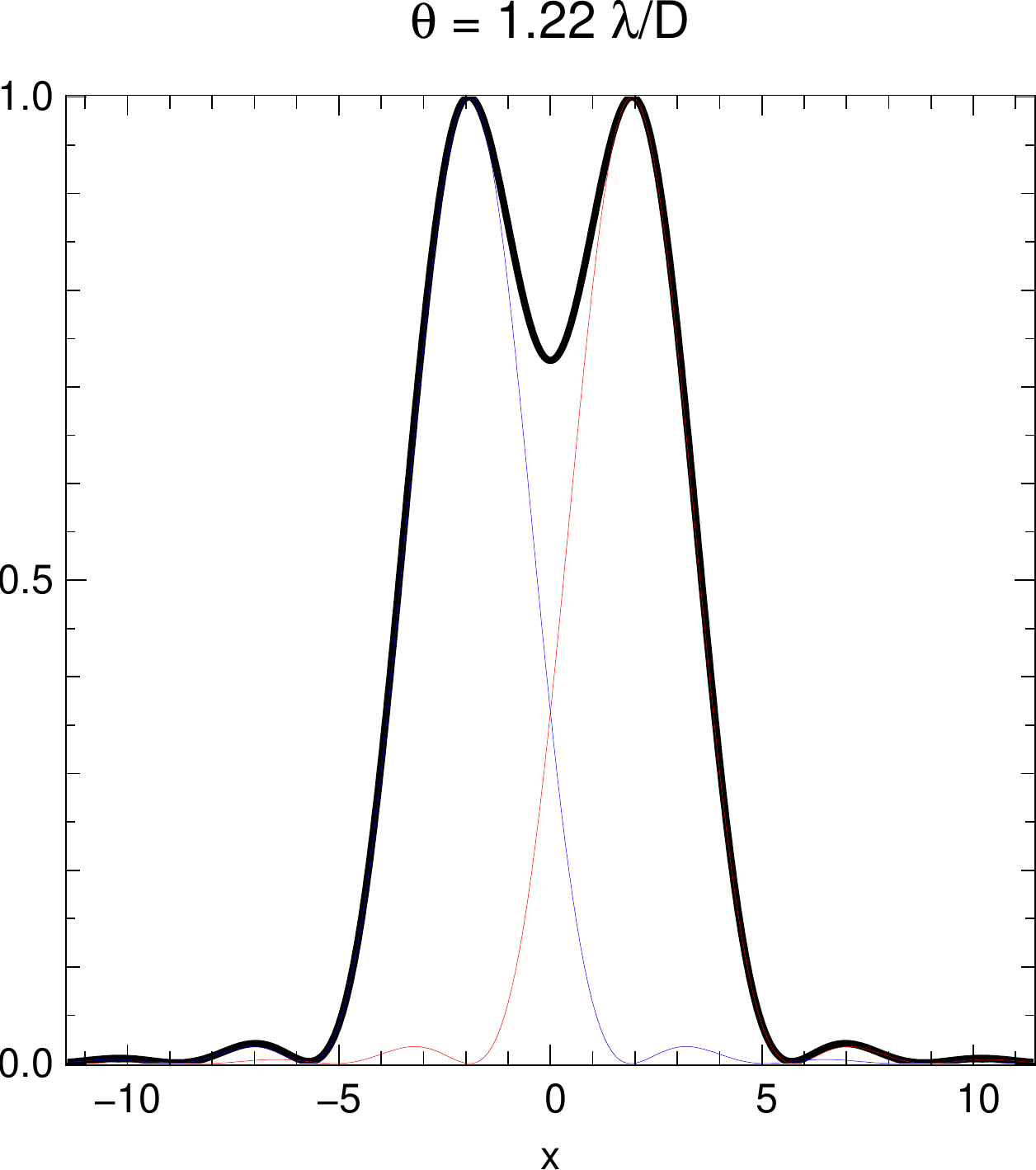} \hspace{.5cm}
  		\centering\includegraphics[width= .25\linewidth]{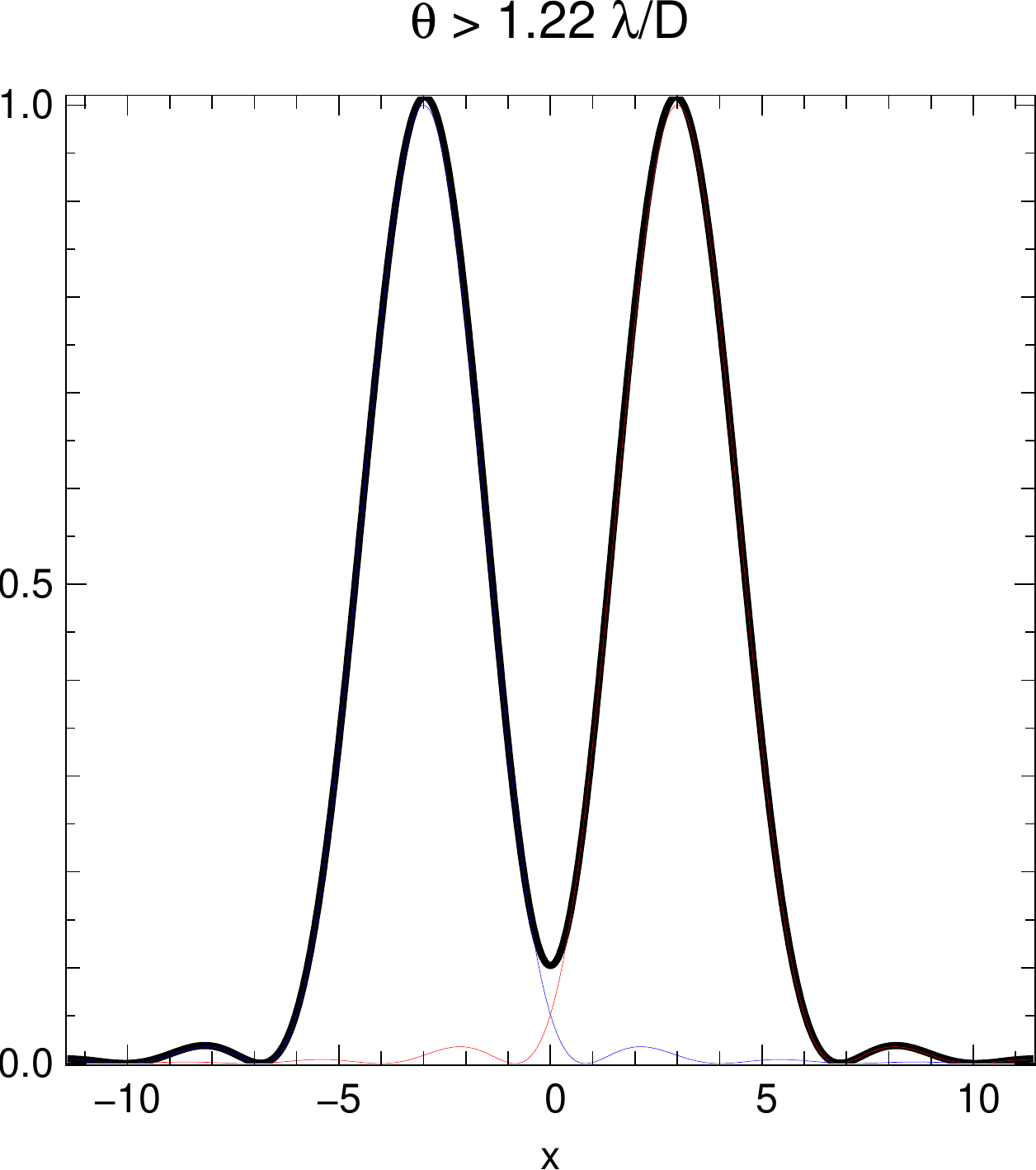}
	\caption[Pouvoir de résolution]{\textbf{Pouvoir de résolution} : images de deux points sources séparés d'un angle $\theta$. Dans l'image de gauche, les deux sources ne sont pas discernables.}
  	\label{image__resolution}
\end{figure}

La capacité d'un télescope à imager à son foyer deux points sources séparées dépend de son diamètre, de la longueur d'onde d'observation et de la séparation entre ces deux points. Le premier anneau de la FEP est défini comme la limite de résolution d'un télescope par Lord Rayleigh. La résolution angulaire en radian a pour expression $R = 1.22 \lambda/D$. Un objet de taille  $\theta\,\geqslant\,R$ sera considéré comme résolu par le télescope. Différents cas sont présentés sur la Fig.~\ref{image__resolution} pour deux points sources séparées d'un angle $\theta$.

\subsection{Fonction d'étalement de point réelle}

\begin{figure}[!p]
	\resizebox{\hsize}{!}{\centering\includegraphics{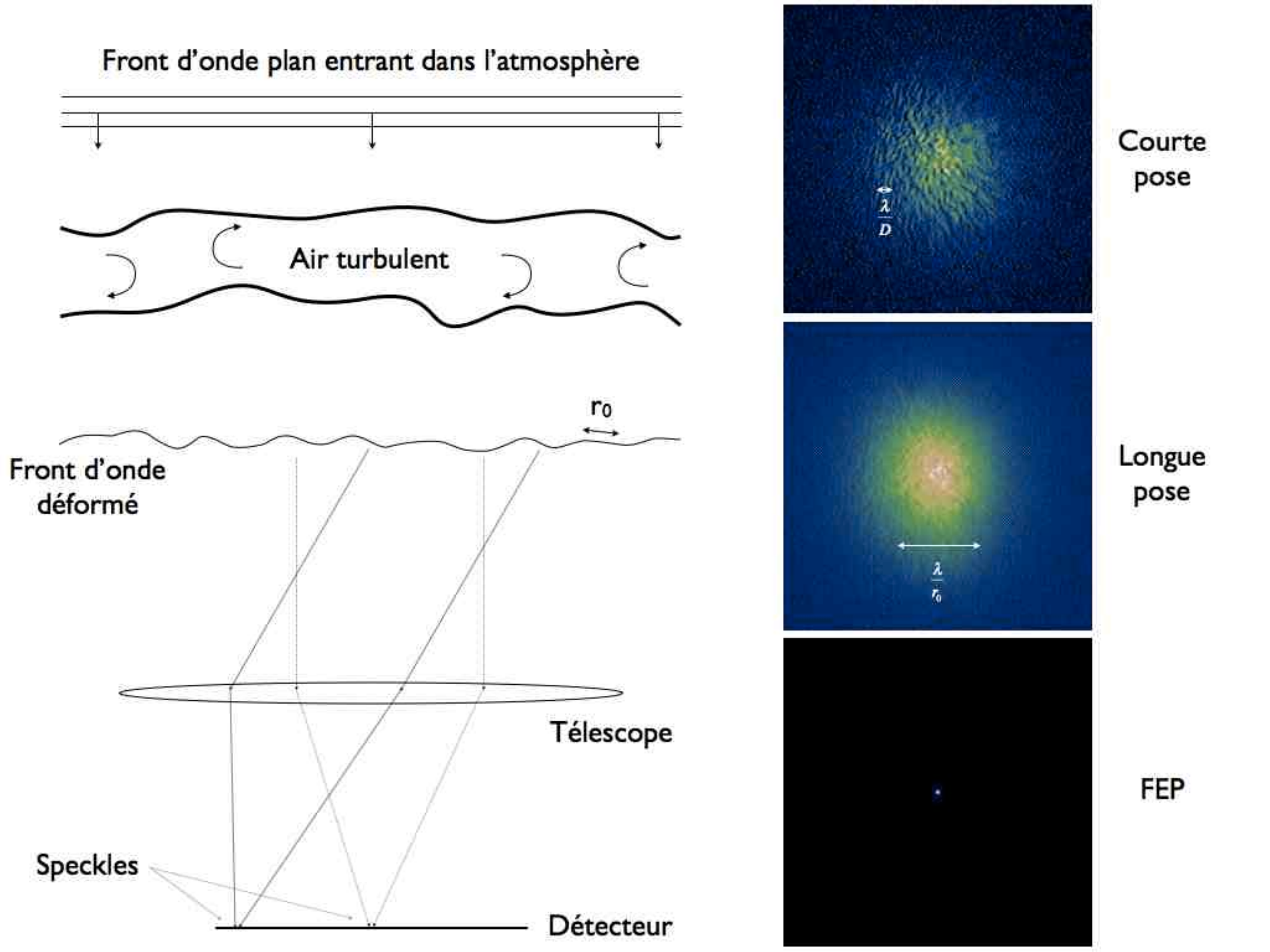}}
	\caption[Déformation du front d'onde]{\textbf{Déformation du front d'onde} : schéma du front d'onde déformé à son passage dans l'atmosphère. Lors d'une longue pose, l'image est la superposition de multiples fronts d'onde. Lors d'une courte pose, la turbulence est presque "gelée", moins d'ondes se superposent.}
  	\label{image__deformation}
\end{figure}

Malheureusement depuis le sol, il est difficile d'atteindre cette résolution à cause de la turbulence atmosphérique. Le front d'onde est déformé par sa traversée de l'atmosphère et n'est plus plan. La turbulence dévie légèrement les rayons lumineux, comme le montre le schéma de la Fig.~\ref{image__deformation}. Il en résulte que chaque point de l'image est la superposition de plusieurs fronts d'onde individuels formant des images distinctes (appelée tavelures ou speckles en anglais). D'un point de vue mathématique, la turbulence se traduit par des variations de phase de l'onde. Plus ces variations sont grandes et plus l'image est dégradée.

La distance caractéristique $r_0$ sur laquelle l'onde incidente reste suffisamment en phase pour produire une figure de diffraction est appelée paramètre de Fried. Pour des bons sites d'observations, ce paramètre est de l'ordre de $5$--$20\,\mathrm{cm}$ dans le visible. La largeur de la FEP, également appelé "seeing", sera alors donnée par $R \sim \lambda/r_0$. Un télescope de taille $D \leqslant r_0$ sera donc limité par la diffraction alors que si $D > r_0$ il sera limité par la turbulence, comme le montrent les images de la Fig.~\ref{image__deformation} pour une longue et courte pose.

\begin{figure}[!p]
	\resizebox{\hsize}{!}{\centering\includegraphics{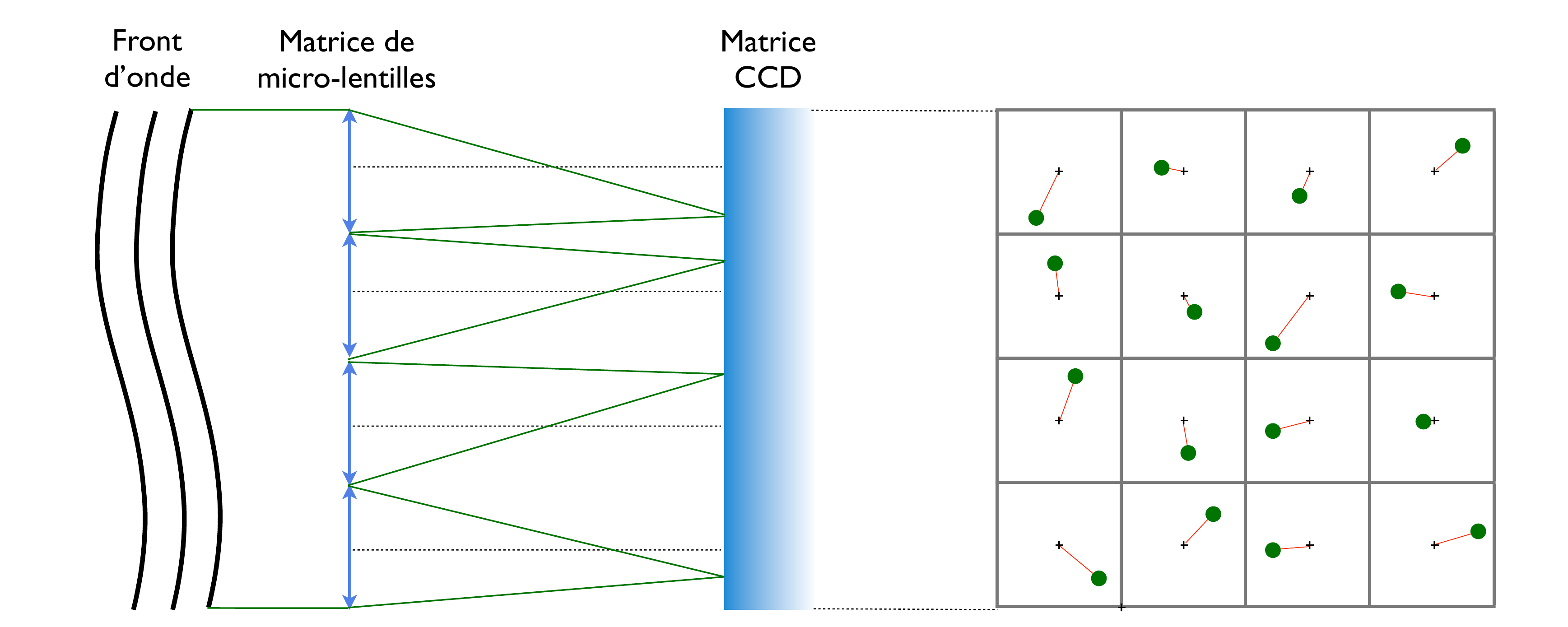}}
	\caption[Principe du Shack-Hartmann]{\textbf{Principe du Shack-Hartmann} : schéma de fonctionnement d'un analyseur de type Shack-Hartmann.}
  	\label{image__SH}
\end{figure}

Dans le modèle de Kolmogorov qui décrit les perturbations du front d'onde introduites par l'atmosphère, ce paramètre a pour expression :
\begin{displaymath}
r_0 = \left( 0.423 \frac{4\pi ^2}{\lambda^2 \cos{z}} \int_0^\infty C^2_n(h) dh \right)^{-\frac{3}{5}} \propto \lambda^{\frac{6}{5}}
\end{displaymath}
où $z$ est la distance angulaire de la source par rapport au zénith et $C^2_n(h)$ le profil de turbulence en fonction de la hauteur $h$. On remarque que $r_0$ augmente avec la longueur d'onde (pour $r_0(0.5\,\mathrm{\mu m}) \sim 10\,\mathrm{cm}$, on aura $r_0(2.2\,\mathrm{\mu m}) \sim 60\,\mathrm{cm}$).

Le seeing est l'un des paramètres le plus important lors des observations car c'est un indicateur de la qualité du ciel. Les meilleurs observatoires ont un seeing de l'ordre de $0.5$--$1\arcsec$ en visible.

\subsection{Shack-Hartmann et miroir déformable}

Les ingrédients nécessaires à l'amélioration du front d'onde sont un analyseur, un calculateur et un système de correction. Il existe différents types d'analyseur de surface d'onde (Shack-Hartmann, de courbure, à pyramide) mais je n'évoquerai ici que le principe d'un analyseur de type Shack-Hartmann (SH) car c'est ce dont est composé le système NAOS.

Un SH décompose le front d'onde en fronts d'ondes élémentaires afin d'en déterminer leur orientation. Son principe est schématisé sur la Fig.~\ref{image__SH}. Au moyen de micro-lentilles, le front d'onde est échantillonné puis imagé sur un sous-ensemble de pixels d'une matrice CCD appelé sous-pupille. La pente locale du front d'onde est estimée en mesurant dans chaque sous-pupille le déplacement de la tache lumineuse par rapport au centre. On remonte ensuite à la forme du front d'onde en approximant ces déformations par des séries de polynômes, généralement sur la base des polynômes de Zernike (également appelés modes, voir annexe~\ref{section__polynome_zenike}). On se doute que plus nous mettrons de sous-pupilles et de modes, meilleure sera la correction. À noter également que le premier mode de ces perturbations (et le plus important) concernant l'inclinaison du front d'onde (tip-tilt) n'est pas corrigé par le SH, mais plutôt par un miroir plan à mouvement rapide placé en amont.

\begin{figure}[!p]
	\centering\includegraphics[width=.8\linewidth]{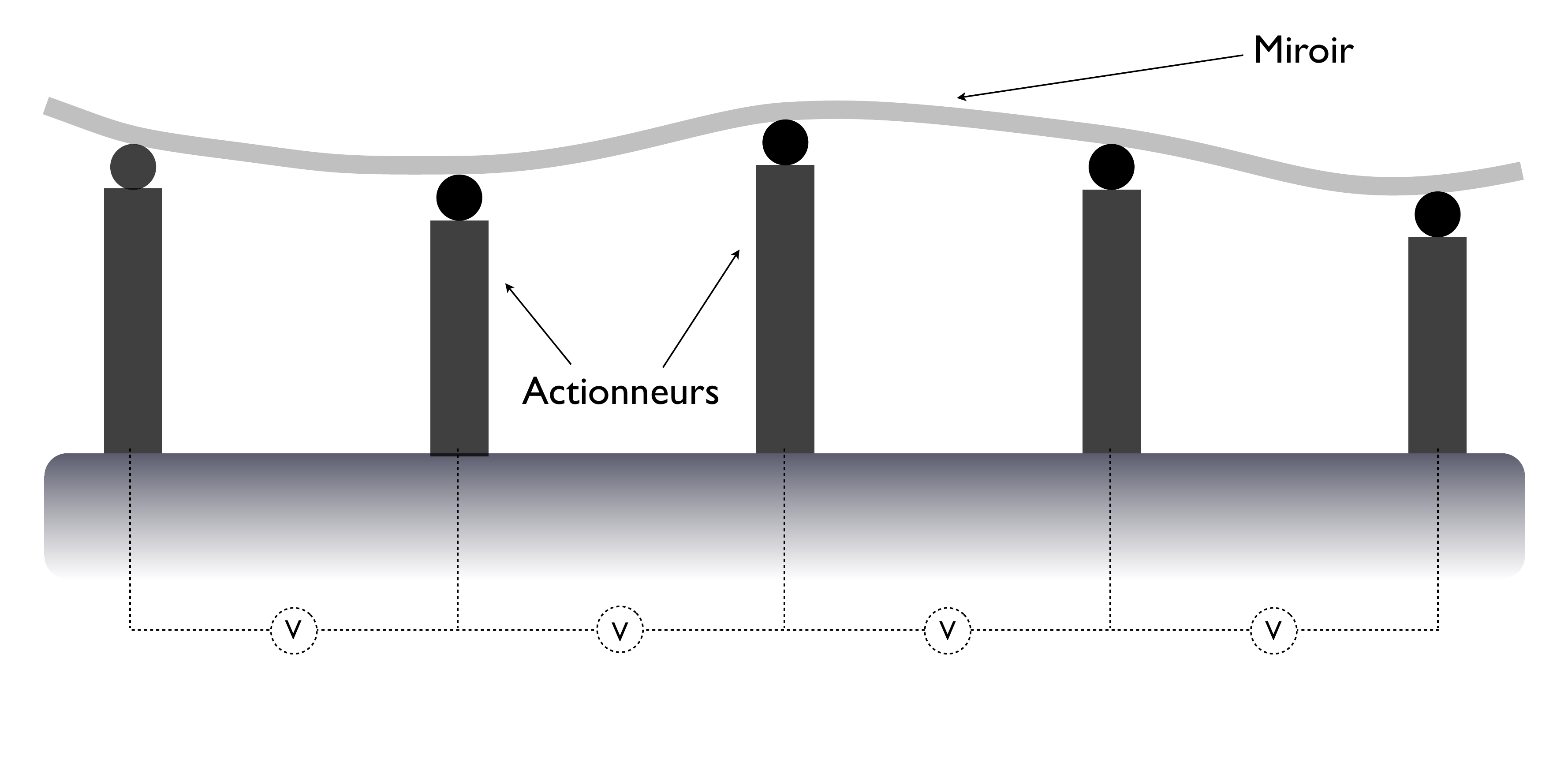}
	\caption[Miroir déformable]{\textbf{Miroir déformable} : schéma du miroir déformable monté sur des actionneurs piézoélectriques pour pouvoir déformer sa surface.}
  	\label{image__DM}
\end{figure}

Une fois le front d'onde connu, il faut le corriger et ce en temps réel. Cette correction n'est pas des plus simples puisqu'elle doit se faire sur un temps très court, de l'ordre de quelques millisecondes, à cause de l'évolution de l'atmosphère. On définit (hypothèse de Taylor) le temps de cohérence $t_0 \sim 0.314\,r_0/\overline{v}$ comme l'intervalle de temps où l'atmosphère n'évolue pas (ou très peu ; $\overline{v}$ représente la vitesse moyenne du vent). La correction doit donc se faire en un temps $t \leqslant t_0$, impliquant un temps de calcul assez court (pour un seeing moyen dans le visible, $r_0(0.55\,\mu\mathrm{m}) \sim 10\,\mathrm{cm}$ et $v \sim 10\,\mathrm{m\,s^{-1}}$ on a $t_0 \sim 3\,\mathrm{ms}$). Cette correction est ensuite appliquée sur un système de miroir déformable, qui a été inventé dans le but d'inverser la forme du front d'onde analysé sous l'effet d'un signal électrique. Comme le montre le schéma de la Fig.~\ref{image__DM}, ce type de miroir est assez flexible et est monté sur des actionneurs piézoélectriques qui permettent la déformation de sa surface. Cela permet de récupérer un front d'onde plan en sortie.

\begin{figure}[!p]
	\resizebox{\hsize}{!}{\centering\includegraphics{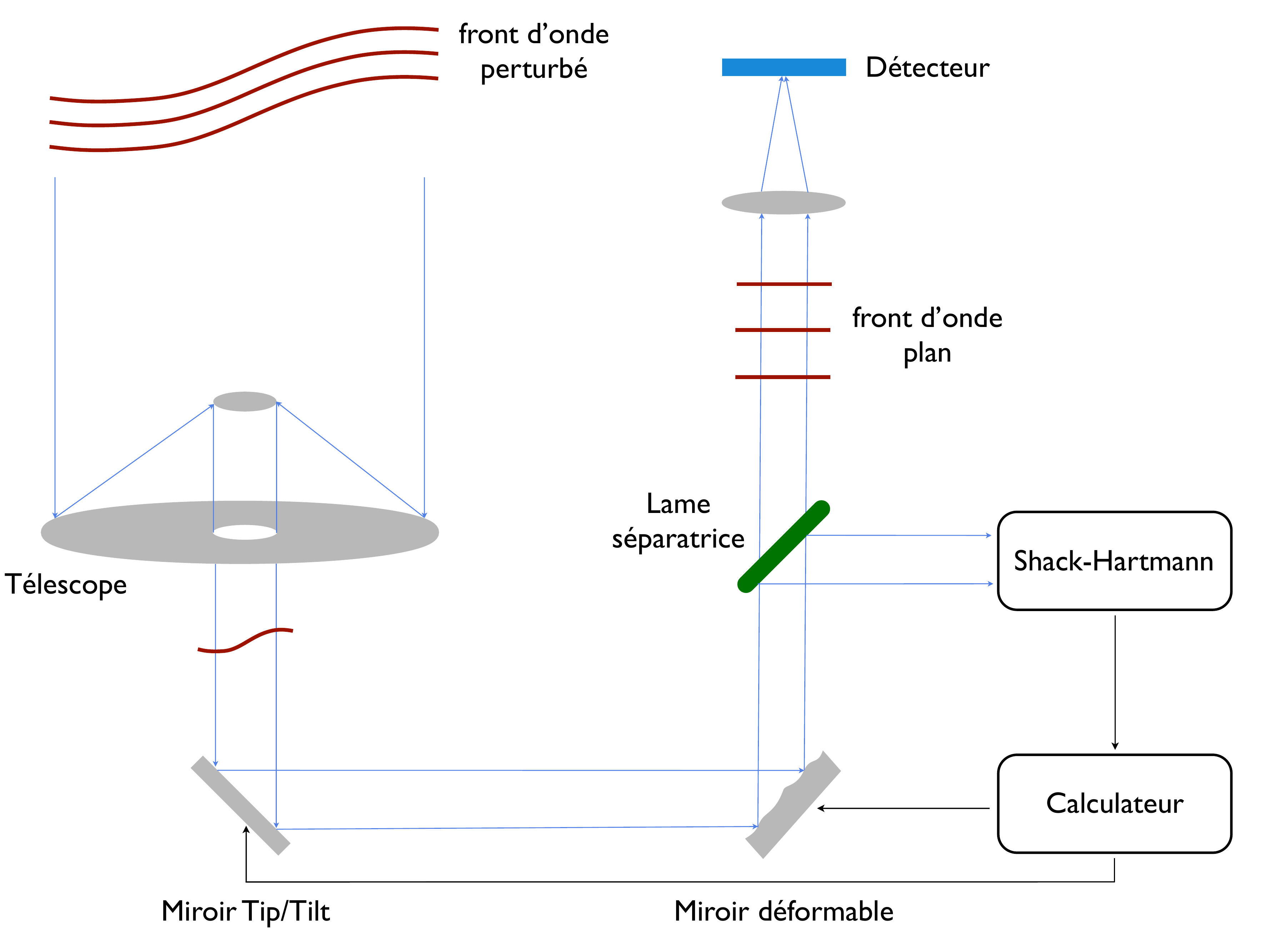}}
	\caption[Schéma de l'optique adaptative]{\textbf{Schéma de l'optique adaptative} : schéma d'un système d'optique adaptative comprenant un miroir déformable et un analyseur de front d'onde de type Shack-Hartmann.}
  	\label{image__principe_optique_adaptative}
\end{figure}

Tous ces systèmes forment l'optique adaptative. Son principe général est schématisé sur la Fig.~\ref{image__principe_optique_adaptative}. La lumière arrive au télescope, passe par un miroir plan, dont la fonction est de corriger du tip--tilt (inclinaison du front d'onde), puis parvient au miroir déformable. Une fraction du flux est ensuite prélevée grâce à une lame séparatrice et envoyée vers l'analyseur. Le front d'onde est reconstruit par un ordinateur qui interpole chaque sous-pupille par des polynômes de Zernike avant de les combiner linéairement. La correction à appliquer est par la suite transmise via des signaux électriques au miroir déformable puis au miroir de tip--tilt. Ce procédé forme un système dit à boucle fermée.


Plus le nombre de polynômes de Zernike sera grand, plus il traduira des aberrations d'ordre élevé et donc des surfaces d'ondes de plus en plus complexes. Mais pour des raisons de temps de calcul, le nombre de polynômes que l'on utilise est limité. Certains ordres élevés ne sont donc pas corrigés, entraînant la formation de speckles résiduels ou d'un halo autour du pic central. Le nombre de modes que l'on peut corriger dépend également du nombre d'actionneurs (et du nombre de sous-pupilles) et sont fonctions de la surface collectrice (télescope) et de $r_0$ par la relation $N \sim (D/r_0)^2$. On utilise également pour une bonne qualité de correction, une fréquence d'échantillonnage temporelle de l'ordre de $\sim 10\,v/r_0$ (soit $1\,\mathrm{kHz}$ pour $r_0(0.55\,\mu\mathrm{m}) \sim 10\,\mathrm{cm}$ et $v \sim 10\,\mathrm{m\,s^{-1}}$). 

\begin{figure}[!p]
	\centering\includegraphics[width=\linewidth]{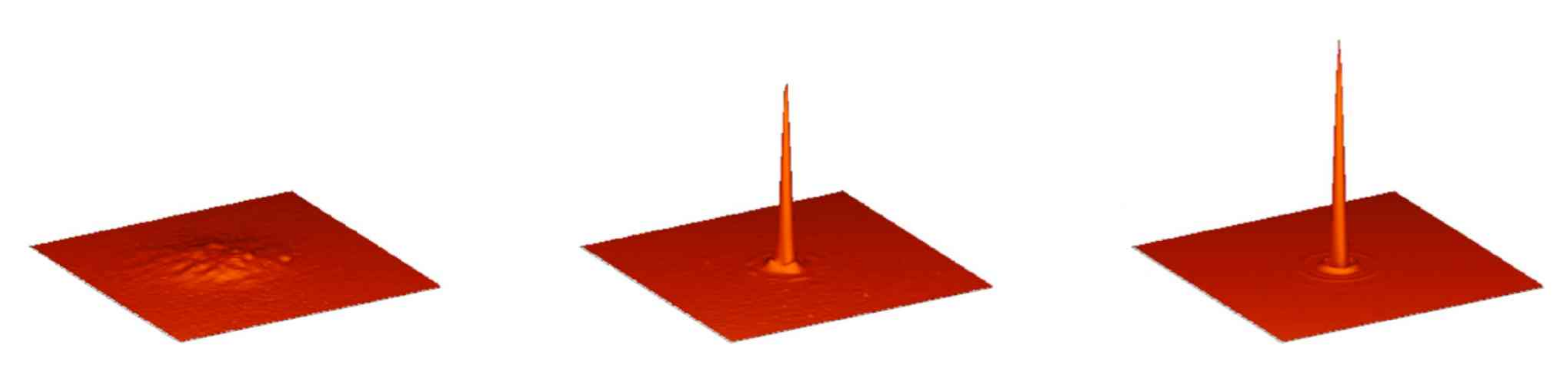}
	\caption[Correction de la FEP par l'optique adaptative]{\textbf{Correction de la FEP par l'optique adaptative} : FEP obtenue par OA en bande $K$ pour un télescope de 8\,m avec $r_0(2.2\,\mu\mathrm{m}) = 1\,\mathrm{m}$, $\overline{v} = 10\,\mathrm{m\,s^{-1}}$, 185 actionneurs et une fréquence d'échantillonnage de $440\,\mathrm{Hz}$. La première image n'est pas corrigée, le Strehl pour les deux autres sont respectivement $0, 64\,\%$ et $100\,\%$ (images simulées par G. Rousset).}
  	\label{image__strehl}
\end{figure}

Pour estimer la performance de l'OA, un paramètre appelé rapport de Strehl ($SR$) a été défini. Il représente le rapport entre le pic d'intensité mesuré par le pic d'intensité théorique. Sur de bons sites d'observations, on peut obtenir $SR~\sim~60\,\%$ à $2\,\mu\mathrm{m}$. On parle également parfois d'énergie cohérente qui correspond à l'énergie contenue dans le pic central sur l'énergie totale. L'énergie cohérente est un bon indicateur du $SR$ car c'est une mesure de l'énergie reconcentrée vers le pic central par l'AO. Des exemples de correction en bande $K$ sont illustrées sur la Fig.~\ref{image__strehl} pour un télescope de $8\,\mathrm{m}$ avec $r_0(2.2\,\mu\mathrm{m}) = 1\,\mathrm{m}$, $\overline{v} = 10\,\mathrm{m\,s^{-1}}$, 185 actionneurs et une fréquence d'échantillonnage de $440\,\mathrm{Hz}$. La première image n'a pas subi de correction, l'image du milieu a été corrigée par OA avec un rapport de Strehl de $64\,\%$ et la dernière image représente la figure de diffraction pour comparaison ($SR = 100\,\%$ par définition).

\begin{figure}[!p]
	\centering\includegraphics[width=.83\linewidth]{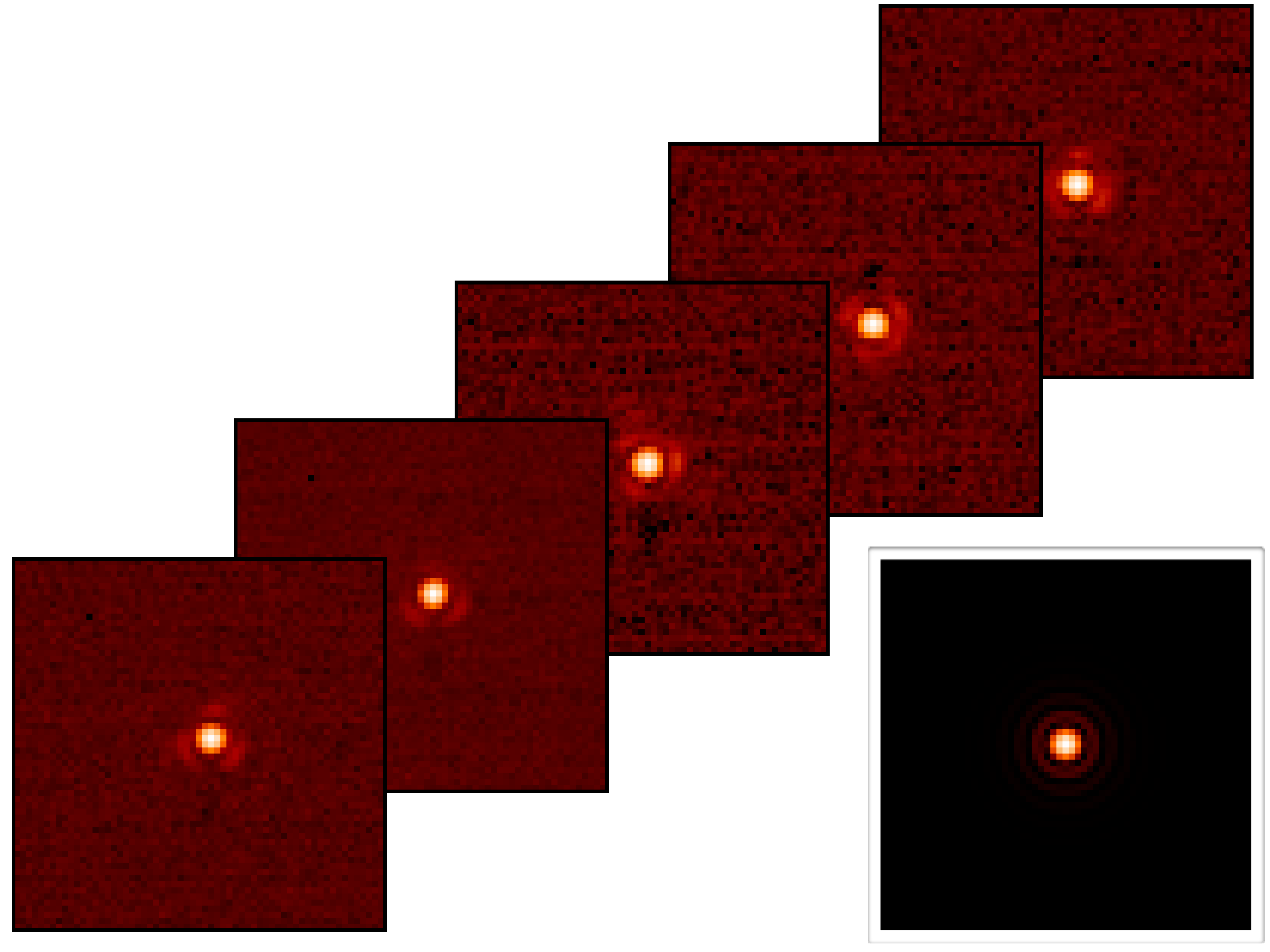}
	\caption[Principe du lucky--imaging]{\textbf{Principe du lucky--imaging} : portion d'images courtes poses ($16\,\mathrm{ms}$) incluses dans un cube de 22~000 images prise au VLT à $8.6\,\mu\mathrm{m}$ (étoile HD~161096, Gallenne et al. 2011 soumis dans A\&A). En bas à droite est représenté la FEP théorique. L'échelle est logarithmique sur toutes les images.}
  	\label{image__shift_and_add}
\end{figure}
\begin{figure}[!p]
	\centering\includegraphics[width=.3\linewidth]{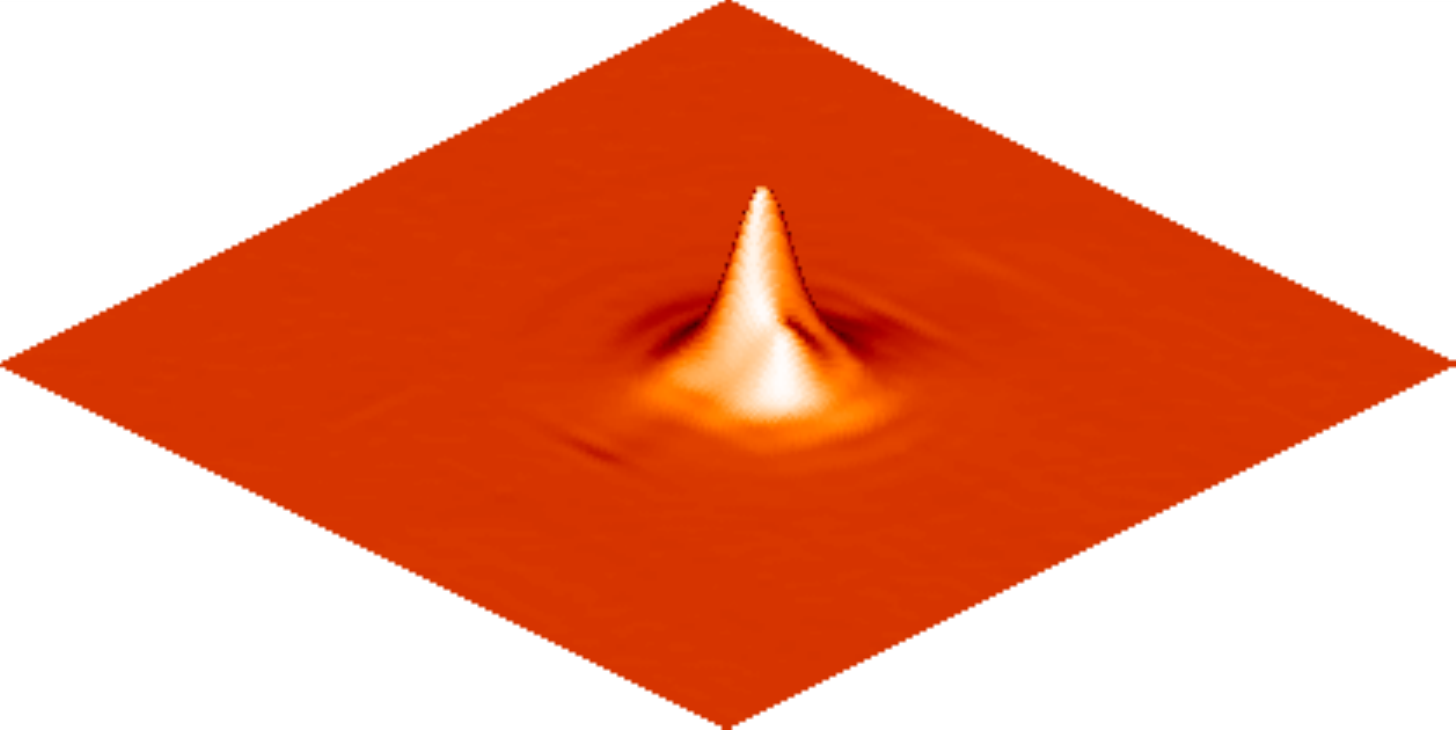} \hspace{.5cm}
	\centering\includegraphics[width=.3\linewidth]{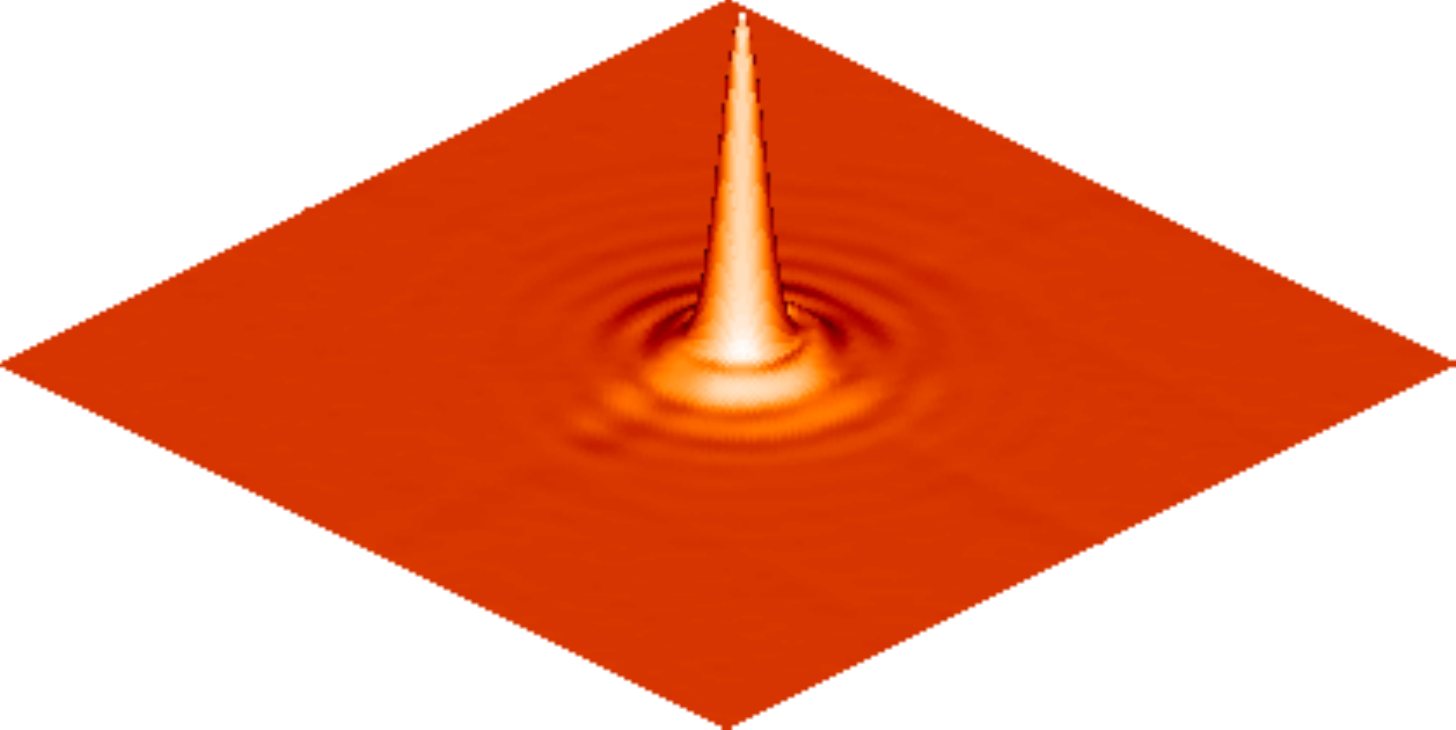} \hspace{.5cm}
	\centering\includegraphics[width=.3\linewidth]{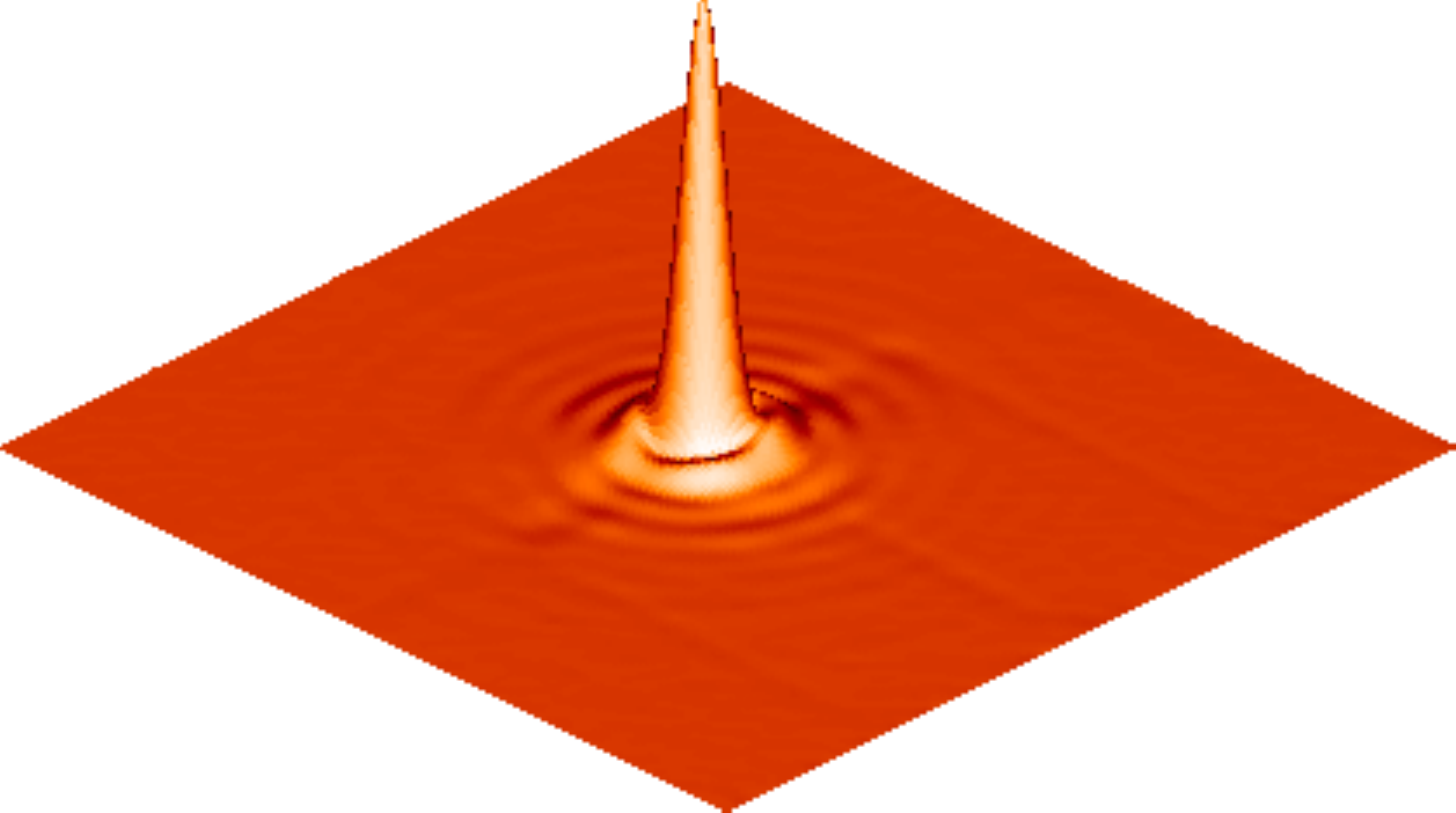}
	\caption[Efficacité du lucky--imaging]{\textbf{Efficacité du lucky--imaging} : image moyenne du cube de la Fig.~\ref{image__shift_and_add}. L'image de gauche est une simple moyenne du cube : $\mathrm{FWHM} = 0.40\arcsec$. L'image centrale représente l'image moyenne du cube recentré sur le pixel le plus brillant : $\mathrm{FWHM} = 0.26\arcsec$. Pour l'image de droite, $50\,\%$ des meilleures images ont été retenues sur le cube recentré avant d'être moyennées : $\mathrm{FWHM} = 0.22\arcsec$.}
  	\label{image__add}
\end{figure}

Pour de bonnes performances, une OA nécessite un objet brillant (à cause du flux qui est divisé dans les sous-pupilles de l'analyseur) et ponctuel (objet brillant et ponctuel = étoile). Parfois les étoiles que l'on souhaite étudier remplissent ces critères et sont donc elles-mêmes utilisables comme références pour la correction du front d'onde. Dans le cas contraire, deux autres solutions sont possibles. Une autre étoile peut être utilisée comme référence mais dans ce cas elle doit être assez proche de l'objet d'étude, de telle sorte qu'elle subisse les mêmes perturbations de front d'onde. On définit un angle limite appelé angle d'isoplanétisme $\theta_0 \sim 0.314\,r_0/\overline{h}$ (dans le cas d'une seule couche turbulente et où $\overline{h}$ représente sa hauteur moyenne), tel que tout objet situé à l'intérieur de cet angle traverse la même atmosphère ($\theta_0 \sim 3\arcsec$ pour $r_0(0.55\,\mu\mathrm{m})  = 10\,\mathrm{cm}$ et $\overline{h} = 2\,\mathrm{km}$). Mais parfois il n'y a pas d'étoiles assez brillantes et proches que l'on puissent utiliser, il faut alors en créer une. Comme nos moyens technologiques (et théorique) ne nous permettent pas de former une étoile à nos besoins, on se contente d'en produire une artificiellement grâce à un laser. Le procédé consiste à pointer une direction proche de l'objet à étudier avec un laser dans le but de stimuler une couche atmosphérique. L'une des couches possibles est celle de sodium, située à environ $90\,\mathrm{km}$ d'altitude, où l'émission stimulée sera utilisée comme source de référence. Cependant utiliser une étoile artificielle à quelques inconvénients. Pour citer seulement un exemple, l'altitude finie de l'étoile artificielle produit un défaut d'anisoplanétisme car son front d'onde n'est plus plan mais sphérique (contrairement à une étoile situé à l'infini). Bien que des solutions existent, une étoile naturelle est toujours préférable comme source de référence.

Les performances d'une OA dépendent de paramètres comme le seeing, la masse d'air, la vitesse de la turbulence, ... Le rapport de Strehl est donc fonction du temps et il n'est pas toujours facile d'obtenir une FEP à la limite de diffraction. L'atmosphère évolue très rapidement (de l'ordre de la milliseconde), sur un temps parfois inférieur au temps d'exposition, empêchant dès lors d'atteindre la résolution angulaire limite du télescope. Une des possibilités est de diminuer le temps d'exposition pour essayer de "figer" l'atmosphère : c'est le principe du lucky--imaging.

\section{Lucky--imaging}
\label{section__lucky_imaging}

Cette technique, proposée pour la première fois par \citet{Bates-1980-03}, permet d'atteindre une résolution proche de la limite de diffraction. Elle consiste à prendre des images instantanément sous forme de cube de données, de recentrer ensuite chaque image du cube par rapport au pixel le plus brillant puis d'en faire la moyenne. Un exemple est exposé sur la Fig.~\ref{image__shift_and_add} pour un télescope de $8.2\,\mathrm{m}$ à $8.6\,\mu\mathrm{m}$ (image de Gallenne et al. 2011 soumis dans A\&A, qui seront présentées au Chapitre~\ref{chapitre__etude_d_exces_infrarouge_par_photometrie}). On remarque que ces images, de temps d'exposition de $16\,\mathrm{ms}$, sont assez proches de la limite de diffraction (présenté sur la même figure). Le rapport signal à bruit final sera augmenté grâce à l'acquisition de milliers d'images. Ce mode d'acquisition permet la sélection des images les moins altérées par l'atmosphère, c'est à dire les images dont presque toute l'énergie est comprise à l'intérieur du pic central. Sur la Fig.~\ref{image__add}, j'expose l'efficacité de cette technique en représentant l'image moyenne sans recentrage ni sélection (image longue pose), puis avec recentrage sans sélection et enfin avec recentrage et sélection de 50\% des meilleures images. Les largeurs à mi-hauteur (FWHM) sont respectivement $0.40\arcsec$, $0.26\arcsec$ et $0.22\arcsec$. En comparant avec la résolution du télescope à cette longueur d'onde de $0.21\arcsec$, on s'aperçoit que cette méthode, également appelée "shift-and-add", apporte un gain considérable en résolution spatiale.

L'inconvénient de cette méthode est son manque de sensibilité. Un temps d'exposition très court implique un flux relativement important pour la détection, et limite par conséquent les objets accessibles à cette technique. Si l'on souhaite un temps d'acquisition plus long, il est possible de combiner cette technique avec l'OA afin de limiter la détérioration de la FEP. Cette combinaison permet de sélectionner en plus des images les moins altérées par la turbulence atmosphérique, les images les mieux corrigées par l'OA, réduisant ainsi le halo créé par les modes non corrigés.

Voyons maintenant une application de l'OA+lucky--imaging à la détection de l'enveloppe de RS~Pup. Les données que j'ai utilisées proviennent du Very Large Telescope (VLT) et plus particulièrement de l'instrument \emph{NACO}. 

\begin{table}[!p]
\centering
\begin{tabular}{cccccc} 
\hline
\hline
Mode 					  	&	Mode de 				&	Mode du 					& 	$\sigma_\mathrm{lec}$	& $\gamma$		& $t_\mathrm{exp}$ min.	\\
instrumental 			&	lecture					&	détecteur					&	(ADU)								& ($\mathrm{e^-/ADU}$)	& 	  (s)								\\
\hline
SW							&	FowlerNsamp		&	HighSensitivity			&	1.3									&	12.1				&	1.7927							\\
SW							&	Double\_RdRstRd	&	HighDynamic				&	4.2									&	11.0				&	0.3454							\\
LW NB imaging			&	Uncorr					&	HighDynamic				&	4.4									&	11.0				&	0.1750							\\
LW Lp imaging			&	Uncorr					&	HighWellDepth			&	4.4									&	9.8				&	0.1750							\\
LW Mp imaging			&	Uncorr					&	HighBackground		&	4.4									&	9.0				&	0.0560							\\
\hline
\end{tabular}
\caption[Caractéristiques du détecteur CONICA]{\textbf{Caractéristiques du détecteur CONICA} : divers modes disponibles de CONICA. $\sigma_\mathrm{lec}$ représente le bruit de lecture et $\gamma$ le gain du détecteur. Le temps d'intégration minimal est également indiqué. SW (pour Short Wavelength) concerne les longueurs d'onde $<~2.5\,\mu\mathrm{m}$ (caractéristiques obtenues d'après le manuel de l'instrument).}
\label{table__conica}
\end{table}

\begin{table}[!p]
\centering
\begin{tabular}{ccccc} 
\hline
\hline
Camera 	&	Échelle 							&	Champ de vue 		& 	Interval spectral		\\
	 			&	($\mathrm{mas/pixel}$)					&	($\arcsec$)			&	($\mu\mathrm{m}$)				\\
\hline
S13			&	13.221 $\pm$ 0.017	&	14$\times$14		&	1.0--2.5				\\
S27			&	27.053 $\pm$ 0.019	&	28$\times$28		&	1.0--2.5				\\
S54			&	54.500 $\pm$ 0.100	&	56$\times$56		&	1.0--2.5				\\
SDI+			&	17.32							&	8$\times$8			&	1.6						\\
L27			&	27.19							&	28$\times$28		&	2.5--5.0				\\
L54			&	54.9								&	56$\times$56		&	2.5--5.0				\\
\hline
\end{tabular}
\caption[Liste des caméras disponibles]{\textbf{Liste des caméras disponibles} : les caméras commençant par "S" concernent les longueurs d'onde $< 2.5\,\mu\mathrm{m}$ (caractéristiques obtenues d'après le manuel de l'instrument).}
\label{table__camera}
\end{table}


\begin{table}[!p]
\centering
\begin{tabular}{cccc} 
\hline
\hline
Taille de fenêtre 		&	$t_\mathrm{exp}$ min. &	$N*t_\mathrm{exp}$ max.	& 	Images perdues	\\
(pixel)	 					&	(s)									&	(s)										&	($\%$)					\\
\hline
1024$\times$1026	&	0.35								&	126									&	20--22				\\
1024$\times$1026	&	0.50								&	126									&	0						\\
512$\times$514		&	0.109							&	508									&	0						\\
256$\times$258		&	0.039							&	2027									&	0						\\
128$\times$130		&	0.016							&	8049									&	0						\\
64$\times$66			&	0.007							&	31711								&	0						\\
\hline
\end{tabular}
\caption[Caractéristiques du mode cube]{\textbf{Caractéristiques du mode cube} : concernent la configuration "Double\_RdRstRd" du détecteur uniquement (caractéristiques obtenues d'après le manuel de l'instrument). $N*t_\mathrm{exp}$ max. représente le nombre maximal d'images enregistrables dans un cube. Le tau de perte d'image est lié à des problèmes électroniques (taille du buffer et vitesse d'écriture sur le disque).}
\label{table__caracteristique_cube}
\end{table}

\section{L'instrument NACO}

L'instrument \emph{NACO} (pour NAos--COnica) est installé depuis fin 2002 au foyer Nasmyth du quatrième télescope du VLT (UT4). Il se compose d'un détecteur infrarouge (CONICA) et d'un système d'optiques adaptatives (NAOS). CONICA est équipée d'un détecteur $1024\times1026$ pixels opérant dans l'intervalle de longueur d'onde $0.8$--$5.5\,\mu\mathrm{m}$. Il peut fonctionner en mode imagerie, coronographie, spectroscopie, polarimétrie ou masquage de pupille. Il est muni à ce jour de 5 filtres à bandes larges, 18 en bandes intermédiaires et 12 en bandes étroites\footnote{Les caractéristiques des filtres sont disponibles sur le site \url{http://www.eso.org/sci/facilities/paranal/ instruments/naco/inst/filters.html}}. Trois modes de lecture peuvent être choisis par l'utilisateur en fonction de l'intensité du fond de ciel (et donc de la longueur d'onde d'observation). Le mode choisi imposera certains autres modes instrumentaux. Diverses caractéristiques qui nous seront utiles dans la suite de ce chapitre et qui concernent ses modes de fonctionnement sont reportées dans la Table~\ref{table__conica}. Cinq caméras sont disponibles en fonction de la longueur d'onde et du champ de vue désiré. La liste est présentée en Table~\ref{table__camera}. 

Le système NAOS quant à lui se compose de deux analyseurs de front d'onde, un dans le visible ($0.45$--$1\,\mu\mathrm{m}$) et un dans l'infrarouge ($0.8$--$2.5\,\mu\mathrm{m}$). Il est doté d'un miroir déformable de 11\,cm monté sur 185 actionneurs et d'un miroir tip--tilt. Les analyseurs de front d'onde sont de type Shack-Hartmann et comprennent de multiples configurations possibles, à choisir en fonction de la magnitude de la source de référence. Dans le cas d'une analyse dans le visible, deux configurations sont disponibles : $14\times14$ et $7\times7$ sous-pupilles. Pour une source brillante, on choisira un échantillonnage de $14\times14$ sous-pupilles alors que pour une source peu brillante, on choisira plutôt $7\times7$. Dans le cas d'une analyse dans l'infrarouge, on a trois configurations possibles : une $14\times14$ sous-pupilles et deux $7\times7$ sous-pupilles avec des champs de vue différents.


Un des modes intéressants de \emph{NACO} est celui du mode cube. Lors d'observations en mode courtes poses, il est possible d'enregistrer chaque image individuelle dans un cube de données (si l'astronome ne demande pas le mode cube, alors l'image fournie sera une moyenne du cube). Ce mode offre 5 tailles de fenêtre possibles pour chaque configuration du détecteur. J'expose les caractéristiques du mode cube dans la Table~\ref{table__caracteristique_cube}, seulement pour la configuration du détecteur utilisée pour les données RS~Pup.

\section{Observation de RS~Puppis avec NACO}
\label{observation_de_rs_puppis_avec_naco}

Les données récoltées en janvier 2009 m'ont permis de mener une étude sur l'environnement circumstellaire de la Céphéide RS~Pup. J'ai pu déterminer un rapport de flux entre l'enveloppe et l'étoile ainsi que la morphologie apparente dans deux bandes photométriques.

Après avoir présenté brièvement cette étoile, je décrirai les observations et la réduction des données. J'exposerai ensuite les deux méthodes d'analyses que j'ai appliquées à ces données et je finirai par les conclusions. Cette section est basée sur un article qui a été publié dans la revue A\&A (Annexe~\ref{article__naco}).

\subsection{RS~Pup}

\begin{figure}[!p]
	\begin{minipage}[h]{.5\linewidth}
  		\centering\includegraphics[width = \linewidth]{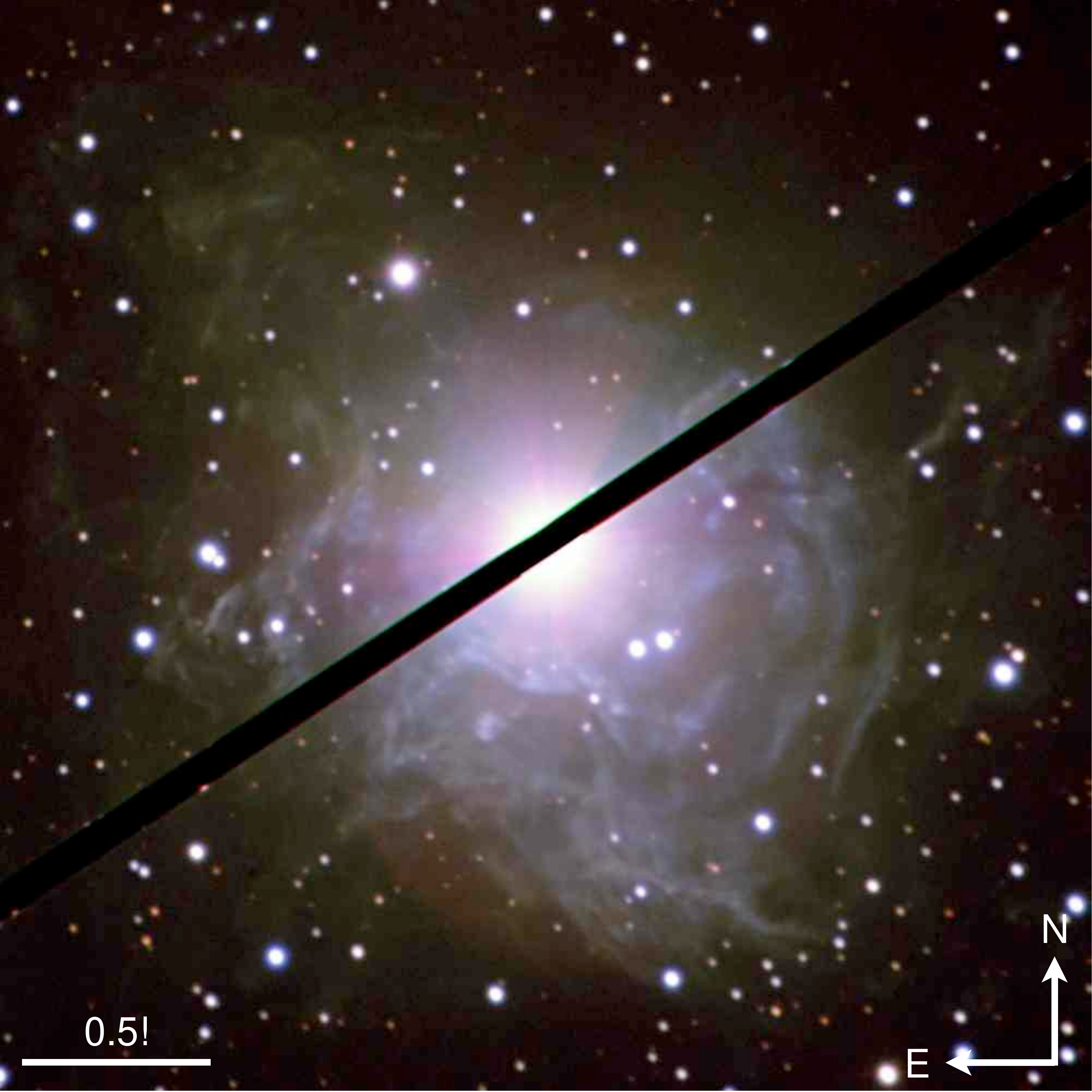}
	\end{minipage}
	\hfill
	\hspace{.2cm}
	\begin{minipage}[h]{.5\linewidth}
	\caption[Image couleur de RS~Pup]{\textbf{Image couleur de RS~Pup} : image composite de la magnifique nébuleuse entourant RS~Pup \citep[images de l'instrument \emph{NTT/EMMI},][]{Kervella-2008-03}. }
  		\label{image__nebuleuse_rs_pup}
	\end{minipage}
\end{figure}

C'est une Céphéide classique de période $P = 41.4\,\mathrm{j}$, bien connue depuis des années pour son immense enveloppe circumstellaire \citep{Westerlund-1961-02,Havlen-1972-01}. RS~Pup est une étoile dix fois plus massive que le Soleil, plus de 200 fois plus grande et en moyenne 15 000 fois plus lumineuse. Il s'agit de la seule Céphéide connue a être entourée d'une grande nébuleuse de poussière. L'origine de cette nébuleuse n'est pas bien connue, serait-ce les restes de la nébuleuse d'origine ou de la matière éjectée par l'étoile elle-même ? Peut-être bien les deux ! Le rayon angulaire de cette nébuleuse est de l'ordre de $1\arcmin$ en bande $V$, ce qui correspond à un rayon linaire d'environ $120~000\,\mathrm{AU}$ \citep[en utilisant $d \sim 2\,\mathrm{kpc}$,][]{Kervella-2008-03}, soit environ 1300 fois le rayon du système solaire. Cette nébuleuse est illustrée sur la Fig.~\ref{image__nebuleuse_rs_pup}. Sa morphologie est très discutée dans la littérature : \citet{Kervella-2008-03} supposèrent une coquille mince et uniforme pour la détermination de la distance par échos de lumière. \citet{Feast-2008-06} a montré qu'un disque incliné concorde également avec les données, et pour finir, une forme plutôt bipolaire est considérée par \citet{Bond-2009-02}. La morphologie de cette nébuleuse est donc un sujet toujours ouvert.

\begin{figure}[!p]
	\begin{minipage}[h]{.55\linewidth}
  		\centering\includegraphics[width = \linewidth]{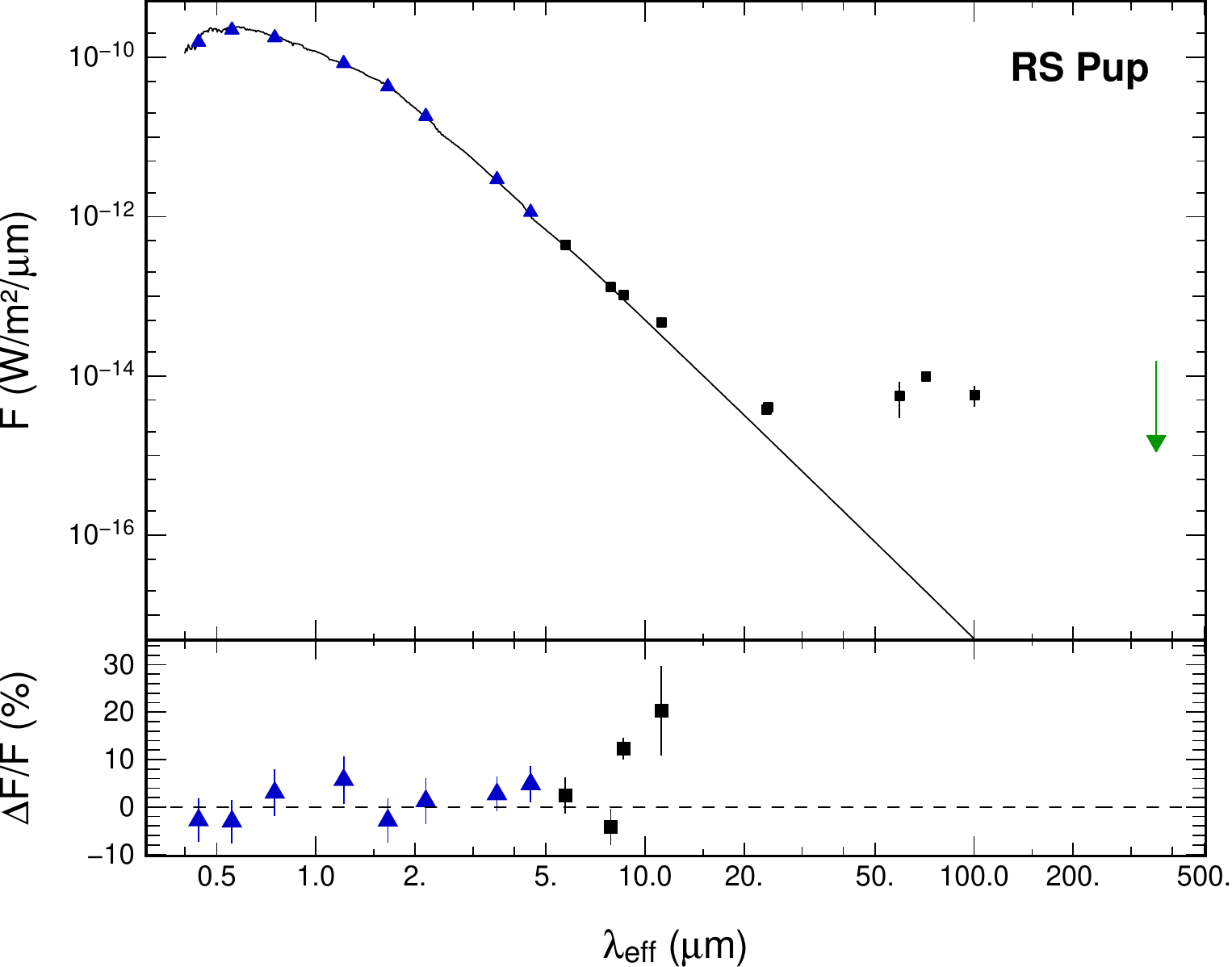}
	\end{minipage}
	\hfill
	\hspace{.2cm}
	\begin{minipage}[h]{.45\linewidth}
	\caption[Distribution d'énergie spectrale de RS~Pup]{\textbf{Distribution d'énergie spectrale de RS~Pup} : spectre synthétique \citep{Castelli-2003-} et mesures photométriques. Le graphique du bas correspond aux résidus de la soustraction des données photométriques au spectre synthétique. Les triangles bleus sont les points utilisés pour l'ajustement du spectre synthétique. La flèche verte indique une limite supérieur à $350\,\mu\mathrm{m}$ que j'ai estimée à partir des données de \emph{APEX/SABOCA}.}
  		\label{image__SED_RS_PUP}
	\end{minipage}
\end{figure}

La plupart des données concernant cette étoile sont photométriques et couvrent des années d'observations. J'ai tracé sur la Fig.~\ref{image__SED_RS_PUP} la distribution d'énergie spectrale moyenne à partir de données récoltées dans la littérature \citep{Kervella-2009-05}. L'ajustement des données par un spectre synthétique \citep{Castelli-2003-} a permis d'estimer un diamètre angulaire moyen $\theta_\mathrm{LD} = 0.93 \pm 0.01\,\mathrm{mas}$ et une température effective moyenne $T_\mathrm{eff} = 5070 \pm 33\,\mathrm{K}$. L'accord entre ce diamètre et $\theta_\mathrm{LD} = 1.02\,\mathrm{mas}$ prédit par \citet{Moskalik-2005-06} est assez bon.

 \begin{table}[!p]
\centering
\begin{tabular}{ccccccccccc} 
\hline
\hline
MJD 				& 	$\phi$	& Filtre 			& Nom 		& $\mathrm{\overline{seeing}}$ &  $\mathrm{AM}$	&	Nbr 	& $\overline{r_0}$	& $\overline{t_0}$ 	 \\
					&				&						&					&	($\arcsec$)						&				&		&	(cm)						&	(ms)						\\
\hline
54~838.021	&		-		& NB\_2.17 	& HD~9362      	& 0.72          					& 1.15 		& 1 	& 9.2						& 3.1						\\
54~838.025	&		-		& NB\_1.64 	& HD~9362 		& 0.77 	         				& 1.15 		& 1 	& 9.0						& 2.9						\\ 
54~838.037	&		-		& NB\_2.17 	& Achernar       	& 0.84 	        					& 1.25		& 1 	& 9.1						& 2.3						\\
54~838.043	&		-		& NB\_1.64 	& Achernar 		& 0.83 							& 1.27 		& 1 	& 9.1						& 2.3						\\ 
54~838.199	&	0.035	& IB\_2.18   	& RS~Pup         	& 0.71\,$\pm$\,0.05 	& 1.03 		& 10	& 13.6\,$\pm$\,0.4	& 3.6\,$\pm$\,0.2	\\
54~838.209	&	0.035	& NB\_1.64 	& RS~Pup         	& 0.68\,$\pm$\,0.04 	& 1.03		& 10 & 13.0\,$\pm$\,0.1	& 3.4\,$\pm$\,0.1	\\	
54~838.223	&		-		& IB\_2.18 	& HD~74417   	& 0.55\,$\pm$\,0.01	& 1.05	 	& 4 	& 19.3\,$\pm$\,0.8	& 5.1\,$\pm$\,0.4	\\
54~838.227	&		-		& NB\_1.64 	& HD~74417		& 	0.54\,$\pm$\,0.01	& 1.04	 	& 4 	& 20.3\,$\pm$\,0.1	& 5.1\,$\pm$\,0.1	\\ 
\hline
\end{tabular}
\caption[Journal des observations NACO]{\textbf{Journal des observations NACO} : le seeing indiqué est la valeur moyenne en bande $V$ mesurée par la station DIMM. AM dénote la masse d'air et Nbr le nombre de cubes enregistrés. $\overline{r_0}$ et $\overline{t_0}$ sont respectivement le paramètre de Fried et le temps de cohérence moyens données par le calculateur de l'AO à $0.55\,\mu\mathrm{m}$.}
\label{table__log_naco}
\end{table}



\subsection{Les observations}

Les observations ont eu lieu dans la nuit du 06 au 07 janvier 2009 sous des conditions atmosphériques convenables. Nous avons utilisé la caméra S13 (Table~\ref{table__camera}) avec deux filtres étroits : NB\_1.64 ($\lambda_0 = 1.644\,\mu\mathrm{m}$ de largeur $\Delta \lambda = 0.018\,\mu\mathrm{m}$) et IB\_2.18 ($\lambda_0 = 2.180\,\mu\mathrm{m}$ de largeur $\Delta \lambda = 0.060\,\mu\mathrm{m}$). Le mode cube a été sélectionné avec des fenêtres respectives de $512\times514$ et $256\times258$ pixels (Table~\ref{table__caracteristique_cube}). Nous avons enregistré 10 cubes de données dans chaque filtre pour la Céphéide RS~Pup et 4 cubes pour l'étoile de référence HD~74417. Chaque cube comprenait 460 images à $1.64\,\mu\mathrm{m}$ et 2000 images à $2.18\,\mu\mathrm{m}$. L'étoile de référence a été observé tout de suite après RS~Pup afin d'être dans les mêmes conditions atmosphériques. Le mode de lecture choisi était "Double\_RdRstRd" (pour un fond de ciel moyen, le détecteur est lu, réinitialisé puis lu une seconde fois), avec le temps d'intégration minimum autorisé pour chaque fenêtre (Table~\ref{table__caracteristique_cube}), soit $109\,\mathrm{ms}$ (pour $512\times514$) et $39\,\mathrm{ms}$ (pour $256\times258$). Le  journal des observations est présenté dans la Table~\ref{table__log_naco}.

Pour comparer ultérieurement le comportement du système d'optique adaptative et les différents artefacts des images RS~Pup, j'ai également utilisé deux autres étoiles observées la même nuit : Achernar et son étoile de référence $\delta$~Phe. Ces données ont été également acquises en mode cube avec la même caméra. Un cube a été obtenu avec le filtre NB\_1.64 et un avec le filtre NB\_2.17 ($\lambda_0 = 2.166\,\mu\mathrm{m}$ de largeur $\Delta \lambda = 0.023\,\mu\mathrm{m}$). Chaque cube de Achernar contient 20~000 images contre 8000 pour $\delta$~Phe. Pour ces observations, la fenêtre du détecteur était de $64\times66$ pixels avec un temps d'intégration minimal de $7.2\,\mathrm{ms}$. La légère différence entre les deux filtres et le temps d'exposition avec RS~Pup n'est pas critique pour notre analyse car je m'intéresse au rapport de flux entre l'enveloppe et l'étoile. De même qu'une fenêtre plus petite diminue le champ de vue, mais je montrerai par la suite que cela n'affecte pas la conclusion.
 
Le premier stade consiste à pré-traiter les données par les étapes habituelles du traitement d'image. J'ai donc pour chaque image de chaque cube corrigé du biais, du champ plat et des mauvais pixels. Le fond de ciel n'a pas été soustrait car il est négligeable à ces longueurs d'onde. J'ai ensuite utilisé deux méthodes d'analyses, l'une utilisant la méthode du "shift-and-add" présentée en Section~\ref{section__lucky_imaging}, et l'autre basée sur une étude statistique du bruit de speckles.

\subsection{Méthode "shift-and-add"}
\label{subsection__methode_shift_and_add}

L'étude commence par l'analyse des images de la Céphéide, puis je comparerai les résultats avec ceux de l'étoile Achernar où l'on ne s'attend pas à détecter d'enveloppe.

\begin{figure}[!p]
		\centering\includegraphics[width=.4\linewidth]{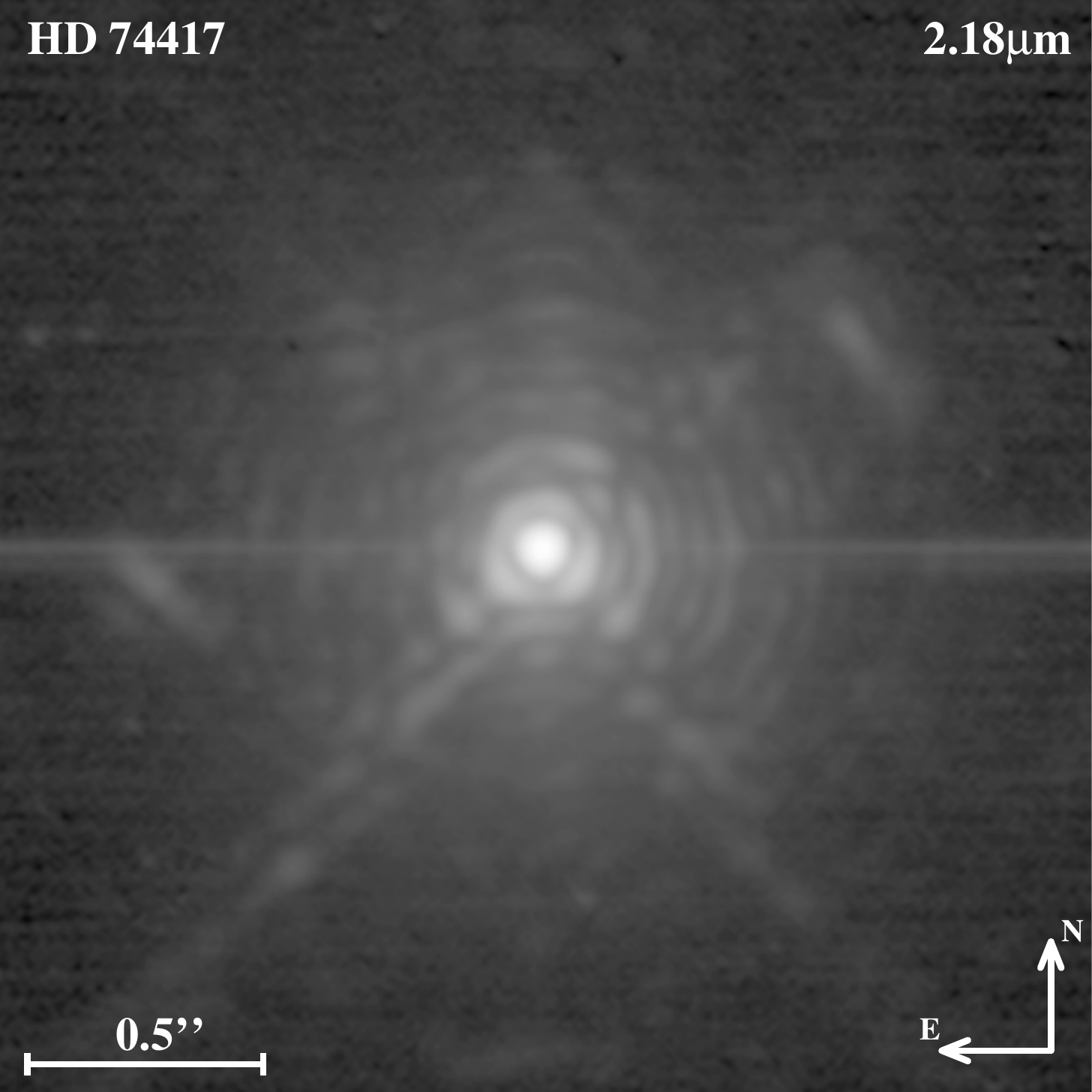} \hspace{.1mm}
		\centering\includegraphics[width=.4\linewidth]{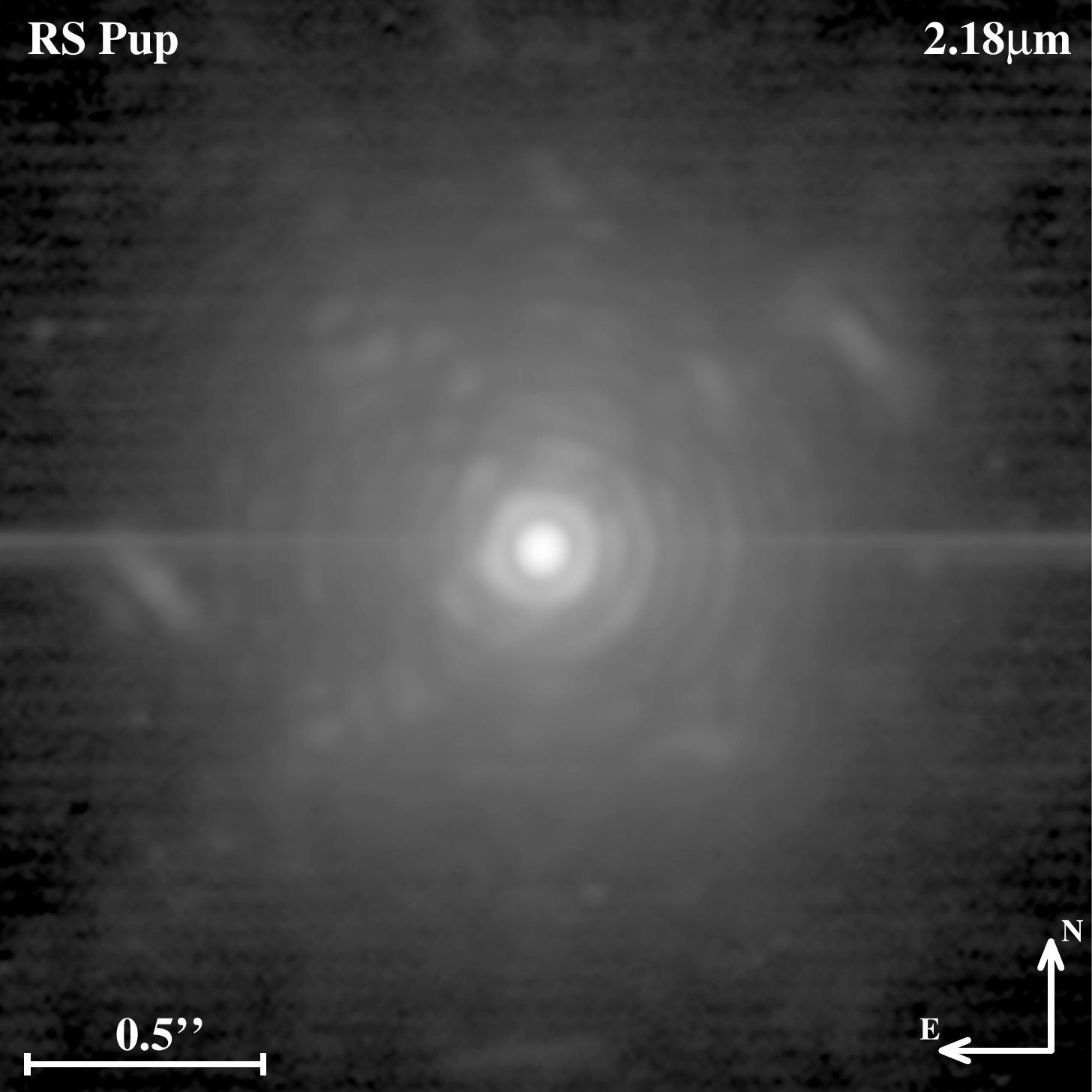}  \\[1.5mm]
		\centering\includegraphics[width=.4\linewidth]{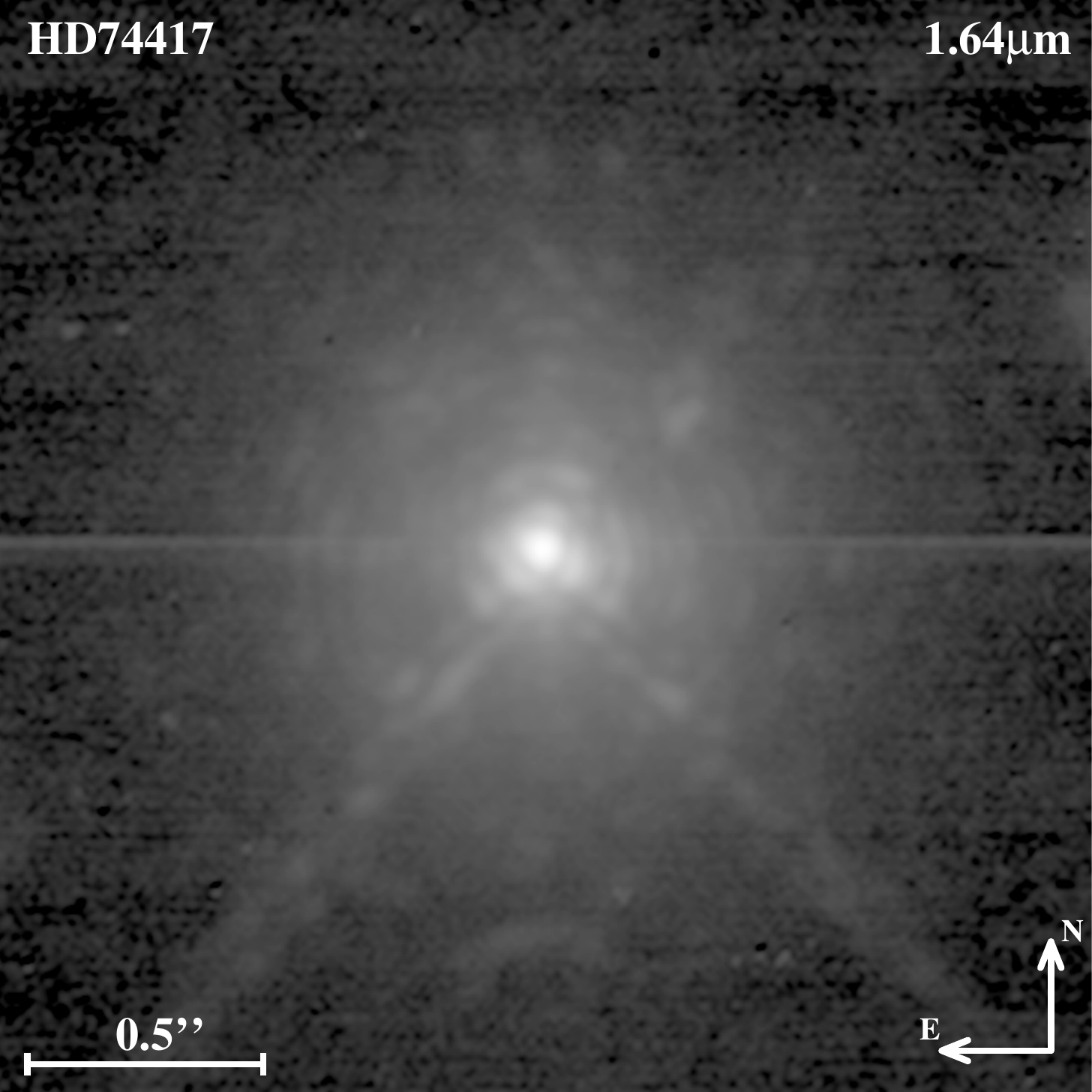} \hspace{.1mm}
		\centering\includegraphics[width=.4\linewidth]{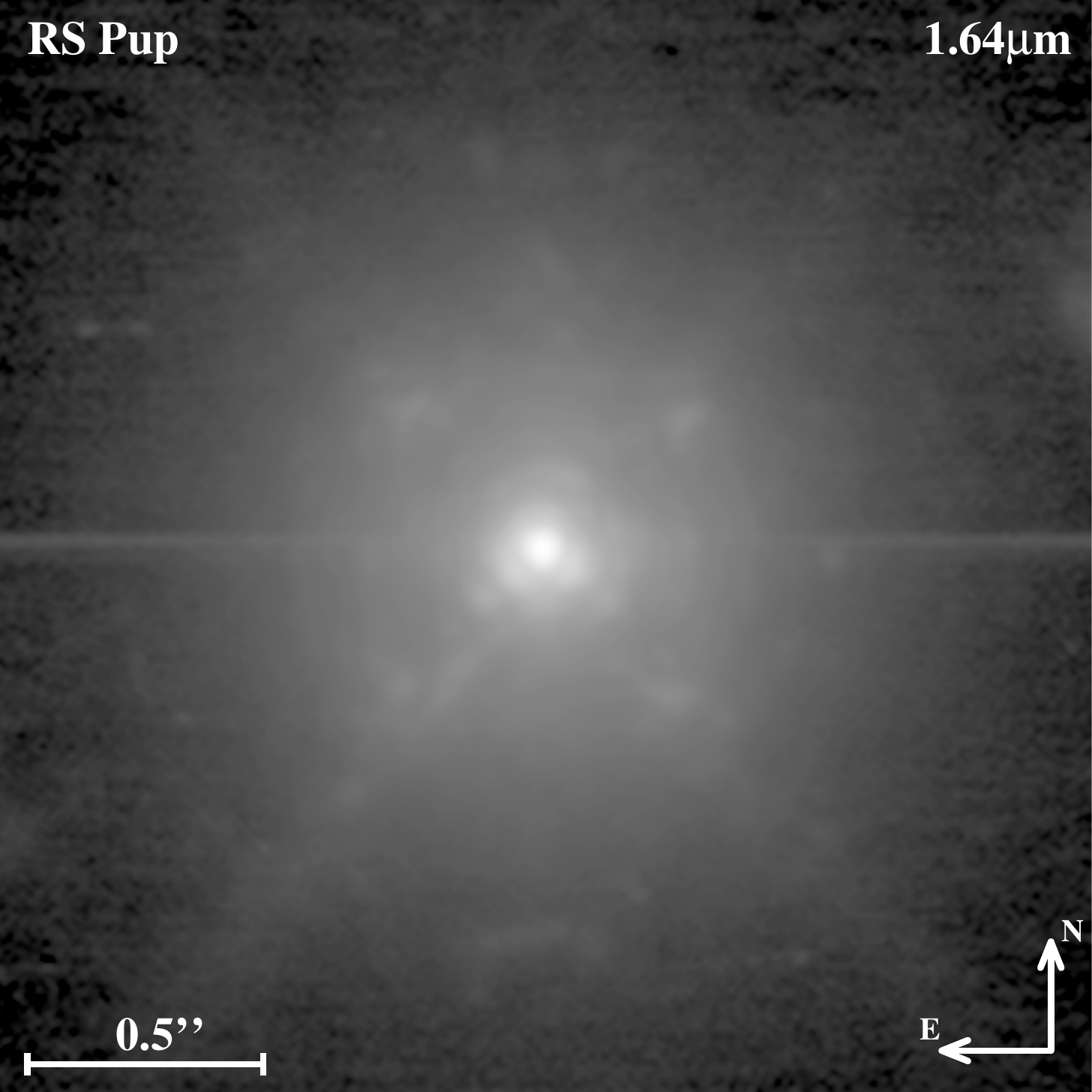}
	\caption[Images moyennes de RS~Pup et HD~74417]{\textbf{Images moyennes de RS~Pup et HD~74417} : les deux images du haut représentent l'étoile de référence et la Céphéide à $2.18\,\mu\mathrm{m}$. En bas, les mêmes étoiles à $1.64\,\mu\mathrm{m}$. L'échelle de niveaux de gris est logarithmique.}
  	\label{image__shift_and_add_final_rs_pup}
\end{figure}

\begin{figure}[!p]
		\centering\includegraphics[width=.4\linewidth]{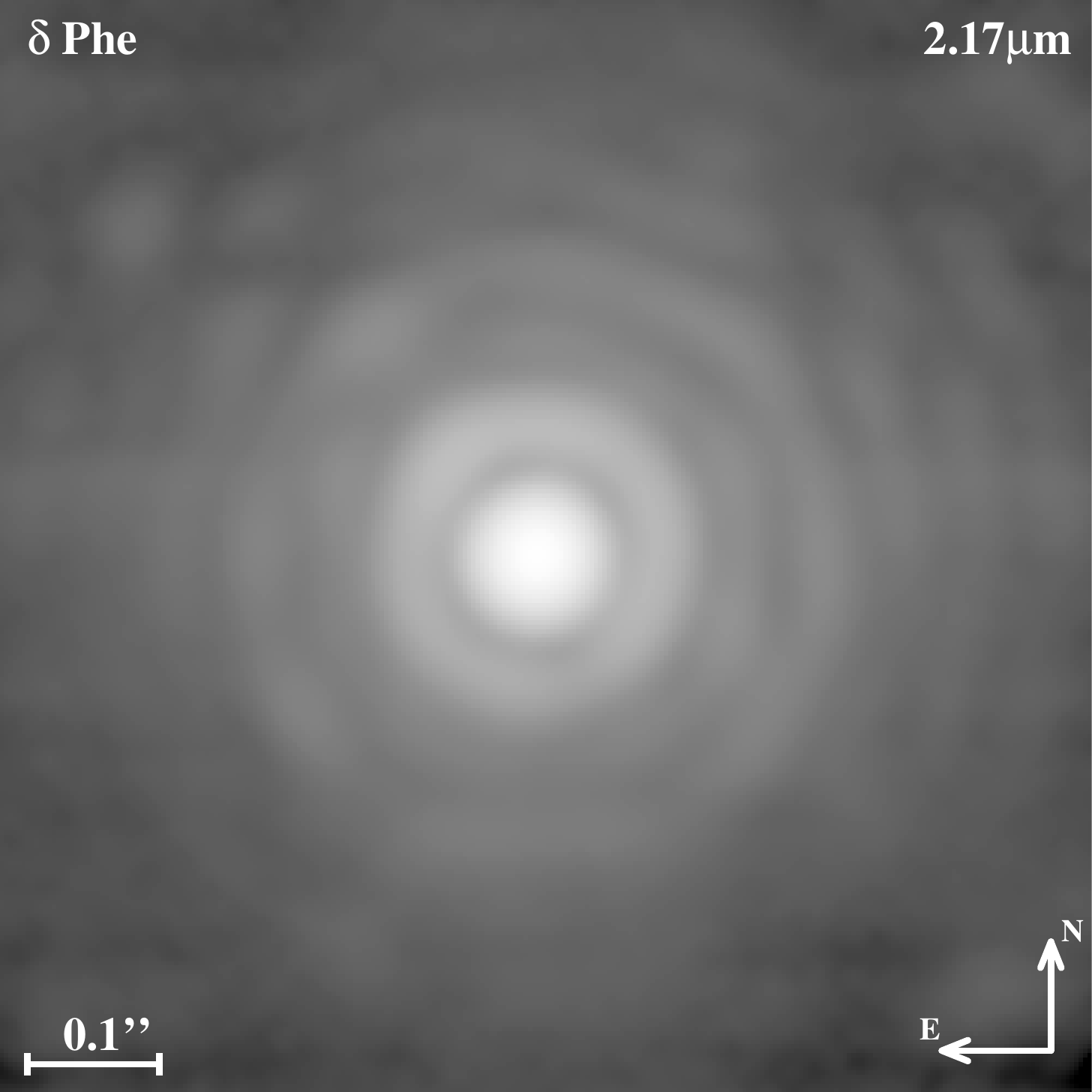} \hspace{.1mm}
		\centering\includegraphics[width=.4\linewidth]{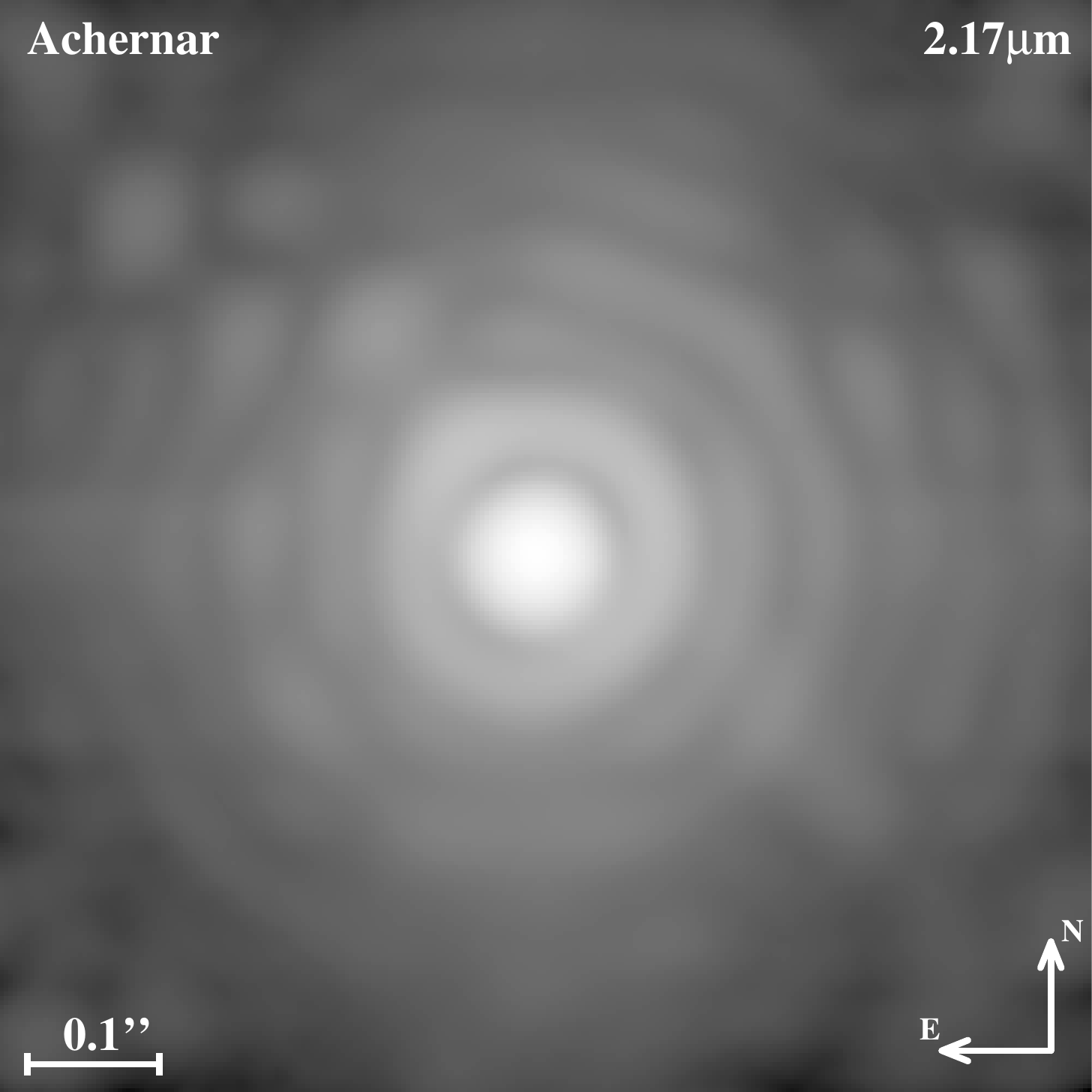} \\[1.5mm]
		\centering\includegraphics[width=.4\linewidth]{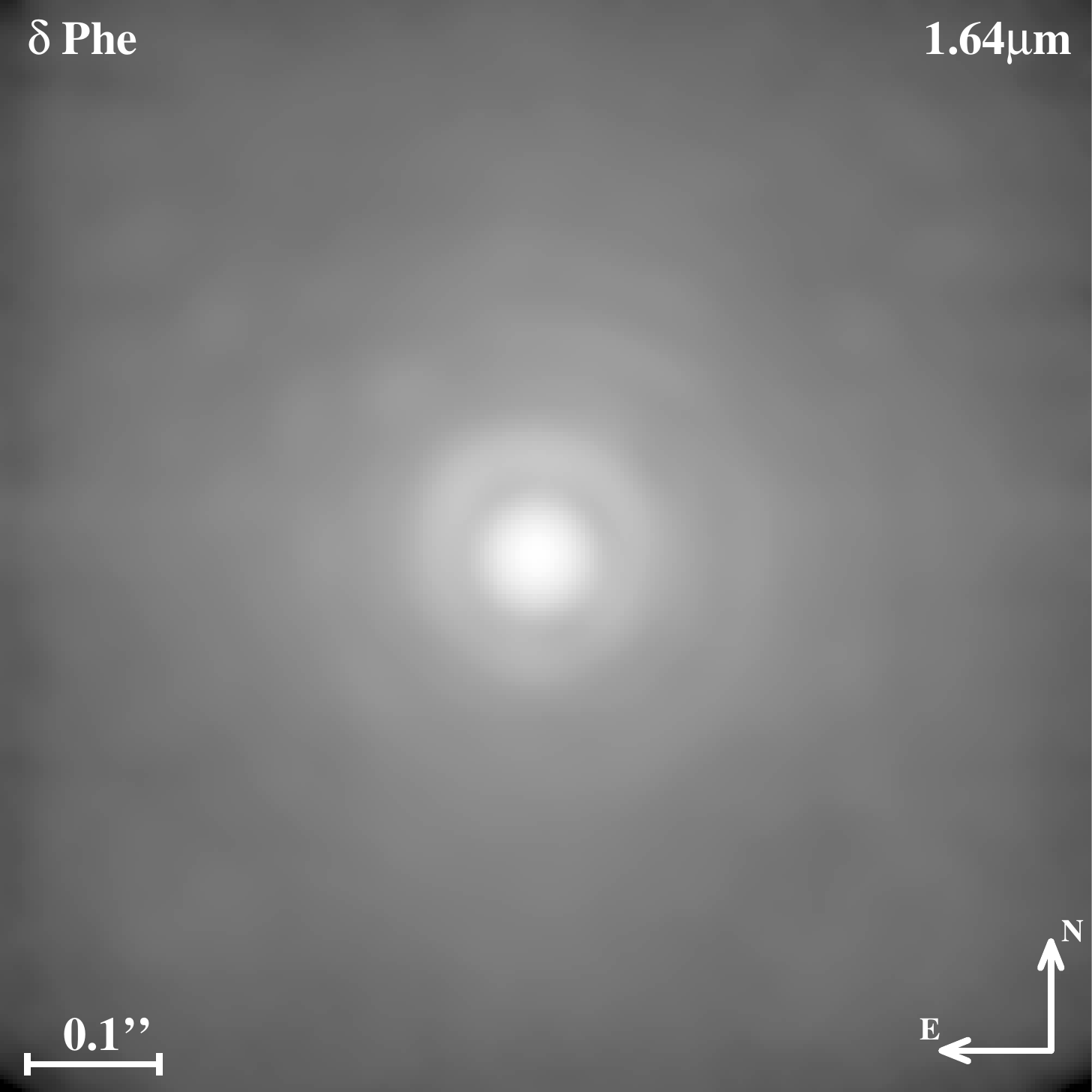} \hspace{.1mm}
		\centering\includegraphics[width=.4\linewidth]{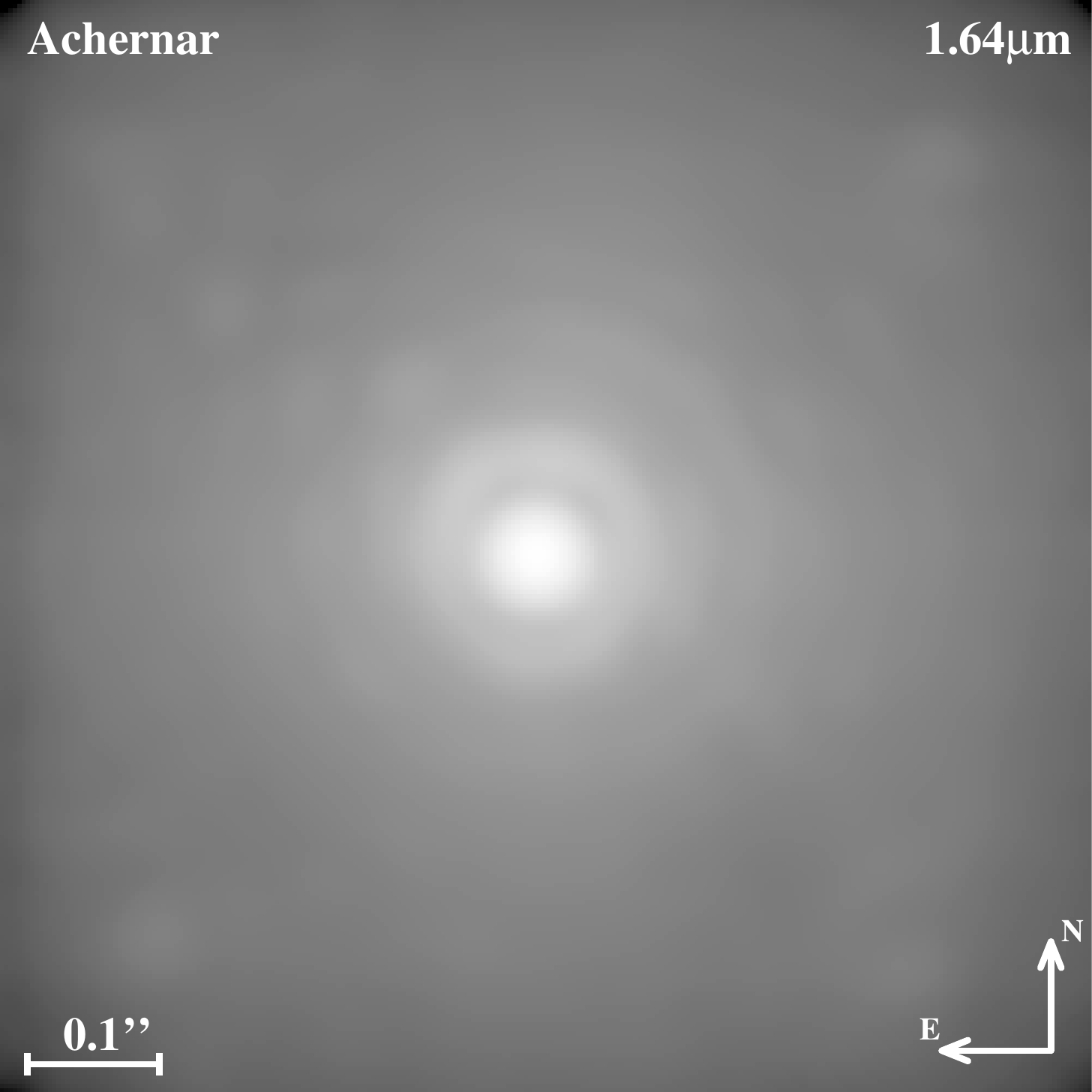}
	\caption[Images moyennes de Achernar et $\delta$~Phe]{\textbf{Images moyennes de Achernar et $\delta$~Phe} : les deux images du haut représentent l'étoile de référence et l'étoile Achernar à $2.17\,\mu\mathrm{m}$. En bas, les mêmes étoiles à $1.64\,\mu\mathrm{m}$. L'échelle de niveaux de gris est logarithmique.}
  	\label{image__shift_and_add_final_achernar}
\end{figure}

Dans un premier temps, j'ai sélectionné dans chaque cube $10\,\%$ des meilleures images en fonction du pixel le plus brillant. J'ai ensuite re-échantillonné chaque image par un facteur 4 en utilisant une interpolation par des splines cubiques\footnote{l'intervalle entre chaque point est interpolé par un polynôme d'ordre 3.} puis recentré en utilisant un ajustement Gaussien. Cela permet d'avoir une précision sur le recentrage de quelques millisecondes d'angle. Chaque cube est ensuite moyenné pour obtenir une image finale. Je présente les résultats de cette méthode sur les Fig.~\ref{image__shift_and_add_final_rs_pup} et \ref{image__shift_and_add_final_achernar} pour RS~Pup, Achernar et leur étoile de référence respective.

\begin{figure}[!p]
	\resizebox{\hsize}{!}{
		\centering\includegraphics{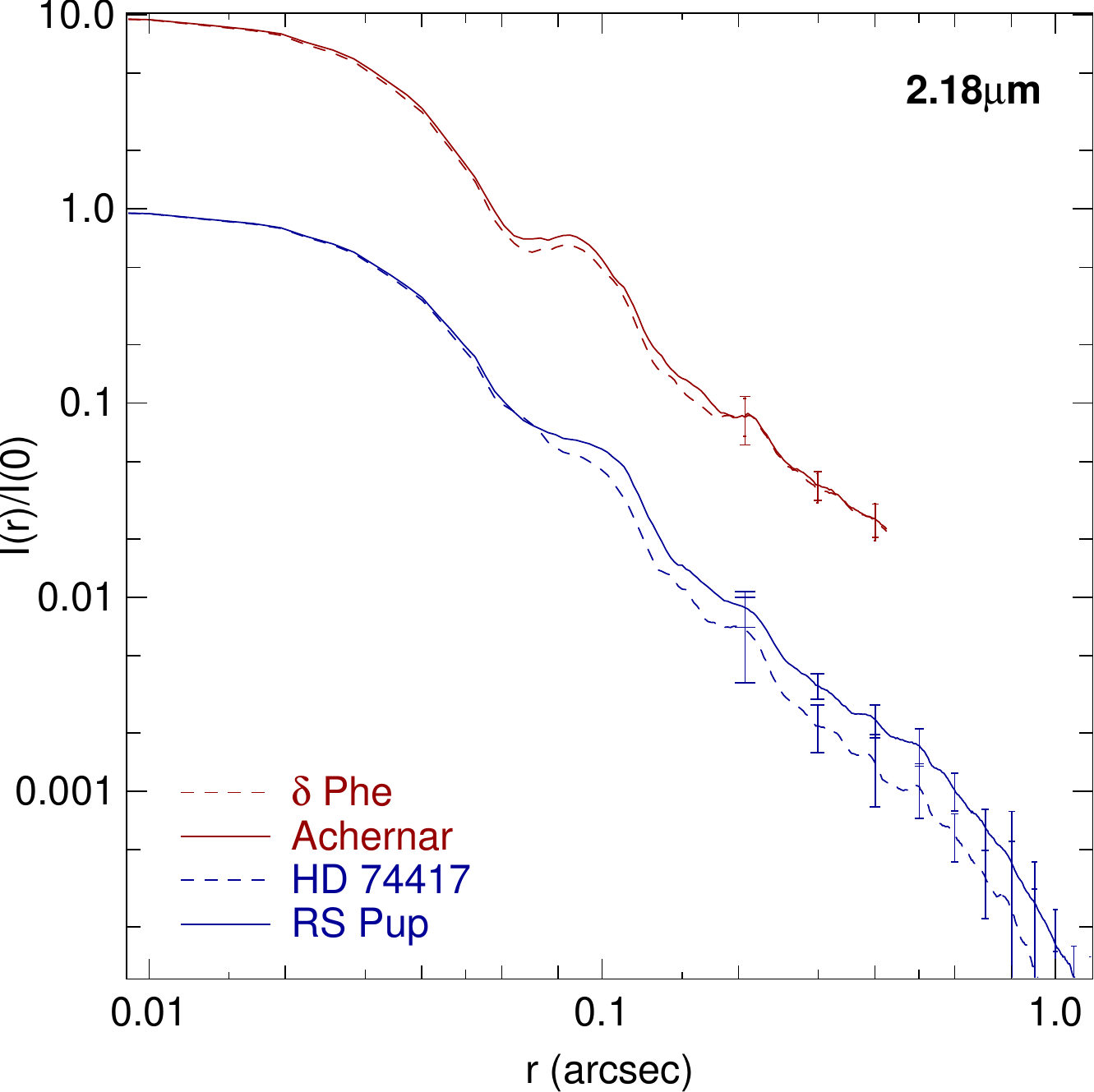} \hspace{.5cm}
		\centering\includegraphics{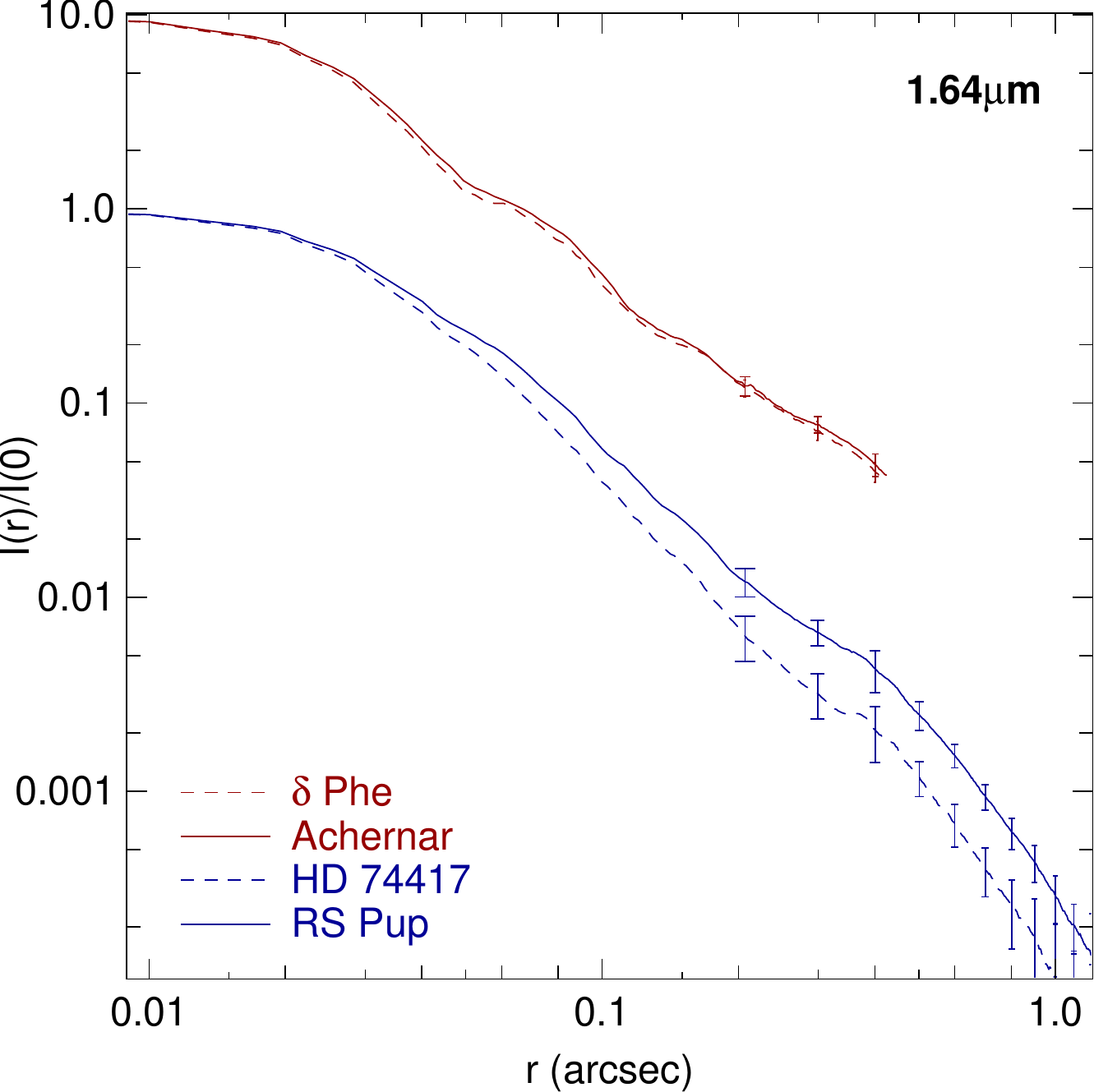}}
	\caption[Profils radiaux]{\textbf{Profils radiaux} : Achernar et $\delta$~Phe sont normalisés à 10 pour la clarté. Les barres d'erreur correspondent à l'écart-type à l'intérieur de chaque anneau et seulement quelques unes sont représentées pour ne pas surcharger le graphe. L'échelle est logarithmique.}
  	\label{image__profile_radiaux}
\end{figure}

On peut déjà remarquer que les anneaux de diffraction sont beaucoup moins visibles sur les images de la Céphéide que pour l'étoile de référence, caractéristique d'une émission étendue autour de RS~Pup (plus faible à $2.18\,\mu\mathrm{m}$). Il est également intéressant d'étudier le profil radial de chaque image, j'ai donc divisé chaque image en plusieurs anneaux élémentaires de largeur d'environ 1 pixel dans lesquels j'ai calculé la médiane et un écart-type. L'image $I(x,y)$ devient donc une fonction $I(r)$ où $r = \sqrt{x^2 + y^2}$. Ces profils sont représentés sur la Fig.~\ref{image__profile_radiaux} où nous voyons clairement une différence entre les courbes. Achernar et $\delta$~Phe ont un profil similaire (courbes grises), cohérent avec une étoile sans émission étendue. Ce n'est pas le cas entre RS~Pup et HD~74417, probablement lié à la présence de l'enveloppe.

Cependant des variations de condition atmosphérique pourraient également nous induire en erreur (à cause du halo de l'OA). Pour écarter cette possibilité, examinons plus en détail ces images, et particulièrement les variations d'énergie encerclée.

\paragraph*{\textcolor{black}{Ajustement du halo}}

L'idée ici est d'ajuster la partie extérieure à la partie centrale, c'est à dire le halo causé par les modes non corrigés de l'OA, et de comparer par la suite les variations de taille du halo avec les variations de l'énergie encerclée.

Utilisons une fonction analytique telle que \citep{Roddier-1981-} :
\begin{equation}
I_{\mathrm{halo}}(r)=f\ \frac{0.488}{\rho^2}\ \left[ 1+\frac{11}{6}\ \left( \frac{r}{\rho}\right) ^{2}\right]^{-\frac{11}{6}}
\label{equation__I_halo}
\end{equation}
où $f$ et $\rho$ sont les paramètres à ajuster. $\rho$ correspond à la largeur à mi-hauteur et est équivalent au seeing pour une image longue pose. $f$ est un paramètre proportionnel au flux.

\begin{figure}[!p]
	\resizebox{\hsize}{!}{
		\centering\includegraphics{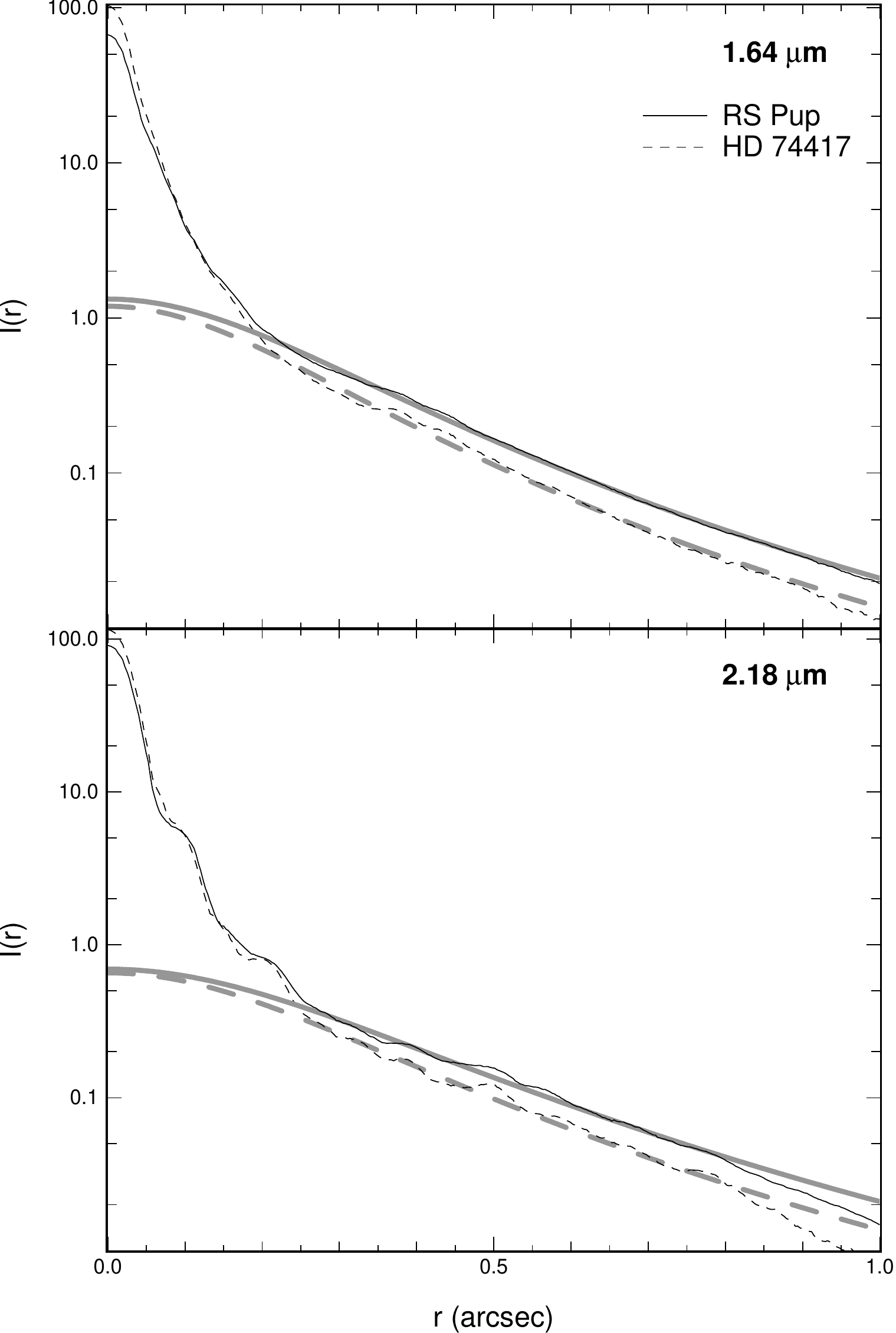} \hspace{.5cm}
		\centering\includegraphics{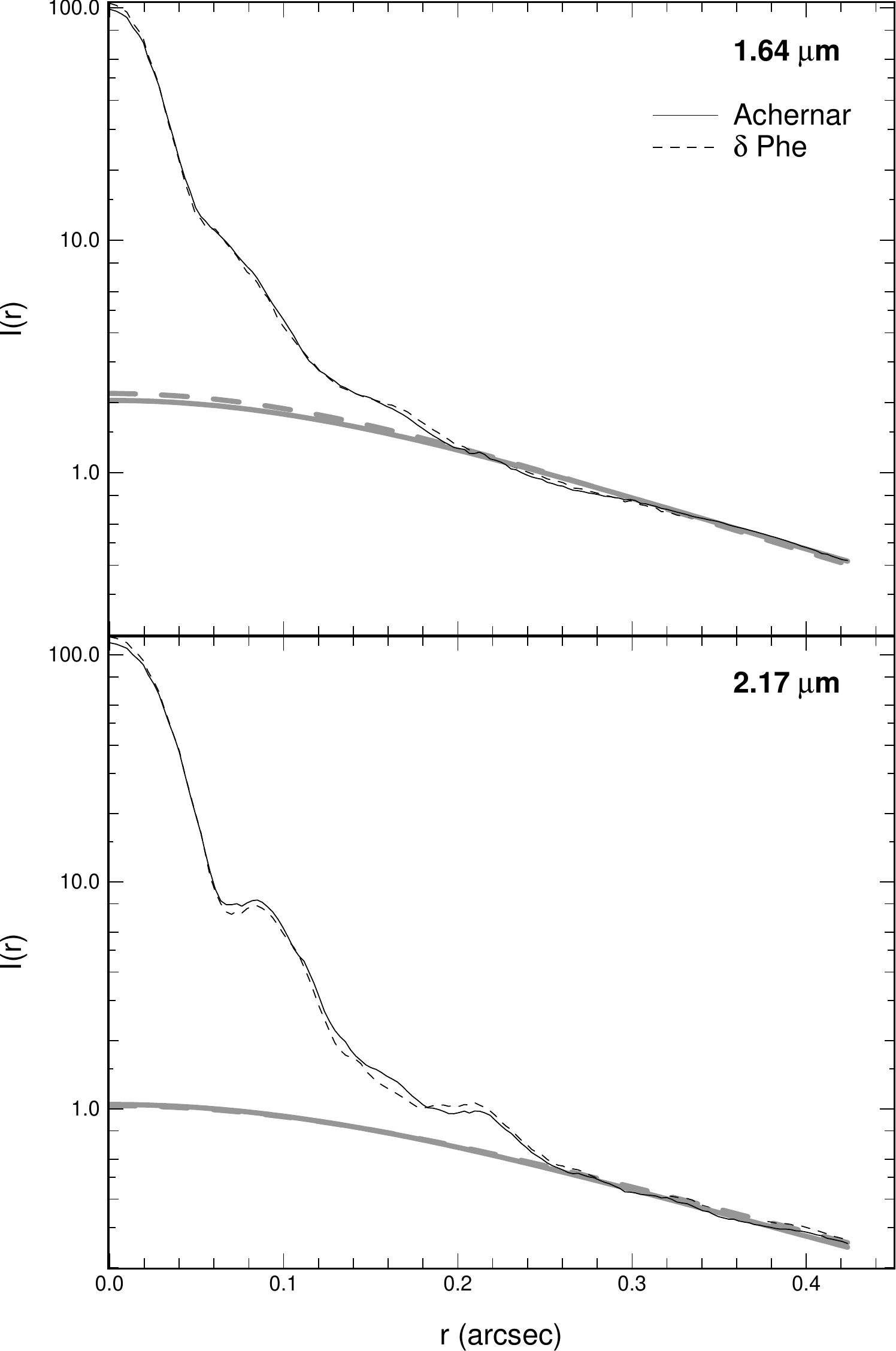}}
	\caption[Ajustement du halo]{\textbf{Ajustement du halo} : les flux sont normalisés au flux total. Les ajustements sont tracés en gris et nos profils radiaux en noir. Les courbes en pointillé représentent les étoiles de référence. L'échelle est logarithmique en ordonnée.}
  	\label{image__profile_radiaux_fit}
\end{figure}

J'ai ajusté l'équation~\ref{equation__I_halo} aux profils radiaux sur l'intervalle $r > 0.26\arcsec$ pour Achernar dans les deux filtres et RS~Pup à $2.18\,\mu$m, et sur l'intervalle $r > 0.20\arcsec$ pour RS~Pup à $1.64\,\mu\mathrm{m}$ (mêmes intervalles respectifs pour leur étoile de référence). Outre ces paramètres, j'ai également évalué l'énergie encerclée normalisée, c'est à dire l'énergie contenue dans le c\oe ur de la FEP (aussi appelé énergie cohérente) divisée par l'énergie totale. Les ajustements sont représentés sur la Fig.~\ref{image__profile_radiaux_fit} (courbes grises) et les paramètres sur la Table~\ref{table__parametre_fit}. Chaque flux est normalisé par le flux total. Nous constatons que le flux dans le halo est différent entre les images de RS~Pup et Achernar. Cela peut être lié à l'émission de l'enveloppe entourant la Céphéide ou à un changement du seeing entre les observations.

Le changement de taille du halo, équivalent à une variation de seeing, peut être évalué par la différence relative des largeurs à mi-hauteur (Table~\ref{table__parametre_fit}). Pour Achernar, la variation de seeing ($\mathrm{\rho_{Achernar}-\rho_{\delta Phe}}$)/$\mathrm{\rho_{\delta Phe}}$, est de l'ordre de $5\,\%$ dans les deux filtres. Comme Achernar ne possède pas d'enveloppe, cette différence est donc attribuée à une variation atmosphérique et/ou un éventuel changement de qualité de correction de l'OA. Pour RS~Pup, cette variation de seeing est un peu plus importante et est de l'ordre de $10\,\%$ dans les deux filtres.

\begin{table}[!p]
\centering
\begin{tabular}{ccccccc} 
\hline
\hline
Filtre 		& Nom	 							& $\rho$ (\arcsec)		& $f$							& $E\mathrm{_{enc}}$	\\
\hline
NB\_1.64 	& RS~Pup 							& 0.46 $\pm$ 0.01 		& 0.58 $\pm$ 0.01 		& 0.327						\\
                	& HD~74417						& 0.43 $\pm$ 0.01	 	& 0.43 $\pm$ 0.01 		& 0.452						\\
                	& Achernar 						& 0.49 $\pm$ 0.01 		& 1.00 $\pm$ 0.01 		& 0.374 						\\
                	& $\delta$~Phe 				& 0.46 $\pm$ 0.01		& 0.97 $\pm$ 0.01 		& 0.376						\\ \hline
IB\_2.18 	& RS~Pup 							& 0.56 $\pm$ 0.01 		& 0.45 $\pm$ 0.01 		& 0.417						\\
                	& HD~74417 					& 0.50 $\pm$ 0.01 		& 0.34 $\pm$ 0.01	 	& 0.517						\\ \hline
NB\_2.17 	& Achernar 						& 0.52 $\pm$ 0.01		& 0.59 $\pm$ 0.01		& 0.495 						\\
                	& $\delta$~Phe 				& 0.54 $\pm$ 0.01		& 0.61 $\pm$ 0.01 		& 0.505 						\\ \hline
\end{tabular}
\caption[Résultats de l'ajustement du halo]{\textbf{Résultats de l'ajustement du halo} : $E\mathrm{_{enc}}$ représente le flux cohérent normalisé au flux total. La fonction analytique $I_{\mathrm{halo}}$ n'est pas normalisée à $f$, c'est à dire $\int I_{\mathrm{halo}}rdr \neq f$.}
\label{table__parametre_fit}
\end{table}

\begin{figure}[!p]
	\resizebox{\hsize}{!}{
		\centering\includegraphics{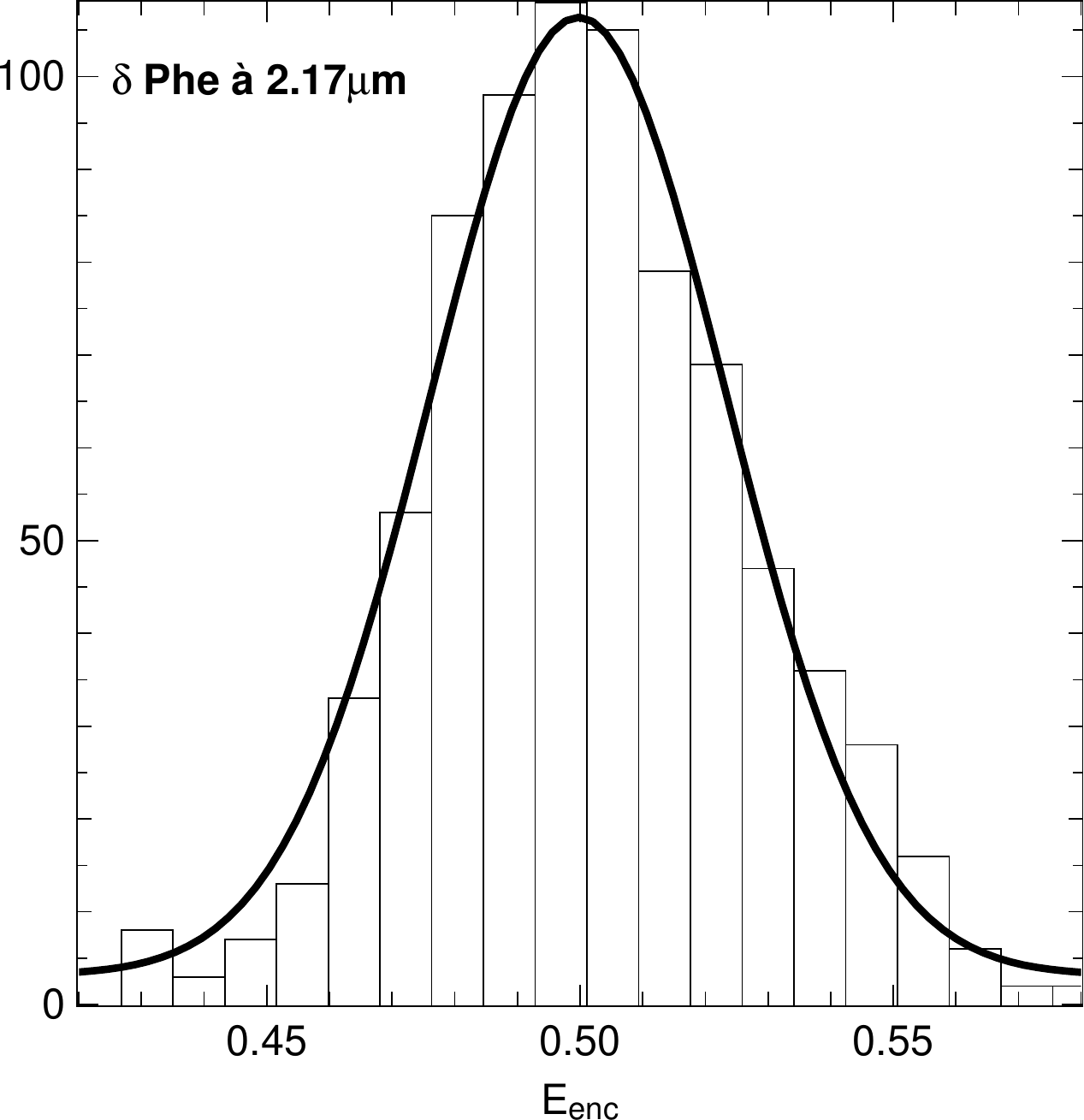} \hspace{1.cm}
		\centering\includegraphics{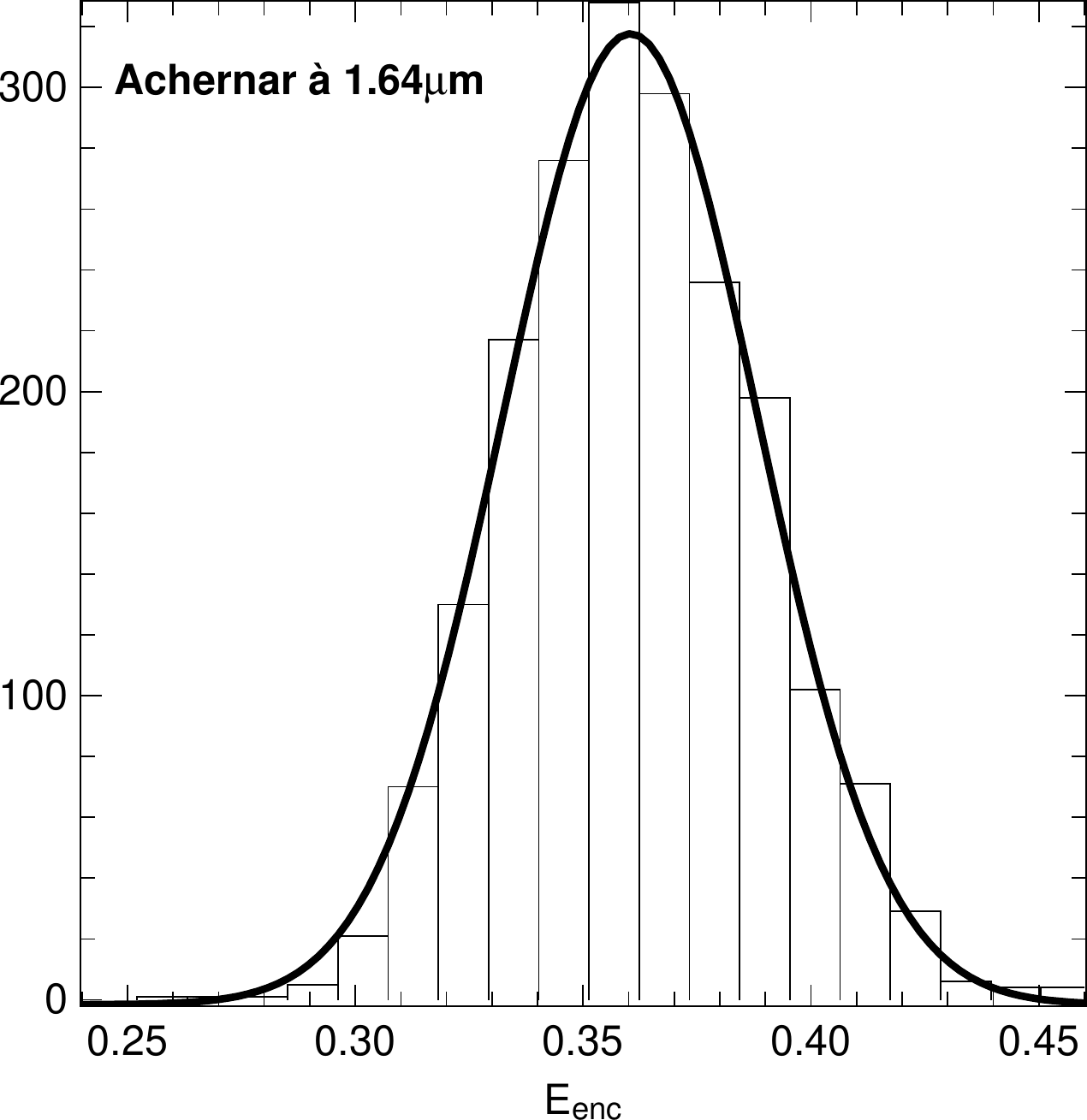}}
	\caption[Distribution de l'énergie encerclée]{\textbf{Distribution de l'énergie encerclée} : histogramme des valeurs de l'énergie encerclée pour $\delta$~Phe à $2.17\,\mu\mathrm{m}$ et Achernar à $1.64\,\mu\mathrm{m}$. La distribution est très proche d'une Gaussienne comme le montre la courbe ajustée en noir.}
  	\label{image__histogramme_energie}
\end{figure}

Aux deux longueurs d'onde, le profil radial normalisé de Achernar est quasiment superposé à celui de $\delta$~Phe, et c'est ce que l'on doit obtenir pour des étoiles non résolues, sans enveloppe et observées dans les mêmes conditions instrumentales et atmosphériques. Si l'on fait le rapport de leur énergie encerclée, on devrait trouver une valeur proche de $100\,\%$. En utilisant les valeurs de la Table~\ref{table__parametre_fit}, je trouve :
\begin{displaymath}
\varepsilon = \frac{E\mathrm{_{enc}(Achernar)}}{E\mathrm{_{enc}(\delta~Phe)}} \sim 98\,\%\quad \mathrm{\grave{a} \ 2.17\,\mu m}
\end{displaymath}
\begin{displaymath}
\varepsilon  \sim 99\,\%\quad \mathrm{\grave{a} \ 1.64\,\mu m}
\end{displaymath}

Comme attendu, les énergies encerclées sont similaires, pas d'émission supplémentaire autour de Achernar n'est détectée, signe que cette étoile ne présente pas d'enveloppe circumstellaire. Nous pouvons estimer une incertitude sur la variable $\varepsilon$ en étudiant ses variations dans les cubes de données. En mesurant $E\mathrm{_{enc}}$ sur chaque image des cubes (les cubes comprenant seulement $10\%$ des meilleures images), j'ai examiné la distribution de l'énergie encerclée et j'ai montré que l'on pouvait faire l'approximation d'une distribution Gaussienne. Deux exemples de distribution sont représentées sur la Fig.~\ref{image__histogramme_energie}. On remarque clairement que cette approximation (les courbes noires et en gras) est assez bonne. De plus les énergies encerclées des deux étoiles ne sont pas corrélées, l'erreur relative a donc l'expression mathématique :

\begin{equation}
\frac{\sigma(\varepsilon)}{\varepsilon} = \sqrt{ \left[ \frac{\sigma(E\mathrm{_{enc}})}{\overline{E\mathrm{_{enc}}}} \right]_{\mathrm{Achernar}}^2 + \left[ \frac{\sigma(E\mathrm{_{enc}})}{\overline{E_{\mathrm{enc}}}} \right]_{\mathrm{\delta~Phe}}^2}
\label{equation__uncertainties}
\end{equation}

Pour Achernar je trouve une erreur relative sur le rapport d'énergie encerclée de l'ordre de $15\,\%$ à $2.17\,\mu\mathrm{m}$ et $11\,\%$ à $1.64\,\mu\mathrm{m}$. 

En appliquant le même procédé pour RS~Pup, je trouve :
\begin{displaymath}
\varepsilon \sim 80 \pm 7\,\%\quad \mathrm{\grave{a} \ 2.18\,\mu m}
\end{displaymath}
\begin{displaymath}
\varepsilon  \sim 72 \pm 9\,\% \quad \mathrm{\grave{a} \ 1.64\,\mu m}
\end{displaymath}

La diminution du rapport d'énergie encerclée est plus importante que celle estimée pour les données Achernar, favorisant l'hypothèse de l'émission de l'enveloppe autour de la Céphéide. De plus cette diminution est plus grande que la variation de seeing mesurée. La chute d'énergie encerclée est $\sim20\,\%$ à $2.18\,\mu\mathrm{m}$ et $\sim28\,\%$ à $1.64\,\mu\mathrm{m}$, alors que la variation de seeing est $\sim10\,\%$. Je considère donc que cette diminution du paramètre $\varepsilon$ pour RS~Pup est causée par l'émission de l'enveloppe.

\paragraph*{\textcolor{black}{Rapport de flux}}

Le but maintenant est d'évaluer le rapport de flux entre l'enveloppe et la photosphère de la Céphéide aux deux longueurs d'onde. Pour cela j'utilise le fait que le rapport d'énergie encerclée théorique doit être $100\,\%$ (comme expliqué précédemment) et qu'une émission supplémentaire au flux de l'étoile aura pour effet de diminuer ce rapport.

L'énergie encerclée peut s'écrire :
\begin{displaymath}
E\mathrm{_{enc}(RS~Pup)}=\frac{F_{\mathrm{coh}} + \alpha F_{\mathrm{env}}}{F + F_{\mathrm{env}}}
\end{displaymath}
où $F$ correspond au flux de l'étoile, $F_{\mathrm{env}}$ au flux de l'enveloppe, $F_{\mathrm{coh}}$ au flux cohérent dans le c\oe ur de la FEP et $\alpha$ la fraction du flux de l'enveloppe contenue dans le c\oe ur. Je suppose que l'enveloppe est beaucoup plus grande que le c\oe ur de la FEP, on a alors $\alpha \ll 1$, et l'équation précédente devient :
\begin{displaymath}
E\mathrm{_{enc}(RS~Pup)}\approx\frac{F_{\mathrm{coh}}/F }{1 + F_{\mathrm{env}}/F}
\end{displaymath}

Cette équation peut être étalonnée avec l'étoile de référence (en supposant que l'énergie encerclée est stable sur \emph{NACO} pendant nos observations) : 
\begin{displaymath}
E\mathrm{_{enc}(HD74417)} = \frac{F_{\mathrm{coh}}}{F}
\end{displaymath}

En faisant le rapport de ces deux équations, nous trouvons le rapport entre le flux de l'enveloppe $F_{\mathrm{env}}$ et celui de la Céphéides $F_\star$ en fonction du rapport d'énergie encerclée $\varepsilon$ :
\begin{displaymath}
\beta = \frac{F_{\mathrm{env}}}{F_\star} = \frac{1 - \varepsilon}{\varepsilon}
\end{displaymath}

En utilisant l'equation~\ref{equation__uncertainties}, l'incertitude sur le rapport de flux est donnée par :
\begin{displaymath}
\sigma(\beta) = \frac{1}{\varepsilon} \frac{\sigma(\varepsilon)}{\varepsilon} = \frac{E\mathrm{_{enc}(HD74417)}}{E\mathrm{_{enc}(RS~Pup)}} \sqrt{ \left[ \frac{\sigma(E\mathrm{_{enc}})}{\overline{E\mathrm{_{enc}}}} \right]_{\mathrm{RS~Pup}}^2 + \left[ \frac{\sigma(E\mathrm{_{enc}})}{\overline{E_{\mathrm{enc}}}} \right]_{\mathrm{HD74417}}^2}
\end{displaymath}

En utilisant les valeurs calculées précédemment, j'ai trouvé que la contribution de l'enveloppe (relative à la photosphère de la Céphéide) est de $38\pm17\,\%$ à $1.64\,\mu\mathrm{m}$ et $24\pm11\,\%$ à $2.18\,\mu\mathrm{m}$. On voit que la fraction de flux émanant de l'enveloppe n'est pas négligeable à ces longueurs d'onde.

Passons maintenant à la géométrie de cette enveloppe. Dans la section suivante, j'examine la morphologie de ce matériel circumstellaire via l'analyse des différents bruits présents dans les images.

\subsection{Étude statistique du bruit de speckles}

Je présente ici une étude qualitative sur la morphologie de l'enveloppe basée sur les propriétés statistiques du bruit. Je commence par examiner les différents bruits avant d'introduire une propriété intéressante du bruit de speckle.

\paragraph*{\textcolor{black}{Méthode d'analyse}}

L'idée générale est de retrouver le flux de l'enveloppe dans les images courtes poses (c'est à dire les images des cubes). Dans le cas d'une étoile non résolue possédant une enveloppe, le flux inter-speckle sera non nul. C'est cette idée que je vais exploiter dans le reste de cette section.

Le bruit de speckle a été étudié par différents auteurs \citep{Racine-1999-05,Canales-1999-,Fitzgerald-2006-01}, et nous savons aujourd'hui que c'est une source de bruit non négligeable. Pour une étoile brillante corrigée par OA, \citet{Racine-1999-05} a même montré que le bruit de speckle domine les autres bruits dans le halo formé par les modes non corrigés d'une OA. L'objectif est de procéder à une analyse similaire grâce à nos cubes de données.

Reprenons les cubes de données comprenant $100\,\%$ des images, non moyennés, mais corrigés du biais, du champ plat et des mauvais pixels. Pour le moment développons le raisonnement avec un seul cube de données et à une seule dimension (le passage 2D à 1D se fait en calculant comme précédemment la médiane dans des anneaux élémentaires). La variance totale dans le cube est donnée par :

\begin{equation}
\sigma^2(r) = \sigma_\mathrm{s}^2(r) + \sigma_{\mathrm{ph}}^2(r) + \sigma_{\mathrm{lec}}^2 + \sigma^2_{\mathrm{strehl}}(r),
\label{equation__variance_bruit}
\end{equation}
où $\sigma_\mathrm{s}^2$ représente la variance du bruit de speckle, $\sigma_{\mathrm{ph}}^2$ la variance du bruit de photon, $\sigma_{\mathrm{lec}}^2$ la variance du bruit de lecture et $\sigma_{\mathrm{strehl}}^2$ la variance du bruit causé par les variations de Strehl (variations de l'énergie dans le c\oe ur). Le bruit de photon du ciel a été omis car à ces longueurs d'onde il est négligeable. Dans la suite de cette section, j'omettrai également le terme lié aux variations de Strehl, car je m'intéresse particulièrement à la partie liée au halo et que ce type de variations se produit uniquement dans le c\oe ur de la FEP.

\begin{figure}[!p]
	\resizebox{\hsize}{!}{
		\centering\includegraphics{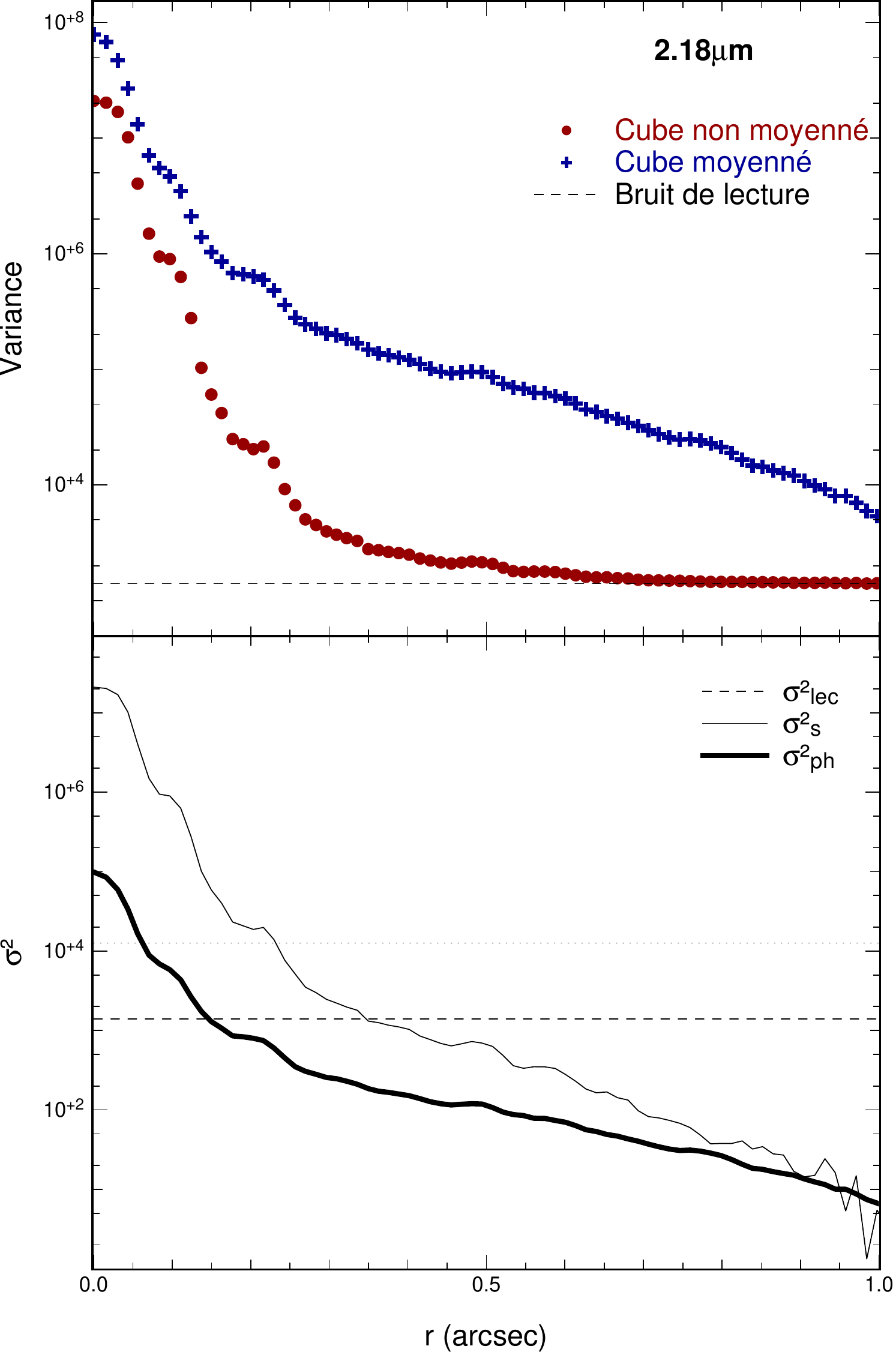} \hspace{.5cm}
		\centering\includegraphics{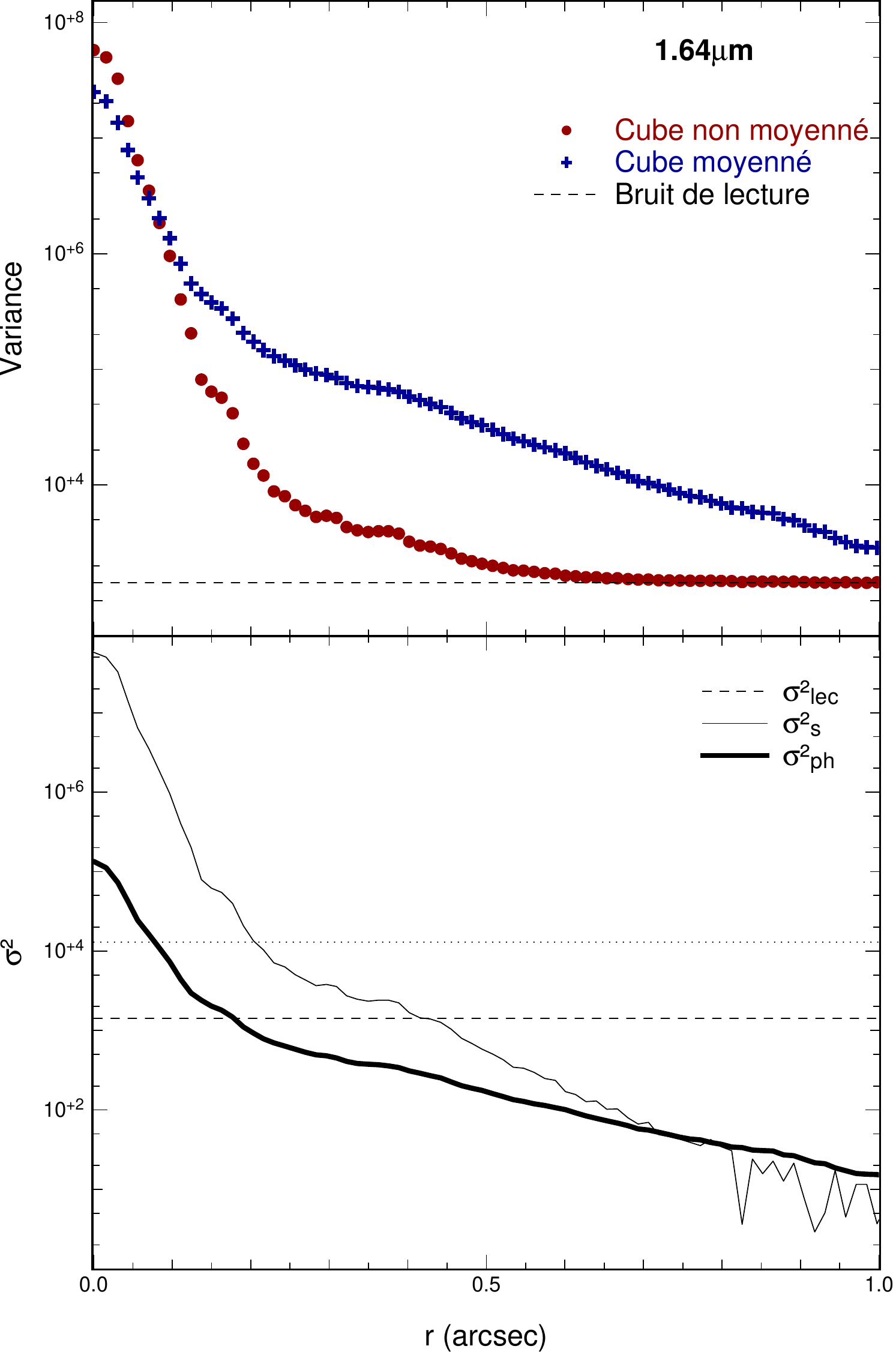}}
	\caption[Variance des différentes sources de bruit]{\textbf{Variance des différentes sources de bruit} : sur les graphiques du haut sont représentés la variance mesurée dans le cube (cercles rouges), la variance du cube moyenné (croix bleues) et le bruit de lecture (tirets noirs). Ces valeurs ont été obtenues pour un cube de l'étoile de référence HD~74417 dans chaque filtre. Sur les graphiques du bas sont représentés la contribution des différents bruits. La courbe en pointillé dénote la limite à $3\sigma$ du bruit de lecture.}
  	\label{image__variance}
\end{figure}

Certains des paramètres de l'équation~\ref{equation__variance_bruit} peuvent être estimés en calculant la moyenne du cube. Par cette opération, le bruit dominant sur l'image moyenne est le bruit de photon. En utilisant les propriétés poissonniennes de ce bruit et en notant $I(r)$ le profil d'intensité, on obtient :
\begin{displaymath}
\sigma_{\mathrm{ph}}^2(r)  = \gamma\,\frac{I(r)}{N},
\end{displaymath}
où $N$ représente le nombre d'images dans le cube et $\gamma$ le gain du détecteur, afin de convertir les différents bruits en unité d'électron ($\gamma = 11\,\mathrm{e^-/ADU}$, d'après la documentation du détecteur). Le bruit de lecture quant à lui est estimé à partir de la variance du cube, par ajustement de la partie extérieure. Je présente un exemple sur la Fig.~\ref{image__variance} pour un seul cube de l'étoile HD~74417, où j'expose la variance mesurée dans le cube non moyenné ($\sigma^2(r)$, cercles rouges), dans le cube moyenné ($\sigma^2_\mathrm{ph}$, croix bleues) ainsi que l'ajustement du bruit de lecture ($\sigma^2_\mathrm{lec}$, tirets noirs). 

Une fois tous ces bruits mesurés, le bruit de speckle peut être estimé en utilisant l'équation~\ref{equation__variance_bruit} :
\begin{displaymath}
\sigma_\mathrm{s}^2(r) = \sigma^2(r) - \sigma_{\mathrm{ph}}^2(r) - \sigma_{\mathrm{lec}}^2,
\end{displaymath}

Sur le graphique du bas de la Fig.~\ref{image__variance}, j'ai représenté l'importance relative des différentes sources de bruit. On constate que le bruit de speckle domine le bruit de photon. Plus précisément, je peux définir deux régions :

\begin{itemize} 
\compactlist
\item $\sigma(r) \simeq \sigma_\mathrm{s}(r)$ pour $ r \lesssim r_\mathrm{1}$, dans cette région, le bruit de speckle domine les autres bruits ($\sigma_\mathrm{s} > 3\,\sigma_{\mathrm{lec}}$)
\item $\sigma(r) \simeq \sqrt{ \sigma_\mathrm{s}^2(r) + \sigma_{\mathrm{lec}}^2} $ pour $r \gtrsim r_\mathrm{1} $
\end{itemize}
avec $r_\mathrm{1} \simeq 0.22\arcsec$ à $1.64\,\mu\mathrm{m}$ et $r_\mathrm{1} \simeq 0.24\arcsec$ à $2.18\,\mu\mathrm{m}$. On remarque que ces rayons limites ont le même ordre de grandeur dans les deux filtres. 

En fait il y a probablement une autre région dans l'intervalle $r < r_\mathrm{1}$, liée aux variations de Strehl affectant principalement le c\oe ur de la FEP. Toutefois je m'intéresse au halo de la FEP, et particulièrement à l'espace inter-speckle et c'est la raison pour laquelle j'ai négligé ce terme de variations auparavant. Comme nous le verrons par la suite, la région d'intérêt correspond à $r \gtrsim r_\mathrm{1} $.

J'introduis maintenant une propriété de la variance du bruit de speckle. J'utilise une formulation de la variance déjà introduite par \citet{Racine-1999-05} mais sous une forme plus générale.

\paragraph*{\textcolor{black}{Fonction invariante}}

\citet{Racine-1999-05} ont fourni une expression intéressante du bruit de speckle, stipulant que sa variance est proportionnelle au carré du flux. Cette propriété implique donc que le rapport $F / \sigma$ est un invariant. Je reprend l'équation~10 de \citet{Racine-1999-05} d'une manière plus générale telle que :
\begin{displaymath}
\sigma_\mathrm{s} = g(r,\lambda,\Delta t,S,...)\,F_\star (\lambda),
\end{displaymath}
où la fonction $g$ est une fonction dépendante de la longueur d'onde $\lambda$, du temps d'intégration $\Delta t$, du rapport de Strehl $S$ et de paramètres atmosphériques.

\begin{figure}[!t]
	\resizebox{\hsize}{!}{
		\centering\includegraphics{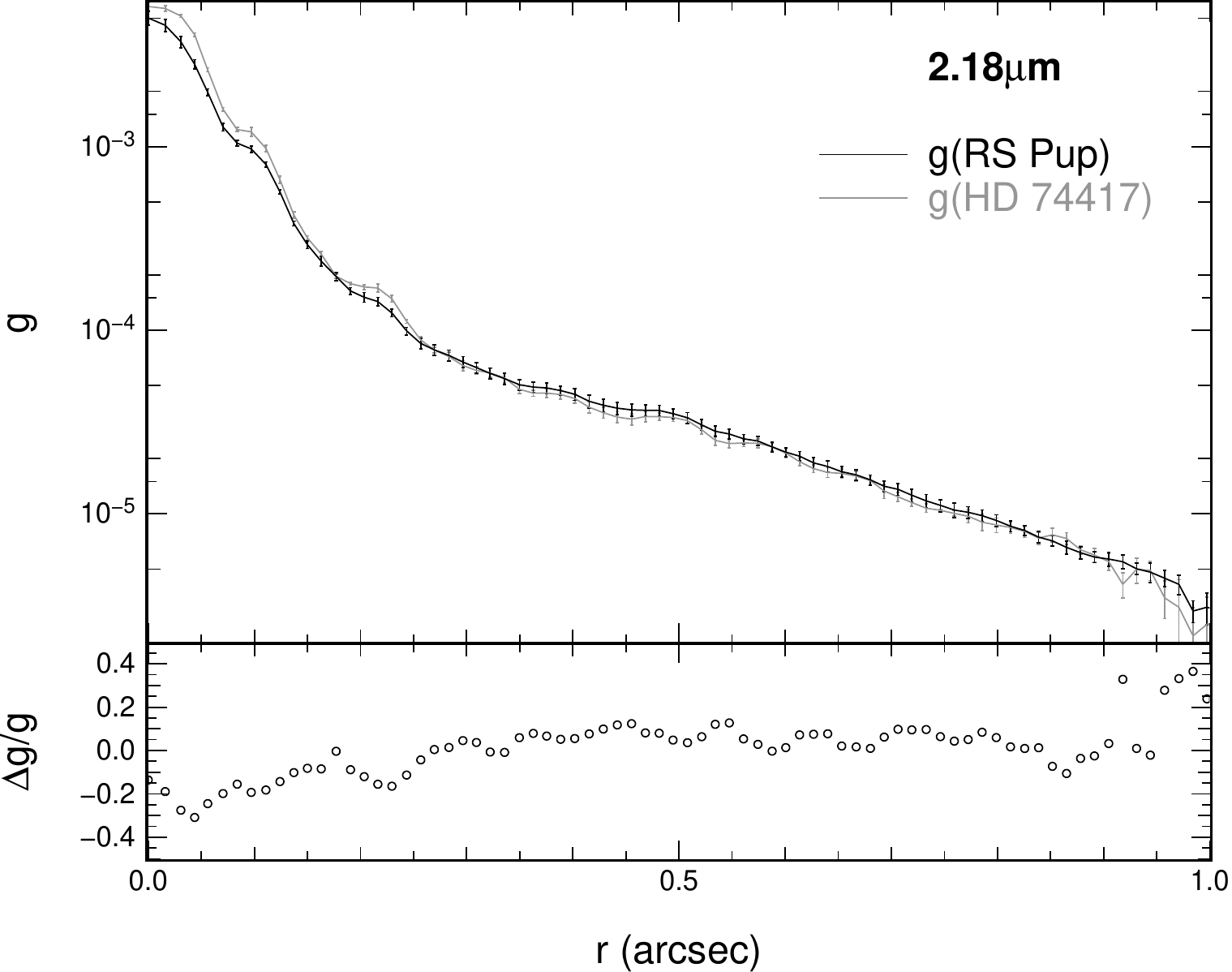} \hspace{.5cm}
		\centering\includegraphics{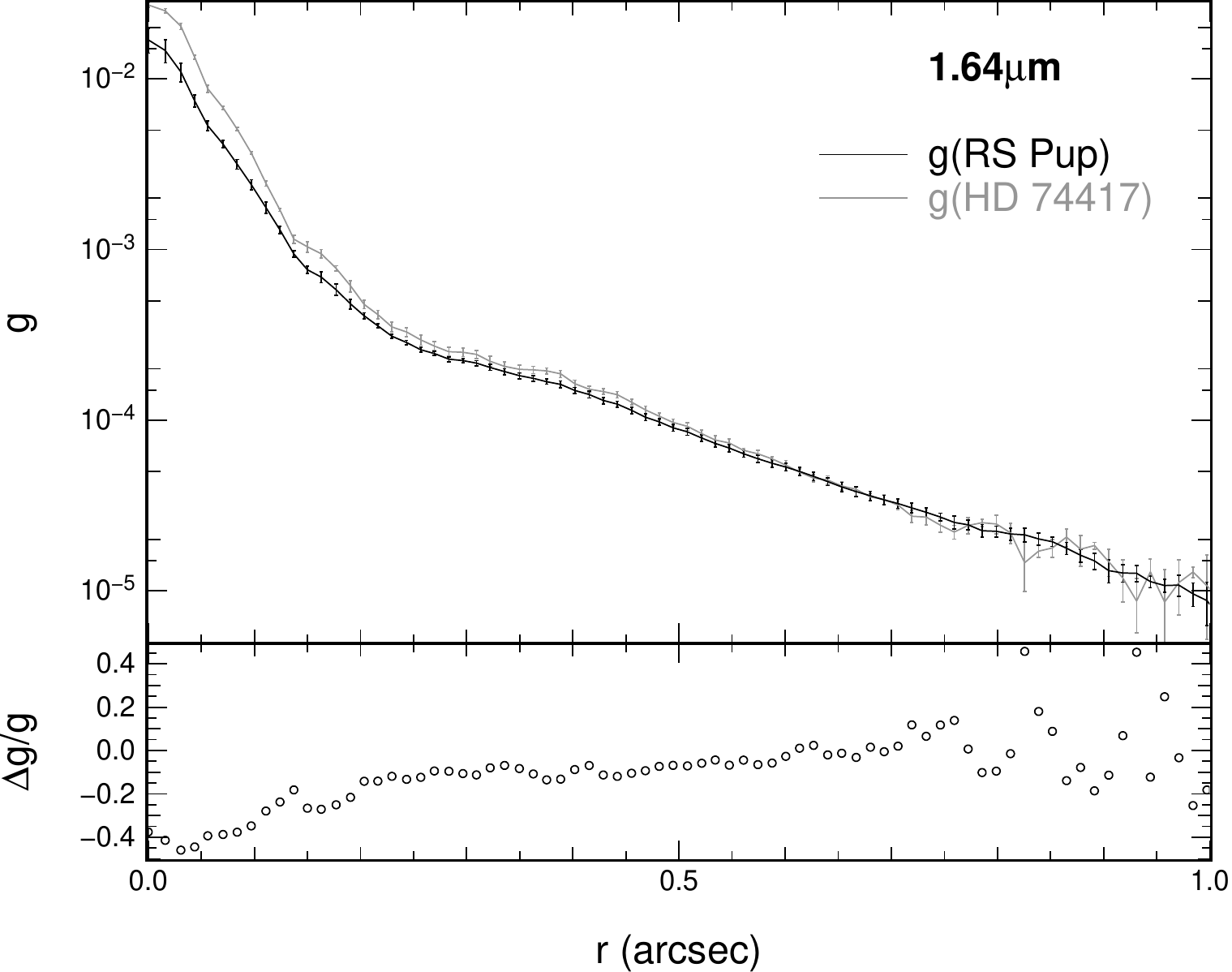}}
	\caption[Fonction invariante]{\textbf{Fonction invariante} : fonction $g$ moyenne pour chaque étoile et pour chaque filtre. Les barres d'erreurs correspondent à l'écart-type estimé dans chaque anneau utilisé lors du profil radial.}
  	\label{image__paramètre_g}
\end{figure}

Commençons par vérifier cette propriété à partir de nos données. Pour chaque étoile, j'estime dans chaque cube et pour chaque pixel la variance (temporelle) que je divise par le flux total moyen. Le résultat est une fonction $g$ en deux dimensions pour chaque cube. Je moyenne ensuite chaque fonction et calcule un profil radial pour chaque étoile (via une moyenne en anneau). Cette fonction est présentée sur la Fig.~\ref{image__paramètre_g} pour la Céphéide et HD~74417. J'ai pris soin de retirer à RS~Pup la contribution de l'enveloppe en utilisant les valeurs calculées dans la Section~\ref{subsection__methode_shift_and_add}. À part dans le c\oe ur, il n'y a pas de différences significatives entre les deux étoiles. Le c\oe ur est le siège des variations de Strehl omises précédemment et que l'on retrouve ici. On peut avoir un ordre de grandeur de ces variations en utilisant, comme précédemment, l'énergie encerclée comme estimateur du rapport de Strehl. Contrairement à la section précédente, je calcule une erreur relative ($\sigma(E_{\mathrm{enc}})/E_\mathrm{{enc}}$) sur $100\,\%$ des images des cubes. Je trouve une erreur relative à $2.18\,\mu\mathrm{m}$ de $\sim15\,\%$ pour RS~Pup et $\sim8\,\%$ pour HD~74417, et respectivement de $\sim17\,\%$ et $\sim9\,\%$ à $1.64\,\mu\mathrm{m}$.

L'invariance totale, c'est à dire pour tout $r$, de la fonction $g$ dépend donc des variations de Strehl. Cependant, comme déjà mentionné plusieurs fois précédemment, ces variations ont lieu dans le c\oe ur de la FEP et peuvent être négligées si on s'éloigne de la partie centrale. C'est là qu'interviennent les deux régions définies précédemment. En se plaçant dans la région $r > r_\mathrm{1}$, on est assez loin du c\oe ur pour ne pas être sensible à ces variations. Dans cette intervalle, l'invariance de la fonction $g$ est vérifiée, avec une différence relative moyenne entre les deux étoiles de $5\,\%$ à $2.18\,\mu\mathrm{m}$ et $8\,\%$ à $1.64\,\mu\mathrm{m}$.

Introduisons maintenant le paramètre $\Gamma(x,y)$, fonction du flux et de l'écart-type dans un cube, tel que :
\begin{displaymath}
\Gamma(x,y) = \frac{F_\star(x,y)}{\sigma(x,y)}
\end{displaymath}

En notant l'indice "cep" et "ref" respectivement pour la Céphéide et HD~74417 et en supprimant la notation $(x,y)$ pour la clarté, on obtient :
\begin{displaymath}
\Gamma_\mathrm{cep} = \frac{F_\mathrm{cep} + F_\mathrm{env}}{\sigma_\mathrm{cep}}\quad \mathrm{et}\quad 
\Gamma_\mathrm{ref} = \frac{F_\mathrm{ref}}{\sigma_\mathrm{ref}}
\end{displaymath}
où $F_\mathrm{cep}$ et $F_\mathrm{ref}$ représentent les flux photosphériques respectifs et $F_\mathrm{env}$ le flux de l'enveloppe autour de la Céphéide. Pour $r > r_\mathrm{1}$ et en définissant $F_\mathrm{ref} = \alpha\,F_\mathrm{cep} = \alpha\,F_\star$, on a :
\begin{displaymath}
\Gamma_{\mathrm{cep}} = \frac{F_\star + F_{\mathrm{env}}}{ \sqrt{ g_\star^2  F_\star^2 + \sigma_{\mathrm{lec}}^2 } }\quad \mathrm{et}\quad 
 \Gamma_{\mathrm{ref}} = \frac{\alpha F_\star}{ \sqrt{ g_\star^2 \alpha^2 F_\star^2 + \sigma_{\mathrm{lec}}^2}}
\end{displaymath}

L'idée maintenant est de soustraire ces deux paramètres :
\begin{displaymath}
\xi = \Gamma_{\mathrm{cep}} -\Gamma_{\mathrm{ref}}= \frac{F_\star + F_{\mathrm{env}}}{ \sqrt{ g_\star^2\,F_\star^2 + \sigma_{\mathrm{lec}}^2 } } - \frac{\alpha\,F_\star}{ \sqrt{ g_\star^2\,\alpha^2 F_\star^2 + \sigma_{\mathrm{lec}}^2}}
\end{displaymath}

Le bruit de lecture peut aisément être estimé, comme je l'ai montré précédemment, et peut donc être soustrait à la variance. On obtient au final un paramètre directement proportionnel au flux de l'enveloppe :
\begin{equation}
\xi = \frac{F_\mathrm{{env}}}{g_\star\,F_\star}
\label{equation__parametre_xi}
\end{equation}

Cette fonction $\xi$ doit donc être nulle s'il n'y a pas d'enveloppe. De plus, comme le dénominateur est un paramètre radial, toute asymétrie de $\xi$ sera lié à une asymétrie de l'enveloppe. Je discuterai de ce point un peu plus loin.

\begin{figure}[!p]
	\resizebox{\hsize}{!}{
		\centering\includegraphics{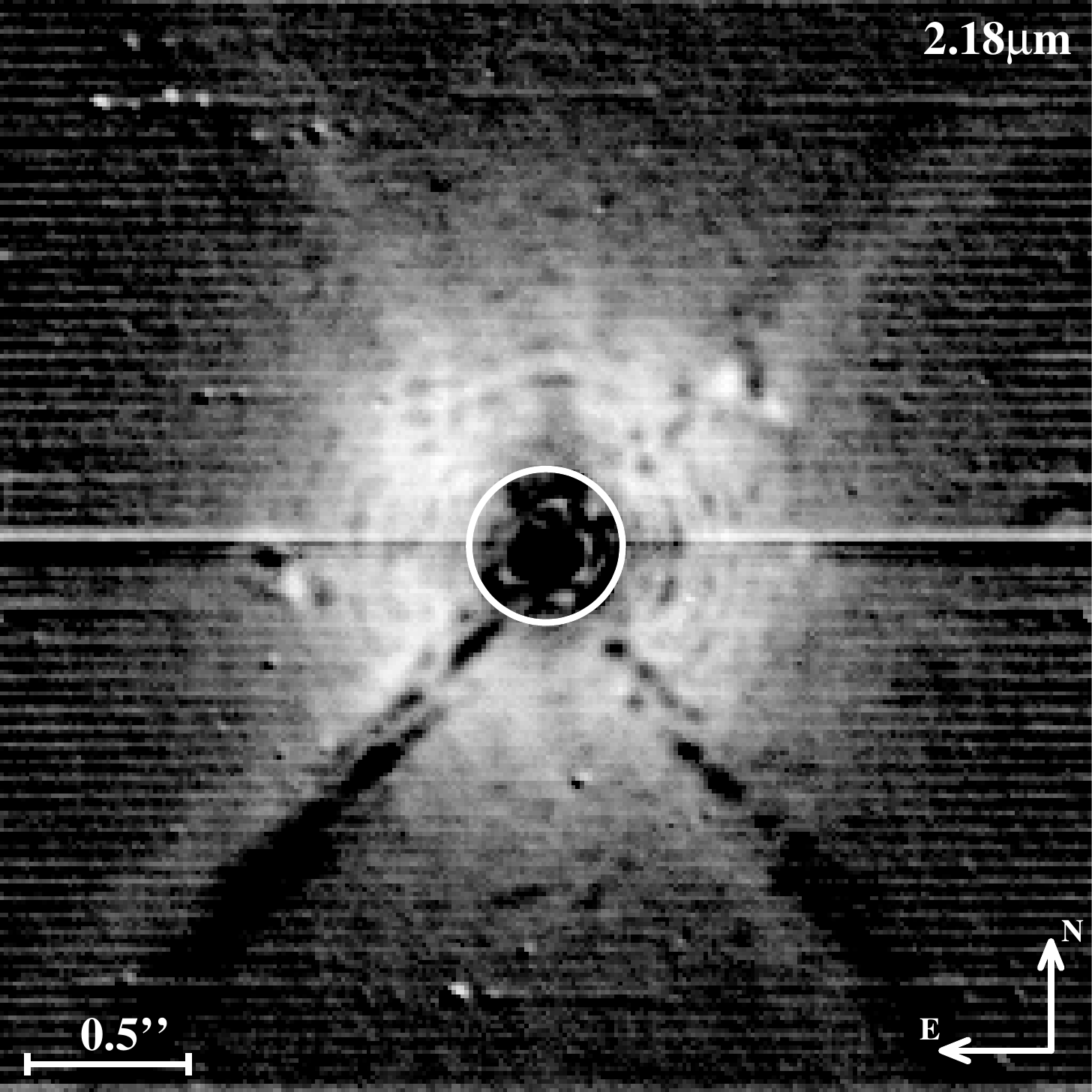} \hspace{.5cm}
		\centering\includegraphics{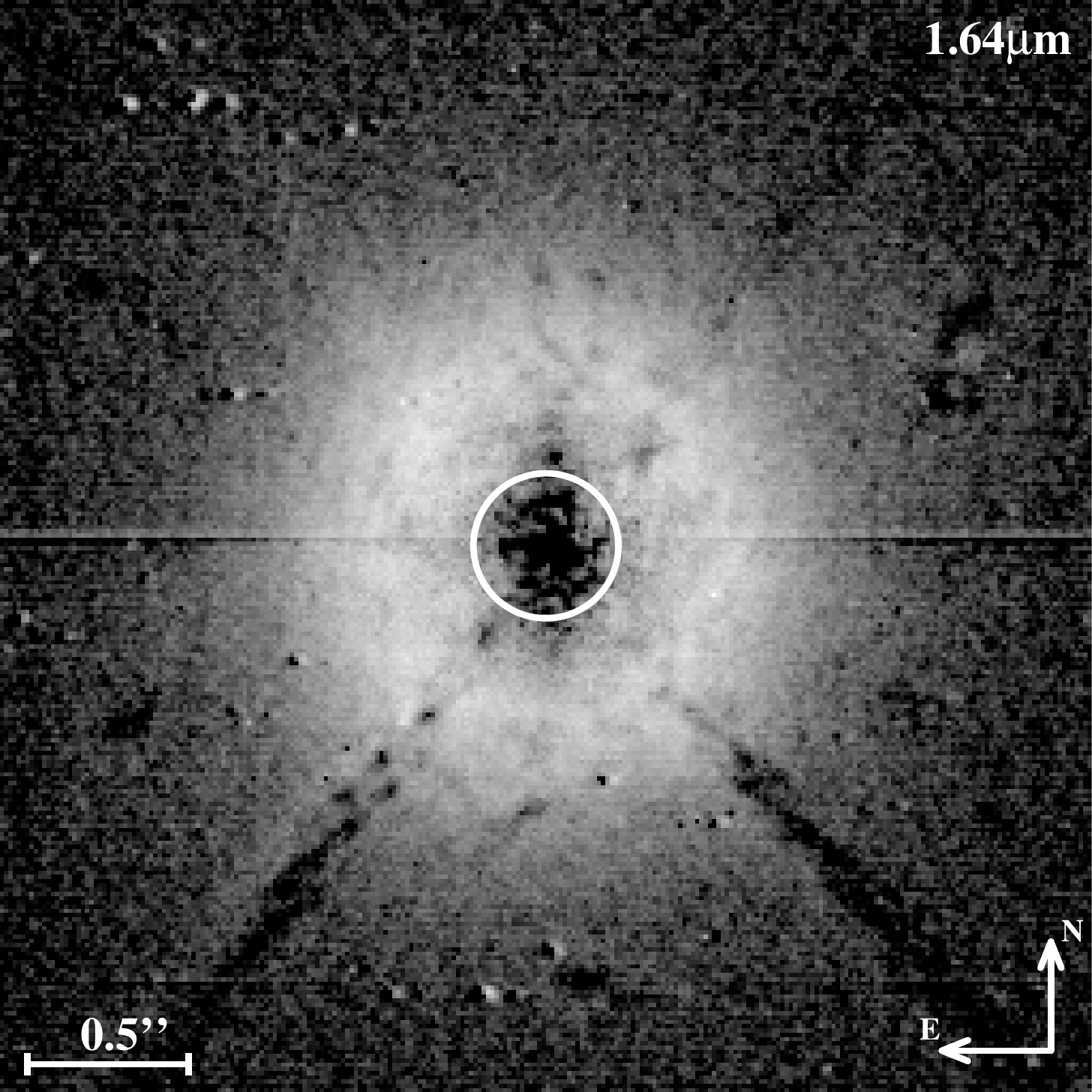}}
	\caption[Paramètre $\xi$]{\textbf{Paramètre $\xi$} : image proportionnelle au flux de l'enveloppe dans les deux filtres (Équ.~\ref{equation__parametre_xi}). Le cercle blanc représente la limite $r = r_\mathrm{1}$. L'échelle d'intensité est logarithmique.}
  	\label{image__parametre_xi}
\end{figure}

Combinons maintenant toutes les données afin d'obtenir une fonction moyenne. Le procédé général d'estimation de $\xi$ pour chaque cube est le suivant : calcul de la variance totale du cube pour chaque pixel, ajustement du bruit de lecture, soustraction du bruit de lecture à la variance, moyenne du cube pour obtenir le flux et effectuer le rapport du flux sur la racine carrée de la variance. J'obtiens ainsi un paramètre $\Gamma$ pour chaque cube de chaque étoile. Pour l'étoile de référence, j'effectue une moyenne :
\begin{displaymath}
\overline{\Gamma_{\mathrm{cal}}(x,y)} = \frac{1}{m} \sum_{i=1}^m \Gamma_\mathrm{cal}(x,y)_\mathrm{i}
\end{displaymath}
où $m = 4$. J'estime ensuite une fonction $\xi$ moyenne telle que :
\begin{displaymath}
\overline{\xi(x,y)} = \frac{1}{n} \sum_{j=1}^n \left[ \Gamma_\mathrm{sci}(x,y)_\mathrm{j}  - \overline{\Gamma_{\mathrm{cal}}(x,y)} \right]
\end{displaymath}
où $n = 10$. J'obtiens ainsi une image moyenne proportionnelle au flux de l'enveloppe. Je présente ces images sur la Fig.~\ref{image__parametre_xi}.

\begin{figure}[!p]
	\resizebox{\hsize}{!}{
		\centering\includegraphics{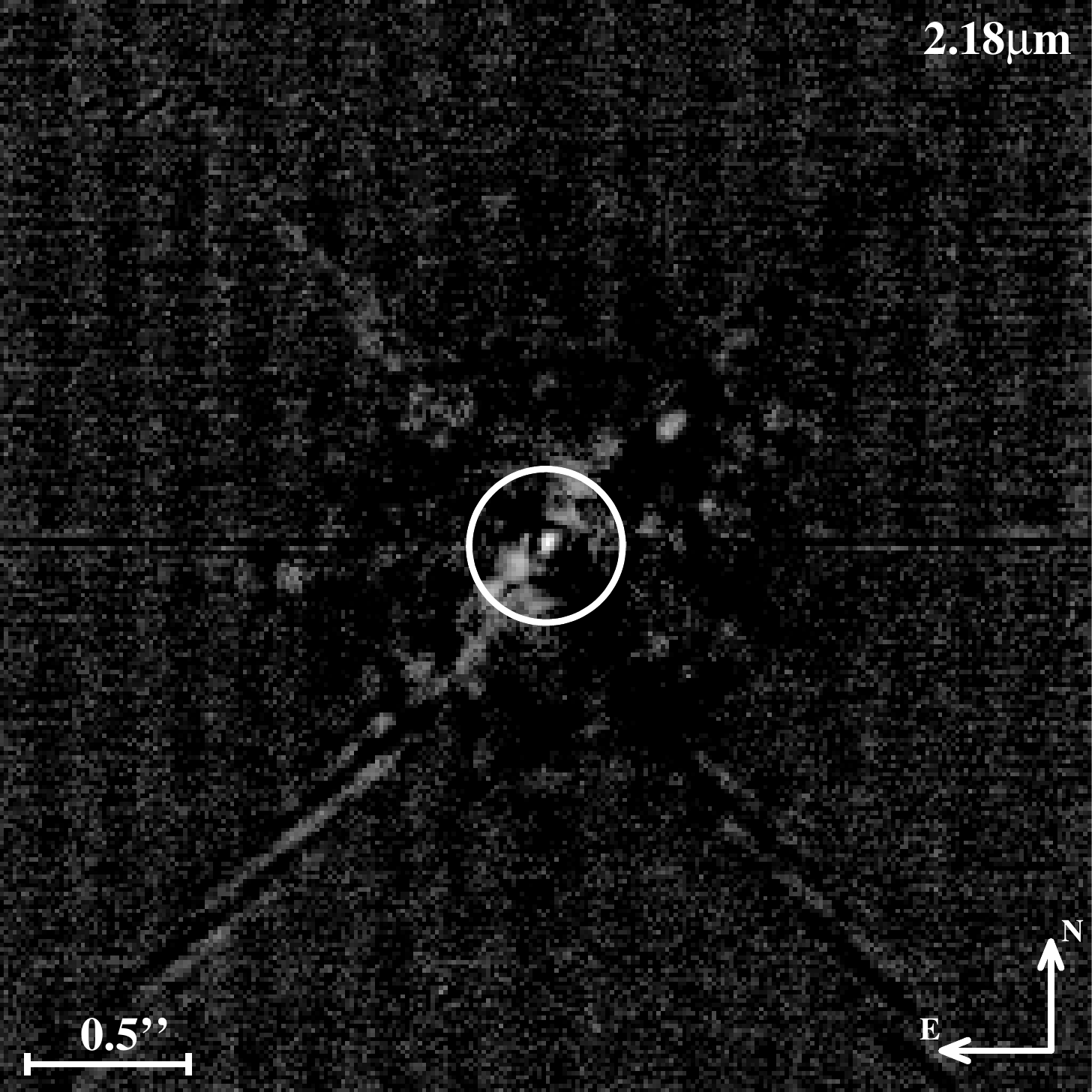} \hspace{.5cm}
		\centering\includegraphics{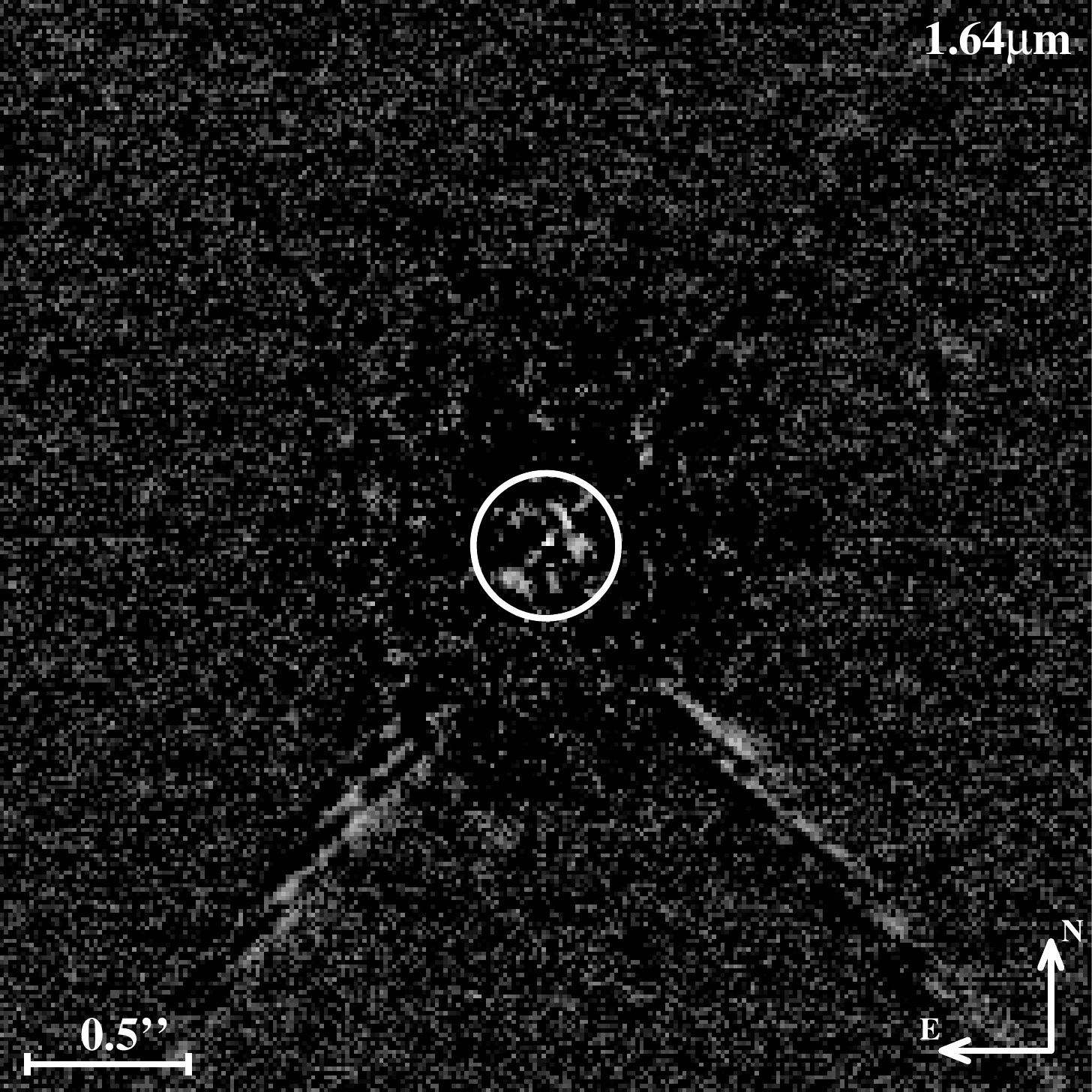}}
	\caption[Test du paramètre $\xi$]{\textbf{Test du paramètre $\xi$} : test de  l'équation~\ref{equation__parametre_xi} pour deux cubes de l'étoile de référence. Le cercle blanc représente la limite $r = r_\mathrm{1}$ définie précédemment. L'échelle d'intensité est logarithmique.}
  	\label{image__parametre_xi_test}
\end{figure}

Pour se convaincre qu'il s'agit bien de l'enveloppe, nous pouvons utiliser l'étoile de référence comme test. Si je calcule le paramètre $\xi$ en utilisant seulement deux cubes de HD~74417, je devrais trouver $\xi\sim0$, excepté peut être au centre où les variations de Strehl dominent. Le résultat est exposé sur la Fig.~\ref{image__parametre_xi_test} pour les deux filtres. On voit clairement, malgré quelques résidus, la différence avec les images RS~Pup. Les cercles blancs correspondent aux mêmes rayons limites définis précédemment. Ce test renforce le résultat antérieur en ce qui concerne la détection de l'enveloppe autour de la Céphéide par cette méthode.

\paragraph*{\textcolor{black}{Morphologie de l'enveloppe}}

On s'attend à ce que l'équation~\ref{equation__parametre_xi} soit nulle s'il n'y pas d'enveloppe. La Fig.~\ref{image__parametre_xi} montre clairement une image positive, indicative d'une détection d'enveloppe. La partie centrale $r < r_\mathrm{1}$ représente la région où la fonction $g$ précédemment définie n'est pas invariante à cause des variations de Strehl. Ces variations sont difficiles à estimer avec précision et par conséquent la région proche de l'étoile ne fournit pas d'information fiable sur l'enveloppe. En revanche la région $r > r_\mathrm{1}$ est tout à fait fiable et révèle la présence d'une enveloppe.

Cette méthode est utile pour étudier la morphologie de l'enveloppe, en particulier pour vérifier s'il y a une symétrie centrale. La Fig.~\ref{image__parametre_xi} semble indiquer une distribution d'intensité uniforme, qui serait compatible avec une enveloppe sphérique ou un disque vu de face (c'est à dire avec un angle d'inclinaison $i\sim90^\circ$).

\begin{figure}[!p]
		\centering\includegraphics[width=.4\linewidth]{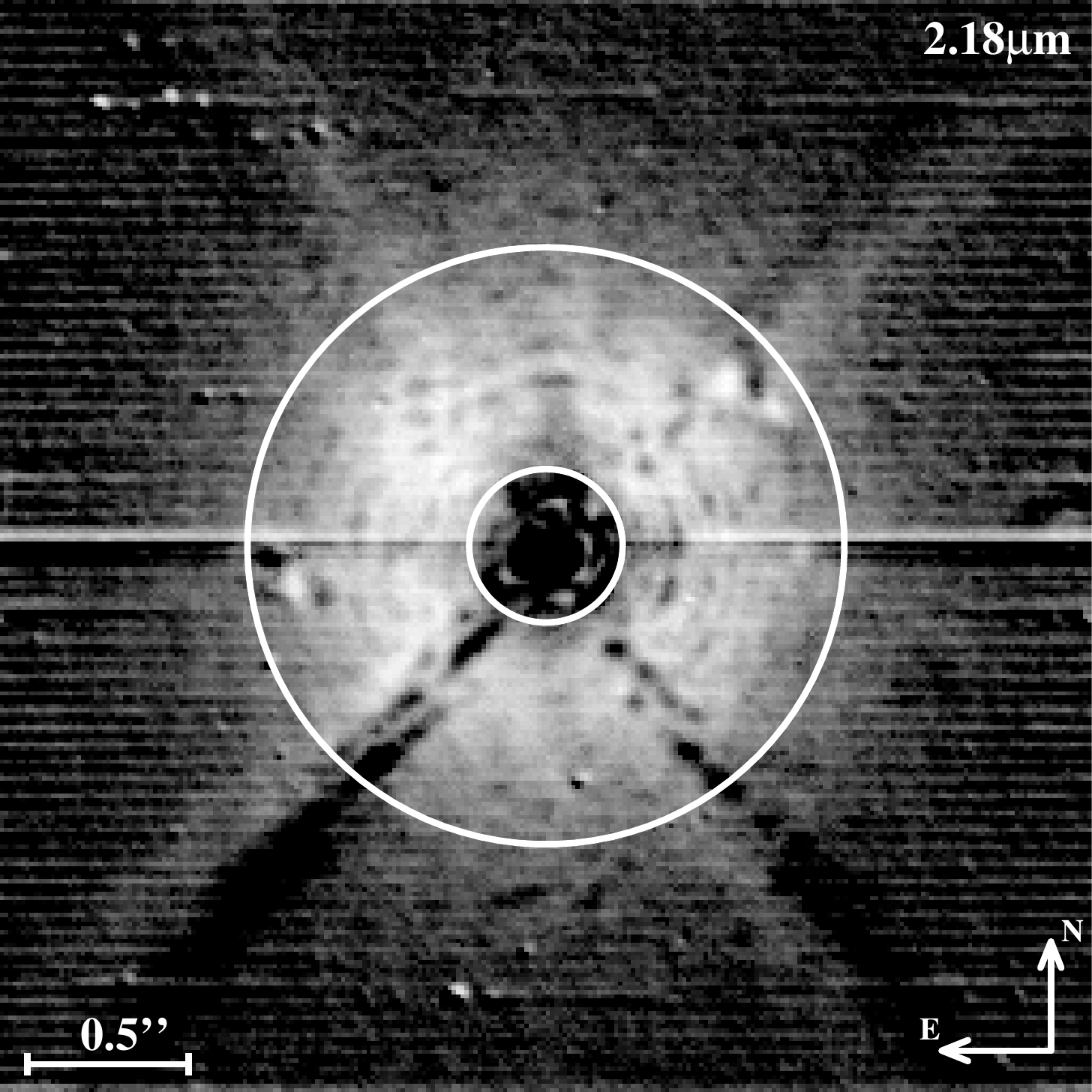} \hspace{.1mm}
		\centering\includegraphics[width=.4\linewidth]{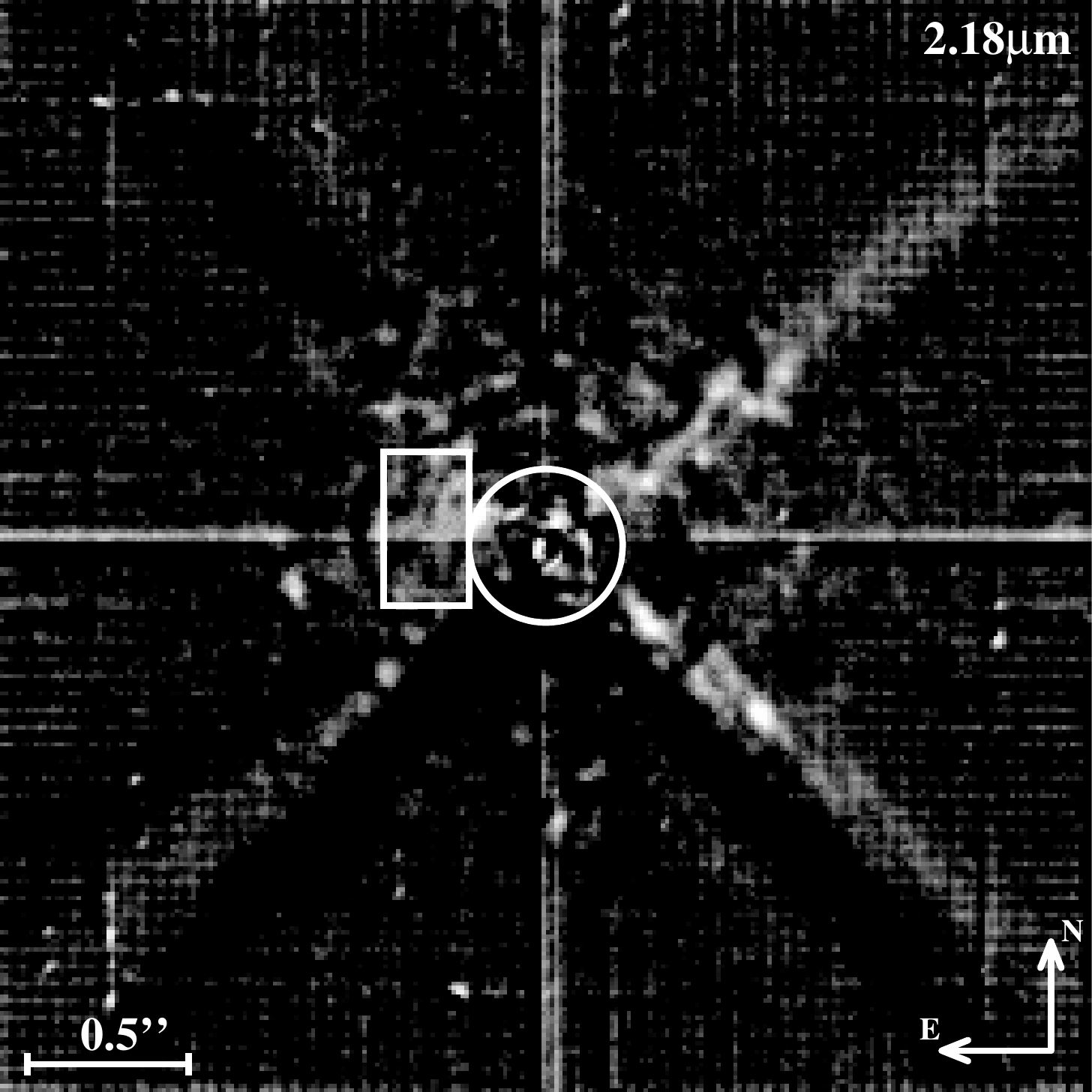} \\[1.5mm]
		\centering\includegraphics[width=.4\linewidth]{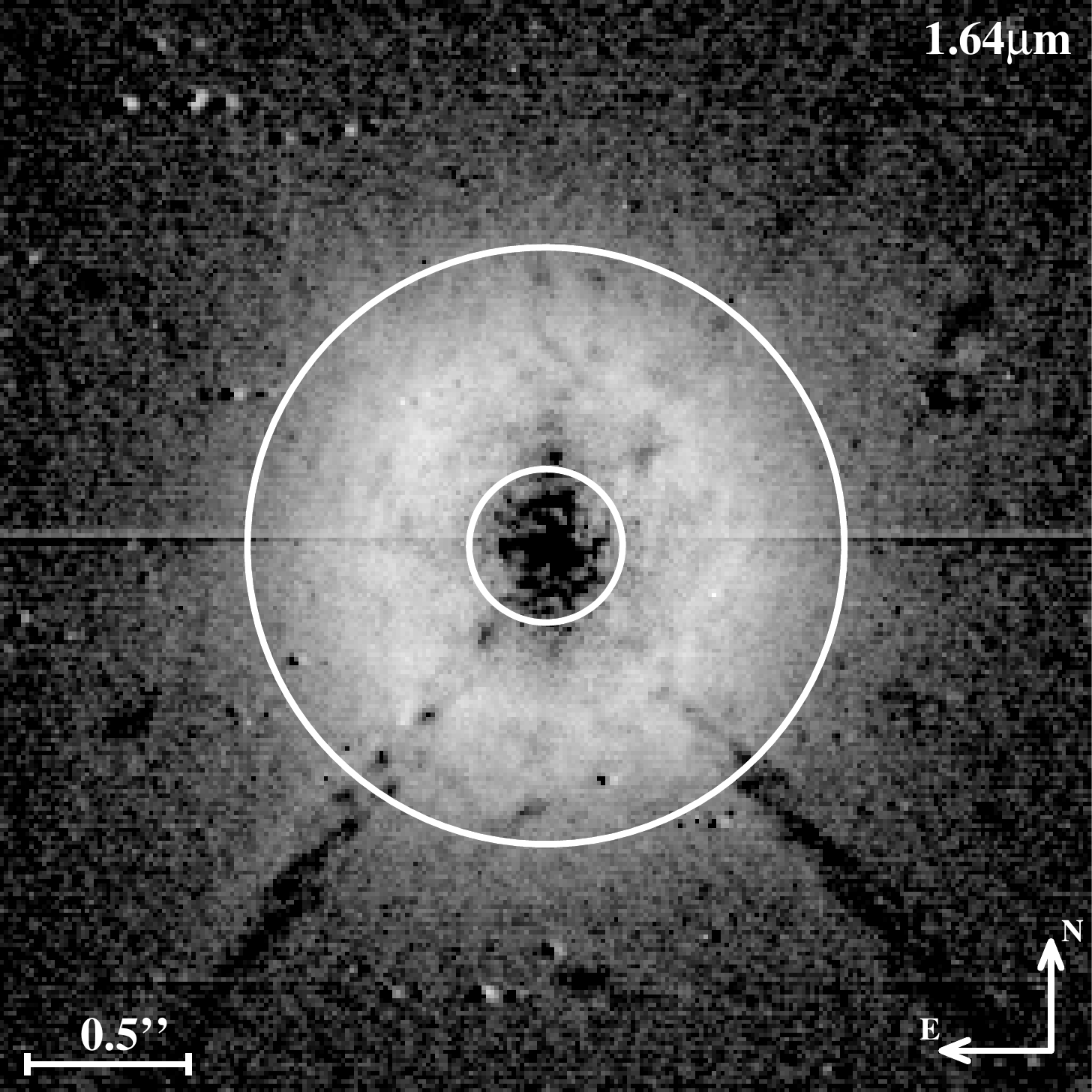} \hspace{.1mm}
		\centering\includegraphics[width=.4\linewidth]{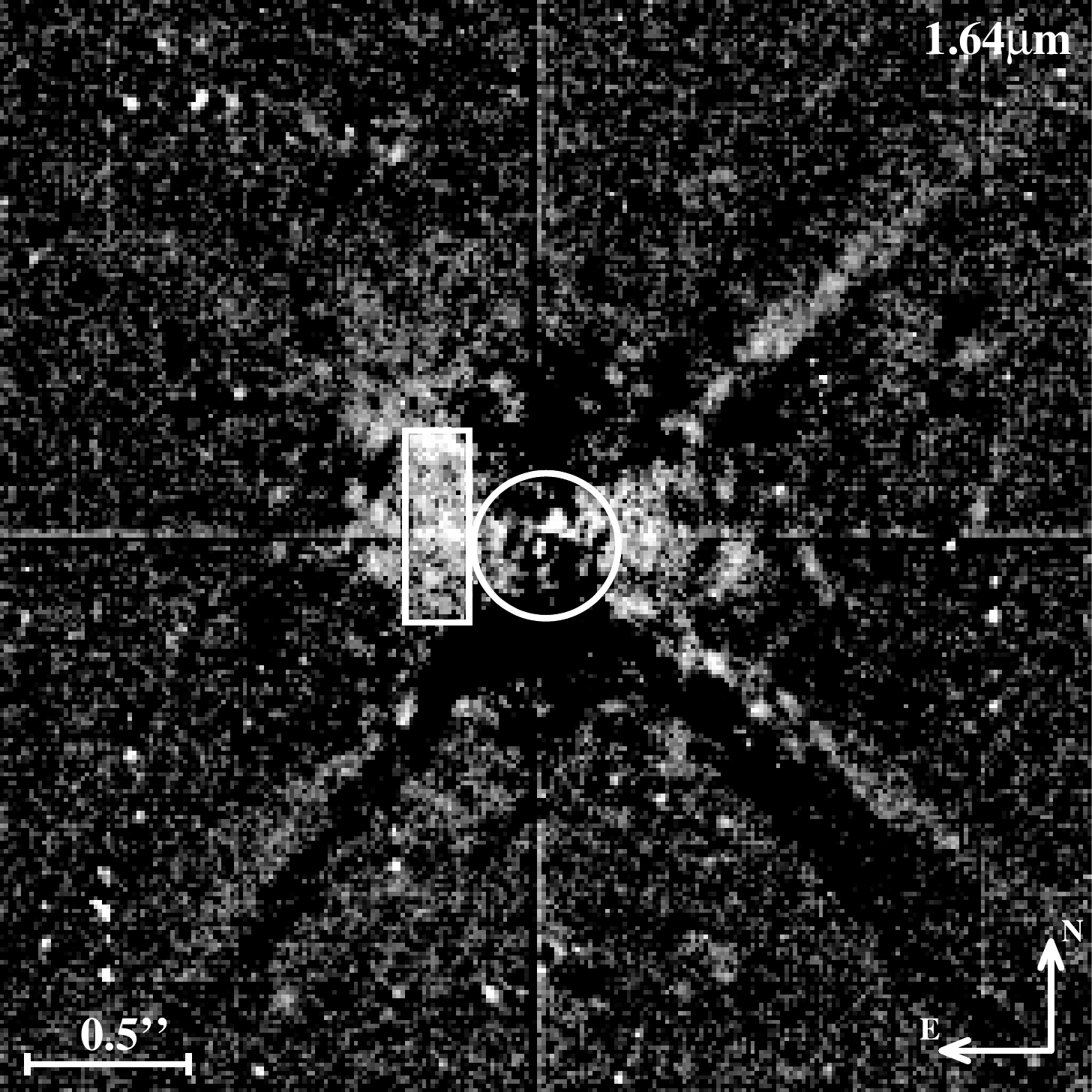}
	\caption[Symétrie de l'enveloppe]{\textbf{Symétrie de l'enveloppe} : estimation du niveau de symétrie en faisant la soustraction des images de la Fig.~\ref{image__parametre_xi} par elles-mêmes tournées de $90^\circ$. La valeur moyenne dans une portion de l'image résiduelle (images 2 et 4) divisée par la valeur moyenne dans un anneau $r_\mathrm{1} < r < r_\mathrm{2}$ de l'image initiale (images 1 et 3) donne le niveau d'asymétrie. L'échelle d'intensité est logarithmique.}
  	\label{image__symetrie}
\end{figure}

Pour estimer un niveau de symétrie, j'utilise une image résiduelle que j'obtiens en faisant la soustraction de l'image de la Fig.~\ref{image__parametre_xi} par elle-même tournée de $90^\circ$. Le résultat de cette différence est présenté sur la Fig.~\ref{image__symetrie}. Dans l'image initiale, je calcule la valeur moyenne à l'intérieur d'une couronne de rayon $r_\mathrm{1} < r < r_\mathrm{2}$ où $r_\mathrm{1}$ est le rayon limite défini précédemment ($0.22\arcsec$ à $1.64\,\mu\mathrm{m}$ et $0.24\arcsec$ à $2.18\,\mu\mathrm{m}$) et $r_\mathrm{2} = 1.10\arcsec$ à $1.64\,\mu\mathrm{m}$ et $r_\mathrm{2} = 0.92\arcsec$ à $2.18\,\mu\mathrm{m}$. Ces rayons sont représentés par des cercles blancs sur la Fig.~\ref{image__symetrie}. Ensuite je calcule une valeur moyenne sur une portion de l'image résiduelle (représentée par un rectangle blanc sur chaque image de la Fig.~\ref{image__symetrie}). Je considère le rapport de ces deux valeurs moyennes comme un estimation du niveau de symétrie (en terme de variation de flux). Je ne détecte pas d'asymétrie (ou variation) plus grande que $7\,\%$ à $1.64\,\mu\mathrm{m}$ et $6\,\%$ à $2.18\,\mu\mathrm{m}$.

\subsection{Conclusion de cette étude}

L'étude de cette Céphéide avec \emph{NACO} a permis d'estimer deux paramètres importants. Le premier concerne le rapport de flux entre l'enveloppe et l'étoile, qui rappelons le est une variable à ne pas négliger dans le cas de mesures de distance dans l'infrarouge. J'ai montré que l'enveloppe avait une émission photométrique de $38\pm17\,\%$ à $1.64\,\mu\mathrm{m}$ et de $24\pm11\,\%$ à $2.18\,\mu\mathrm{m}$, relatif au flux photosphérique. J'ai également trouvé une géométrie centro-symétrique dans les deux bandes, indiquant une morphologie de type sphérique ou disque vu de face.

Cela révèle également la présence d'hydrogène dans l'environnement de la Céphéide. En effet, les filtres NB\_1.64 et IB\_2.18 isolent les transitions électroniques Brackett 12--4 et Brackett 7--4 de l'hydrogène, et l'émission observée correspondrait donc à des raies de recombinaison. Notons également qu'à ces longueurs d'onde, le contraste entre l'étoile et l'enveloppe est diminué à cause de raies d'absorptions de l'hydrogène dans la photosphère de l'étoile, expliquant ainsi l'importante émission relative que l'on mesure. L'émission peut également provenir d'un rayonnement continu de type émission free-free ou encore de la diffusion par les poussières présentes dans l'enveloppe. Toutefois cette contribution doit être de quelques pour cent seulement si l'on se fie à la distribution spectrale d'énergie \citep[Fig.~\ref{image__SED_RS_PUP},][]{Kervella-2009-05}. Par conséquent l'émission observée n'est probablement pas de type continu mais plutôt une émission dans des raies de l'hydrogène.

La méthode originale présentée dans ce chapitre peut être utilisée sur d'autres Céphéides présentant une enveloppe résolue spatialement. Par exemple, \citet{Barmby-2010-11} ont détecté autour de 6 autres Céphéides une probable émission infrarouge résolue, et ce résultat pourrait être confirmé en utilisant ce type d'analyse.

Enfin, il n'est pas pertinent ici de quantifier un biais sur la mesure de diamètre car ces mesures photométriques ont été effectuées dans une bande spectrale étroite, alors que les estimations de diamètre sont généralement en bande large. Avec une bande spectrale plus grande, le rapport de flux sera probablement plus faible car on sera plus sensible à l'émission continu de l'étoile.

\cleardoublepage     

\pagestyle{fancy}
\fancyhf{}
\lhead[\nouppercase{\emph{\thepage}}]{\nouppercase{\emph{\rightmark}}}
\rhead[\nouppercase{\emph{\leftmark}}]{\nouppercase{\emph{\thepage}}}
\newpage

\chapter[Étude d'excès infrarouge par photométrie]{\emph{Étude d'excès infrarouge par photométrie}}
\label{chapitre__etude_d_exces_infrarouge_par_photometrie}

\thispagestyle{empty}

\vspace*{-1cm}

\refbleu  
\textcolor{bleu_chapitre}{\minitoc}
\refnoir  

\section{Introduction}

\malettrine{L}{}a photométrie est l'une des techniques de l'astronomie les plus anciennes. Elle consiste en la mesure de la quantité de lumière reçue des objets astrophysiques. Cette quantité est nommée flux. Les toutes premières mesures photométriques remontent à l'Antiquité où l'astronome grec Hipparque effectua un catalogue d'environ 850 étoiles, à l'\oe il nu. Il définit un système de mesure tel que l'étoile la plus brillante soit de magnitude 1 et la plus faible de magnitude 6. Ce système fut repris ensuite par Ptolémée dans son Almageste et servit de base à l'astronomie pour les siècles qui suivirent.

L'avènement d'outils comme le télescope et les caméras CCD ont permis de faire des progrès considérables en terme de mesures photométriques. Nous pouvons maintenant effectuer des mesures de flux dans diverses gammes de longueur d'onde (ou d'énergie) et étudier la distribution spectrale de l'énergie (SED) des objets célestes. Couplée à d'autres paramètres, la mesure de flux fournit des informations sur d'autres grandeurs physiques, tels que sa luminosité (énergie totale), sa température ou sa taille.

Toutefois, mesurer une quantité absolue, c'est à dire intrinsèque à l'objet céleste, est assez délicat car la quantité de photons voyageant de l'objet jusqu'au détecteur est soumise à divers effets venant compromettre sa mesure. Les poussières situées sur la ligne de visée absorbent et diffusent une certaine quantité de photons, réduisant pas conséquent le flux mesuré. Ce phénomène, connu sous le nom d'extinction interstellaire, est chromatique, diminue avec la longueur d'onde et décroit avec la distance au plan Galactique. Une loi d'extinction interstellaire est généralement appliquée pour corriger cet effet. Le passage des photons dans l'atmosphère entraîne également diverses difficultés. La première concerne la transparence de l'atmosphère à certaines longueurs d'onde. Il existe des "fenêtres atmosphériques" que les rayonnements d'une certaine énergie peuvent traverser, en dehors, l'atmosphère est opaque. Depuis le sol, on est donc restreint à une partie du spectre électromagnétique. La seconde difficulté est le seeing, qui a pour effet une perte en résolution et une mesure de flux plus difficile pour des objets peu brillants. L'utilisation du lucky-imaging introduit dans le chapitre précédent permet d'améliorer ce problème lié à la turbulence. 

À cause de processus atomiques dans l'air, le ciel rayonne dans l'infrarouge et empêche par conséquent la détection d'objets de faible luminosité. La lumière émise par le ciel est donc également une difficulté à la mesure précise du flux à certaines longueurs d'onde. La majeure partie de ce rayonnement peut être supprimée en utilisant la technique du chopping-nodding. Notons également que l'atmosphère est un milieu absorbant et diffusant, causant une perte de photons supplémentaire. La masse d'air traversée doit être minimale pour réduire ces effets.

Lors de ma thèse, j'ai effectué des mesures photométriques pour étudier l'excès IR des Céphéides. Après avoir réduit et analysé les données de l'instrument \emph{VLT/VISIR} aux longueurs d'onde thermiques, j'ai examiné la distribution spectrale d'énergie d'un échantillon de 11 étoiles, où j'ai pu mettre en évidence une émission IR pour certaines d'entres elles. Ce chapitre expose les méthodes et résultats obtenus avec ce jeu de données qui ont été soumis dans la revue Astronomy \& Astrophysics.

Dans un premier temps j'expose quelques notions de base de photométrie stellaire, des principales grandeurs physiques et des obstacles rencontrés par un photon avant sa détection. Je parlerai ensuite de la technique du chopping-nodding utilisée lors des observations \emph{VISIR} et utile à la soustraction du rayonnement du fond de ciel. J'exposerai ensuite les deux types d'analyses effectuées, l'une basée sur l'étude de la SED, et l'autre sur une technique de Fourier pour la recherche d'une émission spatialement étendue. Enfin, je conclurai l'analyse de ces observations par une discussion des résultats obtenues.

\section{Notions de photométrie stellaire}
\label{section__notion_de_photometrie}

\subsection{Intensité, flux et luminosité}

\begin{figure}[!p]
  \centering\includegraphics[width =\linewidth]{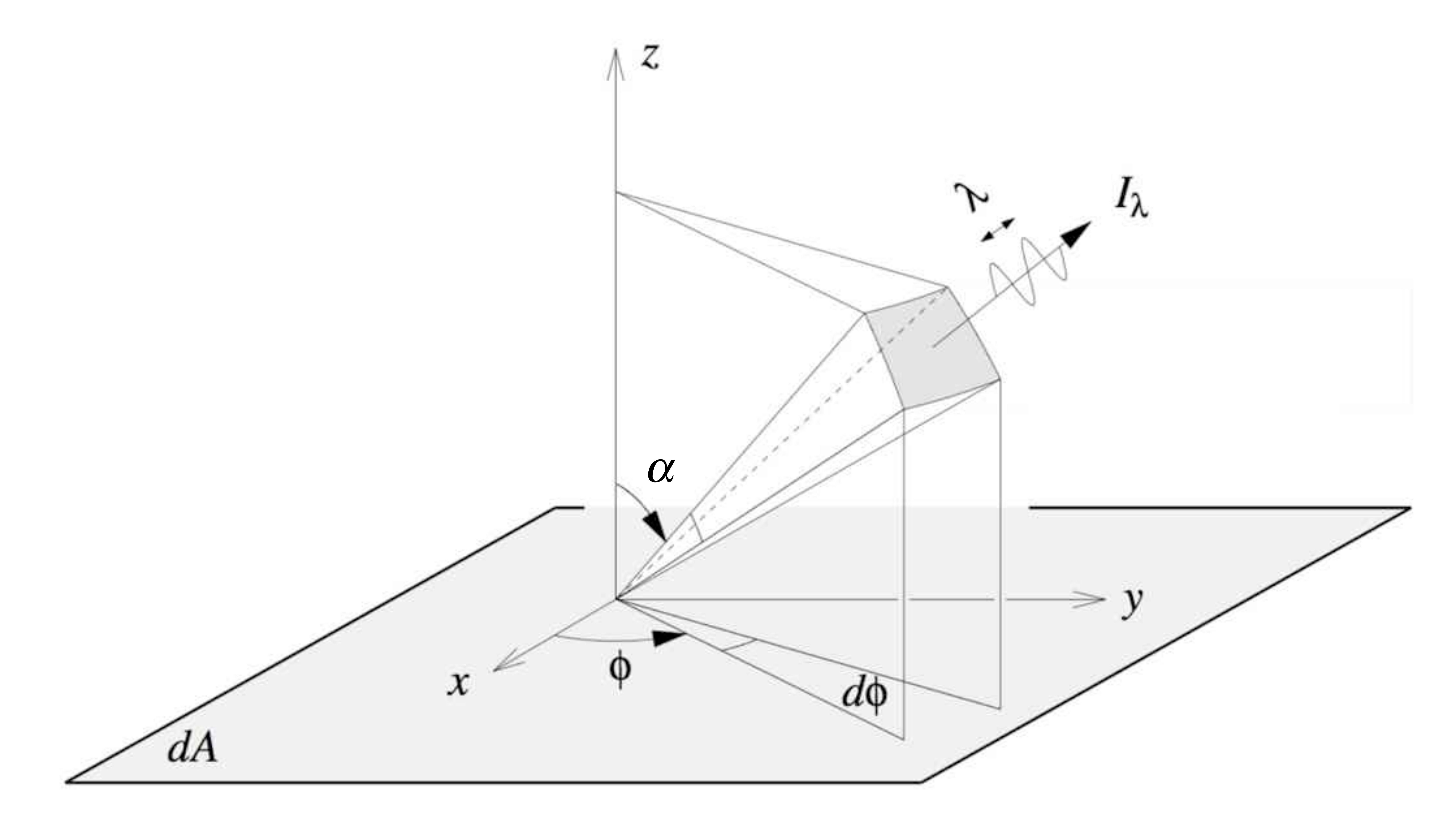}
  \caption[Schéma de définition de l'intensité spécifique à la surface stellaire]{\textbf{Schéma de définition de l'intensité spécifique à la surface stellaire}}
  \label{image__schema_intensite}
\end{figure}

À partir des paramètres définis sur la Fig.~\ref{image__schema_intensite}, un rayon de lumière de longueur d'onde $\lambda$, transportant l'énergie $dE_\lambda$, dans un intervalle spectral $d\lambda$, à travers une surface $dA$, en un temps $dt$, dans la direction $\theta$ et dans un angle solide $d\Omega$, a pour intensité  :
\begin{displaymath}
I_\lambda = \frac{dE_\lambda}{\cos{\alpha}\,dA\,dt\,d\lambda\,d\Omega} \quad (\mathrm{Wm^{-2}\mu m^{-1}sr^{-1}})
\end{displaymath}

Le flux d'énergie monochromatique traversant la surface $dA$ en un temps $dt$ et dans un intervalle spectral $d\lambda$ est défini par :
\begin{displaymath}
F_\lambda = \frac{dE_\lambda}{dA\,dt\,d\lambda} = \int_\Omega I_\lambda\,\cos{\alpha}\,d\Omega \quad (\mathrm{Wm^{-2}\mu m^{-1}})
\end{displaymath}

De plus si la radiation est axisymétrique et en notant $d\Omega = \sin{\alpha}\,d\alpha\,d\phi$ :
\begin{displaymath}
F_\lambda = 2\pi\,\int_\alpha I_\lambda\,\cos{\alpha}\,\sin{\alpha}\,d\alpha \quad (\mathrm{Wm^{-2}\mu m^{-1}})
\end{displaymath}

La quantité d'énergie monochromatique émise par toute la surface stellaire en un temps $dt$ est donnée par :
\begin{displaymath}
L_\lambda = 2\pi\,\int_A \int_\alpha I_\lambda\,\cos{\alpha}\,\sin{\alpha}\,d\alpha\,dA = \int_A F_\lambda\,dA = 4\pi R^2_\star\,F_\lambda
\end{displaymath}
où $R_\star$ correspond au rayon de l'étoile. À une distance $d$ de l'étoile, on a la relation $L_\lambda = 4\pi d^2 F_\lambda$. La luminosité bolométrique, c'est à dire intégrée sur tout le spectre, est donc :
\begin{displaymath}
L = \int_\lambda L_\lambda\,d\lambda = 2\pi\,\int_\lambda \int_A \int_\alpha I_\lambda\,\cos{\alpha}\,\sin{\alpha}\,d\alpha\,dA\,d\lambda = 4\pi R^2_\star\,F \quad (\mathrm{W})
\end{displaymath}

En astronomie, on fait souvent l'approximation que l'étoile rayonne comme un corps noir, c'est à dire qu'elle est en équilibre thermodynamique. L'intensité spécifique obéit alors à la loi de Planck :
\begin{displaymath}
I_\lambda = B_\lambda (T) = \frac{2hc^2}{\lambda^5} \frac{1}{e^{\frac{hc}{\lambda kT}} - 1}
\end{displaymath}

Si on suppose que le rayonnement émis vers l'extérieur par l'astre est isotrope, on obtient :
\begin{displaymath}
F_\lambda = 2\pi\,B_\lambda (T) \int_0^{\pi/2}\,\cos{\alpha}\,\sin{\alpha}\,d\alpha = \pi B_\lambda (T)
\end{displaymath}

En intégrant sur tout le spectre, on trouve la loi de Stefan-Boltzmann $F = \sigma T^4$ et la luminosité intrinsèque $L = 4\pi R^2\sigma T^4$.

\begin{figure}[!p]
  \centering\includegraphics[width =\linewidth]{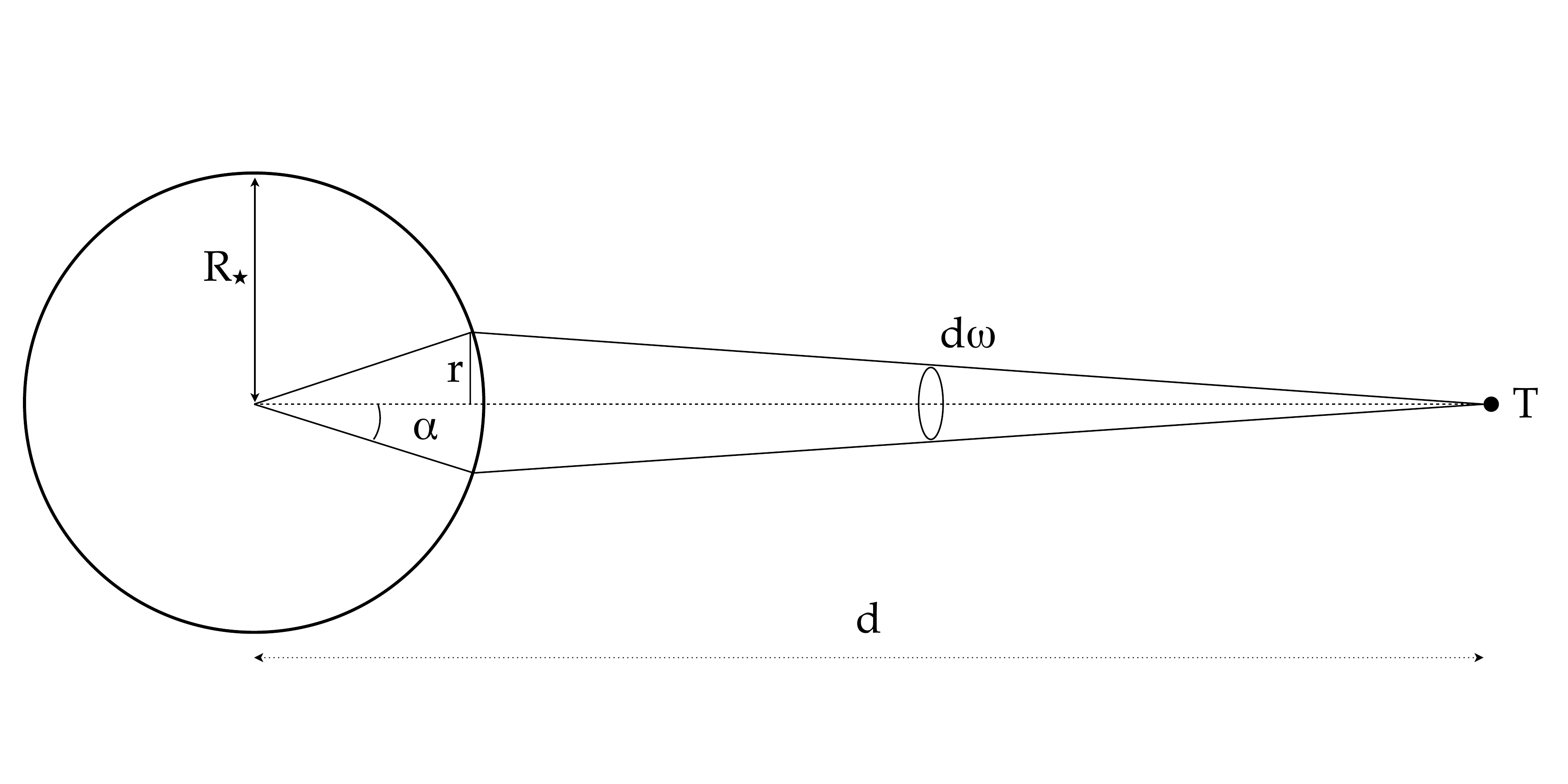}
  \caption[Schéma de définition de l'intensité reçue sur Terre]{\textbf{Schéma de définition de l'intensité reçue sur Terre}}
  \label{image__schema_intensite_2}
\end{figure}

Pour relier le flux sortant de la surface stellaire au flux mesuré sur Terre situé à une distance $d$, il faut intégrer sur l'angle solide que forme l'étoile. À partir des paramètres définis sur la Fig.~\ref{image__schema_intensite_2} on obtient à une longueur d'onde donnée :
\begin{displaymath}
df_\lambda = I_\lambda d\omega = B_\lambda (T) \frac{2\pi r\,dr}{d^2} = \frac{\theta^2}{4} 2\pi B_\lambda (T) \cos{\alpha}\,\sin{\alpha}\,d\alpha
\end{displaymath}
\begin{displaymath}
\Longrightarrow f_\lambda = \int df_\lambda = \pi \frac{\theta^2}{4} B_\lambda (T)
\end{displaymath}
où $\theta \simeq 2R/d$ représente le diamètre angulaire de l'étoile.

\subsection{Magnitude}

Pour représenter le flux d'une étoile reçu sur Terre, on a défini la magnitude apparente. Elle est déterminée par rapport à un flux de référence et à une longueur d'onde donnée :
\begin{displaymath}
m_\lambda = -2.5 \log \left( \frac{f_\lambda}{f_{\mathrm{0},\lambda}} \right)
\end{displaymath}

L'indice de couleur (ou simplement couleur) est défini à partir de la différence de magnitude prises à deux longueurs d'ondes croissantes :
\begin{displaymath}
m_{\lambda_1} - m_{\lambda_2} = -2.5 \log \left(\frac{f_{\lambda_1}}{f_{\lambda_2}} \right)
\end{displaymath}
par exemple $B - V = m_B - m_V$. La couleur donne une idée du spectre de l'étoile.

Pour comparer les astres entre eux, on utilise la magnitude absolue. C'est la magnitude apparente qu'aurait un astre si on le plaçait à une distance de 10 parsecs : 
\begin{displaymath}
m_\lambda - M_\lambda = -2.5 \log\left(\frac{d^2}{10^2}\right) = 5\log d - 5 = \mu
\end{displaymath}
où $\mu$ est appelé module de distance. En mesurant $m_\lambda$ et en connaissant $d$ via l'utilisation d'une autre méthode, on peut estimer $M_\lambda$.


\subsection{Extinction interstellaire}

Durant le trajet jusqu'a la Terre, les photons émis par une étoile sont soumis à divers processus physiques. L'un d'entre eux concerne l'absorption et la diffusion d'une partie des photons par les gaz et les poussières présents sur leur chemin. Cet effet est chromatique, décroissant avec la longueur d'onde, et plus important si l'étoile est localisée dans le plan Galactique. L'étoile parait donc moins brillante qu'elle ne l'est réellement, créant ainsi un biais sur la magnitude mesurée qui se propagera sur les autres grandeurs physiques.

Le flux observé $F_\lambda$ est donc inférieur au flux intrinsèque $F_{0,\lambda}$ et donc $m_\lambda > m_{0,\lambda}$. On définit l'absorption totale telle que :
\begin{displaymath}
m_{0,\lambda} = m_\lambda - A_\lambda
\end{displaymath}
où $A_\lambda$ représente l'extinction à la longueur d'onde $\lambda$. Ce paramètre est généralement relié à un excès de couleur $E(B - V)$ et une loi de rougissement $R_\lambda$:
\begin{equation}
\label{equation__extinction_interstellaire}
A_\lambda = R_\lambda\,E(B - V)
\end{equation}
où $E(B - V) = (B - V) - (B - V)_0$ est estimé en comparant les spectres mesurés à ceux connus pour chaque type spectral et $R_\lambda$ est déterminé à partir des observations.

Ces paramètres ont été mesurés par de nombreux auteurs et diffèrent par rapport à la position de l'étoile (Galactique ou extragalactique).

\subsection{Extinction atmosphérique}

Après des milliers de kilomètres à travers l'espace, les photons font leur entrée dans l'atmosphère. Cette dernière introduit des difficultés supplémentaires en ce qui concerne la mesure de flux en absorbant et diffusant également une fraction du flux incident. L'effet est d'autant plus importante que la couche d'atmosphère traversée est grande. Une étoile située au zénith sera moins affaiblie qu'une étoile identique située à l'horizon. La magnitude observée sera alors :
\begin{equation}
\label{equation__exctinction_atmospherique}
m_\lambda = m_{0,\lambda} + K_\lambda(AM)
\end{equation}
où $K_\lambda(AM)$ dépend de la masse d'air traversée à une longueur d'onde donnée. Cette fonction peut être estimée en observant des étoiles, dont le spectre est connu (étoiles standards), à différentes longueurs d'onde et masses d'air.

\begin{figure}[!p]
  \centering\includegraphics[width =.8\linewidth]{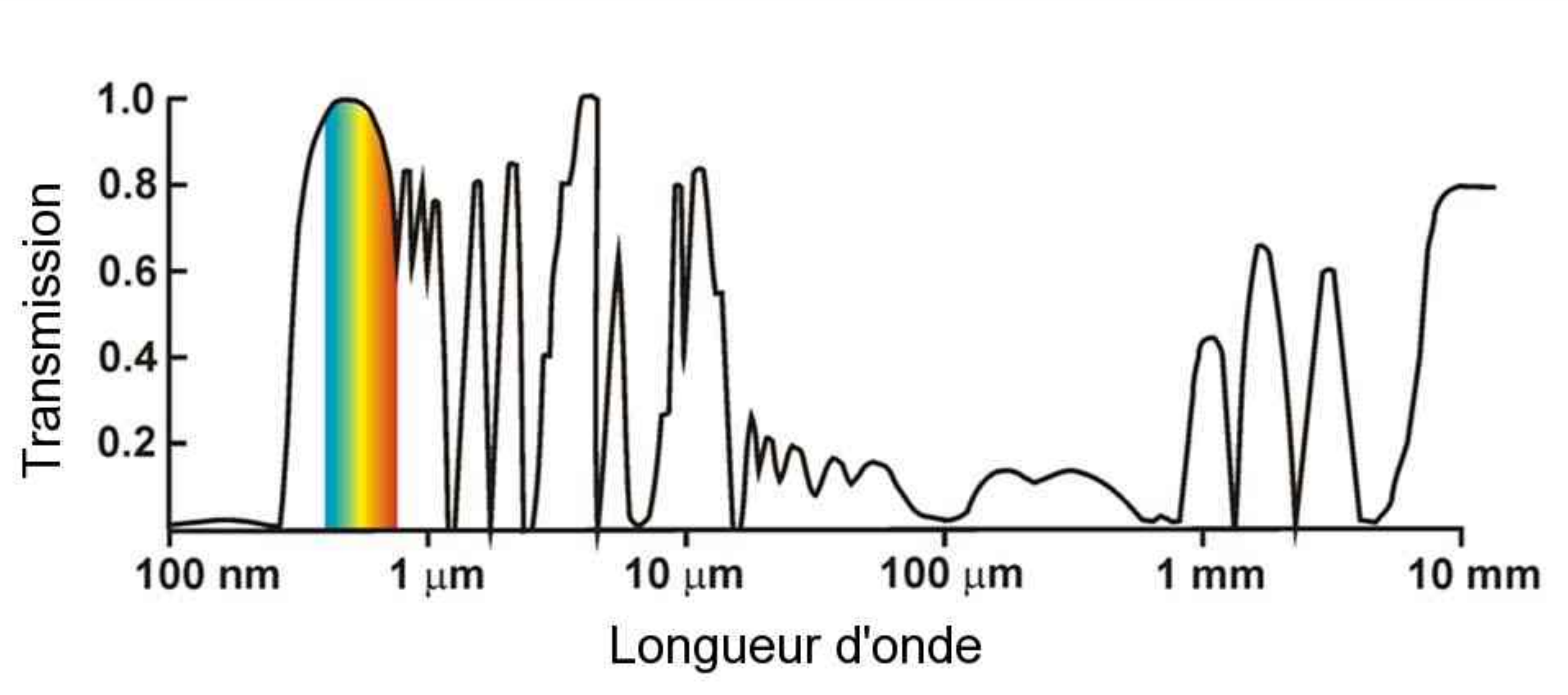}
  \caption[Transmission atmosphérique]{\textbf{Transmission atmosphérique} : transparence de l'atmosphère en fonction de la longueur d'onde.}
  \label{image__transmission}
\end{figure}

Notons également qu'une grande majorité des longueurs d'onde ne franchissent pas l'atmosphère, comme le montre la Fig.~\ref{image__transmission} représentant la transparence de l'atmosphère.

\subsection{Bandes photométriques}

Une fois l'atmosphère traversée, les photons arrivent au télescope qui les oriente vers un instrument de mesure. L'instrument sélectionne généralement certains photons, de même énergie $E_0 = hc/\lambda_0$, grâce à un filtre photométrique. Malheureusement, il n'est pas possible de mesurer un flux monochromatique car les filtres que nous utilisons ont une certaine largeur spectrale $\Delta \lambda$ autour d'une longueur d'onde centrale $\lambda_0$.

\begin{figure}[!p]
  \centering\includegraphics[width =.8\linewidth]{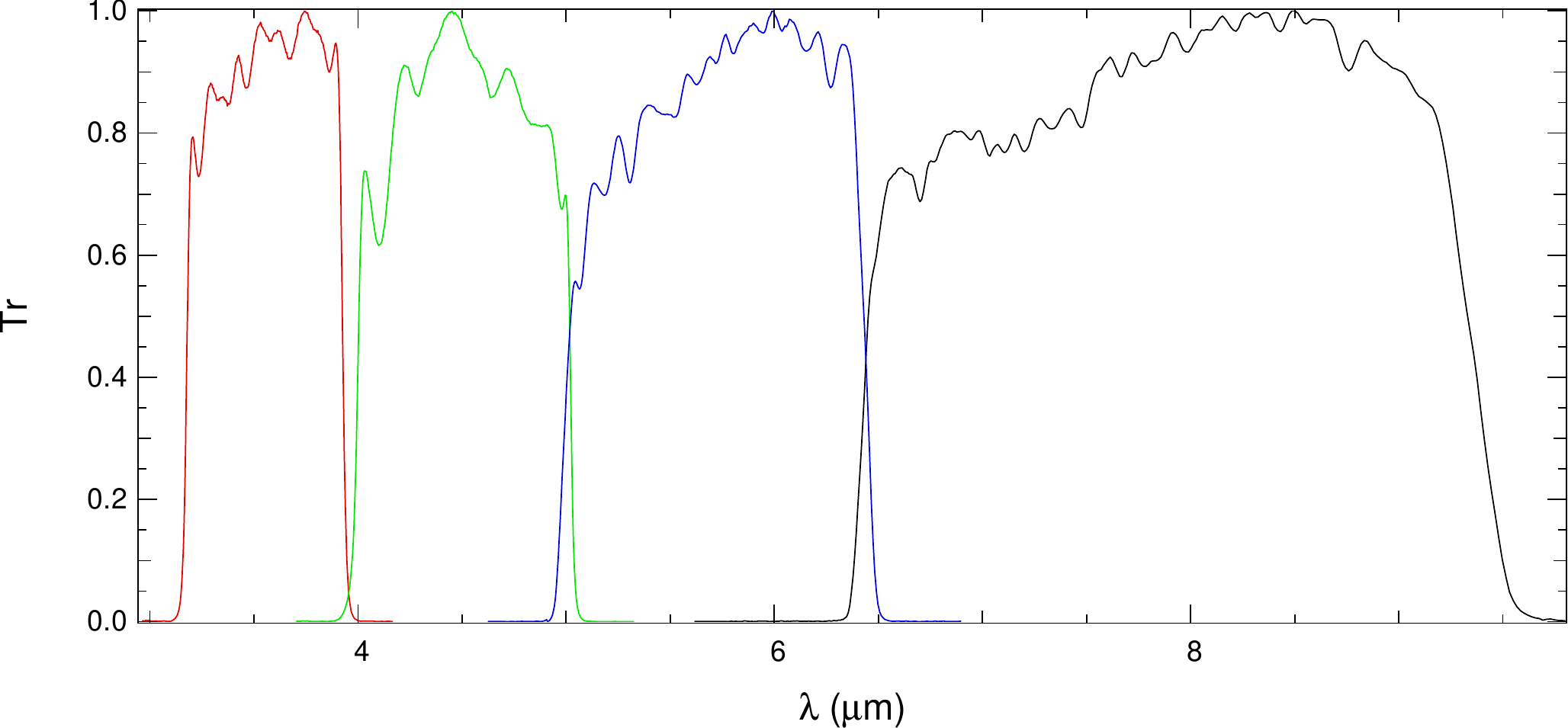}
  \caption[Transmission des filtres IRAC]{\textbf{Transmission des filtres IRAC} : transmission des quatre filtres disponibles de la caméra IRAC de \emph{Spitzer} \citep{Hora-2008-11}.}
  \label{image__filtre_irac}
\end{figure}

Il existe une grande variété de filtres regroupés par bande spectrale et par système photométrique. Les systèmes photométriques sont définis par la largeur de la bande spectrale. Le système le plus familier est le système Johnson $U, B, V, R, I, J, H, K, L, M, N, Q$. Quelques filtres sont présentés dans la Table~\ref{table__filtres}. Lors d'une mesure photométrique, il faut tenir compte non seulement de la largeur du filtre, mais également de sa transmission. Un filtre n'est pas une fonction "porte", comme le montre la Fig.~\ref{image__filtre_irac}, et chaque filtre a sa propre transmission. Le flux mesuré est donc la somme des flux monochromatiques sur la largeur du filtre, pondérée par la transmission du filtre :
\begin{equation}
\label{equation__integration_transmission_filtre}
f_\mathrm{obs}(\lambda) = \frac{\int f(\lambda) T(\lambda) d\lambda}{\int T(\lambda) d\lambda}
\end{equation}

\defcitealias{Bessell-1998-05}{B98}
\defcitealias{Rieke-2008-06}{R08}
\defcitealias{Reach-2005-09}{R05}
\begin{table}[!p]
\centering
\begin{tabular}{cccccc} 
\hline
\hline
Système	&	Bande 	&	$\lambda_0$ 					&	$\Delta \lambda$	& 	$F_0$	& Ref.											\\
 				&				&	($\mu\mathrm{m}$)			&	($\mu\mathrm{m}$)				&	($\mathrm{Jy}$)		& 													\\
\hline
Johnson	&	$U$		&	0.36					&	0.05						&	1800		&	\citetalias{Bessell-1998-05}		\\
Johnson	&	$B$		&	0.44					&	0.09						&	1480		&	\citetalias{Bessell-1998-05}		\\
Johnson	&	$V$		&	0.55					&	0.09						&	3660		&	\citetalias{Bessell-1998-05}		\\
Johnson	&	$R$		&	0.64					&	0.15						&	2970		&	\citetalias{Bessell-1998-05}		\\
Johnson	&	$I$		&	0.80					&	0.15						&	2400		&	\citetalias{Bessell-1998-05}		\\
2MASS		&	$J$		&	1.24					&	0.16						&	1623		&	\citetalias{Rieke-2008-06}		\\
2MASS		&	$H$		&	1.65					&	0.25						&	1075		&	\citetalias{Rieke-2008-06}		\\
2MASS		&	$Ks$		&	2.17					&	0.26						&	676		&	\citetalias{Rieke-2008-06}		\\
IRAC 3.6	&	$3.6$	&	3.6					&	0.76						&	281		&	\citetalias{Reach-2005-09}		\\
IRAC 4.5	&	$4.5$	&	4.5					&	0.99						&	180		&	\citetalias{Reach-2005-09} 		\\
IRAC 5.8	&	$5.8$	&	5.8					&	1.45						&	115		&	\citetalias{Reach-2005-09}		\\
IRAC 8.0	&	$8.0$	&	8.0					&	2.16						&	64		&	\citetalias{Reach-2005-09}		\\
\hline
\end{tabular}
\caption[Bandes passantes et points zéro de quelques filtres photométriques]{\textbf{Bandes passantes et points zéro de quelques filtres photométriques} : le système de référence est donnée pour le système de magnitude Véga. Les références sont \citetalias{Bessell-1998-05} = \citet{Bessell-1998-05}, \citetalias{Rieke-2008-06} = \citet{Rieke-2008-06} et \citetalias{Reach-2005-09} = \citet{Reach-2005-09}.}
\label{table__filtres}
\end{table}

\subsection{Photométrie d'ouverture}
\label{section__photometrie_ouverture}

\begin{figure}[!p]
	\resizebox{\hsize}{!}{
  		\centering\includegraphics{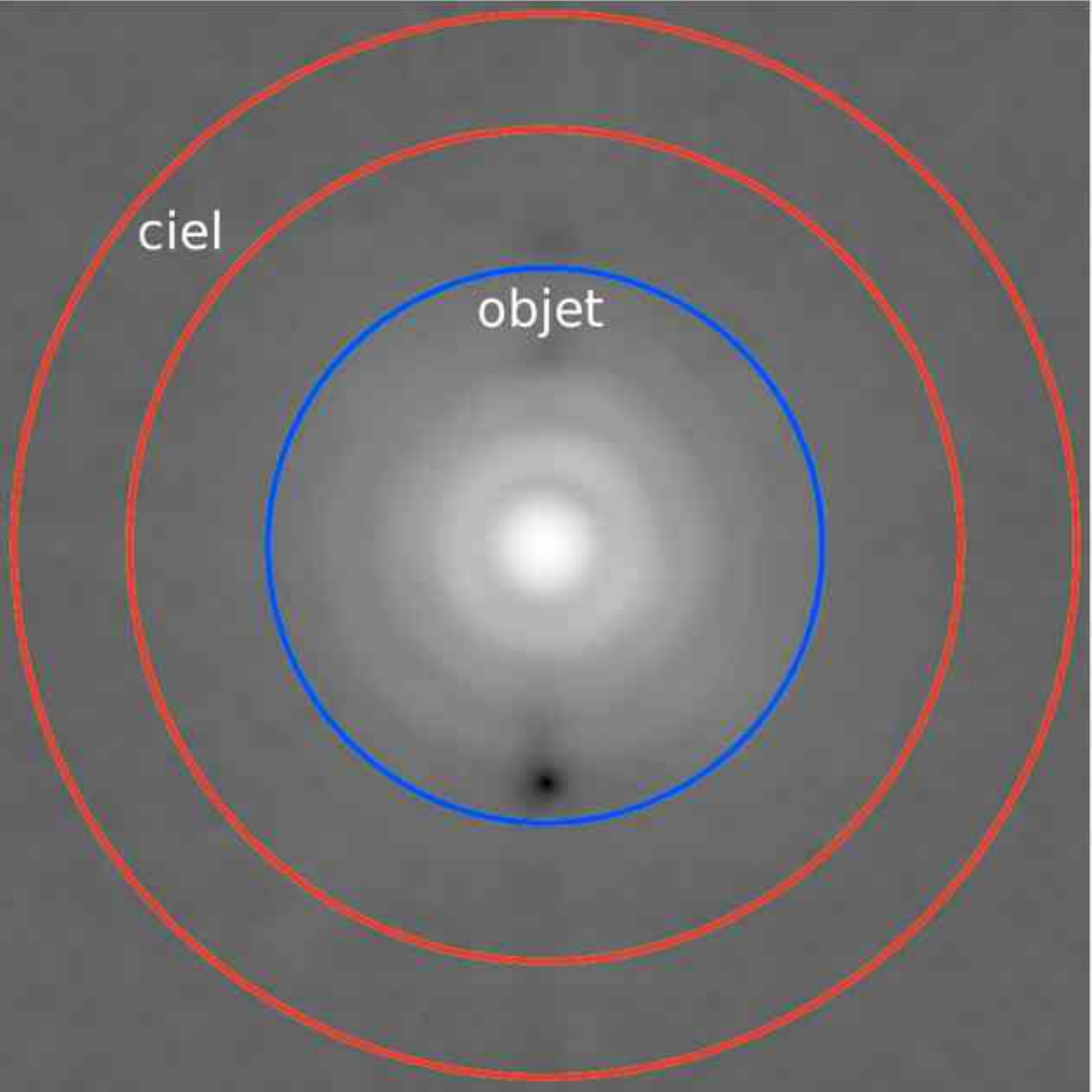}
  		\centering\includegraphics[width =1.2\linewidth]{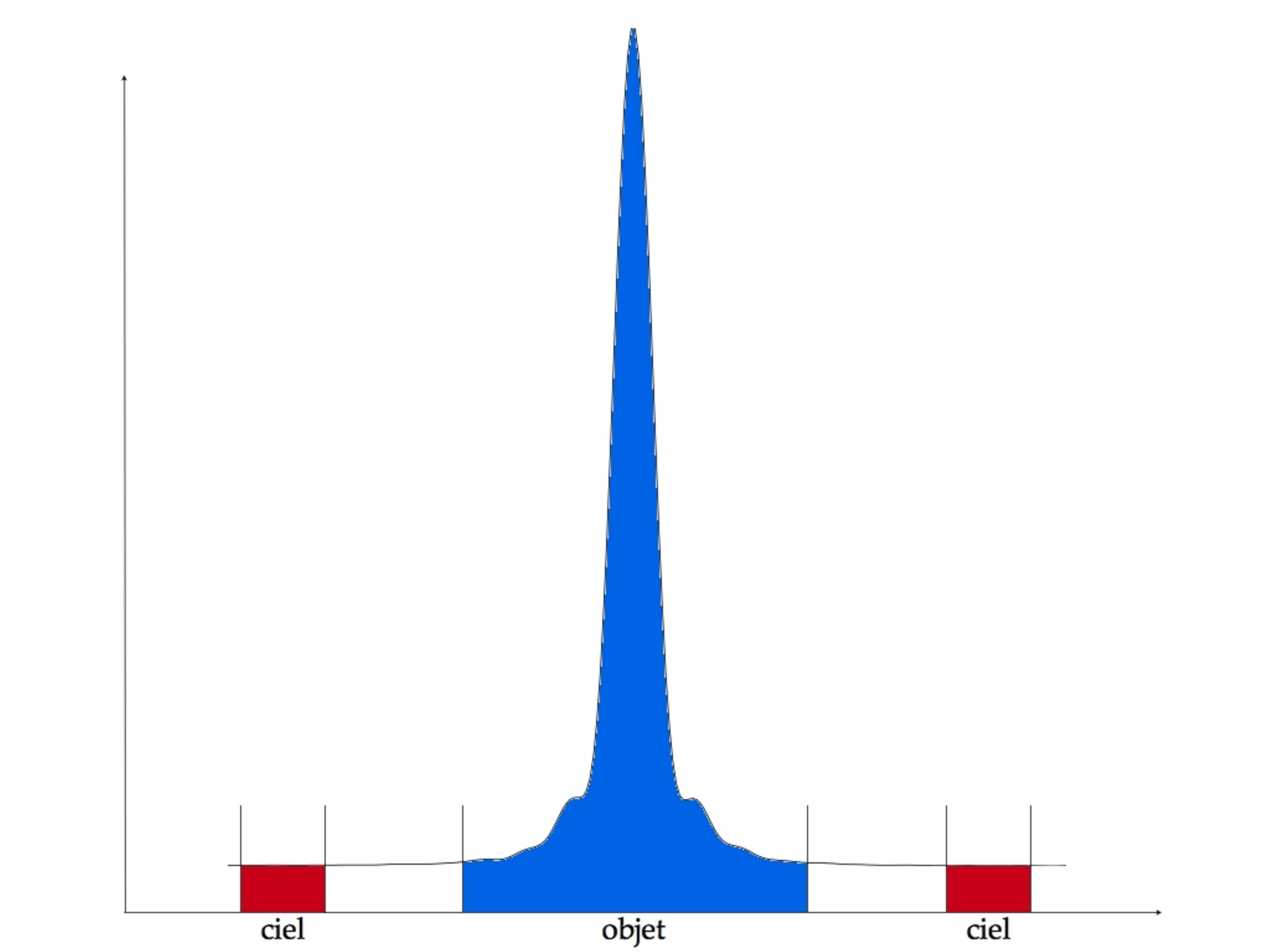}}
  	\caption[Principe de la photométrie d'ouverture]{\textbf{Principe de la photométrie d'ouverture} : image à deux dimensions à droite et à une dimension à gauche. Le but est de soustraire du flux de l'objet (en bleu) l'émission du fond de ciel (en rouge).}
  \label{image__principe_photometrie_ouverture}
\end{figure}

Après la traversée des systèmes optiques et du filtre, les photons atteignent le détecteur CCD et il faut déduire le flux de l'étoile sur l'image. Je ne présente ici que le principe de photométrie d'ouverture car c'est la méthode que j'ai utilisée sur les données \emph{VISIR}. D'autres techniques existent, telles que l'ajustement de FEP, mais je n'en parlerai pas ici. Le processus de photométrie d'ouverture est illustré sur la Fig.~\ref{image__principe_photometrie_ouverture}. Deux ouvertures circulaires sont définies dont la première est centrée sur l'étoile et a un rayon assez large pour inclure tout son flux. La deuxième est également centrée sur l'étoile mais est un anneau dont le rayon intérieur est loin de l'étoile pour ainsi estimer le flux du fond de ciel. Ce dernier, convenablement normalisé au nombre de pixel, est ensuite soustrait au flux du disque central pour obtenir le flux stellaire. La formulation mathématique est la suivante, en notant $I(r,\theta)$ le profil d'intensité sur l'image :
\begin{displaymath}
I(r,\theta) = f_\star\times\mathrm{FEP}(r,\theta) + b
\end{displaymath}
où $f$ représente le flux de l'étoile, FEP la fonction réponse impulsionnelle de l'instrument et $b$ est le bruit du fond de ciel. Le bruit $b_\mathrm{mes}$ que l'on mesure s'effectue sur un anneau de rayons $r_\mathrm{in}$ et $r_\mathrm{ext}$ :
\begin{displaymath}
b_\mathrm{mes} = \frac{1}{\mathcal{A}_\mathrm{a}} \int_0^{2\pi} \int_{r_\mathrm{in}}^{r_\mathrm{ext}} I(r,\theta) rdrd\theta = b + \frac{f_\star}{\mathcal{A}_\mathrm{a}} \int_0^{2\pi} \int_{r_\mathrm{in}}^{r_\mathrm{ext}} \mathrm{FEP}(r,\theta) rdrd\theta 
\end{displaymath}
avec $\mathcal{A}_\mathrm{a} = \pi (r_\mathrm{in}^2 - r_\mathrm{ext}^2)$. Le flux de l'étoile mesuré à l'intérieur d'un disque de rayon $r_\mathrm{d}$ est donc :
\begin{eqnarray*}
&f_\mathrm{mes} = & \int_0^{2\pi} \int_0^{r_\mathrm{d}} (I(r,\theta) - b_\mathrm{mes}) rdrd\theta \\
&\textcolor{white}{f_\mathrm{mes}} = & f_\star \left[ \int_0^{2\pi} \int_0^{r_\mathrm{d}} \mathrm{FEP}(r,\theta) rdrd\theta - \int_0^{2\pi} \int_0^{r_\mathrm{d}} \left( \frac{1}{\mathcal{A}_\mathrm{a}} \int_0^{2\pi} \int_{r_\mathrm{in}}^{r_\mathrm{ext}} \mathrm{FEP}(r,\theta) rdrd\theta \right) rdrd\theta \right] \\
&\textcolor{white}{f_\mathrm{mes}} = & f_\star \left[ \int_0^{2\pi} \int_0^{r_\mathrm{d}} \mathrm{FEP}(r,\theta) rdrd\theta - \frac{\mathcal{A}_\mathrm{d}}{\mathcal{A}_\mathrm{a}} \int_0^{2\pi} \int_{r_\mathrm{in}}^{r_\mathrm{ext}} \mathrm{FEP}(r,\theta) rdrd\theta \right]
\end{eqnarray*}
où $\mathcal{A}_\mathrm{d} = \pi r_\mathrm{d}^2$. Le terme entre crochets est ce que l'on appelle la correction d'ouverture et ne dépend que de $r_\mathrm{d}, r_\mathrm{in}$ et $r_\mathrm{ext}$. Il faut choisir $r_\mathrm{d}$ de tel sorte à englober tout le flux stellaire. L'anneau doit être assez éloigné de la source centrale, tout en évitant d'être contaminé par des étoiles environnantes.

\subsection{Étalonnage photométrique}

L'étalonnage photométrique consiste à convertir le flux mesuré en ADU par le détecteur (Analog to Digital Unit, relié au nombre de photons détectés) dans un système plus conventionnel (Jy, $\mathrm{W/m^2/\mu m}$, ...). L'observation d'étoiles standards, c'est à dire d'étoiles dont le flux absolu est très bien connu, permet cet étalonnage. Cela permet également de corriger des effets de l'extinction atmosphérique.

Toutefois, le choix d'une étoile standard doit vérifier certaines propriétés :

\begin{itemize}
	\compactlist
	\item les coordonnées célestes doivent être proches de celles de l'objet d'étude (même masse d'air).
	\item idéalement, le même $B - V$ que l'objet d'étude.
	\item ne doit pas être une étoile variable.
	\item ne doit pas être une étoile binaire.
	\item le spectre absolu est connu avec une bonne précision.
\end{itemize}

Un fois l'étoile étalon choisie, il faut l'observer dans les mêmes conditions instrumentales que la source d'intérêt, et immédiatement avant ou après afin d'être dans les mêmes conditions atmosphériques. Lors d'une observation, le flux mesuré sur le détecteur pour l'étoile à étudier est :
\begin{displaymath}
f_\mathrm{obs}(\mathrm{sci}) = H f_\mathrm{r\acute{e}el}(\mathrm{sci})
\end{displaymath}
où $H$ représente la fonction de transfert, contenant les conditions instrumentales et atmosphériques. De même pour la source de référence, où l'on peut en déduire H puisque le spectre réel est connu :
\begin{equation}
\label{equation__fonction_transfert}
f_\mathrm{obs}(\mathrm{ref}) = H f_\mathrm{r\acute{e}el}(\mathrm{ref}) \quad \Longrightarrow \quad H = \frac{f_\mathrm{obs}(\mathrm{ref})}{f_\mathrm{r\acute{e}el}(\mathrm{ref})}
\end{equation}
Pour avoir le flux absolu de l'étoile d'intérêt, il suffit donc d'étalonner de la façon suivante :
\begin{equation}
\label{equation__etalonnage_flux}
f_\mathrm{r\acute{e}el}(\mathrm{sci}) =\frac{f_\mathrm{obs}(\mathrm{sci})}{f_\mathrm{obs}(\mathrm{ref})}f_\mathrm{r\acute{e}el}(\mathrm{ref})
\end{equation}

En mesurant le flux à plusieurs longueurs d'onde, on peut reconstituer la distribution spectrale d'énergie. Cependant ces mesures photométriques sont liées au rayonnement total, c'est à dire au spectre continu de l'étoile (processus thermiques où les photons sont émis par désexcitations collisionnelles d'atomes ou de molécules chauffées) et au spectre de raies (interaction entre la matière et le rayonnement). Pour obtenir un spectre plus détaillé, il faut avoir recours à la spectroscopie où des raies en absorption et en émission peuvent être détectées. Toutefois, l'étude du rayonnement continu permet, par exemple, de déterminer la classe spectrale des étoiles, d'estimer une température de surface, ou encore de détecter un excès infrarouge.

\section{Étude d'excès infrarouge}

\begin{figure}[!p]
  	\centering\includegraphics[width =.7\linewidth]{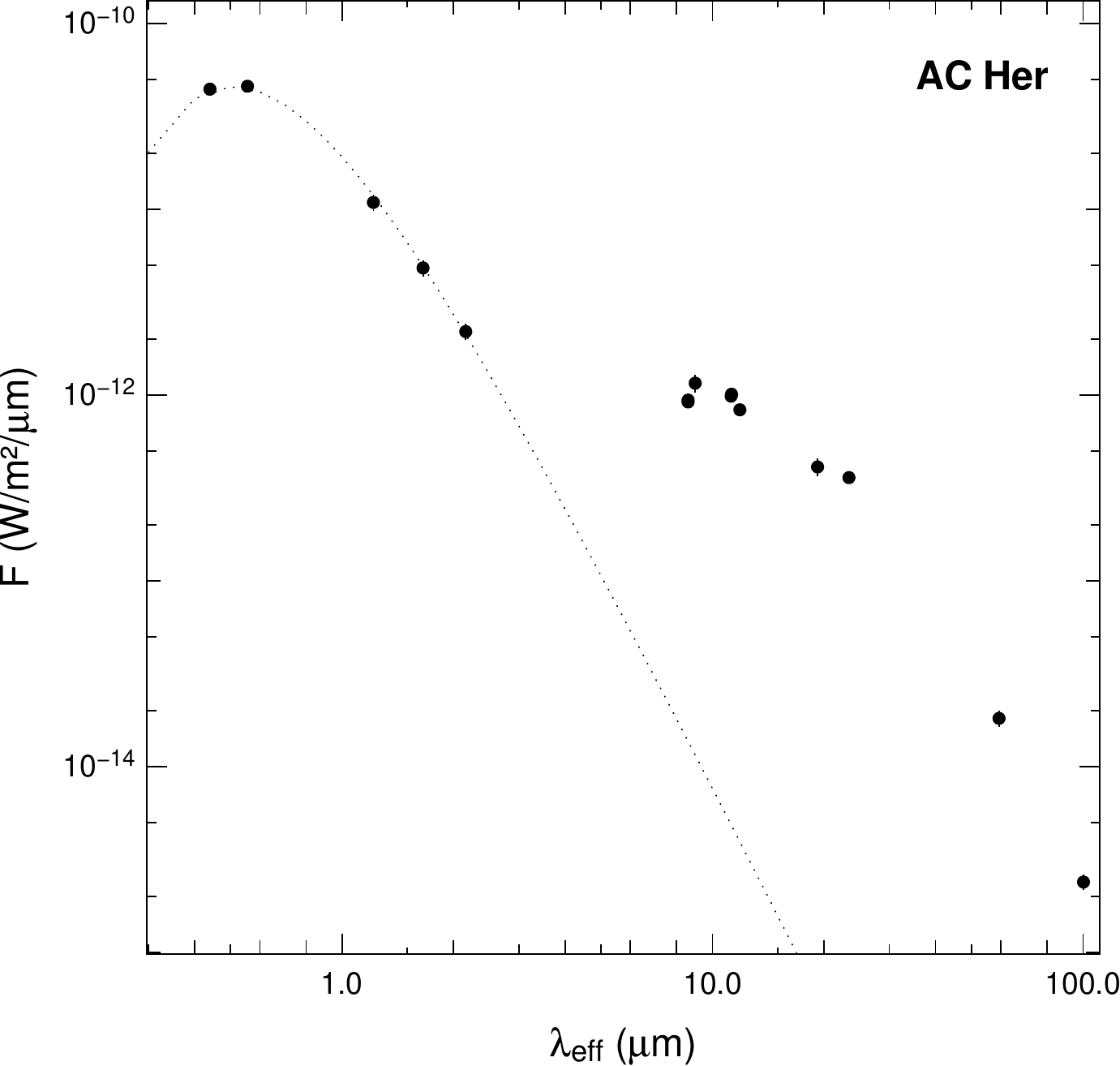}
  	\caption[Exemple d'excès infrarouge avec AC~Her]{\textbf{Exemple d'excès infrarouge avec AC~Her} : distribution d'énergie spectrale de l'étoile AC~Her. Les cercles représentent des données photométriques et la courbe en pointillé est une loi de corps noir symbolisant la distribution d'intensité de la photosphère l'étoile.}
  \label{image__exces_IR_AC_HER}
\end{figure}

Plaçons nous dans le cadre d'une étoile entourée d'une enveloppe circumstellaire. Lorsque l'enveloppe n'est pas résolue spatialement, on se base généralement sur la distribution spectrale d'énergie pour étudier les environnements stellaires. Le rayonnement émis par l'astre central interagit avec les gaz et les poussières présents dans cet environnement. Cette interaction, fortement dépendante de la longueur d'onde et de la taille des composants circumstellaires, se traduit par un excès ou déficit de flux aux longueurs d'onde infrarouges et au-delà. Je présente sur la Fig.~\ref{image__exces_IR_AC_HER} l'exemple de l'étoile AC~Her qui possède un fort excès IR dans sa SED. Les cercles sont des points de mesures photométriques de divers instruments et la courbe en pointillé représente la loi de corps noir symbolisant la distribution d'intensité de la photosphère de l'étoile. Un excès est clairement visible à partir de $10\,\mu\mathrm{m}$ environ, causé par le rayonnement de la matière située autour de l'étoile.

Si l'enveloppe circumstellaire est optiquement épaisse (forte densité), le rayonnement stellaire est complètement absorbé par les particules. Elles réémettent à leur tour leurs propres radiations à des longueurs d'onde plus longues qui sont immédiatement absorbées par les particules voisines, et ainsi de suite. De ce fait, nous ne détectons que l'émission des particules situées à la surface de l'enveloppe. Si le milieu circumstellaire est optiquement mince (faible densité), une partie du flux photosphérique seulement est absorbée puis réémise, causant une émission ou absorption dans le spectre stellaire caractéristique des particules présentes. Dans ce cas l'émission que nous observons provient de toute l'épaisseur de l'enveloppe. Ces émissions créent un flux infrarouge supplémentaire dans la distribution spectrale d'énergie, détectable sous forme d'excès (comme sur la Fig.~\ref{image__exces_IR_AC_HER}). Via l'utilisation de modèles de transfert de rayonnement, il est possible de remonter à certains paramètres physiques de l'enveloppe circumstellaire tels que sa température, sa morphologie, sa taille ou sa composition.

\section{La technique du chopping-nodding}

Les observations dans l'infrarouge ont donc une place importante dans l'étude des environnements circumstellaires. Malheureusement, si l'on souhaite observer depuis le sol à des longueurs d'onde comprises entre $\sim 3$ et $25\,\mu\mathrm{m}$, l'émission du ciel (en plus de l'absorption) est un inconvénient majeur. Vient s'ajouter à cela un rayonnement autour de $10\,\mu\mathrm{m}$ lié à l'émission thermique des divers systèmes optiques. Pour \emph{VISIR}, cette émission de fond (ciel+instrument) est de l'ordre de $3700\,\mathrm{Jy/arcsec}^2$ à $10\,\mu\mathrm{m}$ et $8300\,\mathrm{Jy/arcsec}^2$ à $20\,\mu\mathrm{m}$. En reprenant l'exemple de l'étoile AC~Her, dont le flux du matériel circumstellaire est de $\sim 10^{-12}\,\mathrm{Wm}^{-2}\mu\mathrm{m}^{-1} = 33\,\mathrm{Jy}$ à $10\,\mu\mathrm{m}$, et en supposant que tout le flux est concentré sur une surface circulaire de rayon $0.5\arcsec$, seulement $33/1860 = 1.8\,\%$ du flux reçu sera lié à l'objet astrophysique. Il est donc nécessaire de procéder à la soustraction de cette émission de fond si l'on souhaite faire des mesures de flux précises. On utilise pour cela la technique du chopping-nodding.

\paragraph*{\textcolor{black}{Chopping}}

\begin{figure}[!p]
	\resizebox{\hsize}{!}{
  	\centering\includegraphics{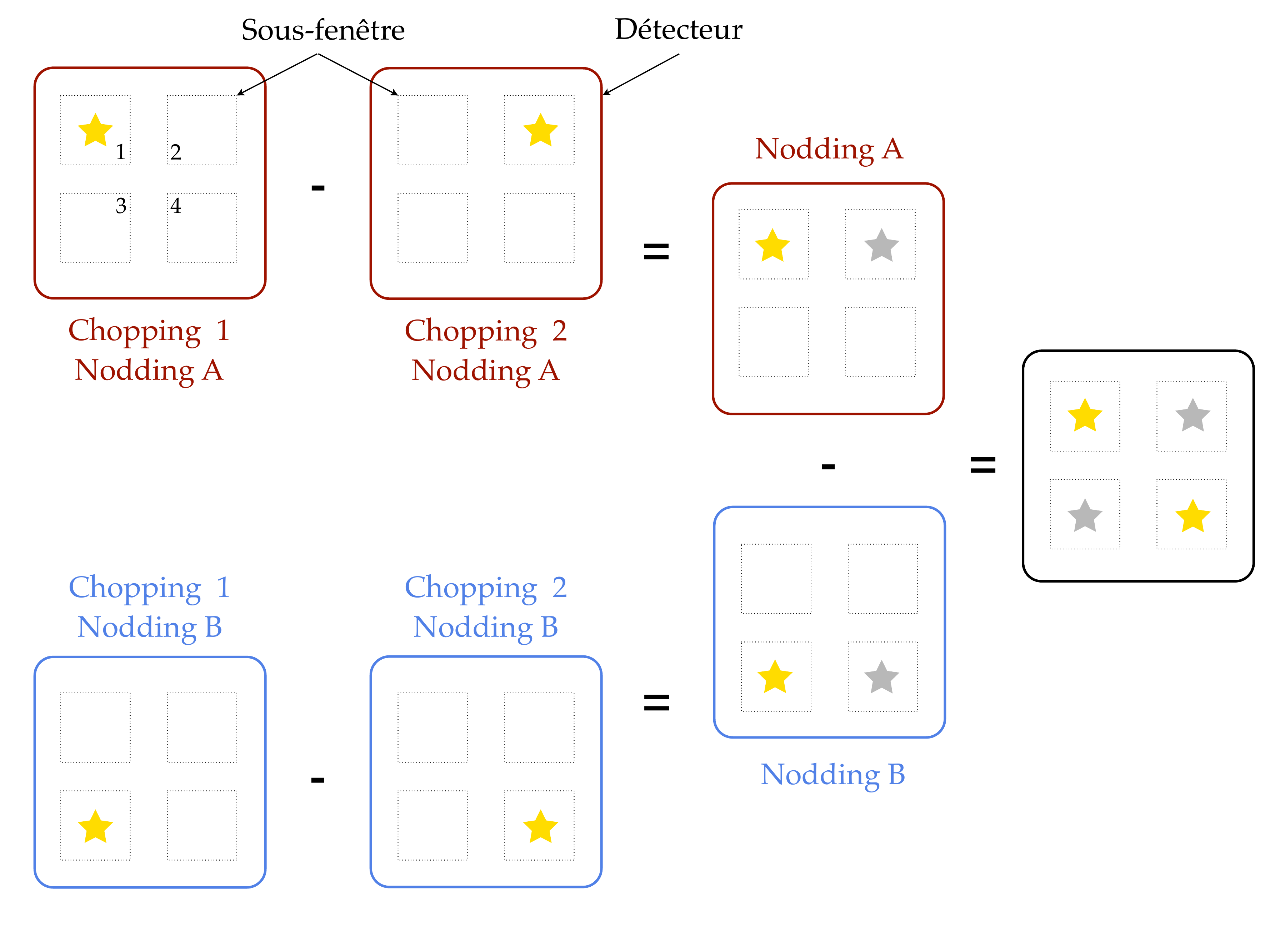}}
  	\caption[Principe du chopping-nodding]{\textbf{Principe du chopping-nodding} : une série d'images alternées est enregistrée en position de nodding A. Le processus est répété en position de nodding B. La soustraction des sous-fenêtres $\mathrm{n^\circ~1}$, $\mathrm{n^\circ~2}$, $\mathrm{n^\circ~3}$ et $\mathrm{n^\circ~4}$ (en pointillé) mène à l'image finale.}
  \label{image__chopping_nodding}
\end{figure}

Le principe consiste à acquérir alternativement et rapidement une image avec et sans l'objet astrophysique dans le champ. On procède généralement de telle sorte que l'objet soit enregistré à deux positions différentes sur le détecteur. Voyons une explication plus en détails en s'aidant de l'illustration de la Fig.~\ref{image__chopping_nodding} et en faisant le raisonnement pour une seule sous-fenêtre, la $\mathrm{n^\circ~1}$ (en haut à gauche). Dans la position chop 1/nod A, le signal mesuré est la somme du flux de l'objet et du fond : 
\begin{displaymath}
\mathrm{S_1 = Obj + ciel_1}
\end{displaymath}
 Dans la position chop 2/nod A, le signal est :
\begin{displaymath} 
\mathrm{S_2 = ciel_2}
\end{displaymath}
En supposant qu'entre les deux acquisitions $\mathrm{ciel_1\sim ciel_2}$ et en soustrayant les deux signaux, on trouve :
\begin{displaymath} 
\mathrm{S_{12} = S_1 - S_2 = Obj}
\end{displaymath}
En pratique $\mathrm{ciel\gg obj}$ et la solution est alors d'acquérir plusieurs centaines d'images alternées à haute fréquence afin d'augmenter le rapport signal à bruit (pour \emph{VISIR} en bande $N$, cette fréquence est de $0.25\,\mathrm{Hz}$). Cette alternance est généralement effectué grâce à un mouvement rapide du miroir secondaire.

\paragraph*{\textcolor{black}{Nodding}}

L'utilisation du chopping supprime une bonne partie de l'émission de fond. Cependant, l'alternance entre chop 1 et 2 implique un différence de chemin optique entre les deux positions, entraînant la présence d'un fond résiduel. Pour compenser, on utilise alors la technique du nodding qui consiste à décaler légèrement et perpendiculairement le champ de vue par rapport à la position précédente et d'appliquer à nouveau du chopping. Le principe est illustré sur la Fig.~\ref{image__chopping_nodding}. En suivant le même raisonnement que précédemment pour la sous-fenêtre $\mathrm{n^\circ~1}$, on a dans la position chop 1/nod A : 
\begin{displaymath} 
\mathrm{S_1 = Obj + ciel_1}
\end{displaymath}
En notant $\mathrm{ciel_2 = ciel_1 + \Delta ciel}$, le signal dans la position chop 2/nod A est :
\begin{displaymath} 
\mathrm{S_2 = ciel_2 = ciel_1 + \Delta ciel}
\end{displaymath}
La soustraction des deux signaux donne :
\begin{displaymath} 
\mathrm{S_{12} = S_1 - S_2 = Obj - \Delta ciel}
\end{displaymath}
On procède maintenant au nodding en décalant légèrement le télescope. Le signal dans la position chop 1/nod B de la sous-fenêtre 1 est :
\begin{displaymath}
\mathrm{S_3 = ciel_1}
\end{displaymath}
En notant $\mathrm{ciel_3 = ciel_1 + \Delta ciel}$, le signal dans la position chop 2/nod B est :
\begin{displaymath} 
\mathrm{S_4 = ciel_3 = ciel_1 + \Delta ciel}
\end{displaymath}
La soustraction des deux signaux donne :
\begin{displaymath} 
\mathrm{S_{34} = S_3 - S_4 = - \Delta ciel}
\end{displaymath}
On procède maintenant à la soustraction des deux positions de nodding :
\begin{displaymath} 
\mathrm{Obj = S_{12} - S_{34}}
\end{displaymath}

La combinaison du chopping et du nodding supprime proprement toute l'émission de fond et le flux de l'objet astrophysique peut être extrait. Toutefois cette technique fonctionne correctement si l'objet est assez brillant afin d'avoir un bon rapport signal à bruit.

Passons maintenant à des données réelles sur lesquelles j'ai appliqué les divers procédés exposés précédemment.

\section{Imagerie de Céphéides avec l'instrument VISIR}
\label{section__imagerie_de_cepheides_avec_visir}

Les observations ont été effectuées le 24 et 25 mai 2008 par une nuit claire sur un échantillon d'étoiles comprenant 8 Céphéides classiques et 3 Céphéides de type II. À partir de mesures photométriques, j'ai étudié la distribution spectrale d'énergie et j'ai pu mettre en évidence un excès infrarouge pour la grande majorité de ces étoiles. J'ai également détecté pour certaine d'entre elles une enveloppe résolue par le télescope, c'est à dire plus large que la FEP du télescope.

Dans un premier temps, j'introduis quelques caractéristiques de fonctionnement de l'instrument \emph{VISIR}, puis je parlerai des observations et de l'algorithme de réduction des données pour l'obtention des images finales. J'exposerai ensuite l'analyse de ces images grâce à l'étude de la SED puis par une technique de Fourier. Enfin, les conclusions de cette étude clôtureront ce chapitre. Cette section reprend les résultats d'un article qui a été soumis à publication dans la revue A\&A.

\subsection{Description de l'instrument}

L'instrument \emph{VISIR} (VLT Imager and Spectrometer for mid-IR) est installé au foyer Cassegrain du troisième télescope du VLT (UT3). Il est composé de deux sous-systèmes : un imageur et un spectromètre. Le premier permet d'obtenir des images à la limite de diffraction (pour de bonnes conditions atmosphériques) à travers 14 filtres allant de $8.59$ à $19.5\,\mu\mathrm{m}$\footnote{Les caractéristiques des filtres sont disponibles sur \url{http://www.eso.org/sci/facilities/paranal/instruments/visir/inst/index.html}}. Deux échelles focales ($75\,\mathrm{mas/pixel}$ et $127$\,mas/pixel) sont disponibles afin d'avoir un bon échantillonnage de la fonction d'étalement de point (FEP) en bande $N$ et $Q$. Le détecteur utilisé comprend  $256\times256$ pixels, ce qui donne un champ de vue de $19.2\arcsec\times19.2\arcsec$ ou $32.5\arcsec\times32.5\arcsec$.

Le spectromètre fournit des spectres en bande $N$ et $Q$ avec trois résolutions spectrales possibles : basse, moyenne et haute (de résolution respective $R = 215$--$390$,  $R = 1800$--$3500$ et $R = 13000$--$32000$ en bande $N$ et pour une largeur de fente de $1\arcsec$). Trois largeurs de fente sont disponibles : $0.4\arcsec, 0.75\arcsec$ et $1\arcsec$. Le détecteur est le même qu'en mode imagerie.

Un mode intéressant quand on utilise l'imagerie est le mode BURST. Ce mode permet d'enregistrer des images très rapidement (donc avec des temps d'exposition très courts), afin d'obtenir des images limitées par la diffraction du télescope (avec un bon seeing). À chaque acquisition, la turbulence atmosphérique est "gelée" et la technique du "lucky-imaging" (présentée au chapitre précédent) peut être utilisée sur les milliers d'images enregistrées.

\subsection{Observations et traitement des données}

L'échantillon de Céphéides classiques a été sélectionné en fonction de divers paramètres : leur brillance, des périodes de pulsation variées et leur diamètre angulaire, afin de pouvoir par la suite confirmer et/ou combiner les résultats avec des mesures interférométriques. L'une des étoiles, Y~Oph, semble avoir la plus brillante enveloppe circumstellaire détectée autour d'une Céphéide \citep{Merand-2007-08} et le but ici est d'explorer cette enveloppe à une plus grande distance de l'étoile.

Les Céphéides de type II sélectionnées ont déjà été étudiées par de nombreux auteurs et sont connues pour avoir un fort excès infrarouge. Ces étoiles seront utilisées pour comparer les propriétés des enveloppes avec les Céphéides classiques et pour valider notre estimation de l'excès IR (basé sur les résultats de travaux précédents sur les types II).

J'ai regroupé dans la Table~\ref{table__cepheide_parametre} quelques paramètres astrophysiques de cet échantillon de Céphéides qui nous seront utiles par la suite. 

\begin{table}[!p]
\centering
\begin{tabular}{cccccccc} 
\hline
\hline
Nom	  & $P$\tablefootmark{a}	& 	$\mathrm{MJD_0}$\tablefootmark{b}	& $<V>$\tablefootmark{c}& $<K>$\tablefootmark{d}	& $\theta$\tablefootmark{e}	& $\pi$\tablefootmark{f}	& Type \\
		  &	(jour)						&						&						&			& 				($\mathrm{mas}$)				& 	($\mathrm{mas}$)		&					\\
\hline
FF~Aql				&  4.4709			& 	41575.928		&	5.37			&	3.49	&	$0.88\,\pm\,0.05$	&	$2.81\,\pm\,0.18$	& I	\\
AX~Cir				&  5.2733			& 	51646.100		&	5.88			&	3.76	&	$0.70\,\pm\,0.06$	&	$3.22\,\pm\,1.22$	& I	\\
X~Sgr				&  7.0128			& 	51653.060		&	4.55			&	2.56	&	$1.47\,\pm\,0.04$	&	$3.00\,\pm\,0.18$	& I	\\
$\eta$~Aql		&  7.1767			& 	36084.156		&	3.90			&	1.98	&	$1.84\,\pm\,0.03$	&	$2.78\,\pm\,0.91$	& I	\\
W~Sgr				&  7.5950			& 	51652.610		&	4.67			&	2.80	&	$1.31\,\pm\,0.03$	&	$2.28\,\pm\,0.20$	& I	\\
Y~Oph				&  17.1242		& 	51652.820		&	6.17			&	2.69	&	$1.44\,\pm\,0.04$	&	$2.04\,\pm\,0.08$	& I	\\
U~Car				&  38.8124		& 	51639.740		&	6.29			&	3.52	&	$0.94\,\pm\,0.06$	&	$2.01\,\pm\,0.40$	& I	\\
SV~Vul				&  45.0121		& 	43086.390		&	7.22			&	3.93	&	$0.80\,\pm\,0.05$	&	$0.79\,\pm\,0.74$	& I	\\
\hline
R~Sct				&  146.50			&	44871.500		&  6.70			&	2.27	&	--								&	$2.32\,\pm\,0.82$	& II	\\
AC~Her				&  75.010			&	35097.300		&	7.90			&	5.01	&	--								&	$0.70\,\pm\,1.09$	& II	\\
$\kappa$~Pav	&	9.0814			&	46683.569		&	4.35			&	2.79	&	$1.17\,\pm\,0.05$	&	$6.00\,\pm\,0.67$	& II	\\
\hline
\end{tabular}
\caption[Quelques paramètres utiles de l'échantillon de Céphéides]{\textbf{Quelques paramètres utiles de l'échantillon de Céphéides} : la période de pulsation $P$, l'époque de référence $\mathrm{MJD_0}$ ($\mathrm{MJD_0}$ = $\mathrm{JD_0}$ - 2400000.5), la magnitude apparente moyenne $<V>$ en bande $V$, la magnitude apparente moyenne $<K>$ en bande $K$, le diamètre angulaire $\theta$ et la parallaxe trigonométrique $\pi$. Type I ou II représente respectivement une Céphéide Classique ou de type II.}
\begin{flushleft}
\tablefoottext{a}{\citet{Feast-2008-06-2} pour $\kappa$~Pav ; \citet{Samus-2009-01} pour les autres.} \\
\tablefoottext{b}{\citet{Samus-2009-01} pour SV~Vul, FF~Aql, $\eta$~Aql, R~Sct et AC~Her ; \citet{Feast-2008-06-2} pour $\kappa$~Pav ; \citet{Berdnikov-2001-12} pour les autres.} \\
\tablefoottext{c}{\citet{Fernie-1995-01} pour les Céphéides classiques ; \citet{Samus-2009-01} pour les types II.} \\
\tablefoottext{d}{\citet{Welch-1984-04} pour FF~Aql et X~Sgr ; \emph{DENIS} pour AX~Cir et W~Sgr ; \citet{Barnes-1997-06} pour $\eta$~Aql ; \citet{Laney-1992-04} pour Y~Oph, U~Car et SV~Vul ; \citet{Taranova-2010-02} pour R~Sct et AC~Her ; \citet{Feast-2008-06-2} pour $\kappa$~Pav.} \\
\tablefoottext{e}{Diamètres angulaires assombris de \citet{Kervella-2004-03a} pour X~Sgr, $\eta$~Aql, W~Sgr et Y~Oph ; \citet{Groenewegen-2007-11} pour FF~Aql ; prédictions des diamètres angulaires de \citet{Moskalik-2005-06} pour AX~Cir ; \citet{Feast-2008-06-2} pour $\kappa$~Pav ; \citet{Groenewegen-2008-09} pour les autres.} \\
\tablefoottext{f}{\citet{Benedict-2007-04} pour FF~Aql, X~Sgr et W~Sgr ; \citet{Hoffleit-1991-} pour U~Car ; \citet{Merand-2007-08} pour Y~Oph ; \citet{Perryman-1997-07} pour les autres.}
\end{flushleft}
\label{table__cepheide_parametre}
\end{table}

\begin{table}[!p]
\centering
\scriptsize
\begin{tabular}{ccccccccc}
\hline\hline
MJD					& $\phi$	& Nom								& Filtre 			& DIT (ms) 	&	$N$			& seeing (\arcsec)		&  AM	&	\#		\\
\hline
54~610.035		&				& HD~89682						&  PAH1		& 16		&	22~500		&	1.2			&	1.26	&	1		\\
54~610.042		&				&	HD~89682					&	PAH2		& 8		&	48~000		&	1.1			&	1.27	&	2		\\
54~610.056		&	0.62		&	U~Car							&	PAH1		& 16		&	22~500		&	1.5			&	1.30	&	3		\\
54~610.064		&	0.62		&	U~Car							&	PAH2		& 8		&	48~000		&	1.1			&	1.32	&	4		\\
54~610.081		&				&	HD~98118					&	PAH1		& 16		&	22~500		&	1.1			&	1.31	&	5		\\
54~610.088		&				&	HD~98118					&	PAH2		& 8		&	48~000		&	1.1			&	1.35	&	6		\\
54~610.104		&	0.65		&	X~Sgr							&	PAH1		& 16		&	22~500		&	1.0.			&	1.61	&	7		\\
54~610.111		&	0.65		&	X~Sgr							&	PAH2		& 8		&	24~000		&	1.0			&	1.53	&	8		\\
54~610.126		&	0.71		&	Y~Oph							&	PAH1		& 16		&	22~500		&	1.2			&	1.63	&	9		\\
54~610.134		&	0.71		&	Y~Oph							&	PAH2		& 8		&	48~000		&	1.8			&	1.55	&	10	\\
54~610.158		&				&	HD~99998					&	PAH1		& 16		&	22~500		&	1.0			&	1.88	&	11	\\
54~610.166		&				&	HD~99998					&	PAH2		& 8		&	48~000		&	1.2			&	2.01	&	12	\\
54~610.182		&				&	HD~124294					&	PAH1		& 16		&	22~500		&	0.9			&	1.12	&	13	\\
54~610.190		&				&	HD~124294					&	PAH2		& 8		&	48~000		&	1.0			&	1.14	&	14	\\
54~610.213		&	0.42		&	W~Sgr							&	PAH1		& 16		&	22~500		&	1.0			&	1.07	&	15	\\
54~610.220		&	0.42		&	W~Sgr							&	PAH2		& 8		&	48~000		&	1.1			&	1.06	&	16	\\
54~610.236		&	0.48		&	R~Sct							&	PAH1		& 16		&	22~500		&	1.1			& 	1.16	&	17	\\
54~610.243		&	0.48		&	R~Sct							&	PAH2		& 8		&	48~000		&	1.1			& 	1.14	&	18	\\
54~610.258		&				&	HD~161096					&	PAH1		& 16		&	22~500		&	0.9			&	1.15	&	19	\\
54~610.266		&				&	HD~161096					&	PAH2		& 8		&	48~000		&	0.8			&	1.15	&	20	\\
54~610.282		&	0.40		&	$\eta$~Aql					&	PAH1		& 16		&	22~500		&	1.0			&	1.22	&	21	\\
54~610.289		&	0.40		&	$\eta$~Aql					&	PAH2		& 8		&	48~000		&	0.8			&	1.20	&	22	\\
54~610.305		&	0.02		&	SV~Vul							&	PAH1		& 16		&	22~500		&	1.0			&	1.72	&	23	\\
54~610.312		&	0.02		&	SV~Vul							&	PAH2		& 8		&	48~000		&	1.1			&	1.69	&	24	\\
54~610.327		&				&	HD~203504					&	PAH1		& 16		&	22~500		&	0.9			&	1.69	&	25	\\
54~610.334		&				&	HD~203504					&	PAH2		& 8		&	48~000		&	0.9			&	1.63	&	26	\\
54~610.349		&	0.14		&	AC~Her							&	PAH1		& 16		&	22~500		& 	0.8			&	1.58	&	27	\\
54~610.357		&	0.14		&	AC-Her							&	PAH2		& 8		&	48~000		&	0.9			&	1.60	&	28	\\
54~610.373		&	0.71		&	Y~Oph							&	PAH1		& 16		&	22~500		&	0.8			&	1.32	&	29	\\
54~610.380		&	0.71		&	Y~Oph							&	PAH2		& 8		&	48~000		&	0.7			&	1.36	&	30	\\
54~610.395		&	0.42		&	$\eta$~Aql					&	PAH1		& 16		&	22~500		&	0.8			&	1.15	&	31	\\
54~610.402		&	0.42		&	$\eta$~Aql					&	PAH2		& 8		&	48~000		&	0.9			&	1.17	&	32	\\
54~610.417		&				&	HD~196321					&	PAH1		& 16		&	22~500		&	0.8			&	1.11	&	33	\\
54~610.424		&				&	HD~196321					&	PAH2		& 8		&	48~000		&	0.7			&	1.12	&	34	\\
54~611.016		&				&	HD~89682					&	PAH1		& 16		&	22~500		&	0.7			&	1.22	&	35	\\
54~611.024		&				&	HD~89682					&	SiC			& 20		&	18~000		&	0.8			&	1.24	&	36	\\
54~611.035		&	0.64		&	U~Car							&	PAH1		& 16		&	22~500		&	0.8			&	1.27	&	37	\\
54~611.042		&	0.64		&	U~Car							&	SiC			& 20		&	18~000		&	0.8			&	1.28	&	38	\\
54~611.054		&	0.26		&	AX~Cir							&	PAH1		& 16		&	22~500		&	0.8			&	1.40	&	39	\\
54~611.061		&	0.26		&  AX~Cir							&	SiC			& 20		&	18~000		&	0.8			&	1.38	&	40	\\
54~611.073		&				&	HD~124294					&	PAH1		& 16		&	22~500		&	0.7			&	1.06	&	41	\\
54~611.081		&				&	HD~124294					&	SiC			& 20		&	18~000		&	0.7			&	1.05	&	42	\\
54~611.093		&	0.27		&	AX~Cir							&	PAH1		& 16		&	22~500		&	0.8			&	1.32	&	43	\\
54~611.101		&	0.27		&	AX~Cir							&	SiC			& 20		&	18~000		&	0.7			&	1.31	&	44	\\
54~611.112		&	0.79		&	X~Sgr							&	PAH1		& 16		&	22~500		&	0.8			&	1.50	&	45	\\
54~611.119		&	0.79		&	X~Sgr							&	SiC			& 20		&	18~000		&	1.1			&	1.44	&	46	\\
54~611.131		&	0.54		&	W~Sgr							&	PAH1		& 16		&	22~500		&	0.8			&	1.43	&	47	\\
54~611.139		&	0.54		&	W~Sgr							&	SiC			& 20		&	18~000		&	0.8			&	1.38	&	48	\\
54~611.151		&				&	HD~161096					&	PAH1		& 16		&	22~500		&	0.8			&	1.49	&	49	\\
54~611.158		&				&	HD~161096					&	SiC			& 20		&	18~000		&	0.8			&	1.43	&	50	\\
54~611.170		&	0.76		&	Y~Oph							&	PAH1		& 16		&	22~500		&	0.7			&	1.27	&	51	\\
54~611.177		&	0.76		&	Y~Oph							&	SiC			& 20		&	18~000		&	0.7			&	1.23	&	52	\\
54~611.189		&	0.48		&	R~Sct							&	PAH1		& 16		&	22~500		&	0.6			&	1.39	&	53	\\
54~611.196		&	0.48		&	R~Sct							&	SiC			& 20		&	18~000		&	0.6			&	1.34	&	54	\\
54~611.207		&				&	HD~161096					&	PAH1		& 16		&	22~500		&	0.6			&	1.21	&	55	\\
54~611.215		&				&	HD~161096					&	SiC			& 20		&	18~000		&	0.6			&	1.19	&	56	\\
54~611.227		&	0.90		&	$\kappa$~Pav				&	PAH1		& 16		&	22~500		&	0.7			&	1.45	&	57	\\
54~611.235		&	0.90		&	$\kappa$~Pav				&	SiC			& 20		&	18~000		&	0.7			&	1.43	&	58	\\
54~611.248		&	0.53		&	$\eta$~Aql					&	PAH1		& 20		&	18~000		&	0.7			&	1.38	&	59	\\
54~611.255		&	0.53		&	$\eta$~Aql					&	SiC			& 25		&	14~400		&	0.7			&	1.33	&	60	\\
54~611.273		&				&	HD~133774					&	PAH1		& 16		&	22~500		&	0.8			&	1.35	&	61	\\
54~611.280		&				&	HD~133774					&	SiC			& 20		&	18~000		&	0.8			&	1.40	&	62	\\
54~611.310		&				&	HD~161096					&	PAH1		& 16		&	22~500		&	0.7			&	1.21	&	63	\\
54~611.317		&				&	HD~161096					&	SiC			& 20		&	18~000		&	0.7			&	1.23	&	64	\\
54~611.340		&	0.15		&	AC~Her							&	PAH1		& 16		&	22~500		&	0.8			&	1.54	&	66	\\
54~611.347		&	0.15		&	AC~Her							&	SiC			& 20		&	18~000		&	0.8			&	1.57	&	67	\\
54~611.365		&	0.04		&	SV~Vul							&	PAH1		& 16		&	22~500		&	0.7			&	1.64	&	68	\\
54~611.372		&	0.04		&	SV~Vul							&	SiC			& 20		&	18~000		&	0.8			&	1.65	&	69	\\
54~611.384		&				&	HD~203504					&	PAH1		& 16		&	22~500		&	0.7			&	1.42	&	70	\\
54~611.392		&				&	HD~203504					&	SiC			& 20		&	18~000		&	0.8			&	1.41	&	71	\\
54~611.403		&	0.62		&	FF~Aql							&	PAH1		& 16		&	22~500		&	0.6			&	1.66	&	72	\\
54~611.411		&	0.62		&	FF~Aql							&	SiC			& 20		&	18~000		&	0.7			&	1.73	&	73	\\
54~611.422		&	0.56		&	$\eta$~Aql					&	PAH1		& 16		&	22~500		&	0.7			&	1.24	&	74	\\
54~611.430		&	0.56		&	$\eta$~Aql					&	SiC			& 20		&	18~000		&	0.7			&	1.27	&	75	\\
\hline 
\end{tabular}
\caption[Journal des observations VISIR]{\textbf{Journal des observations VISIR} : MJD représente le jour Julien modifié, $\phi$ la phase de pulsation de la Céphéide, DIT le temps d'intégration du détecteur, $N$ le nombre total d'images et AM la masse d'air. Le seeing est mesuré dans le visible par la station DIMM.}
\label{table__log_visir}
\end{table}

\begin{sidewaysfigure}
  	\centering\includegraphics[width=.8\linewidth]{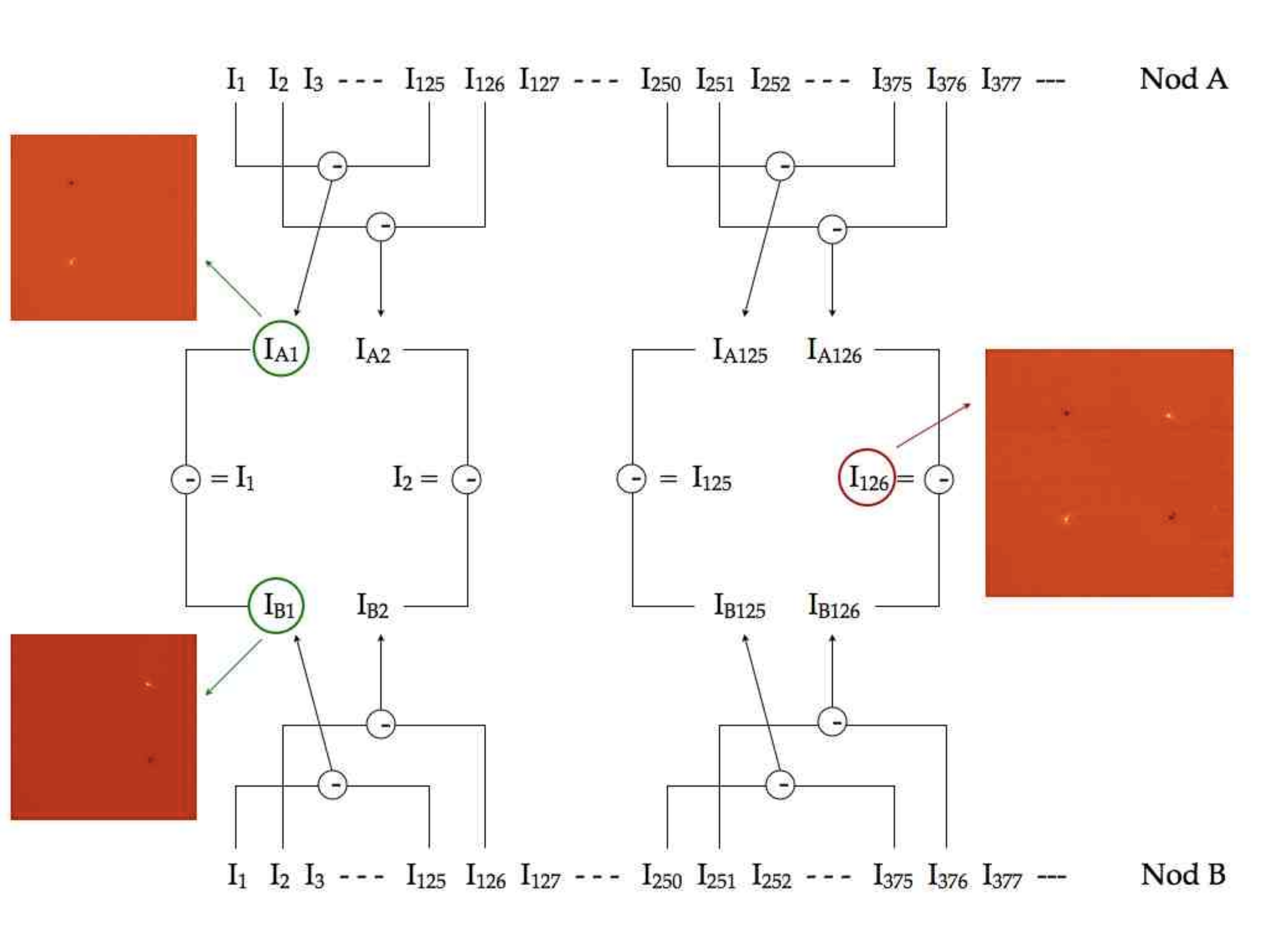}
  	\caption[Algorithme de réduction chopping-nodding]{\textbf{Algorithme de réduction chopping-nodding} : le chopping se fait toutes les 125 images dans cet exemple. Des exemples de données VISIR sont représentés pour chaque étape. Nod A et Nod B sont deux cubes de données séparés.}
  \label{image__algorithme_chop_nod}
\end{sidewaysfigure}

Les données ont donc été obtenues en mode BURST avec chopping/nodding (chop/nod) et à travers 3 filtres photométriques : PAH1, PAH2 et SIC (respectivement $8.59\,\pm\,0.42\,\mu\mathrm{m}$, $11.25\,\pm\,0.59\,\mu\mathrm{m}$ et $11.85\,\pm\,2.34\,\mu\mathrm{m}$). Le journal des observations est présenté dans la Table~\ref{table__log_visir}. Pour chaque Céphéide, une étoile de référence a été observée immédiatement avant ou après, dans les mêmes configurations instrumentales. L'échelle de $75\,\mathrm{mas/pixel}$ a été choisi afin de bien échantillonner la largeur à mi-hauteur de la FEP. 

Les images sont enregistrées sous forme de cube de données où chaque cube représente une position de nodding. Le nombre de cubes correspondant à une série d'acquisition chop/nod est donc un multiple de 2. La séquence d'images enregistrées pour un cube de données (et chaque position de nodding) est la suivante : $m$ acquisitions en position chopping 1 puis $m$ en position chopping 2 et cela répété $n$ fois. Les paramètres $m$ et $n$ dépendent du temps d'intégration minimal et total : $m = 125$ et $n = 30$ en PAH1, $m = 250$ et $n = 16$ en PAH2 et $m = 100$ et $n = 30$ en SIC. Pour une raison inconnue, la position du nodding est hors du détecteur pour les acquisitions \#15, \#17, \#19 à \#22, \#25 à \#28 et \#31 à \#34. Toutefois, il reste suffisamment d'images ($\sim10~000$) pour soustraire correctement le fond. Ce fond résiduel sera supprimé lors de la photométrie d'ouverture. Le chopping et nodding des acquisitions \#23 et \#24 sont hors du détecteur et ne furent pas incluses dans l'analyse. À cause de la faible sensibilité dans le filtre SIC, je n'ai pas pu effectuer le recentrage des images individuelles des cubes \#40 et \#44, ces données n'ont donc pas été analysées. De même pour le cube \#4 dont le signal est très faible, peut être dû au passage d'un nuage.

\paragraph*{\textcolor{black}{Algorithme de traitement des données brutes}}

Les données brutes minimales sont donc deux cubes d'images, un pour chaque position de nodding (\ciap Nod A et Nod B), et chaque cube comprend les deux positions de chopping (\ciap Chop 1 et Chop 2). La première étape pour chaque cube est de soustraire à l'image Chop 1, l'image Chop 2. Les étapes sont représentées sur la Fig.~\ref{image__algorithme_chop_nod} où le chopping se fait tout les 125 images. La première étape est donc $I_{A1} = I_1 - I_{125}$, ... et ce pour chaque image de chaque cube. La seconde étape est de soustraire à chaque image Nod 1, l'image Nod B : $I_1 = I_\mathrm{A1} - I_\mathrm{B1}$, ... Au final on obtient plusieurs images dont chacune comprend quatre images de l'étoile à quatre positions différentes. Ces processus demandent une mémoire vive importante car les cubes de données contiennent des milliers d'images (un cube a une taille $\sim$500Mo), et le traitement d'un couple chop/nod peut prendre quelques minutes.

On découpe ensuite sur chaque image, quatre sous-fenêtres correspondant aux quatre positions (largeur $\sim20\,\mathrm{FWHM}$), que l'on enregistre sous forme de cube d'images. Chaque nouveau cube, qui comprend des milliers d'images ($\sim20~000$), ne contient plus d'émission de fond, et seul le flux de l'étoile est dominant. Pour les données dont le nodding s'est fait hors du détecteur, seule la soustraction du chopping a été effectuée, et un léger fond résiduel est toujours présent.

Chaque image de ce nouveau cube est ensuite corrigée des mauvais pixels avant de sélectionner $50\,\%$ des meilleures en fonction du pixel le plus brillant (en utilisant comme critère du rapport de Strehl). Cette sélection permet de garder les meilleures images uniquement limitées par la diffraction. Ces images subissent par la suite un premier recentrage (d'un niveau de précision de l'ordre du pixel) avant d'être re-échantillonnées d'un facteur 4 via une interpolation par des splines cubiques (la largeur à mi hauteur de la FEP est $\sim 4$ pixels). Un recentrage plus fin (à un niveau de précision de quelques millisecondes d'angle) est ensuite effectué en ajustant une fonction Gaussienne au c\oe ur de la FEP. Finalement, chaque cube est moyenné afin d'obtenir les images moyennes utilisées pour l'analyse.

\begin{figure}[!p]
\centering
\includegraphics[width = .325\linewidth]{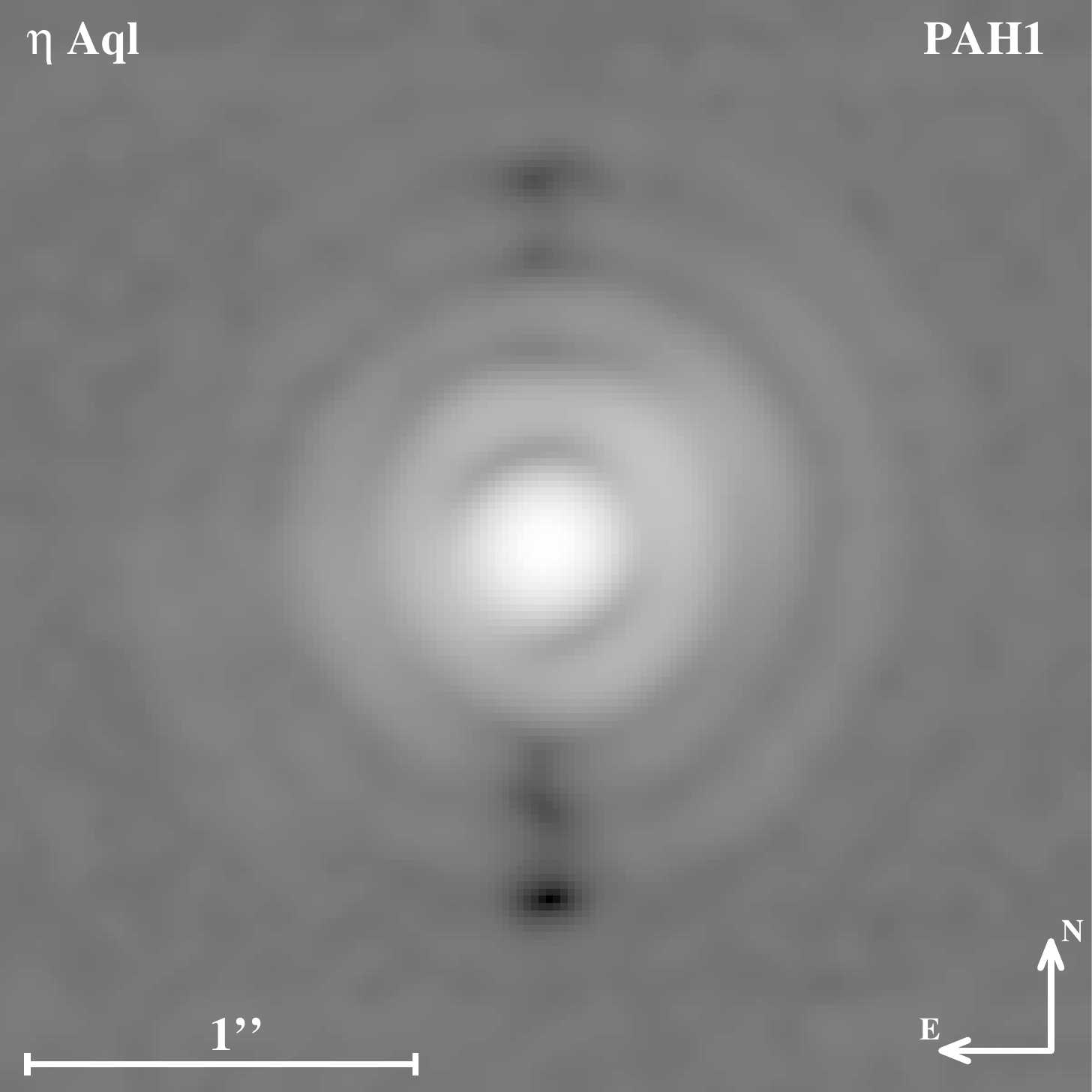}\hspace{.02cm}
\includegraphics[width = .325\linewidth]{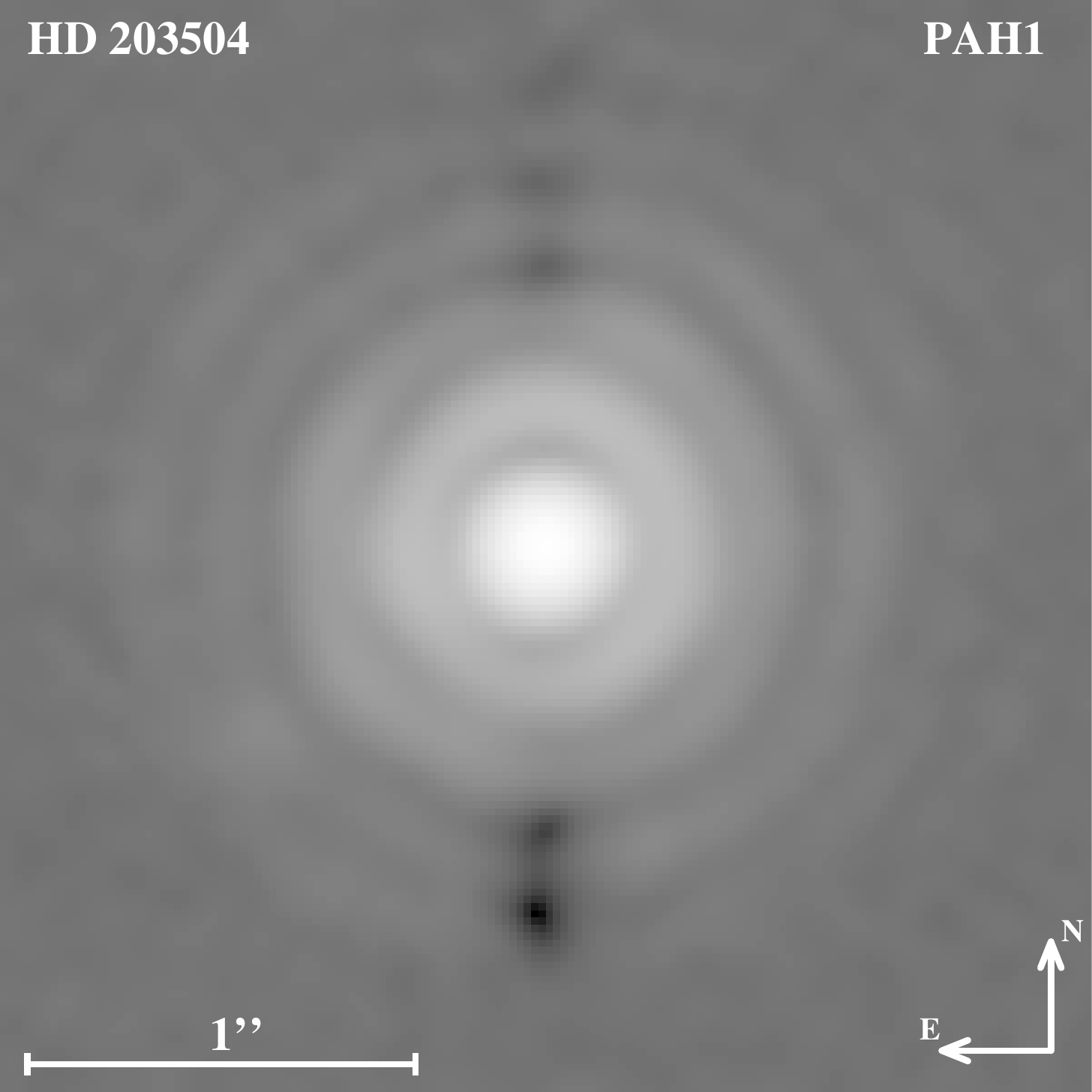}\hspace{.02cm}
\includegraphics[width = .325\linewidth]{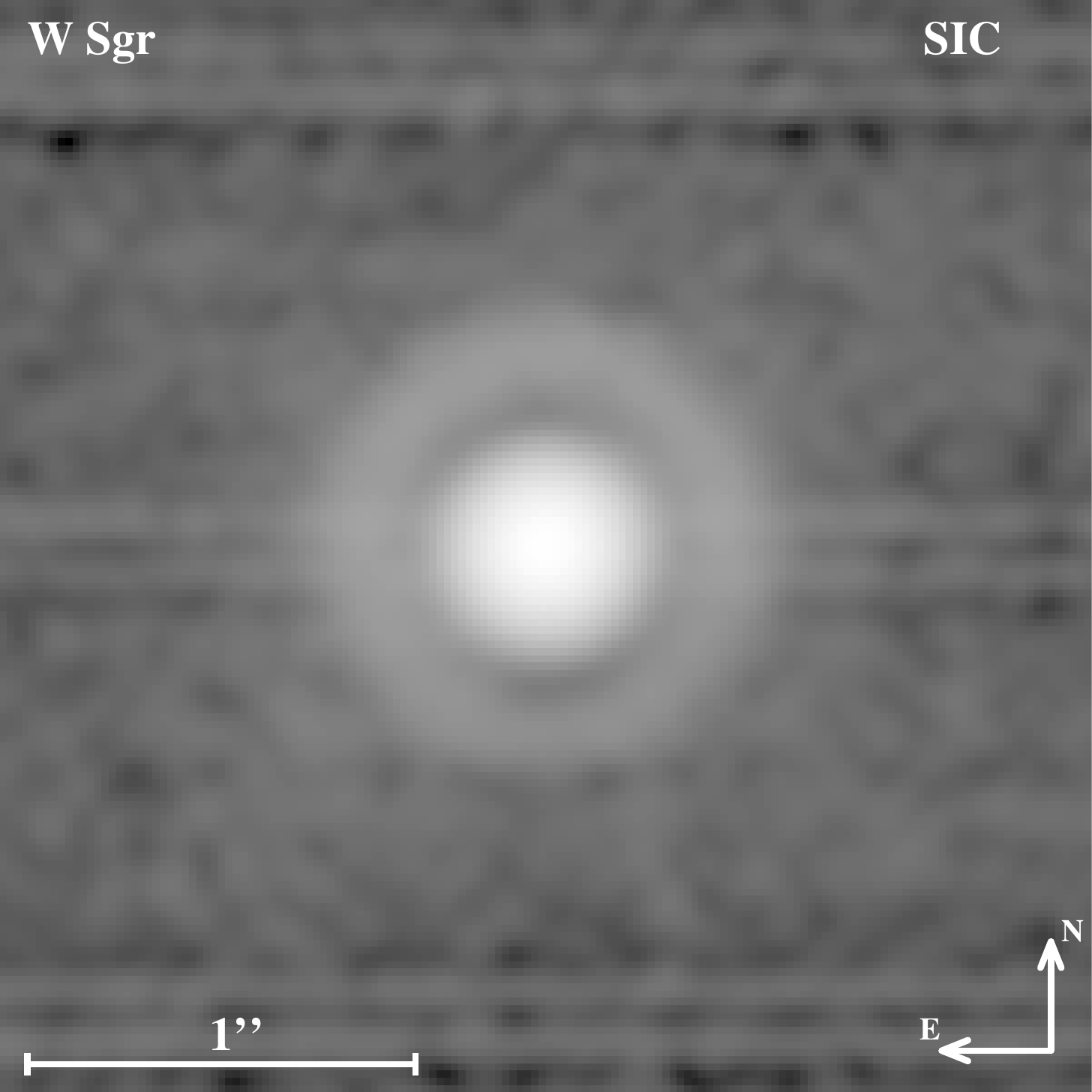}\vspace{.05cm}

\includegraphics[width = .325\linewidth]{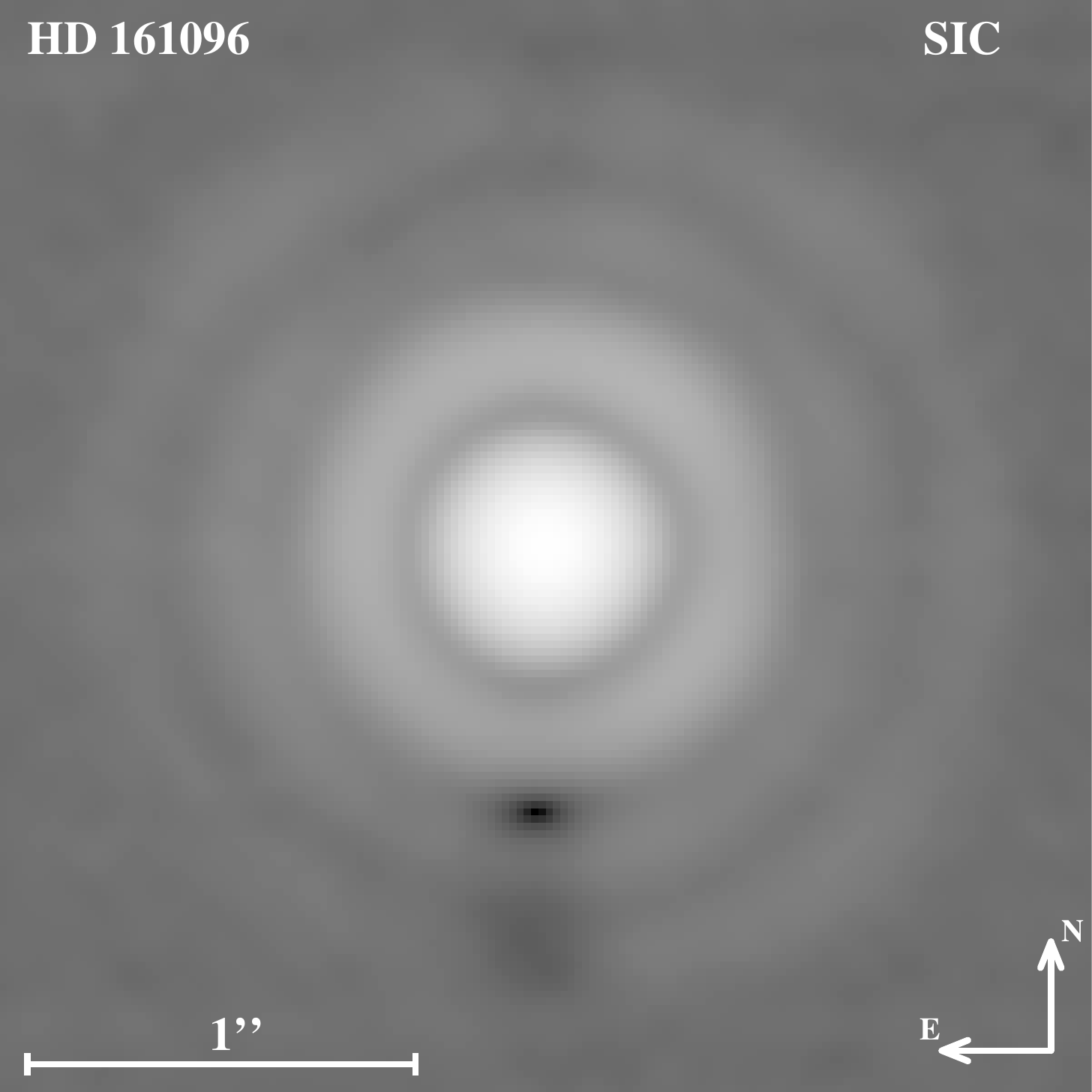}\hspace{.02cm}
\includegraphics[width = .325\linewidth]{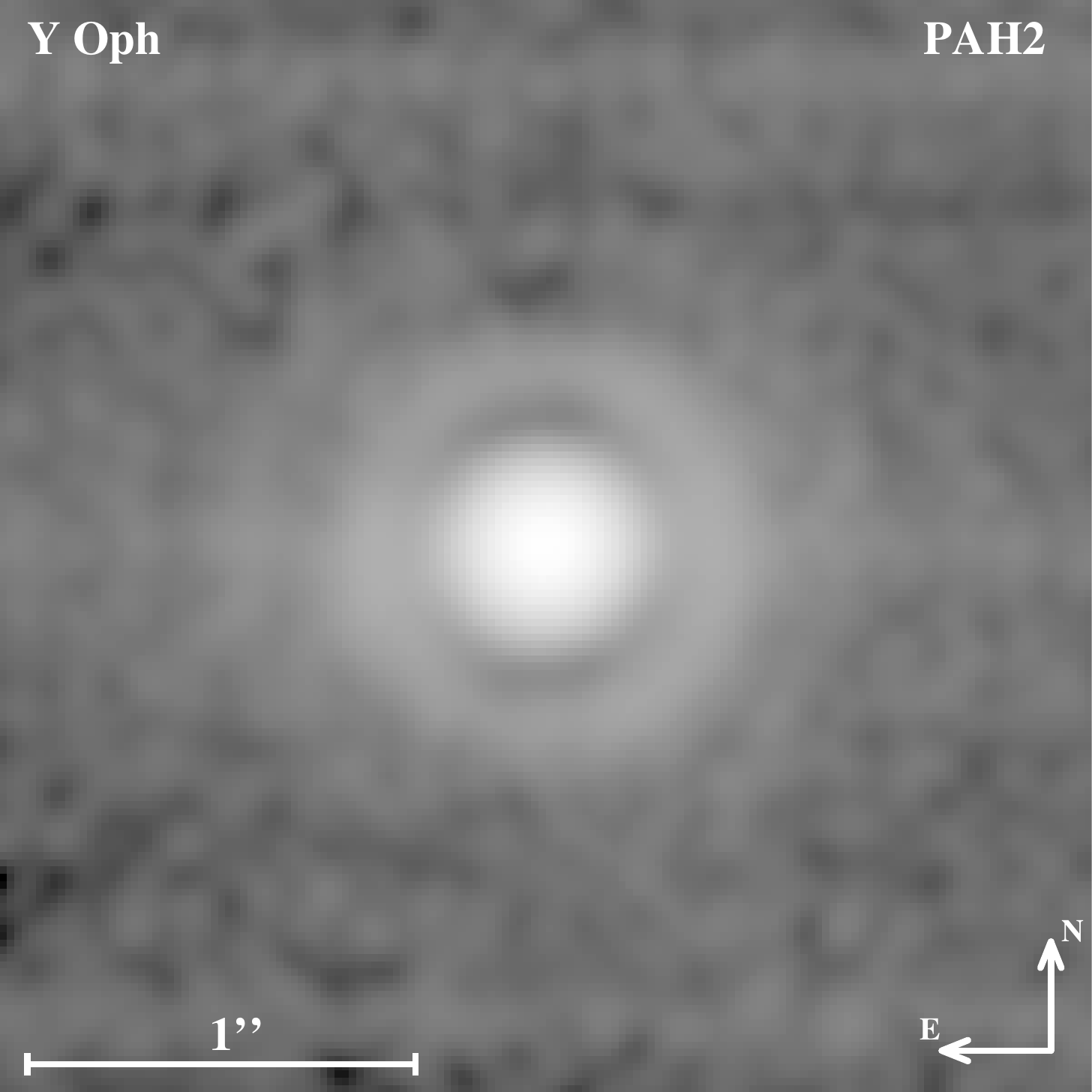}\hspace{.02cm}
\includegraphics[width = .325\linewidth]{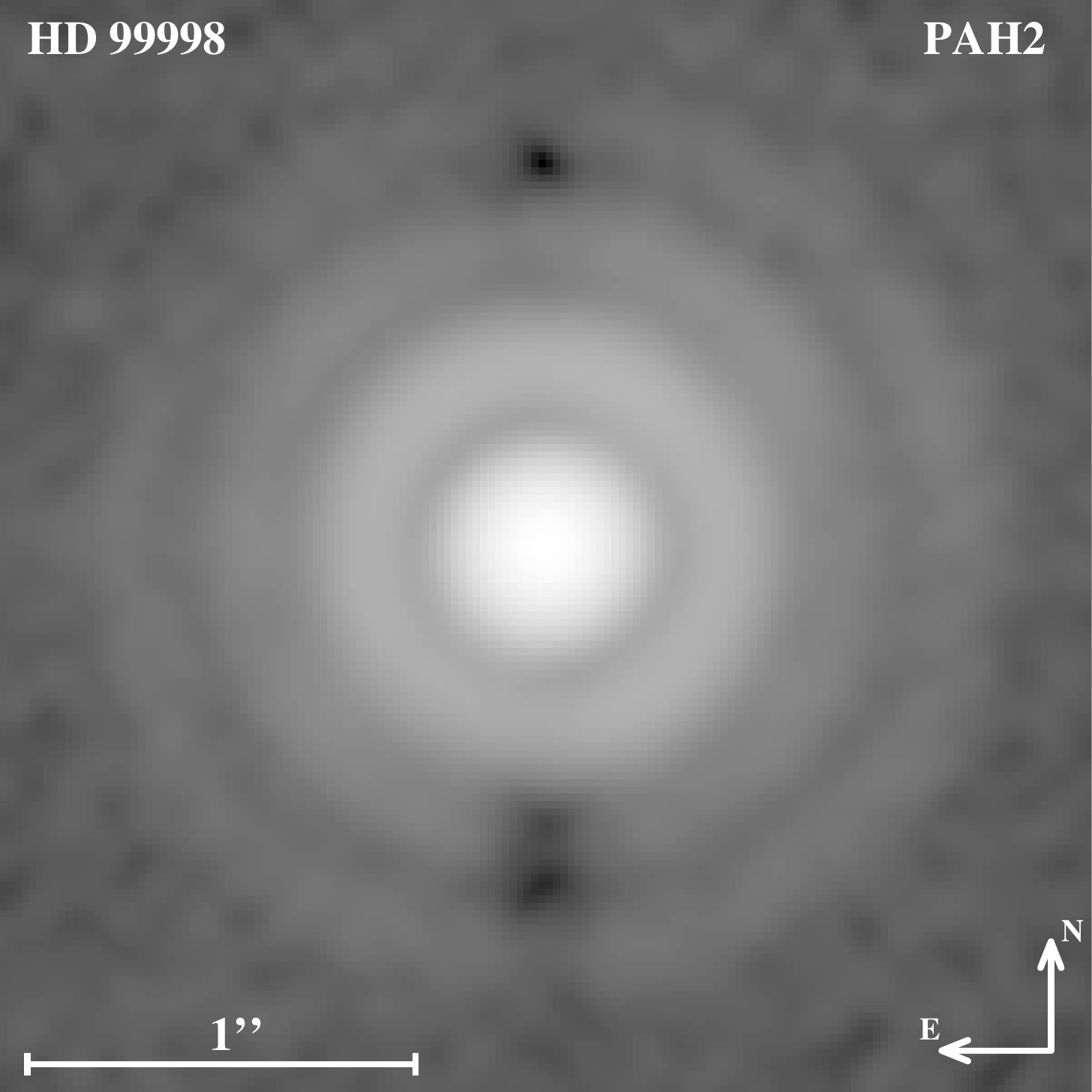}\vspace{.05cm}

\includegraphics[width = .325\linewidth]{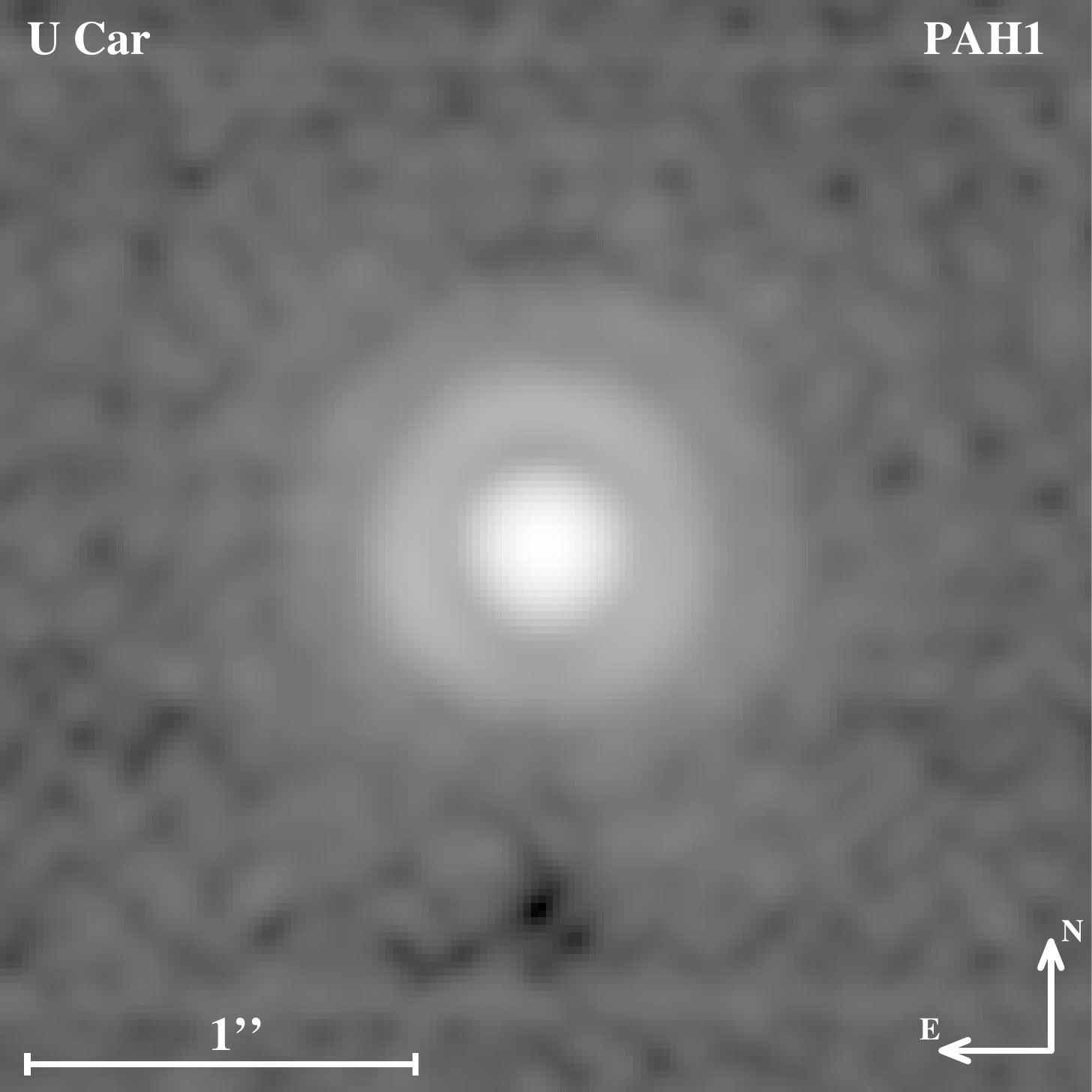}\hspace{.02cm}
\includegraphics[width = .325\linewidth]{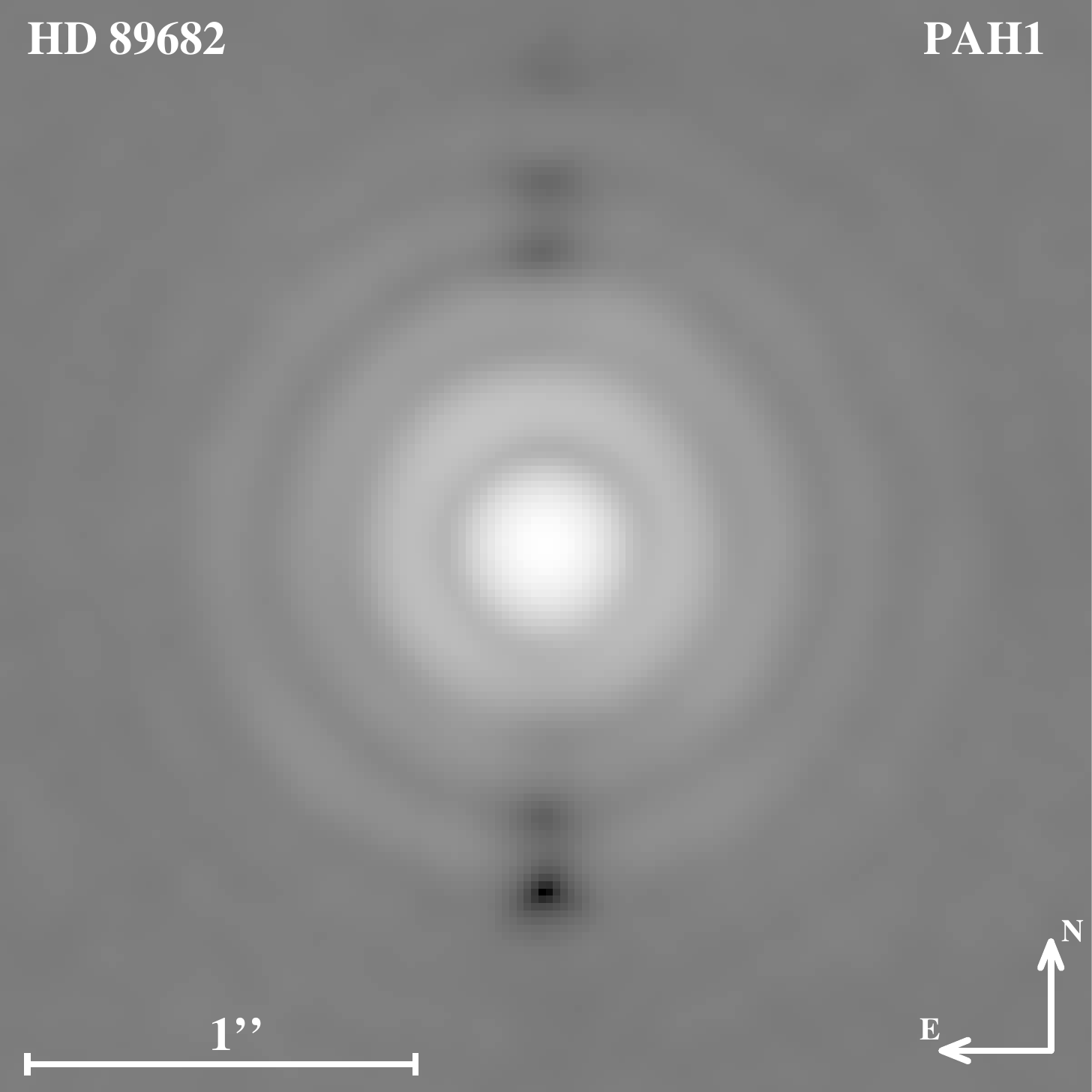}\hspace{.02cm}
\includegraphics[width = .325\linewidth]{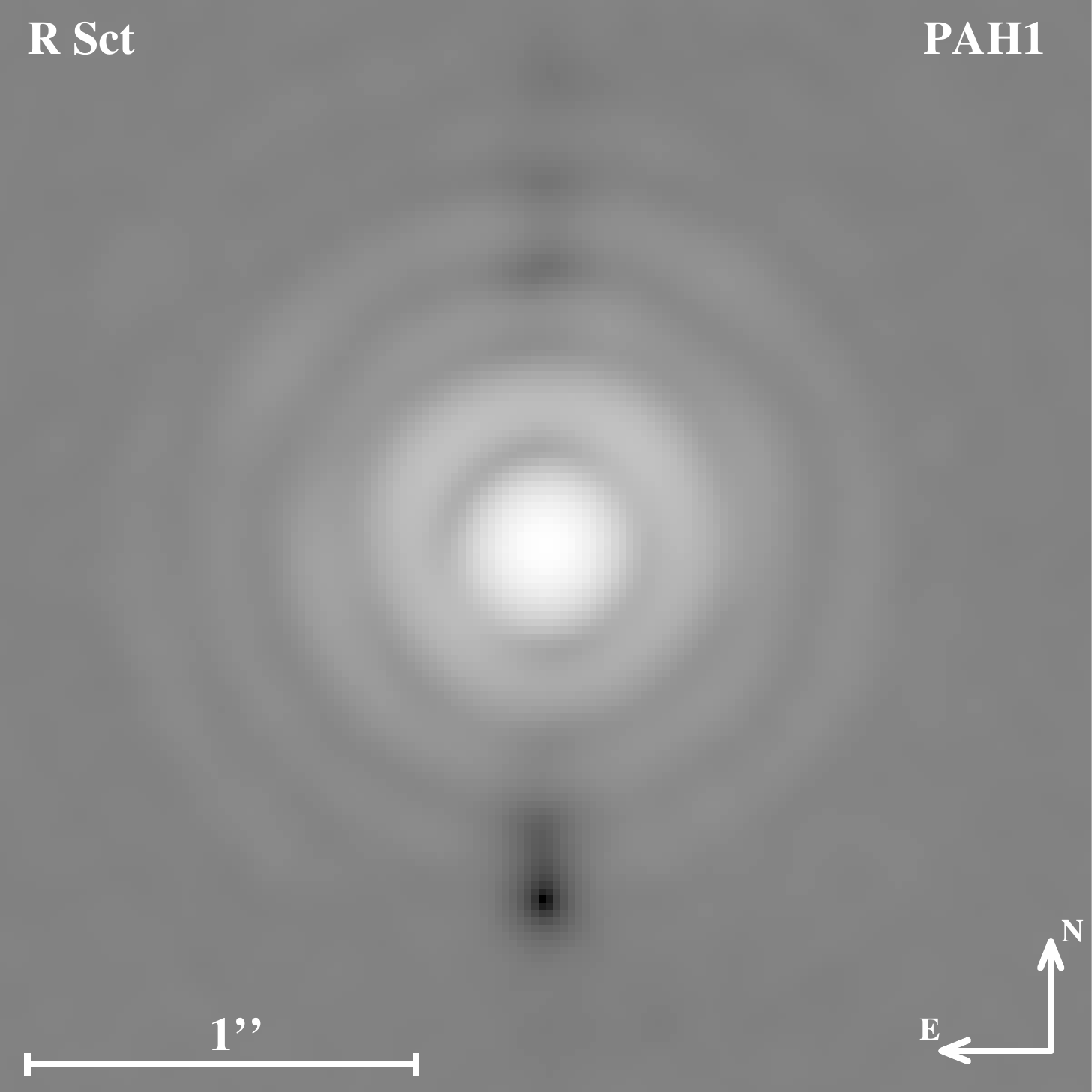}\vspace{.05cm}

\includegraphics[width = .325\linewidth]{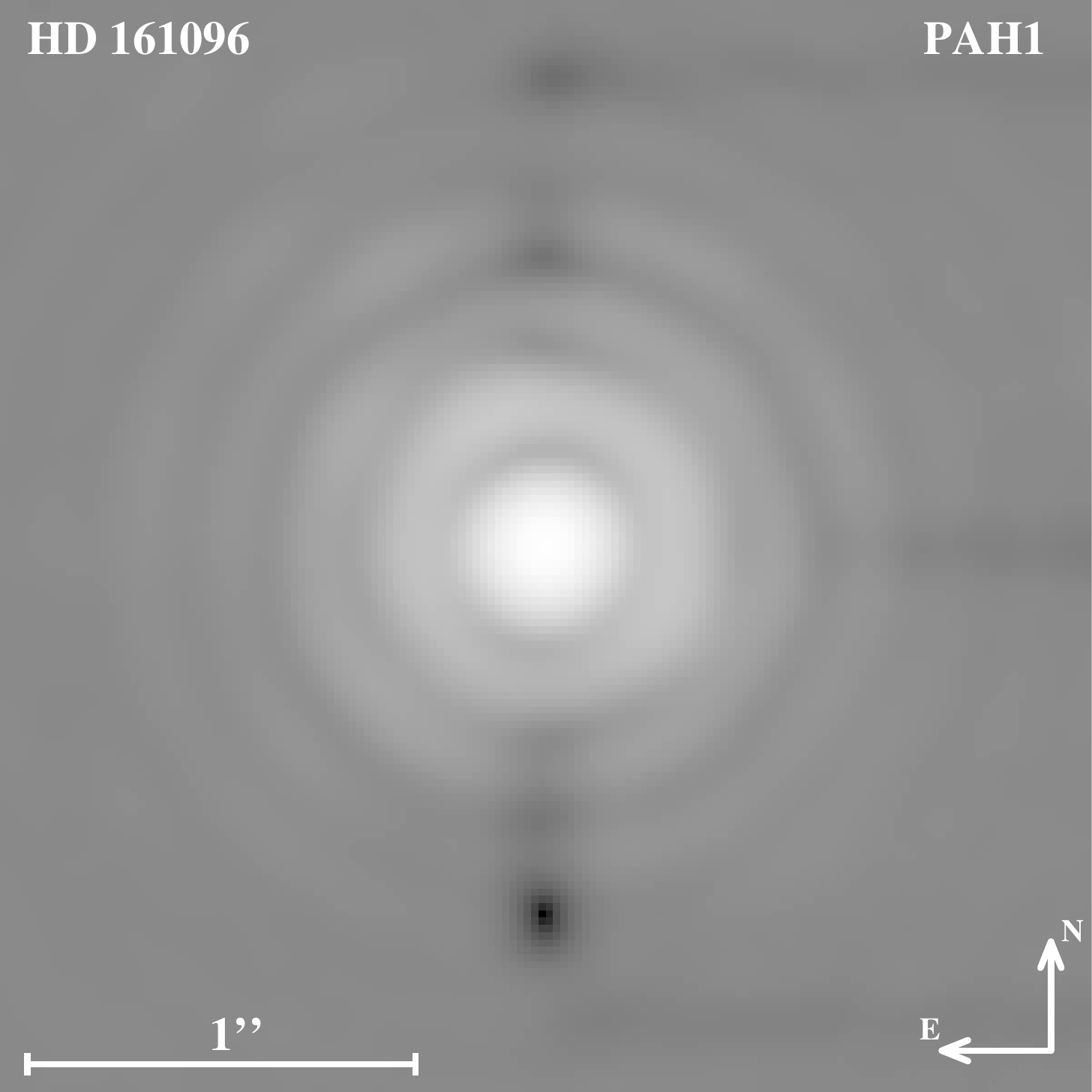}\hspace{.02cm}
\includegraphics[width = .325\linewidth]{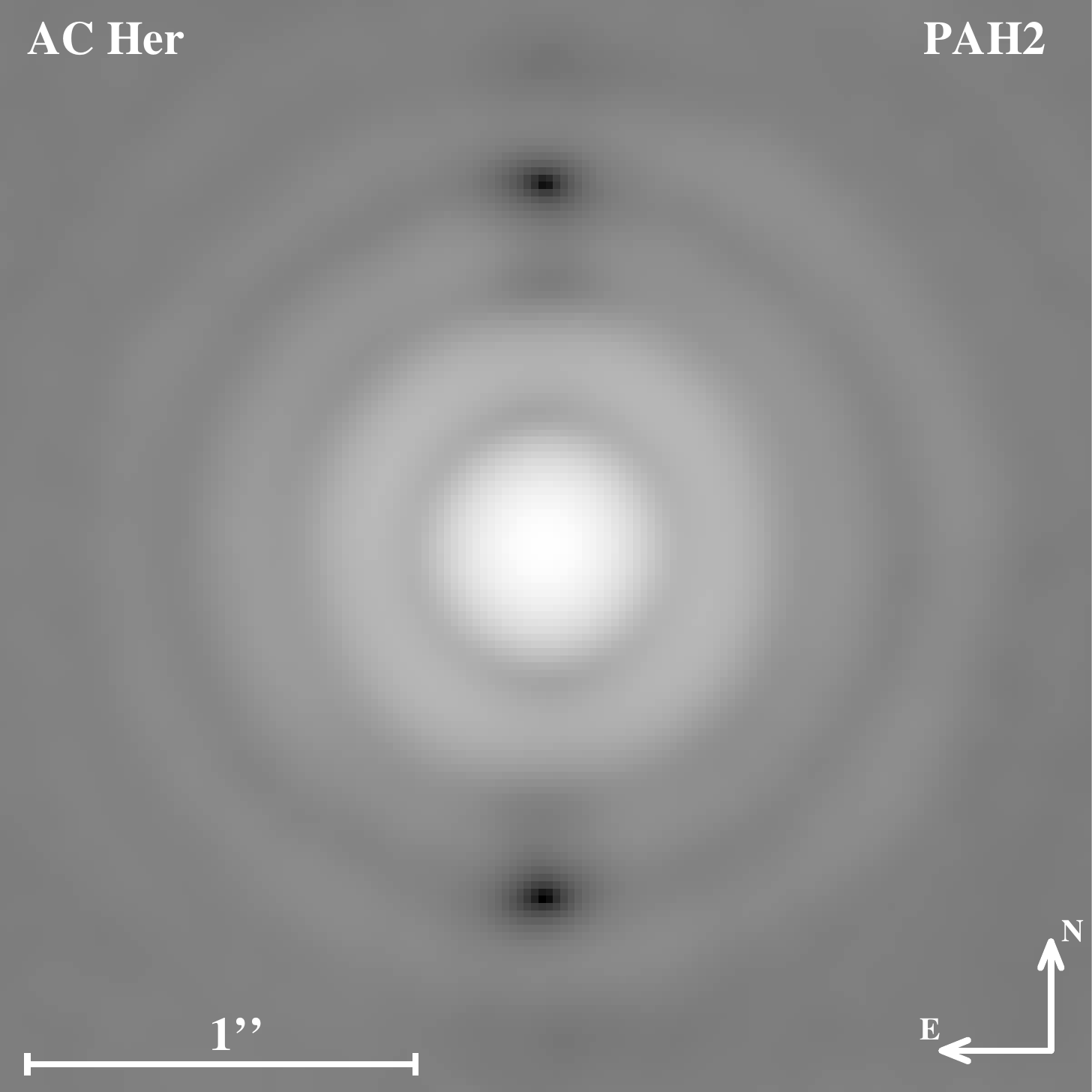}\hspace{.02cm}
\includegraphics[width = .325\linewidth]{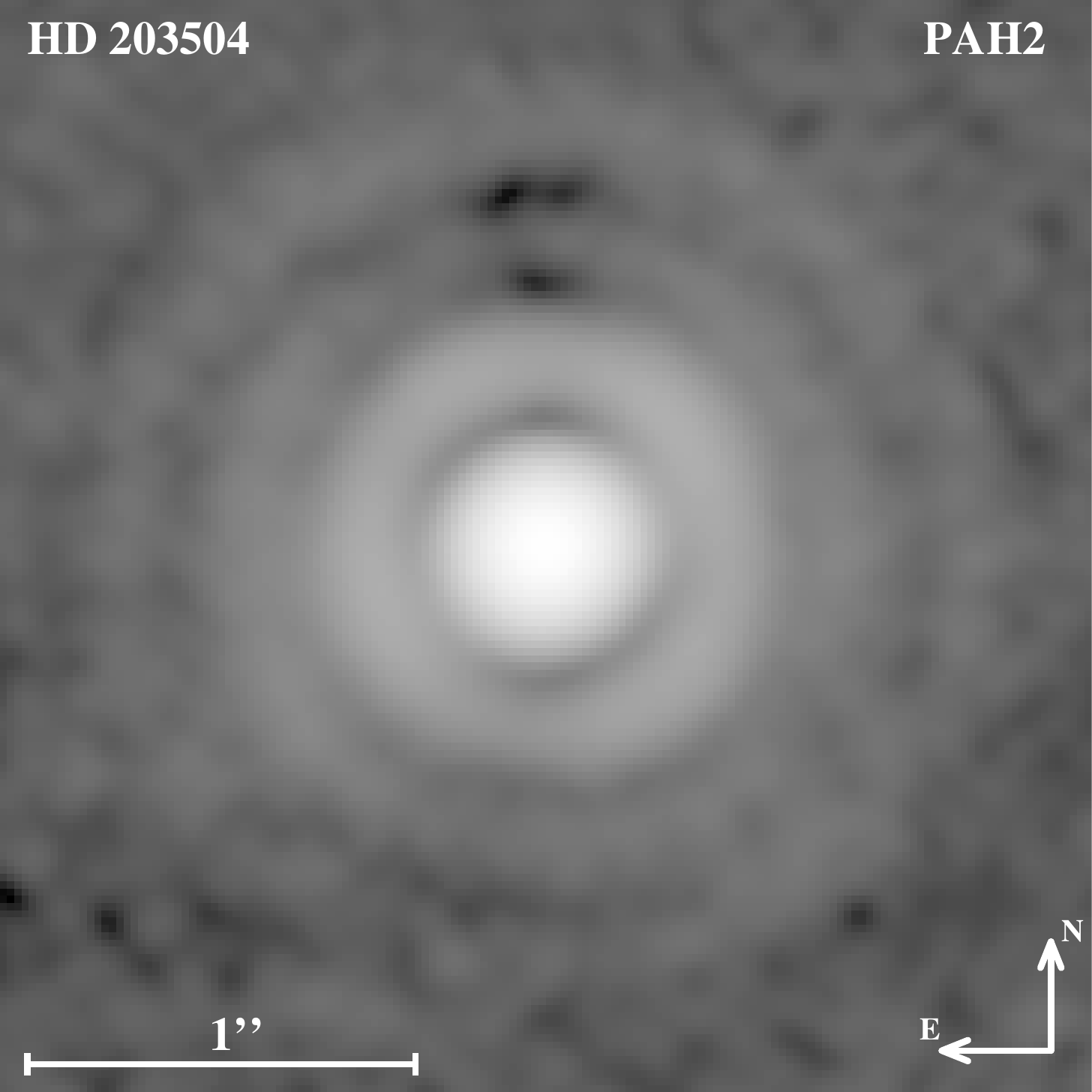}
\caption[Échantillon d'images moyennes]{\textbf{Echantillon d'images moyennes} : l'échelle est logarithmique pour toutes les images.}
\label{image__VISIR_images}
\end{figure}

Ce procédé a déjà été utilisé auparavant et l'efficacité du mode cube par rapport à une longue pose est bien connue \citep{Kervella-2007-11,Kervella-2009-05}. Le gain en résolution grâce au mode cube est présenté en Section~\ref{section__lucky_imaging}. Je présente quelques images moyennes issues de cet algorithme (développé initialement par P. Kervella) sur la Fig.~\ref{image__VISIR_images}.

\subsection{Photométrie}

J'ai effectué sur les images moyennes la photométrie d'ouverture présentée précédemment (Section~\ref{section__photometrie_ouverture}) avec une ouverture $r_\mathrm{d} = 1.30\arcsec$ et une couronne de rayons $r_\mathrm{in} = 1.70\arcsec$ et $r_\mathrm{ext} = 1.90\arcsec$ (grâce au chopping/nodding, le flux dans la couronne est $\sim0$). J'ai utilisé les modèles de spectre de \citet{Cohen-1999-04} que j'ai intégrés sur toute la bande passante des filtres selon l'équation~\ref{equation__fonction_transfert} pour obtenir la photométrie absolue.

Pour améliorer l'étalonnage du flux, j'ai appliqué un facteur de correction lié à la masse d'air traversée pour tenir compte de l'absorption atmosphérique :
\begin{displaymath}
F_\mathrm{corr} = F_\mathrm{obs} \times C(\lambda,\mathrm{AM})
\end{displaymath}
avec $C(\lambda,\mathrm{AM})$ de \citet{Schutz-2005-}:
\begin{displaymath}
C(\lambda,\mathrm{AM}) = 1 + \left[ 0.220 - \frac{0.104}{3}(\lambda - 8.6\,\mu\mathrm{m}) \right] (\mathrm{AM} - 1)
\end{displaymath}

$C(\lambda,\mathrm{AM})$ est bien sûr relié à l'équation~\ref{equation__exctinction_atmospherique} tel que :
\begin{displaymath}
C(\lambda,\mathrm{AM}) = 10^{0.4\,K_\lambda(\mathrm{AM})}
\end{displaymath}

J'ai ensuite estimé l'irradiance des étoiles en utilisant ces flux corrigés dans l'équation~\ref{equation__etalonnage_flux}. Les valeurs sont reportées dans la Table~\ref{table__irradiance_mesure_I} et \ref{table__irradiance_mesure_II}. Les incertitudes comprennent la dispersion statistique de la photométrie (dans les cubes) ainsi que l'incertitude lié à l'étalonnage.

\begin{table}[!p]
\centering
\begin{tabular}{cccccc} 
\hline
\hline
Nom	  					&	    MJD			&	Filtre			& 	Irradiance												& Irradiance				& Excès 		\\
		  					&						&					&	($\mathrm{W/m^2/\mu m}$)				& 	(Jy)						& 	($\%$)			\\
\hline
FF~Aql					& 	54~611.403	&  PAH1		&	$9.23\,\pm\, 0.25\,\times\,10^{-14}$	&	$2.27\,\pm\,0.06$	&	$1.6\,\pm\,2.8$	\\
							& 	54~611.411	&	SiC			&	$2.79\,\pm\, 0.08\,\times\,10^{-14}$	&	$1.28\,\pm\,0.06$	&	$3.7\,\pm\,3.0$	\\
\hline
AX~Cir					& 	54~611.076	&  PAH1		&	$6.97\,\pm\, 0.19\,\times\,10^{-14}$	&	$1.72\,\pm\,0.05$	&	$-0.8\,\pm\,2.9$	\\
\hline
X~Sgr					& 	54~610.104	&  PAH1		&	$2.31\,\pm\, 0.11\,\times\,10^{-13}$	&	$5.69\,\pm\,0.27$	&	$14.1\,\pm\,5.4$	\\
							&	54~611.112	&	PAH1		&	$2.11\,\pm\, 0.10\,\times\,10^{-13}$	&	$5.20\,\pm\,0.25$	&	$3.8\,\pm\,9.8$	\\
							&	54~610.111	&	PAH2		&	$7.82\,\pm\, 0.20\,\times\,10^{-14}$	&	$3.31\,\pm\,0.08$	&	$11.9\,\pm\,2.9$	\\
							&	54~611.119	&	SiC			&	$6.27\,\pm\, 0.15\,\times\,10^{-14}$	&	$2.89\,\pm\,0.07$	&	$5.6\,\pm\,8.1$	\\
\hline
$\eta$~Aql			&	54~610.282	&  PAH1		&	$3.96\,\pm\, 0.14\,\times\,10^{-13}$	&	$9.73\,\pm\,0.35$	&	$0.38\,\pm\,3.6$	\\
							&	54~611.248	&  PAH1		&	$4.30\,\pm\, 0.16\,\times\,10^{-13}$	&	$10.6\,\pm\,0.39$	&	$9.0\,\pm\,4.1$		\\
							&	54~610.289	&	PAH2		&	$1.39\,\pm\, 0.05\,\times\,10^{-13}$	&	$5.89\,\pm\,0.21$	&	$2.6\,\pm\,3.7$		\\
							&	54~611.255	&	SiC			&	$1.28\,\pm\, 0.05\,\times\,10^{-13}$	&	$5.90\,\pm\,0.23$	&	$9.6\,\pm\,4.3$		\\
\hline
W~Sgr					&	54~610.213	&  PAH1		&	$1.70\,\pm\, 0.07\,\times\,10^{-13}$	&	$4.19\,\pm\,0.17$	&	$10.4\,\pm\,4.6$ \\
							&	54~611.131	&  PAH1		&	$1.82\,\pm\, 0.05\,\times\,10^{-13}$	&	$4.48\,\pm\,0.12$	&	$18.2\,\pm\,3.3$	\\
							&	54~610.220	&	PAH2		&	$6.16\,\pm\, 0.36\,\times\,10^{-14}$	&	$2.61\,\pm\,0.15$	&	$16.5\,\pm\,6.8$	\\
							&	54~611.139	&	SiC			&	$5.30\,\pm\, 0.13\,\times\,10^{-13}$	&	$24.4\,\pm\,0.6$		&	$16.2\,\pm\,2.9$	\\
\hline
Y~Oph					&	54~610.126	&  PAH1		&	$2.01\,\pm\, 0.05\,\times\,10^{-13}$	&	$4.95\,\pm\,0.12$	&	$6.8\,\pm\,2.7$	\\
							&	54~611.170	&  PAH1		&	$2.00\,\pm\, 0.05\,\times\,10^{-13}$	&	$4.93\,\pm\,0.12$	&	$6.3\,\pm\,2.7$	\\
							&	54~610.134	&	PAH2		&	$6.62\,\pm\, 0.25\,\times\,10^{-14}$	&	$2.80\,\pm\,0.11$	&	$2.4\,\pm\,3.9$	\\
							&	54~611.177	&	SiC			&	$5.83\,\pm\, 0.15\,\times\,10^{-14}$	&	$2.69\,\pm\,0.07$	&	$4.6\,\pm\,2.7$	\\
\hline
U~Car					&	54~610.546	&  PAH1		&	$1.07\,\pm\, 0.02\,\times\,10^{-13}$	&	$2.64\,\pm\,0.05$	&	$32.1\,\pm\,2.5$	\\
							&	54~611.035	&	PAH1		&	$1.06\,\pm\, 0.05\,\times\,10^{-14}$	&	$2.61\,\pm\,0.12$	&	$30.9\,\pm\,6.2$	\\
							&	54~611.042	&	SiC			&	$2.91\,\pm\, 0.2\,\times\,10^{-14}$	&	$1.34\,\pm\,0.09$	&	$20.9\,\pm\,8.3$	\\
\hline
SV~Vul					&	54~611.365	&  PAH1		&	$8.62\,\pm\, 0.17\,\times\,10^{-14}$	&	$2.12\,\pm\,0.04$	&	$25.0\,\pm\,2.5$	\\
							&	54~611.372	&	SiC			&	$2.35\,\pm\, 0.07\,\times\,10^{-14}$	&	$1.08\,\pm\,0.03$	&	$15.1\,\pm\,3.4$	\\
\hline
\end{tabular}
\caption[Irradiances mesurées par photométrie des Céphéides classiques]{\textbf{Irradiances mesurées par photométrie des Céphéides classiques} : Les flux ont été mesurés sur une ouverture de 1.30\arcsec. L'excès est relatif au modèle photosphérique de l'étoile.}
\label{table__irradiance_mesure_I}
\end{table}

\subsection{Étude de l'évolution atmosphérique}

Grâce aux diverses étoiles étalons observées, il est possible d'étudier qualitativement la transparence du ciel. Pour cela j'utilise l'équation~\ref{equation__fonction_transfert}, la valeur du flux mesuré sur les images et le flux absolu donné par \citet{Cohen-1999-04} pour estimer $H$. La fonction de transfert peut être décomposée telle que $H = T_\mathrm{f}\,T_\mathrm{i}\,T_\mathrm{a}$ où $T_\mathrm{f}$ représente la transmission du filtre, $T_\mathrm{i}$ la fonction de transfert de l'instrument et $T_\mathrm{a}$ la transmission atmosphérique. En faisant l'hypothèse que $T_\mathrm{i}$ est stable pendant les observations, on peut écrire :
\begin{displaymath}
\frac{H}{\overline{H}} \simeq \frac{T_\mathrm{a}}{\overline{T_\mathrm{a}}}
\end{displaymath}

\begin{figure}[!p]
  	\centering\includegraphics[width=.7\linewidth]{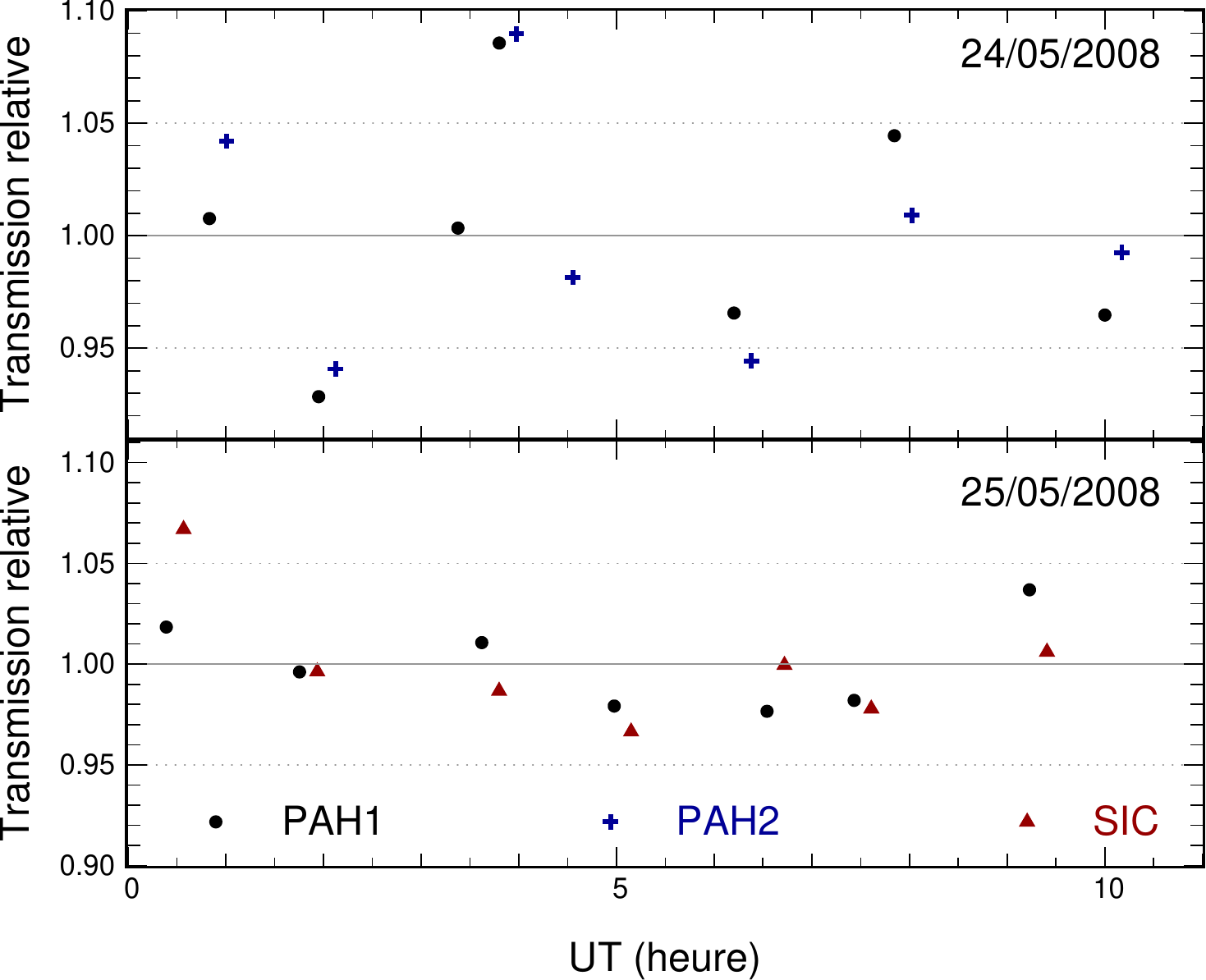}
  	\caption[Transmission atmosphérique relative]{\textbf{Transmission atmosphérique relative} : la courbe en pointillé représente une variation relative de $5\,\%$. La première nuit était assez stable, avec seulement une variation relative de $5\,\%$ en PAH1 et PAH2. La seconde nuit était plus constante avec seulement une variation relative de $3\,\%$ dans les deux filtres.}
  \label{image__transmission_relative}
\end{figure}

La transmission atmosphérique relative est tracée sur la Fig.~\ref{image__transmission_relative} pour les deux nuits. Comme nous pouvons le constater, les nuits étaient très stables avec une variation relative de $5\,\%$ la première nuit et de seulement $3\,\%$ la seconde.

\subsection{Distribution spectrale d'énergie des Céphéides classiques}
\label{subsection__distribution_spectrale_energie_des_cepheides_classiques}

Dans cette section j'étudie la densité d'énergie spectrale des Céphéides classiques afin de mesurer un éventuel excès infrarouge. Pour cela, en plus de nos données autour de $10\,\mu\mathrm{m}$, j'ai collecté dans la littérature des mesures photométriques supplémentaires allant de $\sim0.4\,\mu\mathrm{m}$ à $\sim100\,\mu\mathrm{m}$.

\paragraph*{\textcolor{black}{Choix des données photométriques}}

\begin{figure}[!p]
\centering
\resizebox{\hsize}{!}{\includegraphics{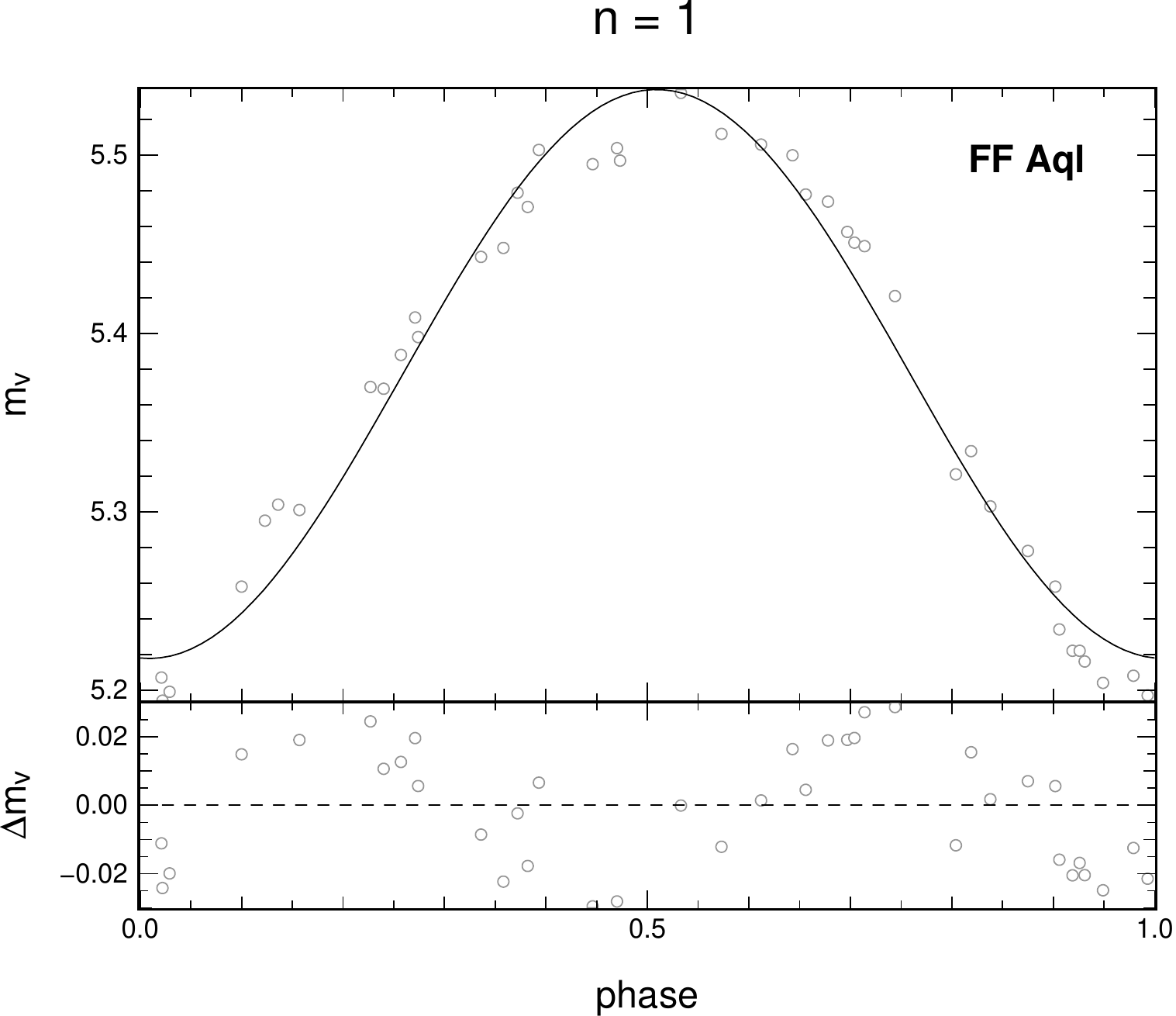}\hspace{.05cm}
\includegraphics{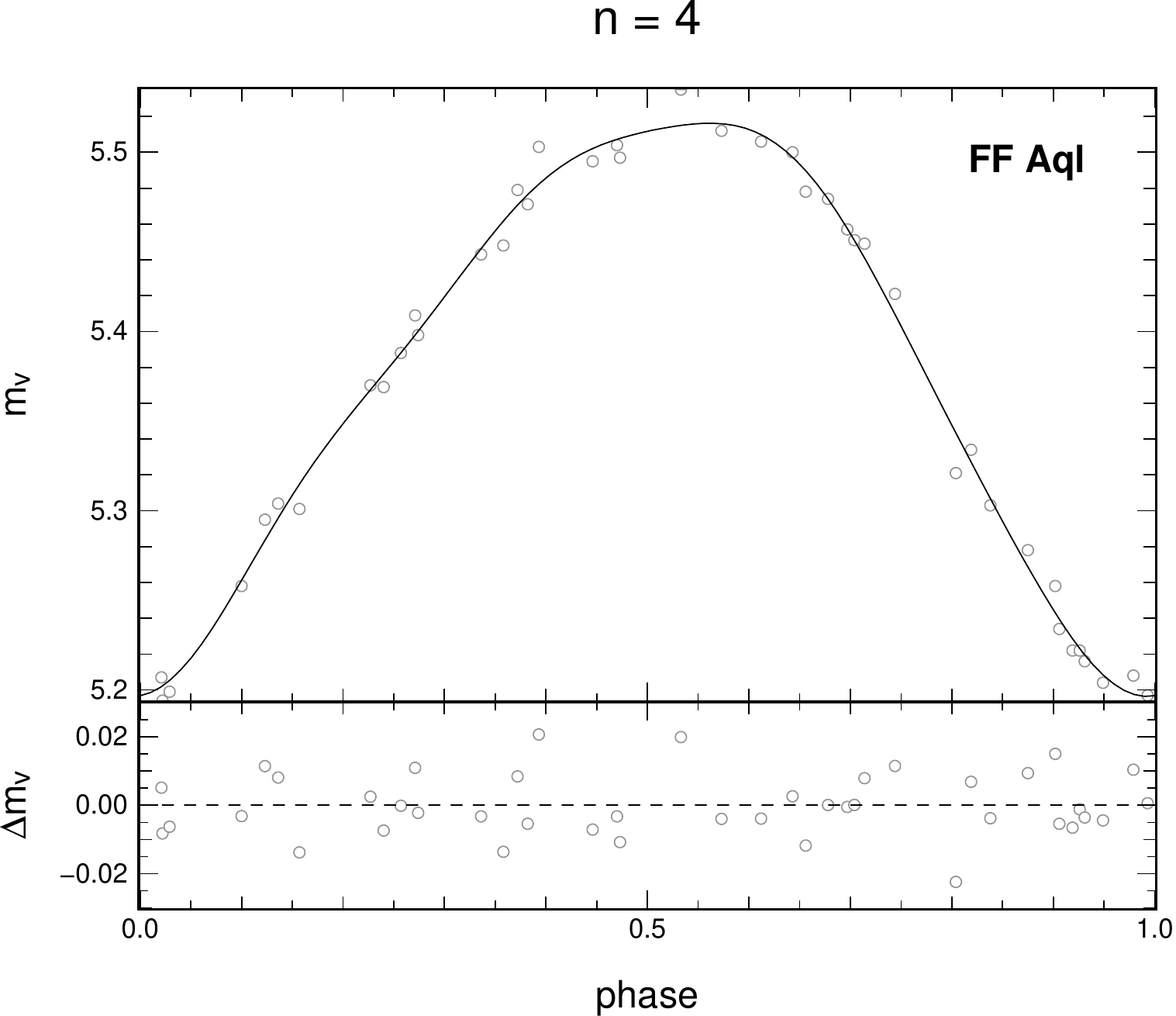}\hspace{.05cm}
\includegraphics{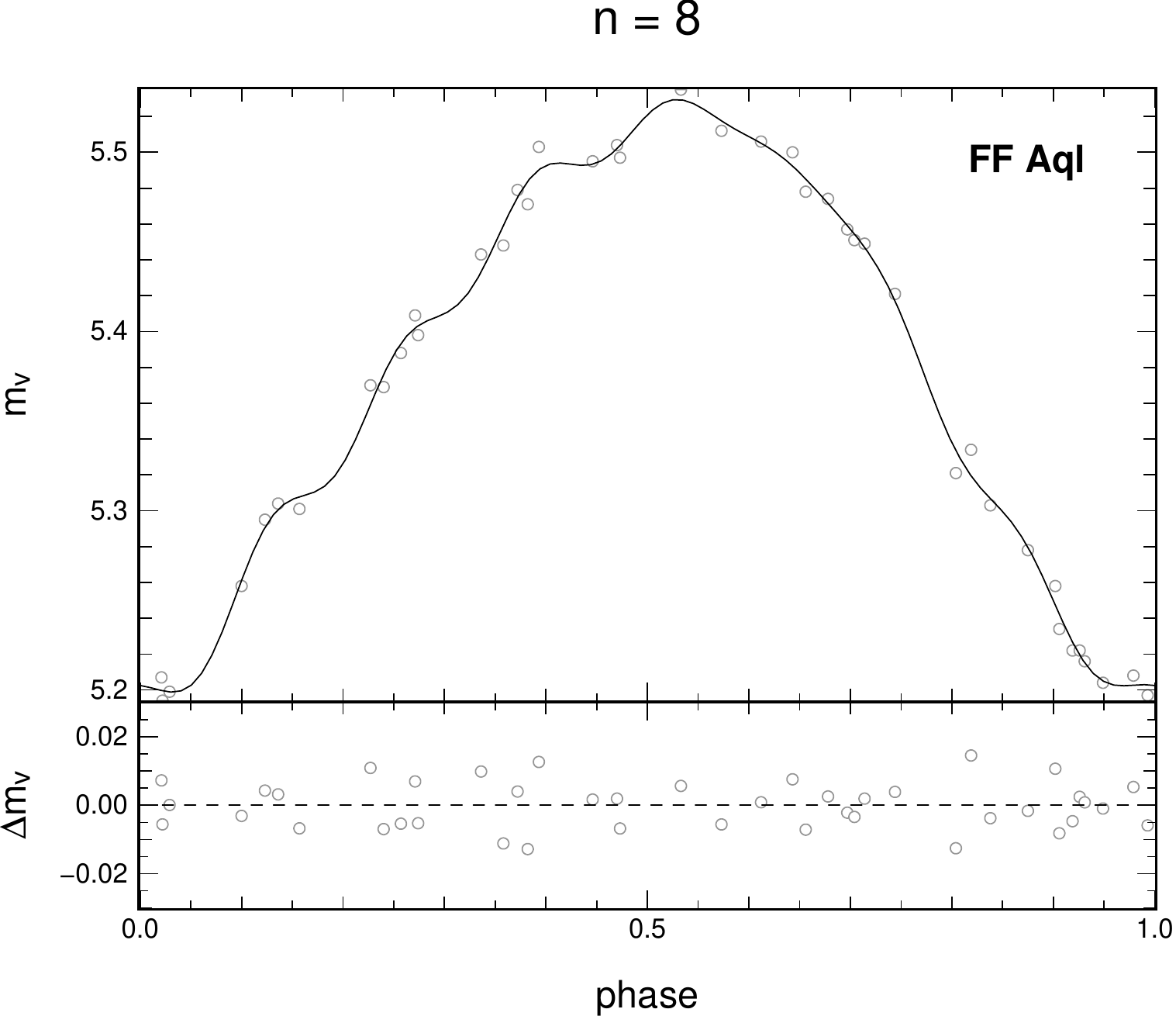}}\vspace{.2cm}
\resizebox{\hsize}{!}{\includegraphics{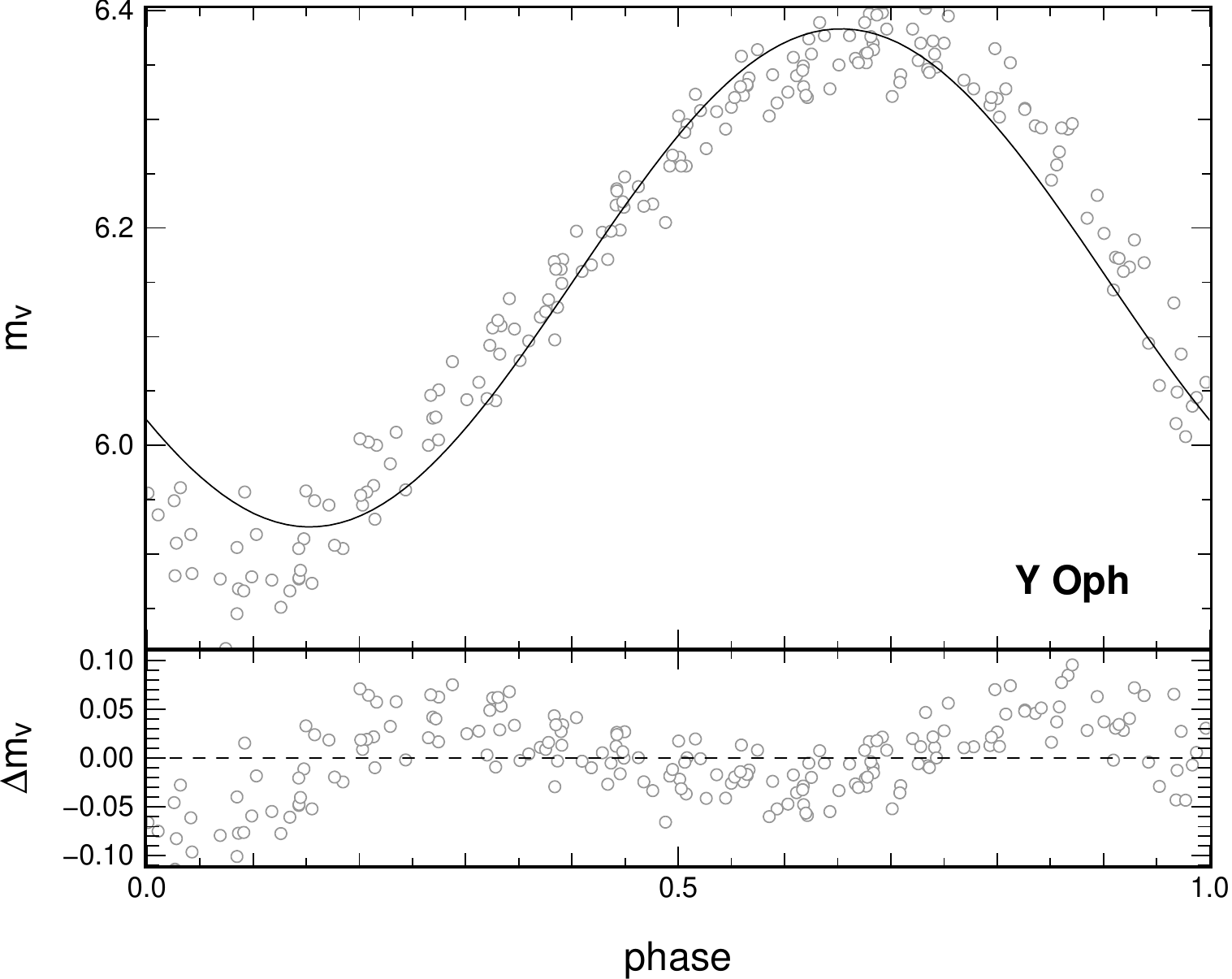}\hspace{.05cm}
\includegraphics{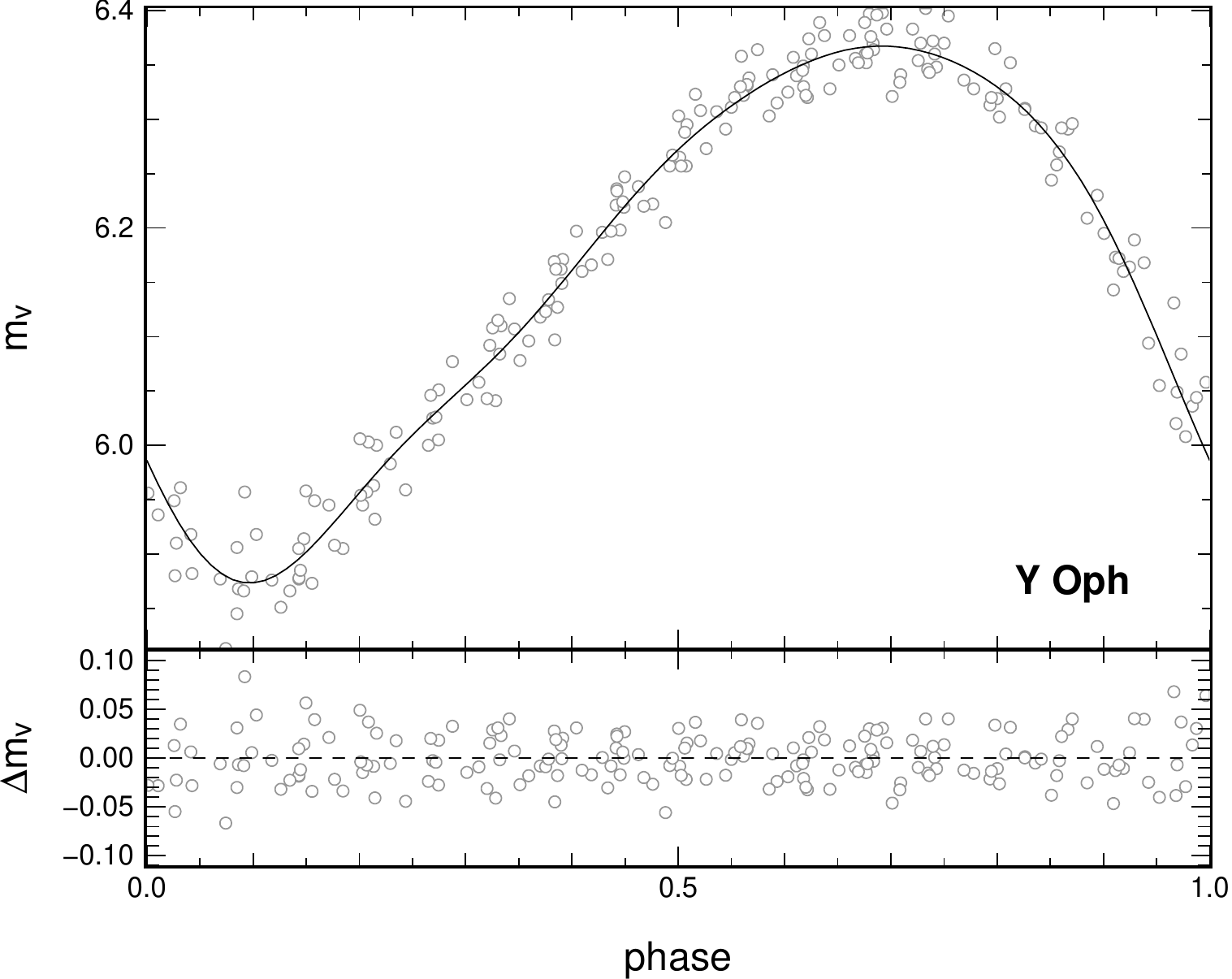}\hspace{.05cm}
\includegraphics{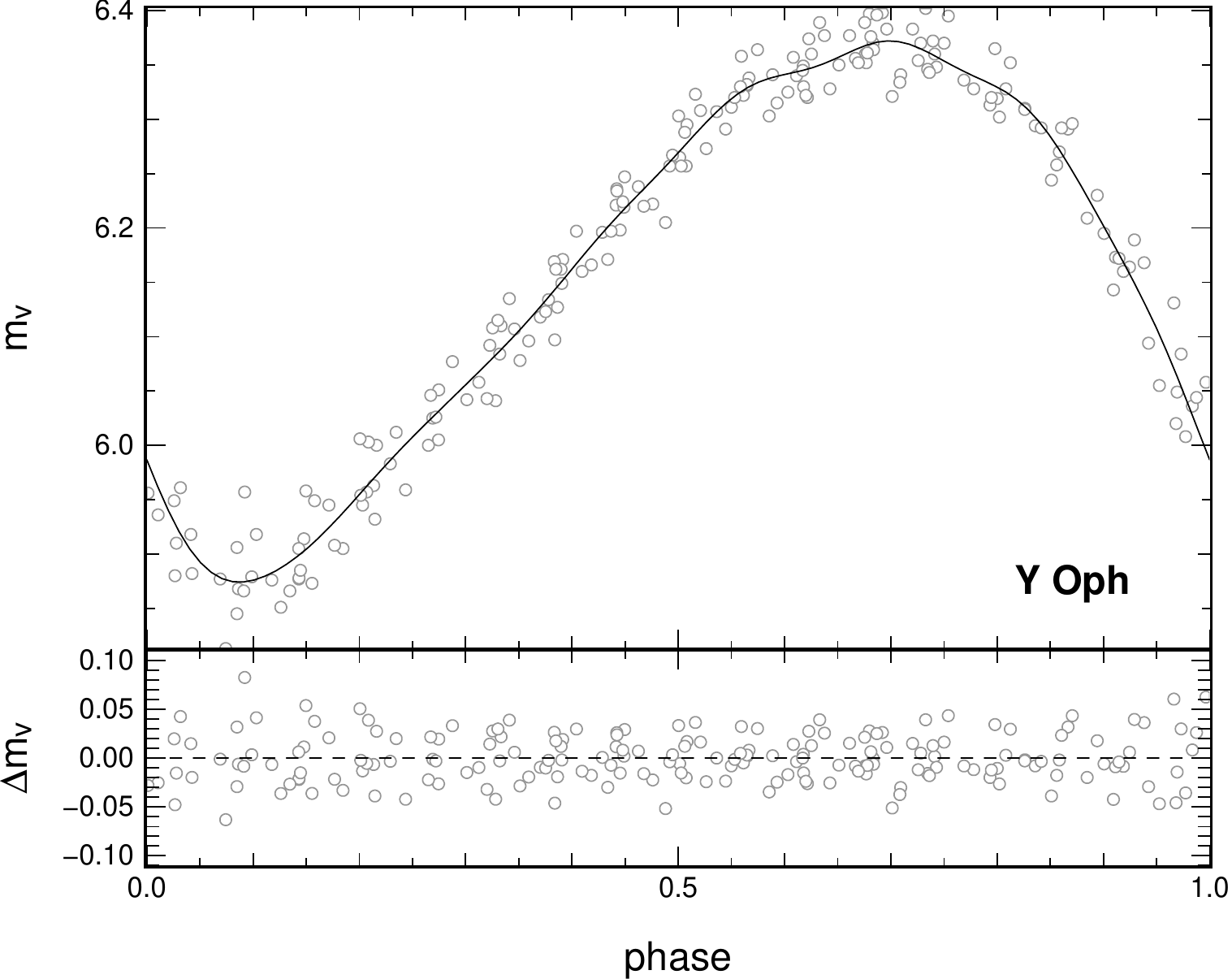}}
\caption[Exemples d'ajustement des courbes de lumière]{\textbf{Exemples d'ajustement des courbes de lumière} : les étoiles prises comme exemple sont FF~Aql et Y~Oph. Les valeurs de l'ordre de l'ajustement sont respectivement de gauche à droite $n = 1, n = 4$ et $n = 8$. Pour FF~Aql, la dispersion moyenne des données par rapport à la courbe ajustée est $\overline{\Delta m_V} = 0.02\,\mathrm{mag}, \overline{\Delta m_V} = 0.009\,\mathrm{mag}$ et $\overline{\Delta m_V} = 0.007\,\mathrm{mag}$, respectivement pour $n = 1, n = 4$ et $n = 8$. Pour Y~Oph, on a $\overline{\Delta m_V} = 0.04\,\mathrm{mag}, \overline{\Delta m_V} = 0.025\,\mathrm{mag}$ et $\overline{\Delta m_V} = 0.025\,\mathrm{mag}$, respectivement.}
\label{image__courbe_lumiere}
\end{figure}

Comme ces étoiles sont pulsantes, il est légitime de penser que la SED varie également avec la phase de pulsation. Il faut donc considérer des mesures photométriques effectuées à une même phase de pulsation. Pour évaluer les magnitudes correspondantes à la phase de nos observations, j'ai utilisé les courbes de lumière disponibles dans la littérature. Pour chaque base de données, j'ai recalculé la phase en utilisant les éphémérides de la Table~\ref{table__cepheide_parametre} puis tracé la magnitude en fonction de la phase. J'ai ensuite appliqué la technique de décomposition de Fourier qui consiste à interpoler la courbe de lumière par une somme de fonctions périodiques \citep[voir par exemple][]{Ngeow-2003-04} :
\begin{displaymath}
m(\phi) = a_0 + \sum_{i=1}^n a_i\,\cos(2\pi i\phi + b_i) 
\end{displaymath}
où $a_i$ et $b_i$ sont respectivement l'amplitude et la phase de Fourier à l'ordre $i$ qu'il faut ajuster et $\phi$ représente la phase de pulsation. Le paramètre $n$ dépend du nombre de points de mesures et de l'amplitude de la courbe de lumière (plus l'amplitude est grande, plus l'ordre à ajuster est grand). Comme le montre la Fig.~\ref{image__courbe_lumiere}, un $n$ trop petit (par exemple $n = 1$) n'ajuste pas correctement les données photométriques. Un $n$ trop grand (par exemple $n = 8$) produit des oscillations supplémentaires non réalistes. J'ai finalement choisi $n = 4$ pour toutes les étoiles, car cette valeur est un bon compromis entre une dispersion faible et pas d'oscillations supplémentaires. J'ai ensuite utilisé les coefficients de l'ajustement pour estimer la magnitude aux phases de pulsation désirées. Cette méthode a été appliquée aux Céphéides FF~Aql (en bande $B, V$ et $R$), AX ~Cir ($B, V$), X Sgr ($B, V, J, H, K$), $\eta$~Aql ($B, V$), W~Sgr ($B, V$), Y~Oph ($B, V$), U~Car ($B, V$) et SV~Vul ($B, V$).

Quand les points de mesures sont espacés de manière très irrégulières, l'ajustement par des séries de Fourier produit également des oscillations non réalistes. Dans ce cas j'ai préféré ajuster les courbes de lumière par une fonction à splines périodiques passant par 4 points flottants \citep[déjà utilisée par][pour interpoler des vitesses radiales]{Merand-2006-}. Plus précisément, ces points de coordonnées $(x_i, y_i)$ sont ajustés de telle sorte que la courbe spline $S_\mathrm{x_i, y_i}(\phi_\mathrm{k})$ minimise :
\begin{displaymath}
\chi^2 = \sum_{k = 1}^{k = N} \frac{[m(\phi_\mathrm{k}) - S_\mathrm{x_i, y_i}(\phi_\mathrm{k})]}{\sigma_k^2}
\end{displaymath}
où $m(\phi_\mathrm{k})$ sont les points de mesures avec leurs incertitudes $\sigma_k$.

\begin{figure}[!p]
\centering
\resizebox{\hsize}{!}{\includegraphics{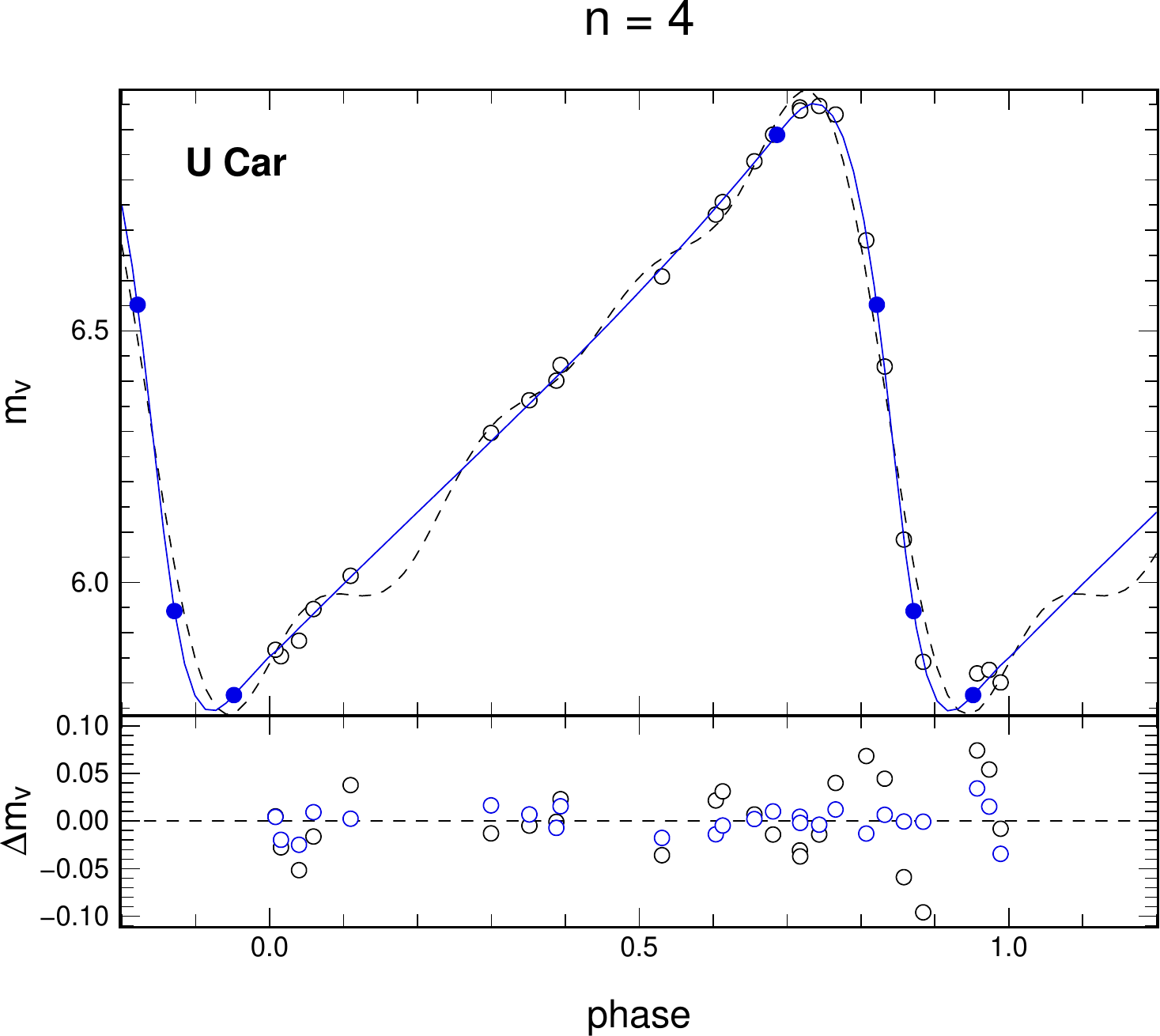}\hspace{.3cm}
\includegraphics{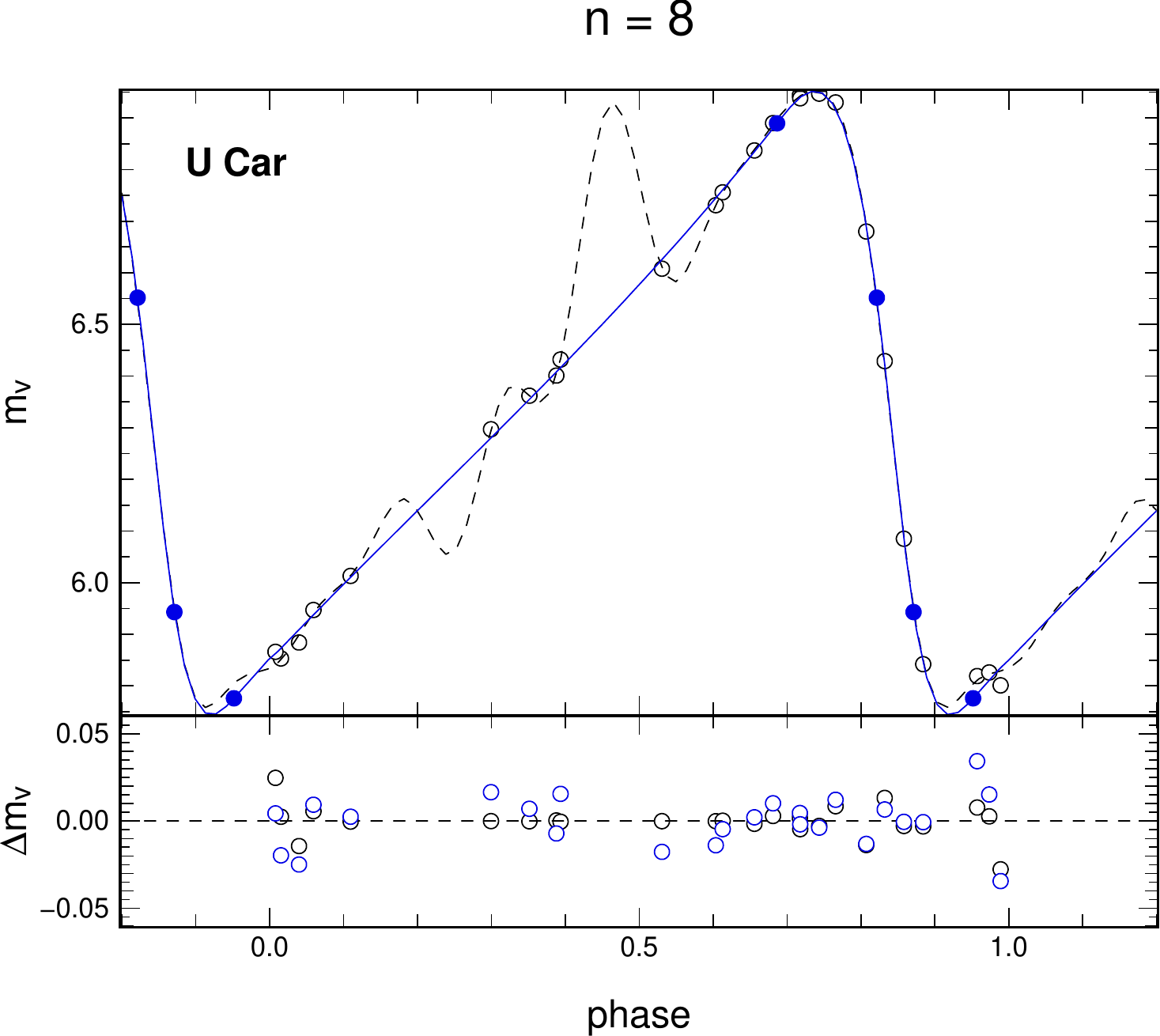}}
\caption[Ajustement de courbe de lumière par des splines périodiques]{\textbf{Ajustement de courbe de lumière par des splines périodiques} : les points en noir sont les mesures et les bleus les points flottants. La courbe en pointillé représente l'ajustement par des séries de Fourier et celle en bleu l'ajustement par des splines périodiques.}
\label{image__courbe_lumiere_spline}
\end{figure}

Un exemple est illustré sur la Fig.~\ref{image__courbe_lumiere_spline} pour la Céphéide U~Car. On voit que pour $n = 4$ ou $n = 8$ (et même $\forall\,n$), l'ajustement de Fourier crée des oscillations non réelles alors que les splines fournissent un ajustement bien meilleur.

L'erreur sur la magnitude est estimée comme étant l'écart quadratique total des valeurs résiduelles (fenêtre du bas de chaque courbe des Fig.~\ref{image__courbe_lumiere} et \ref{image__courbe_lumiere_spline}). L'erreur est considérée la même pour toutes les phases. Ce type d'interpolation a été utilisé pour les Céphéides $\eta$~Aql (en bande $J, H, K$), Y~Oph ($J, H, K$), U~Car ($J, H, K$), SV~Vul ($J, H, K$) et $\kappa$~Pav ($V, J, H, K$).

\begin{figure}[!p]
\centering
\includegraphics[width=.7\linewidth]{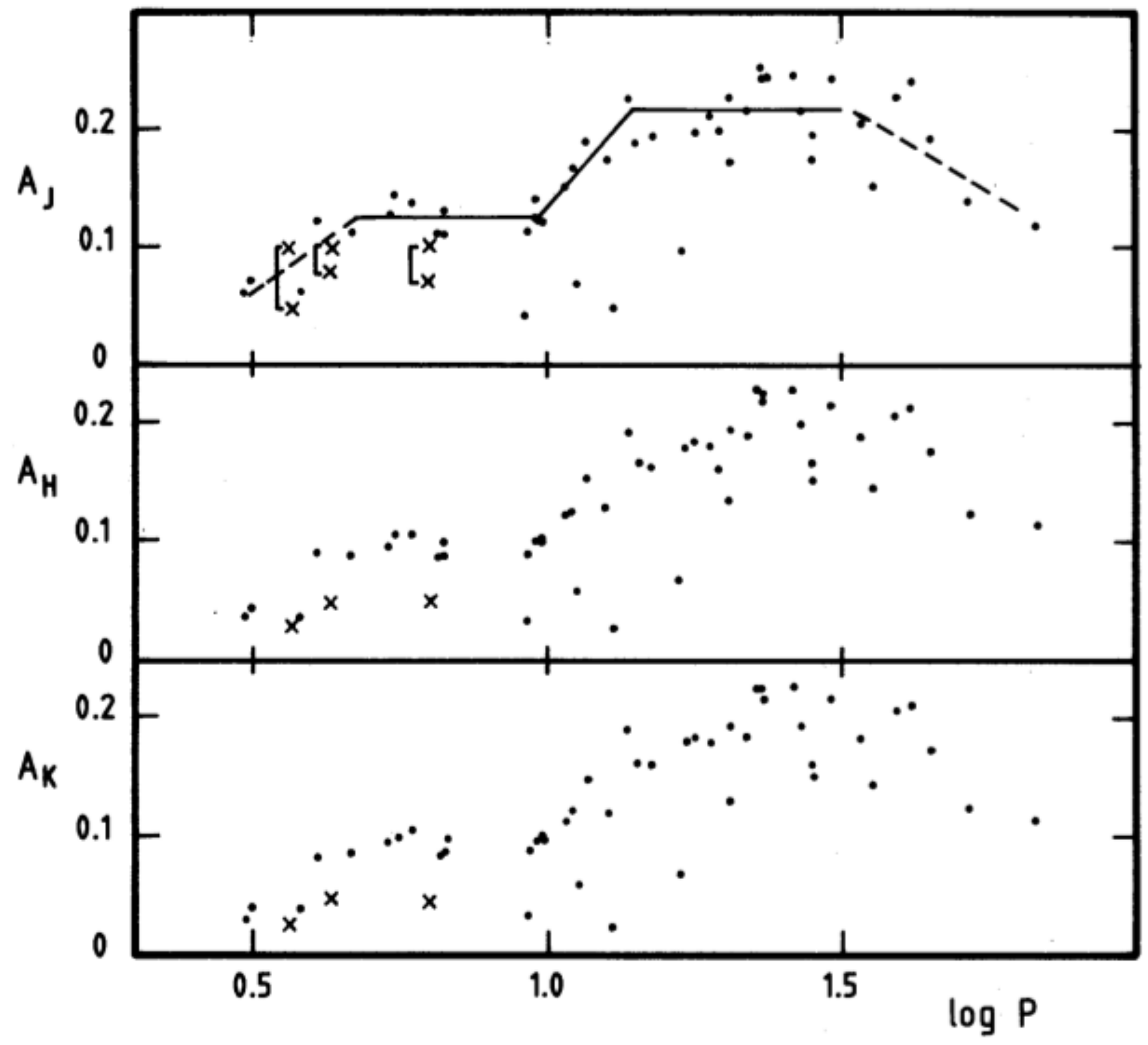}
\caption[Amplitude des courbes de lumière en bande $J, H$ et $K$]{\textbf{Amplitude des courbes de lumière en bande $J, H$ et $K$} : figure tirée de \citet{Laney-1993-01} pour un échantillon de 51 Céphéides Galactiques.}
\label{image__laney}
\end{figure}

L'utilisation des courbes de lumière permet de diminuer l'erreur sur l'estimation de la magnitude à une phase donnée. Cette erreur peut être d'autant plus importante que l'amplitude des variations est grande. Malheureusement, certaines étoiles ne disposent pas de courbes de lumière, ou alors seulement à certaines longueurs d'onde. Parfois il n'est donc pas possible d'estimer la magnitude à une phase donnée avec une bonne précision. Dans un catalogue, seul un ou quelques points de mesures sont disponibles, à une phase pouvant être différente de celle de nos observations \emph{VISIR}. Prenons par exemple l'étoile FF~Aql (Fig.~\ref{image__courbe_lumiere}) dont un catalogue donne $m_V = 5.24\,\pm\,0.03\,\mathrm{mag}$ à la phase $\phi = 0.1$. Sans courbe de lumière, si on utilise cette valeur à la phase $\phi = 0.7$, on ferra une erreur d'environ $0.2\,\mathrm{mag}$ (car d'après la Fig.~\ref{image__courbe_lumiere}, à $\phi = 0.7$ on devrait avoir $m_V\sim5.46\,\mathrm{mag}$). Cette erreur n'est bien sûr pas négligeable. J'ai choisi de raisonner sur l'amplitude des courbes de lumière, $A_\lambda$, pour estimer une incertitude supplémentaire, car elle correspond à l'erreur maximale que l'on peut faire (erreur due à la discordance de phase). Pour cela je me suis appuyé sur les travaux de \citet{Laney-1993-01} qui, pour un échantillon de 51 Céphéides Galactiques, ont tracé l'amplitude des courbes de lumières en bandes $J, H$ et $K$ en fonction de la période de pulsation. Leur figure est reproduite sur la Fig.~\ref{image__laney}. On remarque principalement deux régions :

\begin{itemize}
	\compactlist
	\item $0.5<\log P\leqslant1.0$ où $A_\mathrm{J}\sim0.1\,\mathrm{mag}$
	\item $\log P>1.0$ où $A_\mathrm{J}\sim0.2\,\mathrm{mag}$
\end{itemize}
et ces valeurs diminuent avec la longueur d'onde, avec $A_\mathrm{K}\sim0.08\,\mathrm{mag}$ pour $0.5<\log P\leqslant1.0$ et $A_\mathrm{K}\sim0.15\,\mathrm{mag}$ pour $\log P>1.0$. Pour une étoile donnée, quand il n'existe pas de courbe de lumière et que la magnitude choisie dans un catalogue ne correspond pas à la phase de nos observations, j'ai donc choisi comme incertitudes supplémentaires :

\begin{itemize}
	\compactlist
	\item $\lambda\leqslant3.5\,\mu\mathrm{m}$ : $\sigma_\lambda = 0.1\,\mathrm{mag}$ pour $0.5<\log P\leqslant1.0$ et $\sigma_\lambda = 0.2\,\mathrm{mag}$ pour $\log P>1.0$
	\item $\lambda>3.5\,\mu\mathrm{m}$ : $\sigma_\lambda = 0.05\,\mathrm{mag}$, quelque soit $P$ 
\end{itemize}

Ces incertitudes ont ensuite été sommées quadratiquement aux incertitudes photométriques indiquées dans les catalogues afin d'obtenir des incertitudes totales. Le fait d'utiliser l'amplitude de la courbe de lumière comme erreur maximale sur l'estimation de la magnitude peut surestimer l'incertitude finale, mais cela permettra de distinguer complètement un excès IR d'une erreur de magnitude. D'un autre côté, cela peut également empêcher la détection d'un excès IR.

Certaines de nos Céphéides ont été observées les deux nuits et peuvent par conséquent avoir des phases différentes (spécialement les courtes périodes). J'ai donc considéré pour la SED une phase intermédiaire.

\paragraph*{\textcolor{black}{Modèles de spectres stellaires}}

L'émission photosphérique a été modélisée par les modèles d'atmosphères stellaires obtenus par le code ATLAS9 de \citet{Castelli-2003-}. J'ai choisi une grille de modèles qui a été calculée pour une métallicité de type solaire et une vitesse turbulente de 2\,km.s$^{-1}$. Cette grille comprend des spectres pour 75 valeurs de températures effectives dans l'intervalle $3500<T_\mathrm{eff}<50000$\,K et 11 valeurs de gravité de surface dans l'intervalle $0.0<\log g<5.0$. Elle est ensuite interpolée afin de calculer un spectre quelque soit $T_\mathrm{eff}$ et $\log g$. Ce spectre est par la suite multiplié par l'angle solide que forme la photosphère de l'étoile, $\pi\theta_\mathrm{LD}^2/4$, où $\theta_\mathrm{LD}$ représente le diamètre angulaire du disque assombri. Les modèles sont finalement ajustés aux données photométriques en tenant compte de la réponse spectrale de chaque instrument.
On suppose qu'il n'y a pas d'excès détectable en dessous de $2.2\,\mu\mathrm{m}$ et toutes les longueurs d'onde $\leqslant2.2\,\mu\mathrm{m}$ sont utilisées pour ajuster la température effective et le diamètre angulaire. La gravité de surface n'a pas été ajustée car la photométrie bande large est presque insensible à ce paramètre, sa valeur est donc fixe et choisie dans la littérature, les seuls paramètres à ajuster sont $T_\mathrm{eff}$ et $\theta_\mathrm{LD}$. Notons toutefois que ces variables sont corrélées dans cet ajustement.

\paragraph*{\textcolor{black}{Correction de l'extinction interstellaire}}

J'ai corrigé toutes les mesures photométriques inférieures à $3\,\mu\mathrm{m}$ de l'extinction interstellaire $A_\lambda$ (Equation~\ref{equation__extinction_interstellaire}). J'ai utilisé $R_\lambda$ d'après \citet{Fouque-2003-} et \citet{Hindsley-1989-06} :
\begin{eqnarray*}
&R_\mathrm{B} = &R_\mathrm{V} + 1 \\
&R_\mathrm{V} = &3.07 + 0.28\,(B - V) + 0.04\,E(B - V) \\
&R_\mathrm{R} = &R_\mathrm{V} - 0.97 \\
&R_\mathrm{I} = &1.82 + 0.205\,(B - V) + 0.0225\,E(B - V) \\
&R_\mathrm{J} = &R_\mathrm{V}/4.02 \\
&R_\mathrm{H} = &R_\mathrm{V}/6.82 \\
&R_\mathrm{K} = &R_\mathrm{V}/11 
\end{eqnarray*}
où $B, V, R, I, J, H$ et $K$ représentent les bandes photométriques (Table~\ref{table__filtres}).

L'excès de couleur $E(B - V)$ est tiré de \citet{Fouque-2007-12} pour toutes les Céphéides classiques, excepté AX~Cir qui provient de \citet{Tammann-2003-06}. Pour R~Sct et AC~Her, $E(B - V)$ provient de \citet{Taranova-2010-02} et de \citet{Feast-2008-06-2} pour $\kappa$~Pav. Les flux dont $\lambda>3\,\mu\mathrm{m}$ n'ont pas été corrigés de l'extinction car nous supposons qu'elle est négligeable à ces longueurs d'onde.

\paragraph*{\textcolor{black}{Système photométrique}}

Les mesures photométriques ($B, V, R, I, J, H$ ou $K$) effectuées dans une même bande peuvent être légèrement différentes en fonction des filtres employés pour les observations. Il est nécessaire de corriger de la réponse spectrale de chaque filtre utilisé. Il existe des équations de transformation entre les systèmes les plus utilisés (Johnson, Carter, Glass, ...). Lorsque l'on récolte des données photométriques, il faut veiller à ce que chaque mesure se trouve dans un même système photométrique. Dans notre cas, toutes les données récoltées dans la littérature ont été converties dans le système Johnson, en utilisant les équations de transformation de \citet{Glass-1985-03} et \citet{Leggett-1992-09}.

\subsubsection{FF~Aql}

\defcitealias{Luck-2008-07}{L08}
\defcitealias{Berdnikov-2008-04}{B08}
\defcitealias{Moffett-1984-07}{M84}
\defcitealias{Marengo-2010-01}{M10}
\defcitealias{Ossenkopf-1994-11}{O94}

\begin{figure}[!p]
\centering\includegraphics[width=.7\linewidth]{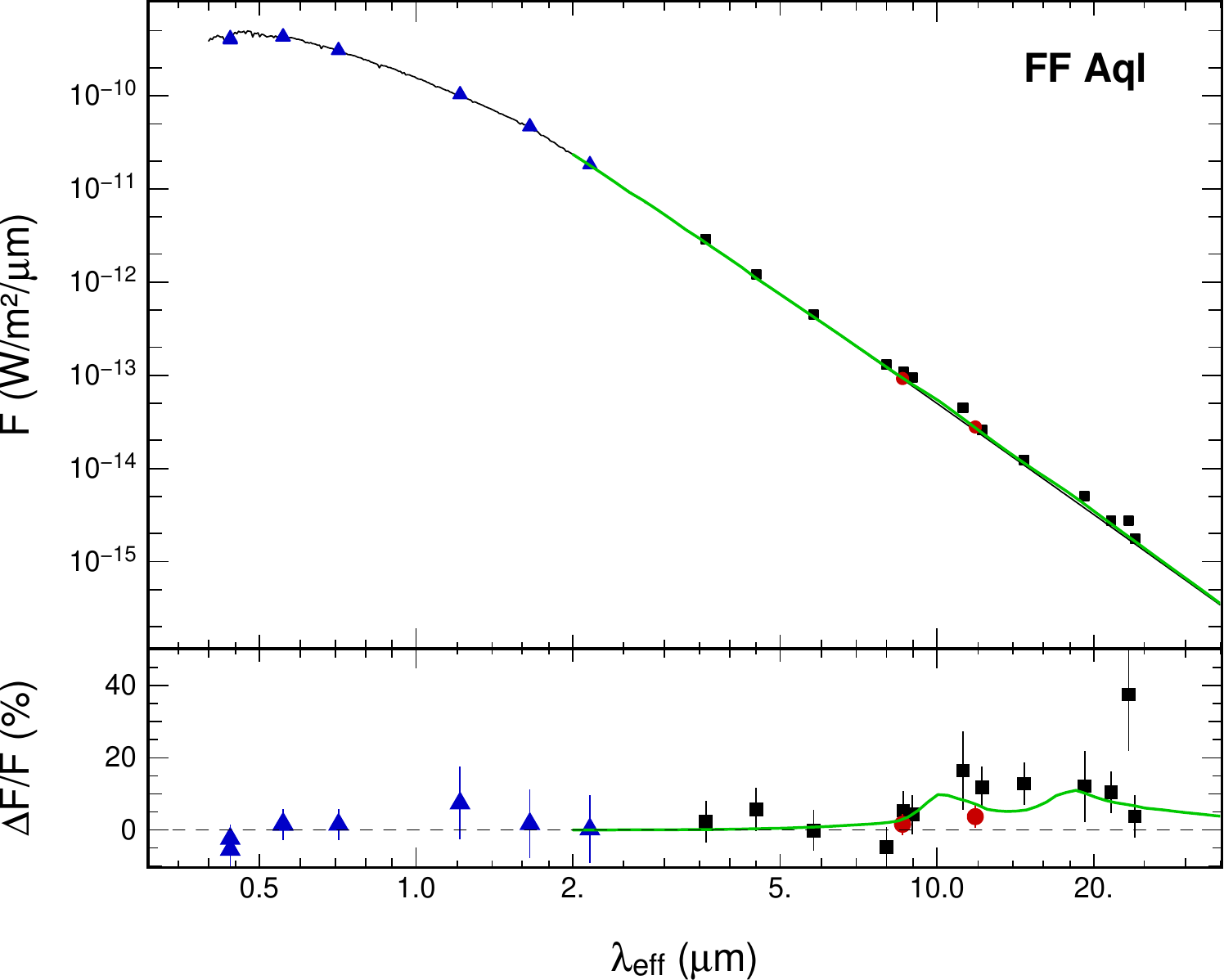}
\caption[Distribution spectrale d'énergie de FF~Aql]{\textbf{Distribution spectrale d'énergie de FF~Aql} : la ligne continue noire représente le spectre synthétique superposé aux points de mesures photométriques. Nos données \emph{VISIR} sont tracées en rouge. Les triangles bleus indiquent les points utilisés pour l'ajustement de la SED. La courbe en vert correspond au modèle de densité de flux. La fenêtre du bas représente la différence de flux relative (relative au modèle photosphérique), intégrée sur la bande passante de chaque filtre.}
\label{image__SED_FF_AQL}
\end{figure}

J'ai sélectionné comme valeur fixe $\log g = 2.05$ d'après \citet[][\ciap \citetalias{Luck-2008-07}]{Luck-2008-07}. La distribution spectrale d'énergie est présentée sur la Fig.~\ref{image__SED_FF_AQL}. Les mesures $B, V$ et $R$ proviennent des courbes de lumière de \citet[][\ciap  \citetalias{Berdnikov-2008-04}]{Berdnikov-2008-04} et \citet[][\ciap \citetalias{Moffett-1984-07}]{Moffett-1984-07}. La photométrie $J, H$ et $Ks$ provient de \citet{Welch-1984-04} et correspond à des valeurs moyennes (et l'écart-type pris comme incertitude). J'ai également rajouté des données des instruments \emph{IRAC} ($3.6, 4.5, 5.8, 8.0\,\mu\mathrm{m}$) et \emph{MIPS} ($24\,\mu\mathrm{m}$), installés sur le télescope spatial \emph{Spitzer} \citep[][\ciap \citetalias{Marengo-2010-01}]{Marengo-2010-01}. Pour finir, j'ai complété avec des données photométriques des télescopes \emph{IRAS} \citep[$12$ et $25\,\mu\mathrm{m}$,][]{Helou-1988-}, \emph{AKARI} \citep[\emph{IRC}: $9$ et $18\,\mu\mathrm{m}$,][]{Ishihara-2010-05} et \emph{MSX} \citep[$8.28, 12.13, 14.65$ et $21.34\,\mu\mathrm{m}$,][]{Egan-1996-12,Egan-2003-}.

L'ajustement de modèle de spectre stellaire est représenté par la courbe en noir sur la Fig.~\ref{image__SED_FF_AQL}. Le résultat des valeurs ajustées est exposé dans la Table~\ref{table__parametre_ajuste}. La température estimée est seulement $3\,\%$ plus petite que celle donnée par \citetalias{Luck-2008-07} ($6062 \pm 43$\,K) à cette phase de pulsation. Le diamètre angulaire est également en excellent accord avec la valeur $0.86 \pm 0.17$\,mas de \citet{Groenewegen-2007-11}.

À partir de la photométrie \emph{VISIR} (Table~\ref{table__irradiance_mesure_I}), je détecte une émission infrarouge de l'ordre de $2\,\%$ dans les filtres PAH1 et SIC. Il semble également qu'un excès plus important soit présent à des longueurs d'onde plus grandes. La fenêtre du bas de la Fig.~\ref{image__SED_FF_AQL} montre bien cette tendance. Cela indique probablement la présence d'une enveloppe autour de cette étoile.

J'ai donc ajusté ensuite une seconde composante suivant un modèle de densité de flux. Ce modèle absorbe uniquement une fraction du flux incident, c'est à dire que je considère le milieu comme étant optiquement mince. Il a comme expression mathématique :
\begin{equation}
\label{equation__corps_gris}
F_\mathrm{d}(\lambda) = \kappa_\lambda\,\beta\,\ B_\lambda(T_\mathrm{d})
\end{equation}
où $B_\lambda(T_\mathrm{d})$ est la fonction de Planck à la température $T_\mathrm{d}$, $\kappa_\lambda$ est l'opacité des poussières et $\beta$ est un paramètre directement proportionnel à la masse de poussières contenue dans l'enveloppe \citep{Li-2005-04} :
\begin{displaymath}
\beta = 2.1\times 10^{-3} \frac{M_\mathrm{d}}{D^2}
\end{displaymath}
avec $D$ la distance de l'étoile en pc et $M_\mathrm{d}$ en masse solaire.

On peut remarquer sur la SED certains pics d'émission autour de $10\,\mu\mathrm{m}$ et $20\,\mu\mathrm{m}$ qui pourraient être caractéristiques de grains de silicate et/ou de carbone. J'ai donc choisi d'utiliser la table l'opacité de \citet[][,sans coagulation des grains, \ciap \citetalias{Ossenkopf-1994-11}]{Ossenkopf-1994-11}, calculée pour ce type de grains avec une distribution en taille de type MRN \citep[][c'est à dire $n \propto a^{-3.5}$ où $a$ est la taille du grain]{Mathis-1977-10}.


La courbe verte de la Fig.~\ref{image__SED_FF_AQL} représente l'ajustement de l'équation~\ref{equation__corps_gris}. Je trouve (Table~\ref{table__parametre_ajuste}) une température de l'enveloppe $T_\mathrm{env} = 539 \pm 52$\,K et un paramètre $\beta = 3.0 \pm 0.7 \times 10^{-19}\,\mathrm{kg\,m^{-2}}$. En utilisant la mesure de distance de la Table~\ref{table__cepheide_parametre}, j'ai pu estimer $M_\mathrm{d} = 1.8 \pm 0.4 \times 10^{-11}\,M_\odot$. La masse totale de l'enveloppe peut être estimée en supposant un rapport gaz/poussière de l'ordre de 100 (valeur typique pour des poussières circumstellaires), soit une masse totale $M_\mathrm{env} = 1.8 \pm 0.4 \times 10^{-9}\,M_\odot$. Nous noterons que certaines bandes photométriques peuvent être surestimées à cause de l'émission du cirrus interstellaire présent autour de cette Céphéide \citep{Barmby-2010-11}.

\subsubsection{AX~Cir}

\defcitealias{Moskalik-2005-06}{M05}

\begin{figure}[!p]
\centering\includegraphics[width=.7\linewidth]{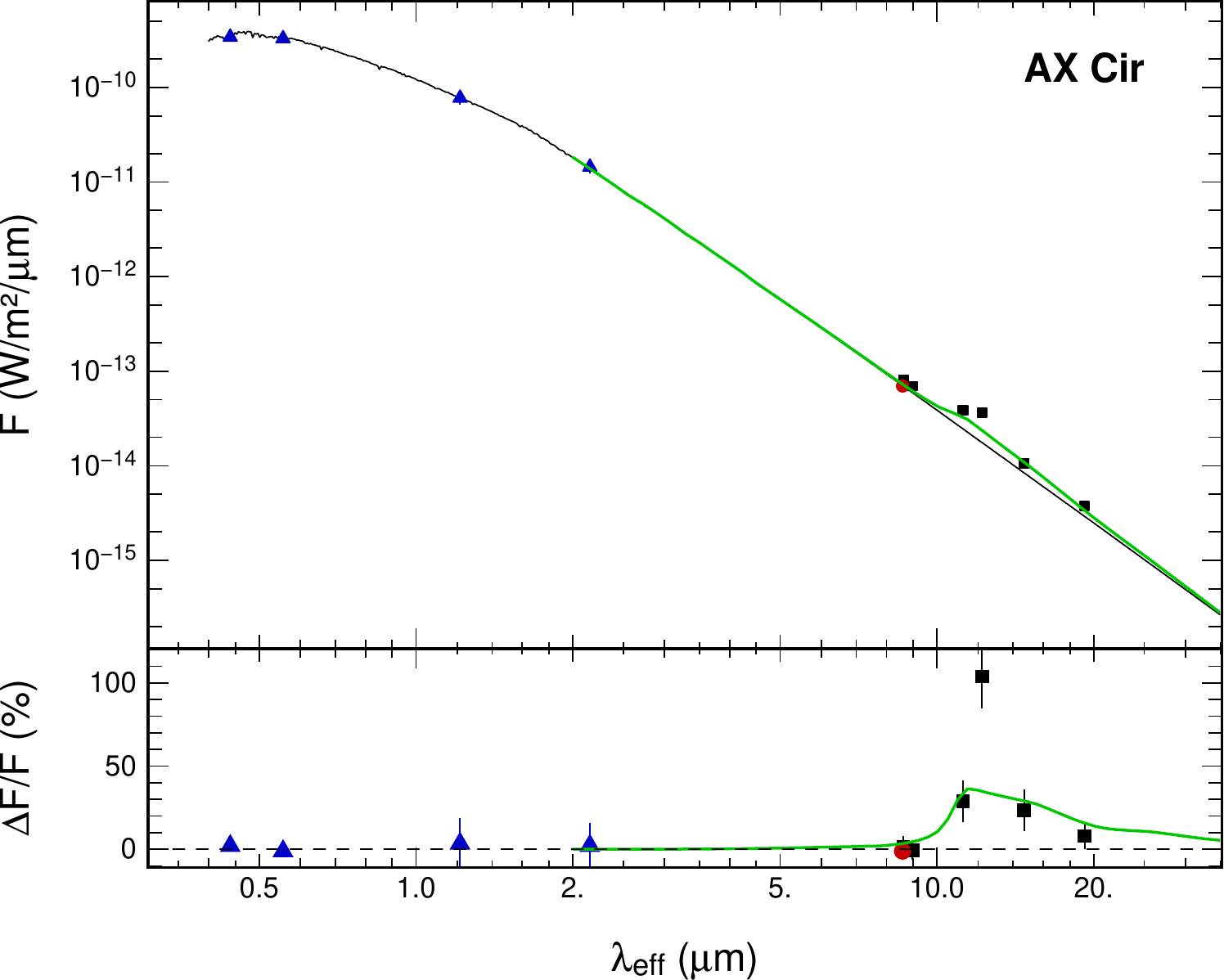}
\caption[Distribution spectrale d'énergie de AX~Cir]{\textbf{Distribution spectrale d'énergie de AX~Cir} : la légende est identique à la Fig.~\ref{image__SED_FF_AQL}.}
\label{image__SED_AX_CIR}
\end{figure}

Cette Céphéide a presque le même type spectral que FF~Aql, j'ai donc choisi de fixer $\log{g} = 2.0$, car il n'y a pas d'estimation de ce paramètre dans la littérature (notons qu'un changement de $\pm0.5$ modifie les valeurs ajustées de seulement $0.7\,\%$). La SED de AX~Cir est exposée sur la Fig.~\ref{image__SED_AX_CIR}. La photométrie $B$ et $V$ provient des courbes de lumière de \citetalias{Berdnikov-2008-04}, les magnitudes en bande $J$ et $Ks$ de l'instrument \emph{DENIS} et les données au-delà de $9\,\mu\mathrm{m}$ proviennent de \emph{MSX} ($8.28, 12.13$ et $14.65\,\mu\mathrm{m}$), \emph{IRAS} ($12\,\mu\mathrm{m}$) et \emph{IRC} ($9$ et $18\,\mu\mathrm{m}$).

Les résultats de l'ajustement sont présentés dans la Table~\ref{table__parametre_ajuste}. Le diamètre angulaire est en accord avec la valeur moyenne prédite par \citet[][\ciap \citetalias{Moskalik-2005-06}]{Moskalik-2005-06} (à un niveau de $7\,\%$, cohérent avec l'amplitude des variations du diamètre angulaire). Il n'existe pas d'estimation de la température dans la littérature permettant une comparaison.

Je ne détecte pas à la longueur d'onde $8.6\,\mu\mathrm{m}$ d'excès IR (voir Table~\ref{table__irradiance_mesure_I}) et ce résultat semble cohérent avec la mesure à $9\,\mu\mathrm{m}$ de \emph{IRC}. Aux autres longueurs d'onde, il pourrait y avoir une émission de l'ordre de $8.5\pm 7.3\,\%$ à $18\,\mu\mathrm{m}$. On peut remarquer que cet excès pique autour de $13\,\mu\mathrm{m}$ et cela pourrait être lié à l'émission de poussières telles que l'oxyde d'aluminium. Je n'ai pas pas trouvé de tables d'opacité dans la littérature pour ce type de grains, j'en ai donc calculé une en faisant quelques hypothèses. J'ai calculé $\kappa_\lambda$ à partir des constantes optiques $m = n + ik$ (calculées en laboratoire) pour de l'aluminium amorphe compact, et en utilisant un modèle simple de sphères homogènes dans la limite de Rayleigh des petites particules. D'après \citet{Bohren-1983-} le coefficient d'absorption dans cette limite est donné par :
\begin{equation}
\label{equation__opacite0}
C_\mathrm{abs} = \frac{6\pi}{\lambda}\,V\,\mathop{\mathrm{Im}}\left( \frac{\varepsilon - 1}{\varepsilon + 2 } \right)
\end{equation}
où $V = 4/3\pi\,a^3$ avec $a$ le rayon de la sphère et $\varepsilon = \varepsilon\arcmin + i\varepsilon\arcsec$. Les constantes optiques sont reliées à $\varepsilon$ par les relations \citep{Bohren-1983-} :
\begin{displaymath}
\varepsilon\arcmin = n^2 - k^2 \qquad \varepsilon\arcsec = 2nk
\end{displaymath}

Puis j'ai calculé l'opacité pour une distribution en taille des particules de type MRN (taille de 5\,nm à 250\,nm comme \citetalias{Ossenkopf-1994-11}) et une densité $\rho = 2.5\,g\,cm^{-2}$ :
\begin{equation}
\label{equation__opacite}
\kappa_\lambda = \frac{\int C_\mathrm{abs}\,a^{-3.5}\,da}{\rho \int V\,a^{-3.5}\,da} 
\end{equation}

Cette opacité a ensuite été utilisée dans l'équation~\ref{equation__corps_gris} pour ajuster l'excès IR en supposant que l'enveloppe est principalement composée d'oxyde d'aluminium. J'ai obtenu $\beta = 1.6 \pm 0.4 \times 10^{-19}\,\mathrm{kg\,m^{-2}}$. La température et la masse totale de l'enveloppe sont présentées dans la Table~\ref{table__parametre_ajuste}. Cependant, des mesures photométriques supplémentaires sont nécessaires pour mieux contraindre ces paramètres.

\subsubsection{X~Sgr}

\defcitealias{Kervella-2004-03}{K04}
\defcitealias{Feast-2008-06-2}{F08}

\begin{figure}[!p]
\centering\includegraphics[width=.7\linewidth]{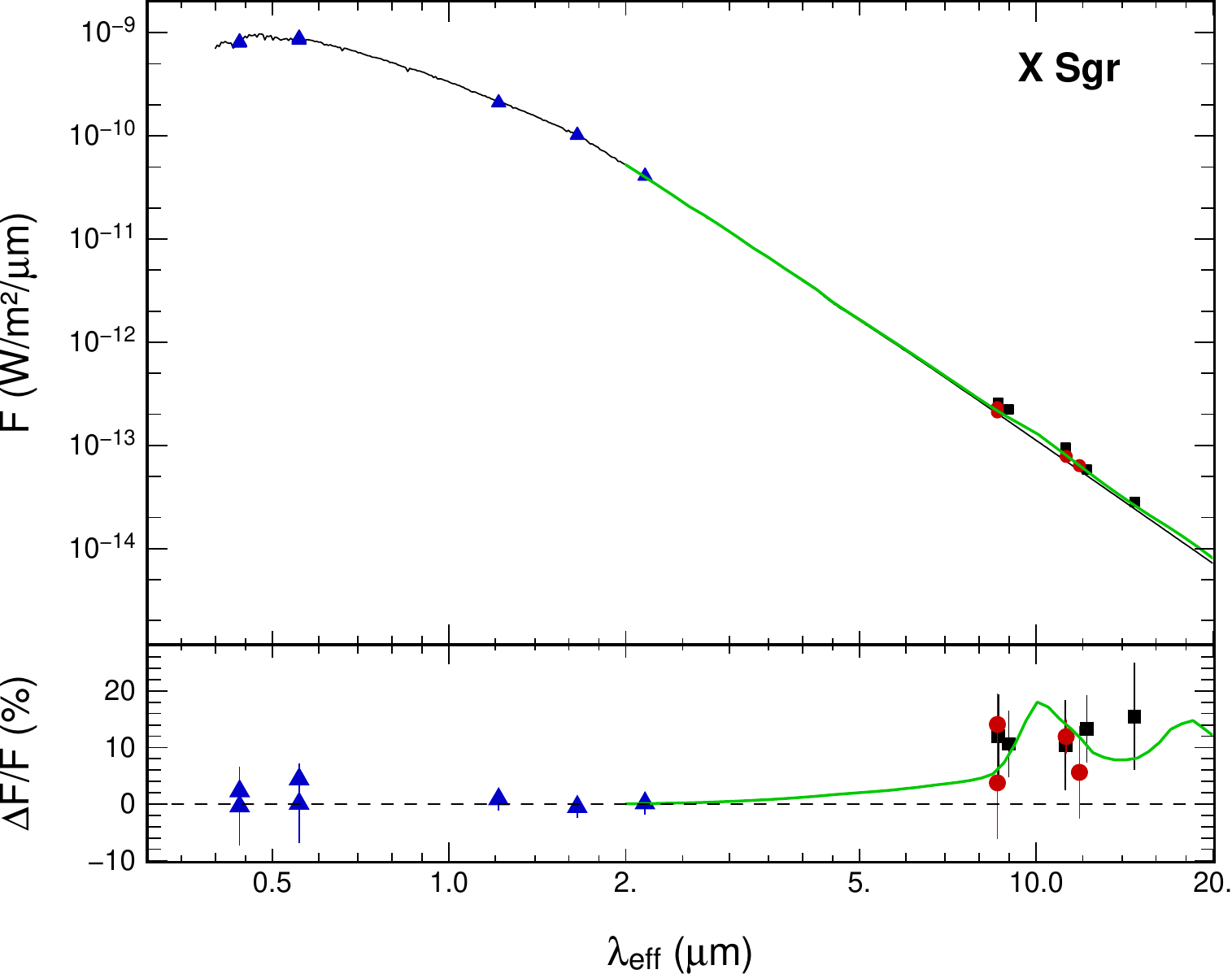}
\caption[Distribution spectrale d'énergie de X~Sgr]{\textbf{Distribution spectrale d'énergie de X~Sgr} : la légende est identique à la Fig.~\ref{image__SED_FF_AQL}.}
\label{image__SED_X_SGR}
\end{figure}

Pour cette étoile, j'ai collecté les données $B$ et $V$ des courbes de lumière de \citet[][\ciap \citetalias{Kervella-2004-03}, données déjà corrigées de l'extinction interstellaire]{Kervella-2004-03} et de \citetalias{Moffett-1984-07}, pour une phase intermédiaire $\phi = 0.72$. Les données $J, H$ et $K$ proviennent de \citet[][\ciap \citetalias{Feast-2008-06-2}]{Feast-2008-06-2}. Enfin, j'ai recueilli les mesures photométriques de \emph{IRC} ($9\,\mu\mathrm{m}$), \emph{MSX} ($8.28, 12.13$ et $14.65\,\mu\mathrm{m}$) et \emph{IRAS} ($12\,\mu\mathrm{m}$). La SED est tracée sur la Fig.~\ref{image__SED_X_SGR}.

La Table~\ref{table__parametre_ajuste} expose les résultats de l'ajustement, pour une même valeur de gravité de surface que précédemment. Je trouve un diamètre angulaire qui est $13\,\%$ et $3\sigma$ plus petit que le diamètre moyen $\theta_\mathrm{LD} = 1.47 \pm 0.04$\,mas mesuré par \citet{Kervella-2004-03a}. Toutefois, l'amplitude de pulsation de cette étoile est de $\sim9\,\%$ en diamètre \citepalias{Moskalik-2005-06}. A contrario, notre estimation est cohérente avec $\theta_\mathrm{LD} = 1.31 \pm 0.12$\,mas évalué à partir de la parallaxe \citep{Benedict-2007-04} et du diamètre linéaire \citepalias{Feast-2008-06-2} à cette phase de pulsation. Notons également que \citetalias{Kervella-2004-03} n'ont utilisé qu'un modèle de disque assombri pour décrire leurs données et que la présence d'une enveloppe engendre une surestimation du diamètre stellaire. C'est l'explication la plus probable à cette divergence.

Nos données \emph{VISIR} indiquent une émission IR de l'ordre de $5$--$15\,\%$ (voir Table~\ref{table__irradiance_mesure_I}). On remarque certaines pics autour de $10\,\mu\mathrm{m}$ qui pourraient être caractéristiques de silicate et/ou carbone. J'ai donc utilisé la table d'opacité de \citetalias{Ossenkopf-1994-11} pour ajuster le modèle de l'équation~\ref{equation__corps_gris}. L'ajustement a donné $\beta = 5.6 \pm 0.9 \times 10^{-19}\,\mathrm{kg\,m^{-2}}$ ainsi que la température et la masse totale de l'enveloppe présentées dans la Table~\ref{table__parametre_ajuste}. Encore une fois, des données à d'autres longueurs seraient utiles pour mieux contraindre ces paramètres.


\begin{table}[!p]
\centering
\begin{tabular}{cccccccc} 
\hline
\hline
Étoile	  			&	$\phi$	&	$\log g$	&	$\theta_\mathrm{LD}$	& 	$T_\mathrm{eff}$	& $M_\mathrm{env}$	& $T_\mathrm{env}$	&	$\chi^2$\\
		  				&				&					&				(mas)				&	(K)						& 	($M_\odot$)				& 	(K)						&				\\
\hline
FF~Aql				&	0.62		&	2.05	&	$0.86 \pm 0.03$	&	$5890 \pm 235$	&	$ 1.8 \pm 0.5 \times 10^{-9}$	&	$539 \pm 53$		&	0.41	\\
AX~Cir				& 	0.27		&	2.00	&	$0.76 \pm 0.03$	&	$5911 \pm 184$	&	$ 7.4 \pm 5.9 \times 10^{-10}$&	$712 \pm 61$		&	0.42	\\
X~Sgr				& 	0.72		&	2.00	&	$1.30 \pm 0.04$	&	$5738 \pm 314$	&	$ 3.0 \pm 0.6 \times 10^{-9}$	&	$703 \pm 52$		&	0.74	\\
$\eta$~Aql		&	0.47		&	1.80	&	$1.86 \pm 0.12$	&	$5431 \pm 498$	&	$ 1.0 \pm 0.7 \times 10^{-8}$	&	$545 \pm 60$		&	1.01	\\
W~Sgr				&	0.48		&	1.70	&	$1.14 \pm 0.06$	&	$5632 \pm 162$	&	$ 1.7 \pm 0.4 \times 10^{-9}$	&	$853 \pm 55$		&	1.45	\\
Y~Oph				&	0.73		&	1.80	&	$1.24 \pm 0.05$	&	$5870 \pm 387$	&	$ 3.9 \pm 0.8 \times 10^{-9}$	&	$1419 \pm 148$	&	1.49	\\
U~Car				&	0.49		&	1.20	&	$0.90 \pm 0.02$	&	$4823 \pm 52$		&	$ 4.4 \pm 2.1 \times 10^{-9}$	&	$746 \pm 94$		&	0.70	\\
SV~Vul				&	0.04		&	1.40	&	$0.76 \pm 0.01$	&	$5744 \pm 144$	&	$ 3.9 \pm 7.4 \times 10^{-8}$	&	$620 \pm 50$		&	1.13 \\
\hline
R~Sct				&	0.48		&	0.00	&	$1.74 \pm 0.06$	&	$4605 \pm 119$	&	--												&	$1486 \pm 335$	&	1.79	\\
						&				&			&								&								&	--												&	$772 \pm 82$		&			\\
AC~Her				&	0.14		&	0.50	&	$0.30 \pm 0.02$ 	&	$5711 \pm 293$	&	--												&	$286 \pm 32$		&	0.91	\\
$\kappa$~Pav	&	0.90		&	1.20	&	$1.09 \pm 0.05$	&	$6237 \pm 119$	&	$9.9 \pm 2.5 \times 10^{-10}$	&	$695 \pm 36$		&	0.30	\\
\hline
\end{tabular}
\caption[Résultats de l'ajustement de la SED]{\textbf{Résultats de l'ajustement de la SED} : $\phi$ représente la phase de pulsation, $\theta_\mathrm{LD}$ et $T_\mathrm{eff}$ sont les paramètres photosphériques ajustés tandis que  $T_\mathrm{env}$ est la température ajustées de l'enveloppe. $M_\mathrm{env}$ est la masse totale (gaz + poussières) de l'enveloppe. $M_\mathrm{env}$ a été estimée en supposant un rapport gaz/poussière\,$\sim 100$. Pour R~Sct, la première ligne dénote la composante compacte. Le $\chi^2$ a été estimé sur l'ajustement du flux photosphérique. $\log g$ est la gravité effective.}
\label{table__parametre_ajuste}
\end{table}

\subsubsection{$\eta$~Aql}

\begin{figure}[!p]
\centering\includegraphics[width=.7\linewidth]{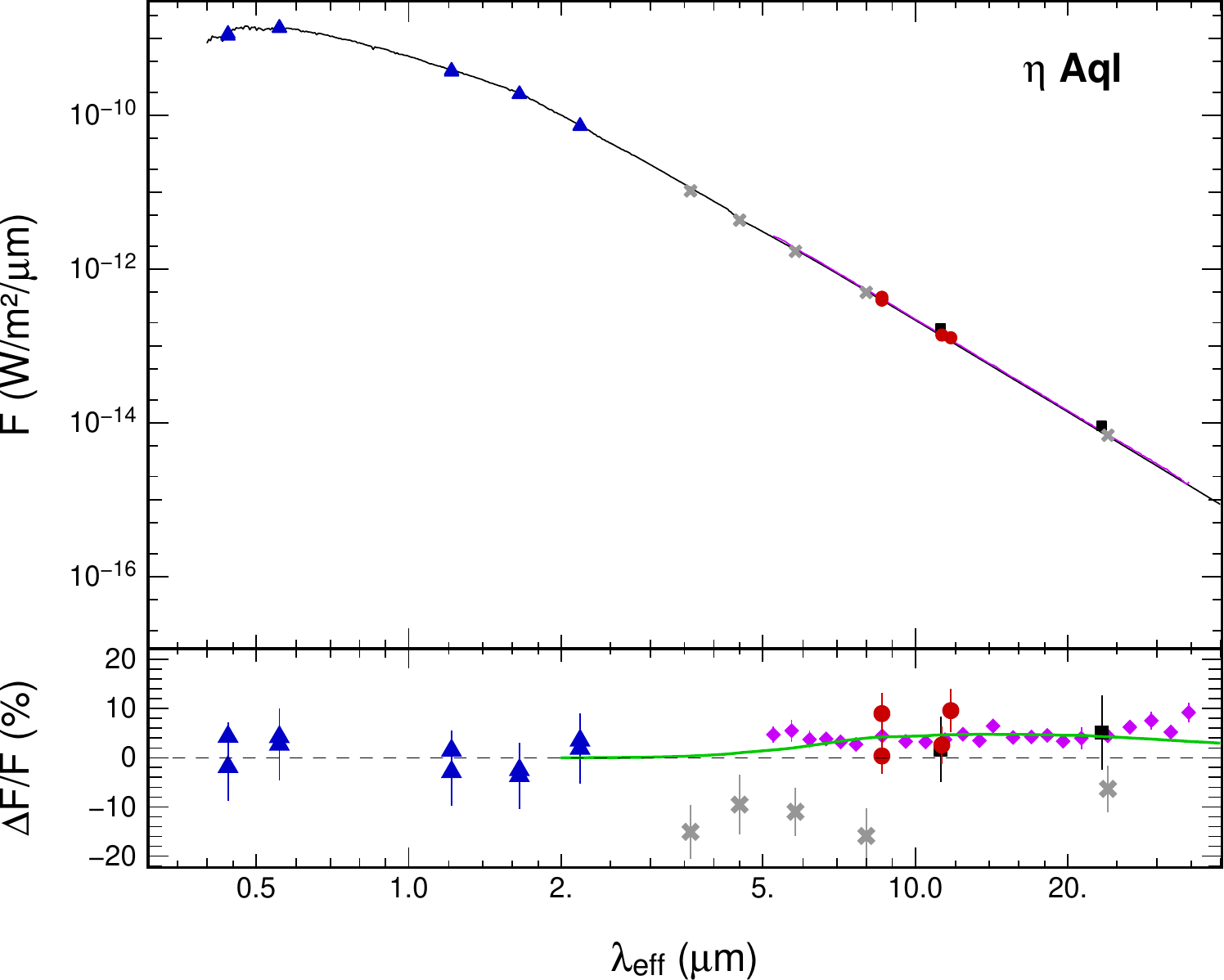}
\caption[Distribution spectrale d'énergie de $\eta$~Aql]{\textbf{Distribution spectrale d'énergie de $\eta$~Aql} : la légende est identique à la Fig.~\ref{image__SED_FF_AQL}. Le spectre \emph{IRS} est superposé en violet. Les points en gris sont les mesures de \citetalias{Marengo-2010-01}.}
\label{image__SED_ETA_AQL}
\end{figure}

J'ai recueilli les données $B, V, J, H$ et $K$ de \citetalias{Kervella-2004-03}, \citetalias{Moffett-1984-07} et \citet{Barnes-1997-06}, à partir de l'ajustement des courbes de lumière pour une phase $\phi = 0.47$. Pour les longueurs d'onde plus grandes, j'ai collecté les flux \emph{IRAS} ($12$ et $25\,\mu\mathrm{m}$) et un spectre \emph{Spitzer/IRS} allant de $5$ à $35\,\mu\mathrm{m}$ \citep{Ardila-2010-12}. La SED est présentée sur la Fig.~\ref{image__SED_ETA_AQL} avec le spectre \emph{Spitzer} en violet. Ce spectre a été mesuré à une phase $\phi = 0.42$, assez proche de notre phase intermédiaire pour ne pas tenir compte de l'incertitude additionnelle liée à la discordance de phase. Dans la fenêtre du bas de la Fig.~\ref{image__SED_ETA_AQL}, seul une valeur tout les $15\,\mu\mathrm{m}$ a été placée pour la clarté du graphe. J'ai également rajouté les mesures de flux avec \emph{Spitzer} de \citetalias{Marengo-2010-01} (en gris sur le graphe), mais elles sont significativement plus faibles que la SED de l'étoile, évaluée à partir des flux visibles et proche infrarouge. Pour cette raison, j'ai préféré ne pas les inclure dans l'ajustement.

L'ajustement du modèle stellaire, avec $\log g = 1.8$ \citep{Luck-2004-07}, donne des valeurs cohérentes (Table~\ref{table__parametre_ajuste}) avec celles d'autres travaux. Le diamètre angulaire est en bon accord avec \citet[$\theta_\mathrm{LD} = 1.87 \pm 0.03\,\mathrm{mas}$,][]{Kervella-2004-03a}, à cette phase de pulsation, et également avec le diamètre moyen de \citet[$\theta_\mathrm{LD} = 1.76 \pm 0.09\,\mathrm{mas}$,][]{Groenewegen-2008-09}. De même, notre température effective est seulement $1\,\%$ plus petite que celle estimée par \citet[$T_\mathrm{eff} = 5508 \pm 40\,\mathrm{K}$][]{Luck-2004-07}.

D'après le spectre \emph{IRS}, $\eta$~Aql semble posséder un faible excès infrarouge, avec un excès de flux relatif (à la photosphère) de $4.7\,\pm\,1.5\,\%$ à $5.3\,\mu\mathrm{m}$ et de $9.2\,\pm\,1.8\,\%$ à $34.7\,\mu\mathrm{m}$. Les mesures photométriques \emph{VISIR} suivent la même tendance avec un ordre de grandeur identique (Table~\ref{table__irradiance_mesure_I}). Il n'y a pas de pics d'émission liés à la présence de silicate, j'ai donc fait l'hypothèse que l'enveloppe est principalement composée de carbone amorphe. Pour le modèle d'enveloppe optiquement mince (Equ.~\ref{equation__corps_gris}), j'ai calculé l'opacité $\kappa_\lambda$ en utilisant le même modèle (sphères homogènes) et les mêmes hypothèses que pour AX~Cir (Equ.~\ref{equation__opacite0} et \ref{equation__opacite}), avec les constantes optiques du carbone amorphe données par \citet{Preibisch-1993-11}. L'ajustement de ce modèle est représenté en vert sur la Fig.~\ref{image__SED_ETA_AQL}, la température et la masse totale de l'enveloppe sont listées dans la Table~\ref{table__parametre_ajuste}.

\subsubsection{W~Sgr}

\begin{figure}[!p]
\centering\includegraphics[width=.7\linewidth]{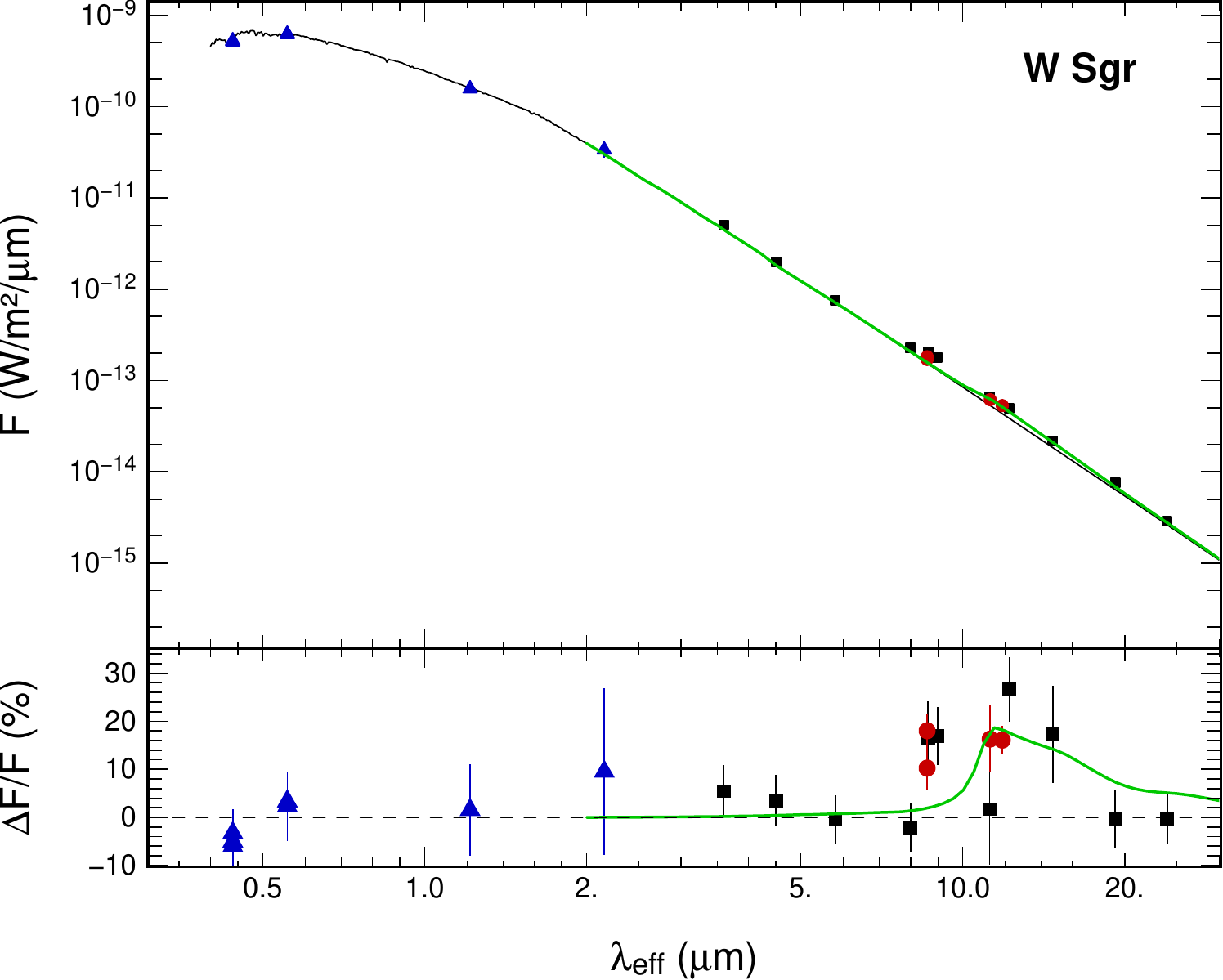}
\caption[Distribution spectrale d'énergie de W~Sgr]{\textbf{Distribution spectrale d'énergie de W~Sgr} : la légende est identique à la Fig.~\ref{image__SED_FF_AQL}.}
\label{image__SED_W_SGR}
\end{figure}

Pour cette étoile, la photométrie $B$ et $V$ provient des courbes de lumière de \citetalias{Kervella-2004-03} et \citetalias{Berdnikov-2008-04} (à une phase intermédiaire $\phi = 0.48$). Les irradiances supplémentaires proviennent des instruments \emph{DENIS} ($J$ et $Ks$), \emph{Spitzer} ($3.6, 4.5, 5.8, 8$ et $24\,\mu\mathrm{m}$ de \citetalias{Marengo-2010-01}), \emph{IRC} ($9$ et $18\,\mu\mathrm{m}$), \emph{MSX} ($8.28, 12.13, 14.65$ et $21.34\,\mu\mathrm{m}$) et \emph{IRAS} ($12\,\mu\mathrm{m}$). La Fig.~\ref{image__SED_W_SGR} représente la SED de W~Sgr.

La température effective donnée par l'ajustement du modèle stellaire (Table~\ref{table__parametre_ajuste}) est en accord à $2\,\%$ avec $T_\mathrm{eff} = 5535 \pm 51\,\mathrm{K}$ de \citet{Luck-2004-07} pour une phase de pulsation similaire et un $\log g = 1.7$ du même auteur. Le diamètre angulaire quant à lui est $15\,\%$ et $2\sigma$ plus petit que celui mesuré par \citetalias{Kervella-2004-03} ($1.31 \pm 0.04\,\mathrm{mas}$) à cette phase de pulsation, alors qu'il est en accord avec le diamètre moyen $\overline{\theta_\mathrm{LD}} = 1.17 \pm 0.11\,\mathrm{mas}$ estimé par \citet{Bersier-1997-04} basé sur la photométrie. L'explication de la divergence avec les mesures interférométriques de \citetalias{Kervella-2004-03} est la même que pour X~Sgr, le diamètre peut être surestimé à cause de la présence de l'enveloppe.

La photométrie \emph{Spitzer} est cohérente avec une distribution de corps noir, comme conclu par \citetalias{Marengo-2010-01}. Au contraire, notre photométrie \emph{VISIR} montre un excès de l'ordre de $15\,\%$ (Table~\ref{table__irradiance_mesure_I}) et la tendance est identique pour les mesures \emph{IRC} (à $9\,\mu\mathrm{m}$) et \emph{MSX} (à $8.28, 12.13$ et $14.65\,\mu\mathrm{m}$). Cela pourrait refléter une composition particulière de la matière présente autour de l'étoile. Comme pour AX~Cir, la SED ne montre pas d'excès autour de $20\,\mu\mathrm{m}$, j'ai donc rejeté la présence de grains de silicate et supposé une enveloppe composée d'oxyde d'aluminium. J'ai utilisé le même modèle d'opacité que AX~Cir pour ajuster aux mesures $>\,3\,\mu\mathrm{m}$ l'équation~\ref{equation__corps_gris}. Le résultat de l'ajustement est listé dans la Table~\ref{table__parametre_ajuste} et représenté en vert sur la Fig.\ref{image__SED_W_SGR}.

\subsubsection{Y~Oph}

\begin{figure}[!p]
\centering\includegraphics[width=.7\linewidth]{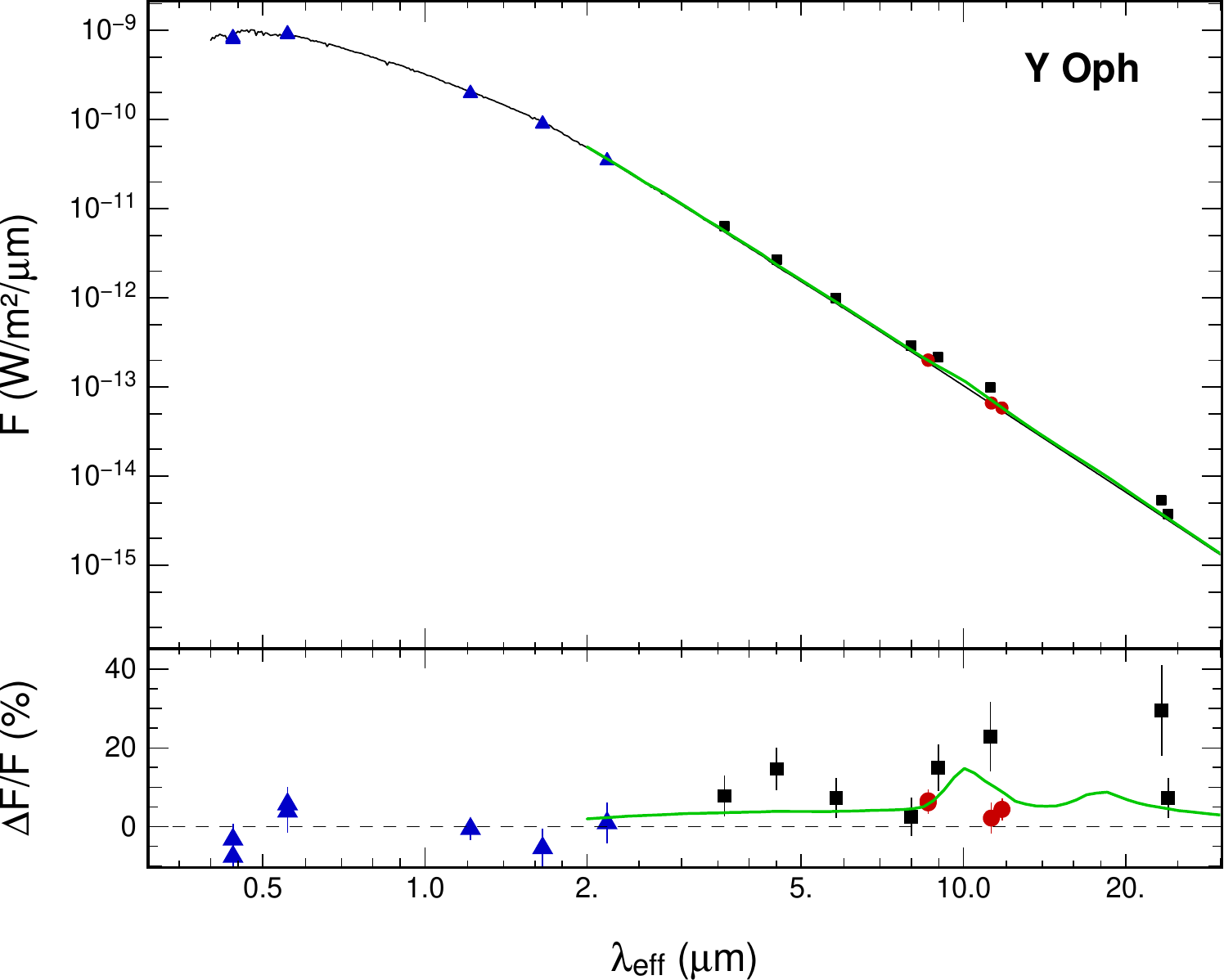}
\caption[Distribution spectrale d'énergie de Y~Oph]{\textbf{Distribution spectrale d'énergie de Y~Oph} : la légende est identique à la Fig.~\ref{image__SED_FF_AQL}.}
\label{image__SED_Y_OPH}
\end{figure}

J'ai estimé la magnitude des bandes $B$ et $V$ à partir des courbes de lumière de \citetalias{Berdnikov-2008-04} et \citetalias{Moffett-1984-07} et la magnitude des bandes $J, H$ et $K$ des courbes de lumière de \citet{Laney-1992-04}, pour une phase intermédiaire $\phi = 0.73$. D'autres mesures photométriques proviennent de \emph{Spitzer} ($3.6, 4.5, 5.8, 8.0$ et $24\,\mu\mathrm{m}$ de \citetalias{Marengo-2010-01}), \emph{IRC} ($9\,\mu\mathrm{m}$) et \emph{IRAS} ($12$ et $25\,\mu\mathrm{m}$). La distribution spectrale d'énergie est tracée sur la Fig.~\ref{image__SED_Y_OPH}.

La Table~\ref{table__parametre_ajuste} expose les résultats de l'ajustement, avec $\log g = 1.8$ de \citetalias{Luck-2008-07}. Ces mêmes auteurs donnent une température effective $T_\mathrm{eff} = 5800 \pm 148\,\mathrm{K}$ qui est seulement $1\,\%$ plus faible que notre estimation. De même pour le diamètre angulaire, qui est en parfait accord avec celui estimé par \citet[$\theta_\mathrm{LD} = 1.24 \pm 0.01\,\mathrm{mas}$,][]{Merand-2007-08}, à la même phase de pulsation.

La photométrie \emph{VISIR} indique un excès IR de l'ordre de $5\,\%$ (Table~\ref{table__irradiance_mesure_I}). Ce résultat est cohérent avec les mesures de \citet{Merand-2007-08} où une enveloppe a été détecté en bande $K$ avec une contribution relative de $5.0 \pm 2.0\,\%$. Comme le spectre présente des pics autour de $10\,\mu\mathrm{m}$ et $20\,\mu\mathrm{m}$, j'ai choisi d'ajusté le modèle de l'équation~\ref{equation__corps_gris} en utilisant la table d'opacité de \citetalias{Ossenkopf-1994-11} pour une enveloppe composée de silicate et de carbone. Le modèle est tracé sur la Fig.~\ref{image__SED_Y_OPH} (en vert). La température ajustée de l'enveloppe et sa masse totale sont présentées dans la Table~\ref{table__parametre_ajuste}. En comparaison des autres Céphéides, cette enveloppe semble plus chaude, indicatif d'une localisation proche de l'étoile et subissant un éventuel chauffage par ses radiations.

\subsubsection{U~Car}

\begin{figure}[!p]
\centering\includegraphics[width=.7\linewidth]{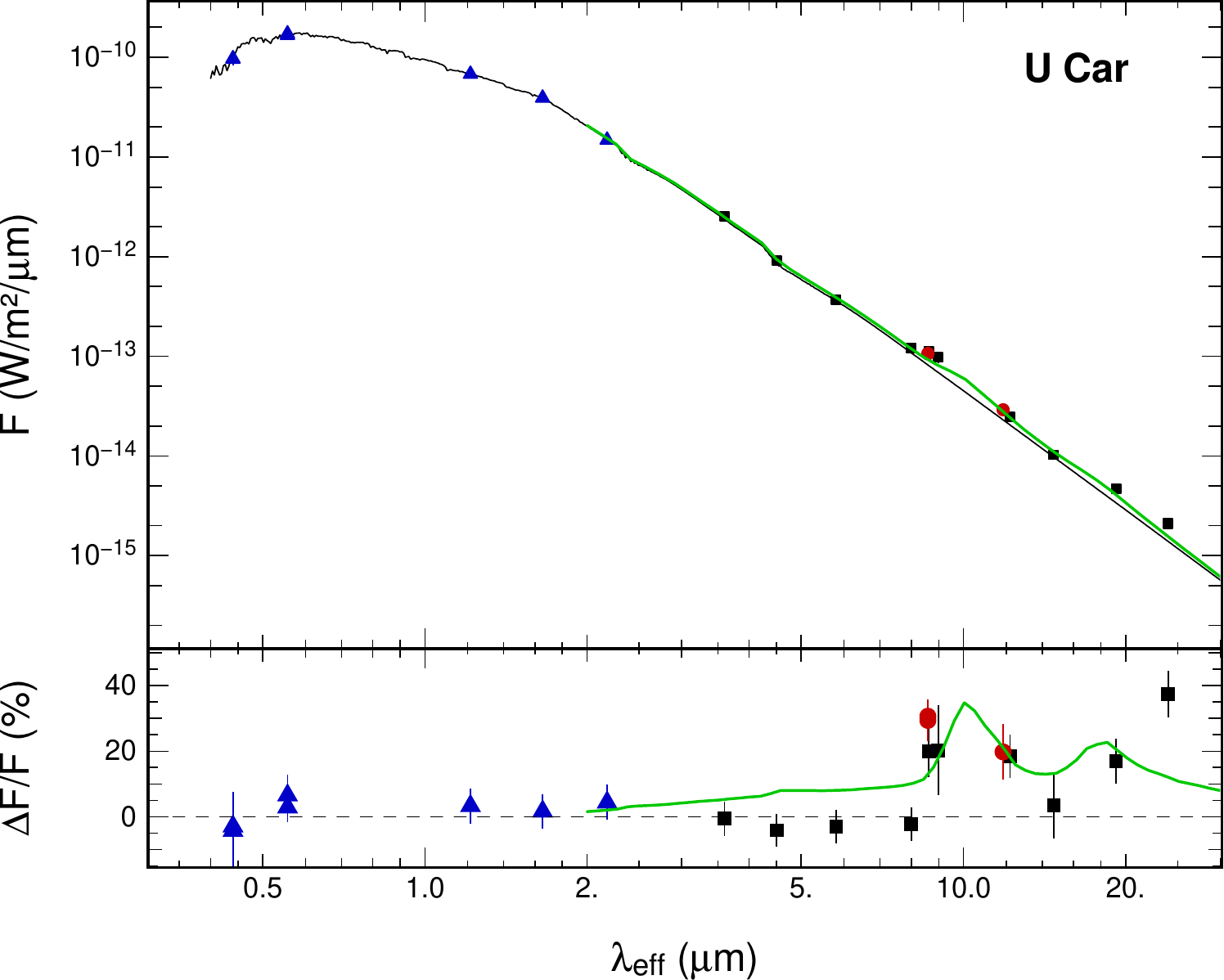}
\caption[Distribution spectrale d'énergie de U~Car]{\textbf{Distribution spectrale d'énergie de U~Car} : la légende est identique à la Fig.~\ref{image__SED_FF_AQL}.}
\label{image__SED_U_CAR}
\end{figure}

Les courbes de lumière de \citetalias{Berdnikov-2008-04} et \citet{Coulson-1985-} ont été utilisées pour estimer la magnitude des bandes $B$ et $V$, à une phase intermédiaire $\phi = 0.63$. J'ai également collecté les données photométriques $J, H$ et $K$ de \citet{Laney-1992-04}, \emph{IRC} ($9$ et $18\,\mu\mathrm{m}$), \emph{MSX} ($8.28, 12.13$ et $14.65\,\mu\mathrm{m}$) et \emph{Spitzer} ($3.6, 4.5, 5.8, 8.0$ et $24\,\mu\mathrm{m}$ de \citetalias{Marengo-2010-01}). La SED est présentée sur la Fig.~\ref{image__SED_U_CAR}.

Les résultats des modèles de \citet{Castelli-2003-} ajustés aux données photométriques sont listés dans la Table~\ref{table__parametre_ajuste}, pour une gravité effective $\log g = 1.2$ d'après \citet[][estimée à une phase $\phi = 0.49$, mais comme indiqué précédemment, la photométrie à bande large est quasiment insensible à ce paramètre]{Romaniello-2008-09}. Mon estimation du diamètre angulaire est cohérente à un niveau de $5\,\%$ (et à $1\sigma$) avec celui estimé par \citet[$0.94 \pm 0.05\,\mathrm{mas}$,][]{Groenewegen-2008-09}. Par contre, la température effective n'est pas en accord avec la valeur moyenne $T_\mathrm{eff} = 5980\,\mathrm{K}$ estimée par \citet{Romaniello-2008-09} à partir des abondances en métaux. Par ailleurs, si on se base sur une relation couleur--température, $\log T_\mathrm{eff} = 3.837 - 0.108\,(B - V)_0$ \citep{Fry-1999-10}, on trouve $T_\mathrm{eff}\sim4800\,\mathrm{K}$, plus cohérente avec mon estimation. Cela rend la valeur de \citet{Romaniello-2008-09} un peu douteuse.

Un important excès IR est mesuré sur les images \emph{VISIR} (Table~\ref{table__irradiance_mesure_I}). C'est également le cas à des longueurs d'onde plus grandes où l'on atteint $37.5 \pm 6.9\,\%$ à $24\,\mu\mathrm{m}$. Pour cette étoile, j'ai également utilisée la table d'opacité de \citetalias{Ossenkopf-1994-11} pour ajuster l'équation~ \ref{equation__corps_gris}. Je présente les résultats de cet ajustement dans la Table~\ref{table__parametre_ajuste} et la Fig.~\ref{image__SED_U_CAR}. Notons toutefois que cet excès pourrait être affecté par l'émission du cirrus interstellaire présent dans la région de cette étoile \citep{Barmby-2010-11}.

\subsubsection{SV~Vul}

\begin{figure}[!p]
\centering\includegraphics[width=.7\linewidth]{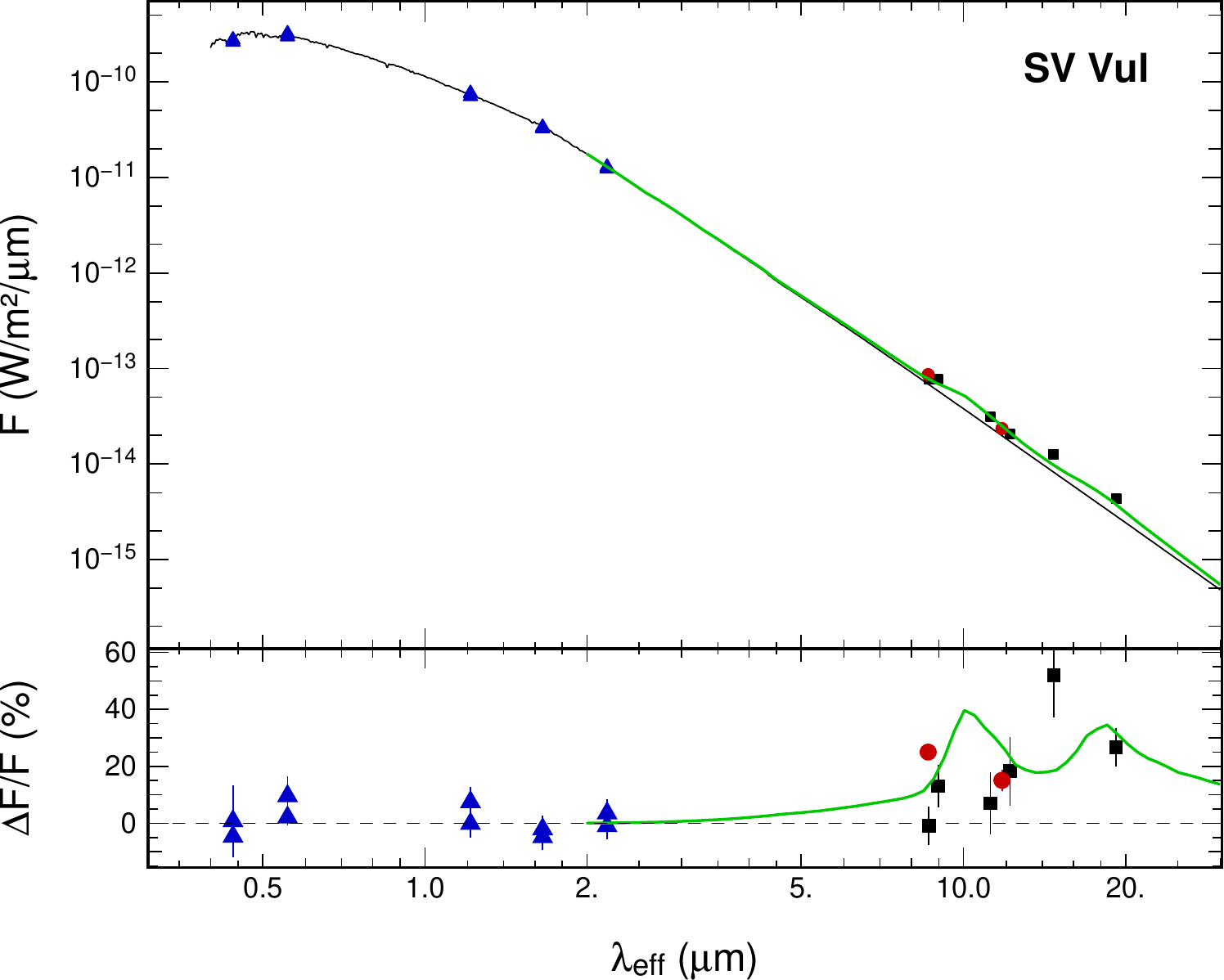}
\caption[Distribution spectrale d'énergie de SV~Vul]{\textbf{Distribution spectrale d'énergie de SV~Vul} : la légende est identique à la Fig.~\ref{image__SED_FF_AQL}.}
\label{image__SED_SV_VUL}
\end{figure}

Toujours à partir des courbes de lumière, j'ai estimé les magnitudes $B$ et $V$ d'après les mesures de \citetalias{Berdnikov-2008-04} et \citetalias{Moffett-1984-07}, et les magnitudes $J, H$ et $K$ d'après les données de \citet{Barnes-1997-06} et \citet{Laney-1992-04}. Les mesures infrarouges supplémentaires proviennent des instruments \emph{IRC} ($9$ et $18\,\mu\mathrm{m}$), \emph{MSX} ($8.28, 12.13$ et $14.65\,\mu\mathrm{m}$) et \emph{IRAS} ($12\,\mu\mathrm{m}$). La distribution spectrale d'énergie de SV~Vul est représentée sur la Fig.~\ref{image__SED_SV_VUL}.

L'ajustement du modèle d'atmosphère stellaire a été effectué en fixant $\log g = 1.4$ d'après \citet{Kovtyukh-2005-01} et les résultats sont exposés dans la Table~\ref{table__parametre_ajuste}. La température effective que nous trouvons est seulement $4\,\%$ et $1.4\sigma$ plus petite que $T_\mathrm{eff} = 5977 \pm 32\,\mathrm{K}$ estimée par \citet{Kovtyukh-2005-01}. Le diamètre angulaire est également en accord avec \citet[][$0.80 \pm 0.05\,\mathrm{mas}$]{Groenewegen-2008-09}, qui est seulement $5\,\%$ plus petit mais à $1\sigma$.

Un excès IR est également détecté dans les filtres PAH1 et SIC (Table~\ref{table__irradiance_mesure_I}). Un excès est aussi visible aux autres longueurs d'onde, jusqu'à $26.7\,\pm\,6.6\,\%$ à $18\,\mu\mathrm{m}$ avec \emph{IRC}. Cela indique sans doute la présence d'une enveloppe circumstellaire autour de cette Céphéide. Pour $\lambda > 3\,\mu\mathrm{m}$, j'ai ajusté le même modèle que précédemment (Equ.~\ref{equation__corps_gris}) avec l'opacité de \citetalias{Ossenkopf-1994-11}. La courbe est présentée sur la Fig.~\ref{image__SED_SV_VUL} (en vert) et les paramètres dans la Table~\ref{table__parametre_ajuste}. L'incertitude sur l'estimation de la masse totale est due à l'incertitude sur la mesure de la parallaxe (Table~\ref{table__cepheide_parametre}). D'autres mesures photométriques, entre $3$ et $8\,\mu\mathrm{m}$ par exemple, sont nécessaires pour mieux contraindre cette émission infrarouge.

\subsection{Distribution spectrale d'énergie des Céphéides de type II}

Dans cette section, les mêmes méthodes que pour les Céphéides classiques sont utilisés. Ces étoiles sont connues pour leur émission infrarouge et nous serviront à valider la méthode et le modèle d'estimation de l'excès IR.

\begin{table}[!p]
\centering
\begin{tabular}{cccccc} 
\hline
\hline
Nom	  					&	    MJD			&	Filtre			& 	Irradiance												& Irradiance				& Excès 		\\
		  					&						&					&	($\mathrm{W/m^2/\mu m}$)				& 	(Jy)							& 	($\%$)			\\
\hline
R~Sct					&	54~610.236	&  PAH1		&	$6.74\,\pm\, 0.16\,\times\,10^{-13}$	&	$16.6\,\pm\,0.4$		&	$134\,\pm\,6$	\\
							&	54~611.189	&  PAH1		&	$7.05\,\pm\, 0.17\,\times\,10^{-13}$	&	$17.4\,\pm\,0.4$		&	$145\,\pm\,6$	\\
							&	54~610.243	&	PAH2		&	$2.56\,\pm\, 0.06\,\times\,10^{-13}$	&	$10.8\,\pm\,0.3$		&	$153\,\pm\,7$	\\
							&	54~611.196	&	SiC			&	$2.24\,\pm\, 0.06\,\times\,10^{-13}$	&	$10.3\,\pm\,0.3$		&	$158\,\pm\,7$	\\
\hline
AC~Her					&	54~610.349	&  PAH1		&	$9.18\,\pm\, 0.25\,\times\,10^{-13}$	&	$22.6\,\pm\,0.6$		&	$8273\,\pm\,228$	\\
							&	54~611.340	&  PAH1		&	$9.40\,\pm\, 0.23\,\times\,10^{-13}$	&	$23.2\,\pm\,0.6$		&	$8473\,\pm\,210$	\\
							&	54~610.357	&	PAH2		&	$10.1\,\pm\, 0.28\,\times\,10^{-13}$	&	$42.8\,\pm\,1.2$		&	$26705\,\pm\,743$	\\
							&	54~611.347	&	SiC			&	$8.33\,\pm\, 0.21\,\times\,10^{-13}$	&	$38.4\,\pm\,1.0$		&	$25586\,\pm\,647$	\\
\hline
$\kappa$~Pav		&	54~611.227	&  PAH1		&	$1.93\,\pm\, 0.05\,\times\,10^{-13}$	&	$4.75\,\pm\,0.12$	&	$22.2\,\pm\,3.2$	\\
							&	54~611.235	&	SiC			&	$5.39\,\pm\, 0.14\,\times\,10^{-14}$	&	$2.48\,\pm\,0.06$	&	$15.3\,\pm\,3.0$	\\
\hline
\end{tabular}
\caption[Irradiances mesurées par photométrie des Céphéides de type II]{\textbf{Irradiances mesurées par photométrie des Céphéides de type II} : Les flux ont été mesurés sur une ouverture de 1.30\arcsec. L'excès est relatif au modèle photosphérique de l'étoile.}
\label{table__irradiance_mesure_II}
\end{table}

\subsubsection{R~Sct}

\begin{figure}[!p]
\centering\includegraphics[width=.7\linewidth]{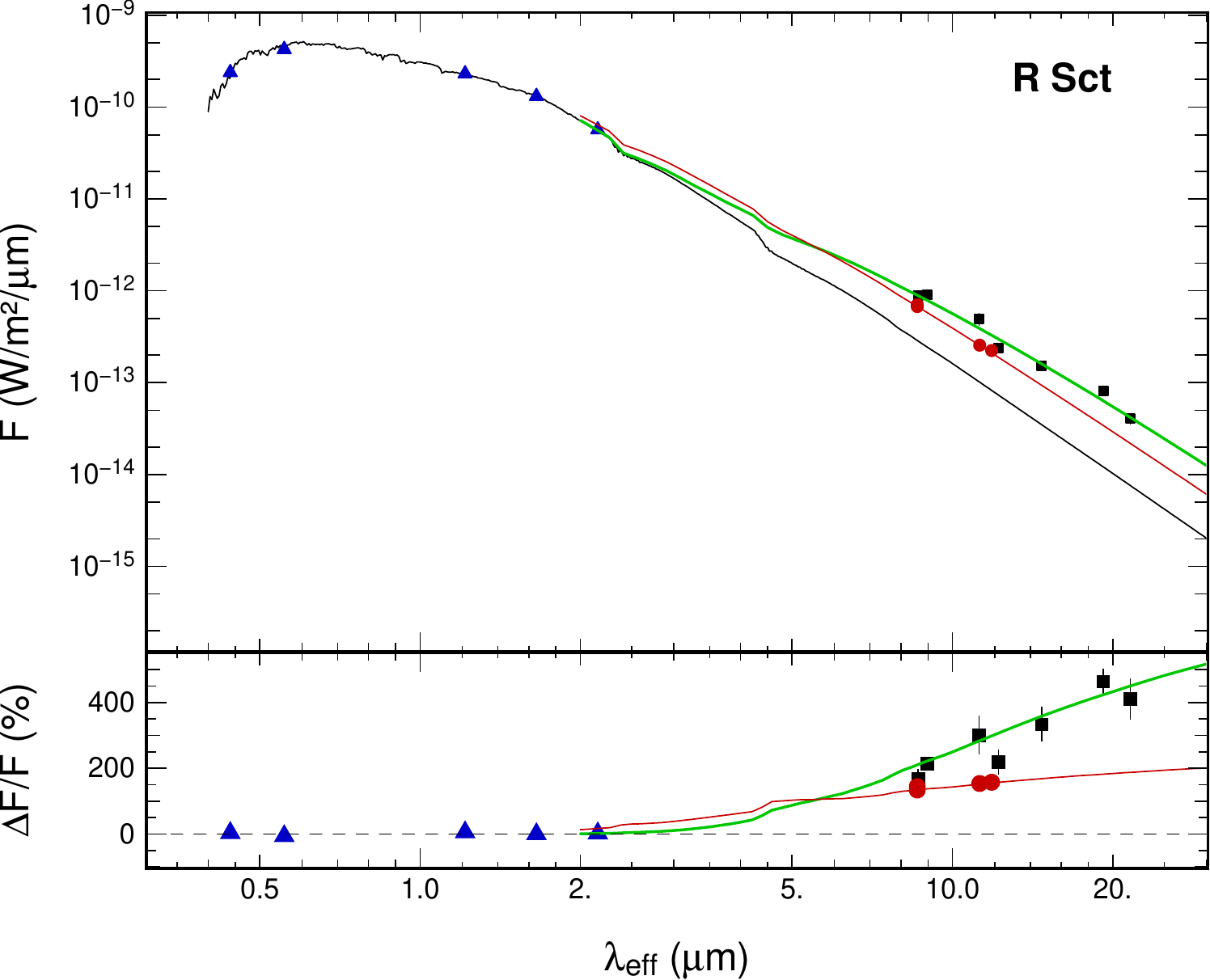}
\caption[Distribution spectrale d'énergie de R~Sct]{\textbf{Distribution spectrale d'énergie de R~Sct} : la légende est identique à la Fig.~\ref{image__SED_FF_AQL}. Deux composantes sont distinguées : la courbe en vert représente le modèle de l'équation~\ref{equation__corps_gris} pour la composante large, tandis que la courbe en rouge représente la composante compacte.}
\label{image__SED_R_SCT}
\end{figure}

\begin{figure}[!p]
\centering\includegraphics[width=.7\linewidth]{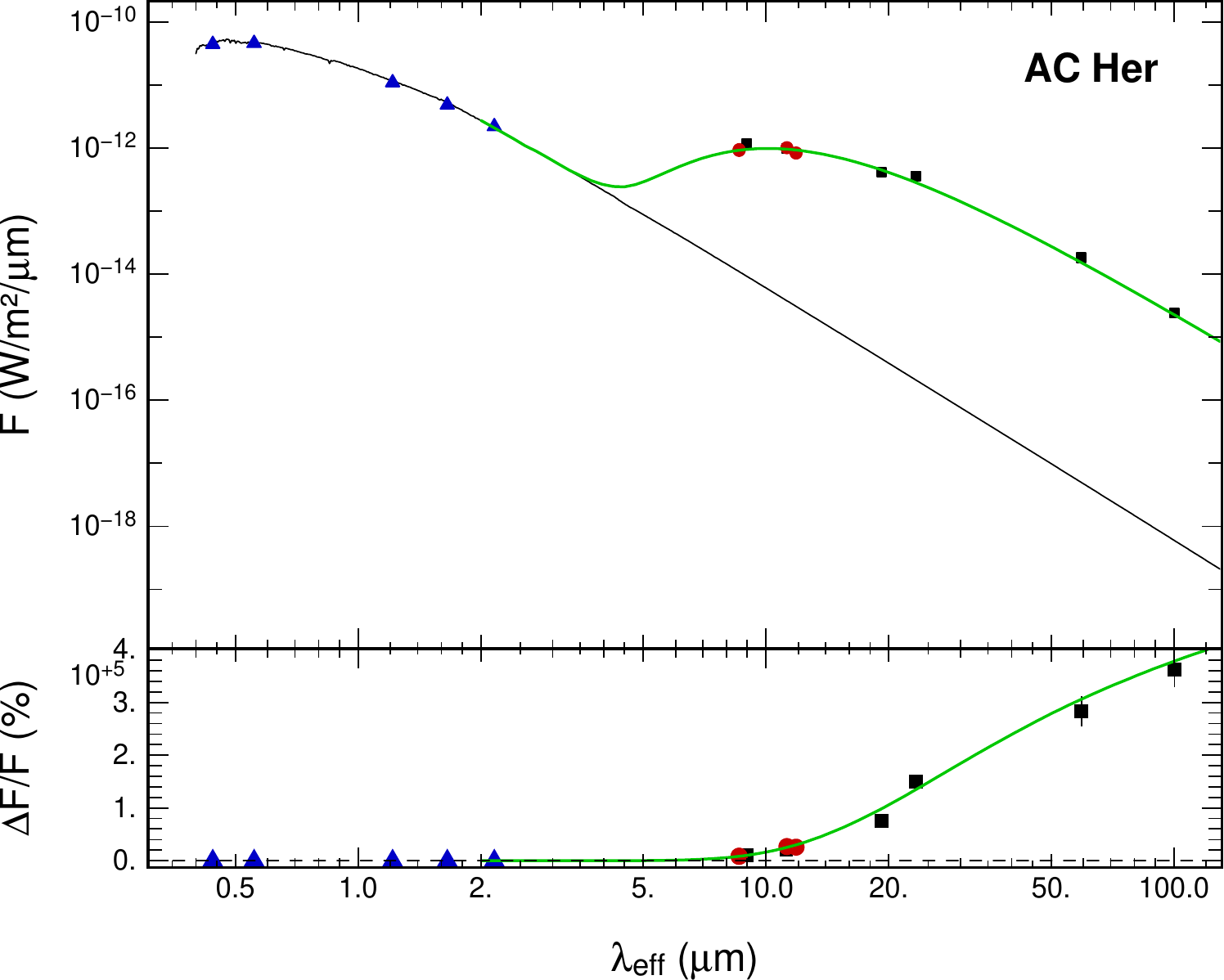}
\caption[Distribution spectrale d'énergie de AC~Her]{\textbf{Distribution spectrale d'énergie de AC~Her} : la légende est identique à la Fig.~\ref{image__SED_FF_AQL}.}
\label{image__SED_AC_HER}
\end{figure}

Cette RV~Tauri est souvent distinguée des autres par sa pulsation irrégulière et sa grande amplitude. Plusieurs auteurs \citep[par exemples][]{Goldsmith-1987-07,Giridhar-2000-03,de-Ruyter-2005-05} ont étudié sa composante circumstellaire, source d'un excès IR bien connu.

J'ai récolté les données photométriques $B, V, J, H$ et $ K$ de \citet{Shenton-1994-07} à la phase de pulsation $\phi = 0.48$. J'ai en plus utilisé les flux mesurés par les instruments \emph{IRC} ($9$ et $18\,\mu\mathrm{m}$), \emph{IRAS} ($12\,\mu\mathrm{m}$) et \emph{MSX} ($8.28, 12.13, 14.65$ et $21.34\,\mu\mathrm{m}$). Je présente la distribution spectrale d'énergie sur la Fig.~\ref{image__SED_R_SCT}.

J'ai ajusté la composante stellaire avec $\log g = 0.0$ donnée par \citet{Giridhar-2000-03}. Comme on pouvait s'y attendre, nous détectons un important excès infrarouge dans nos trois filtres (Table~\ref{table__irradiance_mesure_II}). La température effective évaluée par l'ajustement (Table~\ref{table__parametre_ajuste}) est cohérente à $2\,\%$ avec \citet{Giridhar-2000-03}, qui donnent $T_\mathrm{eff} = 4500\,\mathrm{K}$ (à une phase $\phi = 0.44$ recalculée avec les éphémérides de la Table~\ref{table__cepheide_parametre}). \citet{Shenton-1994-07} trouvent également une température effective de $4500\,\mathrm{K}$ à une phase $\phi = 0.49$. En couplant une estimation du rayon de l'étoile de \citet[$R_\star = 81\,R_\odot$,][]{Shenton-1994-07}, avec la mesure de parallaxe d'Hipparcos ($d = 431\,\mathrm{pc}$), on trouve un diamètre angulaire de $\sim1.79\,\mathrm{mas}$. Cette valeur est en accord à $3\,\%$ avec mon estimation présentée dans la Table~\ref{table__parametre_ajuste}.

Pour la composante circumstellaire, j'ai supposé un milieu optiquement épais. Le flux observé peut alors être ajusté par la fonction suivante :
\begin{equation}
F_\lambda = A\,B_\lambda(T_\mathrm{d})
\label{equation__blackbody}
\end{equation}
où $A$ est un paramètre contenant l'angle solide et l'émissivité de l'enveloppe.

En regardant d'un peu plus près la SED, il semble que mes valeurs soient légèrement plus faibles que les autres, et on pourrait même distinguer deux distributions d'intensités distinctes. Une composante supplémentaire plus compacte est probablement détecté avec \emph{VISIR}. Ceci est une explication plausible car la résolution angulaire avec \emph{VISIR} en bande $N$ est plus grande que celle d'\emph{AKARI} et \emph{IRAS}. J'ai donc décidé d'ajuster deux distributions suivant le modèle de l'équation~\ref{equation__corps_gris} : une composante chaude et compacte détectée par la photométrie \emph{VISIR} et une composante plus grande détectée par les autres données. Ces deux composantes sont représentées sur la Fig.~\ref{image__SED_R_SCT} par la courbe en vert et en rouge. L'ajustement de la composante compacte (courbe rouge) donne $T_\mathrm{env} = 1486 \pm 335\,\mathrm{K}$ (Table~\ref{table__parametre_ajuste}), et est probablement chauffée par les radiations issues de la photosphère de l'étoile. L'autre composante (courbe verte) est plus froide avec $T_\mathrm{env} = 772 \pm 82\,\mathrm{K}$. Notons que \citet{Taranova-2010-02} ont probablement mesuré la composante la plus large car le résultat de leur modélisation donne une température de l'enveloppe de $T_\mathrm{env} = 800 \pm 50\,\mathrm{K}$, compatible avec mon estimation.

\subsubsection{AC~Her}

L'important excès infrarouge causé par l'émission thermique des poussières autour de cette étoile a été étudié par de nombreux auteurs \citep[par exemple][]{de-Ruyter-2005-05}. Comme pour R~Sct, l'objectif est de vérifier que nos mesures photométriques avec \emph{VISIR} sont cohérentes aux autres données.

Les données photométriques $B, V, J, H$ et $K$ proviennent de \citet{Shenton-1992-08}, à la phase $\phi = 0.14$. D'autres mesures de flux ont été collectées d'après les instruments \emph{IRC} ($9$ et $18\,\mu\mathrm{m}$) et \emph{IRAS} ($12, 25, 60$ et $100\,\mu\mathrm{m}$). Ces points sont exposés sur la Fig.~\ref{image__SED_AC_HER}.

Un excès infrarouge est clairement détecté et mes mesures photométriques sont situées au maximum de la densité spectrale d'énergie (Table~\ref{table__irradiance_mesure_II}). Les mêmes modèles que précédemment sont ajustés à cette SED : un modèle photosphérique et un modèle d'enveloppe circumstellaire. Pour le modèle stellaire, j'ai choisi $\log g = 0.5$ d'après \citet{Van-Winckel-1998-08}. Les paramètres stellaires déduits de l'ajustement sont exposés dans la Table~\ref{table__parametre_ajuste}. La température effective est cohérente avec la valeur moyenne $T_\mathrm{eff} \sim 5400\,\mathrm{K}$ de \citet[][en utilisant $T_\odot = 5800\,\mathrm{K}$]{Taranova-2010-02}. Quant au diamètre angulaire, il est également en accord avec $\theta_\mathrm{LD} = 0.29\,\mathrm{mas}$ de \citet[][en utilisant $R = 23.6\,R_\odot$ et $d = 750\,\mathrm{pc}$ de \citealt{Taranova-2010-02}]{Shenton-1992-08}.

L'enveloppe a été ajustée avec le modèle de l'équation \ref{equation__blackbody}, en supposant un milieu optiquement épais (courbe verte de la Fig.~\ref{image__SED_AC_HER}). La température ajustée est listée dans la Table~\ref{table__parametre_ajuste} avec $A = 92.0 \pm 6.9\,\mathrm{mas^2}$. Il est difficile de comparer ces résultats avec d'autres valeurs de la littérature car de nombreuses estimations issues de divers modèles existent. Par exemple, en utilisant un modèle de disque, \citet{Gielen-2007-11} ont estimé une température des poussières de $170\,\mathrm{K}$ avec un rayon interne $R_\mathrm{in} = 50\,\mathrm{mas}$ et externe $R_\mathrm{ext} = 428\,\mathrm{mas}$ (utilisant une distance $d = 1400\,\mathrm{pc}$). D'un autre côté, ces paramètres diffèrent de \citet{de-Ruyter-2005-05} qui ont estimé $R_\mathrm{in} = 22\,\mathrm{mas}$ et $R_\mathrm{ext} = 397\,\mathrm{mas}$, en supposant un modèle de coquille sphérique. \citet{Alcolea-1991-05} ont également ajusté un modèle de coquille sphérique et ont trouvé $R_\mathrm{in} = 76\,\mathrm{mas}$ et $R_\mathrm{ext} = 3330\,\mathrm{mas}$, avec une température d'enveloppe de $360\,\mathrm{K}$. Pour finir, \citet{Close-2003-11} ont étudié AC~Her en imagerie haute-résolution avec optiques adaptatives en infrarouge moyen et n'ont pas détecté de structure plus grande que $200\,\mathrm{mas}$. En conclusion, cette étoile nécessite un effort observationnel important si l'on souhaite contraindre les modèles de SED. Nous verrons dans une prochaine section que j'aboutis à la même limite que \citet{Close-2003-11} en terme de taille angulaire .

\subsubsection{$\kappa$~Pav}

\begin{figure}[!p]
\centering\includegraphics[width=.7\linewidth]{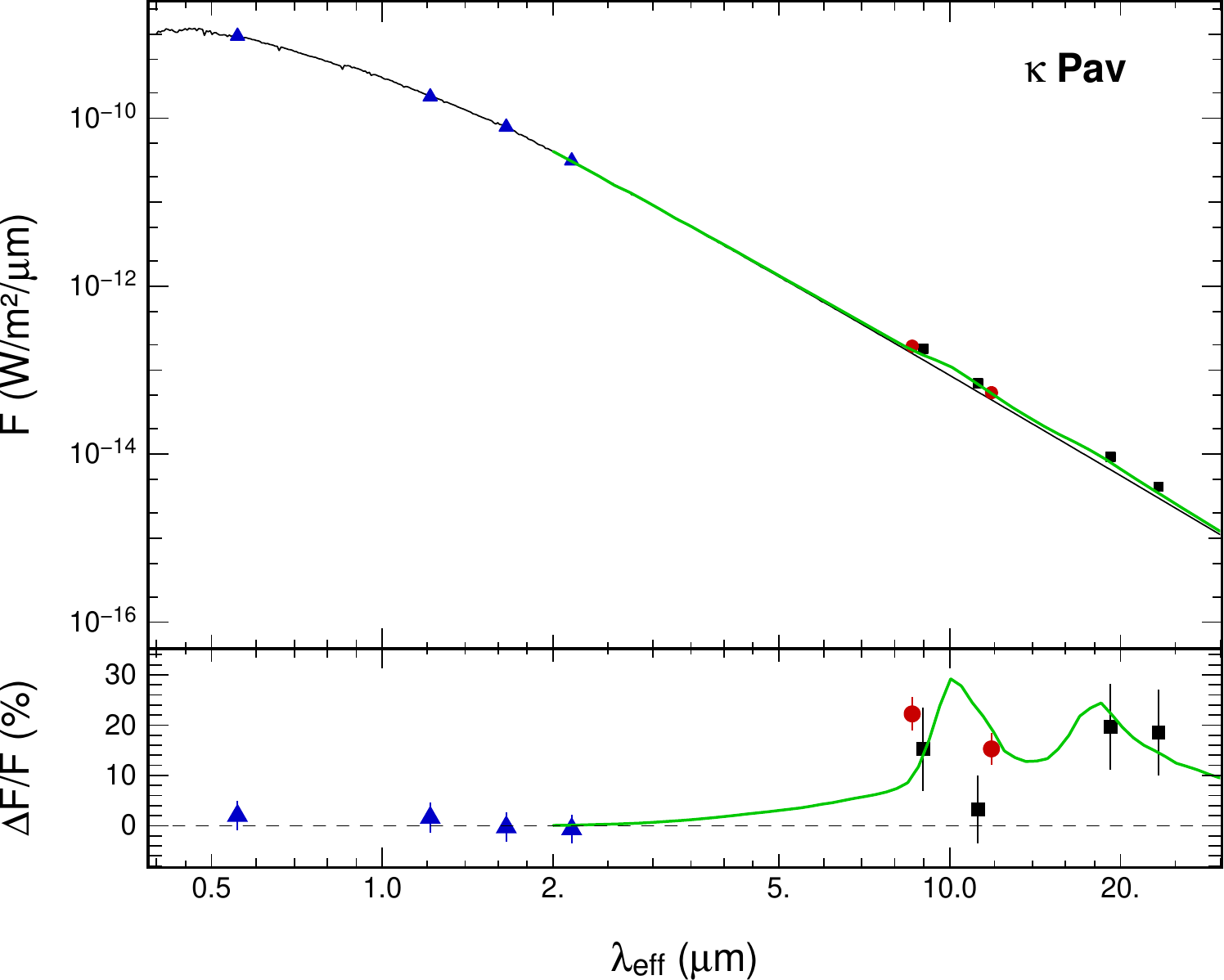}
\caption[Distribution spectrale d'énergie de $\kappa$~Pav]{\textbf{Distribution spectrale d'énergie de $\kappa$~Pav} : la légende est identique à la Fig.~\ref{image__SED_FF_AQL}.}
\label{image__SED_K_PAV}
\end{figure}

Pour cette dernière étoile, j'ai estimé les magnitudes $V, J, H$ et $Ks$ d'après les courbes de lumière de \citetalias{Feast-2008-06-2}. Les flux en infrarouge moyen et lointain ont été mesurés par les instruments \emph{IRC} ($9$ et $18\,\mu\mathrm{m}$) et \emph{IRAS} ($12$ et $25\,\mu\mathrm{m}$). La distribution spectrale d'énergie est illustrée sur la Fig.~\ref{image__SED_K_PAV}.

La gravité effective est fixée à $\log g = 1.2$ d'après \citet{Luck-1989-07} pour l'ajustement du spectre stellaire. Mon estimation du diamètre (Table~\ref{table__parametre_ajuste}) est en très bon accord avec celui estimé par \citetalias{Feast-2008-06-2} ($\theta_\mathrm{LD} = 1.04 \pm 0.04\,\mathrm{mas}$, avec $d = 204 \pm 6\,\mathrm{pc}$ et à la phase $\phi = 0.90$). Par contre, la température effective est $9\,\%$ plus grande que \citet{Luck-1989-07} qui ont estimé $T_\mathrm{eff} \sim 5750\,\mathrm{K}$ (à une phase $\phi = 0.94$ légèrement différente), d'après des mesures d'abondances métalliques. Outre une estimation de température à des phases différentes pouvant expliquer cette différence, il est intéressant de noter que cette étoile est classée comme une Céphéide de type II particulier. Elle est plus brillante et possède une courbe de lumière distincte des types II normaux ayant une même période de pulsation \citep[voir par exemple][]{Matsunaga-2009-08}. La température a pu légèrement varier entre 1980 et 2008.

Un excès infrarouge de l'ordre de $20\,\%$ est également détecté pour cette Céphéide (Table~\ref{table__irradiance_mesure_II}). Cette excès tend à augmenter avec la longueur d'onde, comme le montre la Fig.~\ref{image__SED_K_PAV}, menant à l'hypothèse de la présence d'une enveloppe. Les points de mesures photométriques $> 3 \mu\mathrm{m}$ ont été ajustés par la distribution d'intensité introduite précédemment, c'est à dire que j'ai considéré une enveloppe optiquement mince (Equ.~\ref{equation__corps_gris}). On constate encore quelques caractéristiques de grains de silicate et de carbone dans le spectre, j'ai donc utilisé la table d'opacité de \citetalias{Ossenkopf-1994-11} pour l'ajustement. La masse totale et la température de l'enveloppe sont présentées dans la Table~\ref{table__parametre_ajuste}. Le modèle est tracé en vert sur la Fig.~\ref{image__SED_K_PAV}.

\begin{figure}[!p]
\centering
\includegraphics[width=.7\linewidth]{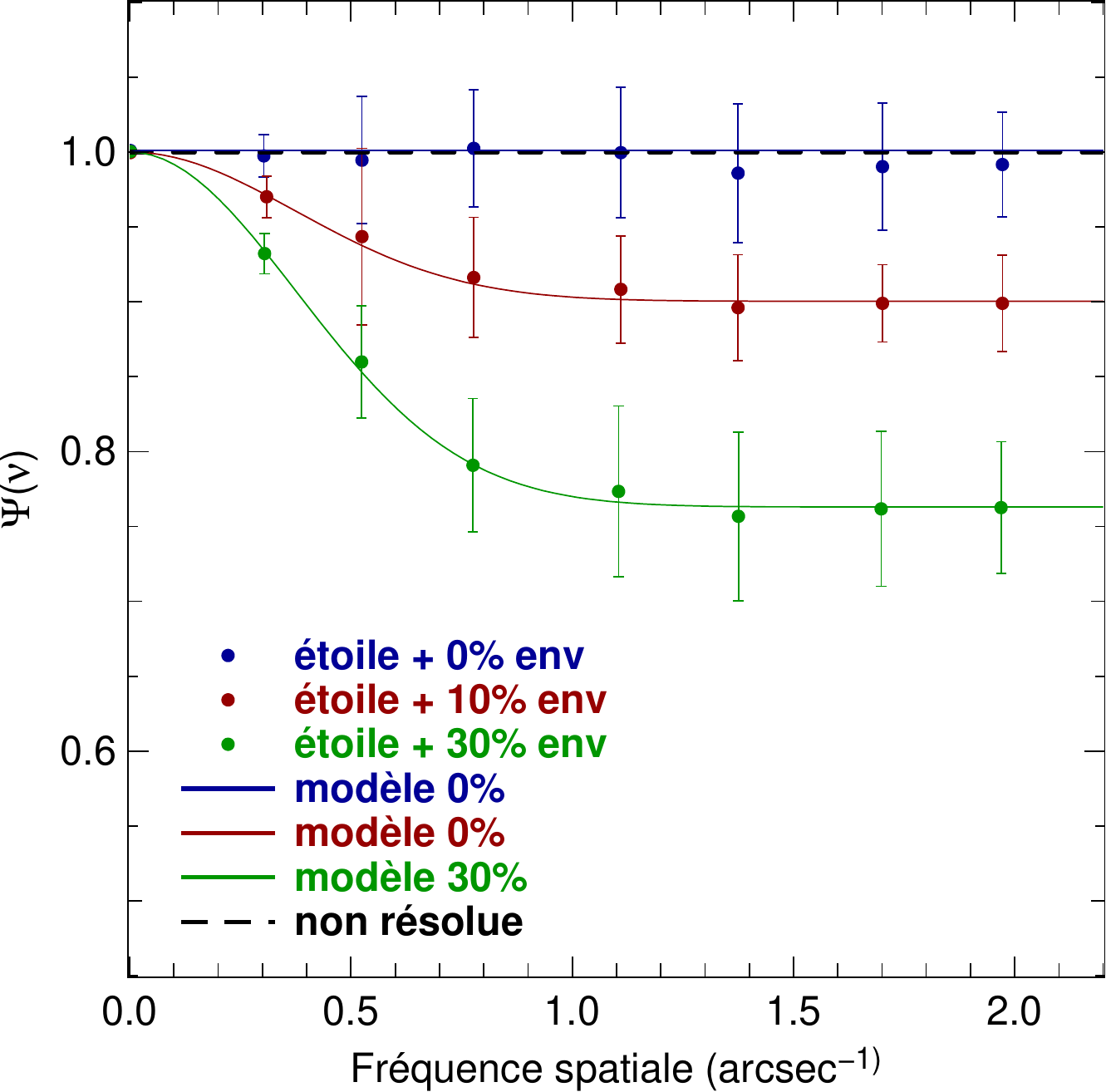}
\caption[Validation de la technique de Fourier]{\textbf{Validation de la technique de Fourier} :  en bleu, la fonction $\Psi (\nu,\rho_\lambda,\alpha_\lambda)$ résultant de l'analyse de Fourier de l'étoile HD~89682 prise comme source scientifique. Pour valider cette technique, nous avons introduit artificiellement deux enveloppes Gaussiennes ayant un flux relatif de $10\,\%$ (en rouge) et de $30\,\%$ (en vert) et de largeur à mi-hauteur $1\arcsec$. La courbe en tirets représente la visibilité d'une source non résolue.}
\label{image__fonction_psi_test}
\end{figure}

Cette section, consacrée aux excès infrarouges, à montrer que sur notre échantillon de 11 Céphéides, 10 présentent une émission infrarouge supplémentaire. Une explication probable serait que cette émission provienne des gaz et poussières présents à l'intérieur d'une enveloppe circumstellaire. Le mécanisme de formation de ces enveloppes pourrait être lié à la perte de masse via le mécanisme de pulsation et des ondes de choc se produisant dans l'atmosphère des Céphéides. Enfin, bien que d'autres données photométriques sont nécessaires, certaines des Céphéides semblent présentées quelques pics d'émission liés à la présence de silicate et de carbone.

\subsection{Technique d'analyse de Fourier}

Procédons maintenant à un autre type d'analyse qui consiste à rechercher une éventuelle émission spatialement étendue. Pour cela j'utilise une technique de Fourier similaire dans son principe à la méthode d'étalonnage utilisée en interférométrie longue base. Cette méthode a déjà été utilisée et validée par \citet{Kervella-2009-05} et \citet{Kervella-2006-07}. Elle est basée sur le théorème de Zernike--Van~Cittert, qui stipule que la fonction de visibilité complexe est égale à la transformée de Fourier de la distribution spatiale d'intensité, mais nous en parlerons plus en détail dans le Chapitre~\ref{chapitre__acces_a_la_haute_resolution_angulaire_interferometrie}.

\paragraph*{\textcolor{black}{Principe}}

Le principe est de diviser le module de la transformée de Fourier de l'image de la Céphéide ($I_\mathrm{cep}$) par celui de l'étoile de référence ($I_\mathrm{ref}$) :
\begin{displaymath}
\Psi (\nu_x,\nu_y) = \left| \frac{ \hat{I}_{\mathrm{cep}}(x,y) }{ \hat{I}_{\mathrm{ref}}(x,y) } \right|
\end{displaymath}
où $\hat{I}$ représente la transformée de Fourier de $I$, $(x,y)$ les coordonnées dans le ciel et $(\nu_x,\nu_y)$ les fréquences spatiales angulaires. Comme indiqué précédemment, cette équation est équivalente à des observations interférométriques qui fournissent des mesures de la transformée de Fourier de la distribution d'intensité de l'objet observé.

Nous nous ramenons à une dimension en calculant la médiane en anneaux (comme présenté dans le Chapitre~\ref{chapitre__imagerie_a_haute_resolution_spatiale_optique_adaptative_et_lucky_imaging}), c'est à dire la médiane sur un rayon de fréquence spatiale $\nu = \sqrt{\nu_x^2 + \nu_y^2}$ pour tout azimut. La fonction $\Psi(\nu)$ ainsi obtenue est équivalente à une visibilité en interférométrie. Les barres d'erreur de cette fonction ont été estimées par la somme quadratique de la dispersion de $| \hat{I}_{\mathrm{ref}} |$ encadrant chaque étoile et de l'écart quadratique de la fonction $\Psi(\nu)$ sur chaque azimut et fréquence spatiale. Une éventuelle asymétrie ne serait pas détectée et serait incluse dans les barres d'erreur.

On définit maintenant un modèle d'étoile non résolue entourée par une enveloppe de forme Gaussienne. En prenant la transformée de Fourier de ce modèle, il est possible de récupérer la distribution d'intensité de l'enveloppe. Ce type de modèle a comme expression mathématique :
\begin{eqnarray*}
\Psi (\nu,\rho_\lambda,\alpha_\lambda)& = &\frac{f_\star V_\star + f_\mathrm{env}V_\mathrm{env}}{f_\star + f_\mathrm{env}} \\
& = &\frac{1}{1 + \alpha_\lambda} \left[ 1 + \alpha_\lambda \exp{(-\frac{(\pi\ \rho_\lambda\ \nu)^2}{4\ln{2}} } \right]
\end{eqnarray*}
où l'enveloppe Gaussienne est définie avec une largeur à mi-hauteur $\rho_\lambda$ et un flux relatif $\alpha_\lambda = f_\mathrm{env}(\lambda)/f_\star(\lambda)$, c'est à dire le rapport entre le flux de l'enveloppe et la photosphère de l'étoile. $V$ représente la visibilité et nous avons posé $V_\star = 1$ puisque les étoiles ne sont pas résolues par le télescope.

\paragraph*{\textcolor{black}{Validation de la technique}}

Pour valider la méthode, j'ai utilisé l'étoile HD~89682 observée avec le filtre PAH1 comme cible scientifique, et les étoiles HD~98118 et HD~124294 comme sources de référence (cela correspond aux observations \#1, \#5, \#35 et \#41 de la Table~\ref{table__log_visir}). Comme le montre la courbe bleue de la Fig.~\ref{image__fonction_psi_test}, la fonction $\Psi$ de HD~89682 est identique à 1, comme attendue pour une source non résolue. Le rapport de flux ajusté est $\alpha = 0.1 \pm 1.1$, alors que la valeur de $\rho$ n'est pas contrainte.

Pour valider cette technique en présence d'une enveloppe de forme Gaussienne, nous avons créé deux images comprenant une source non résolue entourée par une enveloppe de forme Gaussienne. La première image contient une Gaussienne de largeur à mi-hauteur de $1\arcsec$ et de rapport de flux $\alpha = 10\,\%$, tandis que la seconde à un rapport de flux $\alpha = 30\,\%$ pour une même largeur à mi-hauteur. Ces images ont ensuite été convoluées séparément aux deux observations de HD~89682. Une fonction $\Psi$ est calculée pour chaque nuit avant d'être ensuite moyennée. Enfin, on ajuste le modèle Gaussien et l'on obtient les résultats suivant :
\begin{eqnarray*}
\rho & = & 1.00 \pm 0.28\,\arcsec \quad \mathrm{et} \quad \alpha = 11.1 \pm 1.9\,\% \qquad \mathrm{pour\ le\ mod\grave{e}le\ \grave{a}}\ 10\,\% \\
\rho & = & 0.99 \pm 0.13\,\arcsec \quad \mathrm{et} \quad \alpha = 31.1 \pm 4.3\,\% \qquad \mathrm{pour\ le\ mod\grave{e}le\ \grave{a}}\ 30\,\% 
\end{eqnarray*}

Ces paramètres sont en accord à $1\sigma$ avec les valeurs d'entrées, validant ainsi la technique de Fourier. Ces modèles sont tracés sur la Fig.~\ref{image__fonction_psi_test}.

\paragraph*{\textcolor{black}{Application aux Céphéides}}

J'ai appliqué l'ajustement en utilisant une simple minimisation du $\chi^2$ sur toutes les images. Je n'ai pas détecté d'émission étendue pour FF~Aql, $\eta$~Aql, U~Car (pour le filtre SIC), SV~Vul, R~Sct, AC~Her and $\kappa$~Pav. Pour ces étoiles, nous pouvons donner une limite supérieure à l'extension de l'enveloppe circumstellaire basée sur la résolution du télescope, soit $1.22\lambda/D \sim 265\,\mathrm{mas}$.

\begin{figure}[!p]
\centering
\includegraphics[width=8.cm]{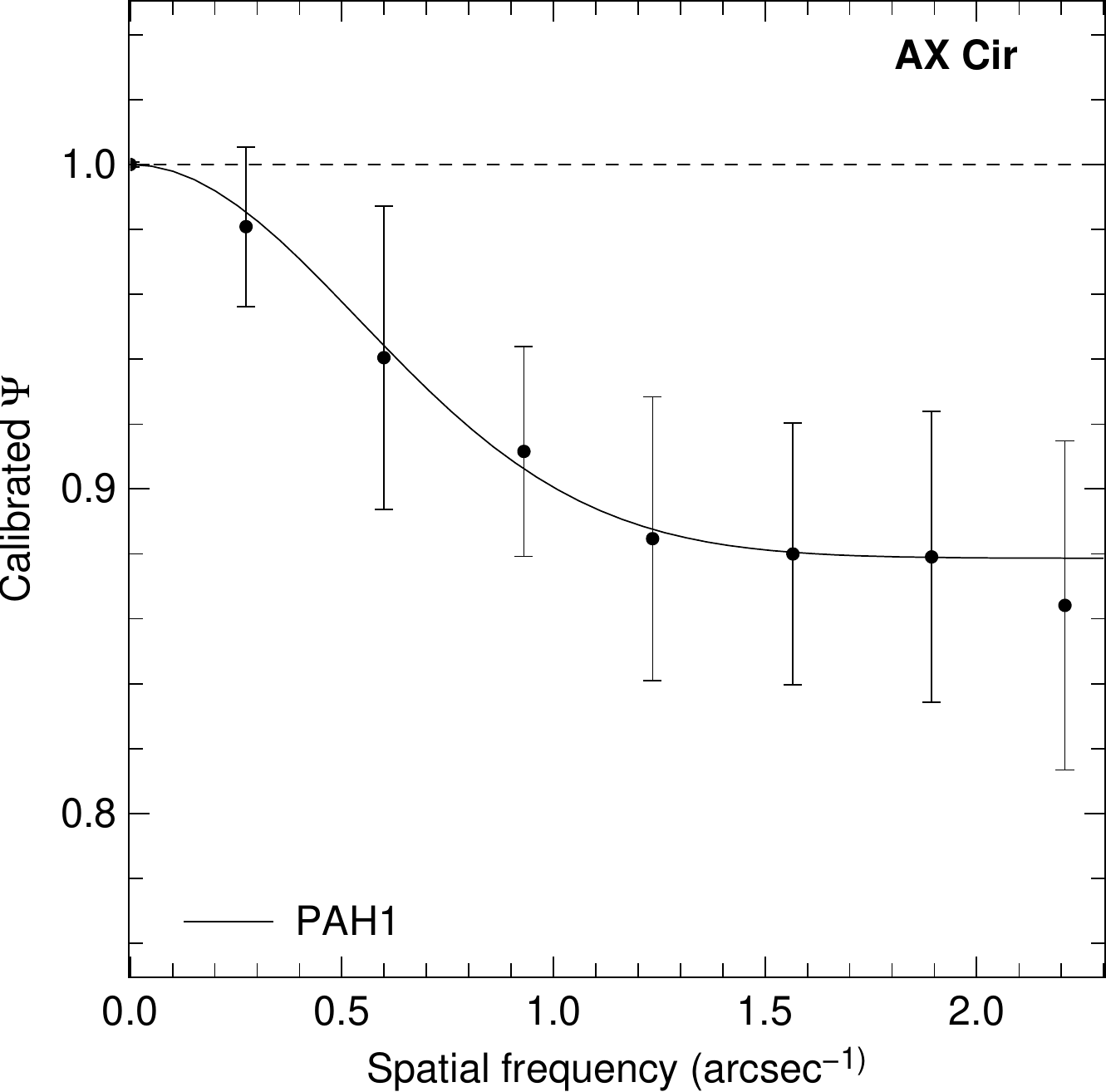}\hspace{.5cm}
\includegraphics[width=8.cm]{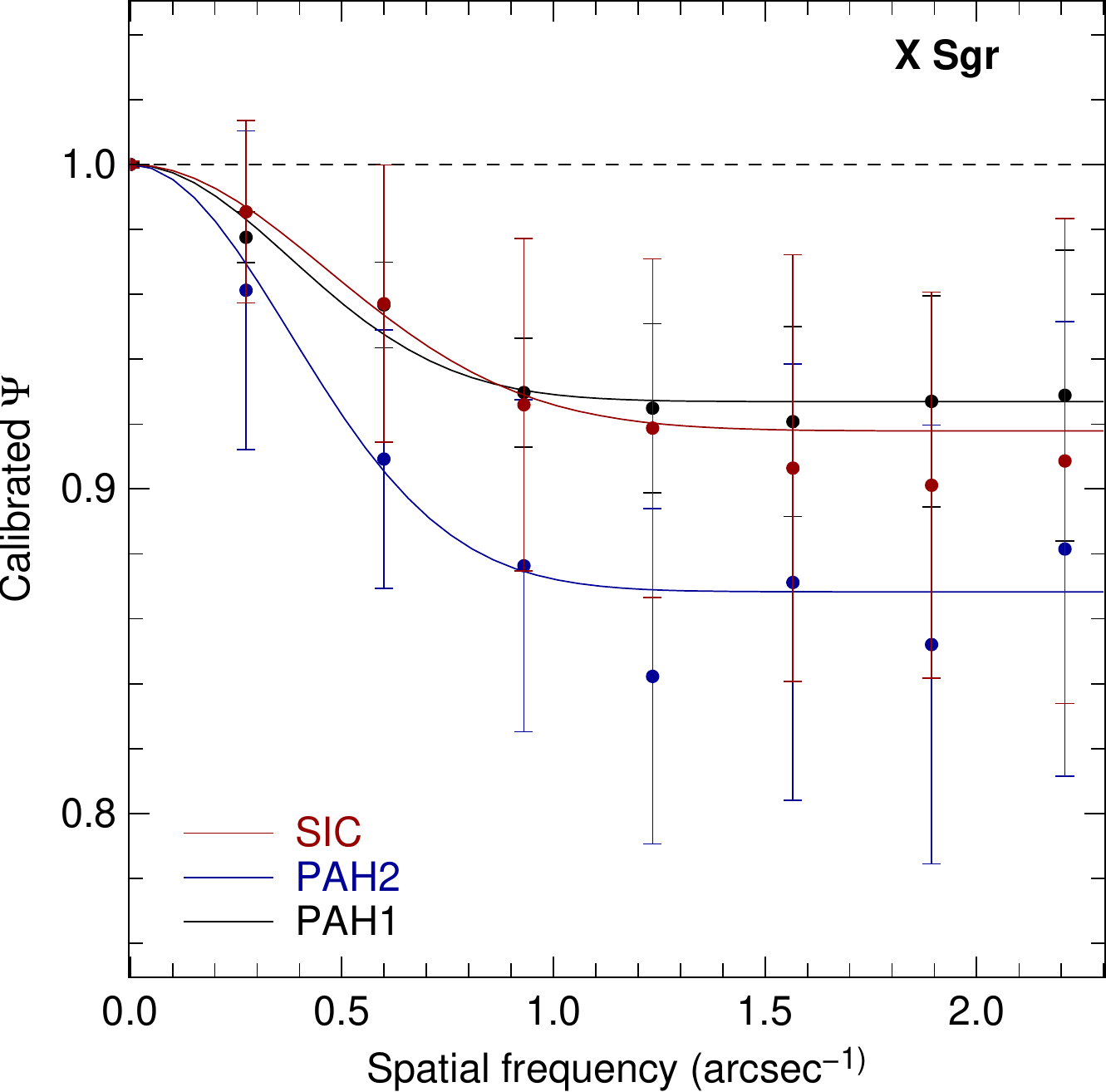}\vspace{.5cm}
\includegraphics[width=8.cm]{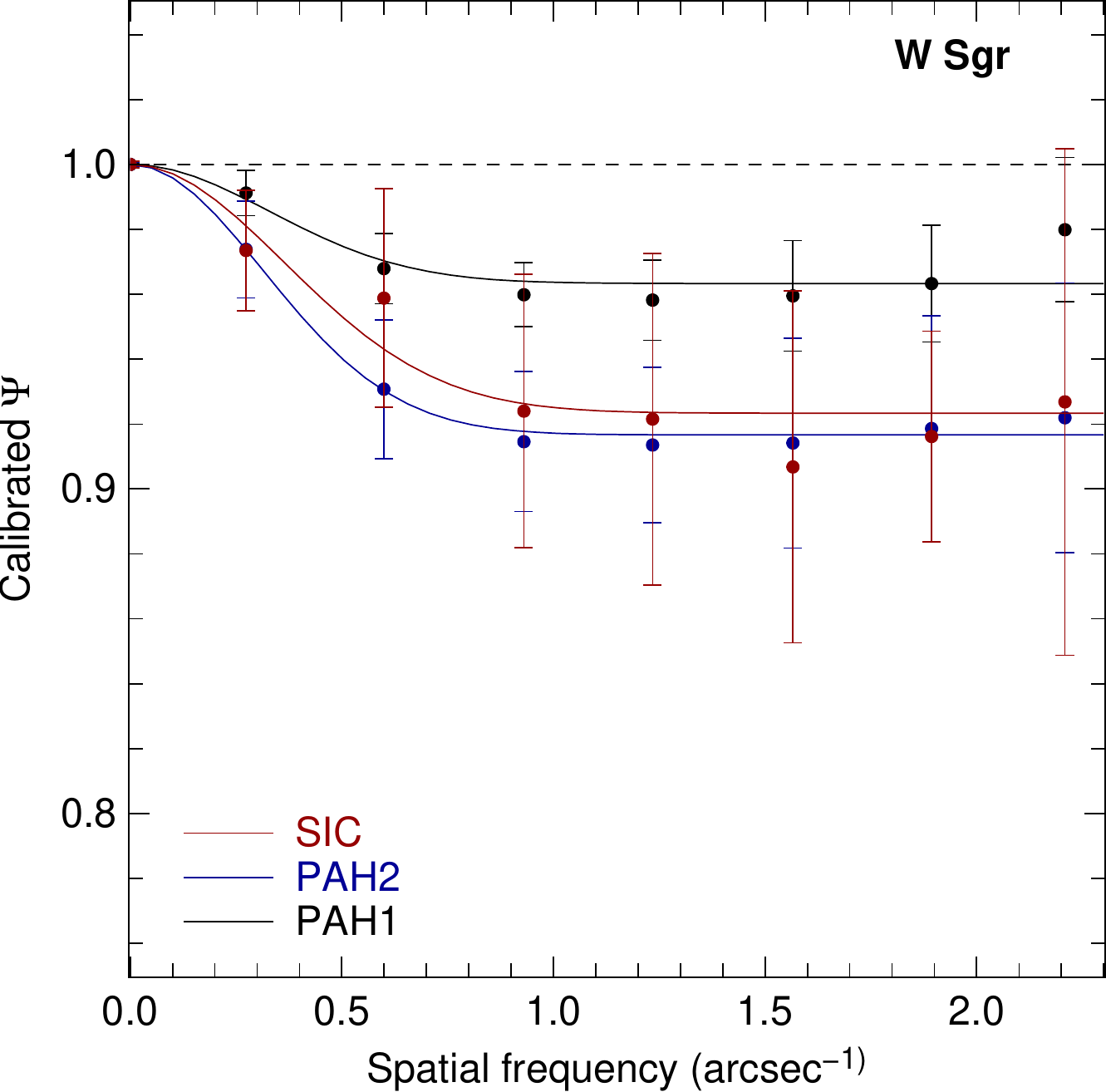}\hspace{.5cm}
\includegraphics[width=8.cm]{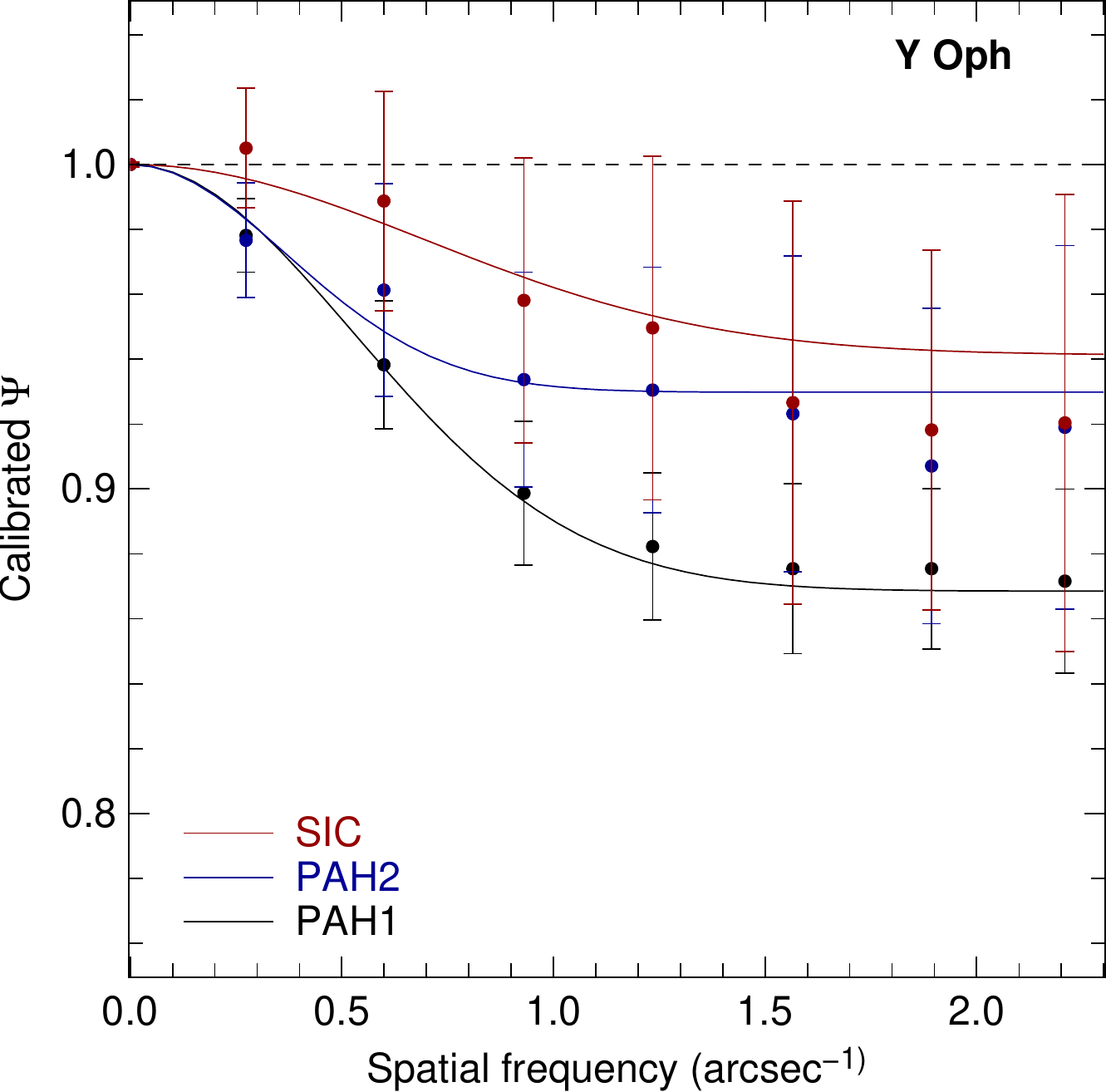}\vspace{.5cm}
\includegraphics[width=8.cm]{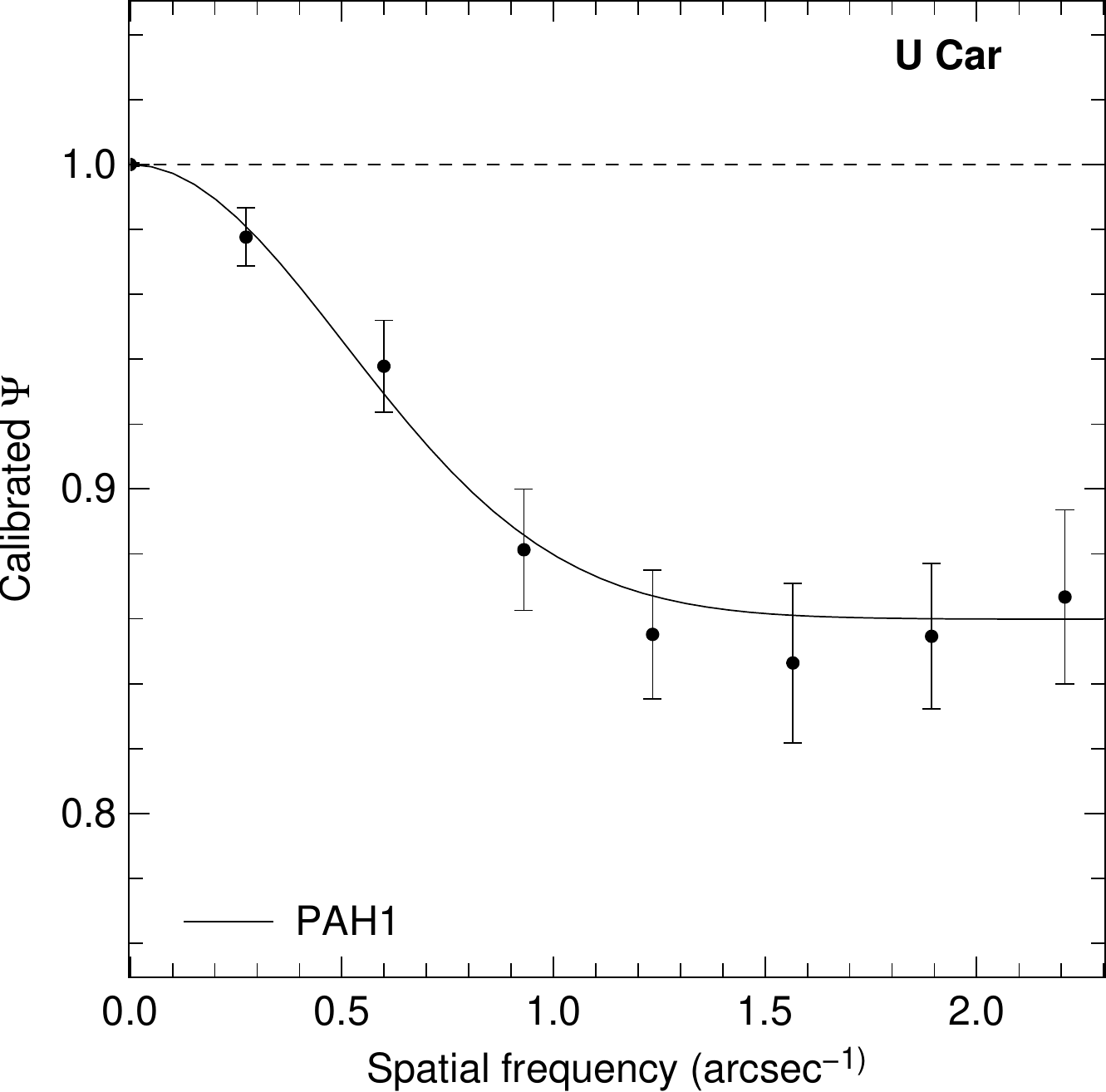}
\caption[Ajustement du modèle de visibilité]{\textbf{Ajustement du modèle de visibilité} : modèle d'étoile non résolue + enveloppe Gaussienne ajusté à la fonction $\Psi (\nu,\rho_\lambda,\alpha_\lambda)$. La courbe en tirets représente la visibilité d'une source non résolue.}
\label{image__fonction_psi}
\end{figure}

Les autres Céphéides quant à elles montrent une composante résolue par le télescope. La fonction $\Psi(\nu,\rho_\lambda,\alpha_\lambda)$ est tracée pour chaque filtre et chaque étoile sur la Fig.~\ref{image__fonction_psi} et les résultats de l'ajustement sont présentés dans la Table~\ref{table__parametre_visibilite_ajuste}. On remarque une perte de la visibilité, représentative de la résolution de l'enveloppe circumstellaire.

À partir de la distance radiale $\rho$ estimée avec cet technique, il est possible d'évaluer une limite minimale pour le tau de perte de masse en raisonnant sur le temps $t$ nécessaire au vent stellaire pour atteindre cette distance. Ce type de raisonnement a déjà été appliqué à la Céphéide $\delta$~Cep par \citet{Marengo-2010-12}. En utilisant la vitesse de libération $v_\mathrm{lib} \sim 100\,\mathrm{km\,s^{-1}}$ \citep{Welch-1988-06} comme vitesse minimale du vent, j'ai trouvé $t = \rho/v_\mathrm{lib} \sim 12$--$23$\,ans. Avec la masse totale de l'enveloppe évaluée dans la Sect.\ref{subsection__distribution_spectrale_energie_des_cepheides_classiques}, j'ai estimé un tau de perte de masse minimale $\dot{M} \sim M_\mathrm{env}/t \sim 7\times10^{-11}\,\mathrm{M_\odot\,yr^{-1}}$ pour AX~Cir et W~Sgr, et $\dot{M} \sim 2\times10^{-10}\,\mathrm{M_\odot\,yr^{-1}}$ pour X~Sgr, Y~Oph et U~Car. Ces valeurs sont comparables à celles prédites par \citet[][variant de $10^{-10}$ à $10^{-7}\,\mathrm{M_\odot\,yr^{-1}}$]{Neilson-2008-09} et à celles mesurées par \citet[][allant de $10^{-10}$ à $10^{-6}\,\mathrm{M_\odot\,yr^{-1}}$]{Deasy-1988-04}.

Enfin, la technique de Fourier donne des valeurs d'excès infrarouge qui sont similaires à celles évaluées à partir des SED présentées dans la Sect.~\ref{subsection__distribution_spectrale_energie_des_cepheides_classiques} (sauf pour AX~Cir). Cette convergence de deux méthodes d'analyses indépendantes nous permet d'avoir une bonne confiance en nos paramètres d'enveloppe que nous avons déterminés.

\begin{table}[!t]
\centering
\begin{tabular}{cccccc} 
\hline
\hline
Nom	  					&	Filtre			& $\rho$ (\arcsec)		& $\alpha$ ($\%$)		\\
\hline
AX~Cir					&  PAH1		&	$0.69 \pm 0.24$		&	$13.8 \pm 2.5$			\\
\hline
X~Sgr					&  PAH1		&	$0.99 \pm 0.23$		&	$7.9 \pm 1.4$			\\
							&  PAH2		&	$0.99 \pm 0.44$		&	$15.2 \pm 3.7$			\\
							&  SiC			&	$0.81 \pm 0.53$		&	$8.9 \pm 3.5$			\\
\hline
W~Sgr					&  PAH1		&	$1.14 \pm 0.39$		&	$3.8 \pm 0.6$			\\
							&  PAH2		&	$1.19 \pm 0.37$		&	$9.1 \pm 1.5$			\\
							&  SiC			&	$1.03 \pm 0.50$		&	$8.3 \pm 2.4$			\\
							\hline
Y~Oph					&  PAH1		&	$0.71 \pm 0.12$		&	$15.1 \pm 1.4$			\\
							&  PAH2		&	$1.02 \pm 0.52$		&	$7.5 \pm 2.3$			\\
							&  SiC			&	$0.54 \pm 0.46$		&	$6.2 \pm 4.1$			\\
\hline	
U~Car					&  PAH1		&	$0.74 \pm 0.10$		&	$16.3 \pm 1.4$			\\
\hline
\end{tabular}
\caption[Résultats de l'ajustement de la visibilité]{\textbf{Résultats de l'ajustement de la visibilité} : $\rho$ et $\alpha$ sont les paramètres de la fonction $\Psi (\nu,\rho_\lambda,\alpha_\lambda)$ et représentent respectivement la largeur à mi-hauteur et le rapport de flux entre l'enveloppe circumstellaire et la photosphère de l'étoile.}
\label{table__parametre_visibilite_ajuste}
\end{table}

\subsection{Discussion}

Les Céphéides de type I et II sont deux classes distinctes d'étoiles pulsantes. Les Céphéides classiques sont connues pour être des étoiles de masse intermédiaire avec une période de pulsation régulière, évoluant comme des étoiles post-séquence principale en phase de combustion centrale de l'hélium. Les types II sont généralement associés à des étoiles de plus faible masse, avec une pulsation irrégulière et étant dans une phase postérieure à la combustion centrale de l'hélium. Par conséquent les enveloppes entourant ces deux classes de Céphéides correspondent à des stades d'évolution différents.

R~Sct et AC~Her ont été intensivement observées et leur fort excès infrarouge, lié à la présence de matière circumstellaire, est interprété comme étant une relique d'une forte perte de masse pendant l'ascension de la branche asymptotique des géantes (AGB). Différents modèles ont été utilisés afin de contraindre la morphologie de cette matière environnante, conduisant à des désaccords dans les résultats obtenus. \citet{de-Ruyter-2005-05} ont interprété leur SED en ajustant un modèle de coquille sphérique. Ils ont trouvé un rayon interne $R_\mathrm{in} = 12.5\,\mathrm{AU}$ et un rayon externe $R_\mathrm{ext} = 224\,\mathrm{AU}$ pour AC~Her et $R_\mathrm{in} = 13.4\,\mathrm{AU}$ et $R_\mathrm{ext} = 5200\,\mathrm{AU}$ pour R~Sct. En utilisant un modèle de disque pour l'étoile AC~Her, \citet{Gielen-2007-11} ont estimé $R_\mathrm{in} = 35\,\mathrm{AU}$ et un rayon externe plus petit, $R_\mathrm{ext} = 300\,\mathrm{AU}$. Pour cette étoile, ces résultats sont en contradiction avec \citet{Close-2003-11} qui ont exclu une émission étendue plus grande que $75\,\mathrm{AU}$, à partir d'observations avec optique adaptative à haut rapport de Strehl. Nous arrivons à la même conclusion avec nos images \emph{VISIR}, uniquement limitées par la diffraction, où nous n'avons pas détecté d'émission étendue plus grande que $R = 100\,\mathrm{AU}$.

Le cas R~Sct est un peu particulier et cette Céphéide est souvent classée comme une exception par de nombreux auteurs. \citet{Alcolea-1991-05} ont utilisé un modèle à deux coquilles sphériques pour interpréter les données en infrarouge moyen et lointain. Ils ont estimé une température pour les plus gros grains de $815\,\mathrm{K}$ ainsi que des rayons $R_\mathrm{in,1} = 30\,\mathrm{AU}$, $R_\mathrm{ext,1} = R_\mathrm{in,2} = 5\,720\,\mathrm{AU}$ et $R_\mathrm{ext,2} = 12\,000\,\mathrm{AU}$ (en utilisant $d = 431\,\mathrm{pc}$ et où les indices $1$ et $2$ représentent respectivement la coquille interne et externe). Peut-être avons nous détecté l'émission de la coquille interne avec nos mesures photométriques à $10\,\mu\mathrm{m}$, en revanche nous n'avons pas détecté d'émission étendue plus grande que $R = 100\,\mathrm{AU}$.

$\kappa$~Pav quant à elle a également été récemment classée par certains auteurs comme une type II particulière à cause de sa courbe de lumière distincte et d'une plus grande brillance en comparaison des autres Céphéides de la même classe avec une même période \citep[voir par exemple][]{Matsunaga-2009-08}. Cette étoile n'est pas au même stade d'évolution que R~Sct et AC~Her puisque dans le diagramme H--R, elle se dirige de la branche horizontale vers la branche asymptotique des géantes. Pendant cette phase, l'étoile subit de nombreux changements, à la fois en son c\oe ur et dans ses couches externes, pouvant mener à une perte de masse via le mécanisme de pulsation et/ou des ondes de chocs. Une active perte de masse passée ou en cours pourrait expliquer l'important excès infrarouge que j'ai détecté.

En ce qui concerne les Céphéides classiques, nous sommes à une échelle spatiale différente des étoiles précédentes puisque les enveloppes détectées jusqu'à présent ont une taille de seulement quelques rayons stellaires. Seules quelques Céphéides sont connues pour posséder une enveloppe circumstellaire \citep[$\ell$~Car, Y~Oph, RS~Pup, Polaris, $\delta$~Cep, S~Mus, GH~Lup, T~Mon et X~Cyg,][]{Kervella-2006-03,Merand-2006-07,Merand-2007-08,Barmby-2010-11}. L'étude effectuée grâce aux données \emph{VISIR} augmente la taille de l'échantillon, avec FF~Aql, $\eta$~Aql, W~Sgr, U~Car et SV~Vul. L'origine de ces enveloppes n'est pas bien comprise à ce jour, leur présence pourrait être liée à une perte de masse de l'étoile. \citet{Kervella-2006-03} ont suggéré que l'excès IR pourrait être dû aux poussières formées par les vents radiatifs. À partir d'un modèle de vents radiatifs, incluant l'effet de la pulsation et des chocs, \citet{Neilson-2008-09} ont conclut que la force radiative n'est pas suffisante à elle seule pour expliquer l'excès observé. Ces auteurs proposent que la perte de masse pourrait également provenir des chocs générés dans l'atmosphère de l'étoile par sa pulsation.

\begin{figure}[!t]
\centering
\includegraphics[width=.7\linewidth]{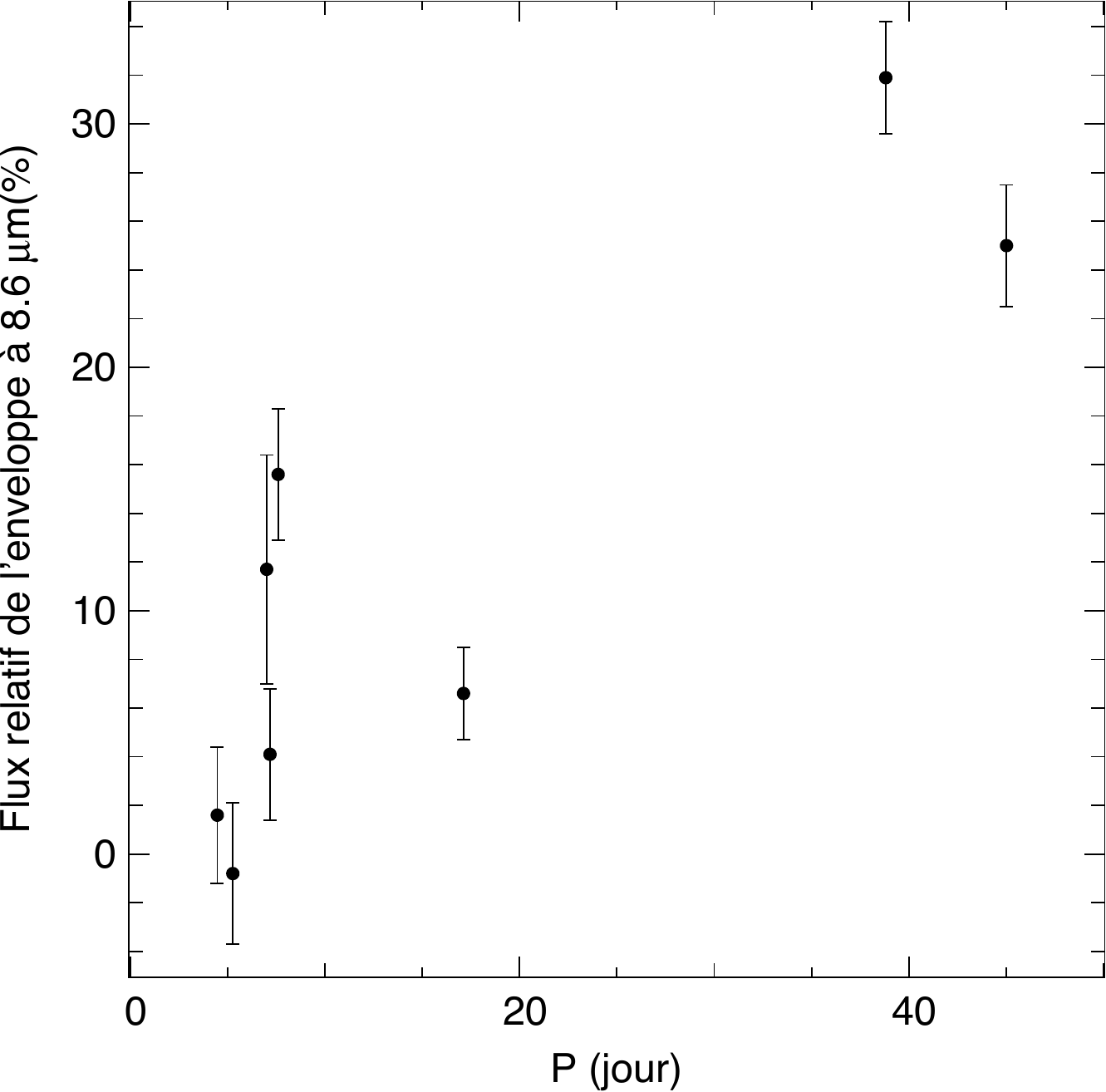}
\caption[Flux relatif de l'enveloppe en fonction de $P$]{\textbf{Flux relatif de l'enveloppe en fonction de $P$} :  valeurs moyennes des excès mesurés lors de la photométrie à $8.6\,\mu\mathrm{m}$ (Table~\ref{table__irradiance_mesure_I}).}
\label{image__flux relatif}
\end{figure}

\citet{Merand-2007-08} ont montré pour des Céphéides classiques une probable corrélation entre la période de pulsation de l'étoile et le flux relatif de l'enveloppe en bande $K$ (relatif à la photosphère). J'ai effectué la même analyse à $8.6\,\mu\mathrm{m}$ (PAH1), et comme le montre la Fig.~\ref{image__flux relatif}, je trouve également une corrélation à cette longueur d'onde. Il semble que les Céphéides de longue période ont un plus grand excès, comme conclu par \citet{Merand-2007-08} en bande $K$. En supposant que cet excès soit lié à une phénomène de perte de masse, cette corrélation montre que les Céphéides de longue période ont une plus grande perte de masse que les courtes périodes, moins massives. Un tel comportement pourrait être expliqué par un champ des vitesses plus important pour les longues périodes, et la présence d'ondes de choc à certaines phases de pulsation \citep{Nardetto-2006-07,Nardetto-2008-10-1}.

L'excès estimé à $8.6\,\mu\mathrm{m}$ varie de 2\,\% à 30\,\%, soit un biais sur la magnitude absolue de 0.03 à 0.4\,mag. Cela implique par exemple que la relation P--L déterminée par \citet{Ngeow-2010-09} à $8\,\mu\mathrm{m}$ pourrait être biaisée. En plus de l'impact sur l'évolution des Céphéides, l'existence de ces enveloppes peut donc avoir un impact sur la mesure de distance des Céphéides dans le domaine infrarouge, en particulier sur les futures mesures avec le JWST.

\cleardoublepage     

\pagestyle{fancy}
\fancyhf{}
\lhead[\nouppercase{\emph{\thepage}}]{\nouppercase{\emph{\rightmark}}}
\rhead[\nouppercase{\emph{\leftmark}}]{\nouppercase{\emph{\thepage}}}
\newpage

\chapter[La très haute résolution angulaire: interférométrie à longue base]{\emph{La très haute résolution angulaire: interférométrie à longue base}}
\label{chapitre__acces_a_la_haute_resolution_angulaire_interferometrie}

\thispagestyle{empty}

\vspace*{-1cm}

\refbleu
\textcolor{bleu_chapitre}{\minitoc}
\refnoir

\section{Introduction}

\malettrine{A}{}u début du XVIIIe siècle, le médecin, égyptologue et physicien britannique Thomas Young découvrit le phénomène d'interférences en plaçant devant une source lumineuse un masque percé de deux petits trous identiques. Il observa sur un écran une alternance de minima et maxima de brillance, le tout contenu dans le profil d'intensité donné par un seul trou. Cette expérience connue maintenant sous le nom d'expérience des trous d'Young a donné naissance à la théorie ondulatoire de la lumière. En 1868, le physicien français Hippolyte Fizeau, déjà reconnu, entre autres pour l'effet Doppler-Fizeau, constata que le contraste de ces minima et maxima, appelés franges d'interférences, diminuait avec l'extension spatiale de la source lumineuse et trouva une relation liant l'écart $d$ entre les deux trous, le diamètre angulaire apparent $\theta$ de la source et le contraste des franges. Il proposa par la suite d'utiliser cette loi pour la mesure de diamètres stellaires. Albert Michelson appliqua cette idée en expérimentant un interféromètre sur le télescope Hooker du Mont Wilson, qui permit en 1920 de mesurer le diamètre angulaire de Bételgeuse.

Au fil du temps, la construction de plus grands collecteurs de photons était indispensable et l'avènement de miroirs de diamètre $D = 8$--$10\,\mathrm{m}$ a permis d'avoir un gain considérable en terme de résolution spatiale et de sensibilité. Cependant, construire de plus grands télescopes pour atteindre une plus haute résolution angulaire est à la fois cher et ambitieux. L'idée est alors d'étendre le principe d'interférométrie des trous d'Young en utilisant deux ou plusieurs télescopes de taille raisonnable, séparés d'une distance $B \gg D$. La résolution est bien meilleure  car elle est maintenant inversement proportionnelle à la base $B$. La première recombinaison réussie est effectuée en 1974 par A. Labeyrie avec l'Interféromètre à 2 Télescopes (I2T) dont la base était de $12\,\mathrm{m}$.

Des progrès technologiques énormes ont été effectués depuis 37 ans et nous disposons aujourd'hui de plusieurs interféromètres longues bases (\emph{VLTI, CHARA, KECK, SUSI, ...}) incluant divers instruments (\emph{FLUOR, MIDI, AMBER, ...}). Il est maintenant possible, via les mesures de franges d'interférences, d'estimer directement un diamètre stellaire (donnant ensuite accès à la température effective), de déterminer l'orbite d'étoiles doubles (donnant ensuite accès aux masses), ou encore de mesurer la variation de diamètre angulaire (donnant accès à la distance). De plus, nous pouvons désormais grâce à la spectro-interférométrie qui disperse les franges dans un intervalle de longueur d'onde, combiner à la fois l'information spectrale et spatiale.

Toutefois, l'interférométrie est également sujette à la turbulence atmosphérique et la recombinaison cohérente de faisceaux provenant de divers télescopes n'est pas une tâche facile. Diverses solutions existent pour essayer de s'affranchir de ces perturbations. On peut par exemple utiliser l'optique adaptative (OA) pour corriger le front d'onde. Les caractéristiques d'une OA ont déjà été exposées dans le Chapitre~\ref{chapitre__imagerie_a_haute_resolution_spatiale_optique_adaptative_et_lucky_imaging} et je n'en reparlerai pas ici. Un autre système très répandu sur les interféromètres est le suiveur de franges. Il permet de corriger des variations du piston atmosphérique (différence de marche aléatoire entre les télescopes) qui ne sont pas corrigées par l'OA. Les variations du piston font se déplacer les franges d'interférences sur le détecteur, ce qui a pour conséquence la dégradation du contraste. La fonction du suiveur de franges est donc de mesurer le déplacement des franges pour ensuite le compenser. Une autre possibilité est d'effectuer des temps de pose très court afin de "geler" les perturbations atmosphériques. Notons toutefois que ce dernier ne corrige pas des variations de piston. La combinaison des ces trois solutions peut être utilisée pour améliorer la stabilité des franges d'interférences. Un autre possibilité consiste à injecter directement la lumière arrivant des télescopes dans une fibre monomode dont la fonction est de transformer les perturbations du front d'onde en variations d'intensité. Seul les variations du piston ne sont pas corrigées.

Dans le cadre de ma thèse, j'ai participé à plusieurs semaines d'observations avec l'instrument \emph{CHARA/FLUOR} \citep[Fiber Linked Unit for Optical Recombination,][]{Coude-du-Foresto-2003-02}, dans le cadre d'un long programme d'étude d'étoiles Céphéides. Le premier objectif de ce programme est de mesurer les variations de diamètre angulaire de Céphéides afin d'appliquer la méthode de Baade-Wesselink présentée au Chapitre~\ref{chapitre__mesurer_l_univers_les_cephéides}. Le second objectif concerne la détection d'enveloppe sur des mesures à courtes bases. Ce programme d'observation s'est terminé en août 2011 et les données sont donc en cours d'analyse. Je présenterai tout de même les premiers résultats qui n'ont pas encore été publiés.

Je présente dans un premier temps, d'une manière non exhaustive, les notions de base d'interférences, quelques modèles utilisés et le principe d'étalonnage. Je rentrerai ensuite dans le vif du sujet par une description de l'instrument \emph{FLUOR}, de l'acquisition des données puis de la méthode de réduction. Pour finir, je présenterai quelques résultats intéressants issus des premières analyses des Céphéides classiques Y~Oph, U~Vul et S~Sge et de la Céphéides de type II R~Sct.

\section{Bases d'interférométrie stellaire}

L'interférométrie est basée sur la nature ondulatoire de la lumière et nous parlerons donc non plus en terme de photons mais en terme d'onde lumineuse. Cette section est basée sur les notes de cours de \citet{Lawson-1999-03}.

\subsection{Point source monochromatique}

\begin{figure}[!p]
\centering
\resizebox{\hsize}{!}{\includegraphics[width = .65\linewidth]{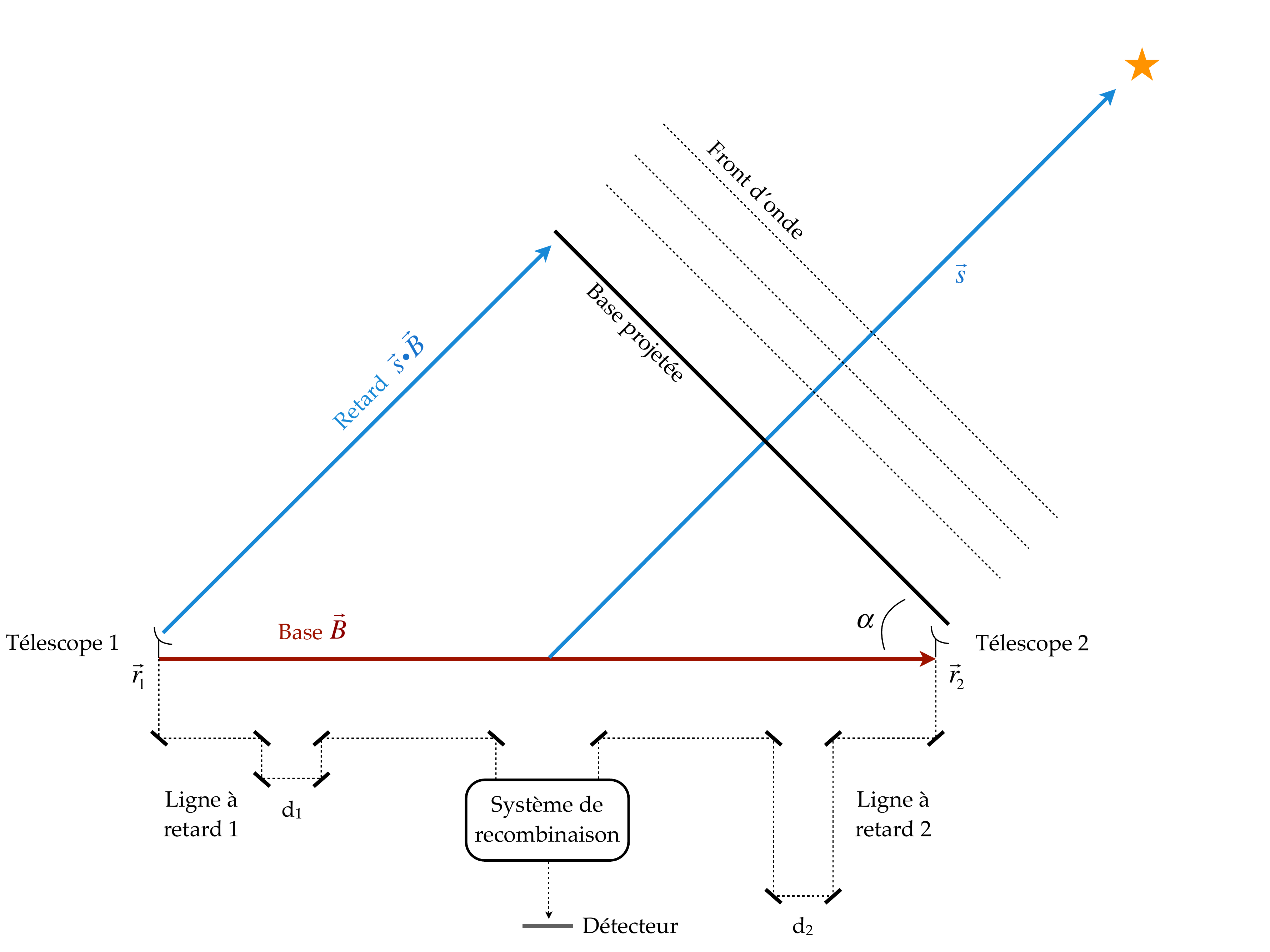}}
\caption[Schéma d'un interféromètre à deux télescopes]{\textbf{Schéma d'un interféromètre à deux télescopes.}}
\label{image__schema_interferometre}
\end{figure}

Considérons le schéma de la Fig.~\ref{image__schema_interferometre} représentant un interféromètre à deux télescopes, séparés d'une distance $\vec{B}=\vec{r_2}-\vec{r_1}$, sur lequel arrive un front d'onde d'une source considérée comme ponctuelle et monochromatique. Dans le cadre de ce manuscrit, cette source sera une étoile et plus précisément une Céphéide située à l'infinie dont on pourra considérer le front d'onde comme plan à son entrée dans l'atmosphère terrestre. Les ondes provenant de chaque télescope observées en un point $(x,y)$ peuvent donc s'écrire :
\begin{displaymath}
S_1(x,y) = A_1(x,y)\,\mathrm{e}^{i(\omega t - \vec{k}.\vec{r}_1 - kd_1)} \qquad \mathrm{et} \qquad S_2(x,y) = A_2(x,y)\,\mathrm{e}^{i(\omega t - \vec{k}.\vec{r}_2 - kd_2)}
\end{displaymath}
où $\omega = 2\pi c/\lambda_0$, $\vec{k} = 2\pi\,\vec{s}/\lambda_0$, ($x,y$) représentent les coordonnées spatiales sur le détecteur, $d_1$ et $d_2$ sont les retards optiques des lignes 1 et 2 (Fig.~\ref{image__schema_interferometre}) et $A_\mathrm{i}$ correspond à l'amplitude de l'onde $i$ (avec $i = 1, 2$).

L'intensité observée sur le détecteur après la recombinaison est le carré de la superposition de ces deux ondes, soit :
\begin{displaymath}
I_\lambda(x,y) = (S_1 + S_2)(S_1^* + S_2^*) = A_1^2 + A_2^2 + 2\,A_1\,A_2 \cos(\vec{k}.\vec{B} + k(d_2-d_1))
\end{displaymath}
où les variables ($x,y$) ont été omises pour la clarté.

En prenant des amplitudes identiques $A_1^2 = A_2^2 = I_0 = \mathrm{cst}$, l'expression précédente devient :
\begin{equation}
 I_\lambda(x,y) = 2\,I_0\,[1 + \cos(kD)] \qquad \mathrm{avec} \qquad D = \vec{s} \cdot \vec{B} + d_2 - d_1
 \label{equation__intensite_monochromatique}
 \end{equation}
 
\begin{figure}[!p]
	\begin{minipage}[h]{.5\linewidth}
  		\centering\includegraphics[width = \linewidth]{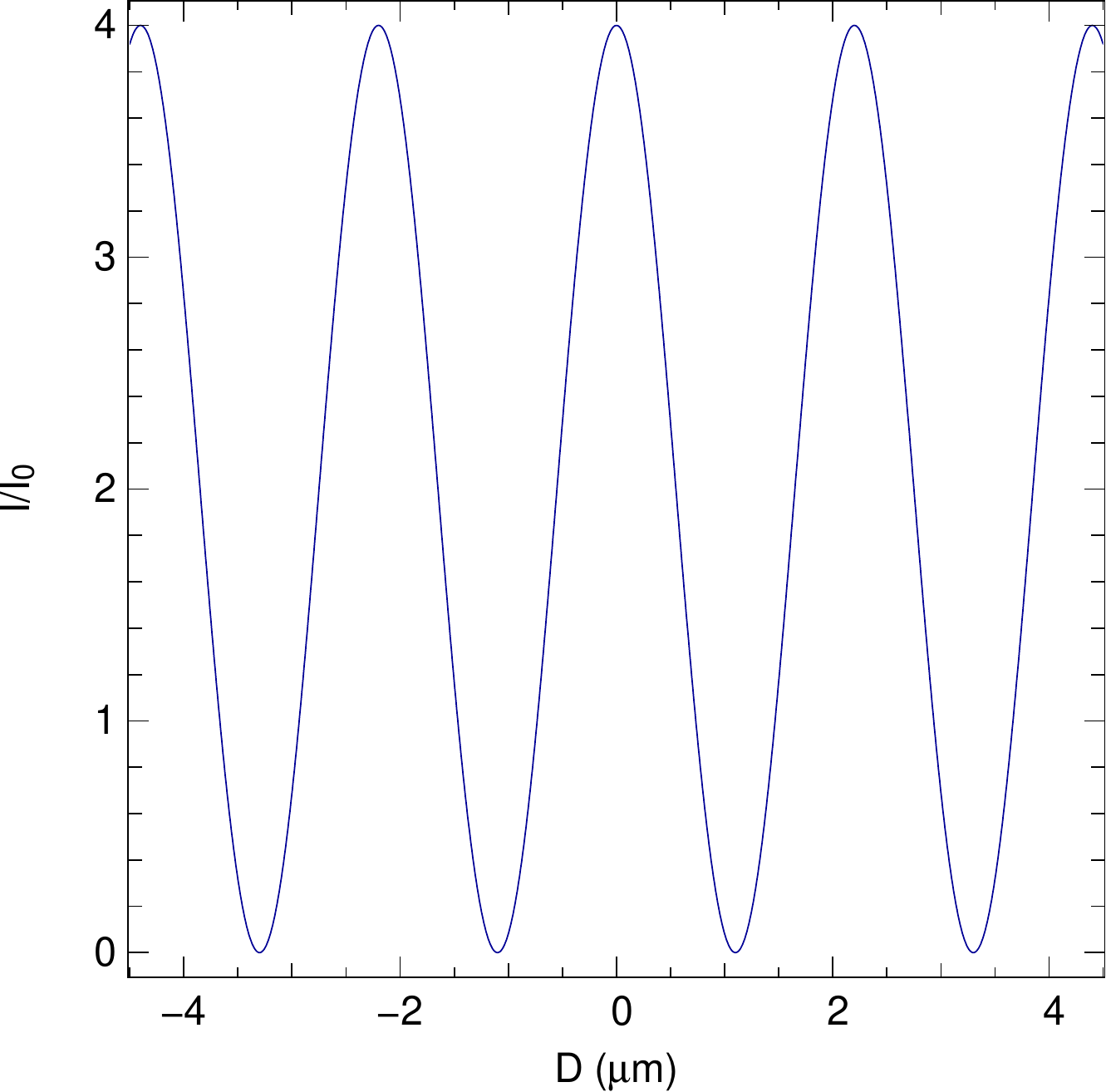}
	\end{minipage}
	\hfill
	\hspace{.2cm}
	\begin{minipage}[h]{.5\linewidth}
		\centering\includegraphics[width = \linewidth]{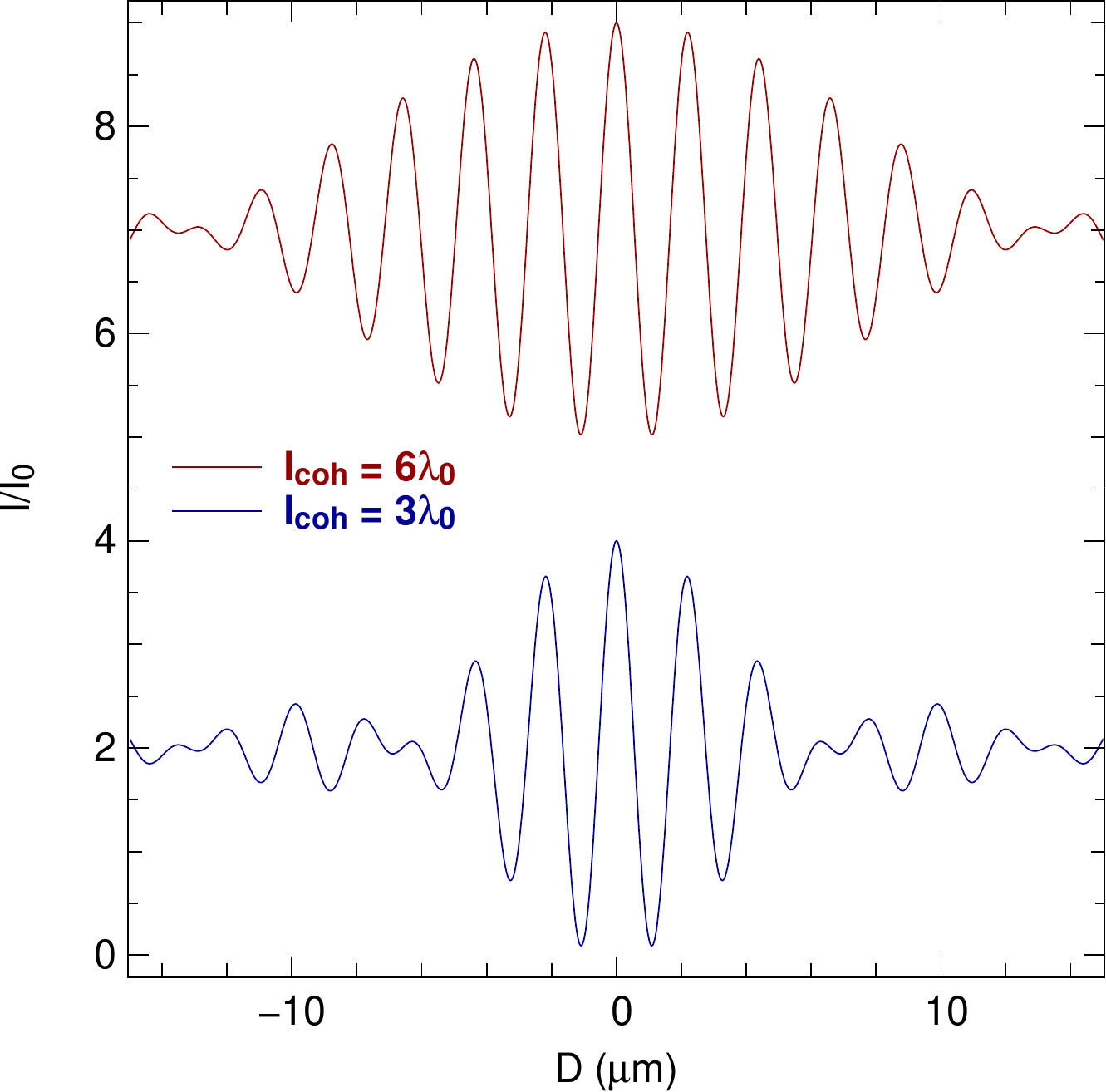}
	\end{minipage}
	\caption[Intensité monochromatique et polychromatique]{\textbf{Intensité monochromatique et polychromatique} : à gauche, intensité monochromatique pour $\lambda = 2.2\,\mu\mathrm{m}$ à une dimension. À droite, intensité polychromatique à une dimension pour $\lambda = 2.2\,\mu\mathrm{m}$ et $l_\mathrm{coh}=3\lambda_0$ et $l_\mathrm{coh}=6\lambda_0$. L'unité en ordonnée est arbitraire.}
  \label{image__intensite_monochromatique}
\end{figure}

 Cette fonction cosinusoïdale est représenté sur la Fig.~\ref{image__intensite_monochromatique} (à gauche) et est formée de minima et maxima, appelés franges d'interférences. Cette série de franges, dont les maxima sont séparés de $\lambda$, est une fonction du retard optique $D$.

\subsection{Point source polychromatique}

En pratique, nous ne sommes jamais dans ce cas idéal monochromatique, mais nous observons plutôt dans une bande spectrale de largeur $\Delta\lambda$. La conséquence est que l'intensité observée est la somme des intensités à chaque longueur d'onde. Par simplicité, supposons que $I_0$ est achromatique et que les ouvertures sont des fentes (première expérience de Young), l'équation~\ref{equation__intensite_monochromatique} devient :
\begin{displaymath}
I(x,y) = \int_{\lambda_0 - \Delta\lambda/2}^{\lambda_0 + \Delta\lambda/2}I_\lambda(x,y)\,d\lambda = 2I_0\,\Delta\lambda\,\left[1 + \frac{\sin \left(\pi D\dfrac{\Delta\lambda}{\lambda_0^2} \right)}{\pi D\dfrac{\Delta\lambda}{\lambda_0^2}}\,\cos(k_0D) \right]
\end{displaymath}

On remarque que les franges sont maintenant modulées par un sinus cardinal, comme exposé sur la Fig.~\ref{image__intensite_monochromatique}, dont la distance caractéristique, appelée longueur de cohérence, est donnée par :
\begin{displaymath}
l_\mathrm{coh}=\frac{\lambda_0^2}{\Delta\lambda}
\end{displaymath}

Le contraste des franges diminue fortement si $D$ est supérieur à $l_\mathrm{coh}$. En pratique, on contrôle $d_2-d_1$ grâce aux lignes à retard de façon à compenser le retard géométrique $\vec{s}\cdot\vec{B}$. Les faisceaux sont ainsi cohérent temporellement et les franges d'interférences sont alors visibles.

\subsection{Source étendue - Théorème de Zernike-Van Cittert}

Un cas supplémentaire apparaît quand l'étoile n'est plus considérée comme ponctuelle mais comme ayant une certaine extension spatiale $\vec{s} = \vec{s_0} + \vec{\Delta s}$ ($\vec{s_0}$ est le vecteur pointant dans la direction de la source et $\vec{\Delta s}$ un léger décalage). Le retard optique devient $D = \vec{\Delta s}\cdot\vec{B}$ (car $\vec{s_0} \cdot \vec{B} + d_2 - d_1 = 0$ grâce aux lignes à retard). Définissons maintenant l'extension spatiale de la source telle que $\vec{\Delta s} = (\alpha, \delta, 0)$ et replaçons nous dans le cas monochromatique par simplicité (le cas monochromatique introduit une modulation des franges). L'intensité observée peut être représentée par une somme infinie de sources ponctuelles incohérentes :
\begin{displaymath}
I_\lambda(\vec{\Delta s},\vec{B}) = \int I_\lambda(\alpha,\delta)\,[1 + \cos(k\vec{\Delta s}.\vec{B})]\,d\alpha d\delta
\label{equ_source_etendue}
\end{displaymath}

En pratique pour mesurer les franges dans le plan pupille (voir Section~\ref{section__interferometrie_plan_pupille_et_image}), on rajoute une différence de marche $\delta_\mathrm{ddm}$ grâce à une autre ligne à retard sur l'un des deux faisceaux pour reconstruire l'interférogramme. De l'équation précédente on a donc :

\begin{displaymath}
I_\lambda(\vec{\Delta s},\vec{B}) = \int I_\lambda(\alpha ,\delta)\,d\alpha d\delta + \int I_\lambda(\alpha ,\delta)\,\cos(k\vec{\Delta s}.\vec{B}+k\delta_\mathrm{ddm})\,d\alpha d\delta
\end{displaymath}

D'un point de vue mathématique, on peut écrire $\cos(k\vec{\Delta s}\cdot\vec{B}+k\delta_\mathrm{ddm}) = \mathop{\mathrm{Re}}[\mathrm{e}^{-ik\vec{\Delta s}\cdot\vec{B}}\,\mathrm{e}^{-ik\delta_\mathrm{ddm}}]$, soit :
\begin{equation}
I_\lambda(\vec{\Delta s},\vec{B}) = \int I_\lambda(\alpha,\delta)\,d\alpha d\delta + \mathop{\mathrm{Re}}\left[ \int I_\lambda(\alpha,\delta )\,\mathrm{e}^{-ik\vec{\Delta s}\cdot\vec{B}}\, \mathrm{e}^{-ik\delta_\mathrm{ddm}}\,d\alpha d\delta \right]
\label{equation__intensite_1}
\end{equation}

On définit la fonction de visibilité complexe telle que :
\begin{displaymath}
\mu(k,\vec{B}) = \int I_\lambda(\alpha,\delta )\,\mathrm{e}^{-ik\vec{\Delta s}\cdot\vec{B}}\,d\alpha d\delta
\end{displaymath}

En introduisant les variables conjuguées $u = B_\mathrm{x}/\lambda$ et $v = B_\mathrm{y}/\lambda$, appelées fréquences spatiales, et en effectuant une normalisation, on obtient :
\begin{equation}
\mu(u,v) = \frac{\int I_\lambda(\alpha,\delta )\,\mathrm{e}^{ -2\pi i(\alpha u + \beta v) }\,d\alpha d\delta}{\int I_\lambda(\alpha,\delta)\,d\alpha d\delta}
\label{equation__visibilite_complexe}
\end{equation}

Cette fonction correspond à la transformée de Fourier (TF) de la distribution de lumière de la source. Ceci nous amène au \textbf{théorème de Zernike-Van Cittert} : \\

\fcolorbox{rouge}{white}{
\begin{minipage}{.9\textwidth}
Pour une source de rayonnement quasi-monochromatique, la fonction de visibilité complexe est égale à la transformée de fourier spatiale de la distribution d'intensité normalisée à l'intensité totale.
\end{minipage}
} \\

Pour une démonstration plus exhaustive de ce théorème, on pourra se référer par exemple à \citet{Lena-1996-}, ou à \citet{Born-1999-10}.

À la vue de cette équation, il est possible de remonter à la distribution spatiale d'intensité par une simple TF inverse. Il faut pour cela un bon échantillonnage du plan $(u,v)$. L'interférométrie échantillonne donc cette fonction de visibilité en apportant une fréquence spatiale pour chaque mesure.

\begin{figure}[!p]
\centering\includegraphics[width = .7\linewidth]{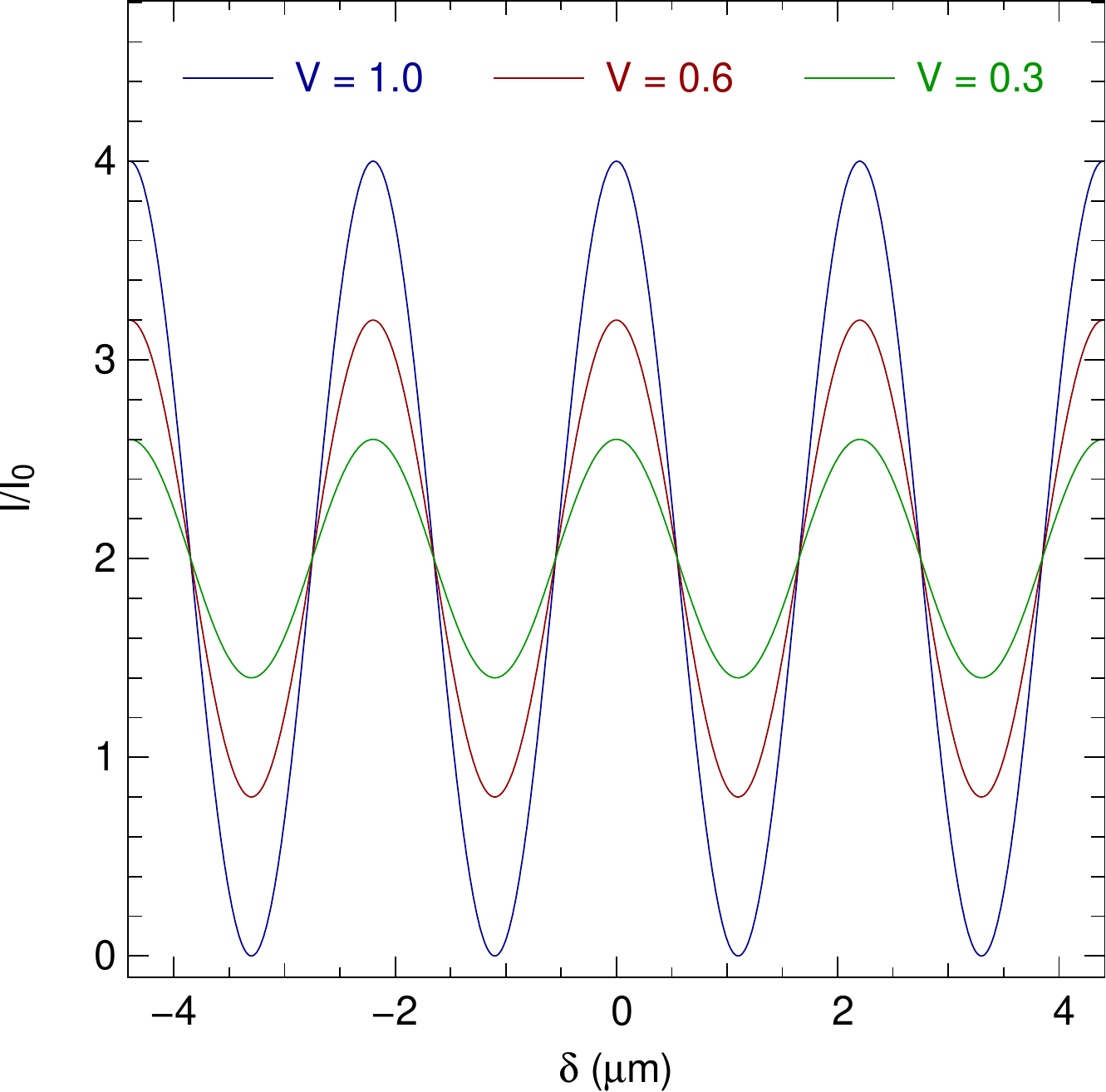}
\caption[Intensité monochromatique observée pour diverses valeurs de visibilité]{\textbf{Intensité monochromatique observée pour diverses valeurs de visibilité} : l'amplitude des franges diminue avec la visibilité.}
\label{image__intensite_visibilite}
\end{figure}

L'intensité mesurée sur le détecteur (Équ.~\ref{equation__intensite_1}) est reliée à la fonction de visibilité par :
\begin{displaymath}
I_\lambda(\vec{\Delta s},\vec{B}) = I_0(1 + \mathop{\mathrm{Re}} [\mu\,\mathrm{e}^{-ik\delta_\mathrm{ddm}}])
\end{displaymath}
où $I_0 = \int I_\lambda(\alpha,\delta)\,d\alpha d\delta$ est l'intensité totale. Le contraste des franges est donc fonction de la taille angulaire de la source et de la séparation entre les deux télescopes. Plus la taille angulaire de la source augmente, plus la visibilité des franges diminue, du fait de la perte de cohérence spatiale. Des exemples pour trois valeurs de visibilité sont exposés sur la Fig.~\ref{image__intensite_visibilite}. On retrouve pour une source ponctuelle, c'est à dire pour $\vec{\Delta s} = 0$, l'équation~\ref{equation__intensite_1} avec un retard optique $\delta_\mathrm{ddm}$.

\section{Interférométrie dans le plan pupille et image}
\label{section__interferometrie_plan_pupille_et_image}

\begin{figure}[!p]
\centering
\resizebox{\hsize}{!}{
\includegraphics{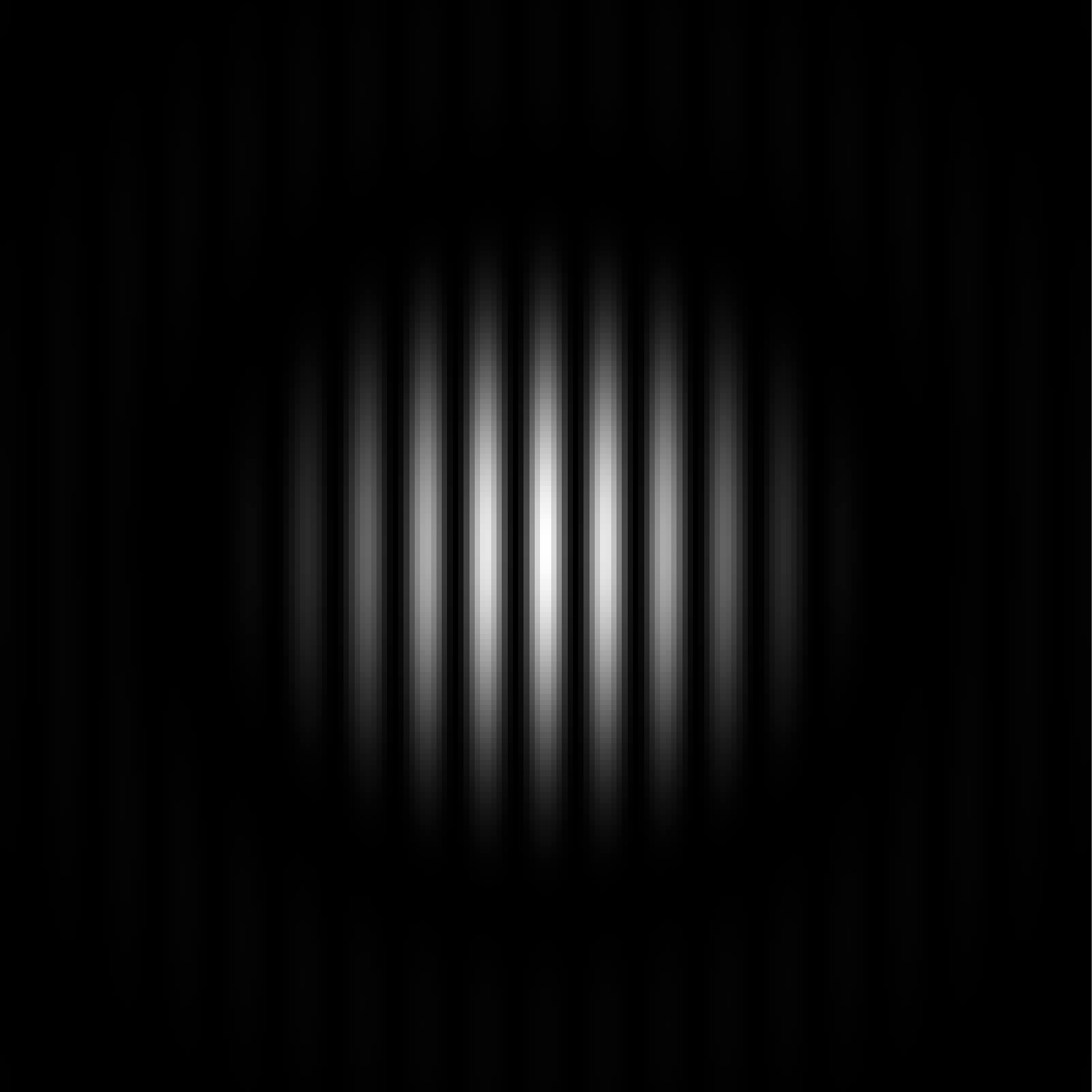}
\includegraphics{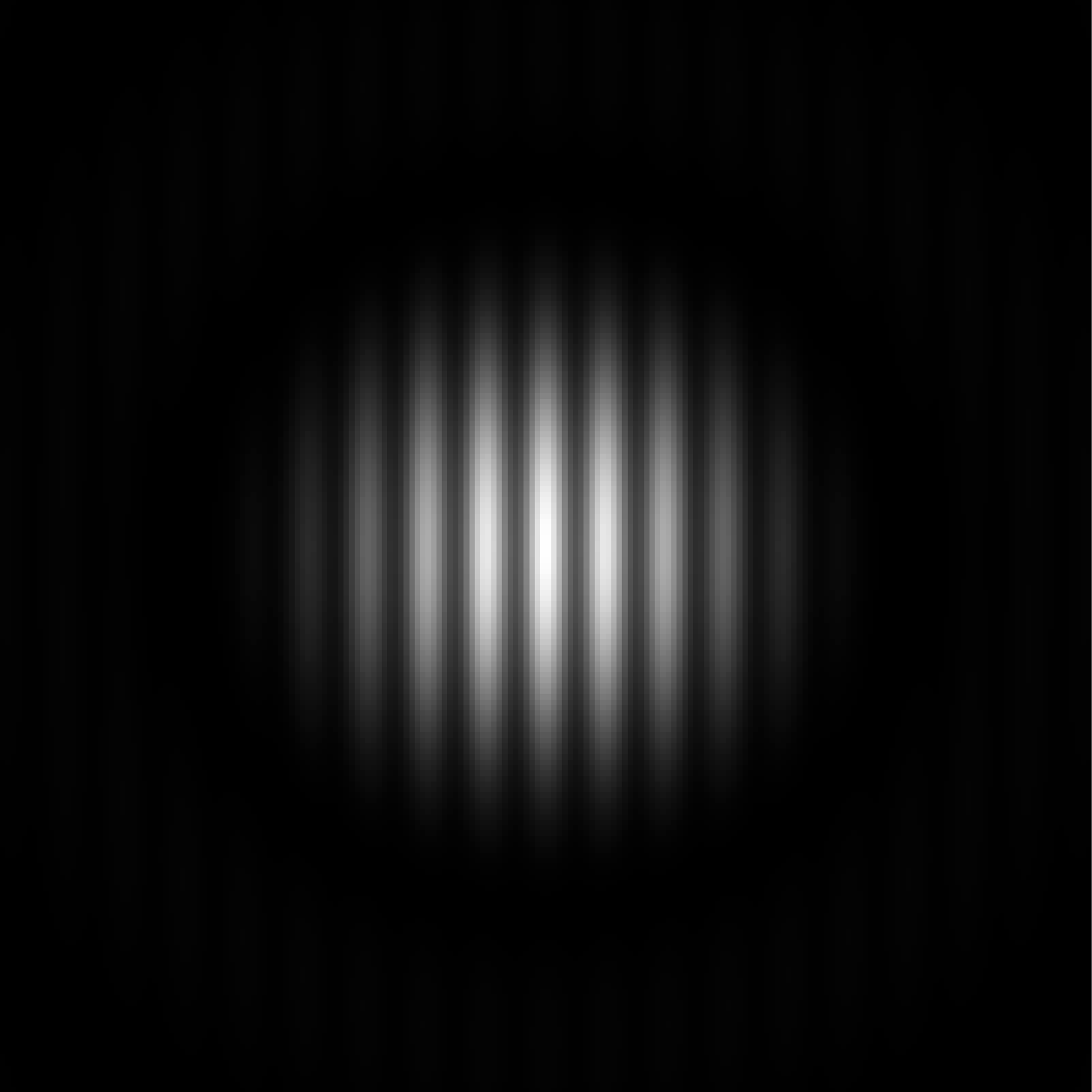}
\includegraphics{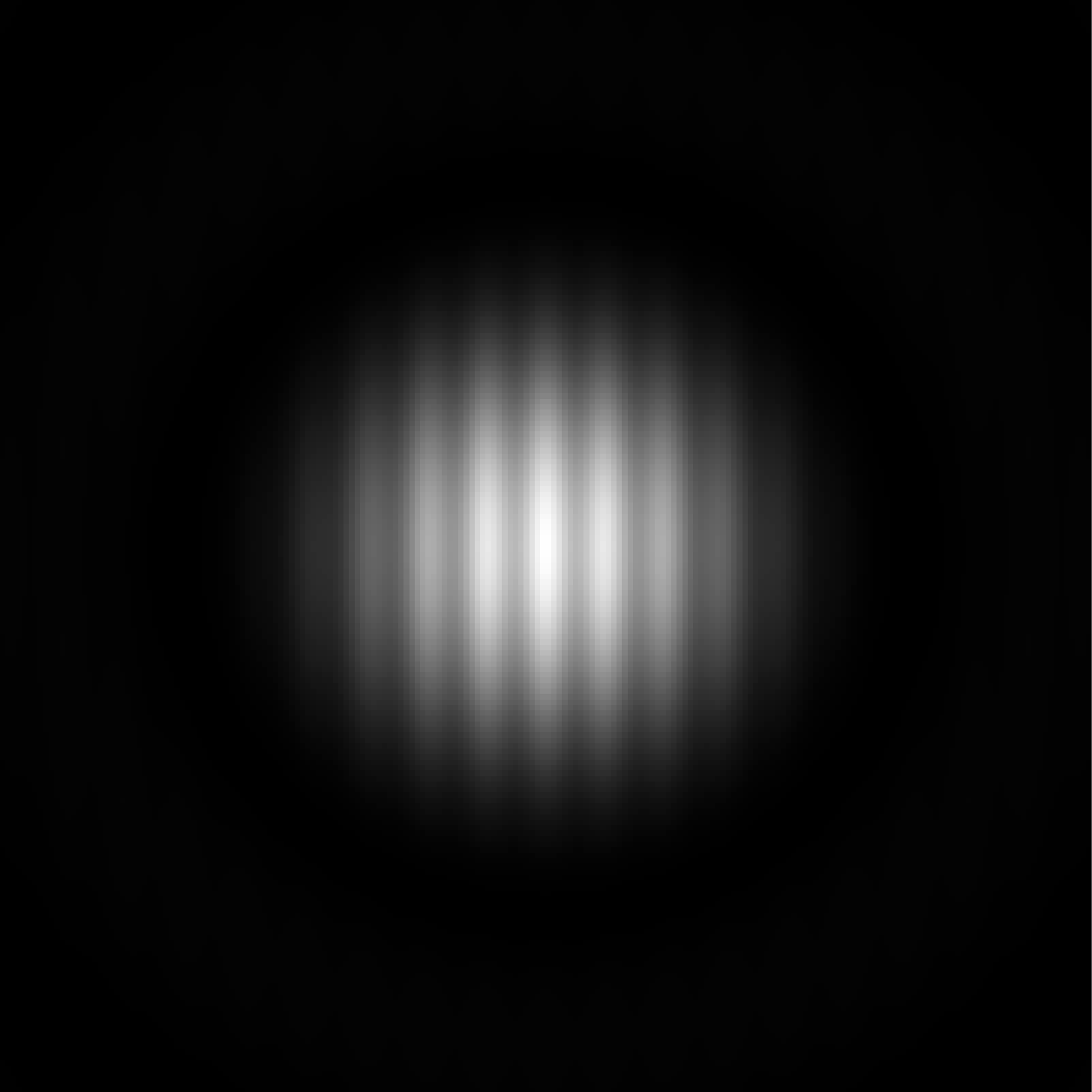}
\includegraphics{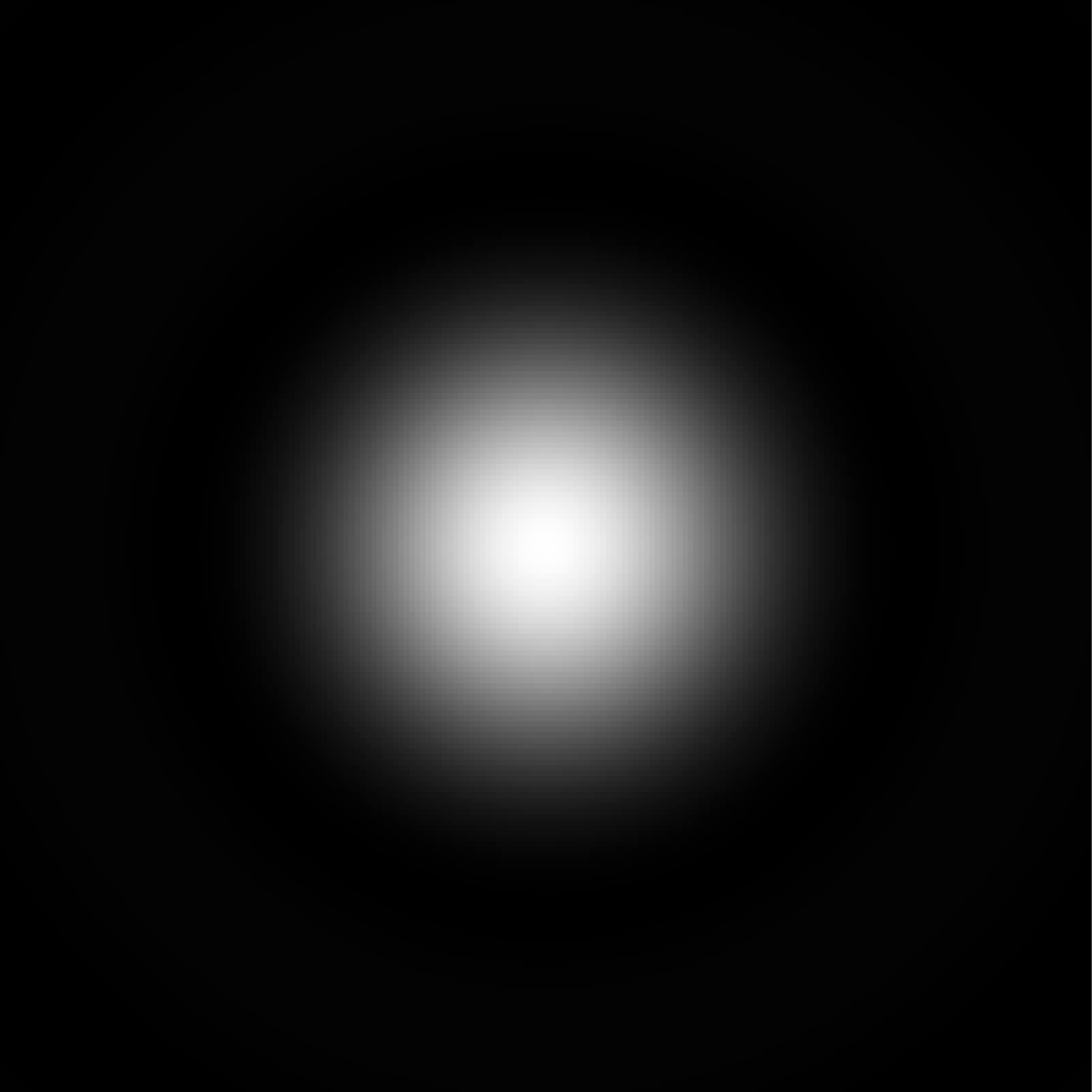}}
\caption[Recombinaison dans le plan image]{\textbf{Recombinaison dans le plan image} : image d'une source au foyer d'un interféromètre. De gauche à droite : $V = 1$ (source non résolue), $V = 0.5$, $V = 0.15$ et $V = 0$.}
\label{image__frange_plan_image}
\end{figure}

Il existe deux manières différentes pour recombiner les faisceaux. Ils peuvent être directement combinés sur le détecteur (plan image), l'image observée est alors une tache d'Airy contenant des franges d'interférences, comme exposé sur la Fig.~\ref{image__frange_plan_image}. Les franges disparaissent (la visibilité diminue) au fur et à mesure que la source est de plus en plus résolue par l'interféromètre.

\begin{figure}[!p]
\begin{minipage}[h]{.245\linewidth}
  \centering\includegraphics[width = \linewidth]{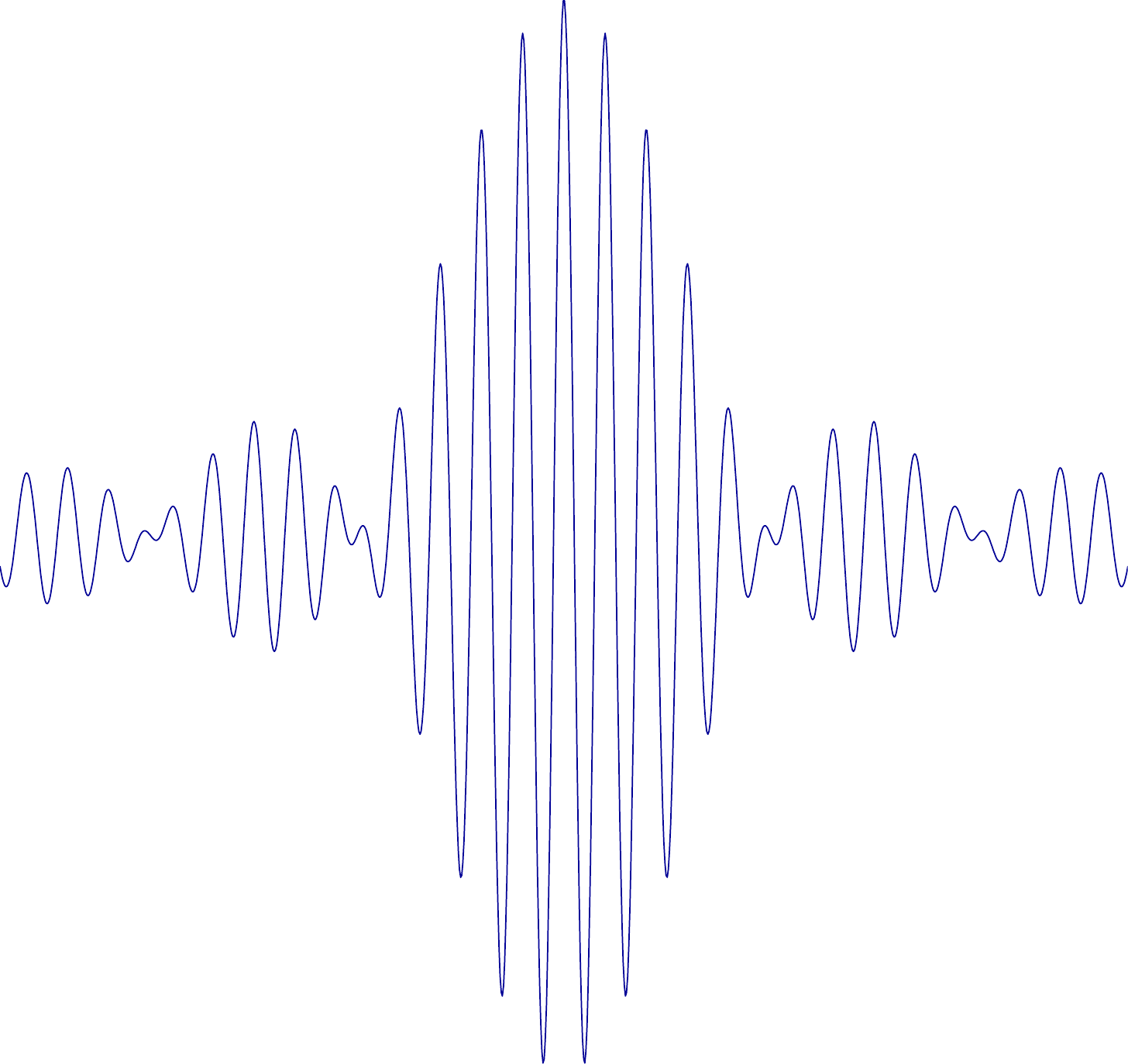}
\end{minipage}
\hfill
\begin{minipage}[h]{.245\linewidth}
  \centering\includegraphics[width = \linewidth]{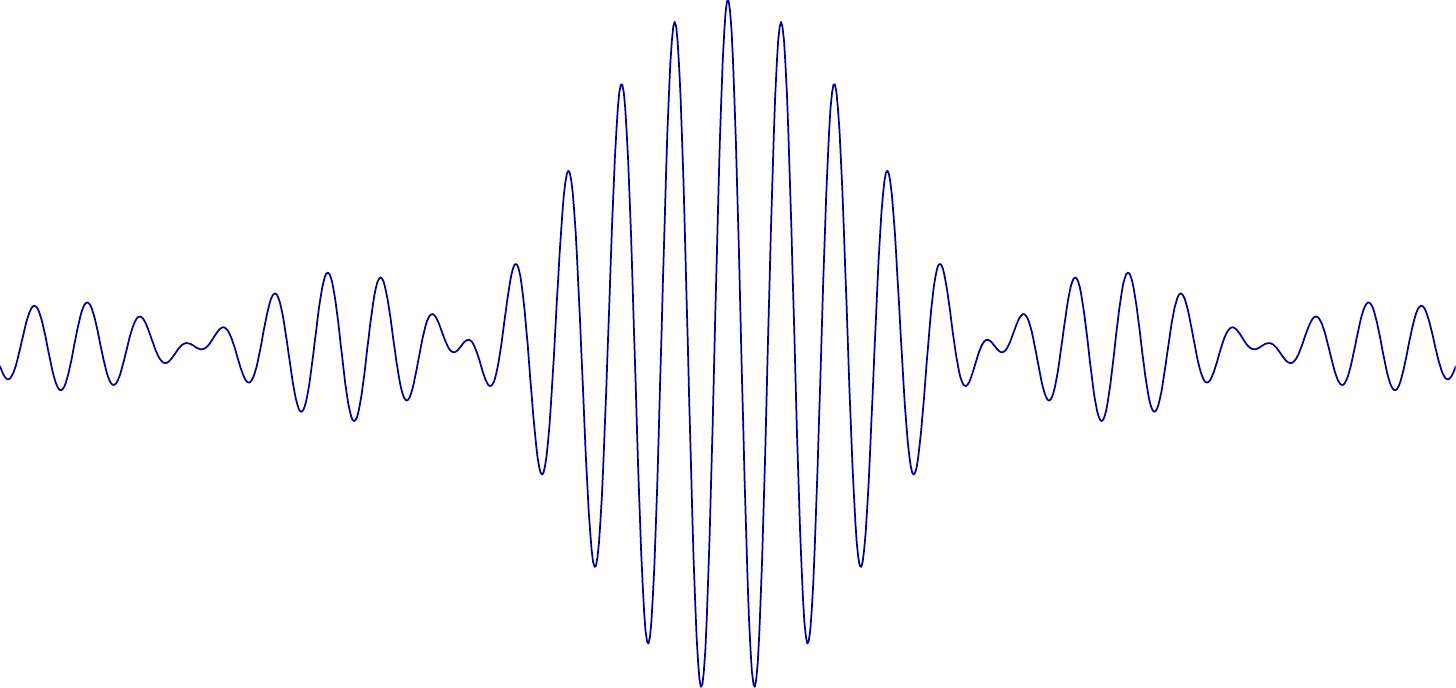}
\end{minipage}
\hfill
\begin{minipage}[h]{.245\linewidth}
  \centering\includegraphics[width = \linewidth]{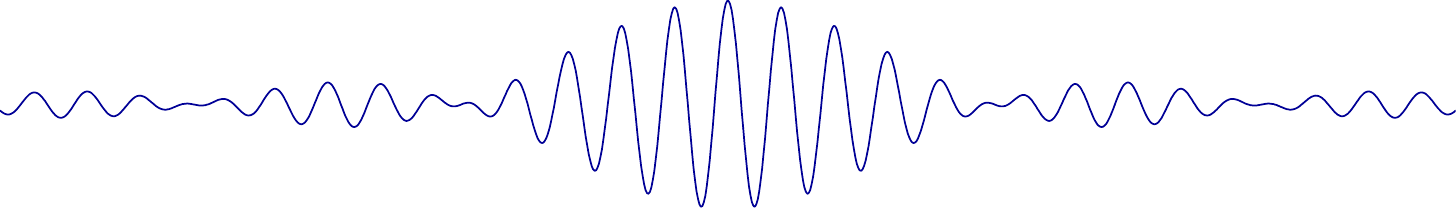}
\end{minipage}
\hfill
\begin{minipage}[h]{.245\linewidth}
  \centering\includegraphics[width = \linewidth]{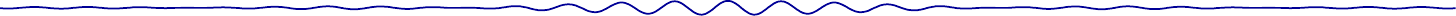}
\end{minipage}
\caption[Recombinaison dans le plan pupille]{\textbf{Recombinaison dans le plan pupille} : Interférogramme obtenu en faisant varier la différence de marche dans un cas polychromatique. De gauche à droite : $V = 1$ (source non résolue), $V = 0.5$, $V = 0.15$ et $V = 0.01$.}
\label{image__frange_plan_pupille}
\end{figure}

L'autre façon, et qui est celle utilisée par l'instrument \emph{FLUOR}, consiste en la superposition des faisceaux dans le plan de \emph{Fourier} (plan pupille), c'est à dire qu'ils sont recombinés avant le détecteur. On fait ensuite varier temporellement la différence de marche pour reconstruire un interférogramme. Le type de motif observé est présenté sur la Fig.~\ref{image__frange_plan_pupille}.

Pour le reste de ce chapitre, on considèrera le cas d'une modulation temporelle des franges d'interférences, telle que réalisée sur \emph{FLUOR}.

\section{Les observables}
\label{section__les_observables}

Il y a deux observables en interférométrie, le module et la phase de la fonction de visibilité complexe (Équ.~\ref{equation__visibilite_complexe}). 

\paragraph*{\textcolor{black}{Module}}

Le module, appelé parfois simplement visibilité, est le paramètre le plus simple à mesurer pour un interféromètre. Il est mesuré en estimant le contraste des franges d'interférences. Cependant, a cause des effets instrumentaux et atmosphériques, la visibilité des franges est différente de la visibilité de l'objet et un étalonnage est nécessaire :
\begin{equation}
| \mu(u,v) | = T\,V(u,v)
\label{equation__mu}
\end{equation}
où $T$ représente la fonction de transfert interférométrique. Elle est déterminée en observant avant et/ou après la source scientifique une étoile de référence dont on connaît à priori sa visibilité. On effectue ensuite l'étalonnage :
\begin{displaymath}
T = \frac{\mu(ref)}{V(ref)} \qquad \Longrightarrow \qquad V = \frac{\mu}{\mu(ref)} V(ref)
\end{displaymath}

Notons qu'en pratique, on travaille le plus souvent avec la quantité $V^2$ car les sources de biais statistiques sont plus facilement corrigibles.

\paragraph*{\textcolor{black}{Phase}}

La phase intrinsèque de l'objet (position des franges) est impossible à mesurer à cause des variations aléatoires atmosphériques. Cependant grâce à diverses techniques, nous pouvons accéder à une partie de l'information de phase. Par exemple en utilisant au minimum trois télescopes, nous pouvons mesurer une quantité appelée clôture de phase, c'est la somme des phases mesurées par chaque paire de télescopes. Cette clôture est indépendante des perturbations atmosphériques. Si les franges sont dispersées spectralement, nous pouvons fixer une longueur d'onde comme phase de référence. On a alors accès à une phase différentielle, donnant une information sur la variation de la phase avec la longueur d'onde.

Dans le reste du manuscrit, je ne parlerai que du module de la visibilité, que je nommerai simplement visibilité, car c'est l'unique quantité mesurée par \emph{FLUOR}.

\section{Plan $(u,v)$}

\begin{figure}[!p]
\begin{minipage}[h]{.5\linewidth}
  \centering\includegraphics[width = \linewidth]{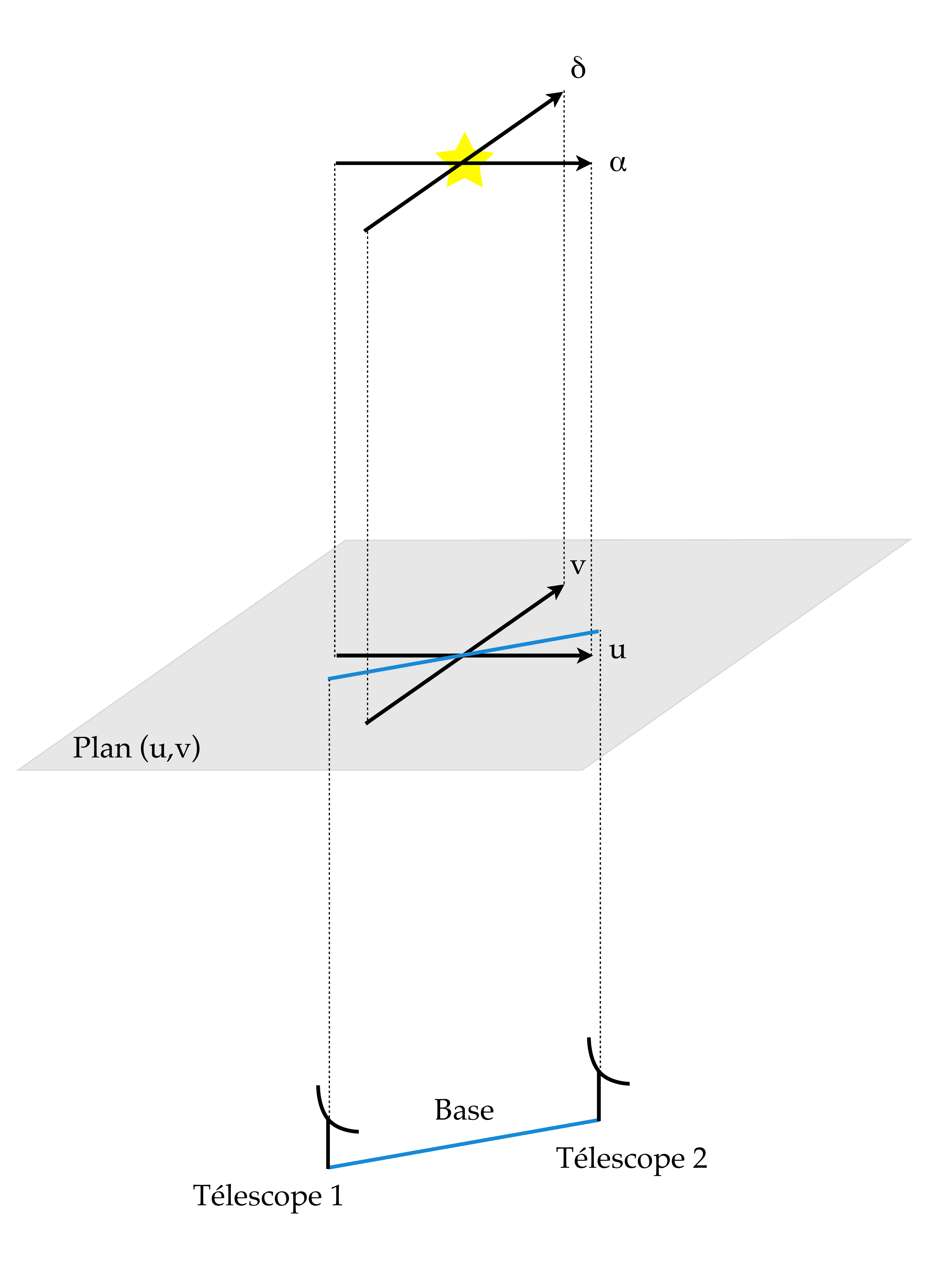}
\end{minipage}
\hfill
\begin{minipage}[h]{.5\linewidth}
\caption[Plan $(u,v)$]{\textbf{Plan $(u,v)$} : à une base donnée, nous n'avons accès qu'à une fréquence spatiale. Il est nécessaire d'échantillonner au maximum le plan $(u,v)$ pour retrouver la distribution d'intensité de la source.}	
\label{image__plan_uv}
\end{minipage}
\end{figure}

La couverture du plan $(u,v)$ est importante si l'on souhaite utiliser les techniques de reconstruction d'image ou examiner la géométrie d"un objet en étudiant l'évolution de la visibilité avec l'angle de projection. Comme exposé sur la Fig.~\ref{image__plan_uv}, une mesure de visibilité à une base donnée donne accès à seulement une fréquence spatiale. Il est donc nécessaire d'avoir une estimation du module et de la phase de la visibilité complexe à plusieurs fréquences spatiales si l'on veut retrouver la distribution d'intensité de la source. Différentes possibilités existent pour couvrir le plan $(u,v)$ :

\begin{itemize}
\compactlist
\item la supersynthèse : on utilise l'effet de rotation de la Terre. Cela est dû au fait que l'étoile ne "voit" pas la base réelle au sol $B$, mais une base $B_\mathrm{p}$ projetée sur le plan normal à la ligne de visée (Fig.~\ref{image__schema_interferometre}). Cette base varie pendant la nuit avec la rotation diurne.
\item spectro-interférométrie : les franges sont dispersées spectralement, on a alors une mesure de visibilité à plusieurs longueurs d'onde.
\item multi-télescopes : l'utilisation de plusieurs télescopes ayant des bases non-redondantes et orientées différemment permet également de "remplir" le plan $(u,v)$.
\end{itemize}

\begin{figure}[!p]
\centering
\resizebox{\hsize}{!}{
\includegraphics{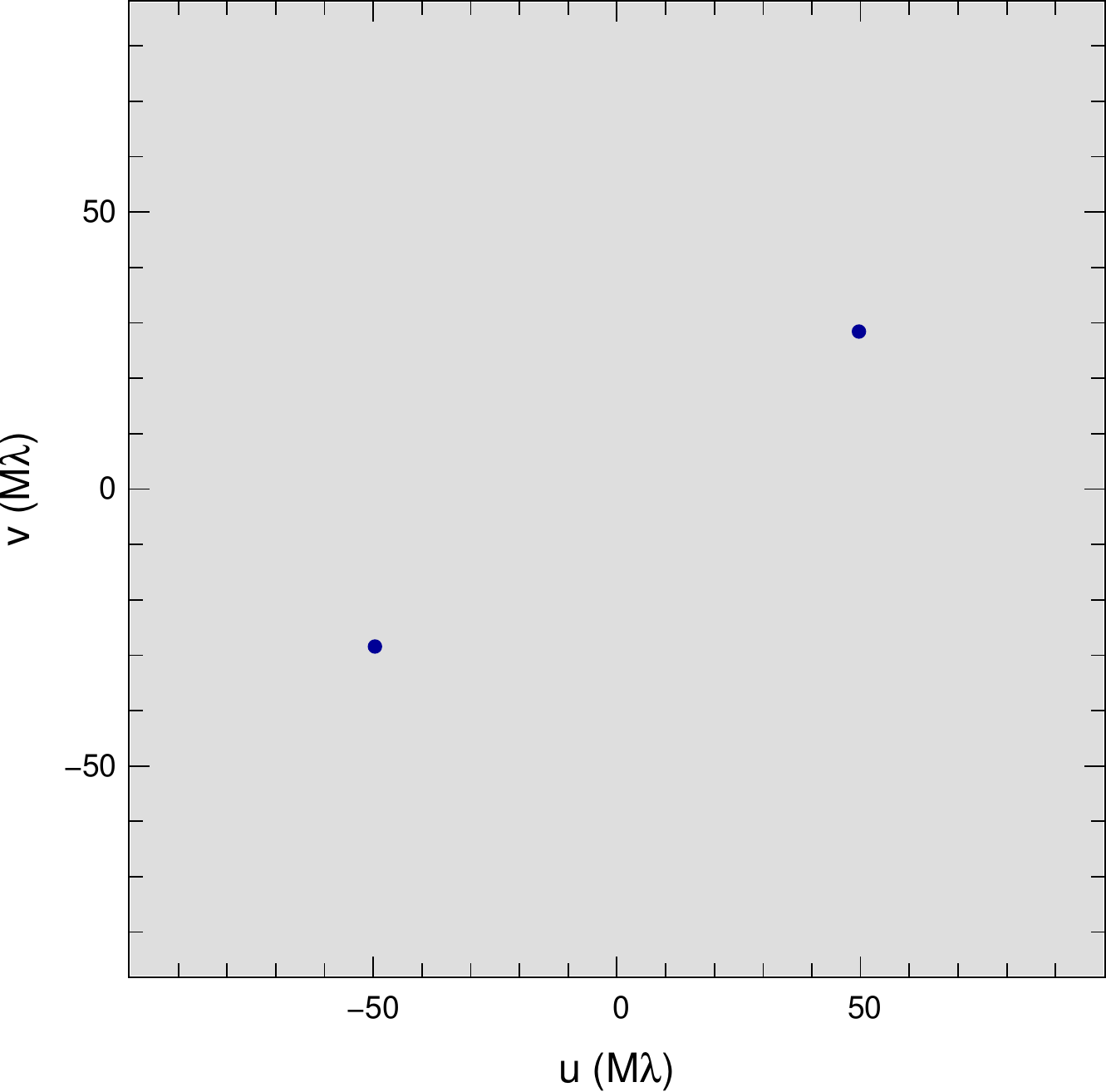}\hspace{.05cm}
\includegraphics{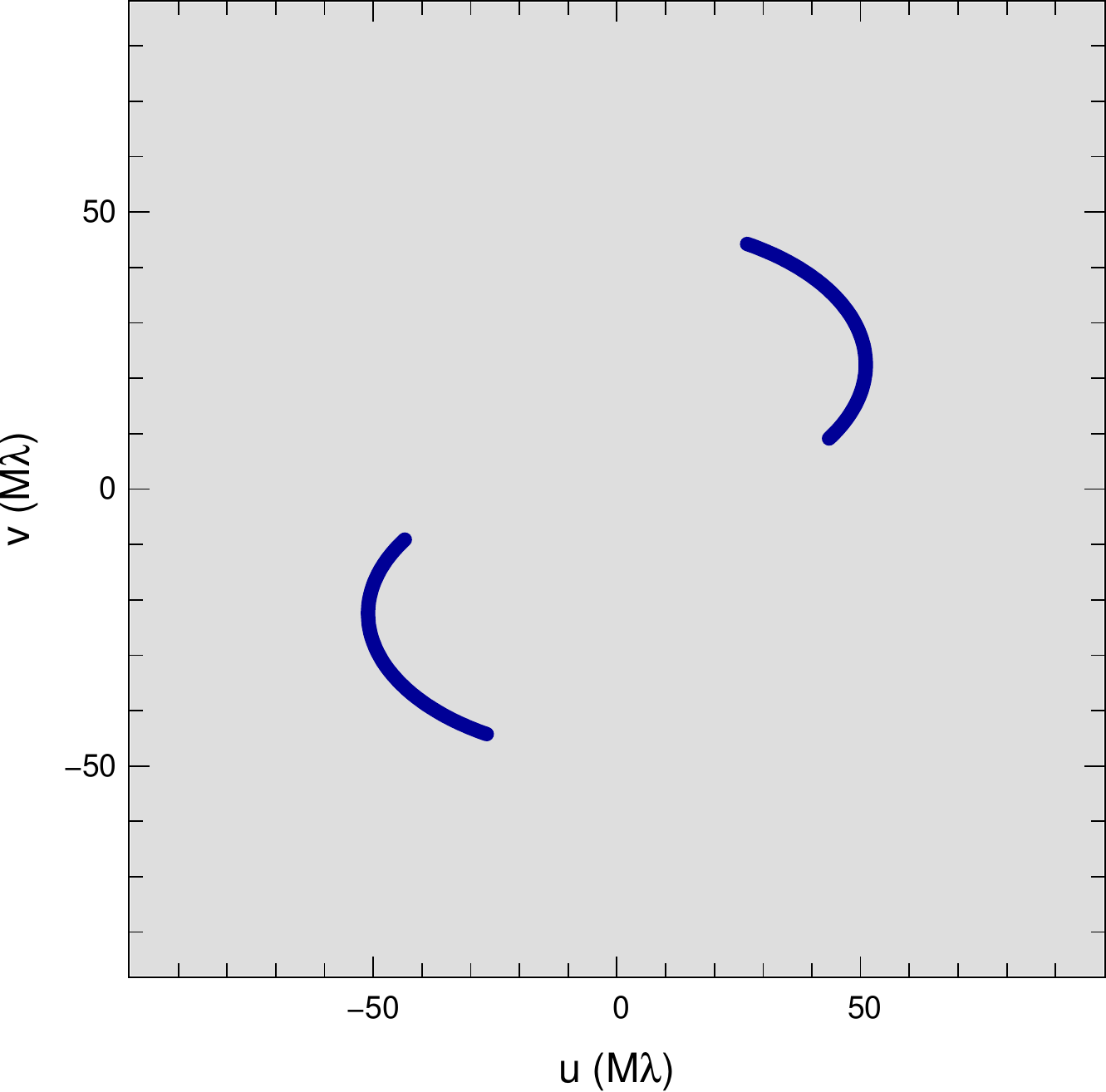}\hspace{.05cm}
\includegraphics{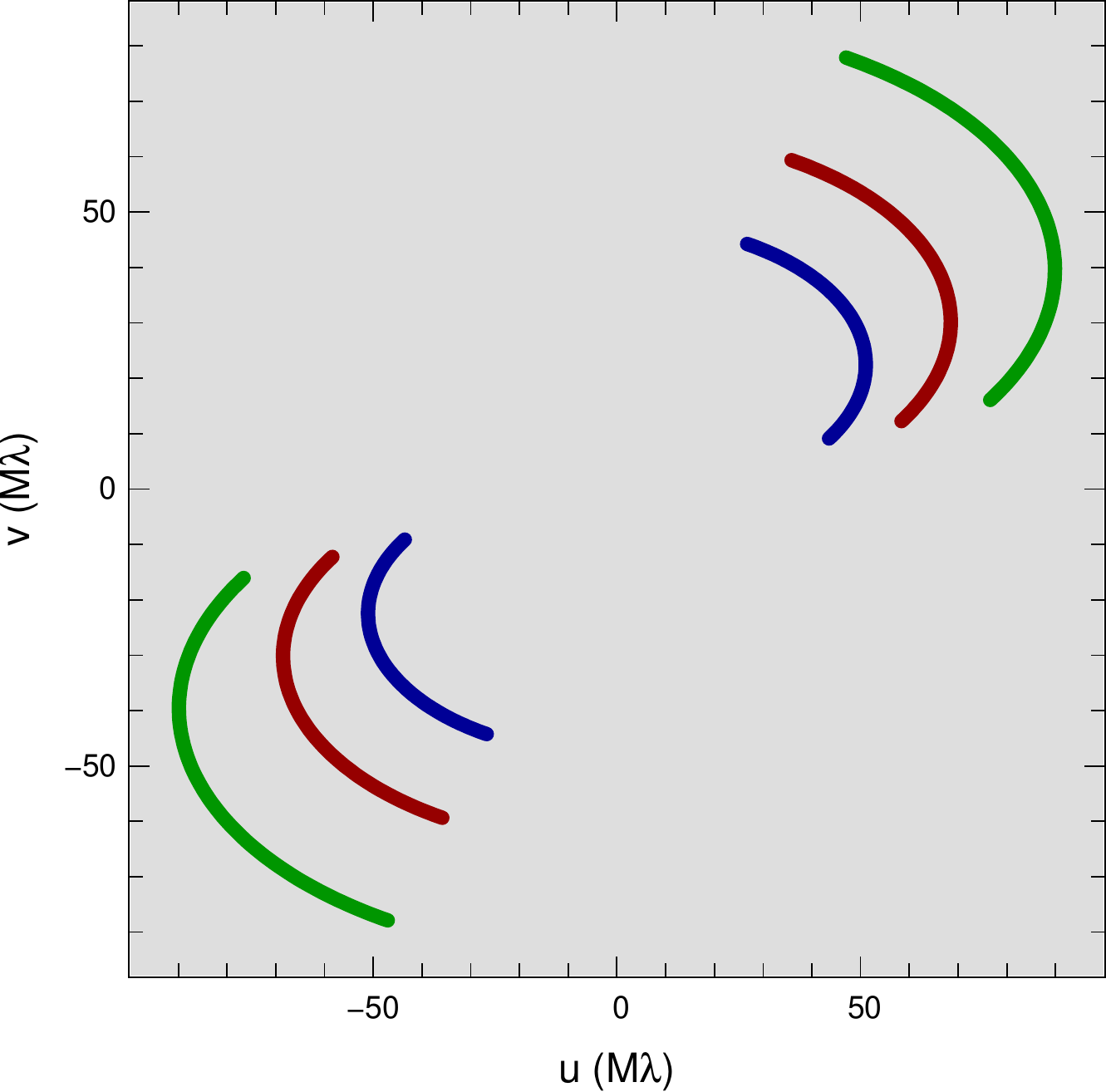}\hspace{.05cm}
\includegraphics{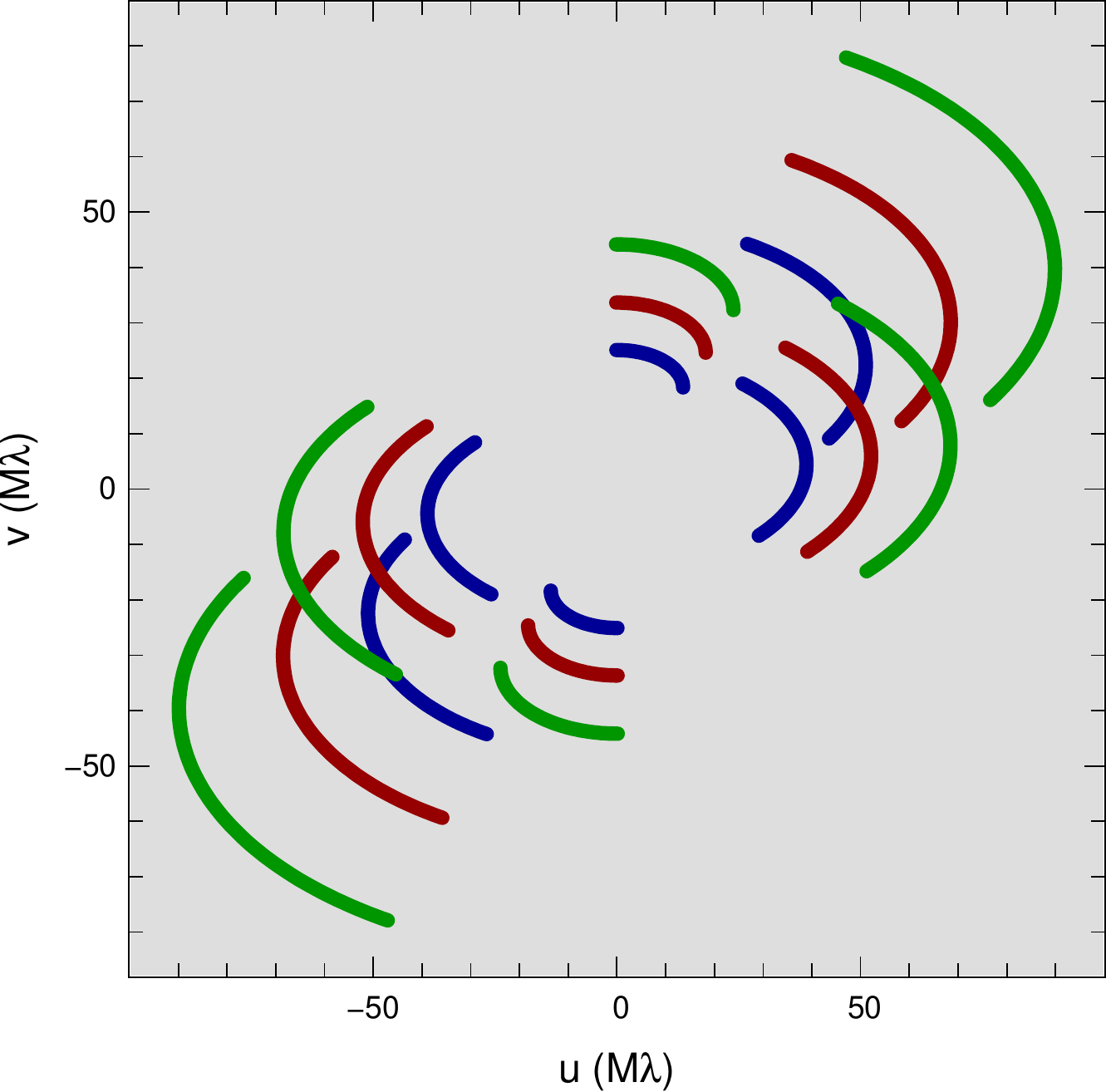}}
\caption[Couverture du plan $(u,v)$]{\textbf{Couverture du plan $(u,v)$} : exemple pour un objet de déclinaison $-34^\circ$ observé avec les télescopes UT1 et UT2 du \emph{VLTI} (base au sol de $130\,\mathrm{m}$). La première image est une mesure de visibilité en bande $K$. La seconde utilise le mouvement diurne pour mesurer plusieurs visibilités (angle horaire $-3\,\mathrm{h} < h < 3\,\mathrm{h}$). La troisième utilise en plus les bandes $J$ (en vert) et $H$ (en rouge). La dernière combine les effets précédent en utilisant 3 télescopes : UT1, UT2 et UT4.}
\label{image__plan_uv_couverture}
\end{figure}

La combinaison de toutes ces possibilités simultanément offrent bien sûr la meilleure couverture. J'expose sur la Fig.~\ref{image__plan_uv_couverture} quelques exemples de remplissage du plan $(u,v)$.

Dans le cadre des mesures \emph{FLUOR}, la reconstruction d'image ne s'applique pas car il ne fonctionne qu'à deux télescopes (pas de mesures de phase) et sans dispersion spectrale (mais une amélioration est prévue prochainement). Cependant, grâce à différentes longueurs de base et angle de projection, il est possible, grâce à de bon modèles de visibilité, d'obtenir certaines informations sur l'objet observé (géométrie, taille, ...).

\section{Modèle de visibilité}

La visibilité d'un objet que l'on observe est liée à ses paramètres intrinsèques (diamètre, rapport de flux, inclinaison, ...). Il est donc important d'utiliser les modèles adéquats pour estimer ces paramètres. Pour cela on suppose que l'objet a une certaine distribution d'intensité, que l'on utilise ensuite dans l'équation~\ref{equation__visibilite_complexe} pour avoir un modèle de visibilité. Dans cette section je présente quelques modèles simples pouvant être appliqués à des mesures de visibilité.

\begin{figure}[!p]
\centering
\resizebox{\hsize}{!}{\includegraphics{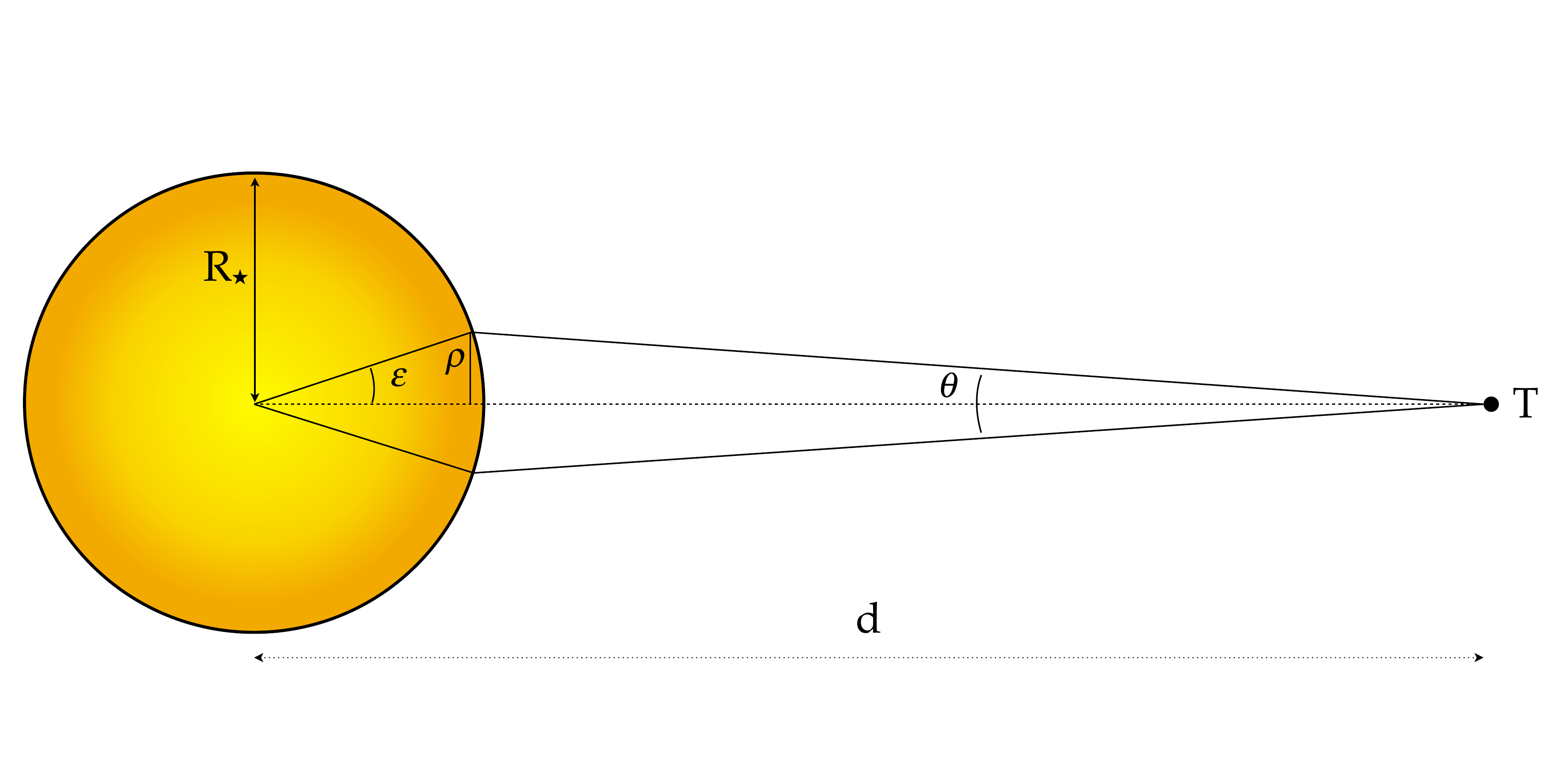}}
\caption[Géométrie d'un objet à symétrie radiale]{\textbf{Géométrie d'un objet à symétrie radiale}}
\label{image__schema_intensite_uniforme}
\end{figure}

Pour les raisons de symétrie présentées sur le schéma de la Fig.~\ref{image__schema_intensite_uniforme}, nous noterons la taille angulaire :
\begin{displaymath}
\theta = \sqrt{\alpha^2 + \delta^2} \sim \frac{2\rho}{d}
\end{displaymath}
où $d$ représente la distance de l'objet et $\rho$ la distance radiale projetée sur la ligne de visée. À partir de maintenant, on se place dans le cas où l'objet observé est une étoile.

\subsection{Source ponctuelle}

C'est le modèle le plus simple pour une étoile non résolue. D'un point de vue mathématiques, on peut écrire :
\begin{displaymath}
I(\alpha,\beta) = I_0\,\delta_\mathrm{D}(\alpha - \alpha_0,\delta - \delta_0)
\end{displaymath}
où $\delta_\mathrm{D}(\alpha - \alpha_0,\delta - \delta_0)$ représente une distribution de \emph{Dirac} en un point $(\alpha_0,\beta_0)$. En utilisant l'équation~\ref{equation__visibilite_complexe}, on obtient : 
\begin{displaymath}
\mu = \mathrm{e}^{-2\pi i(u\alpha_0 + v\beta_0)}
\end{displaymath}

On remarque donc que pour une étoile non résolue, la visibilité vaut 1 ($V = |\mu|$).

\subsection{Disque uniforme}

C'est le modèle le plus simple en ce qui concerne une étoile résolue car l'intensité est uniforme sur toute la surface à une longueur d'onde donnée. La distribution d'intensité est donnée par :
\begin{displaymath}
I(\theta) = I_0\,\delta_\mathrm{D}(\alpha,\delta)\ast\Pi \left(\frac{\theta}{\theta_\mathrm{UD}}\right)
\end{displaymath}
où $\Pi(\frac{\theta}{\theta_\mathrm{UD}})$ est une fonction porte circulaire égale à 1 pour $\theta~\leqslant~\theta_\mathrm{UD} = 2\,R_\star/d$. La transformée de \emph{Fourier} d'une fonction à symétrie radiale est donnée par la transformée de \emph{Hankel} d'ordre 0 :
\begin{equation}
V(u,v) = \left| \frac{\iint I(\alpha,\delta)\mathrm{e}^{-2i\pi (u\alpha + v\delta)}d\alpha d\delta}{\iint I(\alpha,\delta)d\alpha d\delta} \right| \quad \Longrightarrow \quad V(\nu) = \left| \frac{\int_0^1 I(r) J_0(x\,r)r dr}{\int_0^1 I(r) rdr} \right|
\label{equation__visibilite_radiale}
\end{equation}
où $\nu = B_\mathrm{p}/\lambda = \sqrt{u^2 + v^2}$ représente la fréquence spatiale, $x = \pi \theta_\mathrm{UD}\,B_\mathrm{p}/\lambda$ et $r = \rho/R_\star$. Ici on ne s'intéresse qu'à l'amplitude de la fonction visibilité.

En utilisant les propriétés des fonctions de \emph{Bessel}, la visibilité d'un disque uniforme est donnée par :
\begin{displaymath}
V(\nu) = \left| 2\,\frac{J_1(x)}{x} \right|
\end{displaymath}

\begin{figure}[!p]
\centering
\resizebox{\hsize}{!}{
\includegraphics{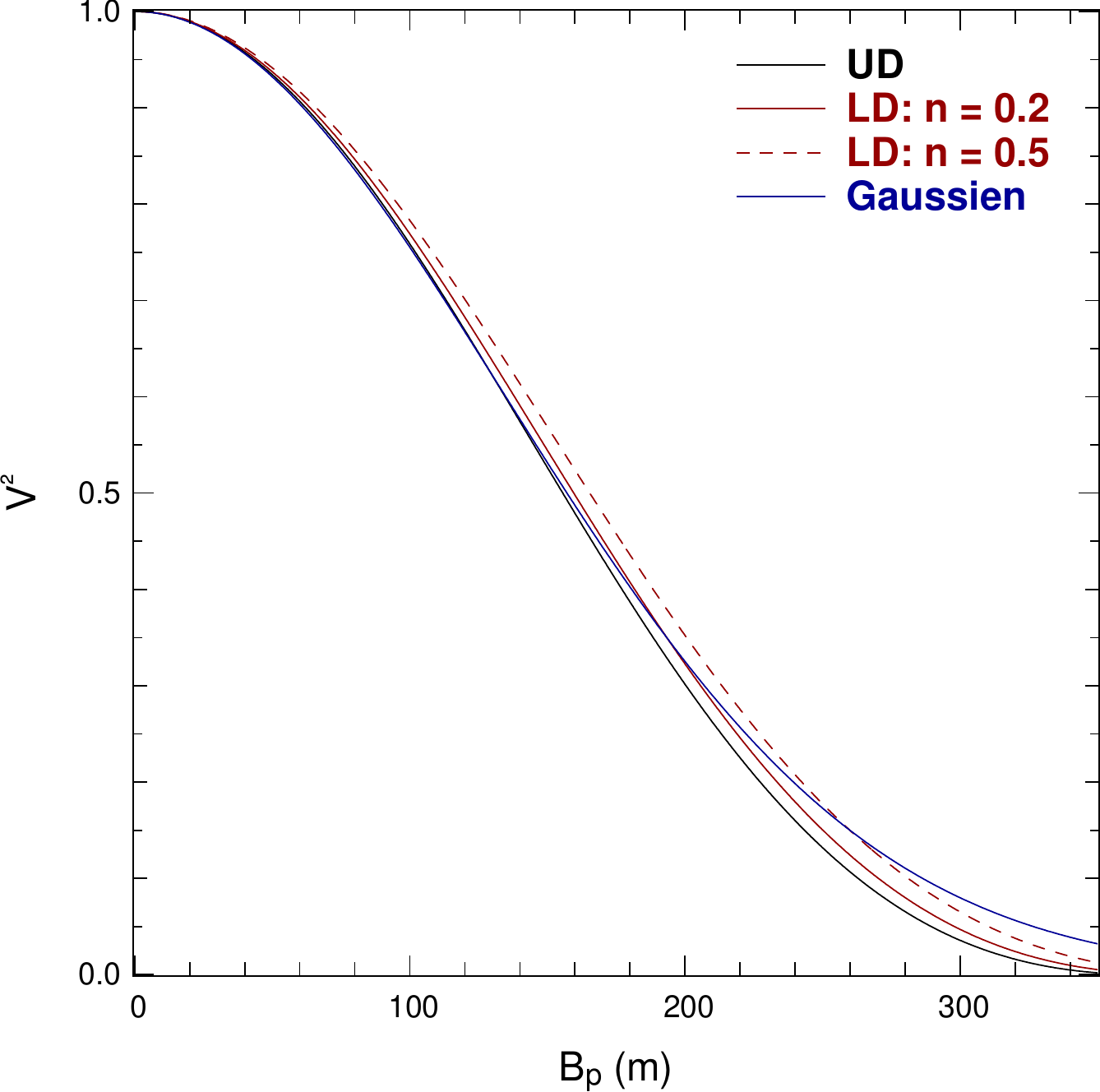}\hspace{.5cm}
\includegraphics{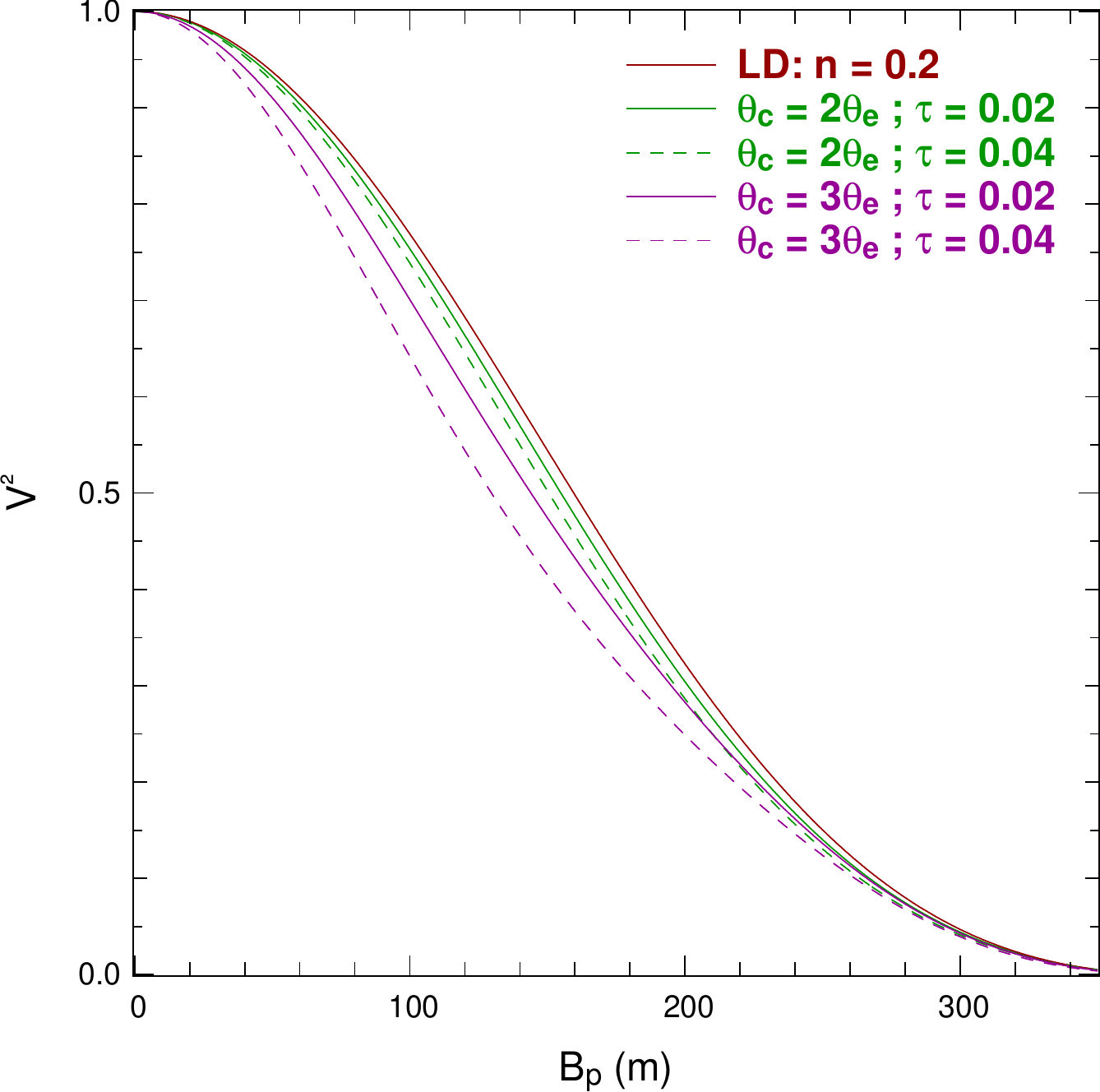}}
\caption[Différents modèles de visibilité]{\textbf{Différents modèles de visibilité} : modèle pour une source de diamètre $1.5\,\mathrm{mas}$ à $\lambda = 2.2\,\mu\mathrm{m}$. En noir, le modèle de disque uniforme, en rouge le disque assombri pour deux valeurs du paramètre $n$, et le modèle gaussien est en bleu. Deux modèles de disque assombri + coquille sphérique sont également représentés avec les valeurs fixes $n = 0.2$ et $\epsilon = 15\,\%$ : $\theta_\mathrm{c} = 2\,\theta_\mathrm{e}$ (en vert) et $\theta_\mathrm{c} = 3\,\theta_\mathrm{e}$ (en violet). $\theta_\mathrm{e} = 1.5\,\mathrm{mas}$ et $\theta_\mathrm{c}$ dénotent respectivement le diamètre angulaire de l'étoile et de la coquille.}
\label{image__modèles_visibilite}
\end{figure}

Cette fonction est représentée sur la Fig.~\ref{image__modèles_visibilite} pour un diamètre angulaire $\theta_\mathrm{UD} = 1.5\,\mathrm{mas}$ et une longueur d'onde $\lambda = 2.2\,\mu\mathrm{m}$.

\subsection{Disque assombri}

C'est un modèle plus réaliste qui prend en compte le phénomène d'assombrissement centre-bord, un effet de projection du disque stellaire où la lumière provenant du centre montre des couches plus profondes et plus chaudes que le limbe. Par sa simplicité, un modèle de distribution d'intensité souvent employé est la loi de puissance \citep{Hestroffer-1997-11} :
\begin{displaymath}
I(\alpha,\delta) = I(0)\,(\cos \varepsilon)^n\ast\delta_\mathrm{D}(\alpha,\delta) \quad \mathrm{avec} \quad n \geqslant 0 \quad \mathrm{et} \quad \cos \varepsilon = \sqrt{1 - r^2}
\end{displaymath}
où $n$ est le paramètre d'assombrissement centre-bord, dépendant de la longueur d'onde d'observation. Il peut être estimé par un ajustement ou fixé à partir de tabulations, par exemple de \citet{Claret-2000-11}, calculées à partir de modèles d'atmosphères stellaires. 

En utilisant l'équation~\ref{equation__visibilite_radiale} et en notant $m = n/2 + 1$, on a pour la visibilité :
\begin{equation}
V_m(\nu) = \left| 2\,m \int (1 - r^2)^{m - 1}\,J_0(xr)rdr \right| \qquad \Longrightarrow \qquad V_m(\nu) = \left| \Gamma(m + 1)\,\frac{J_m(x)}{(x/2)^m} \right|
\label{equation__LD}
\end{equation}
où $x = \pi \theta_\mathrm{LD} B_\mathrm{p}/\lambda$ \citep[][en utilisant l'intégrale des fonctions de \emph{Bessel}, page 487, Équ.~11.4.10]{Abramowitz-1972-}. Cette visibilité est représentée sur la Fig.~\ref{image__modèles_visibilite} pour deux valeurs du paramètre $n$ et un diamètre $\theta_\mathrm{LD} = 1.5\,\mathrm{mas}$ à $2.2\,\mu\mathrm{m}$. On remarque que l'assombrissement centre-bord à pour effet d'élargir le premier lobe (donc diminuer le diamètre apparent) et de réduire la hauteur des lobes suivants. Ce type de modèle à déjà été appliqué avec succès à la Céphéide $\delta$~Cep \citep{Merand-2006-}.

\subsection{Disque Gaussien}

Ce type de modèle peut être utilisé pour représenter une enveloppe optiquement épaisse dans l'environnement de l'étoile. La distribution d'intensité est donnée par :
\begin{displaymath}
I(\alpha,\delta) = I(0)\,\delta_\mathrm{D}(\alpha,\delta)\ast\mathrm{e}^{- \tfrac{4\,\ln 2\,\rho^2}{h^2}}
\end{displaymath}

où $h$ représente la largeur à mi-hauteur de la gaussienne. Cette distribution donne la fonction de visibilité suivante :
\begin{displaymath}
V(\nu) = \mathrm{e}^{-\left( \tfrac{\pi\;B_\mathrm{p}\,h}{4\ln 2\,\lambda} \right)^2}
\end{displaymath}

Ce modèle est illustré sur la Fig.~\ref{image__modèles_visibilite} pour $h = 1.5\,\mathrm{mas}$ à $2.2\,\mu\mathrm{m}$. On remarque notamment que pour des bases courtes, le comportement des courbes est identique à celui d'un disque uniforme.

\subsection{Disque assombri + couronne sphérique}

\begin{figure}[!p]
\centering
\resizebox{\hsize}{!}{\includegraphics{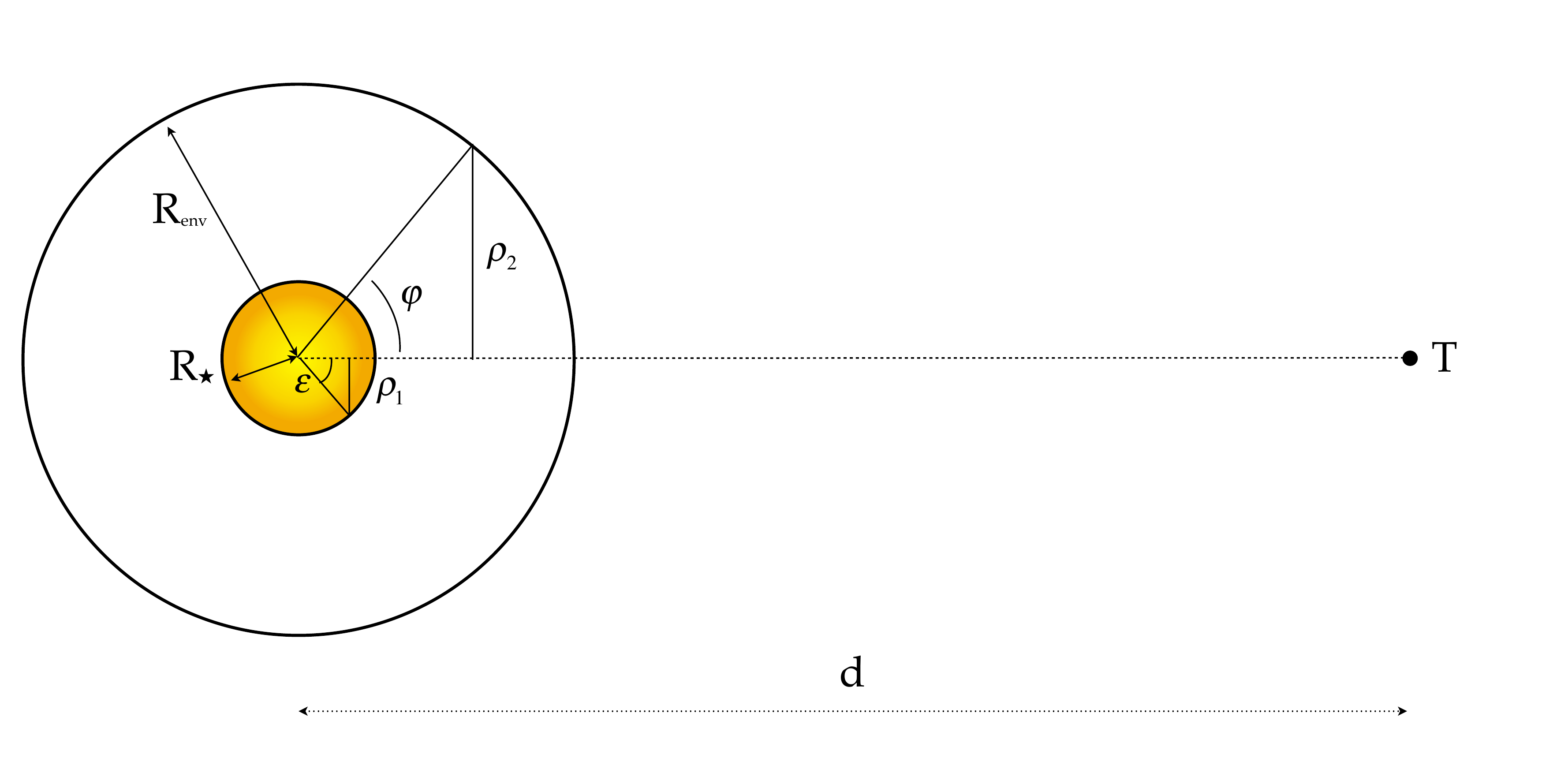}}
\caption[Géométrie d'une étoile avec une couronne sphérique]{\textbf{Géométrie d'une étoile avec une couronne sphérique}}
\label{image__schema_intensite_couronne}
\end{figure}

Ce genre de modèle est utile lorsque la couronne est supposée optiquement mince, c'est à dire qu'elle n'absorbe qu'une partie de la lumière de l'étoile. Suivant le schéma de la Fig.~\ref{image__schema_intensite_couronne}, on a la distribution d'intensité suivante :

\begin{displaymath}
I(\alpha,\delta) = \left\{ 
  \begin{aligned}
     & I_\star\,\mathrm{e}^{-\tfrac{\tau}{\cos \varphi}}\,(\cos \varepsilon)^n + I_\mathrm{env}\,\left( 1 - \mathrm{e}^{-\tfrac{\tau}{\cos \varphi}} \right) \quad \mathrm{pour} \quad \rho_2 \leqslant \rho_1 \\
     & I_\mathrm{env}\,\left( 1 - \mathrm{e}^{-2\tfrac{\tau}{\cos \varphi}} \right) \quad \mathrm{pour} \quad \rho_2 \geqslant \rho_1 \\ 
  \end{aligned}
\right.
\end{displaymath}
où $I_\star$ et $I_\mathrm{env}$ représentent respectivement l'intensité de l'étoile et de l'enveloppe tandis que $\tau$ dénote la profondeur optique de la coquille. En notant $\epsilon = I_\mathrm{env}/I_\star$, $a = \theta_\star/\theta_\mathrm{env}$ et $r = \sin \varepsilon = \sin \varphi/a$, l'équation précédente devient :
\begin{equation}
I(\alpha,\delta) = \left\{ 
  \begin{aligned}
     & I_\star \left[ \mathrm{e}^{-\tfrac{\tau}{\sqrt{1 - (ar)^2}}}\,(1 - r^2)^{n/2} + \epsilon \left( 1 - \mathrm{e}^{-\tfrac{\tau}{\sqrt{1 - (ar)^2}}} \right) \right] \quad \mathrm{pour} \quad \sin \varphi \leqslant \frac{\theta_\star}{\theta_\mathrm{env}} \\
     & I_\star\,\epsilon\left( 1 - \mathrm{e}^{-\tfrac{2\tau}{\sqrt{1 - (ar)^2}}} \right) \quad \mathrm{pour} \quad \sin \varphi \geqslant \frac{\theta_\star}{\theta_\mathrm{env}} \\ 
  \end{aligned}
\right.
\label{equatio__LD_couronne}
\end{equation}

La transformée de \emph{Hankel} est calculée numériquement à partir de ces équations. Sur la Fig.~\ref{image__modèles_visibilite} est représenté ce type de modèle pour différents rapports de taille et de profondeur optique. La présence de la couronne a un effet principalement dans le premier lobe de la fonction de visibilité. Ce type de modèle à déjà été appliqué avec succès à la Céphéide Polaris \citep{Merand-2006-07}. Je l'appliquerai plus tard sur les données de l'étoile Y~Oph.

\subsection{Distinction des différents modèles}

\begin{figure}[!p]
\centering\includegraphics[width=.7\linewidth]{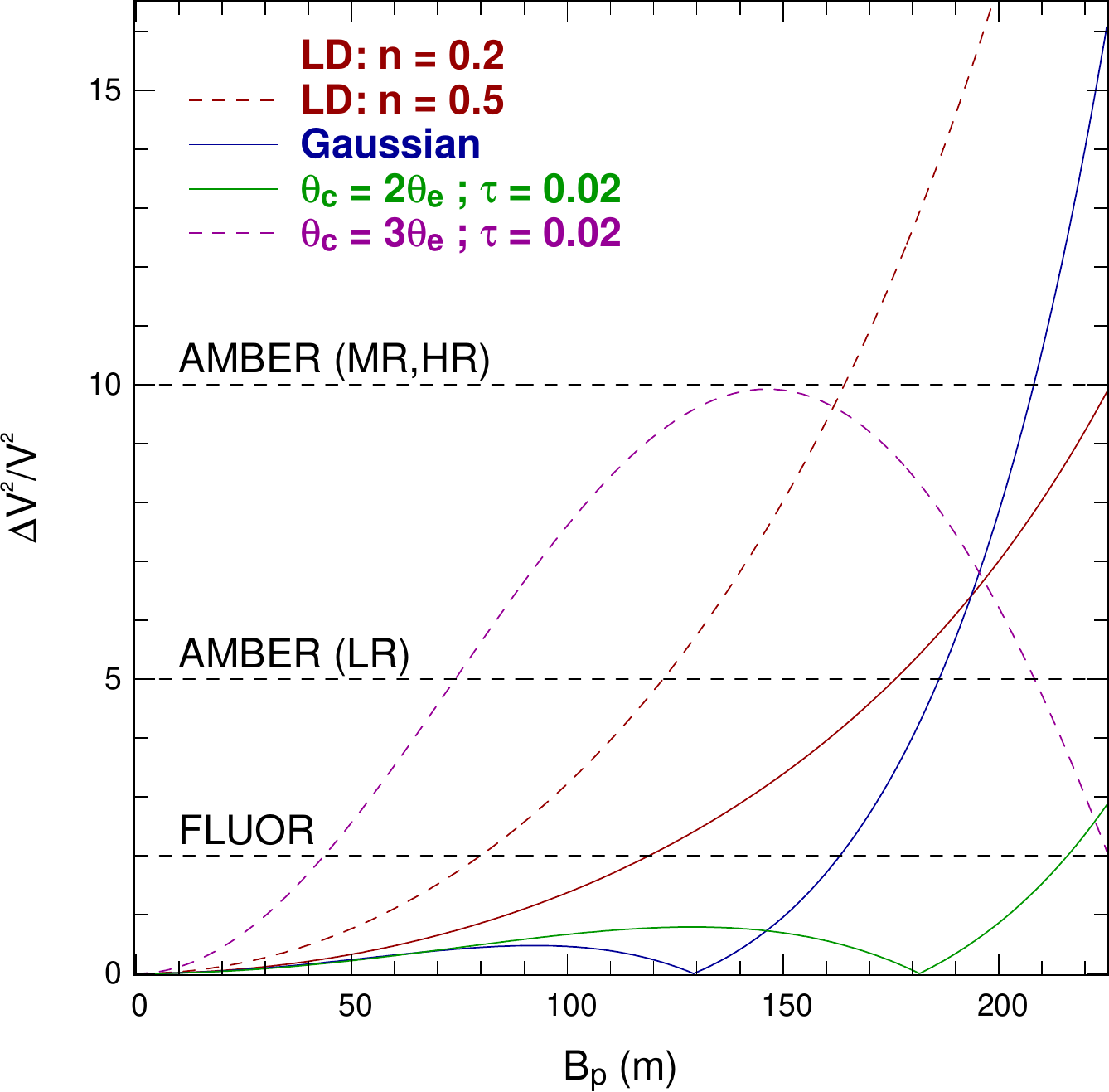}
\caption[Distinction des différents modèles]{\textbf{Distinction des différents modèles} : les légendes et couleurs sont identiques à celles de la Fig.~\ref{image__modèles_visibilite}.}
\label{image__precision_visibilite}
\end{figure}

Tous les modèles illustrés sur la Fig.~\ref{image__modèles_visibilite} sont presque identiques dans le premier lobe. La distinction d'un modèle par rapport un autre dépendra de la précision de chaque instrument. J'ai représenté sur la Fig.~\ref{image__precision_visibilite} la différence relative des différents modèles par rapport au modèle de disque uniforme, soit $(V_\mathrm{mod\grave{e}le}^2 - V_\mathrm{UD}^2)/V_\mathrm{UD}^2$, en fonction de la base projetée.

J'ai utilisé comme instruments de référence \emph{FLUOR} et \emph{AMBER} pour des mesures à $2.2\,\mu\mathrm{m}$ d'un objet ayant un diamètre $\theta_\star = 1.5\,\mathrm{mas}$. On remarque par exemple que \emph{AMBER} en moyenne et haute résolution ne fera pas la différence entre les modèles si on observe avec une base $B_\mathrm{p} < 165\,\mathrm{m}$. À basse résolution, une coquille sphérique de diamètre angulaire de $3\,\theta_\star$ est détectable à partir de $B_\mathrm{p} \simeq 75\,\mathrm{m}$. 

\emph{FLUOR} qui est beaucoup plus précis ($\sigma(V^2)/V^2 \sim 2\,\%$, et peut même atteindre $1\,\%$ avec un bon étalonnage et un haut rapport signal à bruit), détectera la coquille à partir de $45\,\mathrm{m}$ de base, alors que le disque assombri nécessite $B_\mathrm{p} > 80\,\mathrm{m}$. En pratique, le force de l'assombrissement centre-bord est difficile à estimer et nécessite généralement des mesures dans le second lobe de la fonction de visibilité.

\section{Observations de Céphéides avec FLUOR}

Je présente dans cette section quelques premiers résultats des observations de Céphéides avec l'instrument \emph{FLUOR}. Ce long programme d'observation, qui s'est terminé en août 2011, s'est déroulé en trois campagnes d'observations sur deux ans (juin 2010, novembre 2010 et août 2011) auxquelles j'ai pleinement participé. Nous avons pu acquérir plusieurs mesures de visibilité pour 17 Céphéides. Ces données sont toujours en cours de réduction et d'analyse.

Le premier objectif de ce programme est de détecter la pulsation des Céphéides afin d'appliquer la méthode de \emph{Baade-Wesselink}, pour au final obtenir une estimation indépendante de leur distance. Le second objectif est de vérifier la présence ou pas d'une enveloppe autour des étoiles de notre échantillon.

\subsection{Présentation de l'interféromètre CHARA}

\begin{figure}[!p]
	\centering\includegraphics[width = \linewidth]{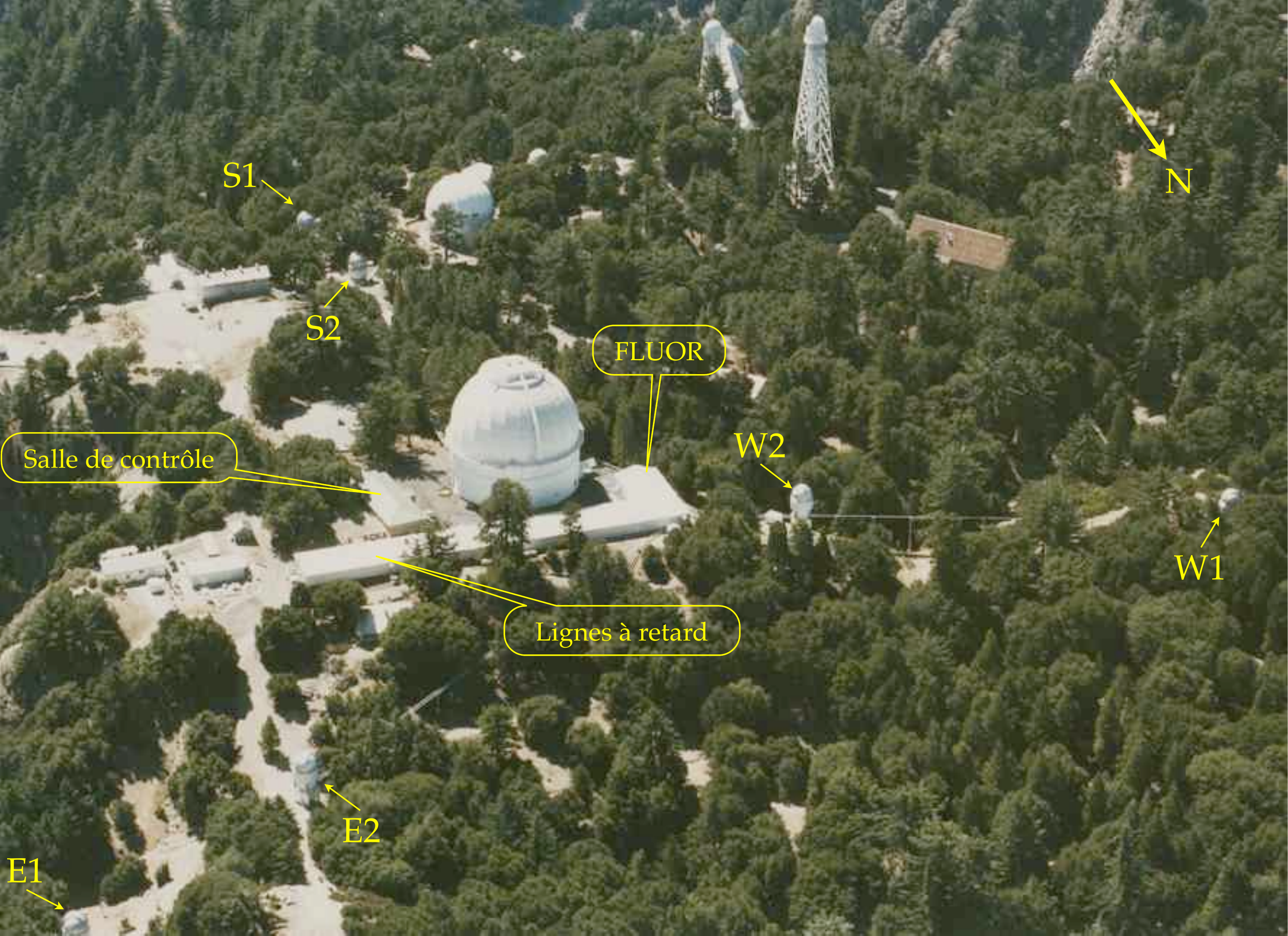}
    \caption[Configuration de l'interféromètre \emph{CHARA}]{\textbf{Configuration de l'interféromètre \emph{CHARA}} : les emplacements des six télescopes de $1\,\mathrm{m}$ sont indiqués par des flèches jaunes (photo de G. Van Belle).}
  		\label{image__chara}
	\end{figure}

L'interféromètre optique et infrarouge \emph{CHARA} (Center for High Angular Resolution Astronomy) est situé au Mont Wilson, près de Los Angeles en Californie. Ce réseau comprend six télescopes fixes de $1\,\mathrm{m}$ de diamètre, disposés dans une configuration en "Y" (non redondance des bases). Il est pleinement opérationnel depuis 2005 et dispose de 15 bases allant de 34 à $331\,\mathrm{m}$. La configuration de cet interféromètre se trouve sur la Fig.~\ref{image__chara} et les longueurs de base au sol sur la Table~\ref{table__chara_base}. Le nom des télescopes représente la direction de la branche sur laquelle ils se trouvent : S pour Sud, E pour Est et W pour Ouest. 

Les miroirs secondaires sont montés sur un système piézo-électrique afin de permettre une correction du tip/tilt. La lumière est ensuite conduite vers les lignes à retard à travers des tubes sous vide (pour réduire les effets de dispersion dans l'air). L'égalisation des chemins optiques entre les deux télescopes est effectuée grâce à deux sous-systèmes. Le premier système, composé de 6 miroirs de repliement (appelés "pop"), est utilisé afin d'introduire un retard fixe. Le second est un système de miroirs montés sur des chariots mobiles permettant une égalisation plus précise et un suivi avec la rotation diurne (je rappelle que pour observer des franges d'interférences, les deux ondes électromagnétiques des deux télescopes doivent être cohérentes temporellement, c'est à dire arriver en même temps à l'instrument de recombinaison). Le réseau \emph{CHARA} peut ensuite alimenter 6 instruments de recombinaison, deux dans le visible et quatre dans l'infrarouge :

\begin{itemize}
\compactlist
\item \emph{MIRC} : acronyme de Michigan IR Combiner, il recombine les faisceaux provenant de 6 télescopes dans l'intervalle de longueur d'onde $1.45$--$2.5\,\mu\mathrm{m}$ à faible résolution spectrale.
\item \emph{CLASSIC} : recombine deux faisceaux en bande $H$ ou $K^\prime$.
\item \emph{CLIMB} : acronyme de CLassic Interferometry on Multiple Baselines. Cette instrument fait interférer 3 télescopes en bande $H$ ou $K^\prime$.
\item \emph{PAVO} : acronyme de Precision Astronomocal Visible Observations, il recombine 3 faisceaux lumineux dans le visible (0.63--0.95\,$\mu\mathrm{m}$).
\item \emph{VEGA} : acronyme de Visible spEctroGraph and polArimeter. C'est un recombinateur 4 télescopes (3 pour le moment) dans la bande spectrale $0.4$--$0.9\,\mu\mathrm{m}$ avec trois résolutions spectrales possibles.
\item \emph{FLUOR} : acronyme de Fiber Linked Unit for Optical Recombinaison, il fait interférer deux faisceaux en bande $K^\prime$.
\end{itemize}

\paragraph*{\textcolor{black}{Note historique}}

Je ne peux pas parler de \emph{CHARA} sans dire quelques mots sur l'histoire de l'observatoire du Mont Wilson, fondé en 1904 par George Ellery Hale. Il abrita de nombreuses découvertes et expériences scientifiques : la première expérience réussite d'interférométrie stellaire par le prix nobel A. Michelson, la détermination de la vitesse de lumière également par A. Michelson, la preuve de l'expansion de l'univers par E. Hubble, où encore la découverte du champ magnétique solaire par G. Hale. Le Mont Wilson est un temple de l'histoire astronomique et rentre encore dans l'histoire aujourd'hui grâce à l'une des premières recombinaisons interférométriques à six télescopes (le 9 juillet 2011 avec l'instrument \emph{MIRC}).

\begin{table}[!p]
	\centering
	\begin{tabular}{cccccc} 
	\hline
	\hline
	Télescopes	&	Base (m)	& PA ($^\circ$)	\\
	\hline
	S1--S2			&	34.07		&	350.29			\\
	E1--E2			&	65.88		&	236.60			\\
	W1--W2		&	107.92		&	99.11			\\
	E2--W2		&	156.27		&	243.23			\\
	S2--W2		&	177.45		&	339.08			\\
	S1--W2		&	210.97		&	340.88			\\
	E1--W2		&	221.82		&	241.27			\\
	E2--S2			&	248.13		&	17.87			\\
	S2--W1		&	249.39		&	317.18			\\
	E2--W1		&	251.34		&	257.73			\\
	S1--W1		&	278.50		&	321.02			\\
	E2--S1			&	278.76		&	14.63			\\
	E1--S2			&	302.33		&	25.70			\\
	E1--W1		&	313.53		&	253.39			\\
	E1--S1			&	330.66		&	22.28			\\
	\hline
	\end{tabular}
  	\caption[Configurations des différents télescopes]{\textbf{Configurations des différents télescopes} : l'angle de projection PA est mesuré par rapport au sud.}
  	\label{table__chara_base}
\end{table}

\subsection{Principe de fonctionnement de FLUOR}

L'instrument \emph{FLUOR} est installé et opérationnel sur \emph{CHARA} depuis 2003 \citep{Coude-du-Foresto-2003-02}. Il permet la recombinaison de faisceaux provenant de deux télescopes en bande $K^\prime$ ($\lambda_0 = 2.15\,\mu\mathrm{m}$ et $\Delta\lambda = 0.3\,\mu\mathrm{m}$), et peut atteindre une précision absolue sur la mesure de visibilité carrée de $1\,\%$.

Le principe de \emph{FLUOR} est d'utiliser une fibre monomode pour filtrer spatialement l'onde entrante. Cela permet de transformer les perturbations du front d'onde liées aux turbulences atmosphériques en fluctuations d'intensité. Ces fluctuations sont mesurables pendant une série d'observations et permettent de corriger les interférogrammes de presque tous les effets de la turbulence, la seule perturbation restante étant le piston.

\begin{figure}[!p]
	\centering\includegraphics[width = .85\linewidth]{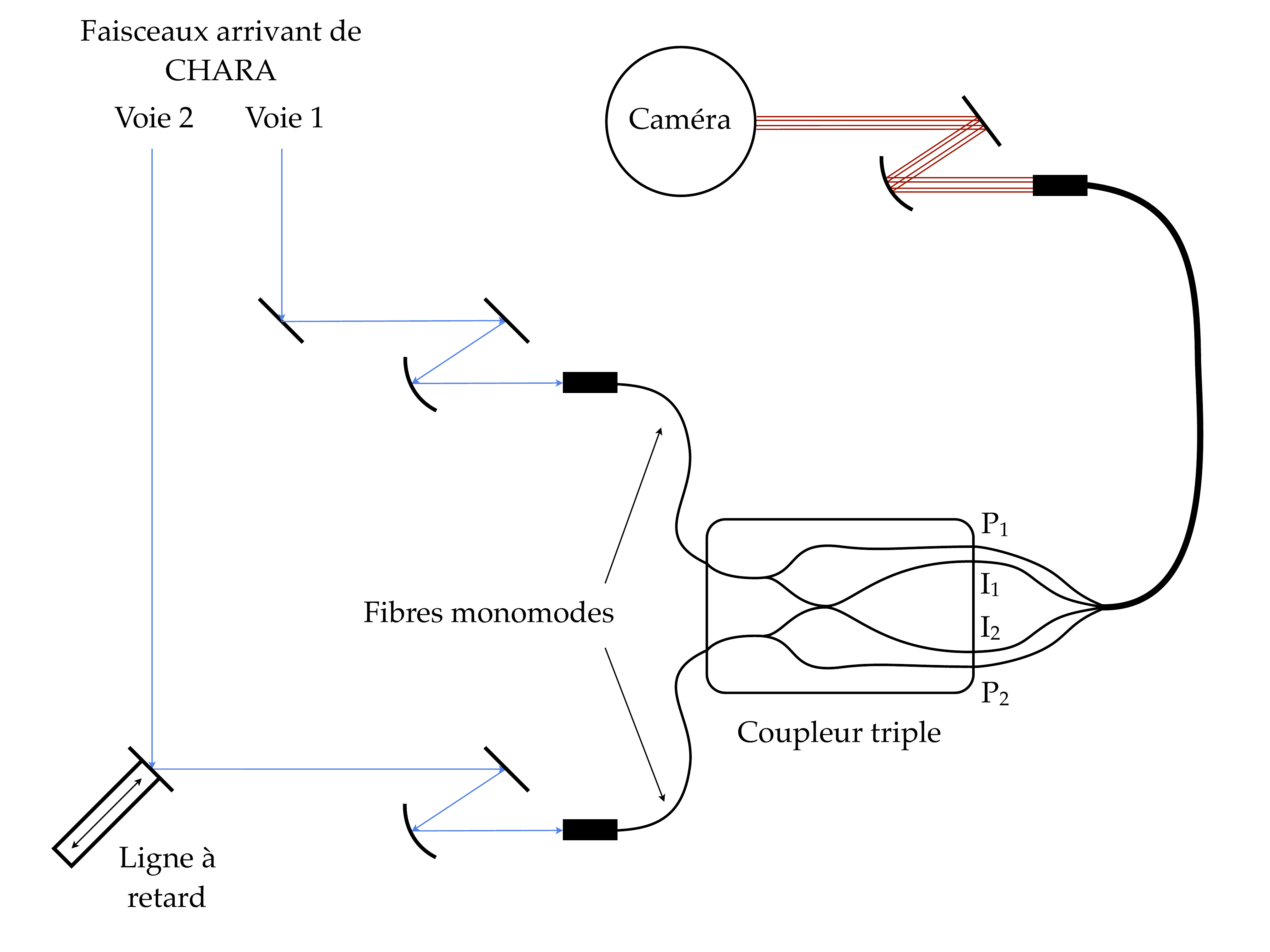}
    \caption[Schéma d'ensemble \emph{FLUOR}]{\textbf{Schéma d'ensemble \emph{FLUOR}} : trajet optique des faisceaux depuis la sortie des lignes à retard jusqu'au détecteur.}
  	\label{image__fluor}
\end{figure}

J'illustre le schéma d'ensemble de \emph{FLUOR} sur la Fig.~\ref{image__fluor}. Les deux faisceaux arrivant des lignes à retard de \emph{CHARA} sont injectés dans les fibres monomodes puis orientés vers le coupleur triple. Ce dernier a pour fonction de prélever les deux signaux photométriques $P_1$ et $P_2$ (nécessaires à la correction des fluctuations d'intensité) et de combiner les faisceaux pour obtenir deux signaux interférométriques $I_1$ et $I_2$ (en opposition de phase). Ces signaux sont ensuite acheminés vers la caméra infrarouge.

Comme la recombinaison se fait dans le plan pupille, les franges doivent être enregistrées en modulant temporellement l'un des faisceaux, c'est à dire en balayant l'interférogramme autour de la différence de marche nulle. C'est la fonction de la ligne à retard interne de \emph{FLUOR}, constituée d'un miroir monté sur un système piézo-électrique permettant une modulation de $\sim 100\,\mu\mathrm{m}$ en $\sim 0.5\,\mathrm{s}$. 

La sensibilité de l'instrument est fixée par la fréquence d'échantillonnage des franges ($500\,\mathrm{Hz}$). La source doit en effet être assez brillante pour avoir un bon rapport signal à bruit en quelques millisecondes. À ce jour, la magnitude limite de \emph{FLUOR} est $K \sim 4$--$5$.

\subsection{Acquisition des interférogrammes}

\begin{figure}[!p]
	\centering\includegraphics[width = .6\linewidth]{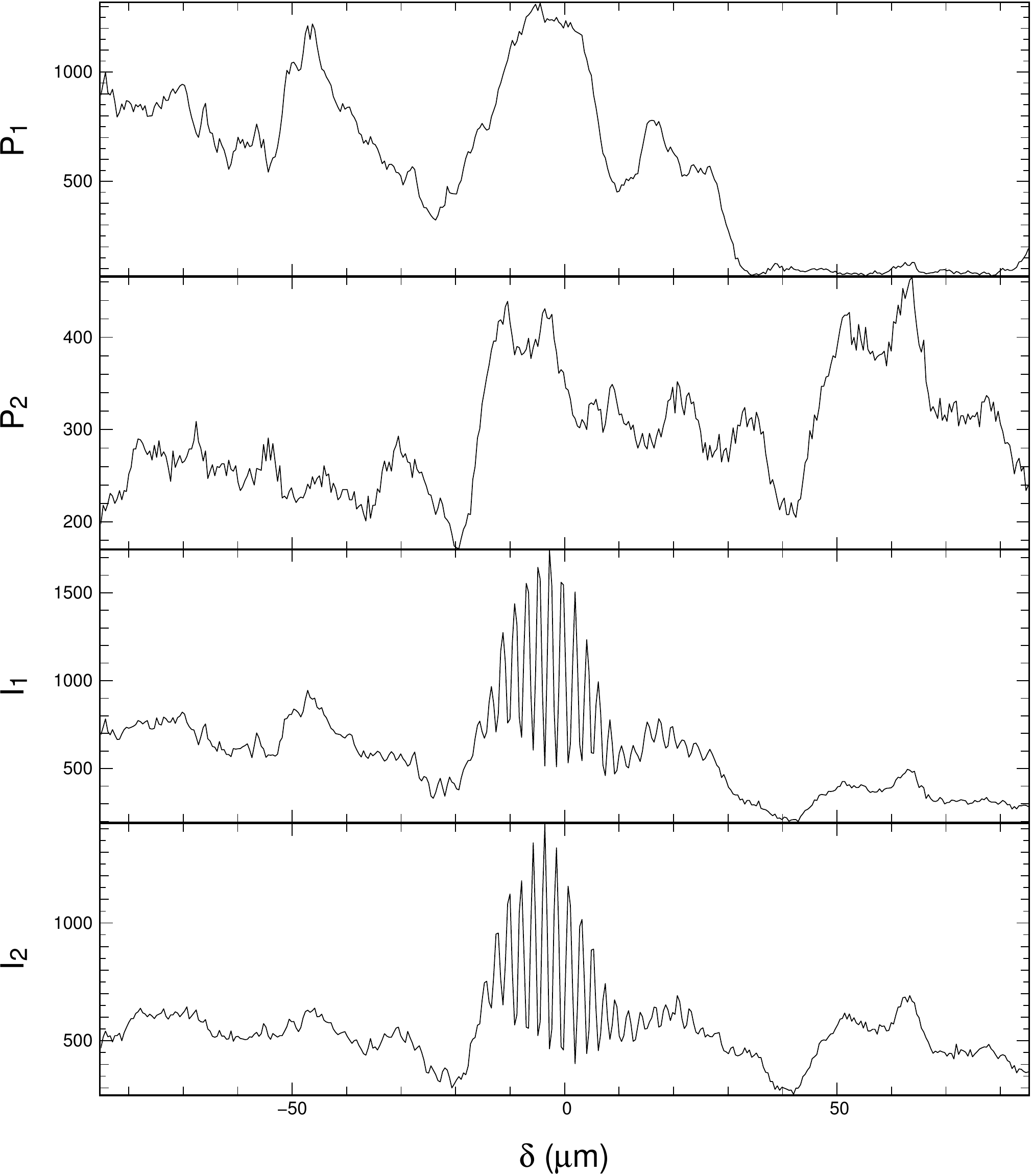}
    \caption[Signaux de sortie]{\textbf{Signaux de sortie} : signaux brutes en fonction de la différence de marche obtenus pour l'étoile Véga. Le flux est en ADU.}
  	\label{image__signaux_sortie_fluor}
\end{figure}

La séquence commence par une acquisition à obturateur fermé de la voie 1, puis 2, puis les deux simultanément, afin de mesurer le courant d'obscurité, l'offset et la matrice des $\kappa$ (que nous verrons dans la section suivante). Chaque interférogramme est ensuite obtenu en balayant la différence de marche $\delta$ de part et d'autre de $\delta = 0$ : c'est le codage temporel. J'expose un exemple des signaux obtenus en sortie de \emph{FLUOR} sur la Fig.~\ref{image__signaux_sortie_fluor}. La séquence se termine par une dernière mesure du courant d'obscurité avec les deux voies fermées.

\begin{figure}[!p]
	\centering\includegraphics[width = .75\linewidth]{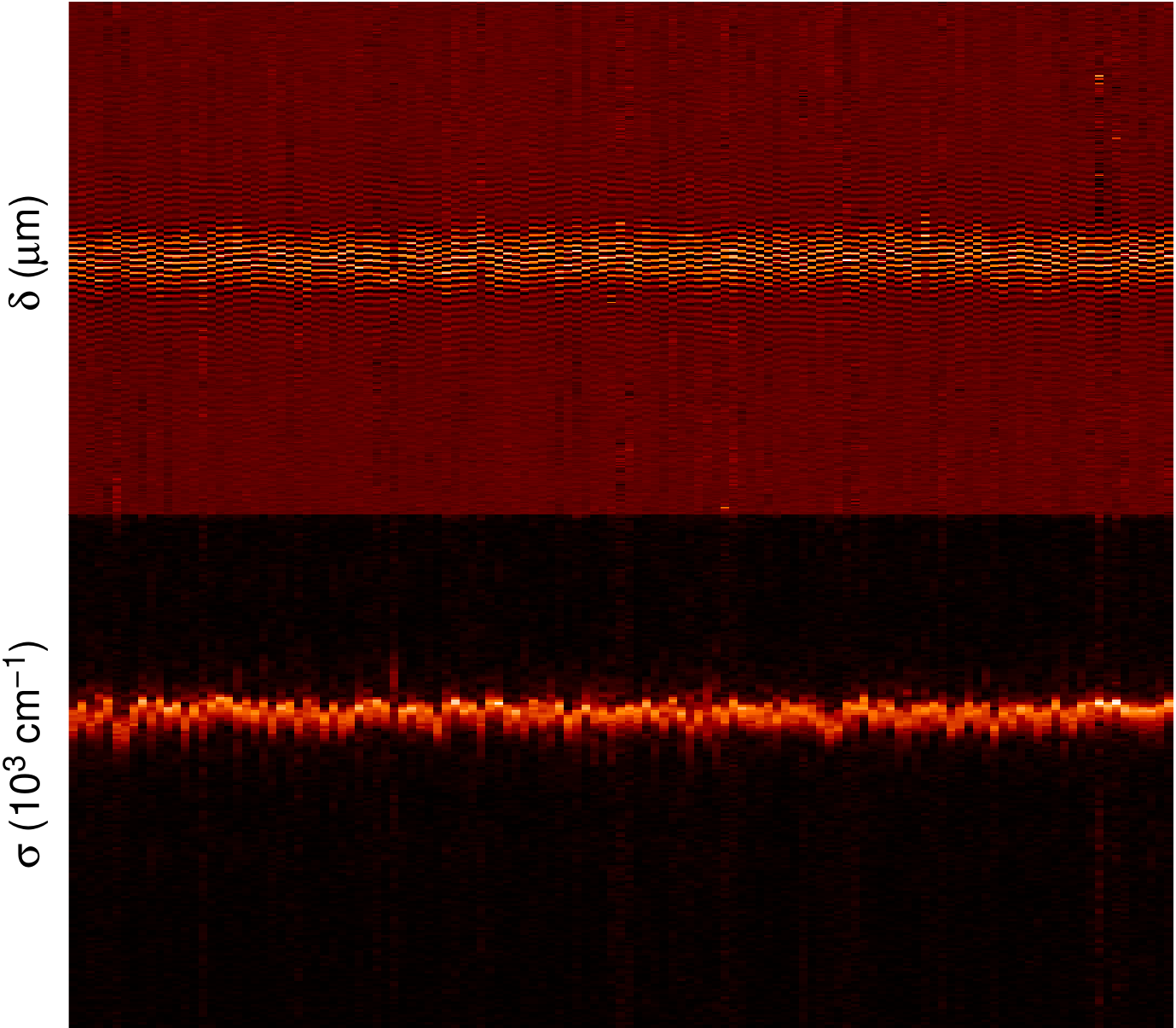}
    \caption[Exemple de série d'acquisition d'interférogrammes]{\textbf{Exemple de série d'acquisition d'interférogrammes} : en haut, franges d'interférences obtenues par soustraction des signaux interférométriques. En bas, puissance spectrale (transformée de \emph{Fourier}) des interférogrammes. Les signaux proviennent de l'étoile VÉGA, présentés sur la Fig.~\ref{image__signaux_sortie_fluor}.}
  	\label{image__scan_fluor}
\end{figure}

Pour chaque étoile, on mesure des centaines d'interférogrammes afin de réduire les erreurs statistiques sur la mesure finale. Un exemple d'une série d'acquisition est présenté sur la Fig.~\ref{image__scan_fluor}

\subsection{Estimation de la visibilité}

La méthode d'estimation des visibilités est exposée en détail par  \citet{Coude-du-Foresto-1997-02} et \citet{Kervella-2004-06}, j'en reprends ici les grandes lignes. J'illustrerai mes propos avec des données obtenues sur l'étoile Véga, car c'est une étoile brillante et la longueur de base utilisée lors des observations était courte (S1--S2), facilitant ainsi l'observation des franges d'interférences. Enfin, le code que j'ai utilisé est celui qui a été développé par \citet{Merand-2006-06}.

L'interférogramme contient une partie basse fréquence, liée aux variations d'intensité dans chaque voie du coupleur, et une partie haute fréquence, liée à la modulation des franges d'interférences. Seule la partie haute fréquence nous intéresse, il faut donc, pour avoir une bonne estimation de la visibilité, corriger ces variations. Les autres sources de bruit que sont le bruit de lecture et de photons doivent également être estimées et supprimées.

\begin{figure}[!p]
	\centering
	\resizebox{\hsize}{!}{\includegraphics{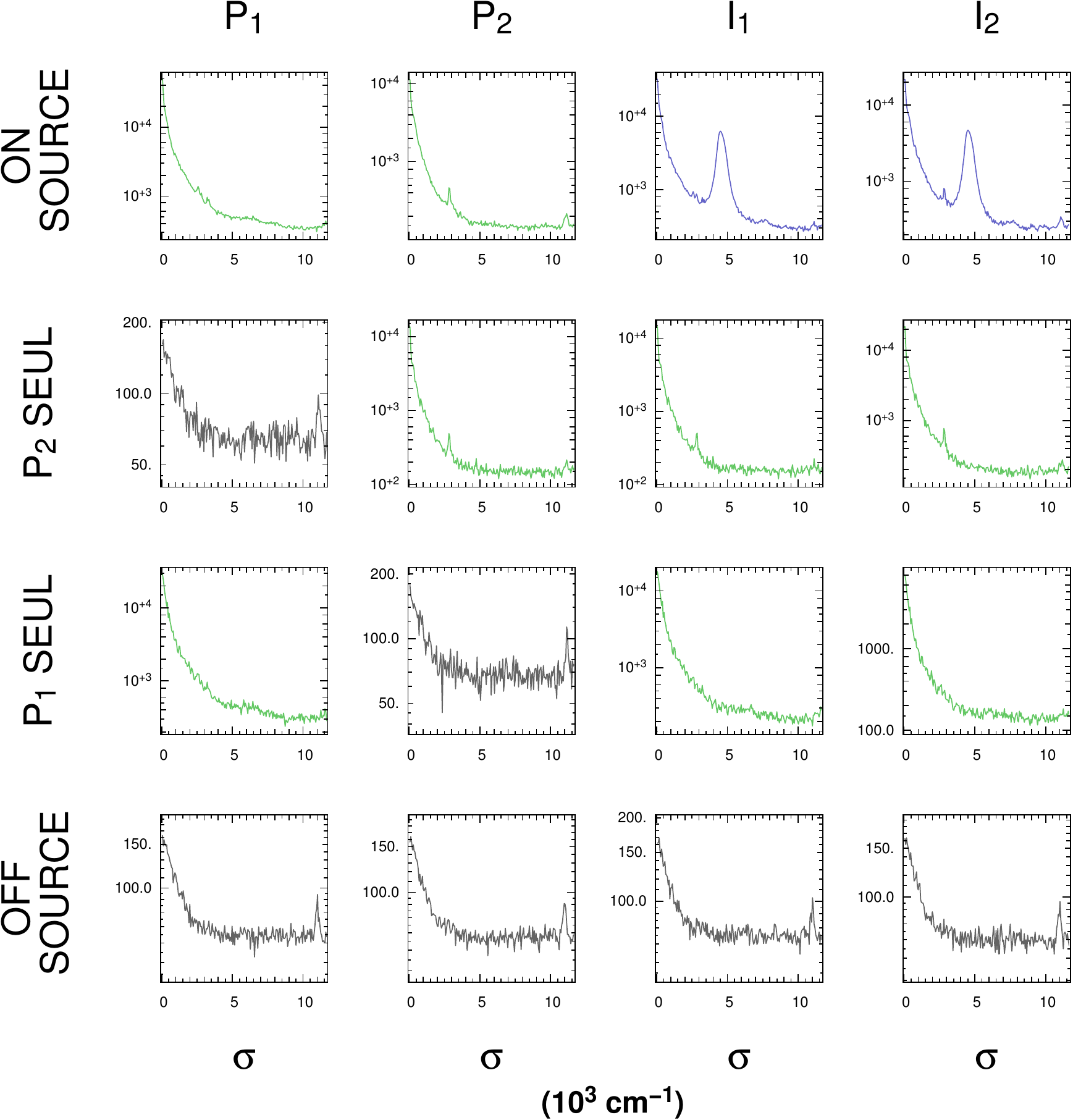}}
    \caption[Spectre de puissance des signaux de sortie]{\textbf{Spectre de puissance des signaux de sortie} : chaque voie est indiquée pour chaque mode d'acquisition. Les données sont celles des Fig.~\ref{image__signaux_sortie_fluor} et \ref{image__scan_fluor} de l'étoile Véga.}
  	\label{image__all_scan_fluor}
\end{figure}

Je présente sur la Fig.~\ref{image__all_scan_fluor}, le spectre de puissance (TF) de toutes les sorties du coupleur avant toute correction pour chaque mode d'acquisition (sur une moyenne de 127 interférogrammes). On remarque clairement à basse fréquence l'effet des fluctuations photométriques. Le pic de franges se situe à $\sigma \sim 5000\,\mathrm{cm}^{-1}$, tandis que l'autre pic plus petit à $\sigma \sim 11000\,\mathrm{cm}^{-1}$ est une caractéristique du bruit de lecture. Le bruit de fond, en moyenne plat pour $\sigma \gtrsim 5000\,\mathrm{cm}^{-1}$, est lié au bruit de photons et le bruit de détecteur.

La première étape consiste à corriger les quatre voies du courant d'obscurité et de l'offset pour estimer le zéro photométrique. Il faut ensuite déterminer la contribution des signaux photométriques aux signaux interférométriques, qui introduit un signal basse fréquence dans l'interférogramme. On cherche alors les coefficients $\overline{\kappa}_\mathrm{ij}$ tels que :
\begin{displaymath}
I_\mathrm{1,BF} = \overline{\kappa}_\mathrm{11}\,P_\mathrm{1} + \overline{\kappa}_\mathrm{12}\,P_\mathrm{2} \quad \mathrm{et} \quad I_\mathrm{2,BF} = \overline{\kappa}_\mathrm{21}\,P_\mathrm{1} + \overline{\kappa}_\mathrm{22}\,P_\mathrm{2}
\end{displaymath}

Pour cela on utilise les signaux obtenus avec obturateur fermé, soient $I_1 = \overline{\kappa}_\mathrm{11}\,P_\mathrm{1}$ et $I_2 = \overline{\kappa}_\mathrm{21}\,P_\mathrm{1}$ pour la voie 1 ouverte et 2 fermée, et $I_1 = \overline{\kappa}_\mathrm{12}\,P_\mathrm{2}$ et $I_2 = \overline{\kappa}_\mathrm{22}\,P_\mathrm{2}$ pour la voie 1 fermée et 2 ouverte. Les coefficients $\overline{\kappa}_\mathrm{ij}$ sont ensuite ajustés par moindres carrés de façon à minimiser $\sum_\delta [I_\mathrm{i,BF} - \overline{\kappa}_\mathrm{i1}\,P_\mathrm{1} - \overline{\kappa}_\mathrm{i2}\,P_\mathrm{2}]^2$, avec $i = 1, 2$ représentant les deux voies interférométriques.

Une fois les coefficients $\overline{\kappa}_\mathrm{ij}$ connus, on déduit la partie modulée de l'interférogramme corrigé en soustrayant la composante basse fréquence et en appliquant un facteur de normalisation :
\begin{displaymath}
I_\mathrm{cor,i}(\delta) = \frac{I_i(\delta) - \overline{\kappa}_\mathrm{i1}\,P_\mathrm{1}(\delta) - \overline{\kappa}_\mathrm{i2}\,P_\mathrm{2}(\delta)}{\sqrt{\overline{\kappa}_\mathrm{i1}\,P_\mathrm{1}(\delta)\,\overline{\kappa}_\mathrm{i2}\,P_\mathrm{2}(\delta)}}
\end{displaymath}

Les deux voies interférométriques sont ensuite soustraites afin de supprimer les fluctuations photométriques résiduelles et le bruit introduit lors de l'étalonnage photométrique :
\begin{displaymath}
I(\delta) = \frac{I_\mathrm{cor,1}(\delta) - I_\mathrm{cor,2}(\delta)}{2}
\end{displaymath}

Par définition, dans le cas monochromatique cette quantité est égale à :
\begin{displaymath}
I_\lambda(\delta) = \frac{1}{2}\,B(\lambda)\,T_\mathrm{r}(\lambda)\,\mu(B,\lambda)\,\cos(2\pi \delta/\lambda + \phi(\lambda)) + S_\mathrm{b}
\end{displaymath}
où $B(\lambda)$ représente le spectre de l'étoile, $T_\mathrm{r}(\lambda)$ la transmission de \emph{FLUOR}, $\mu(B,\lambda)$ le module de la visibilité complexe (Équ.~\ref{equation__mu}), $\phi(\lambda)$ sa phase et $S_\mathrm{b}$ le bruit de lecture et le bruit de photons.

Le module de la visibilité complexe est ensuite estimé à partir de la densité spectrale de puissance (DSP), c'est à dire du carré du module de la transformée de \emph{Fourier} de $I_\lambda(\delta)$ :
\begin{displaymath}
\tilde{I}_\lambda(\delta)^2  = \frac{1}{4}\,B(\lambda)^2\,T_\mathrm{r}(\lambda)^2\,\mu(B,\lambda)^2 + \tilde{S}_\mathrm{b}^2
\end{displaymath}
où le symbole $\ \tilde{}\ $ représente la transformée de \emph{Fourier}.

Toutefois, pour avoir une estimation non biaisée de $\mu^2$, il faut supprimer à la DSP le bruit $\tilde{S}_\mathrm{b}^2 = \tilde{S}_\mathrm{lec}^2 + \tilde{S}_\mathrm{ph}^2$. La DSP du bruit de lecture $\tilde{S}_\mathrm{lec}^2$ est déterminée en utilisant les acquisitions hors-source (HS), c'est à dire avec les mesures obtenues avec les deux obturateurs fermés :
\begin{displaymath}
I_\mathrm{HS,i}(\delta) = \frac{I_\mathrm{HS,i}(\delta) - \overline{\kappa}_\mathrm{i1}\,P_\mathrm{HS,1}(\delta) - \overline{\kappa}_\mathrm{i2}\,P_\mathrm{HS,2}(\delta)}{\overline{\sqrt{\overline{\kappa}_\mathrm{i1}\,P_\mathrm{HS,1}(\delta)\,\overline{\kappa}_\mathrm{i2}\,P_\mathrm{HS,2}(\delta)}}} \quad \Longrightarrow \quad I_\mathrm{HS}(\delta) = \frac{I_\mathrm{HS,1}(\delta) - I_\mathrm{HS,2}(\delta)}{2}
\end{displaymath}
et donc $\tilde{S}_\mathrm{lec}^2 = \tilde{I}_\mathrm{HS}(\delta)^2$. Afin d'éviter de rajouter du bruit aux basses fréquences, on a calculé ici la valeur moyenne (sur tout les mesures) du dénominateur.

Le bruit de photons est déterminé grâce à ses propriétés (bruit blanc) et a donc un spectre uniforme. Sa DSP est estimée en soustrayant à $\tilde{I}_\lambda(\delta)^2,$ sa valeur moyenne à haute fréquence en dehors du pic des franges. Il ne reste plus qu'à estimer la densité spectrale de puissance corrigée $\tilde{I}_\mathrm{corr,\lambda}(\delta)^2 = \tilde{I}_\lambda(\delta)^2 - \tilde{S}_\mathrm{lec}^2 - \tilde{S}_\mathrm{ph}^2$ et d'estimer le module de la visibilité complexe normalisé, en tenant compte de la largeur de bande :
\begin{displaymath}
\mu^2(u,v) = \frac{\int_{\Delta \lambda} B(\lambda)^2\,T_\mathrm{r}(\lambda)^2\,\mu^2(B,\lambda)\,d\lambda}{\int_{\Delta \lambda} B(\lambda)^2\,T_\mathrm{r}(\lambda)^2\,d\lambda}
\end{displaymath}

\begin{figure}[!p]
	\centering
	\resizebox{\hsize}{!}{\includegraphics{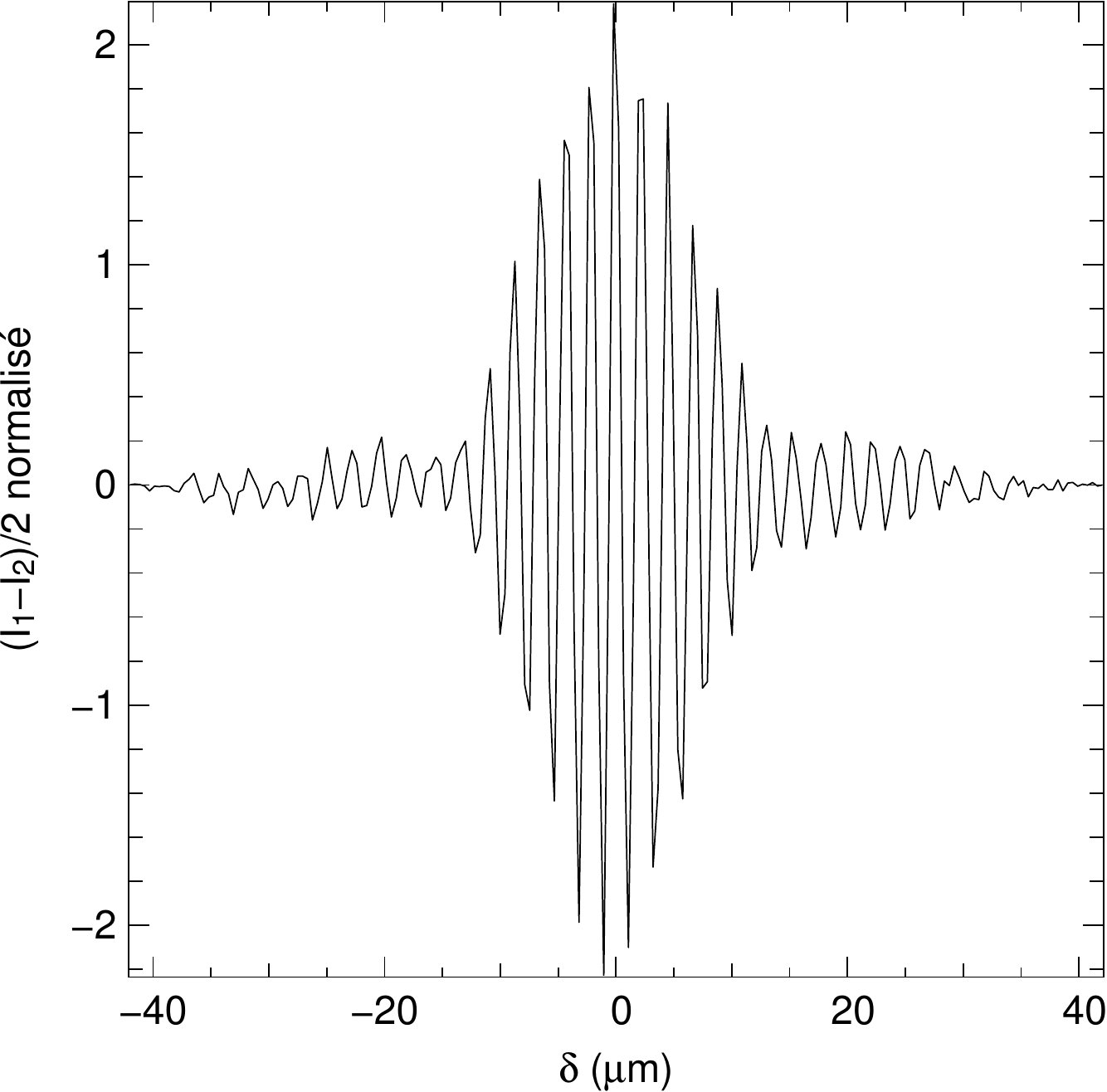} \hspace{.5cm}
	\includegraphics{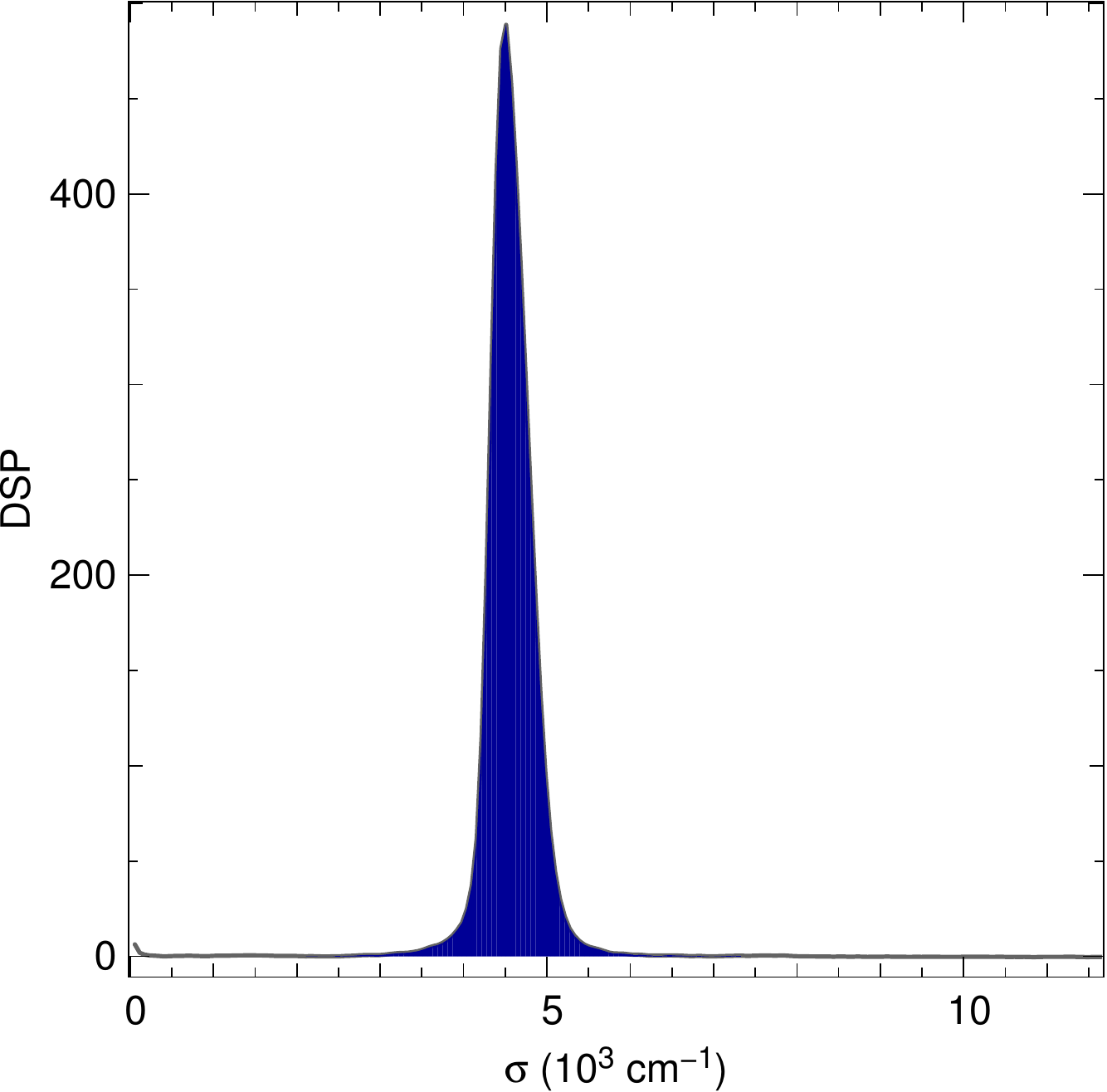}}
    \caption[Densité spectrale de puissance corrigée de l'étoile Véga]{\textbf{Densité spectrale de puissance corrigée de l'étoile Véga}}
  	\label{image__psd_vega}
\end{figure}

La Fig.~\ref{image__psd_vega} illustre les résultats obtenus pour l'étoile Véga. Sur l'image de gauche est représenté le type d'interférogramme obtenu après soustraction des deux voies interférométriques corrigées des variations photométriques. L'image de droite illustre la DSP obtenue sur tous les interférogrammes après corrections. L'intégrale de cette DSP (en bleu) fournira une estimation de $\mu^2(u,v)$.

Le terme de normalisation est appelé facteur de forme et sert à tenir compte de la forme du spectre de la source. Des valeurs typiques de ce paramètres sont données dans \citet{Coude-du-Foresto-1997-02}. Toutefois, il varie peu en bande $K$ d'un type spectral à un autre \citep[][à l'exception des étoiles froides contenant de fortes bandes d'absorption]{Coude-du-Foresto-1997-02}. Notons également que ce facteur a peu d'importance lors de l'étalonnage de la visibilité si la source de référence à le même type spectral que la source d'intérêt scientifique. Cependant, en bande large le facteur de forme n'a pas d'effet notable, j'en donc pas tenu compte dans le code de réduction utilisé.

\subsection{Étalonnage de la visibilité}

Comme expliqué dans la section~\ref{section__les_observables}, le terme obtenu ne correspond pas à la visibilité réelle de l'objet à cause de la transmission atmosphérique et instrumentale, nous mesurons en fait la quantité :
\begin{displaymath}
\mu^2(u,v) = T^2\,V^2(u,v)
\end{displaymath}

En reprenant les notations précédentes :
\begin{displaymath}
V^2(u,v) = \left( \frac{\int_{\Delta \lambda} B(\lambda)^2\,T_\mathrm{r}(\lambda)^2\,\mu^2(B,\lambda)\,d\lambda}{\int_{\Delta \lambda} B_\mathrm{ref}(\lambda)^2\,T_\mathrm{r}(\lambda)^2\,\mu^2_\mathrm{ref}(B,\lambda)\,d\lambda} \right)^2 \left( \frac{\mathcal{F}(\mathrm{ref})}{\mathcal{F}} \right)^2
\end{displaymath}
où $\mathcal{F}^2 = \int_{\Delta \lambda} B(\lambda)^2\,T_\mathrm{r}(\lambda)^2\,d\lambda$ pour la source d'intérêt et de référence.

En pratique, pour un meilleur étalonnage, on observe une étoile étalon avant et après afin d'interpoler la fonction $T^2$ à l'instant de la mesure de l'étoile scientifique.

\subsection{Choix des étoiles étalons}

On ne peut bien sûr pas prendre n'importe quelle étoile comme source de référence. Divers critères de sélection doivent être pris en compte, voici les principaux :

\begin{itemize}
	\compactlist
	\item la source de référence doit avoir un type spectral connu pour l'évaluation du facteur de forme.
	\item elle doit être proche (dans le plan ciel) de la source d'intérêt pour 1) être dans les mêmes conditions atmosphériques et 2) éviter un temps de déplacement des télescopes trop long.
	\item Son diamètre angulaire doit être connu avec la meilleure précision possible pour une base donnée. En effet, l'erreur sur le diamètre de la source de référence se propage sur l'erreur de la visibilité lors de l'étalonnage.
\end{itemize}

\subsection{Estimation de l'erreur sur la visibilité}

L'erreur finale sur la visibilité de l'objet d'intérêt se compose des erreurs statistiques et systématiques. Les erreurs statistiques sont liées à l'estimation du module de la visibilité complexe $\mu^2$ pour l'étoile d'intérêt et de référence. L'erreur statistique de chaque étoile est estimé à partir de la méthode de bootstrapping, c'est à dire  que l'on effectue $n$ tirages aléatoires avec remise de $N$ valeurs afin d'en étudier la dispersion statistique (en général $n > 1000$).

L'erreur systématique est liée à l'estimation de la visibilité théorique des étoiles étalons observées avant et après. En prenant le modèle du disque uniforme, l'erreur sur la visibilité théorique d'une étoile de référence est :
\begin{displaymath}
e_\mathrm{V^2} = 4\,V\,J_2(x)\,\frac{\sigma_{\theta_\mathrm{UD}}}{\theta_\mathrm{UD}}
\end{displaymath}
avec $x = \pi \theta B/\lambda$. On remarque l'intérêt d'avoir une étoile étalon avec une mesure précise de son diamètre. L'erreur systématique finale est évaluée en suivant la méthode de \citet{Perrin-2003-03} qui tient compte d'éventuelles corrélations entre deux étoiles de référence.

L'erreur finale sur la visibilité de l'étoile est finalement la somme quadratique des erreurs statistiques et systématiques.

\subsection{L'échantillon d'étoiles}

Nous avons observé un échantillon de 17 Céphéides avec des longueurs de base allant de $150$ à $331\,\mathrm{m}$. Je ne présenterai ici que les premiers résultats de quelques étoiles uniquement car ces données sont toujours en cours de réduction et d'analyse. J'ai choisi les Céphéides Y~Oph, S~Sge, U~Vul et R~Sct dont le journal de ces observations est présenté dans la Table~\ref{table__log_cepheide_chara}.

Chaque visibilité présentée dans la Table~\ref{table__log_cepheide_chara} a été étalonnée en utilisant des étoiles de référence du catalogue \citet{Merand-2005-04}. Leurs propriétés sont exposées dans la Table~\ref{table__etoile_reference_propriete}.

L'objectif est d'appliquer la méthode de Baade--Wesselink (BW), c'est à dire de déterminer grâce aux variations du diamètre angulaire, la distance de l'étoile et son rayon moyen. Les éléments nécessaires sont des mesures de vitesse radiale (prises dans la littérature), des mesures de diamètre angulaire fournis par \emph{FLUOR} et les éphémérides de l'étoile (période $P$ et époque de référence $T_0$).

\begin{table}[!p]
	\centering
	\begin{tabular}{ccccc} 
	\hline
	\hline
	Nom			&   MJD				&	B$_\mathrm{p}$ (m)	& PA ($^\circ$)	& $V^2$	\\
	\hline
     Y~Oph		&  55353.363 	&  234.04 &  -47.69  	&  $0.2700\pm0.0143$  \\
     				&  55353.385 	&  248.27 &  -49.81  	&  $0.2361\pm0.0125$   \\
     				&  55355.318 	&  286.90 &  35.570  	&  $0.0994\pm0.0050$  \\
     				&  55356.272 	&  310.27 &  39.690  	&  $0.0524\pm0.0070$  \\
     				&  55356.363 	&  257.71 &  27.420  	&  $0.1567\pm0.0191$  \\
     				&  55362.324 	&  271.74	&  31.860  	&  $0.1362\pm0.0056$  \\
     				&  55363.344 	&  248.55	&  79.590  	&  $0.2236\pm0.0106$  \\
     				&	55364.306 	&	154.75	&	69.44 		&	$0.5822\pm0.0200$	  \\	
     				&  55366.248 	&  219.79 &  66.330  	&  $0.3569\pm0.0145$  \\
	\hline
	U~Vul		& 55355.458 	& 316.44 	& 17.53 		& $0.4211\pm0.0156$ \\ 
					& 55361.459 	& 300.15 	& 67.37 		& $0.4457\pm0.0194$ \\ 
					& 55362.489 	& 310.10 	&  5.36 		& $0.4137\pm0.0164$ \\ 
					& 55366.396 	& 220.46 	& 63.25 		& $0.7120\pm0.0143$ \\ 
					& 55778.302 	& 316.09 	& 17.06 		& $0.4162\pm0.0197$ \\ 
					& 55778.402 	& 310.72 	& -7.43 		& $0.4164\pm0.0088$ \\ 
					& 55778.454 	& 318.20 	& -19.72 		& $0.3933\pm0.0152$ \\ 
					& 55779.304 	& 315.30 	& 15.98 		& $0.4192\pm0.0085$ \\ 
					& 55779.329 	& 311.79 	& 10.05 		& $0.4295\pm0.0054$ \\ 
					& 55780.258 	& 322.90 	& 25.02 		& $0.4430\pm0.0064$ \\ 
					& 55780.373 	& 309.48 	& -1.60 		& $0.4521\pm0.0071$ \\ 
					& 55781.341 	& 280.66 	& 61.55 		& $0.5638\pm0.0097$ \\ 
					& 55784.320 	& 278.34 	& -49.49 		& $0.5122\pm0.0098$ \\ 
					& 55786.338 	& 309.65 	&  3.14 		& $0.4202\pm0.0038$ \\ 
					& 55786.431 	& 320.36 	& -12.86 		& $0.4078\pm0.0058$ \\ 
					& 55787.320 	& 310.54 	&  6.89 		& $0.4409\pm0.0106$ \\ 
					& 55790.255 	& 250.12 	& 75.89 		& $0.6293\pm0.0050$ \\ 
	\hline
     S~Sge		& 55366.414 	& 218.76 	& 63.04 		& $0.6763\pm0.0110$ \\ 
					& 55778.279 	& 317.86 	& 25.14 		& $0.4497\pm0.0210$ \\ 
					& 55778.379 	& 301.34 	&  1.71 		& $0.4436\pm0.0262$ \\ 
					& 55779.353 	& 302.93 	&  7.80 		& $0.4472\pm0.0139$ \\ 
					& 55779.375 	& 301.38 	&  2.09 		& $0.4437\pm0.0143$ \\ 
					& 55780.392 	& 301.53 	& -3.13 		& $0.4321\pm0.0099$ \\ 
					& 55781.360 	& 271.36 	& 62.07 		& $0.4896\pm0.0078$ \\ 
					& 55782.340 	& 230.12 	&  4.26 		& $0.6072\pm0.0186$ \\ 
					& 55782.362 	& 229.76 	& -0.89 		& $0.5955\pm0.0196$ \\ 
					& 55784.345 	& 278.21 	& -50.62 		& $0.4766\pm0.0113$ \\ 
					& 55786.384 	& 302.01 	& -5.23 		& $0.4826\pm0.0069$ \\ 
					& 55787.365 	& 301.29 	& -1.15 		& $0.4637\pm0.0081$ \\ 
     \hline
      R~Sct	    &  55366.290 &  220.280  &  66.580  	&  $0.0953\pm0.0027$  \\
	\end{tabular}
  	\caption[Journal des observations de quelques Céphéides]{\textbf{Journal des observations de quelques Céphéides} : MJD représente le jour Julien modifié (MJD = JD - 2~400~000.5), $B_\mathrm{p}$ est la base projetée, PA correspond à l'angle de projection de la base et $V^2$ dénote la visibilité carrée étalonnée.}
  	\label{table__log_cepheide_chara}
\end{table}

\begin{table}[!p]
	\centering
	\begin{tabular}{ccccc} 
	\hline
	\hline
	Étoile de		& Cible		&	Type			& Diamètre de disque	& 	K		\\
	référence		&				&	spectral		& uniforme (mas)			& 			\\
	\hline
    HD~162113 	&  Y~Oph	&	K0~III  		&  $0.904\pm0.012$ 	&  4.5  \\
    HD~169113 	&  				&	K1~III	 	&  $0.897\pm0.011$  	&  3.6   \\
	\hline
	HD~187193 	&  U~Vul	&	K0~II	-III	&  $0.893\pm0.012$  	&  3.9  \\
    \hline
    HD~187193 	&  S~Sge	&	K0~II	-III	&  $0.893\pm0.012$  	&  3.9  \\
    \hline
    HD~169113 	&  R~Sct		&	K1~III	 	&  $0.897\pm0.011$  	&  3.6   \\
     \hline
	\end{tabular}
  	\caption[Paramètres des étoiles de référence]{\textbf{Paramètres des étoiles de référence}}
  	\label{table__etoile_reference_propriete}
\end{table}

\subsubsection{Y~Oph}

\defcitealias{Merand-2007-08}{M07}

Pour cette étoile j'ai suivi le même procédé que \citet[][\ciap \citetalias{Merand-2007-08}]{Merand-2007-08} afin de compléter l'échantillon de points et d'estimer avec une meilleure précision la distance de Y~Oph. Ces auteurs ont déterminé une distance de l'étoile de $491\,\pm\,18\,$pc et un rayon moyen $R = 67.8\,\pm\,2.5\,\mathrm{R_\odot}$. Ils ont également détecté une enveloppe circumstellaire contribuant pour $5\,\pm\,2$\,\% du flux photosphérique. En appliquant un modèle d'étoile entourée d'une couronne sphérique, j'ai pu déterminer une taille angulaire de cette enveloppe. 

\begin{figure}[!p]
  \centering\includegraphics[width = .65\linewidth]{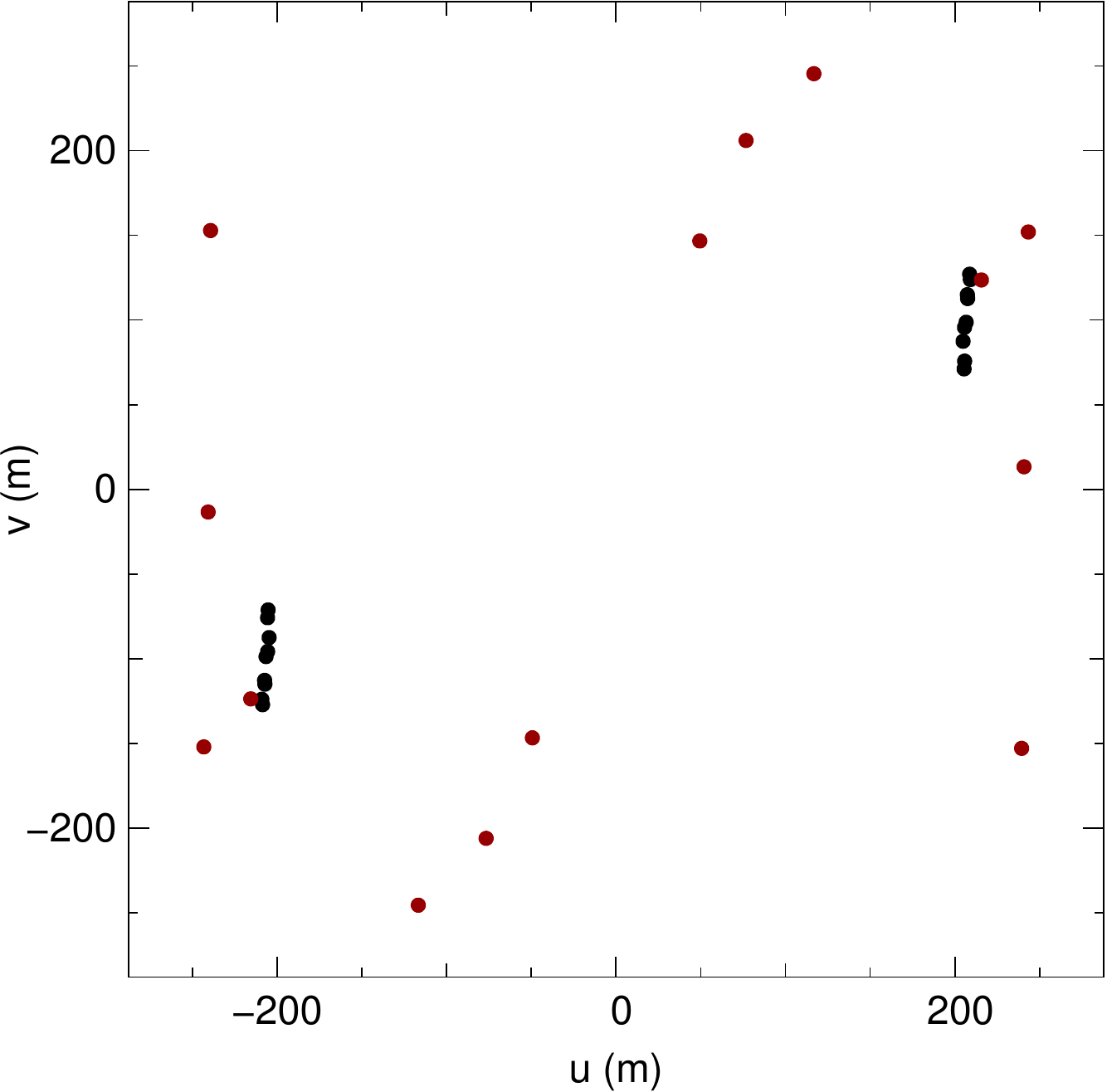}
  \caption[Plan $(u,v)$ des observations de Y~Oph]{\textbf{Plan $(u,v)$ des observations de Y~Oph} : les points en noir représentent les observations de \citetalias{Merand-2007-08} et les rouges nos nouveaux points de mesures.}
  \label{image__plan_uv_y_oph}
\end{figure}

Nous disposons au total de 16 points de visibilité sur cette étoile, je présente ces différentes mesures sur le plan $(u,v)$ de la Fig.~\ref{image__plan_uv_y_oph}.

\paragraph*{\textcolor{black}{Application de la méthode de Baade--Wesselink}}

Pour chaque nouvelle nuit d'observation, j'ai ajusté un modèle simple de disque uniforme sur les points de visibilité de la Table~\ref{table__log_cepheide_chara}. J'ai ensuite calculé une phase de pulsation moyenne par nuit en utilisant les éphémérides de \citet{Fernie-1995-01} et en tenant compte du changement de période de pulsation avec le temps de cette Céphéide ($7.2\pm1.5\,\mathrm{s}\,\mathrm{yr}^{-1}$), grâce à la relation \citep{Fernie-1995-09} :
\begin{displaymath}
	\begin{aligned}
		D =\,& JD - 24400007.720 \\
		E =\,& 0.05839\,D - 3.865\times 10^{-10}D^2 \\
		P =\,& 17.12507 + 3.88\times 10^{-6}E
		\end{aligned}
\end{displaymath}
où $JD$ est la date des observations et $P$ la période de pulsation.

\begin{figure}[!p]
  \centering\includegraphics[width = .7\linewidth]{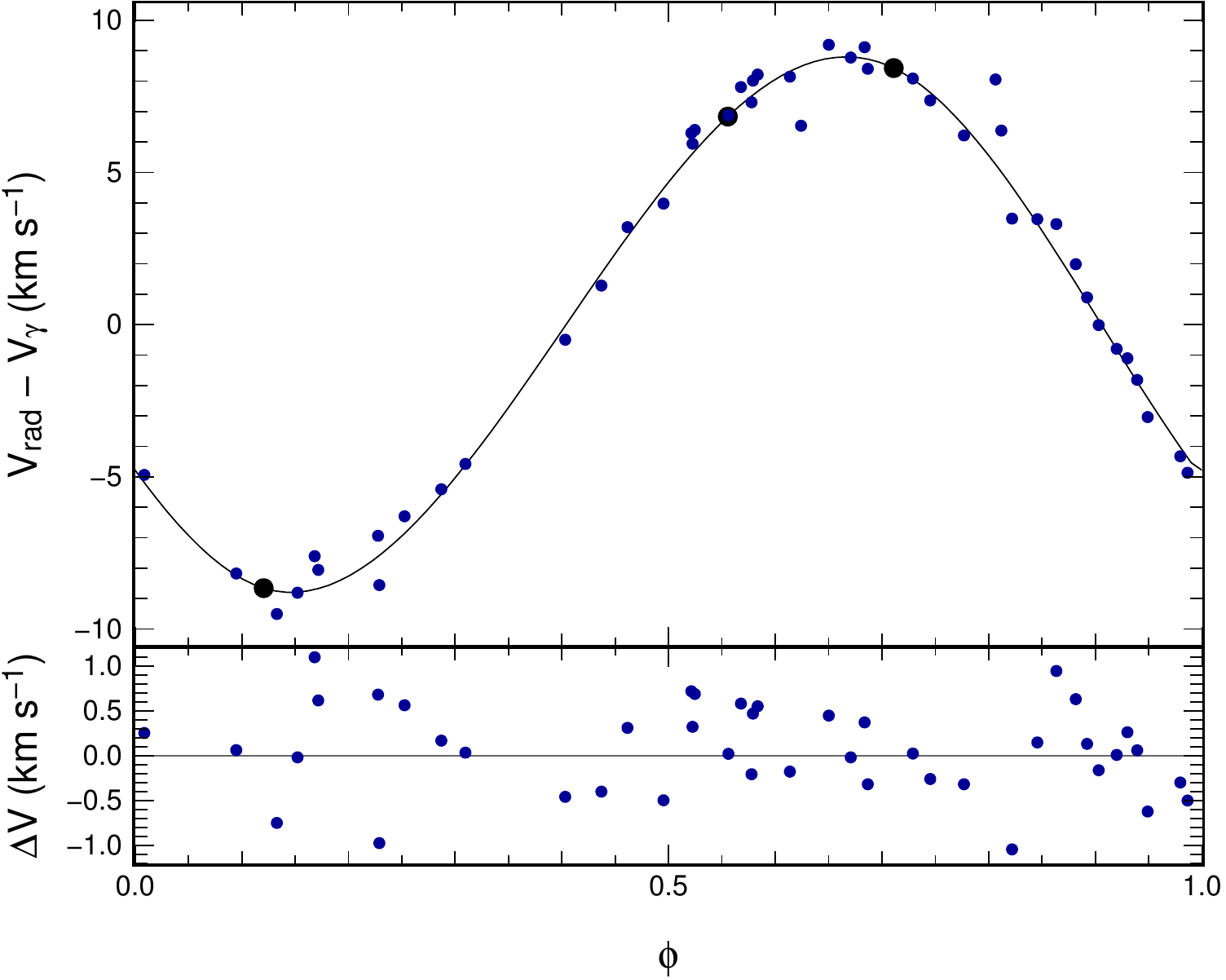}
  \caption[Courbe de vitesse radiale de Y~Oph]{\textbf{Courbe de vitesse radiale de Y~Oph} : la courbe continue est la fonction de splines ajustée sur les trois points flottants illustrés en noir. La vitesse du centre de masse ($V_\gamma = -7.9\pm0.1\,\mathrm{km s}^{-1}$) a été soustraite. Les résidus sont exposés dans la fenêtre du bas.}
  \label{image__vitesse_radiale_y_oph}
\end{figure}

J'ai utilisé les mêmes données de vitesse radiale que \citetalias{Merand-2007-08}, interpolées par une fonction de splines périodiques (comme dans le Section~\ref{section__imagerie_de_cepheides_avec_visir}), afin d'avoir un intervalle identique entre chaque point (pour une meilleure intégration). Les données sélectionnées proviennent de \citet{Gorynya-1998-11} et ont été choisies pour la bonne couverture en phase des points. La décroissance avec le temps de l'amplitude photométrique en bande $B$ et $V$ n'a pas d'effet significatif sur les mesures de vitesse radiale, par conséquent bien que les mesures spectroscopiques et interférométriques sont séparées de plusieurs années, aucune correction n'a été appliquée. La courbe de vitesse radiale interpolée est présentée sur la Fig.~\ref{image__vitesse_radiale_y_oph}.

\begin{figure}[!p]
  \centering\includegraphics[width = .7\linewidth]{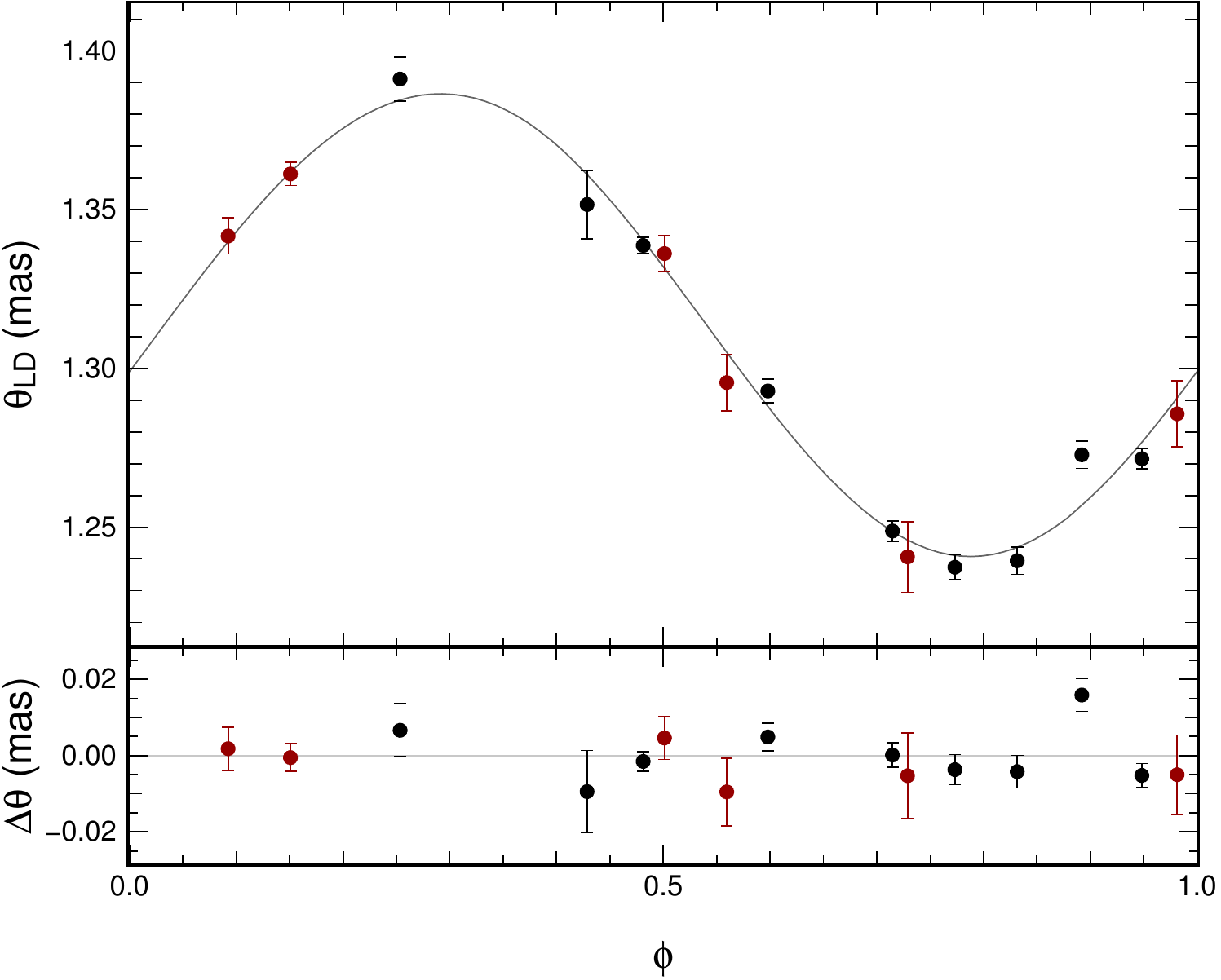}
  \caption[Variation du diamètre angulaire de Y~Oph]{\textbf{Variation du diamètre angulaire de Y~Oph} : nos points supplémentaires sont illustrés en rouge et ceux de \citetalias{Merand-2007-08} en noir. La courbe en continu représente l'équation~\ref{equation__BW}. Les résidus sont exposés dans la fenêtre du bas.}
  \label{image__variation_diametre}
\end{figure}

J'ai ensuite estimé la distance en appliquant la méthode BW sur les mesures de diamètre angulaire de \citetalias{Merand-2007-08} complétées de nos nouveaux points en ajustant la fonction :
\begin{equation}
\theta_\mathrm{UD}(T) - \theta_\mathrm{UD}(0) = -2\,\frac{kp}{d} \int_0^T (v_\mathrm{rad}(t) - v_\gamma)\,dt
\label{equation__BW}
\end{equation}
avec $k = \theta_\mathrm{UD}/\theta_\mathrm{LD}$, $p$ est le facteur de projection (présenté dans le Chapitre~\ref{chapitre__mesurer_l_univers_les_cephéides}), $v_\gamma$ est la vitesse du centre des masses et $d$ dénote la distance. Les paramètres ajustés sont le diamètre moyen $\theta_\mathrm{UD}(0)$ et le coefficient $k\,p/d$. La courbe ajustée du déplacement radial est exposée sur la Fig.~\ref{image__variation_diametre}. Notons qu'un décalage de phase de $0.074\,\pm\,0.003$ entre les nouvelles observations interférométriques et spectroscopiques a été ajusté. La mesure à courte base (154\,m) n'a pas été utilisée car elle est beaucoup moins sensible aux variations du diamètre angulaire, je l'utiliserai par la suite pour l'étude de l'enveloppe circumstellaire.

Ma première approche pour estimer la distance fut d'utiliser le même formalisme et valeurs de paramètres que \citetalias{Merand-2007-08}, c'est à dire que j'ai également supposé $k$ et $p$ constant avec la pulsation (pas de preuves observationnelles du contraire existent pour la moment), le facteur $p = 1.27$ mesuré par \citet{Merand-2006-} pour $\delta$~Cep ($P = 5.37$\,j) a été utilisée pour Y~Oph, et la valeur $k = 1.023$ est adoptée pour tenir compte de la présence de l'enveloppe circumstellaire. Je trouve une distance $d = 491\,\pm\,12$\,pc et un rayon moyen $R = 67.8\,\pm\,1.7\,\mathrm{R_\odot}$. Ces valeurs sont identiques à celles estimées par \citetalias{Merand-2007-08}, avec une précision légèrement meilleure.

Cependant la valeur du facteur $p$ utilisée provient de mesures effectuées sur une autre Céphéide, de période de pulsation différente, et peut ne pas être adapté à Y~Oph. J'ai donc décidé comme seconde approche d'utiliser la relation $p = -0.08_{\pm0.05}\,\log P + 1.31_{\pm0.06}$ de \citet{Nardetto-2009-05}, liant le facteur $p$ à la période de pulsation. Cela donne $p = 1.21 \pm 0.09$ pour Y~Oph. Je trouve finalement $d = 468\,\pm\,12$\,pc et $R = 64.6\,\pm\,1.6\,\mathrm{R_\odot}$. Ces valeurs ne diffèrent pas significativement et sont identiques à $1\sigma$ à celles de \citetalias{Merand-2007-08}.

\paragraph*{\textcolor{black}{Ajustement de l'enveloppe}}

Pour obtenir des paramètres physiques de l'enveloppe, nous avons besoin de plusieurs mesures à courtes bases. Les données de \citetalias{Merand-2007-08} ne contiennent pas de telles mesures et les nouvelles observations n'en contiennent malheureusement qu'une. J'ai alors combiné à ces observations \emph{FLUOR} les mesures de visibilités effectuées par \citet{Kervella-2004-03a} avec l'instrument \emph{VINCI}, fonctionnant dans la même gamme de longueur d'onde, et dont les observations n'ont été réalisées qu'avec des bases relativement courtes ($B_\mathrm{p} = 129$--$139$\,m). Il est maintenant possible de remonter à certains paramètres de l'enveloppe en ajustant un modèle adéquat. 

Malheureusement, dans le cas des étoiles variables, si les mesures n'ont pas été effectuées sur une même nuit ou sur un temps négligeable devant la période de pulsation, la courbe de visibilité varie avec la variation du diamètre angulaire, rendant toute application d'un modèle de visibilité incohérente. Pour corriger de ces variations, j'ai suivi la méthode de \citet{Merand-2006-07} qui consiste à tracer la visibilité mesurée en fonction d'une pseudo-base $B_\mathrm{\theta_0}$ définie telle que $B_\mathrm{\theta_0} = B_\mathrm{p}(\phi)\,\theta_0/\theta (\phi)$, où $\phi$ représente la phase de pulsation de la Céphéide. Le diamètre $\theta (\phi)$ est déterminé pour chaque date d'observation à partir de sa courbe de variation (Fig.~\ref{image__variation_diametre}). Pour une Céphéide sans enveloppe circumstellaire, il en résulte un profil de visibilité $V^2(B_\mathrm{\theta_0})$ d'une étoile de diamètre $\theta_0$.

\begin{figure}[!p]
  \centering\includegraphics[width = .985\linewidth]{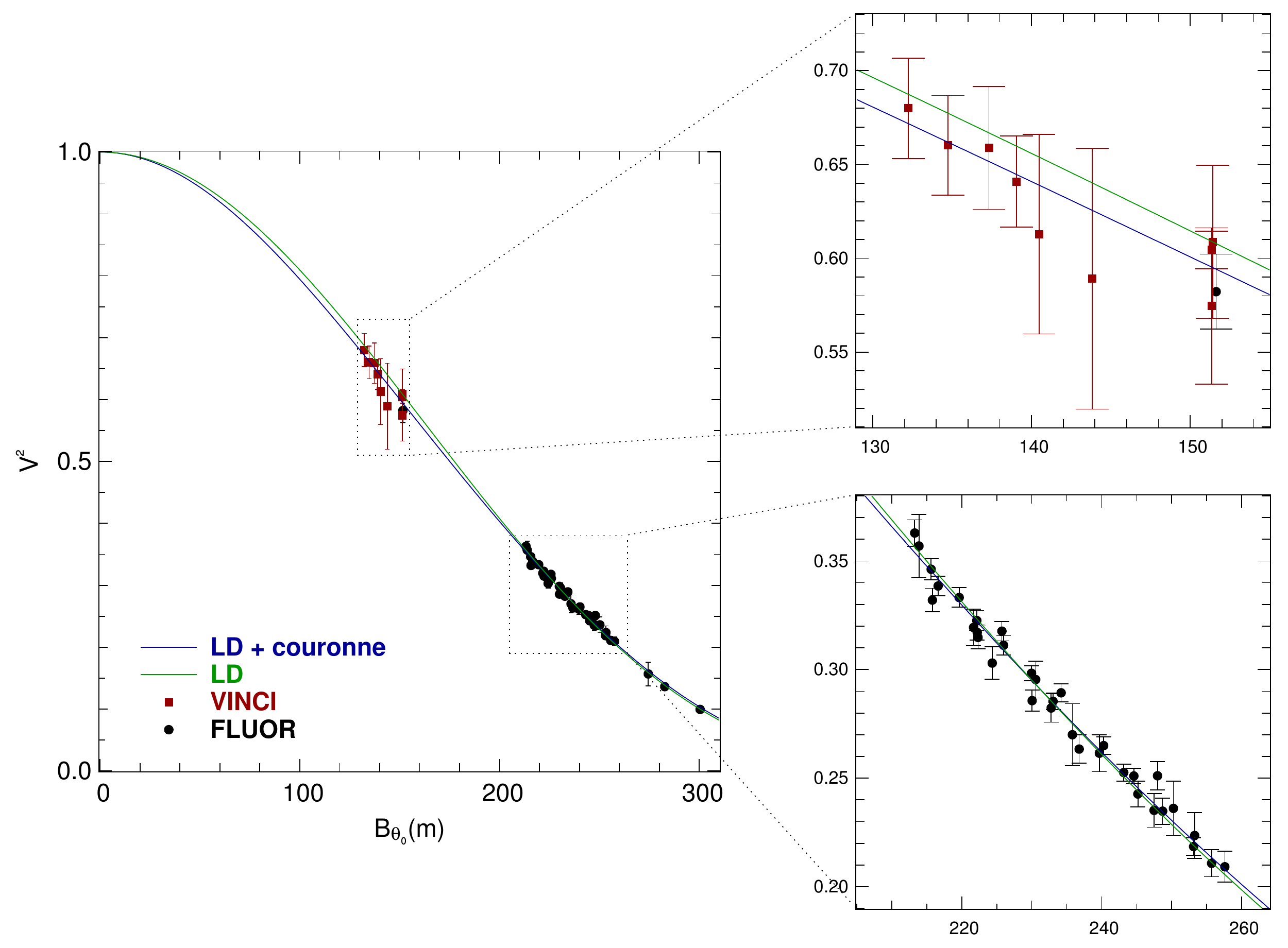}
  \caption[Fonction de visibilité de Y~Oph]{\textbf{Fonction de visibilité de Y~Oph} : la pseudo-base est choisie pour $\theta_0 = 1.28$\,mas. Deux modèles ont été ajustés : un modèle de disque assombri (vert) et un modèle de disque assombri entouré d'une couronne sphérique (bleu).}
  \label{image__visibilite_y_oph}
\end{figure}

La Fig.~\ref{image__visibilite_y_oph} représente la fonction de visibilité de Y~Oph en utilisant le diamètre moyen $\theta_0 = 1.28\,\mathrm{mas}$ (déterminé précédemment) pour la pseudo-base. J'ai commencé par ajuster un modèle de disque assombri (Eq~\ref{equation__LD}). J'ai fixé $n = 0.14$, déterminé à partir des modèles hydrostatiques tabulés de \citet[][pour les paramètres moyens $T_\mathrm{eff} = 5500$\,K, $\log~g = 1.5$, M/H = 0.05 et $v_\mathrm{turb} = 4\,\mathrm{km s^{-1}}$]{Claret-2000-11}. Le seul paramètre libre est le diamètre angulaire. Le résultat est tracé sur la Fig.~\ref{image__visibilite_y_oph} en vert et le diamètre angulaire est exposé dans la Table~\ref{table__visibilite_fit_y_oph}. Le cas où $n$ et $\theta_\mathrm{LD}$ sont laissés libres n'est pas cohérent avec ce que l'on attend pour une Céphéide, car l'étoile serait fortement assombrie et un diamètre angulaire beaucoup trop grand.

À la vue des visibilités légèrement plus faibles mesurées par \emph{VINCI} (en rouge), j'ai décidé d'ajuster un autre modèle, celui d'un disque assombri entouré d'une couronne sphérique (Équ.~\ref{equatio__LD_couronne}). Ce modèle est illustré en bleu sur la Fig.~\ref{image__visibilite_y_oph} où j'ai gardé $n$ et $\epsilon$ fixes \citepalias[$\epsilon = F_\mathrm{cse}/F_\star = 0.05$ d'après][]{Merand-2007-08}. Les autres paramètres laissés libres ($\tau, \theta_\mathrm{LD}$ et $\theta_\mathrm{cse}$) sont exposés dans la Table~\ref{table__visibilite_fit_y_oph}.

On constate que le premier modèle donne un $\chi^2$ réduit légèrement meilleur. Cependant, le second modèle semble plus pertinent car le diamètre de l'enveloppe est mesuré à plus de $3\sigma$. Cela confirme le résultat de \citetalias{Merand-2007-08} concernant la présence d'une enveloppe circumstellaire autour de Y~Oph. Son extension spatiale est également compatible avec la taille typique des enveloppes autour Céphéides mesurée jusqu'à présent ($5.4$\,mas pour $\ell$~Car, $8.6$\,mas pour Polaris et $3.5$\,mas pour $\delta$~Cep).

\begin{table}[!p]
	\centering
	\begin{tabular}{ccccccc} 
	\hline
	\hline
	Modèles					& $\theta_\mathrm{LD}$	&	$n$		&	$\theta_\mathrm{cse}$	&	$\tau$			&	$\epsilon$	&	$\chi^2_r$	\\
	\hline	
     Disque	assombri		&  $1.33\pm0.01$ 	&	\textbf{0.14}	&  --  						&  --							&	--					&	1.04				\\
	\hline	
     Disque	assombri		&  $1.47\pm0.12$ 	&	$1.0\pm0.8$&  --  						&  --							&	--					&	1.01				\\
     \hline
    Disque assombri 		&	$1.32\pm0.01$		&	\textbf{0.14}	&	$4.54\pm1.13$		&	$0.011\pm0.006$	& \textbf{0.05\,$\pm$\,0.02}	&	0.92	\\
      + couronne				&  		&					&   						&  							&						&					\\
     \hline
	\end{tabular}
  	\caption[Résultats de l'ajustement de la visibilité de Y~Oph]{\textbf{Résultats de l'ajustement de la visibilité de Y~Oph} : les paramètres en gras sont gardés fixes. $\tau$ représente la profondeur optique, $\epsilon$ est le rapport de flux entre l'enveloppe et l'étoile, et $n$ correspond au paramètre d'assombrissement. $\theta_\mathrm{LD}$ et $\theta_\mathrm{cse}$ représente respectivement le diamètre de l'étoile et de la couronne sphérique. $\chi^2_r$ dénote le $\chi^2$ réduit.}
  	\label{table__visibilite_fit_y_oph}
\end{table}


\subsubsection{U~Vul et S~Sge}

J'ai utilisé pour l'estimation de la phase de pulsation les éphémérides provenant de \citet{Samus-2009-01} : $P = 7.99068$\,jours et $T_0 = 2444939.58$ pour U~Vul et $P = 8.38209$\,jours et $T_0 = 2442678.792$ pour S~Sge. Le journal de ces observations est exposé dans la Table~\ref{table__log_cepheide_chara}. La magnitude de ces Céphéides classiques en bande $K$ est 4.11 et 3.84 respectivement pour U~Vul et S~Sge.

Les étoiles de référence observées proviennent du catalogue de \citet{Merand-2005-04} et sont présentées dans la Table~\ref{table__etoile_reference_propriete}. J'ai appliqué la méthode de réduction et d'étalonnage présentée précédemment. Notons que ces observations ont été effectuées avec de longues bases ($B_\mathrm{p} > 218$\,m), favorisant la détection de la variation du diamètre angulaire.

\paragraph*{\textcolor{black}{Diamètre angulaire}}

Pour chaque mesure de visibilité, un diamètre de disque uniforme ($\theta_\mathrm{UD}$) a été ajusté. Bien que ce modèle ne soit pas physique, il est particulièrement simple et peut être facilement converti en diamètre de disque assombri ($\theta_\mathrm{LD}$) grâce à un modèle d'atmosphère stellaire. C'est le rôle du facteur $k$ de l'équation~\ref{equation__BW}. Dans le cas d'une loi d'assombrissement (ACB) en loi de puissance $I(\mu) = \mu^\alpha$, on a \citep{Hestroffer-1997-11} :
\begin{displaymath}
k = \frac{\theta_\mathrm{UD}}{\theta_\mathrm{LD}} \approx 1 - 0.1374\,\alpha + 0.0213\,\alpha^2
\end{displaymath}

\begin{figure}[!p]
  \centering\includegraphics[width = .8\linewidth]{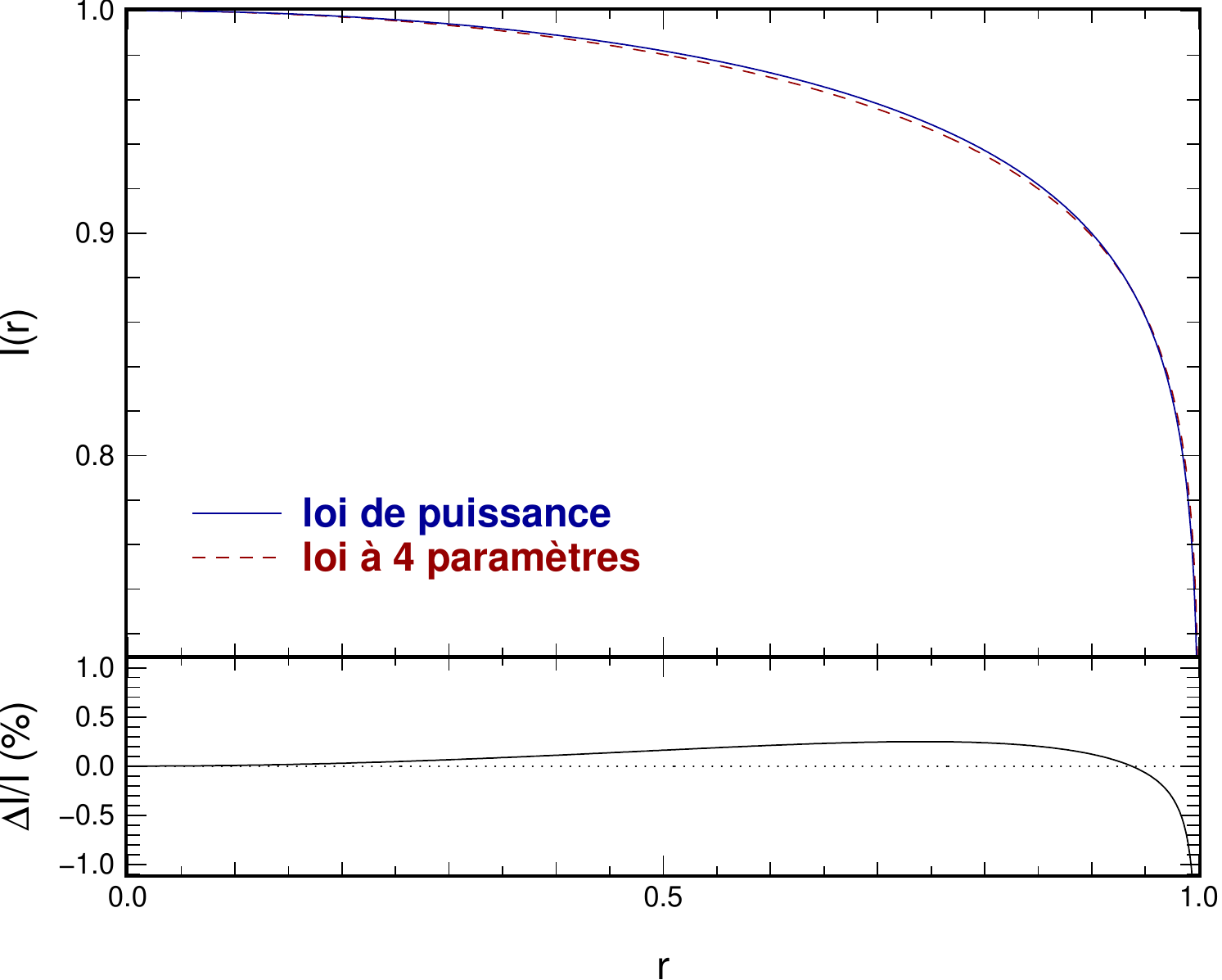}
  \caption[Comparaison du modèle d'ACB à 4 paramètres et de la loi de puissance]{\textbf{Comparaison du modèle d'ACB à 4 paramètres et de la loi de puissance} : le paramètre $r = \rho/R_\star$ est défini sur la Fig.~\ref{image__schema_intensite_uniforme}. Les résidus sont exposés dans la fenêtre du bas.}
  \label{image__comparaison_acb}
\end{figure}

J'ai déterminé $\alpha$ à partir des valeurs tabulées de \citet{Claret-2000-11}.  Cet auteur a utilisé un modèle hydrostatique d'ACB à quatre paramètres dont les valeurs ont été calculées pour diverses bandes photométriques et paramètres stellaires ($T_\mathrm{eff}, \log g, [M/H]$ et $v_\mathrm{turb}$). J'ai donc récolté les quatre paramètres correspondant à la bande $K$ et pour des paramètres stellaires moyens de nos Céphéides : $T_\mathrm{eff} = 5800$\,K, $\log g = 1.8, [M/H] = 0.1$ et $v_\mathrm{turb} = 4\,\mathrm{km~s^{-1}}$ \citep[][identiques pour les deux Céphéides]{Luck-2004-07}. J'ai ensuite ajusté la loi de puissance pour déterminer $\alpha \approx 0.13$. Je compare sur la Fig.~\ref{image__comparaison_acb} la loi à quatre paramètres et la loi de puissance ajustées. On constate la similarité des deux lois dont la différence relative est $< 1\,\%$ jusqu'à $r = \rho/R_\star = 0.99$ (paramètres de la Fig.~\ref{image__schema_intensite_uniforme}). J'ai préféré la loi de puissance car elle est plus simple à utiliser et il n'est pas nécessaire de connaitre les paramètres stellaires.

\paragraph*{\textcolor{black}{Vitesse radiale}}

\begin{figure}[!p]
	\resizebox{\hsize}{!}{
  		\centering\includegraphics{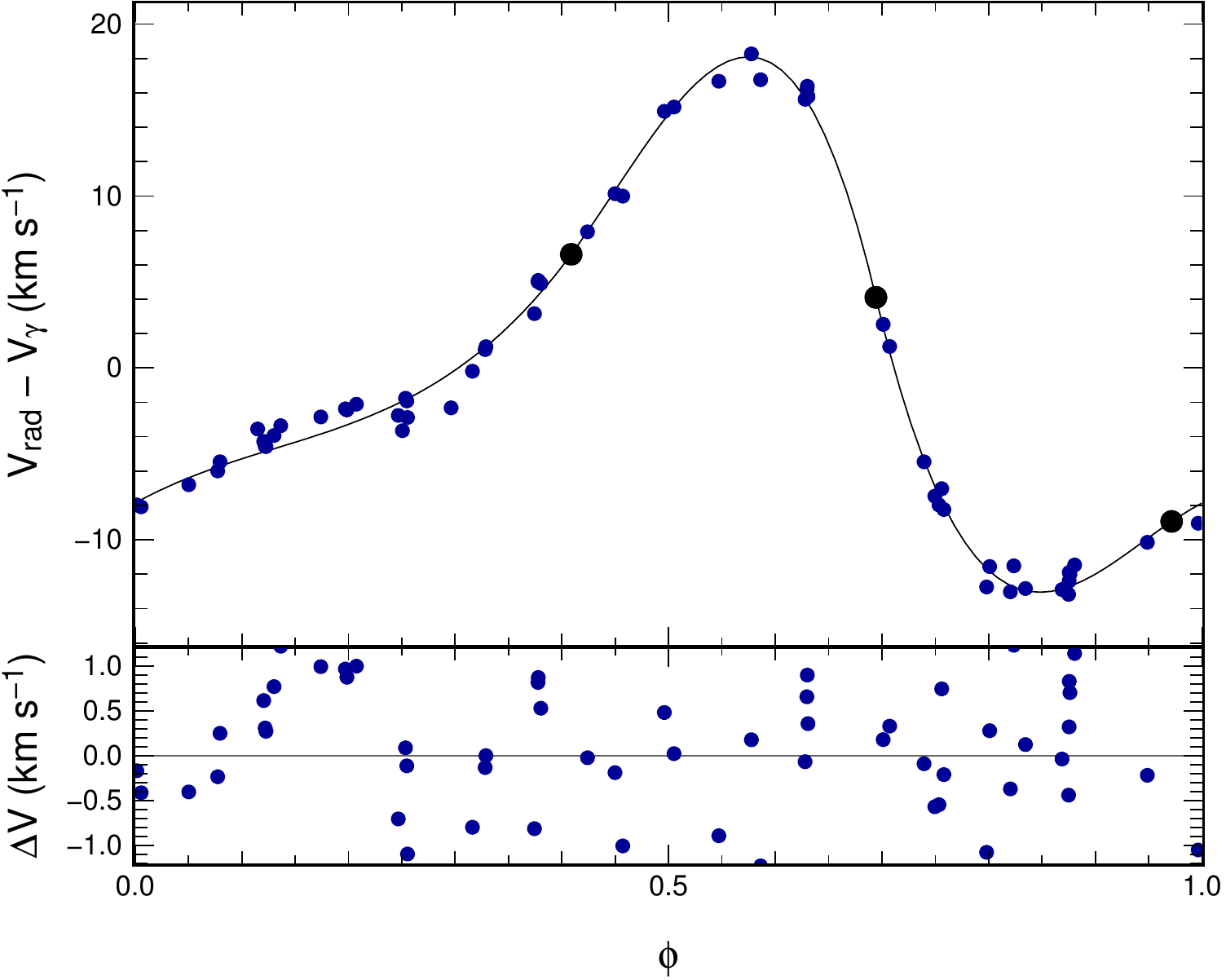} \hspace{.2cm}
		\centering\includegraphics{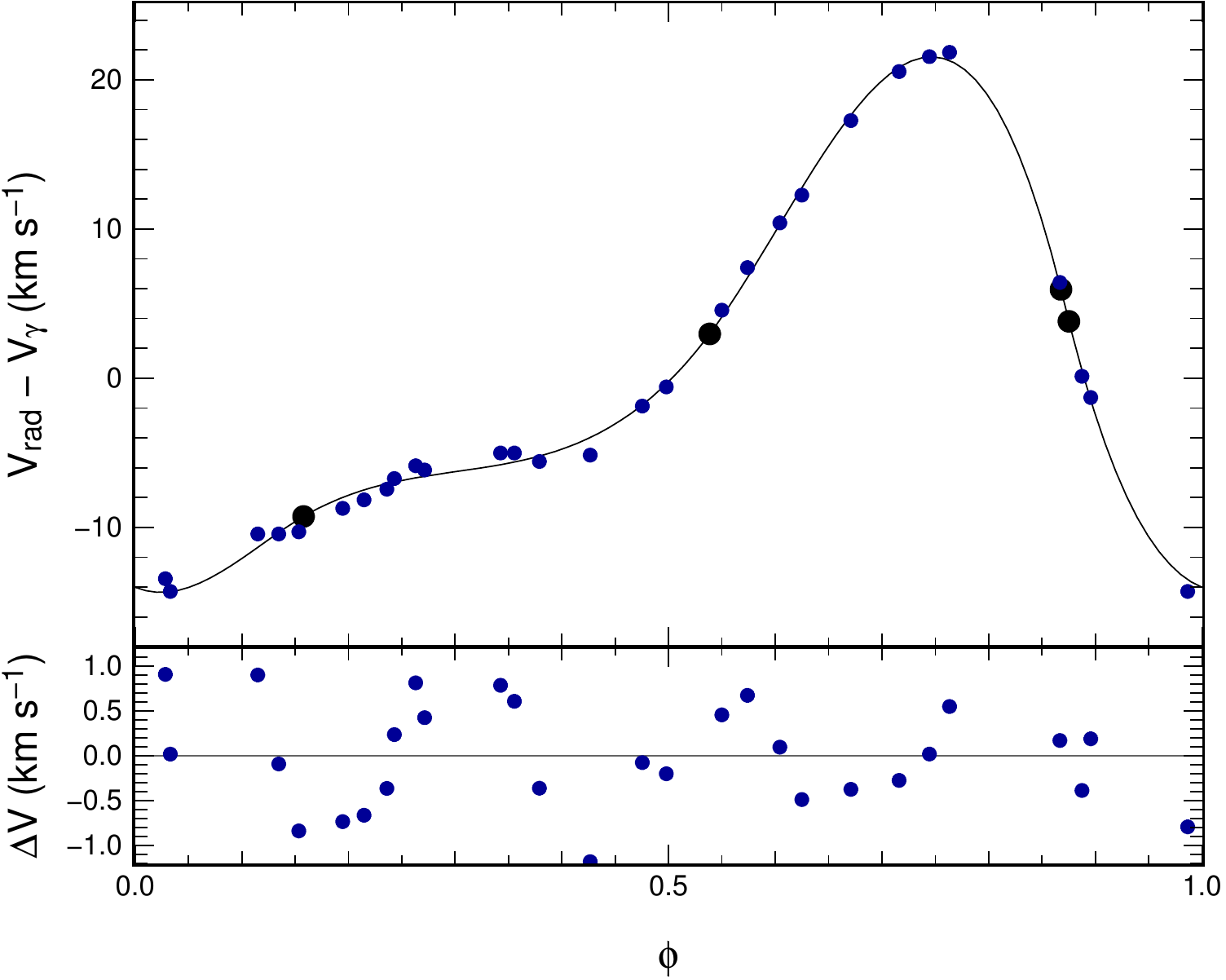}}
	\caption[Courbes de vitesse radiale de U~Vul et S~Sge]{\textbf{Courbes de vitesse radiale de U~Vul et S~Sge} : à gauche, courbe de l'étoile U~Vul corrigée de la vitesse systématique $v_\gamma = -12.9 \pm 0.1\,\mathrm{km~s^{-1}}$. À droite la courbe de S~Sge corrigée de $v_\gamma = -10.3 \pm 0.1\,\mathrm{km~s^{-1}}$. Les fenêtres du bas représentent les résidus.}
  	\label{image__vitesse_radial_uvul_ssge}
\end{figure}

J'ai recueilli les mesures de \citet{Gorynya-1998-11} pour la Céphéide U~Vul. Pour S~Sge, j'ai utilisé les données de \citet{Barnes-2005-02}, dont les vitesses radiales furent déterminées par la méthode de corrélation croisée. Cette étoile étant une binaire \citep[peut être même un système triple,][]{Evans-1993-10}, les données de \citet{Barnes-2005-02} ont l'avantage d'avoir été préalablement corrigées du mouvement propre de l'étoile sous l'effet gravitationnel du compagnon. J'expose sur la Fig.~\ref{image__vitesse_radial_uvul_ssge} les courbes de vitesse radiale interpolées par une fonction de splines périodiques (comme pour Y~Oph).

\paragraph*{\textcolor{black}{Estimation de la distance}}

\begin{figure}[!p]
	\resizebox{\hsize}{!}{
  		\centering\includegraphics{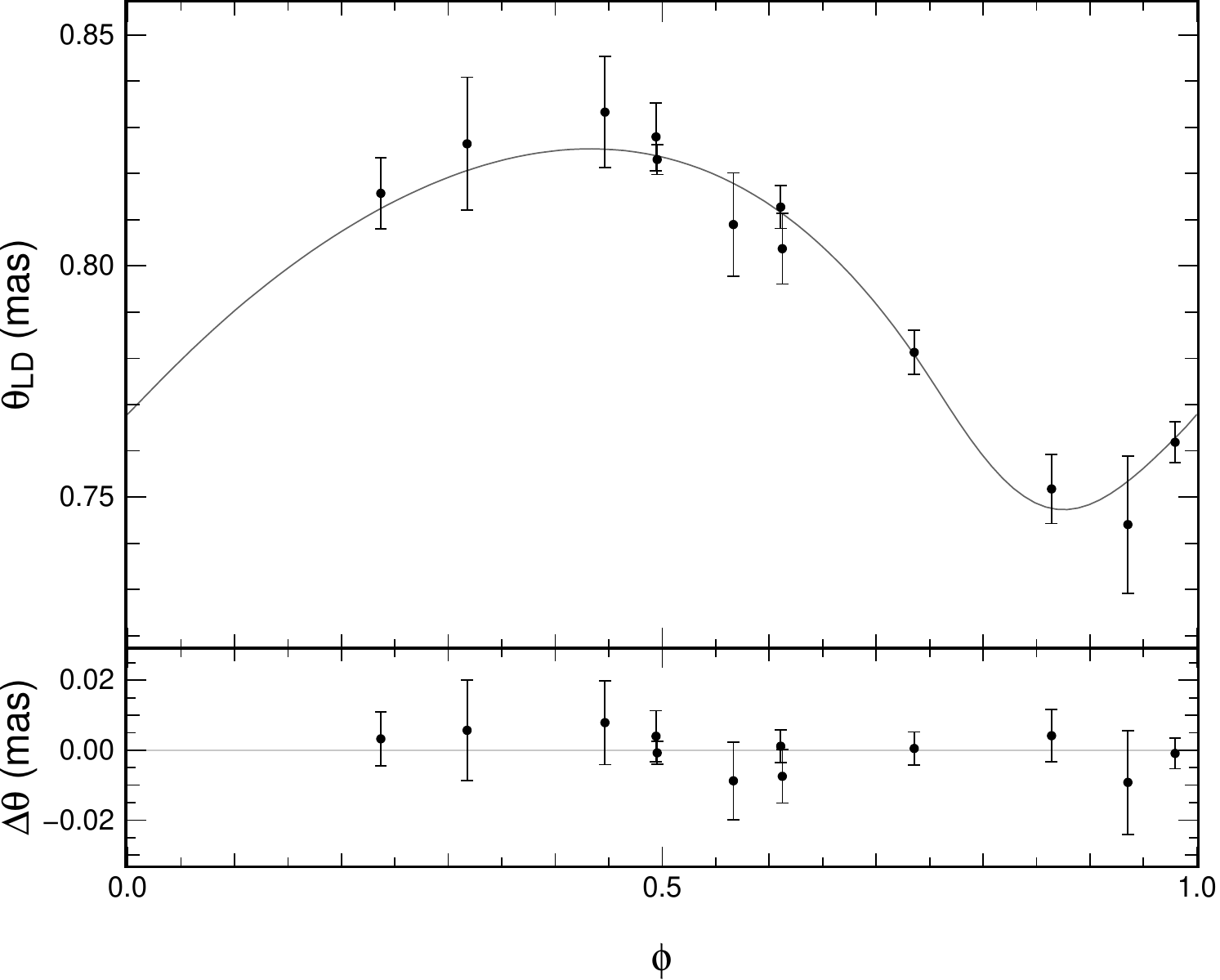} \hspace{.2cm}
		\centering\includegraphics{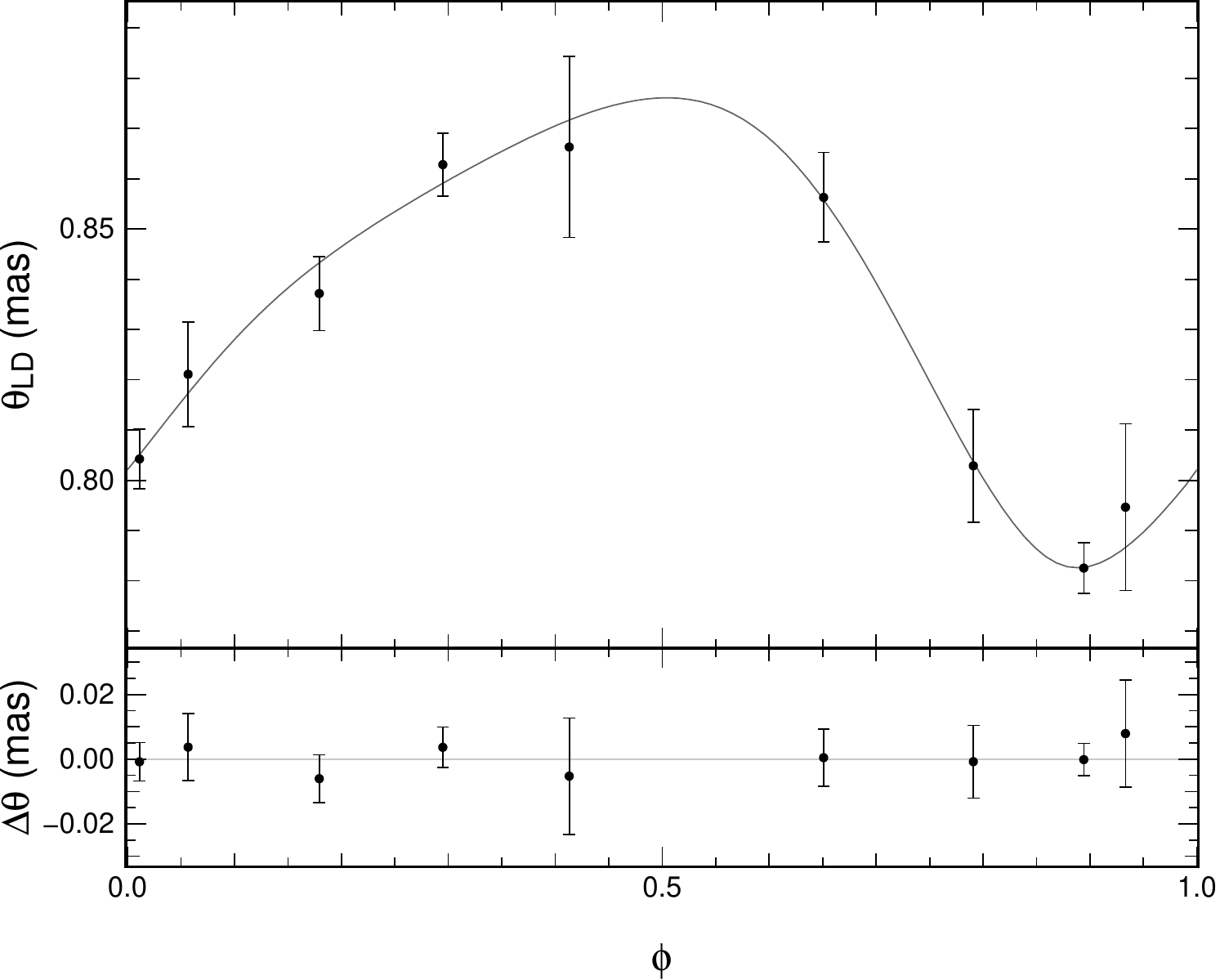}}
	\caption[Courbes de variation du diamètre angulaire de U~Vul et S~Sge]{\textbf{Courbes de variation du diamètre angulaire de U~Vul et S~Sge} : à gauche, l'étoile U~Vul et à droite S~Sge. La courbe continue sur chaque graphe est l'intégration de la courbe de vitesse radiale où j'ai ajusté la distance, le diamètre moyen et le décalage de phase. Les fenêtres du bas représentent les résidus.}
  	\label{image__variation_diametre_uvul_ssge}
\end{figure}

Nous pouvons maintenant appliquer la méthode BW en adoptant une valeur pour le facteur de projection. J'ai choisi d'utiliser la relation linéaire de \citet{Nardetto-2009-05} pour estimer $p = 1.24 \pm 0.08$ pour les deux Céphéides. Les courbes de variation du diamètre angulaire obtenues sont présentées sur la Fig.~\ref{image__variation_diametre_uvul_ssge}. L'ajustement donne pour U~Vul une distance $d = 647\,\pm\,45$\,pc et un rayon moyen $R = 53.4\,\pm\,3.7\,R_\odot$. Pour S~Sge, j'obtiens $d = 661\,\pm\,57$\,pc et $R = 57.5\,\pm\,4.9\,R_\odot$. Notons qu'un décalage de phase entre les observations interférométriques et spectroscopiques de $-0.072\,\pm\,0.026$ pour S~Sge a été autorisé.

La précision de ces résultats dépend fortement de la précision sur la mesure de la visibilité. Ces étoiles étant peu brillantes, il est difficile d'obtenir la même précision que pour Y~Oph. Néanmoins, ils sont compatibles avec les résultats de \citet{Groenewegen-2008-09}, qui estima à partir d'une relation brillance de surface $d = 638\,\pm\,32$\,pc et $R = 50.6\,\pm\,2.6\,R_\odot$ pour U~Vul et $d = 663\,\pm\,22$\,pc et $R = 55.5\,\pm\,1.8\,R_\odot$ pour S~Sge. Ces estimations sont également cohérentes avec les prédictions de \citet{Moskalik-2005-06} : $d = 573\,\pm\,28$\,pc et $R = 56.5\,\pm\,3.2\,R_\odot$ pour U~Vul et $d = 655\,\pm\,31$\,pc et $R = 58.5\,\pm\,3.3\,R_\odot$ pour S~Sge.

\subsubsection{R~Sct}

Je finirai cette étude interférométrique avec cette Céphéide de type II déjà présentée dans le Chapitre~\ref{chapitre__etude_d_exces_infrarouge_par_photometrie}. Le but ici n'est pas d'appliquer la méthode de BW, car elle pulse de façon irrégulière avec une longue période de 146.5 jours, et surtout, nous n'avons qu'un seul point de mesure. Toutefois, un seul point suffit pour avoir une estimation du diamètre angulaire à la phase $\phi = 0.64$, de plus c'est une mesure à visibilité très faible ($V^2 = 0.0953\,\pm\,0.0027$) et donc moins sensible à la présence de matière circumstellaire. 

La visibilité mesurée de R~Sct a été étalonnée en observant l'étoile étalon HD~169113 ($\theta_\mathrm{LD} = 0.816\,\pm\,0.011$\,mas), sélectionnée dans le catalogue \citet{Merand-2005-04}. J'ai ajusté à ce point de visibilité un modèle de disque assombri avec $\alpha = 0.20$, déterminé à partir des valeurs du modèle hydrostatique de \citet{Claret-2000-11} et pour les paramètres stellaire moyens suivants : $T_\mathrm{eff} = 4500$\,K, $\log g = 0.0, [M/H] = -0.4$ et $v_\mathrm{turb} = 3\,\mathrm{km~s^{-1}}$ \citep{Giridhar-2000-03}. 

L'ajustement donne un diamètre $\theta_\mathrm{LD} = 1.81\,\pm\,0.01$\,mas. Il est intéressant de noter que cette valeur est en accord à $4\,\%$ et à $1\sigma$ avec la valeur estimée à partir de la distribution d'énergie spectrale du Chapitre~\ref{chapitre__etude_d_exces_infrarouge_par_photometrie} ($\theta_\mathrm{LD} = 1.74\,\pm\,0.06$\,mas à $\phi = 0.48$).

\subsection{Conclusion de l'étude}

L'interférométrie est un outil puissant pour l'étude des Céphéides, non seulement pour l'étude de l'assombrissement centre-bord ou des enveloppes circumstellaires (pour les plus proches), mais également pour l'estimation de leur distance via la mesure directe des variations de diamètre angulaire. Ces estimations de la distance de manière indépendante ont un rôle important dans l'étalonnage de la relation période--luminosité, comme expliqué dans le Chapitre~\ref{chapitre__mesurer_l_univers_les_cephéides}. Les nouvelles données acquises avec l'instrument \emph{FLUOR} permettront dans les prochains mois d'obtenir un nouvel étalonnage de cette relation P--L.

Grâce à la combinaison des données \emph{FLUOR} et \emph{VINCI}, j'ai pu poursuivre l'étude de \citetalias{Merand-2007-08} concernant l'enveloppe circumstellaire de Y~Oph. En utilisant un modèle d'étoile entourée d'une couronne sphérique, j'ai déterminé un diamètre angulaire moyen photosphérique $\overline{\theta}_\mathrm{LD} = 1.32\,\pm\,0.01$\,mas et une taille angulaire de l'enveloppe de $3.4\,R_\star$. Cette estimation est cohérente avec la taille angulaire typique des enveloppes mesurée jusqu'à présent pour d'autres Céphéides.

Les premiers résultats concernant la distance et le rayon des Céphéides U~Vul et S~Sge sont cohérents avec d'autres valeurs estimées (de manière indirecte) dans la littérature. Ces mesures permettrons d'estimer par la suite, avec les autres Céphéides de notre échantillon, un nouvel étalonnage de la relation période--rayon, période--luminosité et brillance de surface--couleur.

L'interférométrie à grande base, et notamment avec l'instrument \emph{FLUOR}, est devenue l'outil nécessaire à la détermination précise des diamètres angulaires de Céphéides. L'estimation de ces diamètres nous permettra bientôt de réaliser un étalonnage indépendant de la relation P-L et donc, de l'échelle des distances extragalactiques.

\cleardoublepage     

\pagestyle{fancy}
\fancyhf{}
\lhead[\nouppercase{\emph{\thepage}}]{\nouppercase{\emph{\rightmark}}}
\rhead[\nouppercase{\emph{\leftmark}}]{\nouppercase{\emph{\thepage}}}
\newpage


\chapter[Conclusions et perspectives]{\emph{Conclusions et perspectives}}

\thispagestyle{empty}

\vspace*{-1cm}

\refbleu  
\textcolor{bleu_chapitre}{\minitoc}
\refnoir  

\malettrine{J}{ }e termine se manuscrit en présentant mes conclusions sur la détection d'enveloppe circumstellaire et sur la mesure directe de distance par interférométrie. Je parlerai également d'un long programme avec l'instrument \emph{MIDI} du \emph{VLTI} concernant l'étude de l'environnement des Céphéides aux longueurs d'onde thermiques. Je discuterai des perspectives en ce qui concerne l'étude des Céphéides en général grâce au nouveau mode dispersé de \emph{FLUOR}, à l'instrument spatial \emph{GAIA} et au réseau d'antennes ALMA.

\section{Conclusion sur l'étude des enveloppes circumstellaires de Céphéides}

\subsection{RS~Pup}
Grâce à des observations en mode cube avec OA, j'ai pu estimer certaines propriétés sur l'enveloppe de cette Céphéide. Dans un premier temps, en utilisant la technique de "l'imagerie sélective", j'ai déterminé un rapport de flux entre l'enveloppe et l'étoile de $38 \pm 17$\,\% et $24 \pm 11$\,\% aux longueurs d'onde $1.64\,\mu\mathrm{m}$ et $2.18\,\mu\mathrm{m}$ respectivement. L'émission de l'enveloppe a un impact particulier sur l'estimation des distances dans l'infrarouge. Cependant, ces mesures ont été effectuées en bande étroite, le rapport de flux est donc probablement plus faible en bande large, où les diamètres stellaires sont estimés. Ce ne serait donc pas pertinent de quantifier un biais sur la mesure de diamètre à partir de ces estimations.

Dans un second temps, par une analyse statistique originale du bruit dans chaque cube, j'ai pu révéler que cette enveloppe est soit un disque vu de face, soit une enveloppe sphérique uniforme.

\subsection{Les Céphéides observés avec \emph{VISIR}}

Ces données ont permis d'étudier l'excès infrarouge d'un échantillon de 8 Céphéides classiques. Le point important résultant de l'analyse de ces données est que 90\,\% des Céphéides observées présentent dans leur distribution spectrale d'énergie un excès de flux aux longueurs d'onde $\gtrsim 3\,\mu\mathrm{m}$. Cette découverte est assez important et soulève la question sur la présence d'enveloppes circumstellaires autour de toutes les Céphéides. L'excès mesuré à $8.6\,\mu\mathrm{m}$ varie en fonction de l'étoile de 2\,\% à 30\,\%, soit un biais sur l'estimation de la magnitude absolue variant de 0.03 à 0.4\,mag, qui n'est pas négligeable dans l'étalonnage de la relation P--L. De plus, avec quelques hypothèses sur la composition de poussières dans l'enveloppe, j'ai également pu estimer une température et une masse totale de cette enveloppe, de l'ordre de $10^{-8}$--$10^{-10}\,M_\odot$ (Table~\ref{table__parametre_ajuste}).

Une analyse de Fourier a également montré que certaines de nos étoiles ont une enveloppe résolue par le télescope, avec une taille de l'ordre de $1\arcsec$ à $8.6\,\mu\mathrm{m}$ (Table~\ref{table__parametre_visibilite_ajuste}). En utilisant cette distance, j'ai estimé un tau de perte de masse $\sim 10^{-10}$--$10^{-11}\,\mathrm{M_\odot\,yr^{-1}}$, comparables aux valeurs prédites ou mesurées par d'autres auteurs \citep[e.g.][]{Neilson-2008-09,Deasy-1988-04}.

Enfin, j'ai trouvé une corrélation entre l'excès infrarouge à $8.6\,\mu\mathrm{m}$ et la période de pulsation. Cela confirme la même corrélation trouvée par \citet{Merand-2007-08} en bande $K$, les Céphéides de longue période ont un excès IR plus grand que les courtes périodes. Si on suppose que cet excès est lié à la perte masse des Céphéides, les longues périodes perdraient donc plus de masse. Ceci pourrait être expliqué par une dynamique atmosphérique plus importante pour les longues périodes.

\subsection{Y~Oph}

L'étude interférométrique de Y~Oph a permis d'obtenir deux résultats. Le premier est la confirmation des résultats de \citet{Merand-2007-08}, avec une nouvelle estimation de la distance $d = 468 \pm 12$\,pc (non-biaisée par la présence de l'enveloppe) et un rayon $R = 67.8 \pm 1.7\,R_\odot$, en utilisant la méthode de Baade-Wesselink. Le second concerne l'obtention de nouveaux paramètres de l'enveloppe circumstellaire. En couplant des données à longues bases (\emph{FLUOR}) avec des données à courtes bases (\emph{FLUOR + VINCI}), j'ai pu ajuster un modèle de visibilité d'une étoile entourée d'une coquille sphérique optiquement mince. J'ai obtenu un diamètre angulaire $\theta_\mathrm{cse} = 4.53 \pm 1.13\,$mas et une profondeur optique $\tau = 0.011 \pm 0.006$, tout en gardant le rapport de flux $F_\mathrm{cse}/F_\star = 5.0 \pm 2.0$\,\% de \citet{Merand-2007-08}. D'autres mesures à courtes bases et avec une plus grande précision sont nécessaires pour affiner ces résultats.

Cette estimation de la distance, non-biaisée par la présence de l'enveloppe, est particulièrement importante pour utiliser cette Céphéide dans l'étalonnage de la relation P--L.

\section{Les enveloppes avec le nouveau \emph{FLUOR}}

\emph{FLUOR} est actuellement en train d'être amélioré, avec entre autres, un changement des moteurs servant à l'optimisation de l'injection, une nouvelle caméra, et surtout l'installation d'un nouveau mode : le mode dispersif. Le bande $K$ sera maintenant dispersée sur plusieurs canaux spectraux ($R = \lambda/\Delta\lambda \sim 50$). Grâce à cette optimisation, il sera possible d'obtenir en une observation, plusieurs mesures de visibilité simultanées, à différentes fréquences spatiales. L'un des avantages sera de bénéficier de mesures avec la meilleure précision possible ($\sigma(V^2)/V^2 < 2$\,\%, tandis que \emph{AMBER} ou \emph{PIONIER} donnent une précision $\geqslant 5$\,\%). On obtiendra alors des mesures de diamètre angulaire avec une précision jamais atteinte jusqu'à présent.

\begin{figure}[!t]
  \centering\includegraphics[width = .6\linewidth]{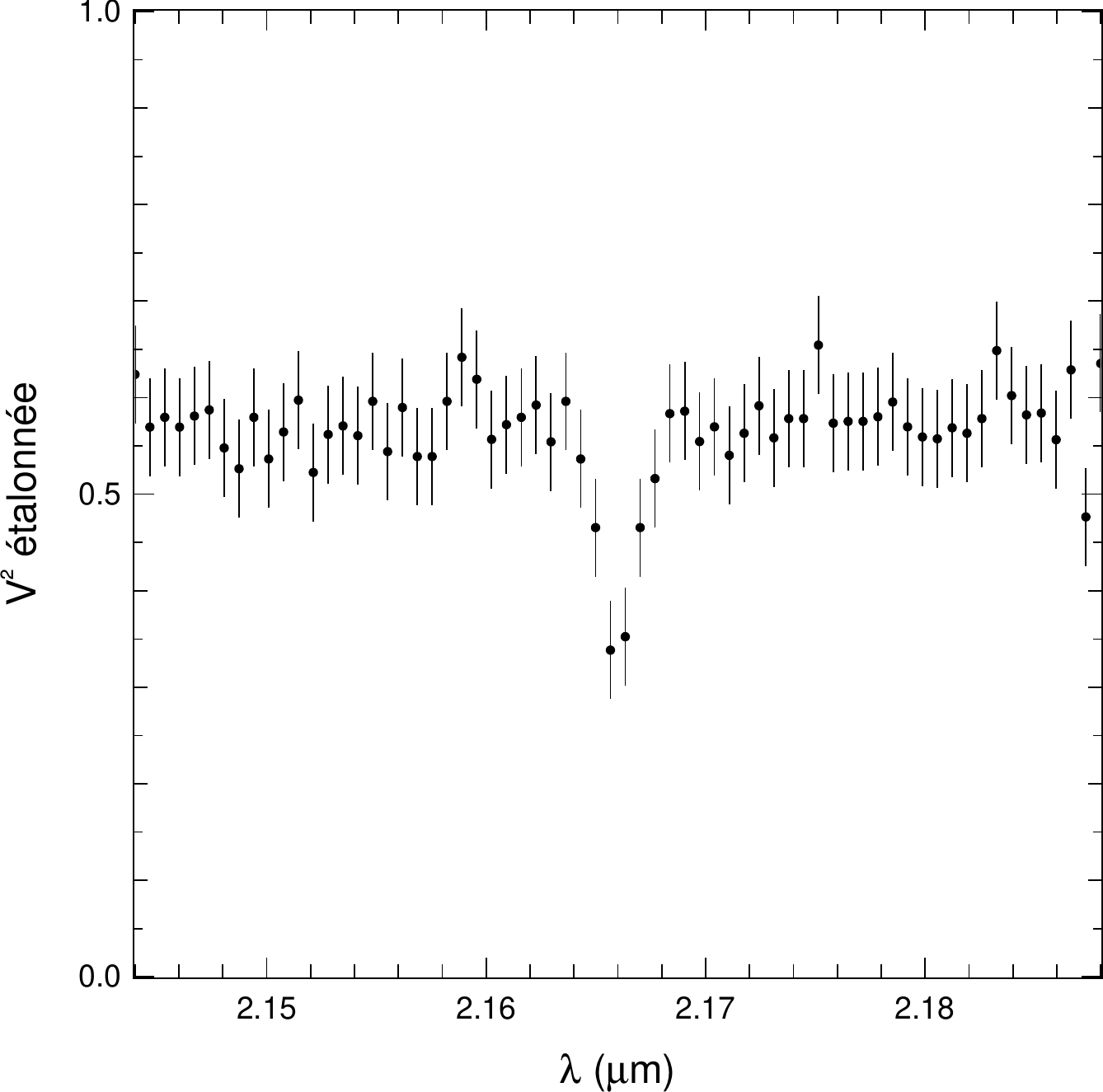}
  \caption[Exemple de visibilité différentielle]{\textbf{Exemple de visibilité différentielle} : la chute de visibilité dans la raie indique que l'objet apparait plus grand à $2.165\,\mu\mathrm{m}$ \citep[travaux de][]{Malbet-2007-03}.}
  \label{image__visibilite_differentielle}
\end{figure}

Dans le cadre de l'étude des enveloppes circumstellaires, la visibilité différentielle sera un paramètre important car elle permet d'étudier la variation de taille de l'objet avec la longueur d'onde. Plusieurs auteurs \citep[par exemple][]{Meilland-2007-03,Malbet-2007-03} ont utilisé la visibilité dans la raie Br$\gamma$ ($2.165\,\mu\mathrm{m}$) pour déterminer les propriétés de l'environnement de certains objets. En effectuant une mesure de la visibilité dans la raie, on peut contraindre la taille angulaire du matériel circumstellaire. Un exemple de ce que l'on pourrait obtenir sur des Céphéides est présenté sur la Fig.\ref{image__visibilite_differentielle}, pour l'objet stellaire jeune MWC 297 \citep{Malbet-2007-03}. La visibilité est estimée de la façon suivante :
\begin{displaymath}
V_\mathrm{Br\gamma} = \frac{F_\mathrm{r}V_\mathrm{r} - F_\mathrm{c}V_\mathrm{c}}{F_\mathrm{r} - F_\mathrm{c}}
\end{displaymath}
où $V_\mathrm{r}$ et $F_\mathrm{r}$ représente la visibilité et le flux dans la raie, $V_\mathrm{c}$ et $F_\mathrm{c}$ correspondent au continuum et 
$F_\mathrm{c} = F_\mathrm{r} + F_\mathrm{Br\gamma}$. Le diamètre estimé à partir de cette visibilité est ensuite comparé à celui dans le continuum (celui d'un disque uniforme par exemple). Le même type de travail peut être envisagé pour l'étude des enveloppes circumstellaires de Céphéides.

Des travaux sur la Céphéides $\ell$~Car utilisant $V_\mathrm{Br\gamma}$ sont actuellement en cours avec des données \emph{AMBER} (Mérand et al. en prép). Un programme d'observation peut être envisagé avec \emph{FLUOR} pour les Céphéides de l'hémisphère nord. Cependant, la résolution spectrale ne sera pas suffisante pour résoudre la raie $Br\gamma$, mais la visibilité différentielle dans le continu apportera probablement des informations utiles sur les enveloppes.

\section{Interférométrie à $10\,\mu\mathrm{m}$ avec \emph{MIDI}}

Un long programme d'observation de 10 Céphéides classiques avec l'instrument \emph{VLTI/MIDI} \citep{Leinert-2003-} est arrivé à terme récemment et les données sont en cours d'analyse. \emph{MIDI} recombine les faisceaux de deux télescopes en bande $N$ (7.6--13.3\,$\mu\mathrm{m}$) avec les résolutions spectrales $R = 30$ et $R = 230$. À cette longueur d'onde, le contraste entre l'étoile et l'enveloppe est réduit et l'utilisation de l'interférométrie permet de résoudre l'environnement proche de l'étoile. \emph{MIDI} est donc particulièrement adapté à la détection d'enveloppe autour des Céphéides.

\begin{figure}[!p]
  \centering\includegraphics[width = .7\linewidth]{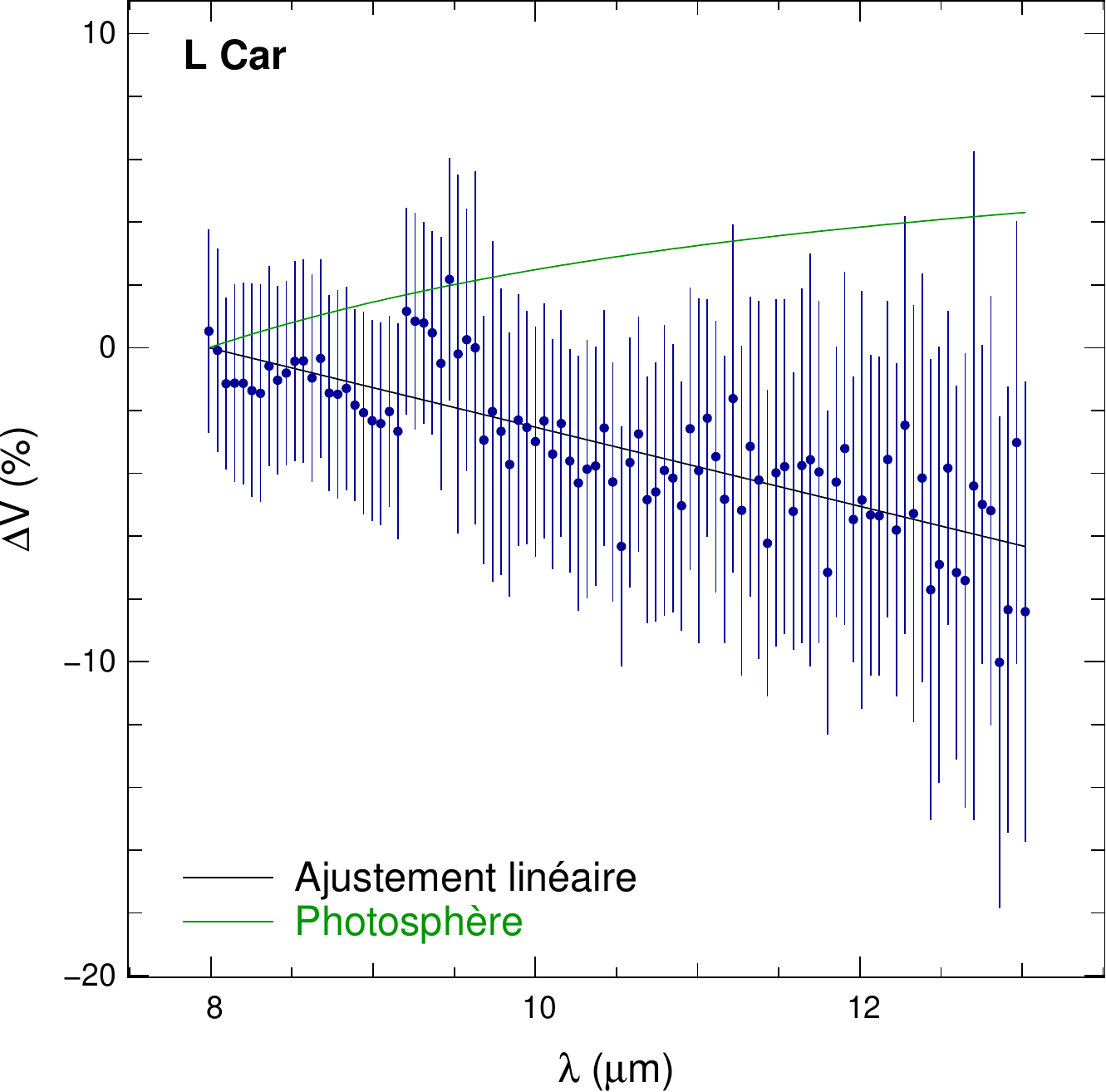}
  \caption[Visibilité différentielle de $\ell$~Car]{\textbf{Visibilité différentielle de $\ell$~Car} : la courbe verte correspond à la visibilité de la photosphère de l'étoile pour un disque assombri de diamètre $\theta_\mathrm{LD} = 3.09$\,mas et une base $B_\mathrm{p} = 123.8$\,m. La courbe noire est un ajustement linéaire des points de visibilité. Les barres d'erreurs indiquées ici correspondent aux erreurs d'origines sur la visibilité absolue estimée par le logiciel de réduction EWS.}
  \label{image__visibilite_midi_l_car}
\end{figure}

Les premières données que j'ai analysées concernaient les étoiles RS~Pup et $\ell$~Car dont l'enveloppe circumstellaire est résolue par \emph{MIDI}. J'expose comme exemple la visibilité différentielle de $\ell$~Car sur la Fig.~\ref{image__visibilite_midi_l_car}. Ces résultats ont été publiés par \citet{Kervella-2009-05} dans une étude comparative avec d'autres instruments infrarouges (Annexe~\ref{article__rs_pup_pierre}).

Il est possible de remonter au rapport de flux entre l'enveloppe et l'étoile grâce à la visibilité différentielle. Elle est définie de la façon suivante :
\begin{displaymath}
\Delta V = \frac{V(\lambda)}{V(8\mu\mathrm{m})} - 1
\end{displaymath}

Pour une étoile entourée d'une enveloppe, la visibilité a la forme suivante :
\begin{displaymath}
V(\lambda,B) = \frac{F_\star(\lambda)V_\star(\lambda,B) + F_\mathrm{env}(\lambda)V_\mathrm{env}(\lambda,B)}{F_\star(\lambda) + F_\mathrm{env}(\lambda)} = \frac{V_\star(\lambda,B)}{1 + \alpha (\lambda)}
\end{displaymath}
où $F_\star$ et $F_\mathrm{env}$ représentent respectivement le flux photosphérique et le flux de l'enveloppe, et $V_\star$ et $V_\mathrm{env}$ les visibilités respectives. On a également défini $\alpha = F_\mathrm{env}/F_\star$. Pour l'exemple de $\ell$~Car, on a supposé que l'enveloppe est résolue par l'interféromètre, soit $V_\mathrm{env}(\lambda,B) = 0$. La photométrie Spitzer a $8\,\mu\mathrm{m}$ pour cette étoile indique un très faible excès IR, on peut alors présumer que $\alpha (8\,\mu\mathrm{m}) = 0$. L'expression de la visibilité différentielle prend donc la forme suivante :
\begin{displaymath}
\Delta V(\lambda,B) = \frac{\Delta V_\star(\lambda,B) - \alpha(\lambda)}{1 + \alpha(\lambda)} \qquad \Longrightarrow \qquad \alpha (\lambda) = \frac{\Delta V_\star - \Delta V}{1 + \Delta V}
\end{displaymath}

$\Delta V_\star$ est estimée à partir d'un modèle de visibilité de l'étoile pour un disque assombri $\theta_\mathrm{LD} = 3.09$\,mas et une base $B_\mathrm{p} = 123.8$\,m (à la date des observations), tandis que $\Delta V$ est estimée à partir d'un ajustement linéaire sur les mesures. Cela donne à $13\,\mu\mathrm{m}$ un rapport de flux $\alpha = 10.5 \pm 1.0$\,\% pour la Céphéide $\ell$~Car. 

La visibilité absolue de \emph{MIDI} a une précision de 10--15\,\%, alors que la même quantité différentielle atteint quelques pour cent. C'est pour cela  qu'il est préférable pour la détection d'enveloppe d'utiliser la visibilité différentielle. Le même type d'analyse est envisagé pour tout l'échantillon de Céphéides afin de sonder l'environnement proche.

\section{Les enveloppes à plus grandes longueurs d'onde: ALMA, Hershel}

\subsection{ALMA}

L'observation de Céphéides aux longueurs d'onde millimétriques ou radio est un domaine quasi inexploré. Il serait intéressant d'étudier les enveloppes dans cette gamme de longueurs d'onde car cela fournirait des informations sur l'évolution et l'histoire de la perte de masse. Quand le matériel circumstellaire chaud se refroidit, de la poussière peut se former à une distance de quelques rayons stellaires. Pour la détecter, la haute résolution angulaire est nécessaire et à ces longueurs d'onde, seule l'interférométrie donne accès à une résolution suffisante. \emph{ALMA} nous permettra d'observer avec la résolution nécessaire l'émission continue de poussières froides et des raies moléculaires. Ce réseau interférométrique couvrira les longueurs d'onde $0.42\,\mathrm{mm} < \lambda <  3$\,mm avec une longueur de base pouvant atteindre 16\,km, soit une résolution angulaire maximale $\sim 10$\,mas. Il possèdera 64 antennes de 12\,m et 12 de 7\,m qui permettront une excellente couverture du plan $(u,v)$ pour la reconstruction d'images.

\begin{figure}[!p]
  \centering\includegraphics[width = .6\linewidth]{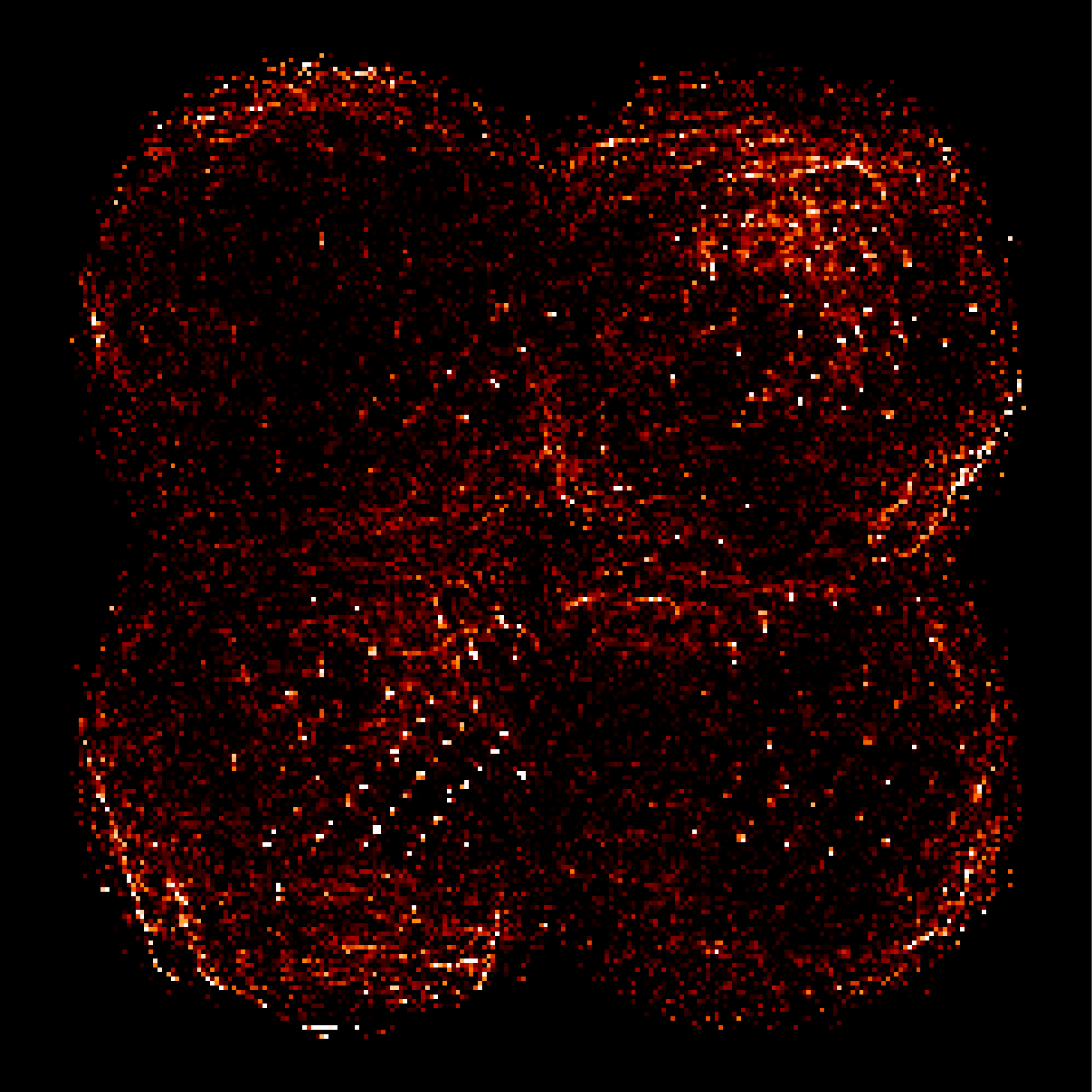}
  \caption[Image à $\lambda = 0.35$\,mm de RS~Pup]{\textbf{Image à $\lambda = 0.35$\,mm de RS~Pup}.}
  \label{image__saboca}
\end{figure}

La première cible envisagée pour des observations avec \emph{ALMA} serait la Céphéide RS~Pup car cette étoile est entourée d'une grande nébuleuse ($\sim 2\arcmin$ en bande $V$), faisant d'elle une cible idéale pour sonder l'environnement froid. Sur cette étoile, nous avons déjà effectué des observations avec une seule antenne à 0.35\,mm avec \emph{APEX/SABOCA} dans le cadre du programme de vérification scientifique. J'ai réduit les données en utilisant le logiciel miniCRUSH \citep{Kovacs-2008-08} afin d'obtenir une image du matériel circumstellaire. Malheureusement notre temps d'observation obtenu ayant été diminué d'un facteur 4, nous n'avons pas obtenu la sensibilité nécessaire pour détecter l'enveloppe (lors de la mise en service de l'instrument, il a été déterminé que la sensibilité était 10 fois meilleure que prévu, et notre temps d'observation est passé de 4h à 1h). Je présente l'image obtenue sur la Fig.\ref{image__saboca}. J'ai pu néanmoins grâce à ces données estimer une limite maximale $F_{0.35\mathrm{mm}}< 65$\,Jy, en estimant le niveau de détection ($3\sigma$) dans l'image. 

\emph{ALMA} dans sa configuration finale offrira la sensibilité nécessaire à la détection de l'enveloppe. J'expose dans la Table~\ref{table__alma_sensibilite} la sensibilité attendue dans les bandes disponibles à la prochaine période. On peut estimer le temps d'exposition nécessaire pour une détection à $10\sigma$ de l'enveloppe à $\lambda = 0.4$\,mm avec \emph{ALMA}. En ajustant une distribution de corps noir à la distribution spectrale d'énergie de RS~Pup, j'estime le flux de l'enveloppe attendu $F_{0.4\mathrm{mm}} \sim 1.6\times10^{-16}\,\mathrm{W\,m^2\,\mu m^{-1}} \simeq 6.8$\,Jy. En considérant une taille de l'enveloppe de $2.5\arcmin \times 2.5\arcmin$, cela donne une sensibilité nécessaire d'environ 0.14\,mJy/beam (avec la base minimale de 18\,m). En utilisant le calculateur de sensibilité de \emph{ALMA}\footnote{\url{http://almascience.eso.org/call-for-proposals/sensitivity-calculator}} pour 16 antennes de 12\,m, sans faire de mosaïques et un mode de polarisation double (et une largeur de bande de 7.5\,GHz par polarisation pour des observations du continuum), cette sensibilité peut être atteinte avec un temps d'observation de 10\,min. L'enveloppe de RS~Pup sera donc détectable avec \emph{ALMA}.

\begin{table}[!p]
	\centering
	\begin{tabular}{cccc} 
	\hline
	\hline
	bandes	&	f (GHz)			& 	$\lambda$ (mm)	&	Sensibilité (mJy)	\\
	\hline
		3		&	84--116		&	2.6--3.6				&	0.17--0.53		\\
	    6		&	211--275		&	1.1--1.4				&	0.32--0.38		\\
		7		&	275--373		&	0.8--1.1				&	0.38--3.4			\\
		9		&	602--720		&	0.4--0.5				&	17.3--25.9		\\
	\hline
	\end{tabular}
  	\caption[Sensibilité de \emph{ALMA}]{\textbf{Sensibilité de \emph{ALMA}} : valeurs estimées en utilisant le calculateur de sensibilité de \emph{ALMA} avec 16 antennes de 12\,m, un temps d'exposition de 1h sans mosaïque et un mode de polarisation double (observation du continuum)}
  	\label{table__alma_sensibilite}
\end{table}

\begin{figure}[!p]
  \centering\includegraphics[width = .7\linewidth]{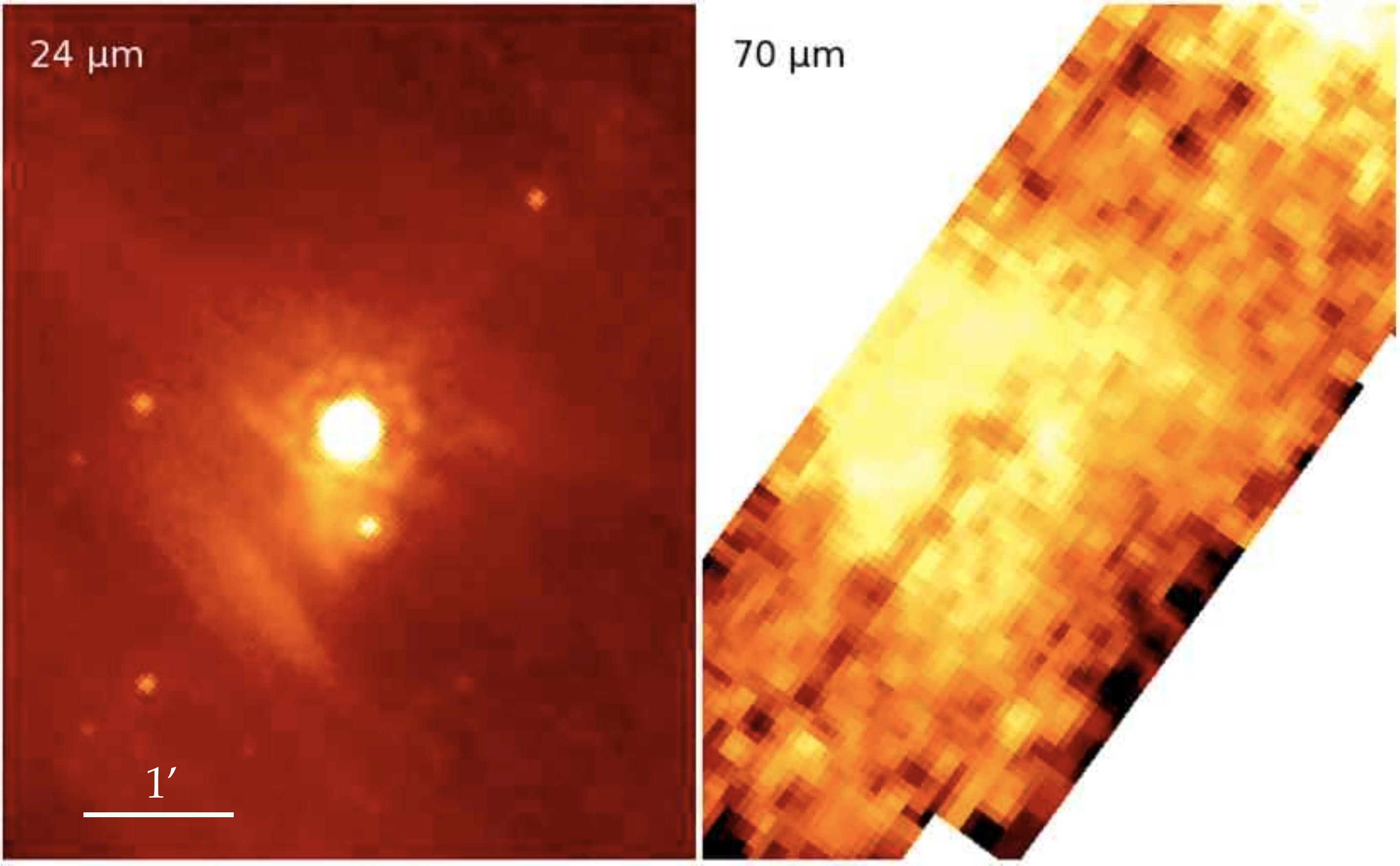}
  \caption[Image 24\,$\mu\mathrm{m}$ et 70\,$\mu\mathrm{m}$ de $\delta$~Cep]{\textbf{Image 24\,$\mu\mathrm{m}$ et 70\,$\mu\mathrm{m}$ de $\delta$~Cep} : observations du télescope \emph{Spitzer} avec l'instrument \emph{MIPS} \citep{Marengo-2009-01}.}
  \label{image__spitzer_delta_cep}
\end{figure}

\subsection{Herschel}

L'infrarouge lointain est également un domaine peu exploré des Céphéides. Le télescope spatial \emph{Herschel} complètera les mesures aux longueurs d'onde $55\,\mu\mathrm{m} < \lambda < 672\,\mu\mathrm{m}$. Il est équipé de trois instruments : \emph{PACS}, qui fonctionne soit en imagerie soit en spectrométrie grand champ dans deux bandes comprises entre $55\,\mu\mathrm{m} < \lambda < 210\,\mu\mathrm{m}$, \emph{SPIRE}, fonctionnant également soit en imagerie soit en spectrométrie, mais dans trois bandes dans l'intervalle $194\,\mu\mathrm{m} < \lambda < 672\,\mu\mathrm{m}$, et \emph{HIFI}, un spectromètre haute résolution dans deux bandes comprises entre $157\,\mu\mathrm{m} < \lambda < 625\,\mu\mathrm{m}$.

Des observations de RS~Pup et $\delta$~Cep sont actuellement en cours avec \emph{PACS} et \emph{SPIRE} en mode imagerie. Les images de l'instrument \emph{PACS} auront une meilleure résolution que \emph{Spitzer/MIPS} et permettront de contraindre plus précisément la morphologie de l'enveloppe. Les observations de \citet{Marengo-2009-01} de $\delta$~Cep avec le télescope spatial \emph{Spitzer} à 24\,$\mu\mathrm{m}$ et 70\,$\mu\mathrm{m}$ (présentées sur la Fig.\ref{image__spitzer_delta_cep}) montrent une extension angulaire comparable à RS~Pup. Ces deux étoiles ayant des périodes différentes, une étude simultanée permettra de vérifier une éventuelle corrélation avec la pulsation.

La température et la masse  de l'enveloppe de poussière pourront être déterminées grâce aux données \emph{SPIRE}. Étant donnée une sensibilité des deux instruments de l'ordre de 4\,mJy pour un temps d'exposition de 1\,h (sans mosaïque) et un flux attendu $F_{670\mu\mathrm{m}} \sim 3$\,Jy, l'enveloppe de poussières de ces Céphéides devrait être détectée. L'interprétation de ces données en utilisant une code de transfert radiatif nous donnera une meilleure compréhension sur la structure en densité et en température de ces enveloppes.

\section{Conclusion sur l'estimation des distances des Céphéides}

L'estimation des distances des Céphéides de manière indépendante est indispensable à un bon étalonnage de la relation P--L. L'interférométrie longues bases permet maintenant de mesurer directement la variation de diamètre angulaire et d'appliquer la méthode interférométrique de Baade-Wesselink pour estimer la distance. Pour une bonne précision sur la distance, les instruments de recombinaison doivent fournir une précision absolue sur la mesure de diamètre de l'ordre du pour cent. \emph{FLUOR} atteint cette précision et permettra, grâce aux nouvelles données acquises récemment, une estimation des distances assez précises pour les étoiles de l'hémisphère nord. Dans l'hémisphère sud, l'instrument \emph{PIONIER}, qui est l'équivalent de \emph{FLUOR} à 4 télescopes, atteindra cette précision très prochainement. Cependant la faible longueur des bases du \emph{VLTI} restreint le nombre de Céphéides pour lesquelles on peut détecter la pulsation. 

Ces estimations de distance via la mesure de diamètres angulaires souffrent encore de biais qu'il reste à quantifier précisément. L'une des sources de biais est la présence d'enveloppes circumstellaires qui peuvent biaiser la mesure de distance de quelques pour cent. Le facteur $p$ est également une source de biais, il n'existe qu'une seule mesure de ce paramètre dans la littérature \citep{Merand-2006-} et elle est généralement utilisée pour les autres Céphéides. Le même procédé que \citep{Merand-2006-} peut être appliqué aux autres Céphéides dont la mesure de distance existe \citep{Benedict-2007-04}. Pour le moment, l'utilisation de la relation $p$--P de \citet{Nardetto-2009-05} déterminée à partir de l'étude d'une raie métallique photosphérique semble plus appropriée. L'ACB est probablement la source de biais la plus difficile à estimer car sa mesure nécessite des mesures de visibilité dans le second lobe. Cela implique des bases supérieures à 330\,m ou des observations à des longueurs d'onde plus courtes. Des instruments tels que \emph{VEGA} ou \emph{PAVO}, installés sur \emph{CHARA} et fonctionnant dans le visible, permettraient de telles mesures pour les Céphéides.

\section{Mesures de parallaxe avec GAIA}

Le satellite \emph{GAIA} mesurera des parallaxes trigonométriques d'un milliard d'objets dans notre Galaxie avec une précision pouvant atteindre 10\,$\mu$as à $V = 15$ (soit la taille angulaire d'une pièce de 1 euro sur la Lune). Il est attendu pour élargir l'échantillon de Céphéides avec des mesures de parallaxes précises et ainsi améliorer l'étalonnage de la relation P--L. \emph{GAIA}  fournira des données astrométriques, photométriques et spectroscopiques, dont la précision attendue est présentée dans la Table~\ref{table__precision_gaia}. Il fera une étude sans précédent sur tout le ciel jusqu'à des étoiles de magnitude 20. Notons toutefois qu'il reste sensible à l'extinction interstellaire pour les mesures photométriques.

\begin{figure}[!p]
  \centering\includegraphics[width = .7\linewidth]{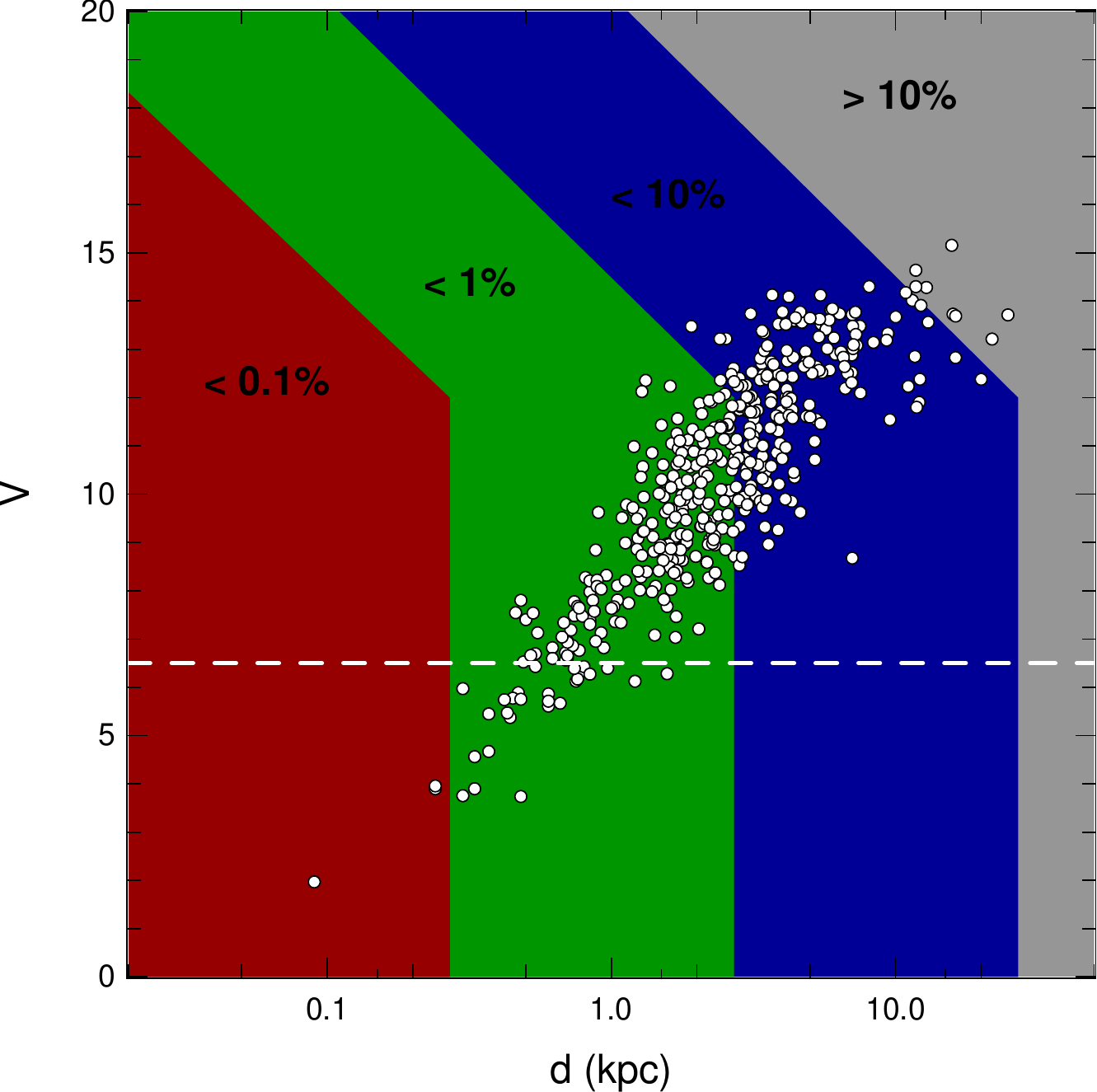}
  \caption[Précision de \emph{GAIA} pour les Céphéides]{\textbf{Précision de \emph{GAIA} pour les Céphéides} : diagramme \citet{Lindegren-2005-10} auquel j'ai rajouté l'échantillon de Céphéides provenant de \citet{Berdnikov-2000-04}. À chaque zone colorée correspond une zone de précision. On constate qu'une grande partie des Céphéides sera mesurée avec une précision $< 1$\,\%. La courbe en tirets représente la limite en magnitude, les étoiles plus brillantes que $V \sim 6.5$ ne seront pas observables.}
  \label{image__precision_gaia}
\end{figure}

Dans le cadre des Céphéides, il produira les courbes de lumière ($\sim 25$ bandes comprises entre $0.33\,\mu\mathrm{m} < \lambda < 1\,\mu\mathrm{m}$) et de vitesse radiale avec la précision nécessaire pour permettre une bonne description de paramètres physiques tels que la luminosité et la température. Je présente sur la Fig.~\ref{image__precision_gaia} un diagramme $(V-\sigma_\pi)/(\pi-d)$. J'ai repris le diagramme de \citet{Lindegren-2005-10} pour y placer l'échantillon de 455 Céphéides étudié par \citet{Berdnikov-2000-04}. On constate que la distance d'environ 235 Céphéides sera connue avec une précision $< 1$\,\%. Cela devrait permettre d'obtenir une estimation du point zéro de la relation P--L avec une précision inédite.

Ces mesures de distance combinées à des mesures interférométriques de variation de diamètre angulaire pourraient fournir, grâce à la méthode de Baade-Wesselink, une estimation de facteur $p$ pour un grand nombre de Céphéides. Malheureusement \emph{GAIA} ne devrait pas être capable d'observer des étoiles plus brillantes que $V \sim 6.5$, soit une magnitude $K \sim 4.5$, et pour le moment, la magnitude limite des mesures interférométriques est de l'ordre de $K \sim 5$. La combinaison des mesures spatiales et terrestres ne sera possible que pour une dizaine  de Céphéides seulement.

\begin{table}[!p]
	\centering
	\begin{tabular}{c|l} 
	\hline
	\hline
	Astrométrie		&	$-\; 7\,\mu$as pour $V < 10$ 									\\
	($V < 20$)		&	$-\; 12-25\,\mu$as pour $V \sim 15$ 						\\
							&	$-\; 100-300\,\mu$as pour $V \sim 20$					\\
	\hline
	Photométrie		&	$-\; 8-20$\,mmag pour $V \sim 15$								 \\
	($V < 20$)		&																				 	\\
	\hline
	Spectroscopie	&	$-\; < 1\,\mathrm{km\,s^{-1}}$ pour $V \sim 13$ 		\\
	($V < 16-17$)	&	$-\; < 15\,\mathrm{km\,s^{-1}}$ pour $V \sim 16$ 	\\
 	\hline
	\end{tabular}
  	\caption[Précision attendue de \emph{GAIA}]{\textbf{Précision attendue de \emph{GAIA}}}
  	\label{table__precision_gaia}
\end{table}

\section{Conclusion générale}

L'avènement de nouveaux instruments de haute résolution spatiale et spectrale apportera une caractérisation plus complète des enveloppes : leur taille, leur géométrie, leur température ou leur composition chimique. Mon travail de thèse a montré que la présence d'enveloppes est peut être une phénomène global, car toutes les Céphéides observées jusqu'à présent semblent en posséder une, favorisant ainsi l'hypothèse de la perte de masse. Leur détection reste cependant difficile sans l'usage d'instruments à haute résolution angulaire. 

Les objectifs de la thèse ont été atteint en utilisant plusieurs techniques d'observations à diverses longueurs d'onde. Une solution observationnelle a été présentée au Chapitre~\ref{chapitre__imagerie_a_haute_resolution_spatiale_optique_adaptative_et_lucky_imaging} : avec des observations en mode cube et grâce à une analyse statistique originale, il est possible de détecter et d'extraire la morphologie des enveloppes de Céphéides si elle sont résolues spatialement.

Une corrélation entre l'excès en infrarouge thermique et la période de pulsation de l'étoile à été trouvée. Cette corrélation montre que les Céphéides de longues périodes ont un excès plus important que les courtes périodes. L'hypothèse actuelle est que les longues périodes perdent plus de masse via le mécanisme de pulsation et des ondes de choc présentes dans l'atmosphère. Des modèles d'atmosphères stellaires incluant la pulsation et la perte de masse sont nécessaires pour valider cette hypothèse.

Enfin, les excès présentés au Chapitre~\ref{chapitre__etude_d_exces_infrarouge_par_photometrie} varie de 2\,\% à 30\,\% à $8.6\,\mu\mathrm{m}$, et engendrent un biais sur l'étalonnage de la relation P--L via l'estimation de la magnitude absolue. Ce biais varie de quelques pour cent à quelques dizaines de pour cent à cette longueur d'onde et n'est donc pas négligeable. De plus, la corrélation trouvée précédemment implique que ce biais n'est pas constant et dépend de la période de pulsation. Les mesures sur les Céphéides de longue période, qui sont les plus brillantes et donc observables à de plus grandes distances, présentent un biais plus important. Ceci souligne d'autant plus l'importance de l'étude des enveloppes de Céphéides.

La connaissance des distances dans l'univers est primordiale pour la connaissance des paramètres physiques des astres. Ces mesures de distances se font grâce à un échafaudage de techniques où les Céphéides y ont une place centrale. L'étude des Céphéides en général et de leur enveloppe circumstellaire en particulier sera d'autant plus importante et intéressante avec l'arrivée de nouveaux instruments de haute résolution, tels que \emph{GAIA} ou \emph{Gravity}.


\cleardoublepage  

\pagestyle{fancy}
\fancyhf{}
\lhead[\nouppercase{\emph{\thepage}}]{\nouppercase{\emph{Références}}}
\rhead[\nouppercase{\emph{Références}}]{\nouppercase{\emph{\thepage}}}
\newpage

\makeatletter
\renewcommand{\@biblabel}[1]{}			
\makeatother

\bibliographystyle{aa}
\addcontentsline{toc}{chapter}{Références}




\appendix
\cleardoublepage  

\pagestyle{fancy}
\fancyhf{}
\lhead[\nouppercase{\emph{\thepage}}]{\nouppercase{\emph{Annexes~\thesection}}}
\rhead[\nouppercase{\emph{Annexes~\thesection}}]{\nouppercase{\emph{\thepage}}}
\newpage

\renewcommand*\thesection{A.\arabic{section}}
\renewcommand{\thefigure}{\thesection}

\chapter*{\textcolor{bleu_chapitre}{\emph{Annexes}}}
\addcontentsline{toc}{chapter}{Annexes}
\addtocontents{lof}{\protect\addvspace{10pt}}%
\addtocontents{lot}{\protect\addvspace{10pt}}%

\thispagestyle{empty}

\section{Évolution des étoiles de masse intermédiaire}
\label{annexe__evolution_des_etoile_de_masse_intermediaire}

\begin{figure}[!p]
    \centering\includegraphics[width = .67\linewidth]{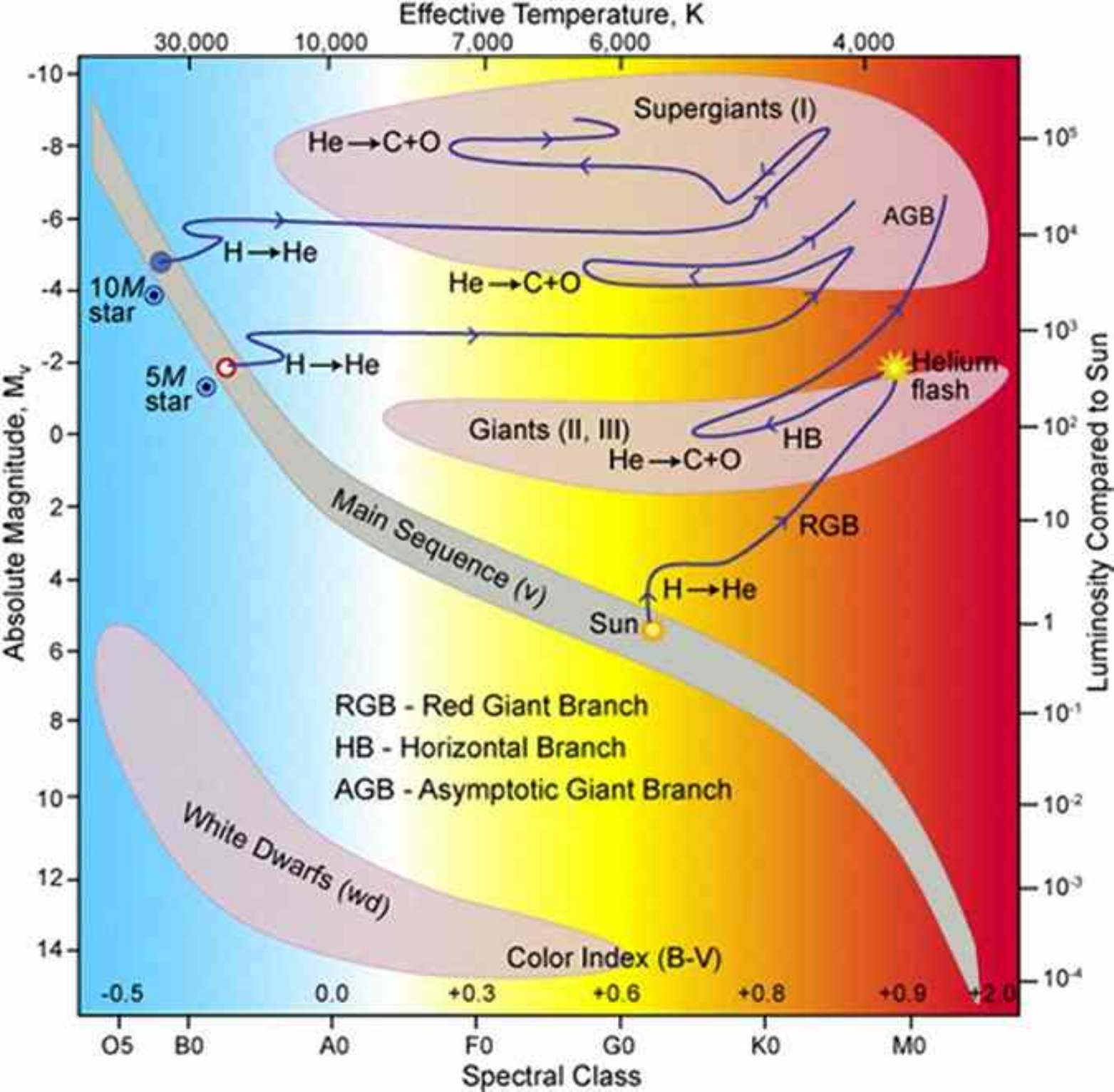}
    \caption[Évolution stellaire]{\textbf{Évolution stellaire} : diagramme H--R où sont présentées quelques phases de l'évolution stellaire.}
    \label{hr_diagram_annexe_A}
\end{figure}

Je présente ici d'une manière non exhaustive l'évolution d'une étoile de masse intermédiaire dans le diagramme H--R. Le but est d'avoir une idée des différentes phases de l'évolution stellaire, du stade séquence principale au stade final. Sur la Fig.~\ref{hr_diagram_annexe_A} sont présentés quelques trajets d'évolution d'une étoile en fonction de sa masse, incluant certaines phases par lesquelles passent l'étoile durant son évolution.

Après sa formation, une étoile se trouve sur ce que l'on appelle la séquence principale (MS), fusionnant l'hydrogène en hélium en son c\oe ur. L'étoile est alors en équilibre hydrostatique, contre-balançant la force de gravité grâce à la pression thermique interne générée par les réactions nucléaires. En fonction de sa masse, l'étoile sera à une position différente dans la MS et plus elle est massive, plus court sera le temps passé dans cette phase. L'épuisement de l'hydrogène au centre entraine la contraction du c\oe ur, l'étoile quitte la séquence principale et commence son ascension dans la branche des géantes rouges (RGB). 

L'hydrogène fusionne maintenant en couche dans l'atmosphère de l'étoile, le rayon externe augmente à cause de la pression radiative tandis que le noyau est entrain de s'effondrer. Quand la température centrale atteint $\sim 10^8\,\mathrm{K}$ la fusion de l'hélium s'enclenche, l'étoile atteint une nouvelle phase d'équilibre et entre dans la branche horizontale (HB). L'hélium fini par s'épuiser au c\oe ur et commence à fusionner en couche avec l'hydrogène. L'étoile grossit à nouveau et recommence son ascension vers la phase des géantes en suivant une trajectoire presque similaire à la RGB. L'étoile se trouve désormais dans la branche asymptotique des géantes (AGB). L'hélium et l'hydrogène qui brulent en couche donnant lieu à des instabilités dans l'atmosphère et entrainent des éjections de masse. En même temps, le c\oe ur se contracte et l'hélium fusionne en carbone et oxygène. Une étoile de masse intermédiaire n'atteint pas la température nécessaire pour continuer la fusion en d'autres éléments plus lourd et le noyau forme un gaz dégénéré stoppant la contraction. Elle finie par expulser ses couches extérieures formant ainsi une nébuleuse planétaire dont l'objet central est une naine blanche.

\section{Potentiel de gravitation dans un système binaire}
\label{section__potentiel_de_gravitation_dans_un_systeme_binaire}

\begin{figure}[!p]
\resizebox{\hsize}{!}{
    	\centering\includegraphics[width = \linewidth]{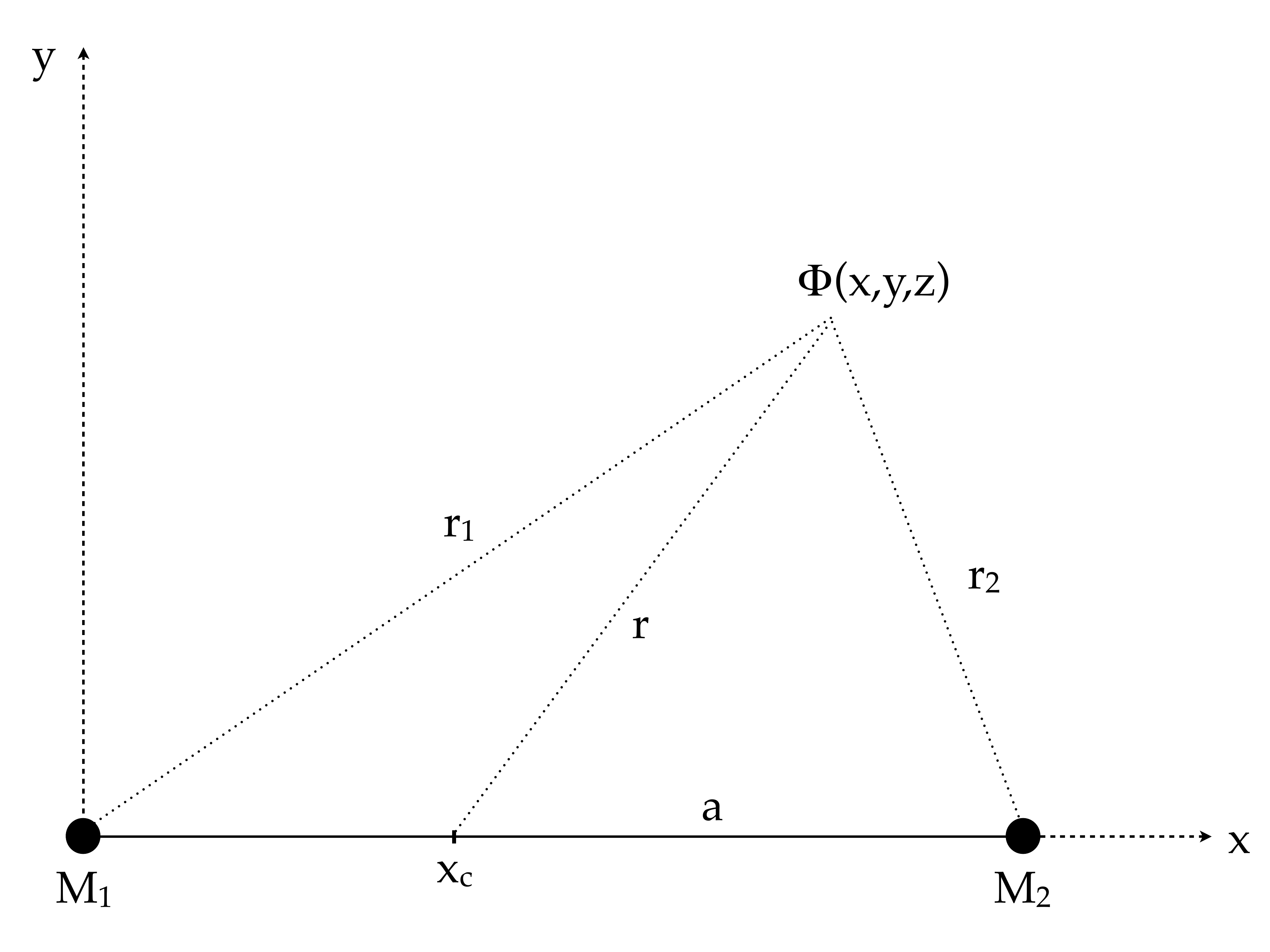}
      	\centering\includegraphics[width = \linewidth]{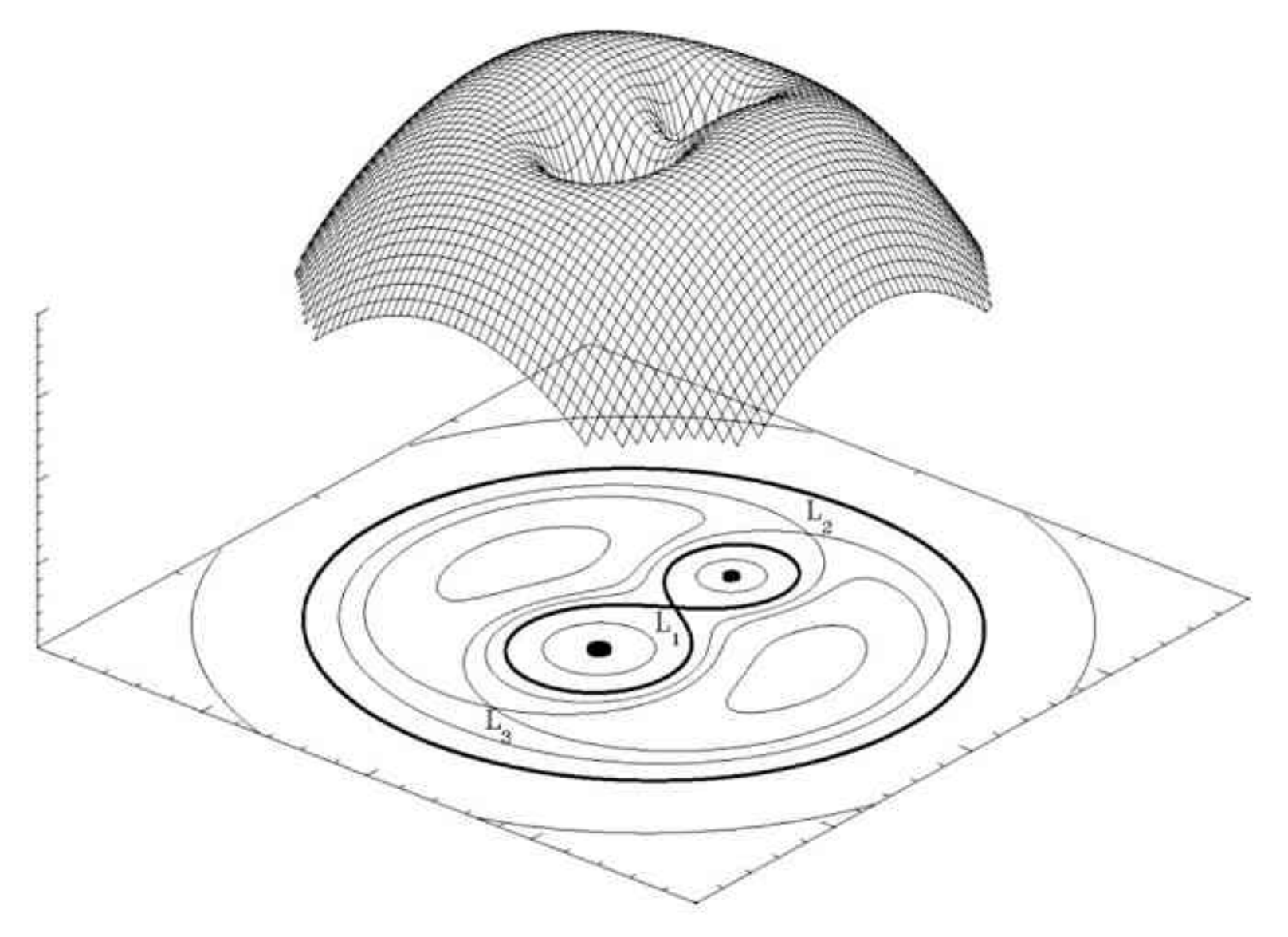}}
    \caption[Potentiel de roche]{\textbf{Potentiel de roche} : à gauche, géométrie utilisée pour le système binaire. À droite, potentiel de gravitation dans un système binaire pour un rapport de masse de 0.5.}
    \label{image__potentiel_de_roche_annexe_B}
\end{figure}

Supposons deux astres ponctuels en orbite circulaire, de masse $M_1$ et $M_2$ placés suivant le schéma de la Fig.~\ref{image__potentiel_de_roche_annexe_B} . Le potentiel total est la somme des potentiels gravitationnels et du potentiel centrifuge :
\begin{displaymath}
\phi = -\frac{GM_1}{\Vert \vec{r}_1 \Vert} - \frac{GM_2}{\Vert \vec{r}_2 \Vert} - \frac{1}{2}\Vert \vec{\omega} \times \vec{r} \Vert^2
\end{displaymath}
où $\vert \vec{\omega} \vert^2 = G(M_1 + M_2)/a^3$, $r_1^2 = x^2 + y^2 + z^2$, $r_2^2 = (x - a)^2 + y^2 + z^2$ et $r^2 = (x - x_c)^2 + y^2 + z^2$.

Les courbes équipotentielles sont définies par $\phi = cste$ et décrivent les surfaces de Roche, visibles sur la Fig.~\ref{image__potentiel_de_roche_annexe_B} (à droite). Il existe des points particuliers du plan orbital, dits points de Lagrange ($L_1, L_2, L_3, ...$) où l'orbite d'un troisième corps est stable (ceux sont les 5 solutions de l'équation $\nabla \phi = 0$). Les lobes de Roche sont les surfaces équipotentielles liées gravitationnellement à un seul astre. Si un astre remplit son lobe de Roche, représenté par une trajectoire en "8" sur la Fig.~\ref{image__potentiel_de_roche_annexe_B} (à droite), il perdra de la masse via le point $L_1$ sous l'effet de l'attraction gravitationnelle du second astre.

\section{Polynômes de Zernike}
\label{section__polynome_zenike}

Ces polynômes permettent de décrire les aberrations des systèmes optiques. L'expression la plus employée est celle donnée par \citet{Noll-1976-03}. Les polynômes sont définis en coordonnées polaires ($r, \phi$) en fonction du degré radial $n$ et azimutal $m$ par :

\begin{equation*}
Z_n^m (r,\phi) = 
\begin{cases}
   	  \sqrt{2(n + 1)} R_n^m(r) \cos (m\phi) 	& \text{pour $j$ pair et $m \neq 0$} \\
      \sqrt{2(n + 1)} R_n^m(r) \sin (m\phi)  & \text{pour $j$ impair et $m \neq 0$} \\
      \sqrt{2(n + 1)} R_n^0(r)  & \text{pour $m = 0$}
\end{cases}
\end{equation*}

\noindent avec $j = n (n + 1)/2 + m + 1$ et

\begin{displaymath}
R_n^m(r) = \sum_\mathrm{k=0}^\mathrm{(n-m)/2} \frac{(-1)^\mathrm{k}(n - k)!}{k!\,[(n + m)/2 - k]![(n - m)/2 - k)]!}r^\mathrm{n - 2k}
\end{displaymath}

\begin{figure}[!p]
\resizebox{\hsize}{!}{\centering\includegraphics[width = \linewidth]{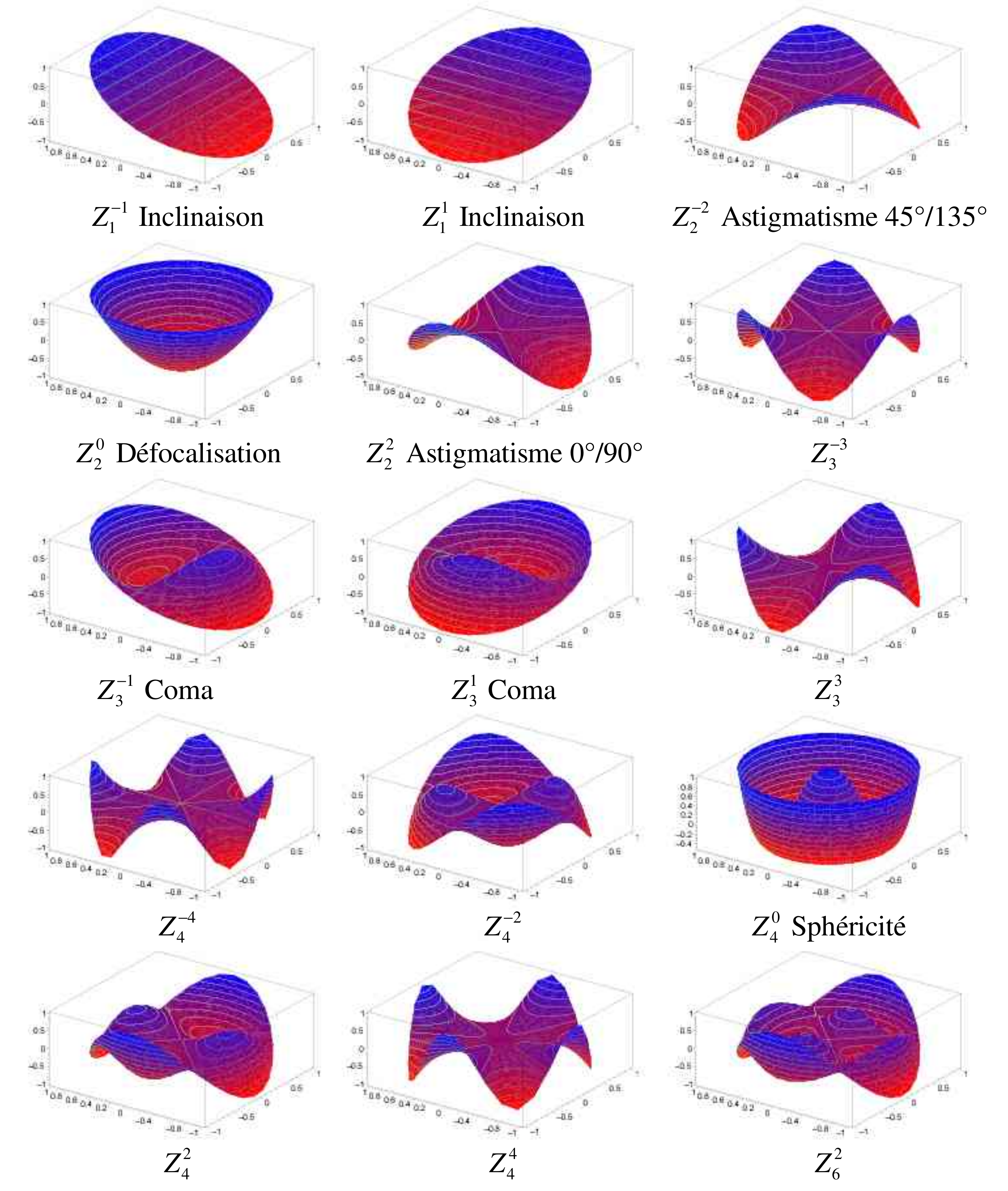}}
    \caption[Polynômes de Zernike]{\textbf{Polynômes de Zernike}}
    \label{image__polynome_zenike}
\end{figure}

La Fig.~\ref{image__polynome_zenike} illustre quelques polynômes, dont les abérrations optiques les plus communes.

\cleardoublepage
\section{Article : Spatially extended emission around the Cepheid RS~Puppis in near-infrared hydrogen lines - Adaptive optics imaging with VLT/NACO}
\label{article__naco}

Article publié dans la revue \emph{Astronomy \& Astrophysics}, volume 527, numéro A51.

\includepdf[pages=-]{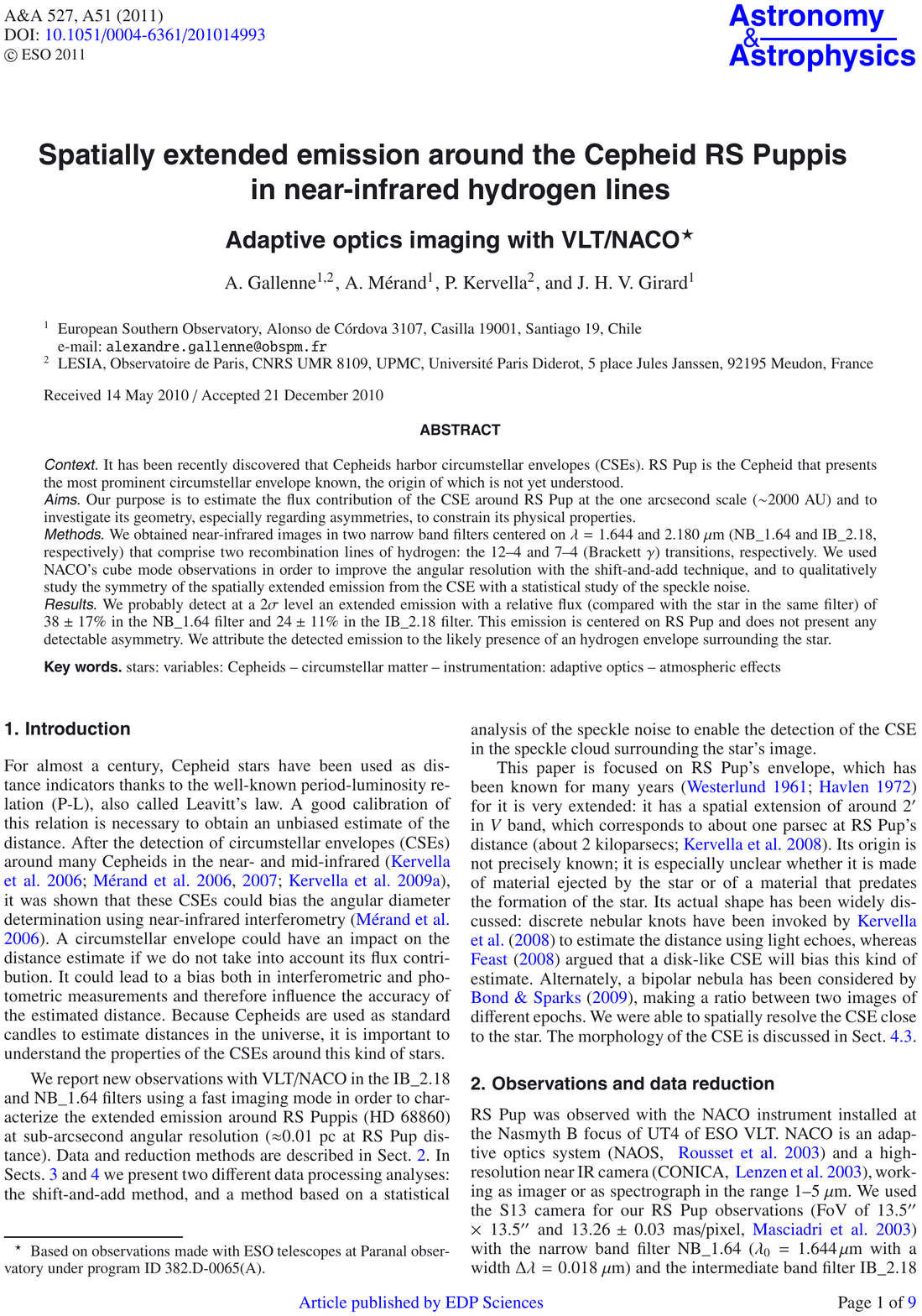}

%
%

\cleardoublepage
\section{Article : The circumstellar envelopes of the Cepheids $\ell$~Carinae and RS~Puppis. Comparative study in the infrared with Spitzer, VLT/VISIR, and VLTI/MIDI}
\label{article__rs_pup_pierre}

Article publié dans la revue \emph{Astronomy \& Astrophysics}, volume 498, page 425.

\includepdf[pages=-]{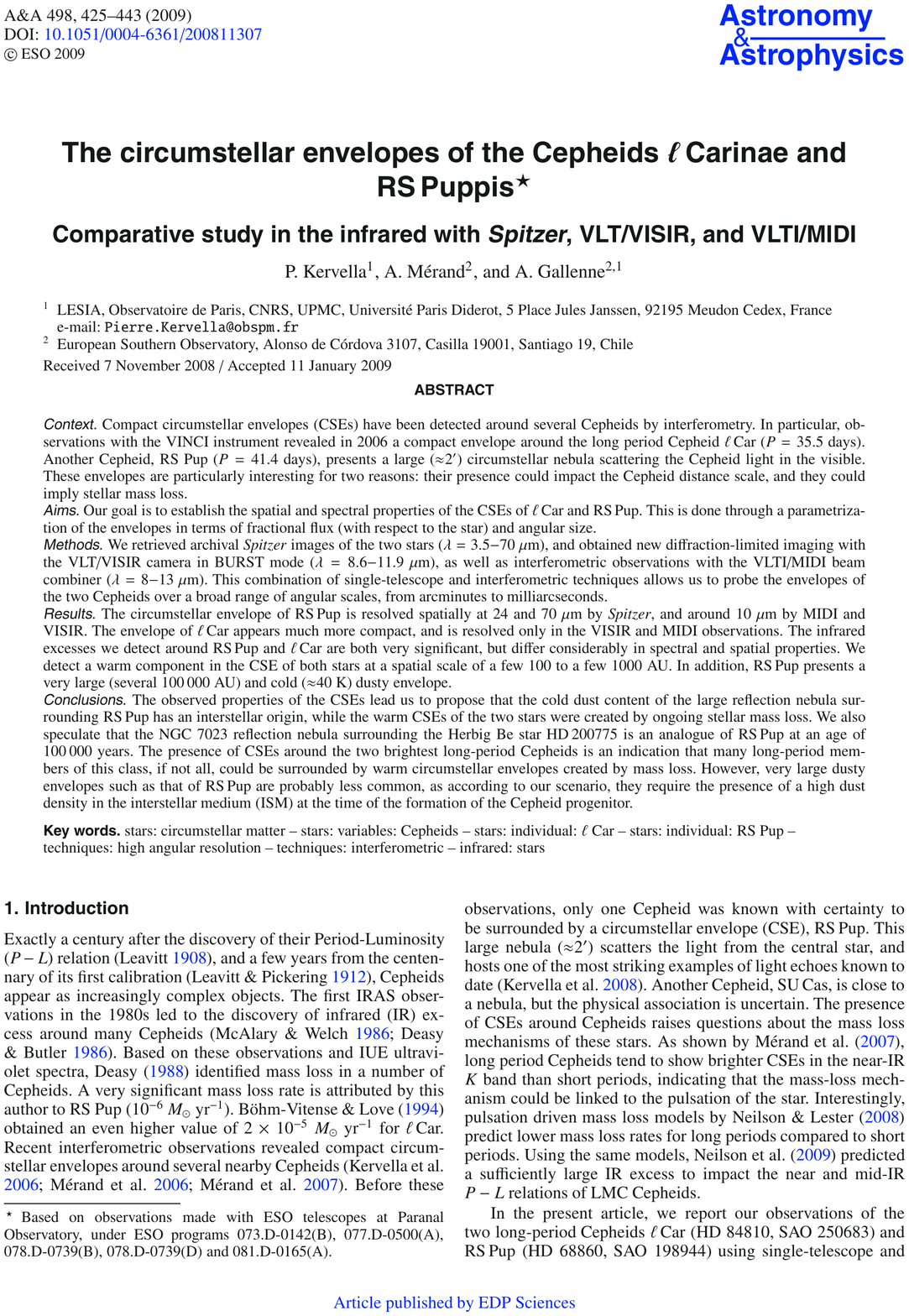}


\cleardoublepage  
\thispagestyle{empty}		

\section*{\textcolor{bleu_chapitre}{\Huge\textbf{\emph{Abstract}}}}
\vspace*{.75cm}

\thispagestyle{empty}

Since more than a century, the period--luminosity relation of Cepheid stars is a fundamental step of the cosmological distance ladder. However, the distance estimates from this law are only accurate to $\sim 5$\,\% and this uncertainty is mainly due to its calibration. The enhancement of this calibration requires an accurate and independent determination of the distances of a sample of nearby Cepheids.

Until recently, Cepheids were considered as devoid of circumstellar material. In 2005, interferometric observations with \emph{VLTI/VINCI} and \emph{CHARA/FLUOR} revealed the existence of a circumstellar envelope (CSE) around some Cepheids. This surrounding material is particularly interesting for two reasons: it could have an impact on the distance estimates and could be linked to a past or on-going mass loss. The use of Baade-Wesselink methods for independent distance determinations could be significantly biased by the presence of these envelopes.

Although their observations are difficult because of the high contrast between the photosphere of the star and the CSE, several observation techniques have the potential to improve our knowledge about their physical properties. In this thesis, I discuss in particular high angular resolution techniques that I applied to the study of several bright Galactic Cepheids.

First, I used adaptive optic observations with \emph{NACO} of the Cepheid RS~Puppis, combined with an observing mode called "cube", in order to deduce the flux ratio between the CSE and the photosphere of the star in two narrow bands of central wavelength $\lambda = 2.18\,\mu\mathrm{m}$ et $\lambda = 1.64\,\mu\mathrm{m}$. In addition, thanks to the cube mode, I could carry out a statistical study of the speckle noise and inspect a possible asymmetry.

Secondly, I analysed \emph{VISIR} data to study the spectral energy distribution of a sample of Cepheids. These diffraction--limited images enabled me to carry out an accurate photometry in the $N$ band and to detect an IR excess linked to the presence of a circumstellar component. On the other hand, applying a Fourier analysis I showed that some components are resolved.

I then explored the $K^\prime$ band with the recombination instrument \emph{FLUOR} for some bright Cepheids. Thanks to new set of data of Y~Oph, I improved the study of its circumstellar envelope. Using a ring-like model for the CSE, I assessed an angular size of $4.54\,\pm\,1.13$\,mas and an optical depth of $0.011\,\pm\,0.006$. For two other Cepheids, U~Vul and S~Sge, I applied the interferometric Baade--Wesselink method in order to estimate their distance. I found $d = 647\,\pm\,45$\,pc and $d = 661\,\pm\,57$\,pc , respectively for U~Vul and S~Sge, as well as a linear radii $R = 53.4\,\pm\,3.7\,R_\odot$ and $R = 57.5\,\pm\,4.9\,R_\odot$ respectively.

\vspace*{.5cm}
\noindent \emph{\textbf{Keywords :} adaptive optics, lucky--imaging, interferometry, Cepheids, circumstellar envelopes, distances, Baade--Wesselink, pulsation}

\newpage

\section*{\textcolor{bleu_chapitre}{\Huge\textbf{\emph{Résumé}}}}
\vspace*{.75cm}

\thispagestyle{empty}

Depuis plus d'un siècle, la relation période--luminosité (P--L) des étoiles Céphéides est un échelon fondamentale de l'échelle des distances cosmologiques. Cependant, l'estimation des distances à partir de cette loi n'est précise qu'à $\sim 5$\,\% et cette incertitude est principalement due à son étalonnage. L'amélioration de cet étalonnage nécessite une détermination précise (de manière indépendante de la relation P--L) de la distance des Céphéides proches.

Jusqu'à récemment, les Céphéides étaient considérées comme dépourvues de matériel circumstellaire. En 2005, des observations interférométriques \emph{VLTI/VINCI} et \emph{CHARA/FLUOR} ont révélé l'existence d'enveloppe circumstellaire autour de certaines Céphéides. Ce matériel environnant est particulièrement intéressant pour deux raisons : il pourrait avoir un impact sur l'estimation des distances et pourrait être lié à une perte de masse passée ou en cours. L'utilisation de la méthode de Baade--Wesselink classique pour la détermination indépendante des distance pourrait être significativement biaisée par la présence de ces enveloppes.

Bien que leurs observations soient difficiles à cause du fort contraste entre la photosphère de l'étoile et l'enveloppe circumstellaire, plusieurs techniques d'observations ont le potentiel d'améliorer notre connaissance sur leurs propriétés physiques. Dans ce manuscrit, je discute en particulier des techniques de haute résolution angulaire que j'ai appliqué pour l'étude de plusieurs Céphéides Galactiques.

Dans un premier temps j'ai utilisé des observations de la Céphéide RS~Puppis en imagerie par optique adaptative avec \emph{NACO}, couplée à un mode d'observation dit "cube", pour déduire le rapport de flux entre l'enveloppe et la photosphère de l'étoile dans deux bandes étroites centrées sur $\lambda = 2.18\,\mu\mathrm{m}$ et $\lambda = 1.64\,\mu\mathrm{m}$. De plus grâce au mode cube, j'ai également pu effectuer une étude statistique du bruit de speckle me permettant d'étudier une éventuelle asymétrie.

Dans un second temps, j'ai analysé des données \emph{VISIR} pour étudier la distribution d'énergie spectrale d'un échantillon de Céphéides. Ces images, qui sont limitées par la diffraction, m'ont permis d'effectuer une photométrie précise dans la bande $N$ et de mettre en évidence un excès infrarouge lié à la présence d'une composante circumstellaire. D'autre part en appliquant une analyse de Fourier j'ai montré que certaines de ces composantes sont résolues.

J'ai ensuite exploré la bande $K^\prime$ avec l'instrument de recombinaison \emph{FLUOR} pour certaines Céphéides brillantes. Grâce à de nouvelles données sur l'étoile Y~Oph, j'ai approfondi l'étude de son enveloppe circumstellaire. En utilisant un modèle d'étoile entourée d'une couronne sphérique, j'ai déterminé une taille angulaire de $4.54\,\pm\,1.13$\,mas et une profondeur optique de $0.011\,\pm\,0.006$. Pour deux autres Céphéides, U~Vul et S~Sge, j'ai appliqué la méthode de Baade--Wesselink afin d'estimer une première mesure directe de leur distance. J'ai trouvé une distance de $d = 647\,\pm\,45$\,pc et $d = 661\,\pm\,57$\,pc , respectivement pour U~Vul et S~Sge, ainsi qu'un rayon linéaire moyen $R = 53.4\,\pm\,3.7\,R_\odot$ et $R = 57.5\,\pm\,4.9\,R_\odot$ respectivement.

\vspace*{.5cm}
\noindent \emph{\textbf{Mots clés :} optique adaptative, imagerie sélective, interférométrie, Céphéides, enveloppes circumstellaires, distances, Baade--Wesselink, pulsation}

\end{document}